\documentclass{aa}  

\usepackage{txfonts}
\usepackage{natbib}
\usepackage{graphicx}
\usepackage{xcolor}
\usepackage{hyperref}
\usepackage{array}
\usepackage{subfigure}
\usepackage{enumitem}
\usepackage{comment}
\usepackage{appendix}
\usepackage{multicol}
\usepackage{multirow}
\usepackage{longtable}
\usepackage{threeparttable}
\usepackage{booktabs}
\usepackage{supertabular}
\usepackage{setspace}
\usepackage{geometry}
\usepackage{hyperref}
\usepackage{float}
\usepackage{afterpage}
\usepackage{hypcap}

\usepackage{amsmath}
\usepackage{txfonts}
\newcommand{\eri}[1]{\textcolor{cyan}{#1}} %Changes from Eduardo Ros
\defcitealias{BlandfordKonigl}{BK79}
\defcitealias{Plavin2019}{PL19}

\begin{document} 

\title{Time variability of the core-shift effect in the blazar 3C\,454.3}
\titlerunning{Time variability of the core-shift effect in the blazar 3C\,454.3} 

\author{Wara Chamani\inst{1,2}\fnmsep\thanks{E-mail address: wara.chamani@aalto.fi}
          \and
          Tuomas Savolainen\inst{1,2,3}\fnmsep\thanks{E-mail address: tuomas.k.savolainen@aalto.fi}
          \and
          Eduardo Ros\inst{3}
          \and
          Yuri Y. Kovalev\inst{4,5,3}
          \and
          Kaj Wiik\inst{6}
          \and
          Anne L\"ahteenm\"aki\inst{1,2}
          \and
          Merja Tornikoski\inst{1}
          \and
          Joni Tammi\inst{1}
  }

 \institute{Aalto University Mets\"ahovi Radio Observatory, Mets\"ahovintie 114, FI-02540 Kylm\"al\"a, Finland 
              %\email{wara.chamani@aalto.fi}
            \and
            Aalto University Department of Electronics and Nanoengineering, PO Box 15500, FI-00076 Aalto, Finland               
            \and
            Max-Planck-Institut f\"ur Radioastronomie,
            Auf dem H\"ugel 69, D-53121 Bonn, Germany
            \and
            Lebedev Physical Institute of the Russian Academy of Sciences,
            Leninsky prospekt 53, 119991 Moscow, Russia
            \and
            Moscow Institute of Physics and Technology, Institutsky per.~9, Dolgoprudny, Moscow region, 141700, Russia
            \and
            Department of Physics and Astronomy, FI-20014 University of Turku, Finland}

\date{\today}
\date{Received XXX ; accepted YYY}

% \abstract{}{}{}{}{} 
% 5 {} token are mandatory
 
\abstract
% {context.}
% {Aims.}
% {methods.}
% {REsults}
% {Conclusions.}
{Measuring and inferring key physical parameters of the jets in Active Galactic Nuclei (AGN) requires high resolution very long baseline interferometry (VLBI) observations. Using VLBI to measure a so-called core shift effect is a common way of obtaining estimates of the jet magnetic field strength -- certainly a key parameter for understanding jet physics. The VLBI core is typically identified as the bright feature at the upstream end of the jet, and the position of this feature changes with the observed frequency, $r_\mathrm{core} \propto \nu^{-1/k_r}$. Due to the variable nature of AGN, flares can cause variability of the measured core-shift. In this work, we investigated the time variability of the core-shift effect in the luminous blazar \object{3C\,454.3}. We have employed self-referencing analysis of multi-frequency (5, 8, 15, 22$-$24, and 43\,GHz) Very Long Baseline Array (VLBA) data covering nineteen epochs from 2005 until 2010. We found significant core shift variability ranging from 0.27 to 0.86\,milliarcseconds between 5\,GHz and 43\,GHz. These results confirm the core-shift variability phenomenon observed before. Furthermore, we also found time variability of the core-shift index, $k_r$, which was typically below one, with an average value of $0.85 \pm 0.08$ and a standard deviation of $0.30$. Values of $k_r$ below one were found both during flaring and quiescent states. Our results indicate that commonly assumed conical jet shape and equipartition conditions do not always hold simultaneously. Still, these conditions are typically assumed when deriving magnetic field strengths from core shift measurements, which can lead to unreliable results if $k_r$ significantly deviates from unity. Therefore, one should verify that $k_r = 1$ actually holds before using core shift measurements and the equipartition assumption to derive physical conditions in the jets. When $k_r = 1$ epochs are selected in the case of \object{3C\,454.3}, the magnetic field estimates are indeed quite consistent, even though the core shift varies significantly with time. Subsequently, we estimated the magnetic flux in the jet of \object{3C\,454.3} and found that the source is in the magnetically arrested disk state, which agrees with earlier studies. Finally, we found a good correlation of the core position with the core flux density, $r_\mathrm{core}\propto S_\mathrm{core}^{0.7}$, which is consistent with increased particle density during the flares. 
%We argue that magnetic field strengths and particle densities cannot be reliably measured when $k_r$ significantly deviates from one. 
%Hence, the expression for the magnetic field at one parsec, $B_\mathrm{1pc}$ in equipartition with conical jet holds strictly as long as $k_r$ is unity. 

% \ericom{The abstract should provide the likely reason of the time variability of the core-shift, this is what people will look when reading it.} 
% \ykcom{I second it.}
% \ericom{The journal encourages (see \href{https://www.aanda.org/for-authors/author-information/paper-organization}{here}) the use of structured abstracts, altough it is not mandatory, but the logical structure is desired.} 
% \textit{Wara: I hope it is fixed now.}
% \ykcom{The abstract is still not structured. Again, no strict requirement to do that but useful.}

}

  % conclusions heading (optional), leave it empty if necessary 
   

   \maketitle
%
%------------------------------------------------------------------

\section{Introduction}  

The core-shift effect is an observational feature of synchrotron-emitting relativistic jets in Active Galactic Nuclei (AGN). The so-called core shift is the change of the radio core's distance from the central engine as a function of frequency. First observations of the phenomenon were reported by \cite{MarcaideShapiro1984}, and since then it has been detected often in multi-frequency very-long-baseline interferometry (VLBI) images of compact extragalactic sources \citep[e.g.,][]{Kovalevetal2008,OSullivanGabuzda2009, Sokolovskyetal2011, Hada2011, Pushkarevetal2012, Fromm2015}. 
%being the transition region between optically thick and thin emission. 
The physical nature of the VLBI core is still a matter of debate and it has been suggested that at least at mm-wavelengths it might be due to a standing shock in the jet  \citep{Marscher2010}. 
In such a case one would not expect to see a significant frequency-dependency of the core position. 
On the other hand, following the prediction of \citet[][hereafter \citetalias{BlandfordKonigl}]{BlandfordKonigl}, the core is the region where the jet becomes self-absorbed at a given frequency $\nu_\mathrm{obs}$, i.e., the optical depth to synchrotron self-absorption, $\tau$, becomes approximately one. In the \citetalias{BlandfordKonigl} model, the core's position along the jet, $r_\mathrm{core}$,
varies with frequency as $r_\mathrm{core} \propto \nu_\mathrm{obs}^{-1/k{_r}}$ because of the magnetic field strength and particle density gradients in the jet. The \citetalias{BlandfordKonigl} model assumes a freely expanding, supersonic, narrow, conical jet with a constant half-opening angle, $\phi$, and with a constant Lorentz factor $\Gamma$. %Along the jet the different radiating cores have different distances with respect to the jet's apex. 
Within the radiating cores the magnetic field strength $B$ and particle number density $N$ are assumed to be uniform and constant. Both decay with the distance $r$ along the jet as: $B=B_{1}\left( r_{1}/r \right)^m$ and $N=N_{1}\left({r_{1}}/{r}\right)^{n}$, where $B_{1}$ and $N_{1}$ are the values at the distance of $r_{1} = 1$\,pc from the apex of the jet. Assuming energy equipartition between particle and magnetic energy densities, the indices should be scaled such that $n=2m$. Choosing $m=1$ and $n=2$ results in $k_r = 1$ irrespective of the optically thin spectral index of the emission \citep{Konigl1981, Lobanov1997}. Based on these previous models, a recent study further showed that in microquasars with moderately relativistic flow speeds and significant jet opening angles the core position can also vary with the inclination angle and different magnetic field configurations \citep{Sharma2022}.

Several studies have confirmed $r_\mathrm{core} \propto \nu_\mathrm{obs}^{-1}$ \citep{Lobanov1997, Hirotani2005, OSullivanGabuzda2009, Fromm2010, Fromm2013b,Sokolovskyetal2011, Hada2011, Mohan2015, Fromm2015, Pushkarev2018} and the amount of core-shift together with the equipartition assumption has been frequently used to infer jet magnetic field strengths \citep[e.g.,][]{Pushkarevetal2012, Voitsik2018, Plavin2019, Chamani2021}. However, there is very little knowledge of the stability of the index $k_r$ over time or how this affects the inferred magnetic field strength. Knowing the jet magnetic field strength is important since it is one of the key parameters to test the results from the general relativistic magneto-hydrodynamic (GRMHD) simulations on formation and collimation of jets \citep[e.g.,][with references therein]{Tchekhovskoy2015}. For example, these works show that when enough magnetic flux is available for accretion, the magnetic flux surrounding the supermassive black hole may reach a saturation point and a magnetically arrested disk (MAD) develops \citep{Narayan2003}. MAD accretion behaves very differently from the standard weakly magnetized disk, and it is able to launch very powerful relativistic jets in simulations \citep{Tchekhovskoy2011,McKinney2012}. 
\cite{Zamaninasab2014} used core-shift measurements to infer magnetic fluxes in 76 jets in blazars and radio galaxies from the MOJAVE survey and they concluded that the high-power jets indeed appear to result from MAD accretion. Core-shift-inferred magnetic fields can therefore play a major role in our understanding of accretion and ejection in AGN. Therefore, it is essential to investigate how reliable these measurements are. 

\begin{figure*}[!t]
\centering
   \subfigure[]
    {
    \includegraphics[width=0.32\textwidth]{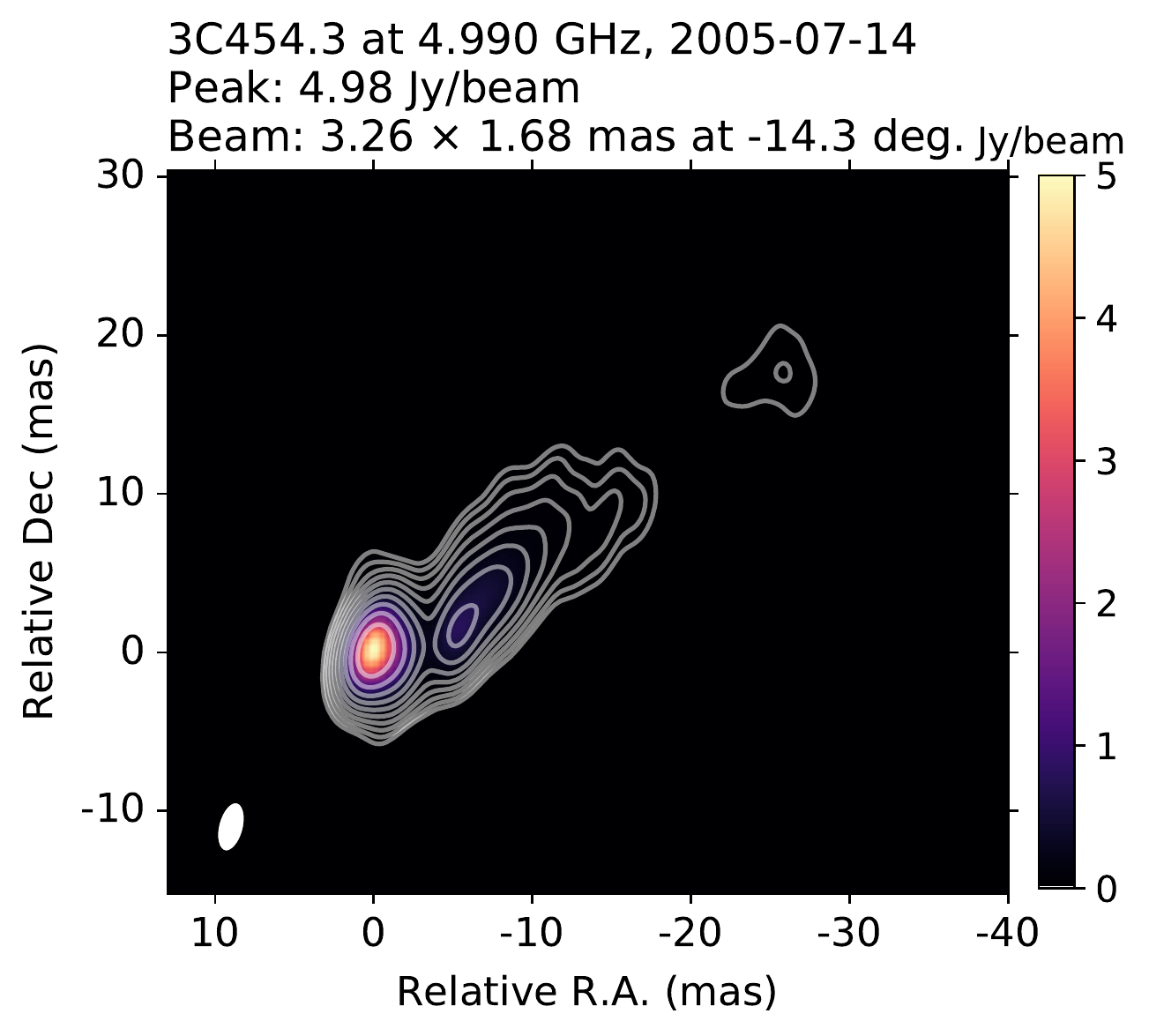}
    }
     \subfigure[]
    {
        \includegraphics[width=0.32\textwidth]{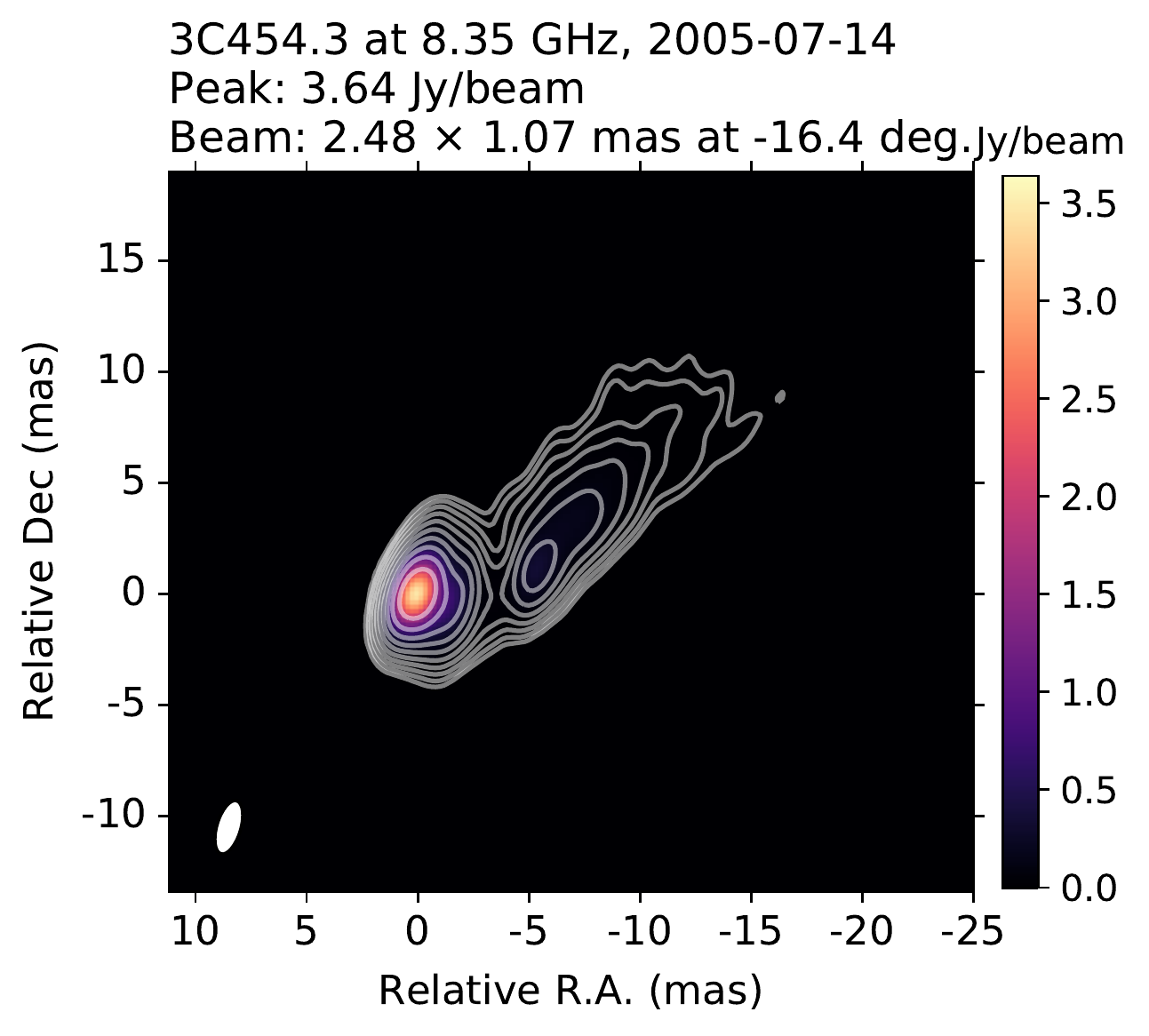}
    }
    \subfigure[]
    {
        \includegraphics[width=0.32\textwidth]{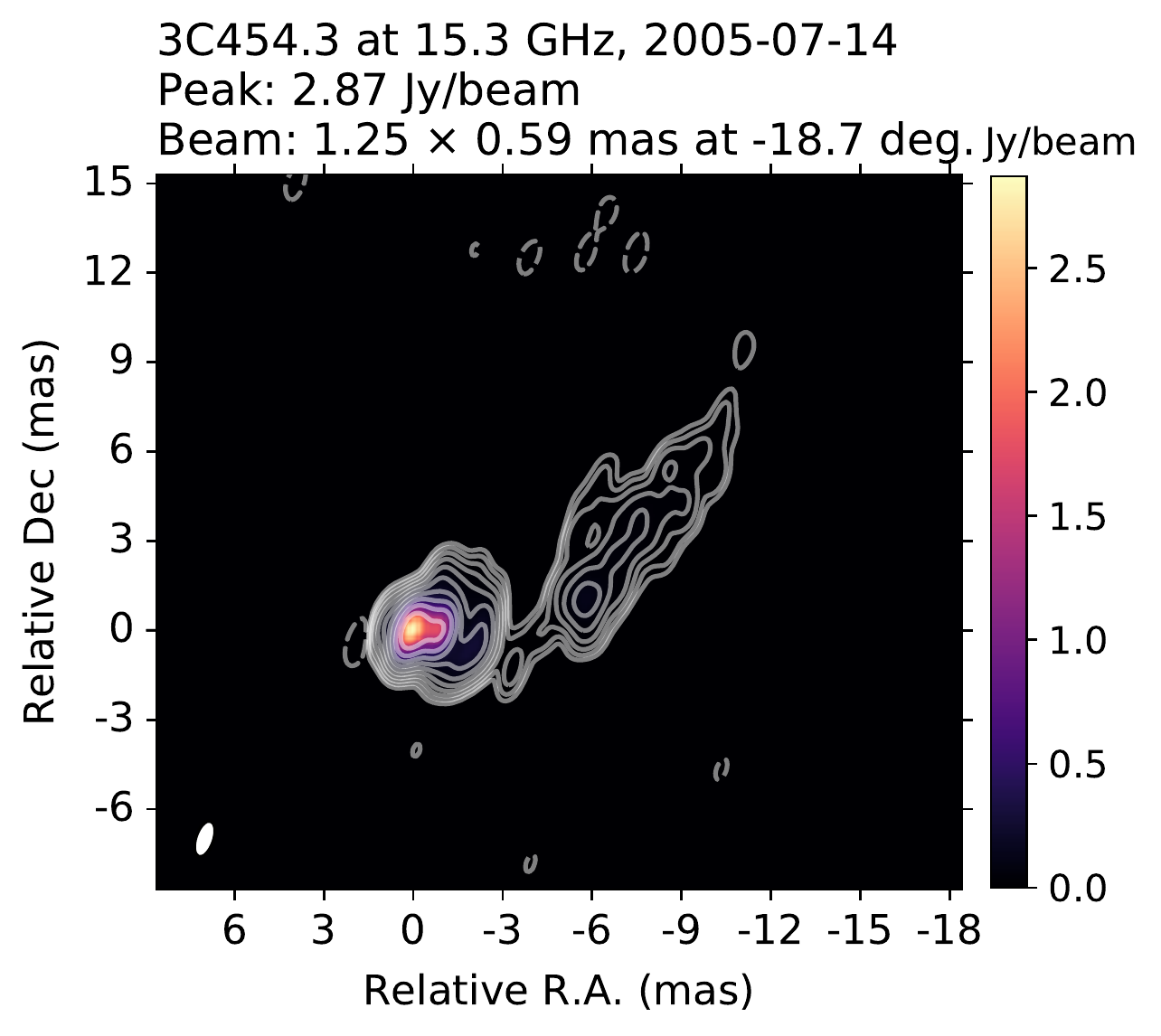}
    }
    \subfigure[]
    {
        \includegraphics[width=0.32\textwidth]{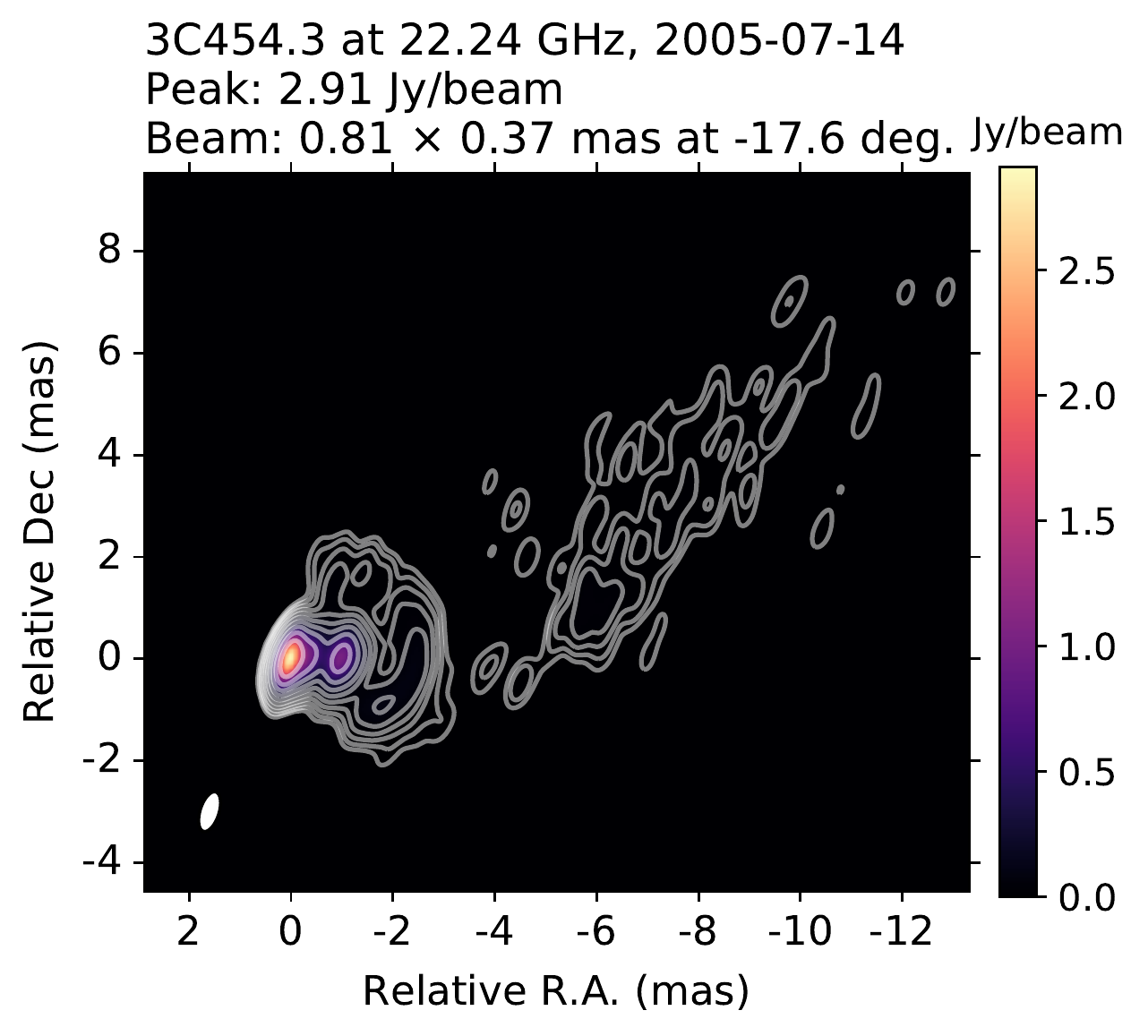}
    }
    \subfigure[]
    {
        \includegraphics[width=0.32\textwidth]{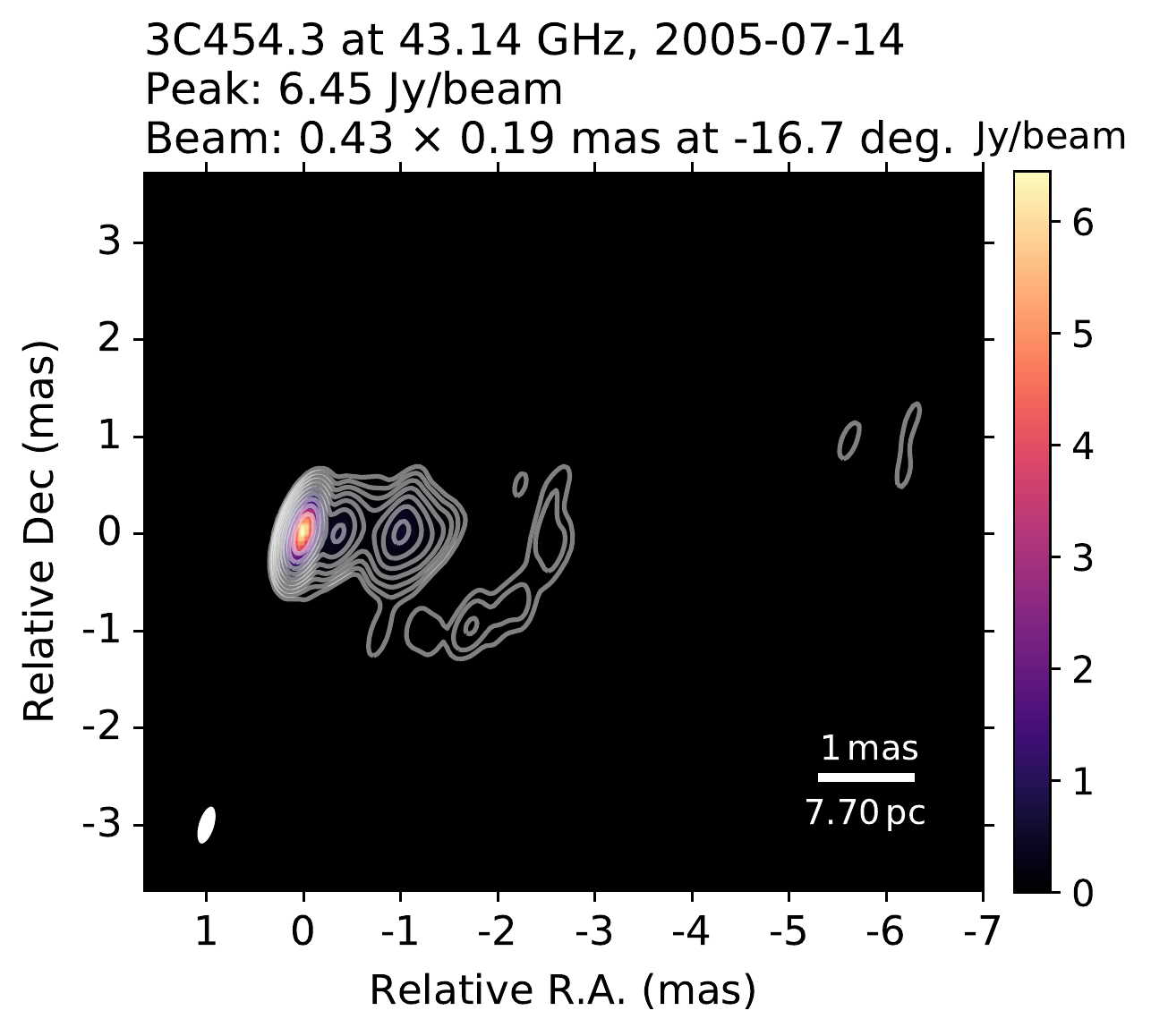}
    }
     \caption{CLEAN images of \object{3C\,454.3} on 
     2005-07-14 of the following frequency bands: a) C, b) X, c) U, d) K and e) Q  with contours at -0.1\%, 0.1\%, 0.2\%, 0.4\%, 0.8\%, 1.6\%, 3.2\%, 6.4\%, 12.8\%, 25.6\%, and 51.2\% of the peak intensity at each image. The interferometric beam (ellipse) is displayed on the bottom-left corner of each image. The rms noise levels from the lowest to the highest frequency are 0.11, 0.20, 0.35, 0.30 and 0.38\,mJy/beam.}
    %  \ericom{Please keep the same dimensions for the different figures, in the way they are presented they look strange.  Each black box should have the same dimensions in the page.  Actually, the labels in the figures can be put either in the caption or be tabulated. }}
    \label{multifreqimag}
\end{figure*}

A number of previous works have reported core-shift and magnetic field measurements in a diversity of sources \citep[e.g.,][]{OSullivanGabuzda2009, Sokolovskyetal2011, Pushkarevetal2012}, however, these studies have been limited to single epoch observations. Recently, a compelling multi-epoch investigation on the core shift on an extensive AGN data set having more than ten observing epochs has been reported by \citet[][hereafter \citetalias{Plavin2019}]{Plavin2019}. Their results exhibit a significant variability of the core-shift up to 1\,mas for the frequency pair of 2\,GHz and 8\,GHz on time scales of years. Although the latter shows for the first time substantial fluctuation of the core position, dedicated observations at three or more frequencies are necessary to understand the potential variability of the functional form of the core-shift, which could tell about deviations from the equipartition or effects of the propagating flares on the physical conditions in the jet. Ideally, such investigations should be done with multi-frequency and multi-epoch observations of a representative sample of sources. Examples of such monitoring efforts targeting individual sources include, e.g., \object{3C\,273} \citep{Savolainen2008b}, \object{3C\,345} \citep{Lobanov1999}, \object{CTA\,102} \citep{Fromm2013b}, and \object{PKS\,2233-148} \citep{Pushkarev2018}.

Due to the variable nature of AGN, strong outbursts could potentially cause time variability of the core-shift effect and possibly disturb the relation $r_\mathrm{core} \propto \nu^{-1}$ as discussed by \citep{Kovalevetal2008,Fromm2013b, Fromm2015,Plavin2019}. As a consequence, magnetic field strength estimations and astrometry could be significantly affected. Thus, blazars being conspicuous flaring sources are attractive targets to study the variability of the core shift and explore the stability of the magnetic field strength. 

% \textbf{3C454.3 emits a high relativistic jet pointing towards us in a small viewing angle. Early observations of 3C454.3 at very low frequencies ranging from 326 MHz to 5 GHz have resolved unambiguously its core-jet morphology (see \cite{Altschuleretal1995} with references therein).}

An intriguing study case is the very luminous and variable flat-spectrum radio quasar \object{3C\,454.3} (\object{B\,2251+158}, \object{4C\,+15.76}) also known as the \textit{Crazy Diamond} located at the redshift $z=0.859$ \citep{Jackson1991}. 3C454.3 emits a powerful relativistic jet pointing towards us at a small viewing angle. It has been catalogued as one of the most luminous astronomical objects and the brightest extragalactic gamma-ray source. The source exhibits a typical flat-spectrum radio-loud quasar (FSRQ) features such as non-thermal emission and variability across the whole electromagnetic spectrum. Many multi-wavelength observing campaigns of \object{3C\,454.3} ranging from radio to gamma-rays have been performed over the years. The source started an extraordinary flaring behaviour in the early 2005 with the first strong peak in 2006, the second peak in the mid-2008, and the largest peak in 2010 as seen in the Mets\"ahovi Radio Obsevatory single-dish observations at 37\,GHz (see Figure~\ref{2-plots}) and in the 15\,GHz Owens Valley Radio Telescope (OVRO) observations \citep[see e.g.,][]{Sarkar2019}. This period corresponds to the major multi-wavelength (radio, mm, optical, X-rays and gamma-rays) flaring events during 2005 and 2006 \citep[e.g.,][]{Remillard2005, Giommi2006, Villata2006, Pian2006, Jorstadetal2010}, and in 2007$-$2010 \citep[e.g.,][]{Ghisellini2007,Raiteri2008, Vercelloneetal2009, Donnarumaeral2009, Abdo2009, Vercelloneeatlal2010, Jorstadetal2010, Raiteri2011, Jorstadetal2013} with extraordinarily bright gamma-ray flares in the early December 2009 \citep{Bonnoli2011, Paccianietal2010} and in November 2010 \citep{Abdo2011}. \object{3C\,454.3} continuously flared until 2011 and then entered a quiescent state in 2012 lasting until 2014. After 2014, small flaring events have been registered although they are not strong compared to the period prior to 2012 \citep{Sarkar2019}.

% Long-term observations in the continuum in combination with VLBI observations have been extensively reported \citep[e.g.,][]{Jorstadetal2010, Jorstadetal2013}

Recent studies based on multi-frequency flux density monitoring observations and the motion of superluminal components observed in 3C\,454.3 have suggested the existence of a supermassive binary black hole system in this source \citep{Volvach2021, Qian2021}. \citet{Volvach2021} performed harmonic analysis using radio, optical, and gamma-ray flux density monitoring covering a period from 1966 until 2020 to establish the precession and orbital periods for what they argue correspond to a close black hole binary system. They report a possible period of 14\,yr based on the light curve analysis. In another study, a double precessing jet scenario has been proposed for 3C454.3, where two groups of superluminal knots are ejected at different directions that, according to the authors' model, originate from two different jets precessing with the same period of 10.5\,years \citep[][with references therein]{Qian2021}. The physical origin of the observed variability in 3C\,454.3 might be explained in the context of the precessing jet scenario as suggested by \cite{Volvach2021, Qian2021}. However, other mechanisms, such as fluctuations in the magnetic field in the inner disk, variations in the accretion flow, or development of shocks or instabilities in the jet can also be potential origins of its variability. 
%On the other hand, the existence of a black hole binary system in 3C\,454.3 is a phenomenon that requires important attention for gravitational wave observations of this variable and extraordinary blazar, which exhibits a single-jet.

In the view of its variable nature, \object{3C\,454.3} is an excellent target for studying whether also the core-shift effect is variable in this source. Previously, \cite{Pushkarevetal2012} and \cite{Kutkinetal2014} have reported core shift measurements of the jet in \object{3C\,454.3} at altogether three epochs, but only the measurement in  \cite{Kutkinetal2014} had enough frequencies to measure $k_r$. Additionally, radio light curve based core shift estimations have been reported by \cite{Mohan2015}. However, these data were not enough to study the time variability of the core-shift and $k_r$.

We present here results from three multi-wavelength Very Long Baseline Array (VLBA) monitoring programs of \object{3C\,454.3} that we have carried out between 2005 and 2010, as well as archival VLBA data. With this rich data set, we aim to test how stable the core-shift magnitude, the $k_r$ index, and the inferred magnetic field strengths are over time in \object{3C\,454.3}. We measure the frequency-dependent shifts of the core position in a frequency range spanning from 5\,GHz to 43\,GHz along with the core spectrum and core-shift vector directions. We present a direct measurement of the time variability of the $k_r$ index for the first time. %\textbf{in 3C454.3}. 
We study how the variable source flux density affects the core shift and investigate the role played by the flaring events. Using the core-shift measurements, jet's magnetic field strength at 1\,pc is estimated for $k_r=1$ and $k_r \neq 1$ cases. We also test whether the source is in a MAD state during the observed epochs. Finally, we explore the connection of core-shift magnitude and $k_r$ with core flux densities, the relation of $k_r$ with the jet position angle, and the possible correlation of magnetic field strength at 1\,pc with core-flux density when $k_r=1$.

The paper is outlined as follows: In Section 2, we describe the observations and data calibration; in Section 3, we describe the analysis method; in Section 4, we present our results. Discussion and the summary are in Sections 5 and 6, respectively. We adopt a cosmology with $\Omega_m= 0.27$, $\Omega_\Lambda = 0.73$ and $H_0 = 71\,$km\,s$^{-1}$\,Mpc$^{-1}$ \citep{Komatsu2009}. The source is at a luminosity distance of 5.49\,Gpc and at this distance the scale is $7.70\,$pc\,mas$^{-1}$. Throughout the paper the spectral index $\alpha$ is defined as $S_\nu\propto \nu^\alpha$, where $\nu$ is the observed frequency and $S_\nu$ the flux density.

%Departures from the core-shift effect could be possible due to radiative cooling (mention other processes). To study changes in the $k_{r}$ index of the core shift effect power law $r\propto \nu^{-1/k_{r}}$ for a conical jet shape (Blandford and K\"{o}nigl 1979) are necessary for testing how stable the core-shift effect is over the time. 

% Hovatta et al 2013, spectra index maps

%- To calculate the jet magnetic power?,
%\cite{Nalewajko2014} the jet magnetic power at 1pc can be calculated as $L_B=\frac{c}{4}(\Gamma \theta_j)^2 (1pc)^2 B'^2_{1pc}$  which in cgs units is equivalent to $7.14 \times 10^{46} (\Gamma \theta_j)^2B'^2_{1pc} [erg s^{-1}]$.

\begin{table}[]
\centering
\begin{threeparttable}
\caption{VLBA observations of \object{3C\,454.3} used in this work.  
}
\label{t1}
\begin{tabular}{@{}cccc@{}}
\hline
\hline \noalign{\smallskip}
Epoch & Date & \multicolumn{1}{c}{\begin{tabular}[c]{@{}c@{}}Frequency-bands\\ (GHz)\end{tabular}} & \begin{tabular}[c]{@{}c@{}}Project\\ Code\end{tabular} \\ \hline \noalign{\smallskip}
1 & 2005-05-19 & C, X, U, K, Q & BS157A \\
2 & 2005-07-14 & C, X, U, K, Q & BS157B \\
3 & 2005-09-01 & C, X, U, K, Q & BS157C  \\
4 & 2005-12-04\tnote{1} & C, X, U, K, Q & BS157D \\
5 & 2006-08-03\tnote{2} & C, X, U, K, Q & BW086A \\
6 & 2006-10-03\tnote{3} & C, X, U, K, Q & BW086B \\
7 & 2006-12-04 & C, X, U, K, Q & BW086C \\
8 & 2007-01-26 & C, X, U, K, Q & BW086D \\
9 & 2007-04-26 & C, X, U, K, Q & BW086E \\
10 & 2007-06-16 & C, X, U, K, Q & BW086F  \\
11 & 2007-07-25\tnote{4} & C, X, U, K, Q & BW086G \\
12 & 2007-09-13\tnote{5} & C, X, U, K, Q & BW086H  \\
13 & 2008-01-03\tnote{6} & C, X, U, K, Q & BW086I \\
14 & 2008-12-07 & C, X\tnote{*}, U, K\tnote{**}\,, Q  & BO033  \\
15 & 2009-09-22 & C, X, U, K, Q & S2087BA  \\
16 & 2009-10-22 & C, X, U, K, Q & S2087BB \\
17 & 2009-12-03 & C, X, U, K, Q & S2087BC \\
18 & 2010-01-18\tnote{7} & C, X, U, K, Q & S2087BD  \\
19 & 2010-02-21 & C, X, U, K, Q & S2087BE \\ \hline
\end{tabular}
\begin{tablenotes}
\item[1] Missing antenna(s): Brewster. 
% \ericom{You would save space if you just give codes for antennas missing, e.g., $\star$ for MK, $\dagger$ for BR, etc.  As the notes are now, they take a lot of space.}
\item[2] Missing antenna(s): Mauna~Kea. 
\item[3] Missing antenna(s): Brewster missing for half of the observation due a focus/rotation mount failure.
\item[4] Missing antenna(s): Kitt~Peak. 
\item[5] Missing antenna(s): Hancock and St.~Croix.
\item[6] Missing antenna(s): Hancock and St.~Croix; Also, Brewster missing for half of the observation due to snow problems.
\item[7] Missing antenna(s): Hancock. 
\item[*] X-band is split in lower (Xl: 7.9\,GHz) and higher (Xh: 8.9\,GHz).
\item[**] K-band is split in lower (Kl: 21.8\,GHz) and higher (Kh: 24\,GHz).
\end{tablenotes}
\end{threeparttable}
\end{table}

\section{Observations and Data Calibration}

\subsection{Observations}

\subsubsection{Very Long Baseline Array}

We analyzed altogether nineteen epochs of VLBA data dating from 2005 until 2010 and comprising 5 different frequency bands; C band (5\,GHz), X band (8\,GHz), K$_\mathrm{U}$ band (15\,GHz; hereafter 'U' band), K band (22-24\,GHz), and Q band (43\,GHz). These observations include both our own multi-frequency monitoring programs\footnote{The first one of these programs (BS157; PI T.~Savolainen), was triggered by the major multi-wavelength flaring event in May 2005 \citep{Villata2006}. The second program (BW086; PI K.~Wiik) was a  continuation of the monitoring started with BS157. The third program (S2087; PI Y.~Y.~Kovalev) was triggered by the \textit{Fermi}-LAT detected $\gamma$-ray flare in September 2009. The latter monitoring also covered the large $\gamma$-ray outburst of December 2009 \citep{Ackermann2010}.} as well as archival data\footnote{Public data from the project BO033 (PI S.~P.~O'Sullivan) was downloaded from the NRAO archive.}. Table~\ref{t1} lists the data with all the epochs and frequencies used in this work. All ten VLBA antennas were scheduled at all the epochs and any antennas that were taken out of the observation for any reason are given in the table.  

The multi-frequency VLBA observations were quasi-simultaneous, i.e., all the bands were observed in every run. Individual 4$-$6 minute-long scans at different bands were interleaved in order to maximize the $(u,v)$ coverage. The observing runs in the programs BS157 and BW086 lasted for 10\,hrs each and included CTA\,102 and 1749+096 as calibrators for bandpass, polarization leakage, and absolute EVPA. In BS157, the total integration time per epoch on \object{3C\,454.3} was 49\,min at C, X and U bands, 68\,min at K band and 89\,min at Q band. In BW086, more time was spent on CTA\,102 and the total integration times on \object{3C\,454.3} were 36\,min at C, X, and U bands, 40\,min at K band, and 54\,min at Q band. The recording was made with dual circular polarization, $4 \times 8$\,MHz sub-bands (IFs) per polarization, and two-bit digitization, which gave a total recording rate of 256\,Mbps. Standard frequency setups for continuum observations were used -- except for the X band, which had the sub-bands spread across the 500\,MHz filter in order to facilitate a Faraday rotation measurement. The BS157 and BW086 programs also included observations at 86\,GHz, but since the data quality at this band is highly variable due to tropospheric phase fluctuations and antenna pointing issues, and since 86\,GHz was not observed in S2087, we do not include these data in the current analysis. 

In the program S2087, 8-hr long observing sessions were used with BL~Lac scheduled as a calibrator. The total integration times on \object{3C\,454.3} ranged from 45\,min at C and X bands to 82\,min at Q band. The recording setup was again dual circular polarization, $4 \times 8$\,MHz sub-bands per polarization, and two-bit digitization. The K band centre frequency was moved to 23.8\,GHz, away from the waterline, in order to improve the continuum sensitivity. Data from all three programs were correlated at the VLBA correlator in Socorro.

Parts of the data from these three programs have been published earlier: \cite{Zamaninasab2013} presented multi-frequency VLBA polarimetry analysis of \object{3C\,454.3} based observations made on 2005-05-19 and 2009-09-22, and \citet{Fromm2013a,Fromm2013b,Fromm2015} analyzed the kinematics, spectra, and core-shift properties of the calibrator CTA\,102 based on the data taken between 2005-05-19 and 2007-04-26. 
%\textit{Wara: Tuomas, could you please specify here the exact dates?}

%\ericom{Please unify the way of referring to dates. Following the journal recommendations (see section 3.5) \href{https://www.aanda.org/doc_journal/instructions/aadoc.pdf}{here} this should be done either as 2005-May-19 or 2005-05-19. Check tables later.}

\subsubsection{Mets\"ahovi Radio Observatory total flux density monitoring}
%\textbf{More text will be added here?, waiting for Anne's reply}\\

The 37\,GHz observations were made with the 13.7\,m diameter Mets\"ahovi radio telescope. A typical integration time to obtain one flux density data point is between 1200 and 1400\,s. The detection limit of the telescope at 37\,GHz is on the order of 0.2\,Jy under optimal conditions. Data points with a signal-to-noise ratio < 4 are handled as non-detections.

The flux density scale is set by observations of \object{DR\,21}. Sources \object{NGC\,7027}, \object{3C\,274} and \object{3C\,84} are used as secondary calibrators. A detailed description of the data reduction and analysis is given in \citep{Teraesrantaetal}. The error estimate in the flux density includes the contribution from the measurement rms and the uncertainty of the absolute calibration.

In our study period from 2005 to 2010, \object{3C\,454.3} went through multiple large outbursts which are well visible in the 37\,GHz flux density curve shown in the top panel of Figure~\ref{2-plots}. These outbursts peaked in the early 2006, mid-2008, and early 2010. Furthermore, our VLBA observation epochs (marked by red arrows) coincide with the different phases of these flares. These include the strong rising flare from 2005-05-19 to the short-lived plateau until 2005-12-04, and the transition from a quiescent phase from 2006-08-03 to the formation of the moderate flare in 2008-01-03. The post-flare of 2008 coincides with our VLBA data in 2008-12-07, and the rising part of the major flare in 2009$-$2010 includes the epochs from 2009-09-22 until 2010-02-21.

\begin{figure*}[t]
\centering
    \subfigure[]
    {
        \includegraphics[width=0.4\textwidth]{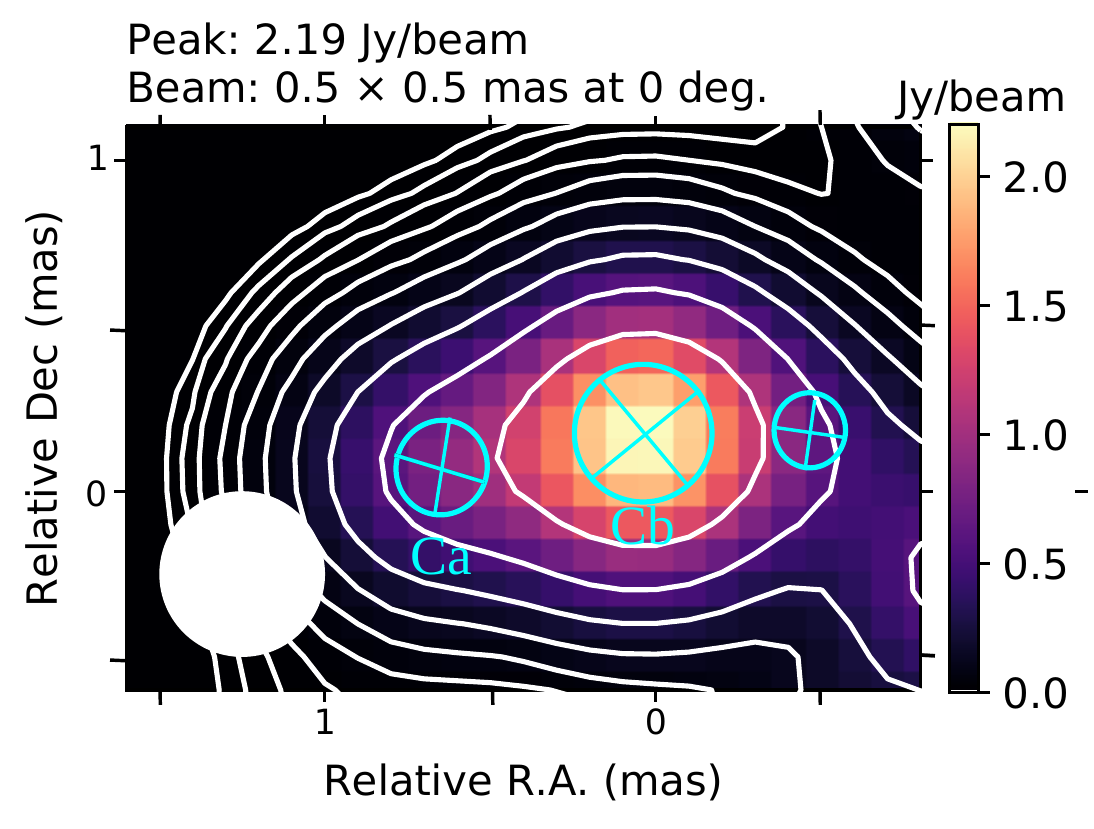}
        
    }
    \hspace{2cm}
     \subfigure[]
    {
        \includegraphics[width=0.4\textwidth]{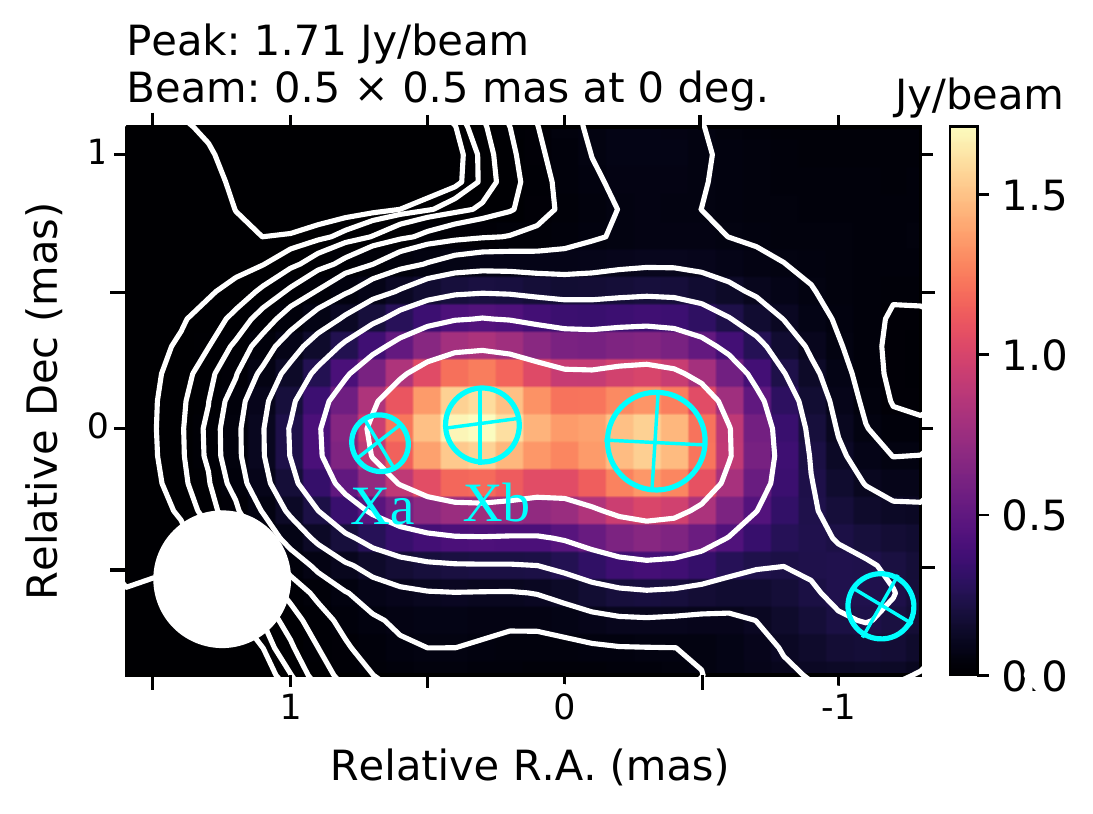}
        
    }
    \caption{Super-resolved CLEAN images of \object{3C\,454.3} on 2005-07-14 with contours at 0.1\%, 0.2\%, 0.4\%, 0.8\%, 1.6\%, 3.2\%, 6.4\%, 12.8\%, 25.6\%, and 51.2\% of the peak intensity at each image. 2D Gaussian components in the core region marked with the cyan color. a) At the C-band (5\,GHz), the features are denoted as `Ca' for the upstream-most feature and `Cb' for the downstream feature or bright component. b) Similarly, at the X-band (8.4\,GHz), the components are marked as `Xa' and `Xb'.} 
    \label{core-low-freq}
\end{figure*}

\begin{figure*}[t]
\centering
   \subfigure[]
    {
        \includegraphics[width=0.45\textwidth]{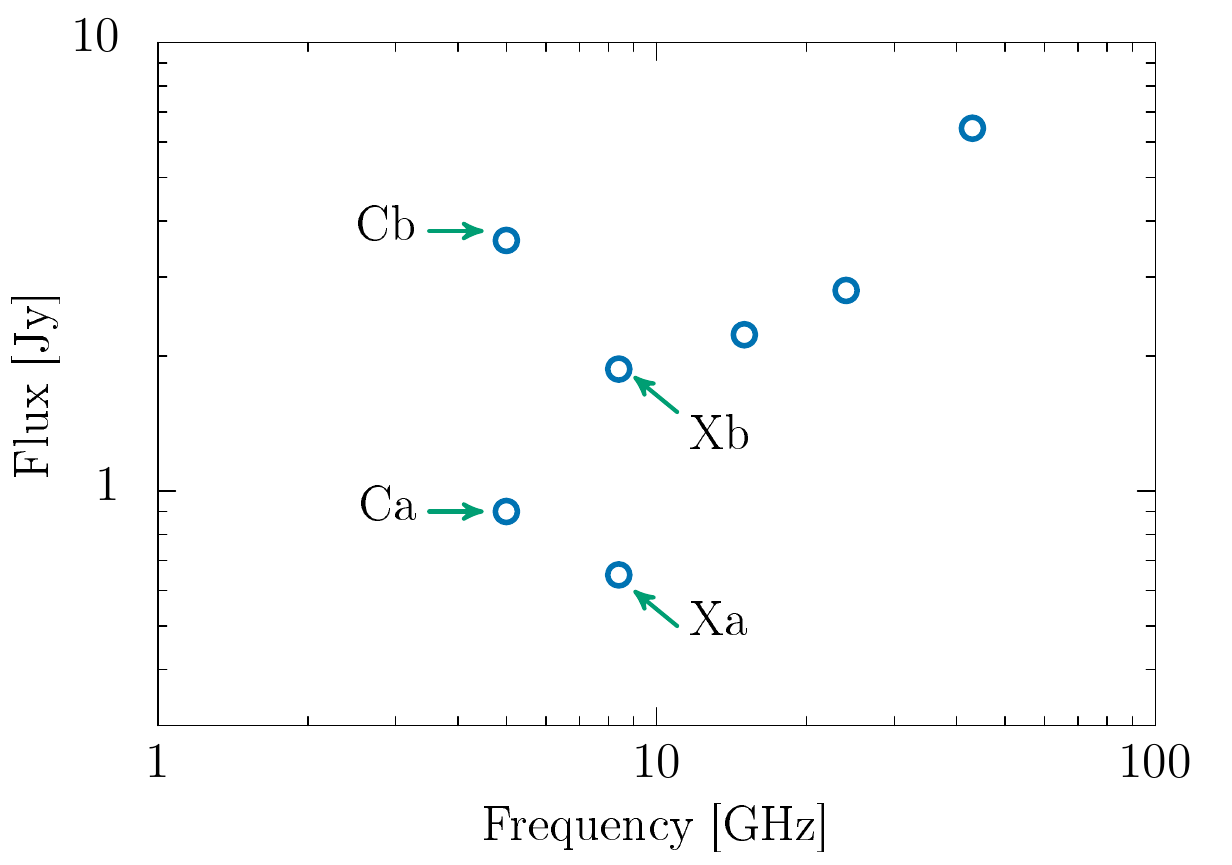}
        
    }
     \subfigure[]
    {
        \includegraphics[width=0.5\textwidth]{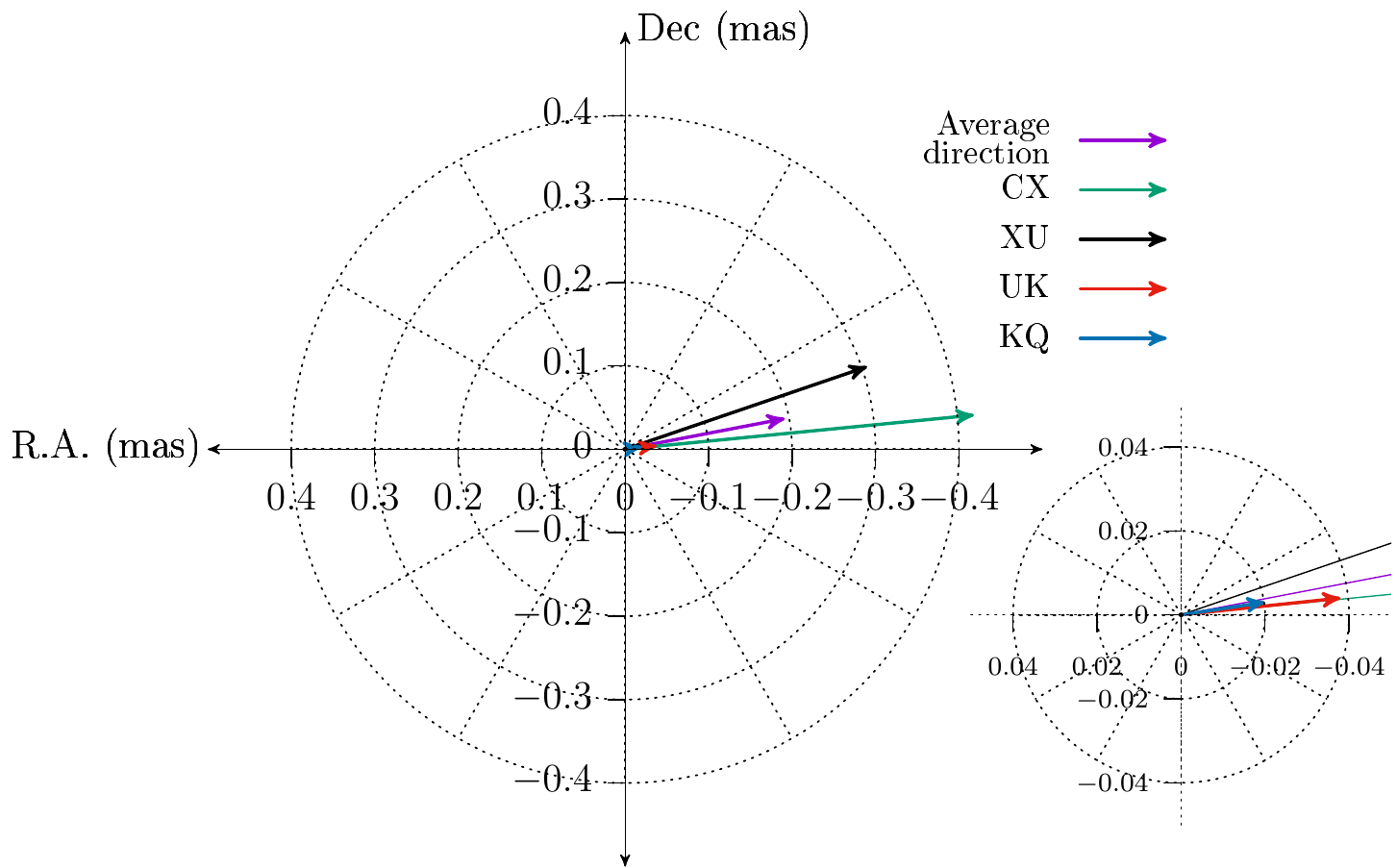}
        
    }
    \caption{(a) Radio spectrum (5\,GHz to 43\,GHz) of core and jet features in \object{3C\,345} in 2005-07-14. The flux densities of the upstream-most components are denoted by `Ca' and `Xa' and the components downstream by `Cb' and `Xb'. (b) Core-shift vectors of each frequency pair and average vector direction in a polar grid. The dotted lines are given at intervals of 30{\degr}. The inset plot at the bottom right corner shows the core-shift vector pairs for the high-frequency pairs since these are very small. All vectors  consistently point in the large-scale jet direction towards the West (see Figure~\ref{multifreqimag}). The `true' cores are features `Cb' and `Xb' at the C and X-bands, respectively, since they produce core-shift vectors whose direction is consistent with the jet direction. Comparisons of the core-shift vectors produced by the other C/X component choices are displayed in Appendix~\ref{appendix:choices}.}
    \label{spectrum-csvectors}
\end{figure*}

\subsection{Calibration and imaging of the VLBA data}

We have calibrated the VLBA data using standard methods implemented in the Astronomical Image Processing System ($\tt AIPS$) software package \citep{Greisen2003}. The procedure followed the one described in \citet{Lister2009a} except that we typically did not use pulse calibration tones to align the phases across the IFs, since the pulse calibration phases often showed unexpected jumps that may be related to the frequent band changes in our observing schedules. Instead, we performed fringe fitting of a single scan of a bright source -- either a calibrator or the target itself -- and used the results of the global solution to correct single-band delays and align the phases across the IFs. We also always carried out the full global fringe fitting of the whole observation solving for delays, rates and phases with solutions intervals roughly set according to the expected coherence times at the individual bands. Atmospheric opacity correction of the \textit{a priori} visibility amplitudes was performed for the data taken at U, K and Q bands. We note that the MOJAVE team has calibrated and imaged a significant fraction of our U band observations as a part of their effort to image archival VLBA data in order to increase the temporal sampling of the MOJAVE program sources \citep{Lister2018}. If already processed U band data was available from the MOJAVE archive\footnote{\url{https://www.cv.nrao.edu/MOJAVE/}}, it was used instead of re-doing the calibration.

Imaging (using the standard CLEAN algorithm) and self-calibration (of both amplitudes and phases) of the VLBA data sets were performed with the $\tt DIFMAP$ package \citep{Shepherd1997}. After an initial round of imaging and self-calibration, any significant antenna-based gain errors were identified and a single correction factor per antenna per experiment was applied in $\tt AIPS$, if necessary. Then a new round of imaging and self-calibration was performed in $\tt DIFMAP$ -- this time normalizing the amplitude self-calibration solutions to unity in order to prevent the flux scale from wandering. The resulting amplitude calibration accuracy is estimated to be generally $\sim 5$\% at C, X and U bands and $\lesssim 10$\% at K and Q bands in accordance with previous studies \citep{Savolainen2008, Sokolovskyetal2011, Lister2018}. We used multiple visibility weighting schemes in imaging, going from super-uniform to uniform and finally to natural weighting as the model improved. The final set of CLEAN images was produced with natural weighting and an example of a set of multi-frequency images at one epoch is shown in Figure~\ref{multifreqimag}. The images for all the epochs are presented in Appendix~\ref{fullVLBAimages}. 

We note presence of the arc-like structure $\sim  2$\,mas downstream of the core. This feature has been previously reported by \citet{Britzen2013} and \citet{Zamaninasab2013}. The latter authors modelled the structure as a shock wave and showed that the structure exhibits a frequency-stratification in its thickness that follows the expectation for a thin particle acceleration layer -- such as shock wave -- from which the accelerated electrons are advected away losing their energy to synchrotron radiation. The arc is visible in our VLBA images mostly at 15\,GHz and 22$-$24\,GHz (until late 2009). At 43\,GHz, a partial arc -- with the southern side brighter than the northern one -- is visible until mid-2007.

%\tscom{Check how large image sets should be presented in A\&A. It might be that you need to have it as an Appendix.} We note that polarization data from these experiments will be presented elsewhere.

We highlight here that we dropped out the epoch 5 (2006-08-03) from further analysis due to Mauna Kea, the antenna that gives the longest baselines, not taking part in the observation (see Table~\ref{t1}). However, the images are still presented in  Appendix~\ref{fullVLBAimages}. The core-shift measurement for epoch 13 (2008-01-03) is still presented but was dropped out from the subsequent analysis. There are several missing antennas (Hancock, St.~Croix, Brewster partly) at this epoch, which results in a poor $(u,v)$ coverage and a degraded resolution such that the core cannot be well resolved at the low frequencies. This leads to ambiguities and a very large core-shift value that is unlikely to be real. The epochs 4, 6, 11, and 18 also had one (at least partially) failed antenna and epoch 12 had two failed antennas, but we kept these data in the analysis since their core-shift measurements did not indicate serious issues.  However, one should keep in mind that the $(u,v)$ coverage is degraded at these epochs.

%On the other hand, even though epochs 6, 11, 12 and 18 miss up to two antennas, this did not prevent the measurement of core shifts for the whole analysis presented in the following sections.

\begin{figure*}[ht]
\centering
  \subfigure[]
    {
    \includegraphics[width=5cm, height=5cm]{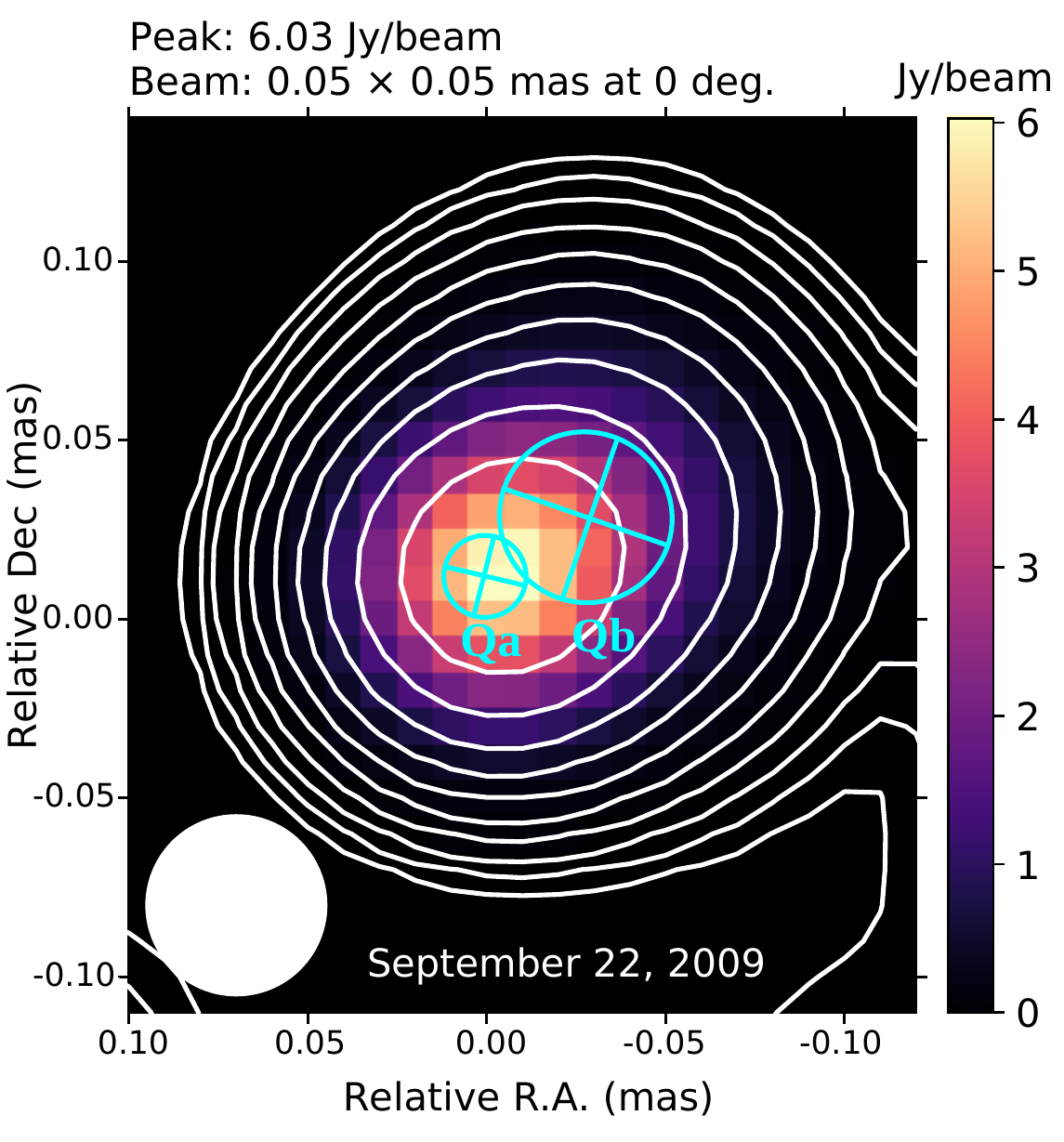}
        
    }
     \subfigure[]
    {
        \includegraphics[width=5cm, height=5cm]{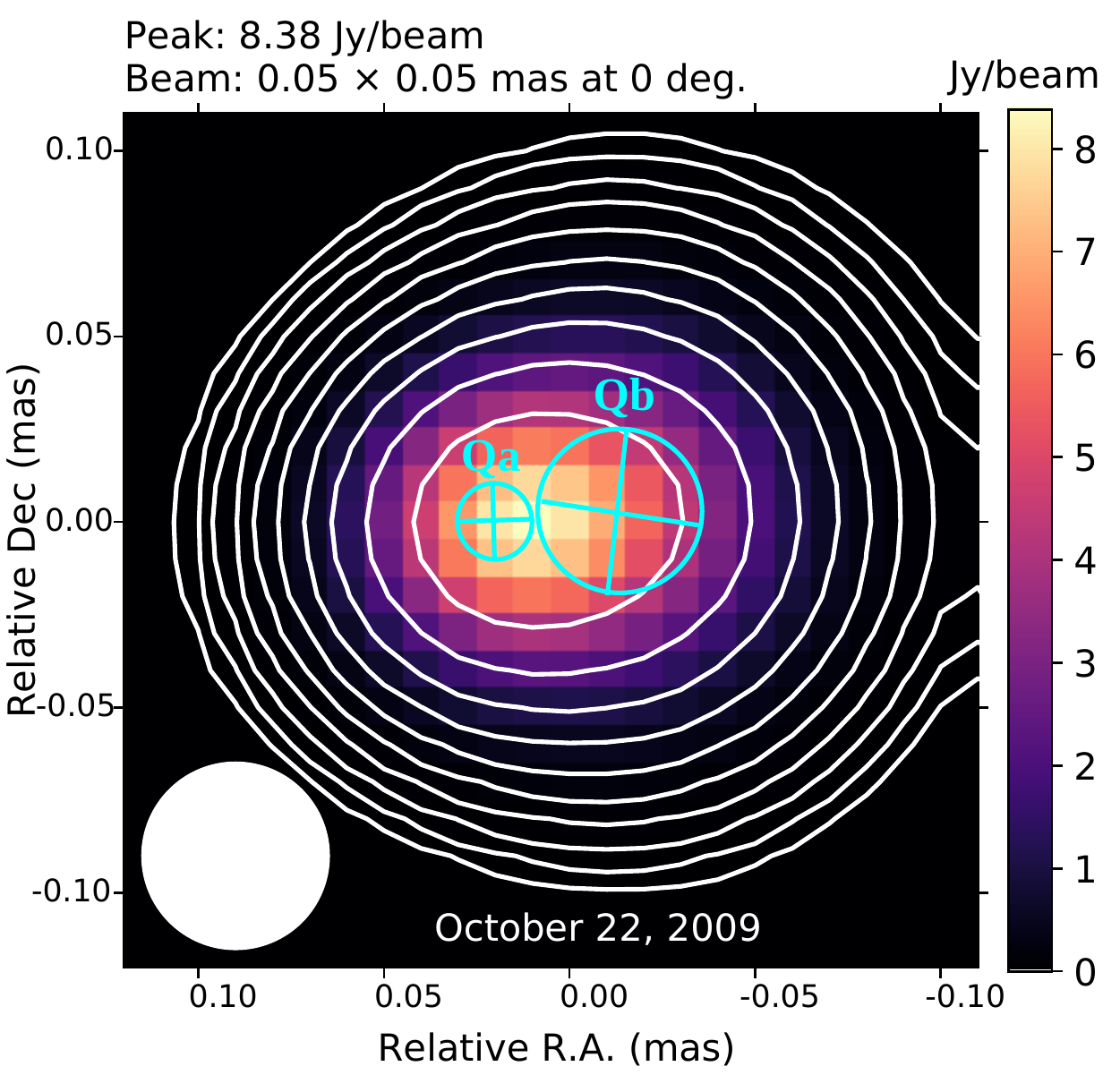}
    }
    \subfigure[]
    {
    \includegraphics[width=5cm, height=5cm]{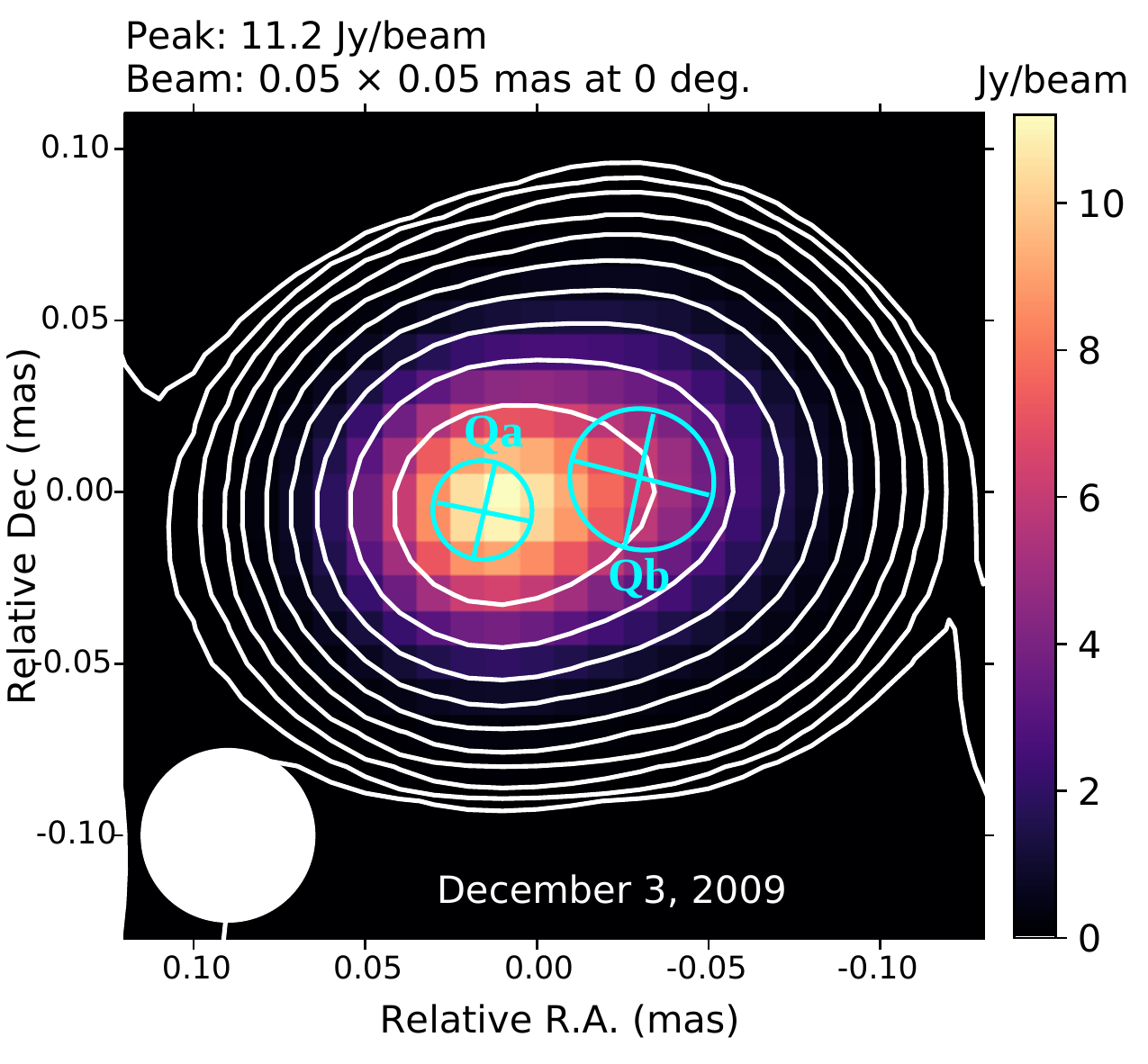}
    }
    \subfigure[]
    {
    \includegraphics[width=5cm, height=5cm]{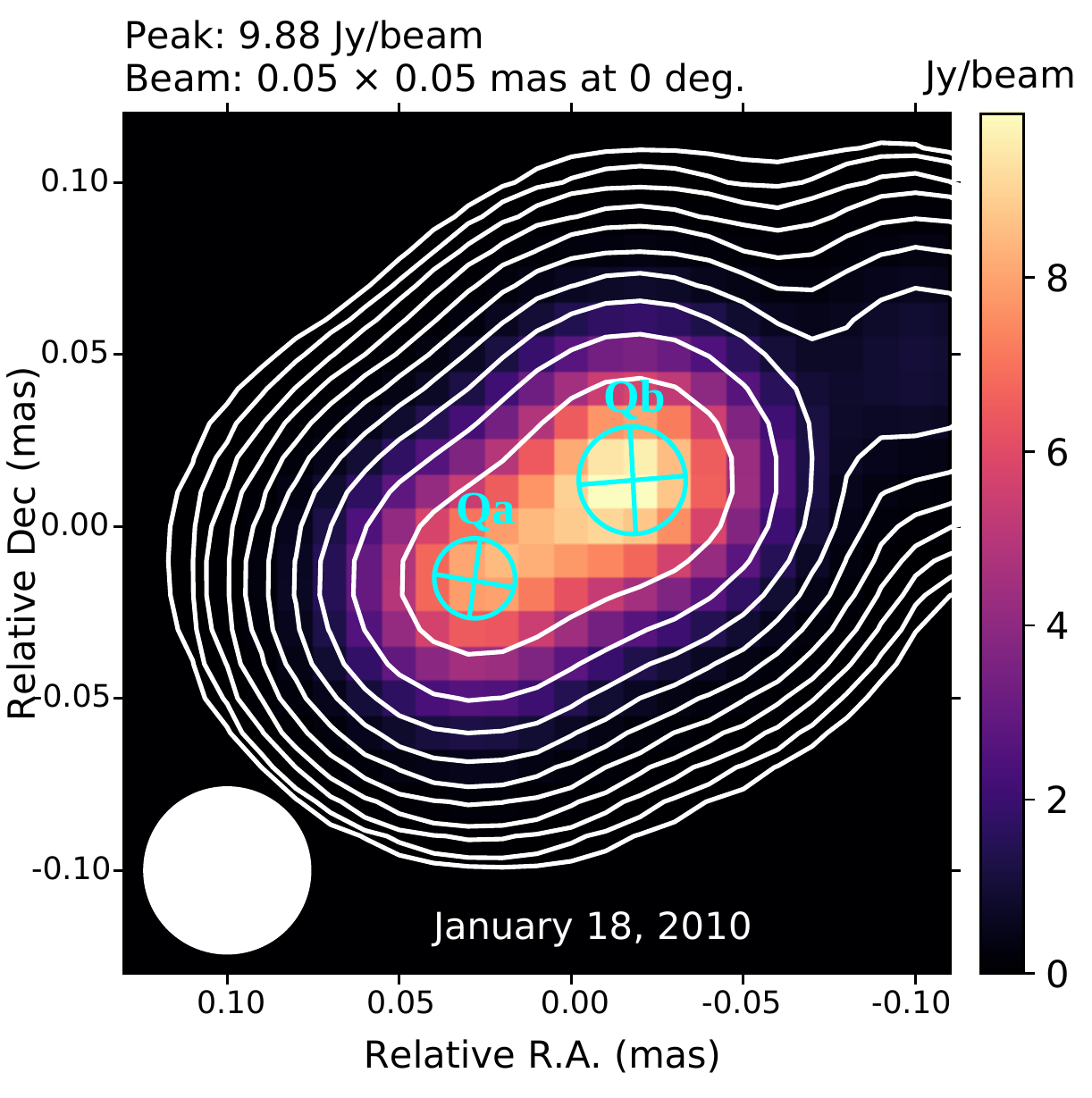}
    }    
    \subfigure[]
    {
    \includegraphics[width=5cm, height=5cm]{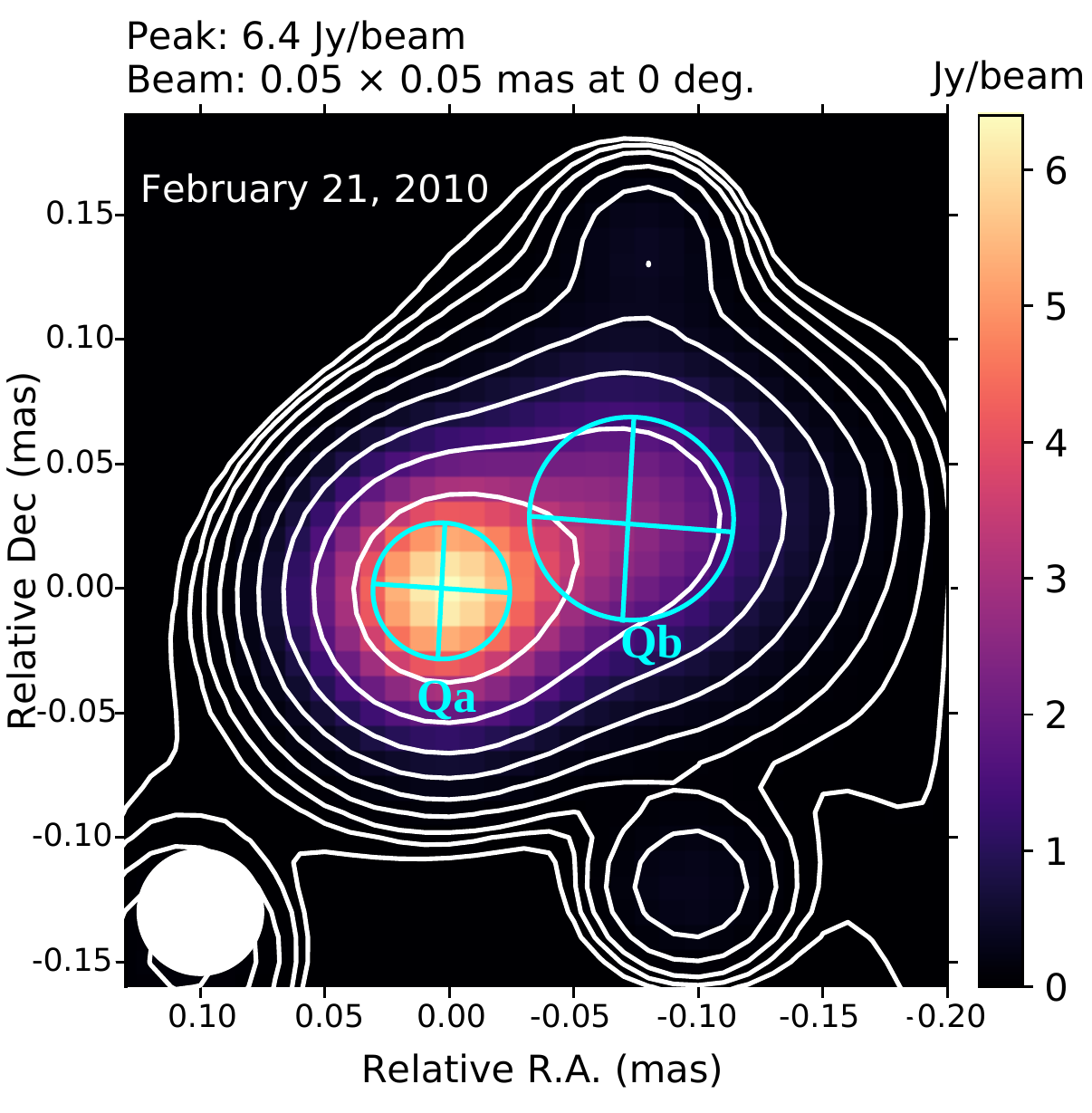}
    }
     \caption{Super-resolved images of \object{3C\,454.3} at 43\,GHz for five different epochs from 2009 to 2010. The contours are given at 0.1\%, 0.2\%, 0.4\%, 0.8\%, 1.6\%, 3.2\%, 6.4\%, 12.8\%, 25.6\%, and 51.2\% of the peak intensity at each image. Each map uses a circular convolving beam of 0.05\,mas and the components are displayed by cyan colour. The restoring beam (circular) is displayed on the bottom-left corner of each image. The rms noise level is in a) 1.1, b) 2.4, c) 4.1, d) 7.7 and e) 2.9\,mJy/beam. The core is marked by `Qa' and `Qb' denotes a bright, recently ejected component that is moving downstream.}
    \label{43GHzcore}
\end{figure*}

\section{Data analysis}

\subsection{2D cross-correlation of the images}

In order measure the core shift, it is necessary to first align the images made at different frequencies. This is non-trivial since the absolute source positions are lost in the phase self-calibration procedure during imaging. \object{3C\,454.3} exhibits an extended parsec-scale jet structure at all the observed frequencies (Figure~\ref{multifreqimag}). The extended structure allows one to align the images by matching the optically thin parts of the jets that are assumed to have frequency-independent structures. We do this by cross-correlating the optically thin jet regions in the CLEAN images at adjacent frequencies. The 2D cross correlation technique has been widely used to align multi-frequency VLBI images \citep[e.g.,][]{Walkeretal2000, Kovalevetal2008, CrokeGabuzda2008, OSullivanGabuzda2009, Hovattaetal2012, Pushkarevetal2012,  Kutkinetal2014, Kravchenko2016, Plavin2019, Chamani2021}. 

In order to apply the image cross-correlation analysis, we first created CLEAN images at the different frequency pairs with a pixel size of 1/20 of the minor axis of the beam of the higher frequency image and then the pair of images were convolved with the same restoring beam size (of the lower frequency one). Next, the image alignment was performed for adjacent pairs of frequencies, i.e. CX, XU, UK, KQ, using a similar procedure as described in \citet{Pushkarevetal2012}. Here, e.g., the CX notation means that the alignment of the C-band image was done with respect to the X-band image. Finally, to obtain a statistical average of the image shift with an associated uncertainty, we performed the alignments ten times per frequency pair selecting slightly different optically thin features every time, similarly to the analysis in \citet{Chamani2021}. The procedure involved five alignments with matched common $(u,v)$ range images and five full $(u,v)$ range images. The resulting spectral index maps have both: common and full $(u,v)$ range of a frequency pair. A common $(u,v)$ range means that the lower limit of the $(u,v)$ distance for both images is the lower limit of the high frequency, and the upper limit coincides with the upper limit of the low frequency. Spectral index maps with a common $(u,v)$ range should be used in any analysis of the extended, optically thin emission, since the missing short $(u,v)$ spacings at the higher frequency can result in artificial steepening of the spectrum of the extended emission features. The common $(u,v)$ range spectral index images are displayed in Appendix~\ref{fullSImaps}.

%so we employ similar process as described in \cite{Kutkinetal2014}. A good and detailed description of the whole process can be be found in the previous work. To evaluate the goodness of the alignments we obtain and inspect the spectral index maps.

\subsection{Visibility plane model-fitting}

In order to accurately measure the core position, we fitted simple models of the source structure directly to the visibilities. We used the (fully self-) calibrated visibility data and model-fitted the core with 2D Gaussian components consisting of circular components at each frequency. To model the core, we first removed CLEAN components from the nuclear region in the jet (around the brightest pixel) leaving the extended emission unchanged. In general, the area removed was one-beam size for all frequencies, however at the low frequencies (C, X bands) the area had to be increased in order to obtain better Gaussian fits similarly to the approach performed by \citet{Homan2021}. The model-fits were performed minimizing the reduced $\chi^2$ with a Levenberg-Marquardt non-linear least-squares fitting algorithm implemented in $\tt DIFMAP$ so that a value as close to one as possible was reached. Then, we combined the core coordinates (from the map centre) with the image shifts measured by 2D cross-correlation to estimate the core-shift vectors by following the method described in \cite{Pushkarevetal2012}. We calculated, in particular, the core-shift vectors for the adjacent frequency pairs: CX, XU, UK, and KQ.

A typical VLBI image often has the brightest feature at the upstream end of the jet. Such structure is usually assumed to be the 'core' of the jet; however, in \object{3C\,454.3}, at some epochs, the core region requires several components, reflecting a complex nuclear structure in this source. The emergence of such features has been noticed at 2.8\,cm (10.7\,GHz) already in the 1980s by \cite{Pauliny-Toth1987}. Their study suggests that the additional features near the core may be stationary shocks. Another alternative includes the possibility of the blending of the core with the jet \cite[e.g.,][]{Kovalevetal2008, Sokolovskyetal2011, Algaba2019}. Hence, the presence of several features can lead to the misidentification of the true core, and the identification of the core (i.e., $\tau = 1$ surface) is non-trivial. This turned out to be an issue at the two lowest frequencies (5 and 8\,GHz). An example of this is displayed in Figure~\ref{core-low-freq}, where a moderately bright emission feature is seen \emph{upstream} of the brightest emission at the C and X-bands. 
The two components in the core region are denoted by `Ca' and `Cb' at the C-band and `Xa' and `Xb' at the X-band. The flux of each component is marked in the core spectrum displayed in Figure~\ref{spectrum-csvectors}a. 

In order to correctly identify the core at each frequency, we calculate the CX and XU core-shift vectors by taking the following combinations: Ca/Xa, Ca/Xb, Cb/Xa, and Cb/Xb. In many cases, it is expected that only one of the combinations will lead to both the CX and XU core-shift vectors pointing in the expected jet direction (towards West) and resulting in a single power-law frequency dependency. For example, we found that the unique combination Cb-Xb leads to correct CX and XU core-shift directions in the 2005-07-14 epoch as shown in Figure~\ref{spectrum-csvectors}b. The other combinations produce CX and/or XU core shift vectors that do not align with the jet direction. These combinations are displayed in Appendix~\ref{appendix:choices}. The correct combination is also verified by the resulting power-law fits. A fit is considered acceptable as long as the core shifts are not too large to look unrealistic (which can be driven by the wrong choice of the core at the low frequencies). Hence, this approach allows us to identify the most likely core component at each image.

Since the core identification plays a crucial role in accurate core-shift measurements and consequently the estimation of the core-shift index $k_r$, we employed the criteria described above as a sanity check for the observing epochs where the core identification at the low frequencies is unclear -- these are specified in Table~\ref{appendix:core-shift-values}. The model fitting of the core for the U, K and Q bands in general works well with one single circular Gaussian. However, an emerging new component was identified at the Q band in our data sets in 2009 and 2010. The ejection of a new emission feature is possibly related to the strong flaring event in 2009. Although we do not aim to study the jet kinematics of \object{3C\,454.3} in this paper, we discuss this specific ejection in the following section, since it significantly affects the core shift measurements.

\begin{figure}[t]
\centering
\includegraphics[width=0.5\textwidth]{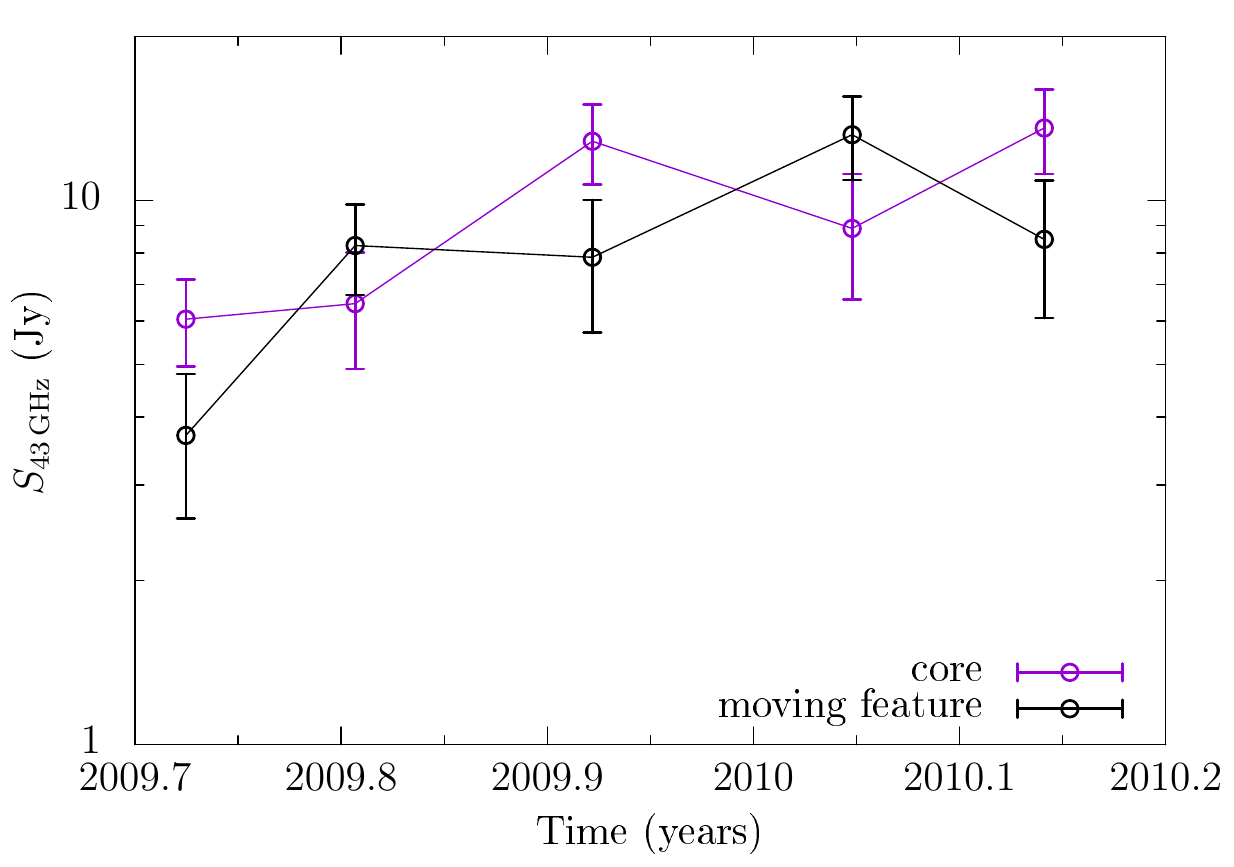}
\caption{Flux density at 43\,GHz of the core and the moving feature downstream as a function of time. The error bars include the uncertainty in the flux scale which is $\sim 10$\%.}
\label{core-featurefluxes}
\end{figure}

\begin{figure}[t]
\centering
\includegraphics[width=0.4\textwidth]{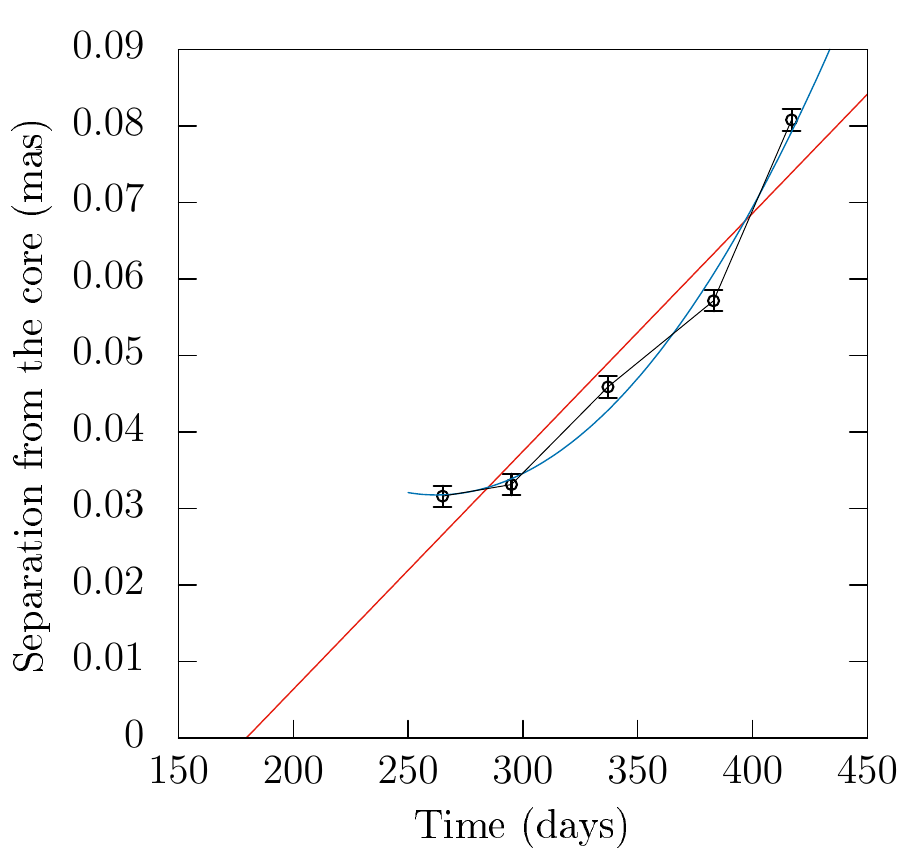}
\caption{Separation of the moving feature (see Figure~\ref{43GHzcore}) from the core as a function of time at 43\,GHz. The first day is counted from 2009-01-01. The black circles indicate the observations on 2009-09-22, 2009-10-22, 2009-12-03, 2010-01-18, and 2010-02-21. The red and blue curves indicate a linear and accelerating fitting functions, respectively.}
\label{blob-kinematics}
\end{figure}

\subsection{Ejection of a moving feature downstream at 43\,GHz in the period of 2009$-$2010} \label{Sect:ejection}

In 2009$-$2010, the nuclear region at 43\,GHz was modeled first with one circular Gaussian, but the fit became better and more stable with two circular components, which indicated that a emission feature had been ejected from the core. Figure \ref{43GHzcore} shows the two resolved, bright components marked by cyan colour. 
%\ericom{the figure appears before it is referred in the text} 
The core is labelled `Qa' and the moving feature downstream `Qb'. The feature follows an east-west trajectory. The flux density curves of each component are shown in Figure~\ref{core-featurefluxes}. Since the core and the moving feature are so close to each other they can "swap flux" in model-fitting. Due to this, the flux ratio of the two components is uncertain, and the formal flux uncertainties of the fit (shown in Figure~\ref{core-featurefluxes}) likely underestimate the errors. Finally, we note that the moving feature tracked from September 2009 until February 2010 may be cross-identified the knot 'K09' in by \cite{Jorstadetal2013}. 

Figure~\ref{blob-kinematics} displays the feature separation from the core as a function of time. The errors of the core separation are small since both components are very bright. The uncertainties in the component position were derived from Gaussian model-fitting by moving the component by small amounts around the best-fit position, fixing the component position, and finding the set of remaining parameters that minimize the $\chi^2$. The errors are found by requiring $\Delta \chi^2 = \chi^2 - \chi^2_\mathrm{min} < C^{\alpha}_{p}$, where $\chi^2_\mathrm{min}$ is the minimum $\chi^2$ corresponding to the best-fit model and $C^{\alpha}_{p}$ is the critical value of $\chi^2$ distribution for $p$ degrees of freedom and a significance level $\alpha$ (we use $\alpha = 0.32$ corresponding to 1-$\sigma$ errors). For details, see \citet{Chamani2021}.

A linear function was fitted to the data to follow the trajectory of the moving feature. 
The fitting model gives $d = -(0.06 \pm 0.02) + (3 \pm 0.5)\cdot 10^{-4}\,t$, where $d$ is the separation (mas) from the core and $t$ the time (days from 2009-01-01). 
%\ericom{error bars?}
Hence, the feature has a proper motion of ($0.11 \pm 0.02$)\,mas\,yr$^{-1}$ corresponding to an apparent superluminal speed of $\beta_\mathrm{app} = 5.0 \pm 1.0$. 
The new feature's extrapolated ejection time ($T_0$) is $179 \pm 66$\,days corresponding to 2009-06-28, or equivalently to $2009.49 \pm 0.18$. 
On the other hand, fitting an accelerating model gives 
$d = (0.17 \pm 0.08) -(1 \pm 0.5)\cdot10^{-3}\,t + (2 \pm 0.7)\cdot10^{-6}\,t^2$, 
where the acceleration of the feature is ($0.26 \pm 0.09$)\,mas\,year$^{-2}$.
% \ericom{check IAU conventions, year can be "a", or shorter, also yr}
We note that in \cite{Jorstadetal2013} the new component `K09' is tracked only from January 2010 on, and it moves at ($0.21 \pm 0.02$)\,mas\,year$^{-1}$ corresponding to $\beta_\mathrm{app}=9.6\pm0.6$ and $T_0=2009.86\pm0.05$ (using 24 epochs). 
In their study, they estimated two acceleration components both along and perpendicular to the jet given by ($0.10 \pm 0.01$) and ($0.13 \pm 0.02$)\,mas\,year$^{-2}$, respectively. Additionally, the flux density of the moving feature in early 2010 is comparable to the flux density of `K09' shown in Figure~13 in \cite{Jorstadetal2013}. The separation from the core is below 0.1\,mas and is similar to our observed values. Hence, we conclude that 'K09' is likely the same feature that we detected in late 2009 and early 2010. If this is the case, the moving feature has experienced acceleration in order to explain the factor of two difference in the average speeds.

\begin{figure}[t]
\centering
    \subfigure[]
    {
        \includegraphics[width=0.95\linewidth]{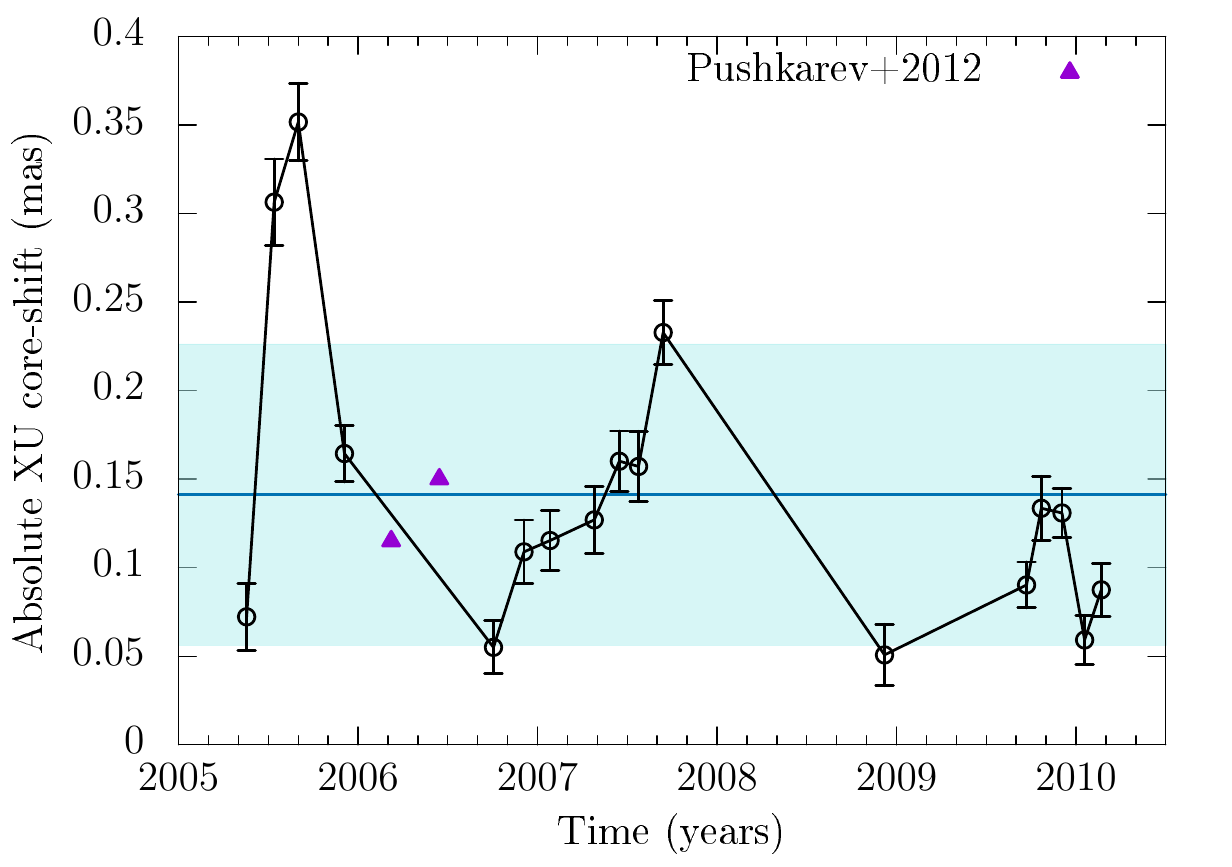}
    }
    \subfigure[]
    {
        \includegraphics[width=0.95\linewidth]{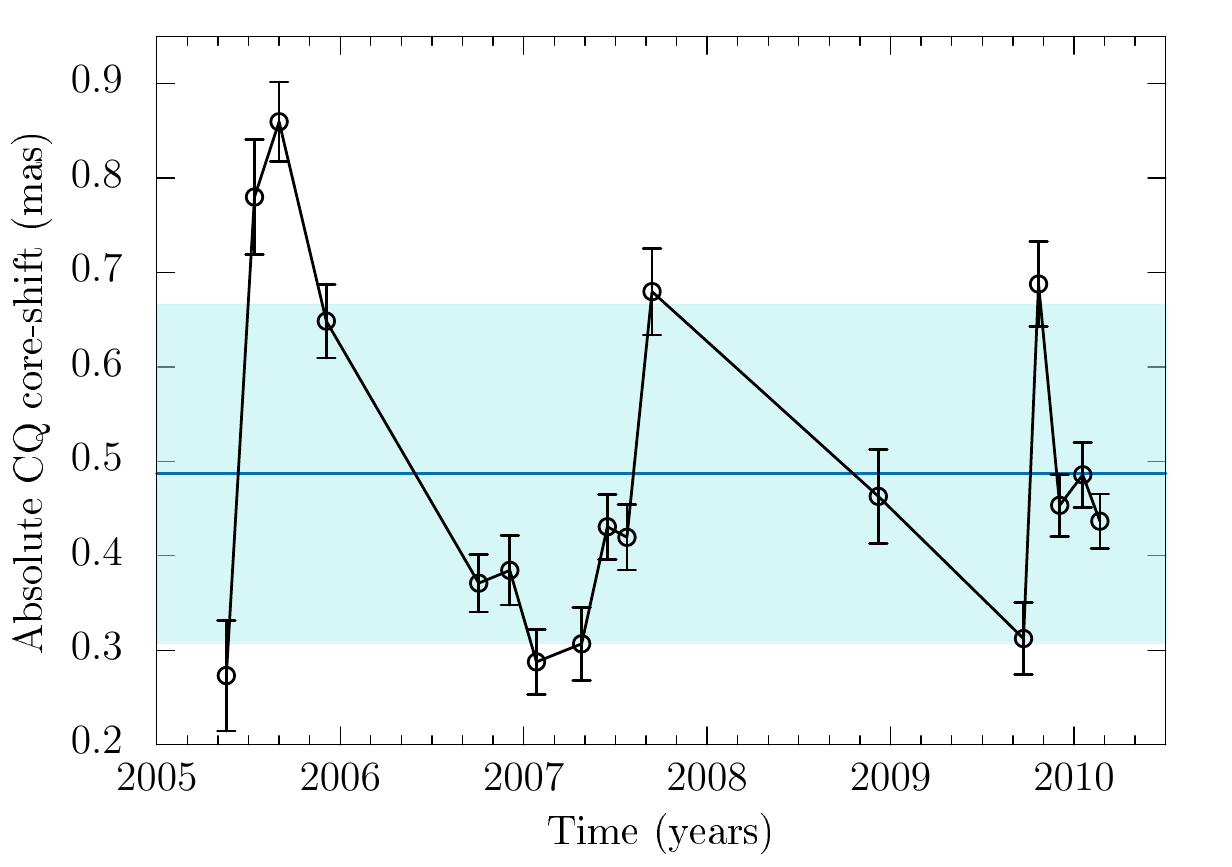}
    }
     \caption{Variability of the core shift in \object{3C\,454.3}. a) Core shift between the X (8\,GHz) and U (15\,GHz) bands versus time. The purple triangles represent the core shifts derived by \citet{Pushkarevetal2012}. b) Core shift between the C (5\,GHz) and Q (43\,GHz) bands versus time. In both figures the mean value is represented by the blue line and $\pm\sigma$ is displayed by the shaded cyan background. The epoch 2008-01-03, which had three missing antennas, was dropped out from this plot as an outlier with a CQ core shift of $1.64 \pm 0.04$\,mas. Zero core-shift physically means that the cores at two frequencies coincide at the same location.} 
    \label{cs-variable}
\end{figure}

\section{Results}

In previous sections, we described our method to identify the core and the estimations of the core shifts for adjacent frequency pairs. Here we present and analyse the resulting core shifts.

\subsection{Core-shift measurements and variability}

We list in Appendix~\ref{appendix:core-shift-values}, Table~\ref{coreshift-table} the core-shift vectors with components in Right Ascension (R.A.) and Declination (Dec) as well as the core-shift absolute values and projected absolute values. The projected absolute values were obtained by projecting the core-shift vectors onto the average vector direction in a similar fashion as described in Section~4.1 in \citet{Chamani2021}. This is done to diminish the effect of random errors on core-shift vector directions, and it is implicitly assumed that the jet is straight at any given epoch. The total positional uncertainties of the core shifts (for both coordinates) are estimated as
\begin{equation}
\label{eq:cserror}
\Delta \theta_\mathrm{total,core-shift}=\sqrt{\Delta \theta^2_\mathrm{2DCC}+\Delta \theta^2_\mathrm{core-position}}\,,
\end{equation}
where $\Delta \theta_\mathrm{2DCC}$ is the error obtained from the 2D image cross-correlation analysis, and $\Delta\theta_\mathrm{core-position}$ is the error of the core position obtained from the 2D Gaussian model-fitting. For the latter we employed the same method as used by \citet{Chamani2021} and also described in Sect.~\ref{Sect:ejection}.

By following the method and notation for the core identification described in the previous section (see Figures \ref{core-low-freq} and \ref{spectrum-csvectors}), we present in Appendix~\ref{appendix:core-shift-results} the core spectrum, the core-shift vectors, and the core-shift power-law fit for each epoch. Using the absolute (projected) core-shifts of Table~\ref{appendix:core-shift-values}, we fitted the data with power-law curves using 43\,GHz as the reference frequency,
\begin{equation}
\label{eq:csfitting-function}
    \Delta r=a\,(\nu_\mathrm{GHz}^{-1/k_r}-43^{-1/k_r}),
\end{equation}
where $\Delta r$ represents the core-shift in mas, $\nu$ the observing frequency, $a$ is a fitting parameter, and $k_r$ is the core-shift power-law index. The fitting parameters are summarized in Table~\ref{fits-parameters}.

We found, in general, that all core-shift vectors point towards the west-north-west direction, which agrees well with the jet direction downstream of the \object{3C\,454.3} jet. Changes in the direction towards the south-west are seen in some core-shift vectors, for instance in Figures~\ref{CSepoch1}, \ref{CSepoch14}, \ref{CSepoch15}, \ref{CSepoch16}, \ref{CSepoch17}, \ref{CSepoch18}, and \ref{CSepoch19}. 
Major flaring events have been registered for instance in May 2005 and from late 2009 to early 2010, which might have affected the direction of some core-shift vectors.

We note that the core shift at the U-band is off the fitting curve in the observation on 2008-12-07; thus, we dropped out 15\,GHz from the fit at this epoch. The data set without the U-band data point yields a better fit at this epoch (see Figure~\ref{CSepoch14}). 
A similar case is seen in the observation on September 22, 2009 (see Figure~\ref{CSepoch15}). Furthermore, the outburst starting in late 2009 appears to affect the core shift measurements of the last five epochs. 
In these observations, the core-shift vectors changed the direction significantly towards the south-west, and the functional form of the core-shift changed significantly with $k_r$ increasing to values above one, and the typical core-shift behaviour became in some cases distorted. This effect is evident on 2009-12-03, where the core-shift for the K band increased by a factor of two from the previous epoch. 
As a consequence, the fit became poor, deviating greatly from a power-law (see Figure\ref{CSepoch17}c). Hence, we did not include this epoch for the further analysis described in the next subsections. Even though the core shifts are not greatly distorted on 2010-01-18 and 2010-02-21, the uncertainties are large and the fits exhibit considerable deviations from $k_r=1$ (see Figures \ref{CSepoch18}d and \ref{CSepoch19}d).

\begin{figure}[t]
 %\centering
 \includegraphics[width=0.48\textwidth]{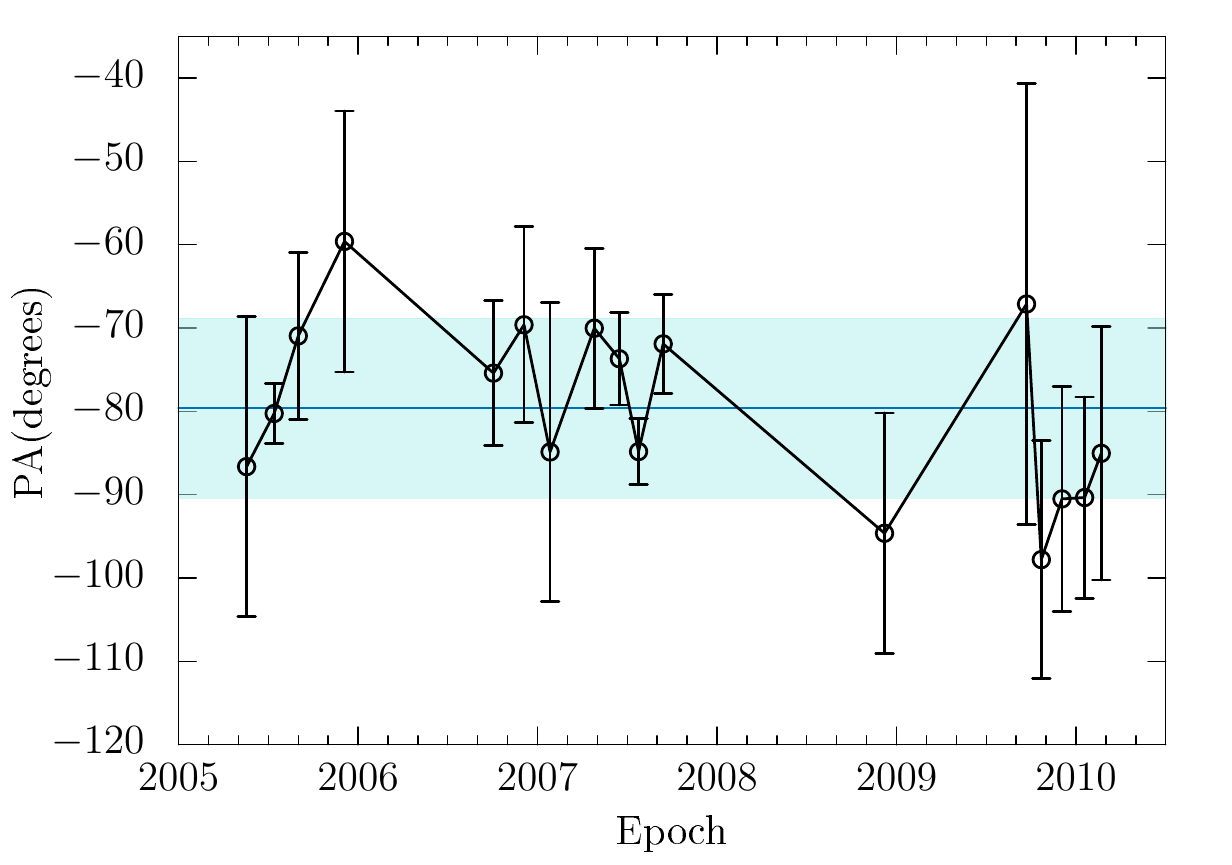}
  \caption{Time variability of the jet position Angle (PA) from the  model fitting to the core. The PA values (black open circles) represent the average PA per epoch using all core-shift vectors per frequency pair (CX, XU, UK, KQ).  The blue line is the mean PA for the whole data set. The area in light cyan represents $\pm 1\sigma$.}
% \tscom{The PA values come from the core shift vectors, right? It is not clear now and should be clarified. Remove the "from the circular model fitting to the core", which just makes this confusing. Eduardo was asking if the PA is the orientation of an elliptical component fitted to the core, but that is not the case, right? The PA comes from the core shift vectors, right?}
 \label{PAvar}
\end{figure}

\eri{
\begin{table}[]
\centering
\begin{threeparttable}
\caption{Core-shift power-law fit parameters. $k_r$ is the core-shift index, $a$ is a fitting parameter. Rounded values are given here.} 
%\ericom{Indicate in the caption or in the note below the meaning of $k_r$, $a$, and $epsilon$.  Note that when an error bar is provided starting with 1, a further digit is needed, namely, $0.70\pm0.10$ but not $0.7\pm0.1$; it is correct $0.7\pm0.2$ or $0.70\pm0.08$.  I modified this table (original copy above in the latex file) to get the $\pm$ aligned for each column}
\label{fits-parameters}
%\begin{tabular}{@{}c{2.2cm}c{2.5cm}c{2.2cm}c{2.2cm}@{}}
\begin{tabular}{@{}ccr@{\,$\pm$\,}lr@{\,$\pm$\,}lr@{\,$\pm$\,}l@{}}
\hline \noalign{\smallskip}
\hline \noalign{\smallskip}
Epoch & Date  & \multicolumn{2}{c}{$k_{r}$} & \multicolumn{2}{c}{$a$} \\ \hline  \noalign{\smallskip}
1& 2005-05-19  & 0.70  & 0.10 & 2.6   & 1.1  \\
2& 2005-07-14  & 0.45  & 0.05 & 29.4  & 12.5  \\
3& 2005-09-01  & 0.52  & 0.07 & 19.6  & 9.2    \\
4& 2005-12-04  & 0.70  & 0.10 & 7.2   & 2.7   \\
6& 2006-10-03  & 1.0   & 0.3  & 1.7   & 0.8  \\
7& 2006-12-04  & 0.67  & 0.07 & 4.4   & 1.3 \\
8& 2007-01-26  & 0.65  & 0.06 & 3.8   & 1.1 \\
9& 2007-04-26  & 0.80  & 0.10 & 2.8   & 1.0   \\
10& 2007-06-16 & 0.60  & 0.10 & 5.9   & 2.5  \\
11& 2007-07-25 & 0.69  & 0.05 & 4.5   & 0.8  \\
12& 2007-09-13 & 0.90  & 0.10 & 4.2   & 0.8  \\
%13& 03.01.2008 & 1.1 & 0.2 & 7.8   & 2.6 & 1.4 & 0.1 \\
14& 2008-12-07  & 1.4 & 0.4\tnote{(a)} & 1.7 & 0.6   \\
  & 2008-12-07  & 1.1 & 0.1\tnote{(b)} & 2.3 & 0.4   \\
15& 2009-09-22  & 1.1 & 0.6\tnote{(c)} & 1.7 & 1.2  \\
  & 2009-09-22  & 0.8 & 0.3\tnote{(d)} & 2.4 & 1.7  \\
16& 2009-10-22  & 1.1 & 0.3 & 3.4 & 1.4    \\
17& 2009-12-03\tnote{$\dagger$}  & \multicolumn{2}{c}{---}  &\multicolumn{2}{c}{---}\\
18& 2010-01-18 & 1.7 & 0.7 & 1.6 & 0.5  \\
19& 2010-02-21 & 1.3 & 0.3 & 1.7 & 0.5 \\
\hline \noalign{\smallskip}
\end{tabular}
\begin{tablenotes}
\item [] The 13th observing epoch (2008-01-03) was dropped from the analysis.
\item [(a)] Using all frequency bands, see ~\ref{CSepoch14}c.
\item [(b)] Without the U band, see Figure~\ref{CSepoch14}e.
\item [(c)] Using all frequency bands, see ~\ref{CSepoch15}e.
\item [(d)] Without the U band, see Figure~\ref{CSepoch15}f.
\item [$\dagger$] No good power-law fit, see Figure~\ref{CSepoch17}d.
\end{tablenotes}
\end{threeparttable}
\end{table}
}

We plot the core-shift variability in Figure~\ref{cs-variable}. The plots show the absolute values of the XU and CQ core shifts as a function of time. For a comparison, we included the XU core-shift values measured by \cite{Pushkarevetal2012}. Their values lie well within the range of our measurements between 0.05\,mas and 0.34\,mas. The mean XU core shift is $0.14 \pm 0.02$\,mas with a standard deviation, $\sigma$, of 0.09\,mas. %The majority of the data points are within $1\sigma$, but nearly 12\% of the data lie strictly within $2\sigma$. 
In order to test whether the XU core shift is variable given the measurement uncertainties, we set up a null hypothesis stating that the core shift does not vary and remains stable around the mean value and calculated the $\chi^2$ statistic,
\begin{equation}
\label{eq:chisquare-test}
 \chi^2=\sum_{i}^{N} \frac{\left (\Delta r_{(\nu_1\nu_2)i}-\overline{\Delta r_{\nu_1\nu_2}} \right )^2}{{\sigma'_i}^2},
\end{equation}
where $\Delta r_{(\nu_1\nu_2)i}$ represents the core-shift of the $i$th observation between two frequencies, $\nu_1$ and $\nu_2$ ($\nu_2>\nu_1$), $\sigma'_i$ is the core-shift error, $\overline{\Delta r_{\nu_1\nu_2}}$ is the mean core-shift and $N(=17)$ is the number of observations. At the 99.9$\%$ confidence level for $N-1$ degrees of freedom (d.o.f), $\chi_\mathrm{critical}^2$ = 39.25 is the critical value for rejecting the null hypothesis. For XU core shift ($\Delta r_\mathrm{XU}$), $\chi^2$ results in 312.5, which is above the critical value, confirming that the XU core shift significantly varies with time. 

A similar analysis was also performed for the CQ core-shifts ($\Delta r_\mathrm{CQ}$) (see Figure~\ref{cs-variable}b) where the mean value is $0.49 \pm 0.05$\,mas with a standard deviation of 0.18\,mas. The core shifts vary between 0.27\,mas and 0.86\,mas. There is one extreme core shift value, $1.64 \pm 0.04$\,mas, on January 3, 2008. Since this observation misses both St.~Croix and Hancock for the whole observation, as well as Brewster for half of the observation, the resulting poor $(u,v)$ coverage makes this result of suspect and we drop it from our analysis and from Figure~\ref{cs-variable}. The $\chi^2$ test for the CQ core shift gives 279.02 (for $N=17$), above the critical level. Thus, the CQ core shift also shows significant variability. Overall, the variability curves of both the XU and CQ core shifts exhibit similar trends, although there is no one-to-one correspondence.

Additionally, we studied the mean core shift direction variability, which represents approximately the overall inner jet direction (PA) in the core region. The mean jet PA was measured by averaging the direction of the core-shift vectors of all frequency pairs.
%PA$_\mathrm{CX}$, PA$_\mathrm{XU}$, PA$_\mathrm{UK}$ and PA$_\mathrm{KQ}$. 
The PA variability plot is presented in Figure~\ref{PAvar}. 
For the study period, the mean jet PA is $-80^{\circ} \pm 3^{\circ}$, $\sigma \sim 10.8$, $\chi^2 = 13.66$  (for $N=17$).
%\eri{$\chi^2 = 13.66$}. 
%\ericom{The use of normal likelihood, not reduced by dividing for the degrees of freedom is confusing for the reader, since the critical value is every time different throughout the text.}
The latter is below the critical value ($\chi_\mathrm{critical}^2 = 39.25$) indicating no significant jet PA variability. That means that even if there is significant core shift variability, the jet vector direction stayed relatively stable in our study period. While the changes in jet PA are not statistically significant, it is interesting to note an increasing trend in the PA from mid-2005 to early 2006, corresponding to a major mm-wavelength flare, and then more or less stable PA from late 2006 until late 2007. Possibly related to the strong outburst, PA decreased to lower values from late 2009 to early 2010. 

%The jump seen on September 22, 2009, is due to the arbitrary change of the KQ core shift-vector direction pointing very close to North (Figure~\ref{CSepoch15}d). \tscom{Considering the large errors, I am not sure this "jump" needs to be mentioned?}

%(Note: we could take the mean PA by dropping out the KQ vector and we get approximately -86 degrees so the jump will go down.)

\begin{figure}[t]
\centering
 \includegraphics[width=0.5\textwidth]{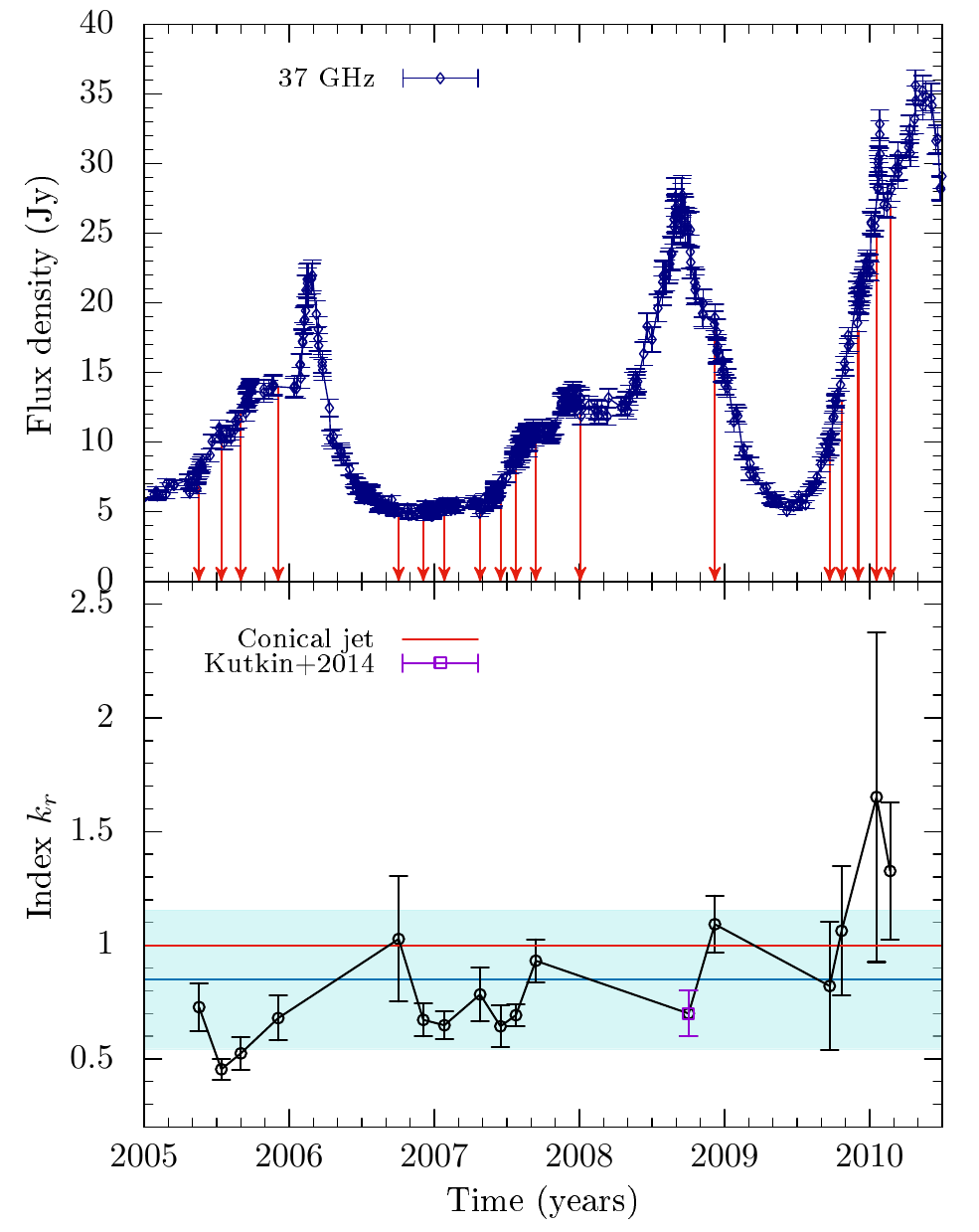}
 \caption{Top: Total flux density at 37\,GHz for \object{3C\,454.3} observed at the Mets\"{a}hovi Radio Observatory. 
 The red arrows show the time correspondence with our VLBA observations. 
 Bottom: Core-shift power-law $k_r$ index versus time. The measured $k_r$ values are represented by black open circles. The purple square, $0.7\pm 0.1$, represents the value  obtained by \cite{Kutkinetal2014}. The observation epoch on 2009-12-03 is not included. The blue dashed line shows the mean for the whole data set. The area in the light cyan represents $\pm 1\sigma$. The red line represents the ideal BK79 conical jet with $k_r=1$}.
 \label{2-plots}
\end{figure}

\begin{table}[t]
\centering
\begin{threeparttable}
\caption{Physical properties of \object{3C\,454.3} }
\label{properties}
\begin{tabular}{@{}lcl@{}}
\hline
\hline \noalign{\smallskip}
Name & Parameter & Value \\ \hline \hline \noalign{\smallskip}
Redshift \tnote{(a)} & $z$ & 0.859  \\
Luminosity distance \tnote{(b)} & $D_\mathrm{L}$ & 5.49\,Gpc \\
Doppler factor \tnote{(c)} & $\delta$ & 24.6 $\pm$ 4.5  \\
Lorentz factor \tnote{(c)} & $\Gamma$ & 15.6 $\pm$ 2.2  \\
Viewing angle \tnote{(c)} & $\theta$ & 1.3$^\circ$ $\pm$ 1.2$^\circ$  \\
Half-opening angle \tnote{(c)} & $\theta_\mathrm{j}$ & 0.8$^\circ$ $\pm$ 0.2$^\circ$ \\
Black hole mass \tnote{(d)} & $M_\mathrm{BH}$ & 4.9 $\times10^{8}$\,$M_\mathrm{\odot}$  \\
Accretion disk luminosity \tnote{(d)} & $L_\mathrm{acc}$ & 7.2 $\times10^{46}$\,erg s$^{-1}$ \\\hline
\noalign{\smallskip}
\end{tabular}
\begin{tablenotes}
\item[(a)] Observed by \cite{Sargent1988, Jackson1991}.
\item[(b)] Using the calculator of \cite{Wright2006}.
\item[(c)] Global jet parameters  adopted from \cite{Jorstad2005}. We assumed the parameters to remain constant during the period 2005-2010.
\item[(d)] Values adopted from \citet[][with references therein]{Zamaninasab2014}.
%\ericom{BH is not defined in the text} 
\end{tablenotes}
\end{threeparttable}
\end{table}

\subsection{Variability of the core-shift index $k_{r}$}

Core-shift indices $k_{r}$ are listed in Table~\ref{fits-parameters}, and the variability plot of $k_{r}$ is displayed in the bottom panel of Figure~\ref{2-plots}. For comparison, we included the result obtained \cite{Kutkinetal2014} for a single observation. We find that $k_{r}$ varies in the following range: $0.45<k_{r}<1.7$, with an average value of $0.85 \pm 0.08$ and with a standard deviation $\sigma=0.30$. We note that we dropped out the observation on 2009-12-03 since the fitting result gives an unreasonably  large $k_r$ value that is due to the ongoing flare significantly increasing the KQ core-shift value, while the lower frequencies are not affected (see Figure~\ref{CSepoch17}d). It is important to point out that $k_r$ indices in the period of 2009 to 2010 have large uncertainties due to the strong flare affecting the core-shift measurements (see Figures~\ref{CSepoch15} to \ref{CSepoch19}). It appears that the flare increases the distance between the 24 and 43\,GHz cores. This may be due to several reasons, including increased particle density and/or magnetic field strength due to the flare that increase the synchrotron opacity and move the 24\,GHz core downstream. It is also possible that the blending of the newly ejected component with the core at 24\,GHz drags the apparent core position downstream. In any case, the power-law dependencies measured during this period are disrupted by the flare and the measured $k_r$ values are affected.

Similarly to the variability test of the core-shift, we set up a null hypothesis that $k_{r}=1$ and does not vary over time. We also included in the analysis the result from \citep{Kutkinetal2014}. The chi-square test resulted in $\chi^2 = 146.50$, which is above the critical value (for $N=17$). If we exclude the epochs from 2009 to 2010, the mean of $k_r$ is $0.74 \pm 0.05$ with $\sigma=0.18$, and $\chi^2 = 67.16$, which is still above the corresponding critical value of $\chi_\mathrm{critical}^2 = 32.91$ (12\,d.o.f). Hence we reject the null hypothesis and conclude that $k_{r}$ is indeed also a variable parameter. These results could indicate that the jet in \object{3C\,454.3} does not always stay in equipartition or does not follow the classical Blandford \& K\"{o}nigl jet model. The last explanation also includes the possibility that the core is not strictly a $\tau= 1$ surface at our highest observing frequencies during the strong flares.

The variability plot of $k_r$ is displayed in the bottom panel of Figure~\ref{2-plots}. $k_r$ indices below 1 can occur either during flaring (2005 to 2006) or quiescent episodes (2007). These results indicate that the jet in \object{3C\,454.3} cannot be described with the ideal \citetalias{BlandfordKonigl} model with equipartition and conical jet assumptions even during the quiescent period.  On the other hand, $k_r$ values close to one within the uncertainties are found in a quiescent state (late 2006 and late 2007), post-flare (December 2008) and during flaring periods (late 2009 to early 2010). 

\subsection{Core position}

The core position or distance from the central engine, $r_\mathrm{core}$, at frequency $\nu$ is given by 
\begin{equation}
\label{eq:rcore}
    r_\mathrm{core}(\nu_\mathrm{GHz})=\frac{\Omega_{r\nu}}{\nu_\mathrm{GHz}^{1/k_r}\,\mathrm{sin\,\theta}} [\mathrm{pc}],
\end{equation}
where $\Omega_{r\nu}$ is the so-called core offset and $\theta$ is the source's viewing angle which is a fixed constant parameter here \citep{Lobanov1997}. $\Omega_{r\nu}$ is formulated as
%The uncertainty in \theta is omitted since it enlarges the r_core error by a factor of 3.
\begin{equation}
\label{eq:omega}
    \Omega_{r\nu}=4.85\cdot 10^{-9} \frac{\Delta r_\mathrm{\nu_1\nu_2}\,D_L}{(1+z)^2} \frac{\nu^{1/k_r}_{1}\nu^{1/k_r}_{2}}{\nu^{1/k_r}_{2}-\nu^{1/k_r}_{1}} [\mathrm{pc\cdot GHz^{1/k_r}}],
\end{equation}
where $D_{L}$ denotes the luminosity distance, $\Delta r_\mathrm{\nu_1\,\nu_2}$ is the core-shift measured in mas at two frequencies, and $z$ the redshift. 
%The values for $D_{L}$ and redshift are listed in Table~\ref{properties}.

To measure $r_\mathrm{core}$ at each frequency, we employed the mean value of $\Omega_{r\nu}$ per epoch given by ($\Omega_{r,CX}$+$\Omega_{r,XU}$+ $\Omega_{r,UK}$+ $\Omega_{r,KQ}$+$\Omega_{r,CQ}$+ $\Omega_{r,XQ}$+$\Omega_{r,UQ}$)/7. The uncertainties of $r_\mathrm{core}$ were obtained via error propagation of Equation~\ref{eq:omega}, and the details are presented in Appendix~\ref{appendix:error-prop}. We use $r_\mathrm{core}$ for a further analysis presented in Section~\ref{Sect:correlation}.

\subsection{Magnetic field parameters}

The magnetic field at 1\,pc from the jet apex, $B_\mathrm{1pc}$, can be estimated from the measured core shift as long as  the energy equipartition between the magnetic field and radiating particles at the radio core holds \citep{Lobanov1997,Hirotani2005}. The expression for $B_\mathrm{1pc}$ as used in several previous works \citep[e.g.][]{Hirotani2005,OSullivanGabuzda2009, Zamaninasab2014} misses one $(1+z)$ term as shown by \citet{Zdziarskietal2015}. We take into account this factor and use the corrected version,
\begin{equation} 
\label{eq:B1eq}
    B_\mathrm{1pc} \approx 0.025 \left [\frac{\sigma_\mathrm{rel}\,\Omega_{r\nu}^{3k_r}\,(1+z)^3} {\delta^2\, \theta_\mathrm{j}\, \mathrm{sin}^{3k_{r}-1}\theta}     \right ]^{\frac{1}{4}}[\mathrm{G}],
\end{equation}
where $\sigma_\mathrm{rel}$ is the ratio of magnetic and particle energy densities and taken as unity, $\delta$ is the Doppler factor, and $\theta_\mathrm{j}$ the jet's half opening angle. The physical properties of \object{3C\,454.3} relevant for the magnetic field parameter estimations are summarized in Table~\ref{properties}. We adopted from \citet{Jorstad2005} the average values for $\delta$, $\Gamma$ and $\theta$. We note that \citet{Jorstadetal2010, Jorstadetal2013} obtained $\delta$, $\Gamma$ and $\theta$ for several ejected components named 'K1', 'K2', 'K3', 'K09' and 'K10'. \citet{Jorstadetal2010} shows that 'K1', 'K2' and 'K3' were ejected in 2005.5, 2007.49, and 2007.93, respectively. Component K09 as described in section~3.3 was ejected in mid-2009 and K10 appeared to be ejected at the end of 2010 according to \citet{Jorstadetal2013}. Components 'K1', 'K2', and 'K09' have measured $\delta$, $\Gamma$ and $\theta$ values which are similar and in good agreement with the parameter values measured by \citet{Jorstad2005}. Since these appear to be relatively stable at least for our period of study, we kept them as constants. Component 'K3' does have a significantly higher Lorentz factor and a smaller viewing angle, and consequently significantly higher $\delta$. However, we consider 'K3' here as a likely outlier since it faded away very close to the core ($<0.2$\,mas) \citep{Jorstadetal2013} compared to the other components.

\begin{figure}[!t]
\centering
    \includegraphics[width=0.5\textwidth]{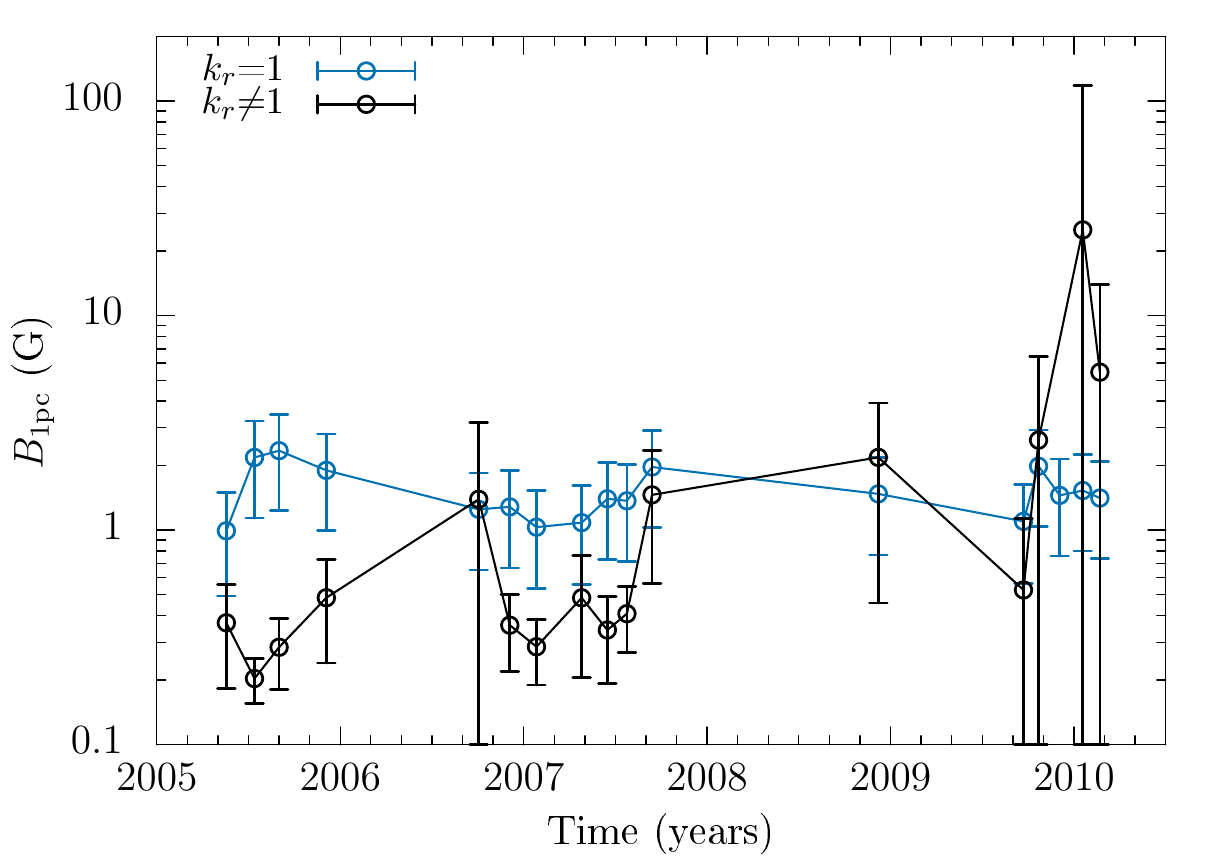}
     \caption{Magnetic field $B_\mathrm{1pc}$ at one parsec by employing the CQ core-shift pair %\eri{(see section FILL in text)}.
     We use here the (projected) core-shift absolute values. 
     The blue open circles represent the results assuming $k_r$=1. 
     The black open circles represent the results using measured $k_r$ values.}
    \label{B-field}
\end{figure}

To estimate $B_\mathrm{1pc}$, we employ the CQ core-shift per epoch together with the $k_r$ measurements. The uncertainties of $B_\mathrm{1pc}$ are evaluated by error-propagating equation~\ref{eq:B1eq} and taking into account the uncertainties in $\delta$, $\theta$ and $\theta_\mathrm{j}$ (see Appendix~\ref{appendix:error-prop}). 
%Uncertainties in $\delta$, $\theta$ and $\theta_\mathrm{j}$ are omitted, since these parameters are assumed to be constants across all the observations. This means that the absolute uncertainty in $B_\mathrm{1pc}$ is slightly larger than what is shown in Figure~\ref{B-field}. \ts{The contribution to the total uncertainty in $B_\mathrm{1pc}$ from $\delta$, $\theta$ and $\theta_\mathrm{j}$ ranges from 7\% for $k_r = 0.5$ to 30\% for $k_r = 1.7$.}
The $B_\mathrm{1pc}$ values as a function of time are shown in Figure~\ref{B-field}. The results show an evident discrepancy of $B_\mathrm{1pc}$; meaning that there is up to an order of magnitude of difference in $B_\mathrm{1pc}$ between $k_r=1$ and $k_r\neq1$. Furthermore, $B_\mathrm{1pc}$ varies more than two orders of magnitude when $k_r\neq 1$. These results suggest that reliable $B_\mathrm{1pc}$ estimations can be obtained only as long as $k_r=1$, which is understandable remembering the assumptions underlying Equation~\ref{eq:B1eq}, i.e., an ideal Blandford \& K{\"o}nigl conical jet and equipartition. If we cannot reasonably assume that these hold, then resulting $B_\mathrm{1pc}$ values are of suspect.

We assume $B_\mathrm{1pc}$ estimations to be more reliable when our measured $k_r$ indices are near one and also consistent within the error bars. From our results shown in Figure~\ref{B-field}, there are only four observations satisfying such a condition. Thus using $k_r=1$, the magnetic fields for the four epochs are displayed in Table~\ref{4epochs-Bfield}. In this table, the weighted mean $B_\mathrm{1pc}$ is $(1.5 \pm 0.4)$\,G which is comparable with the value of 1.1\,G obtained by \cite{Pushkarevetal2012}. We note that \cite{Hu2021} made recent estimations of the magnetic field in 3C\,454.3 based on the spectral fitting of six SEDs with a one-zone leptonic jet model concentrated around the time of the strong gamma-ray flares in 2009. For the two epochs they can constrain the distance of the emission region from the central engine, they obtain 0.7$-$1.0\,G at the distance of 0.3$-$0.4\,pc, i.e., their inferred magnetic field strength is smaller by a factor of about five compared to what one would expect based on our results, if the magnetic field strength scales as $r^{-1}$.
%We note that their magnetic field values agree with our values (in equipartition) only within the error bars. }

\begin{table}[t]
\centering
\caption{$B_\mathrm{1pc}$ values using $k_r=1$. The epochs selected have measured $k_r$ indices near one.}
\begin{tabular}{cc}
\hline\hline \noalign{\smallskip}
Epoch & $B_\mathrm{1pc}$(G) \\ \hline\hline \noalign{\smallskip}
 2006-10-03 & 1.3 $\pm$ 0.6  \\
 2007-09-13 & 2.0 $\pm$ 0.9 \\
 2008-12-07 & 1.5 $\pm$ 0.7 \\
 2009-10-22 & 2.0 $\pm$ 0.9  \\ \hline \noalign{\smallskip}
\end{tabular}
\label{4epochs-Bfield}
\end{table}

We further use the observations listed in Table ~\ref{4epochs-Bfield} to estimate the jet's magnetic flux in \object{3C\,454.3}. We examined whether the magnetic flux value is consistent with the source having developed a magnetically arrested disk (MAD). We adopted the formula for jet magnetic flux derived by \citet{Zdziarskietal2015}, who considered the observed condition of $\Gamma \theta_\mathrm{j} \sim 0.1$ and $\Gamma \theta \neq 1$ for blazars and radio galaxies, respectively, instead of $\Gamma \theta_\mathrm{j} = 1$ assumed in the original \citet{Zamaninasab2014} paper. The updated formula for the magnetic flux used by \citet{Chamani2021} is
\begin{equation}
\label{eq:mflux-jet}
     \Phi_\mathrm{jet} = 8 \times 10^{33} f(a)\, [1+\sigma]^{1/2} \left[\frac{M_\mathrm{BH}}{10^{9} \mathrm{M}_{\odot}}  \right ] \left [\frac{B_\mathrm{1pc}}{\mathrm{G}} \right ]\, \hspace{0.2cm} [\mathrm{G \,cm^{2}}],
\end{equation}
with $a$ the black hole (BH) spin and $f(a)=\frac{1+(1-a^2)^{1/2}}{a}$. Assuming that \object{3C\,454.3} hosts a fast spinning BH ($a = 1$), $f(a)=1$. Setting the jet magnetization parameter as $\sigma=(\Gamma\,\theta_\mathrm{_j})^2$, the  $\Phi_\mathrm{jet}$ values are readily obtained. We compare the measured $\Phi_\mathrm{jet}$ values with the predicted magnetic flux threading the BH, $\Phi_\mathrm{BH}$, at the magnetically arrested disk (MAD) state:
\begin{equation}
\label{eq:mflux-MAD}
 \Phi_\mathrm{BH}= 2.4\times10^{34} \left [\frac{\eta}{0.4}  \right ]^{-1/2} \left [ \frac{M_\mathrm{BH}}{10^9\,\mathrm{M}_{\odot}} \right ]\left [\frac{L_\mathrm{acc}}{1.26\times 10^{47}\mathrm{\,erg\,s}^{-1}}  \right ]^{1/2}\,
\end{equation}
given in units of G\,cm$^2$ by following e.g., \citet{Tchekhovskoy2011} and \citet{Zamaninasab2014}. The parameters $\eta$, $L_\mathrm{acc}$ indicate the radiative efficiency of the accretion disk and the accretion disk luminosity. 

\begin{figure}[t]
\centering
    \includegraphics[width=0.52\textwidth]{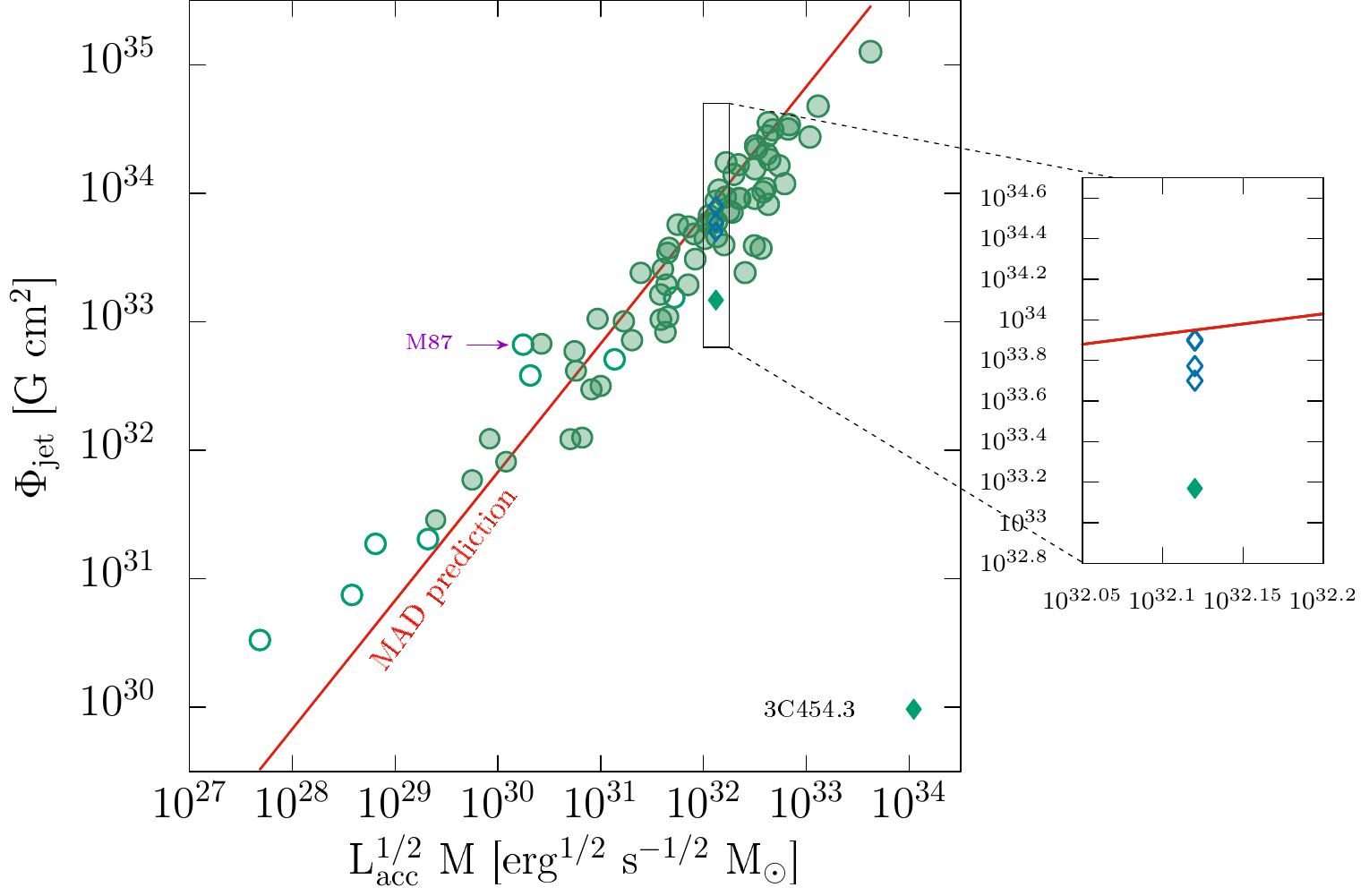}
    \caption{Measured $\Phi_\mathrm{jet}$ versus L$_\mathrm{acc}^{1/2} \mathrm{M}$ for radio galaxies (open circles) and blazars (filled circles) shown in green color. The plot is adapted from \citet{Zamaninasab2014} and it uses the corrections from \citet{Zdziarskietal2015} (see also \citet{Chamani2021}). The blazar \object{3C\,454.3} is marked with the green diamond. The plot on the right zooms in the values only for \object{3C\,454.3} for four epochs. The light-blue diamonds represent the $\Phi_\mathrm{jet}$ values using  $B_\mathrm{1pc}$ with measured CQ core-shifts and with $k_r=1$. If \object{3C\,454.3} harbours a maximally rotating ($a=1$) black hole, then the source appears to remain near the magnetically arrested disc state in the four observing epochs.}
    \label{MADplot}
\end{figure}

Figure~\ref{MADplot} displays the $\Phi_\mathrm{jet}$ values for \object{3C\,454.3} for the four epochs that have $k_r = 1$. The plot of the jet's magnetic flux as a function of the accretion luminosity and BH mass was adapted  from \citet{Zamaninasab2014} (who assumed $\Gamma\theta_\mathrm{j}=1$). We used the corrections given for radio galaxies ($\Gamma\theta_\mathrm{j} \neq 1$) and blazars ($\Gamma\theta_\mathrm{j}=0.13$) presented in \citet{Chamani2021}. Our result appears to be very close to the predicted MAD limit, even closer than the measurement given in \citet{Zamaninasab2014}.

\begin{figure}[]
\centering
    \includegraphics[width=0.46\textwidth]{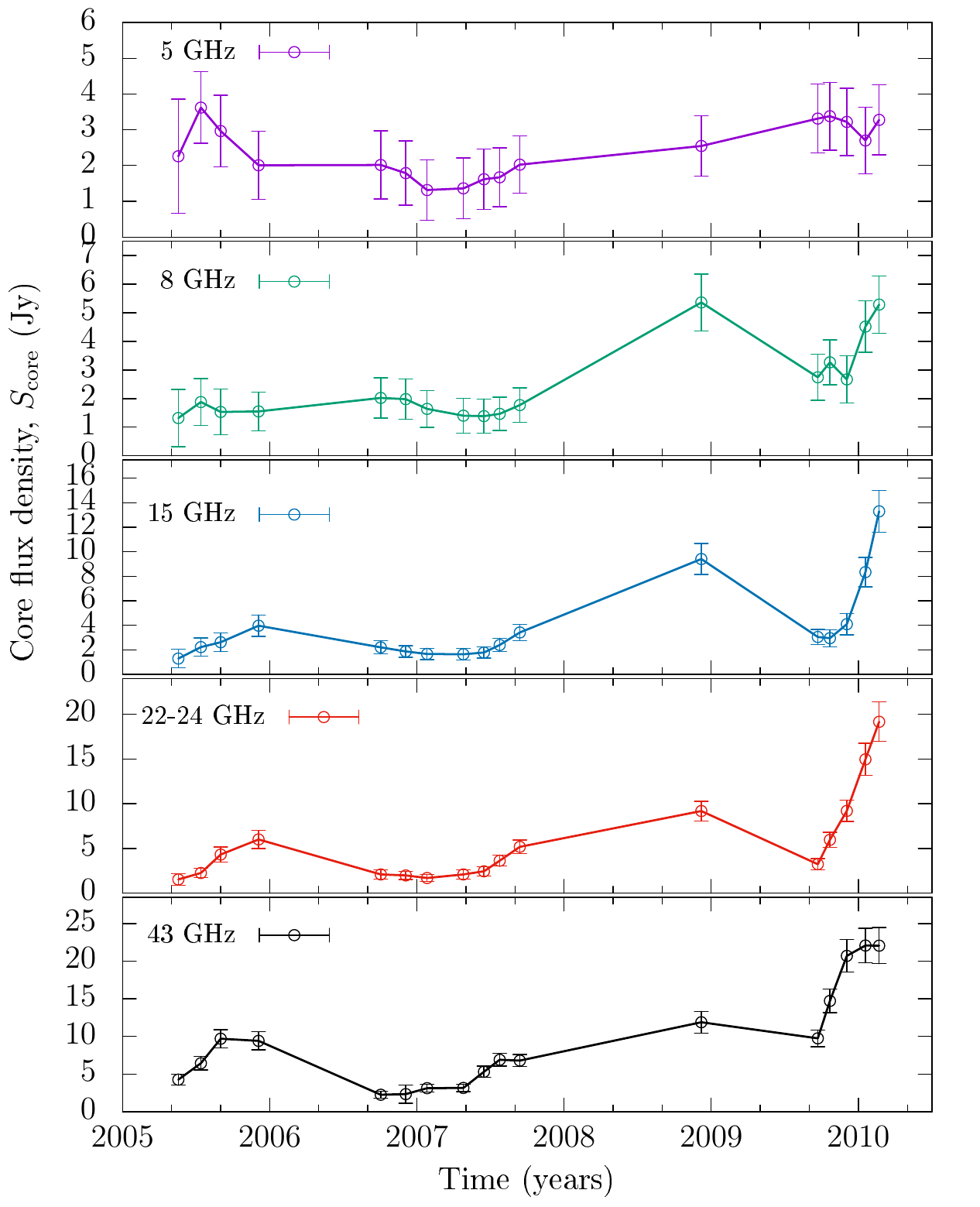}
    \caption{VLBA core flux density of \object{3C\,454.3} at 5, 8, 15, 22$-$24, and 43\,GHz estimated from Gaussian model fitting as described in section~3.2. The error bars include the uncertainty in the flux scale which is 10$\%$.}
    \label{corelightcurves}
\end{figure}

%------ core-shift versus core flux density figures below -----

\begin{figure}[t]
\centering
\includegraphics[width=0.5\textwidth]{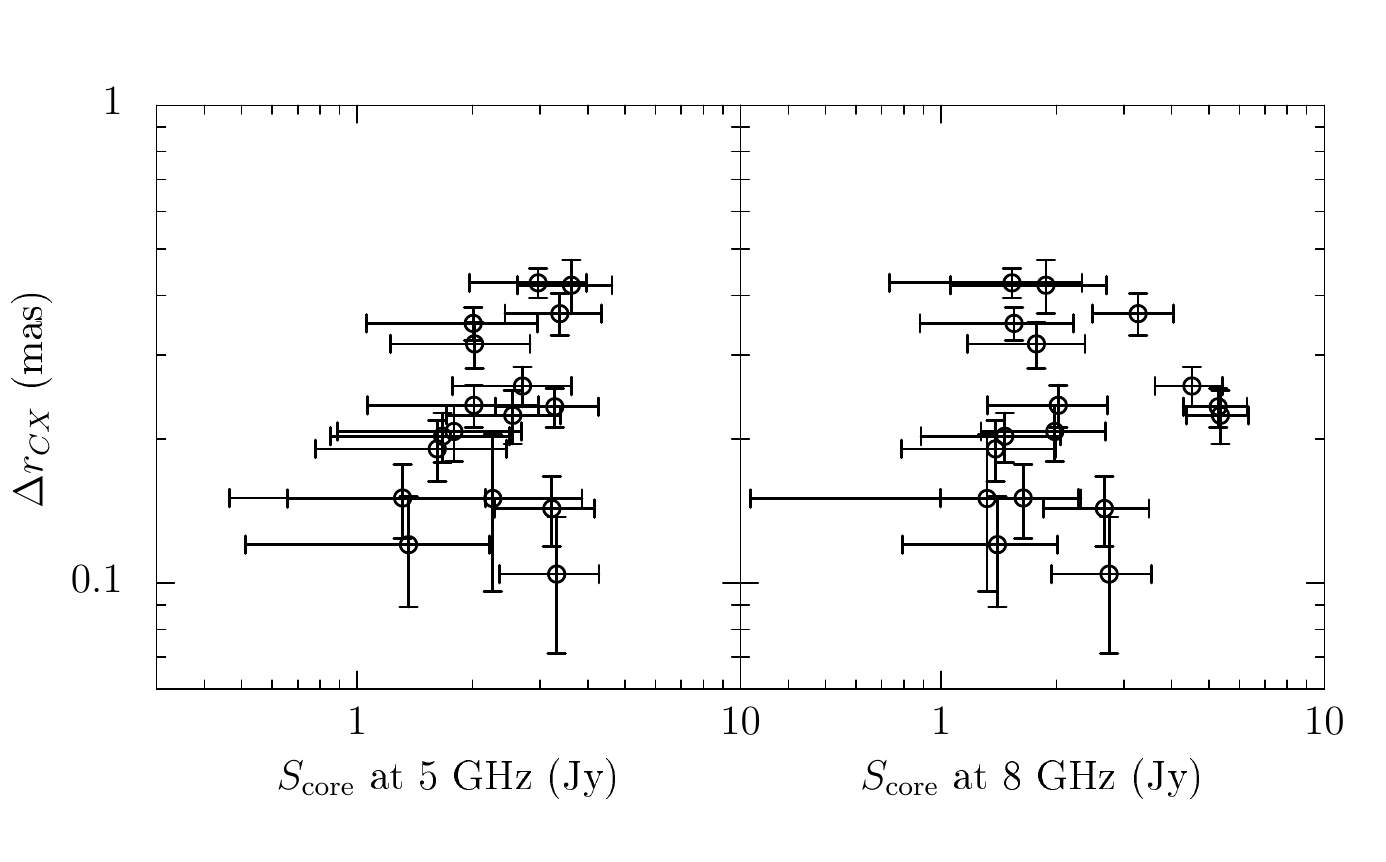}
    \caption{CX core-shifts versus core flux density at different frequencies. No correlations are found with the core flux at 5 and 8\,GHz}. 
    \label{CXcs-flux}
\end{figure}

\begin{figure}[t]
\centering
\includegraphics[width=0.5\textwidth]{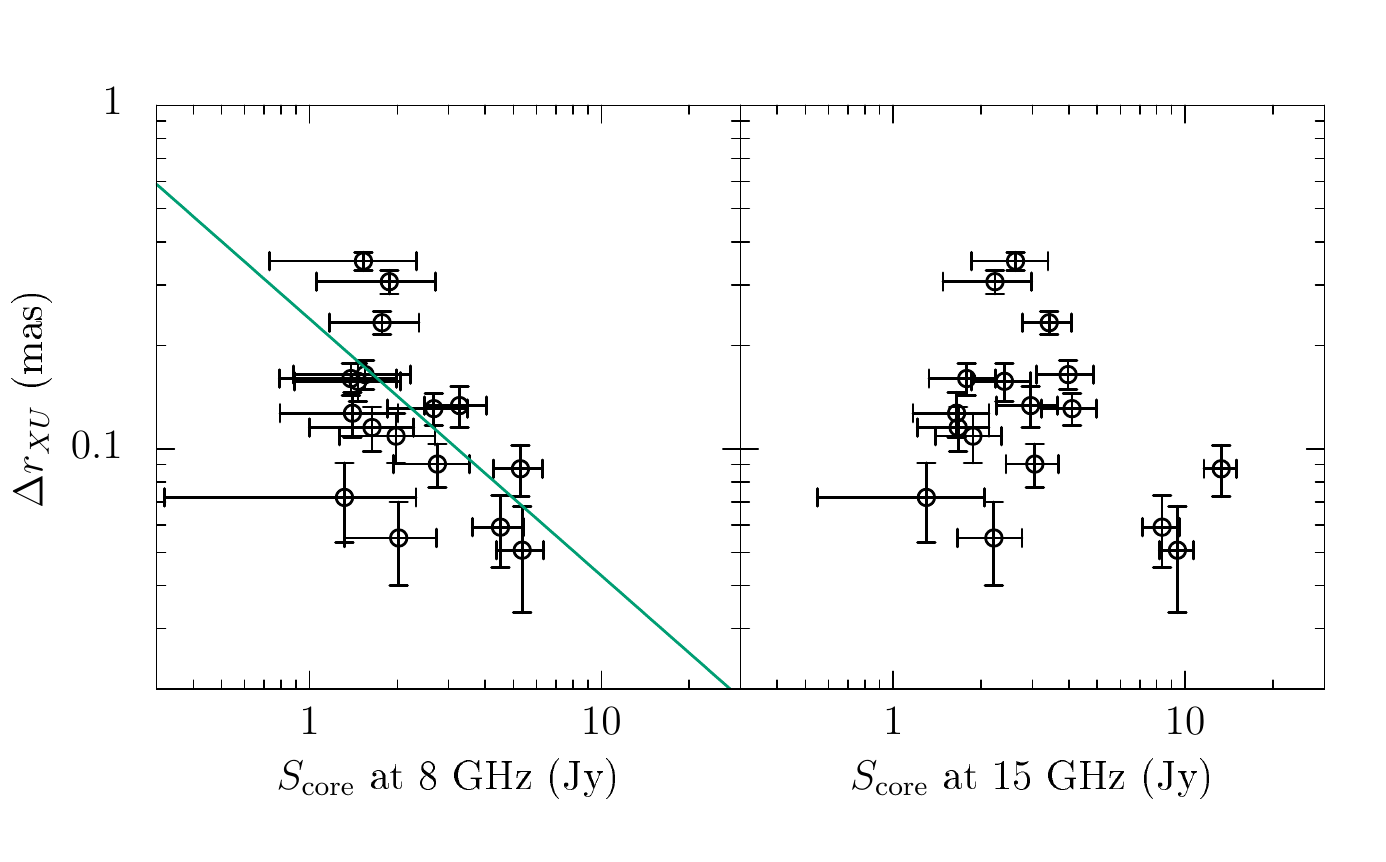}
    \caption{XU core-shifts versus core flux density at different frequencies. A negative correlation with the core flux at 8\,GHz is found but not at 15\,GHz. The green line indicates a power-law fit, see Table~\ref{correlations} for further details.}
    \label{XUcs-flux}
\end{figure}

\begin{figure}[t]
\centering
\includegraphics[width=0.5\textwidth]{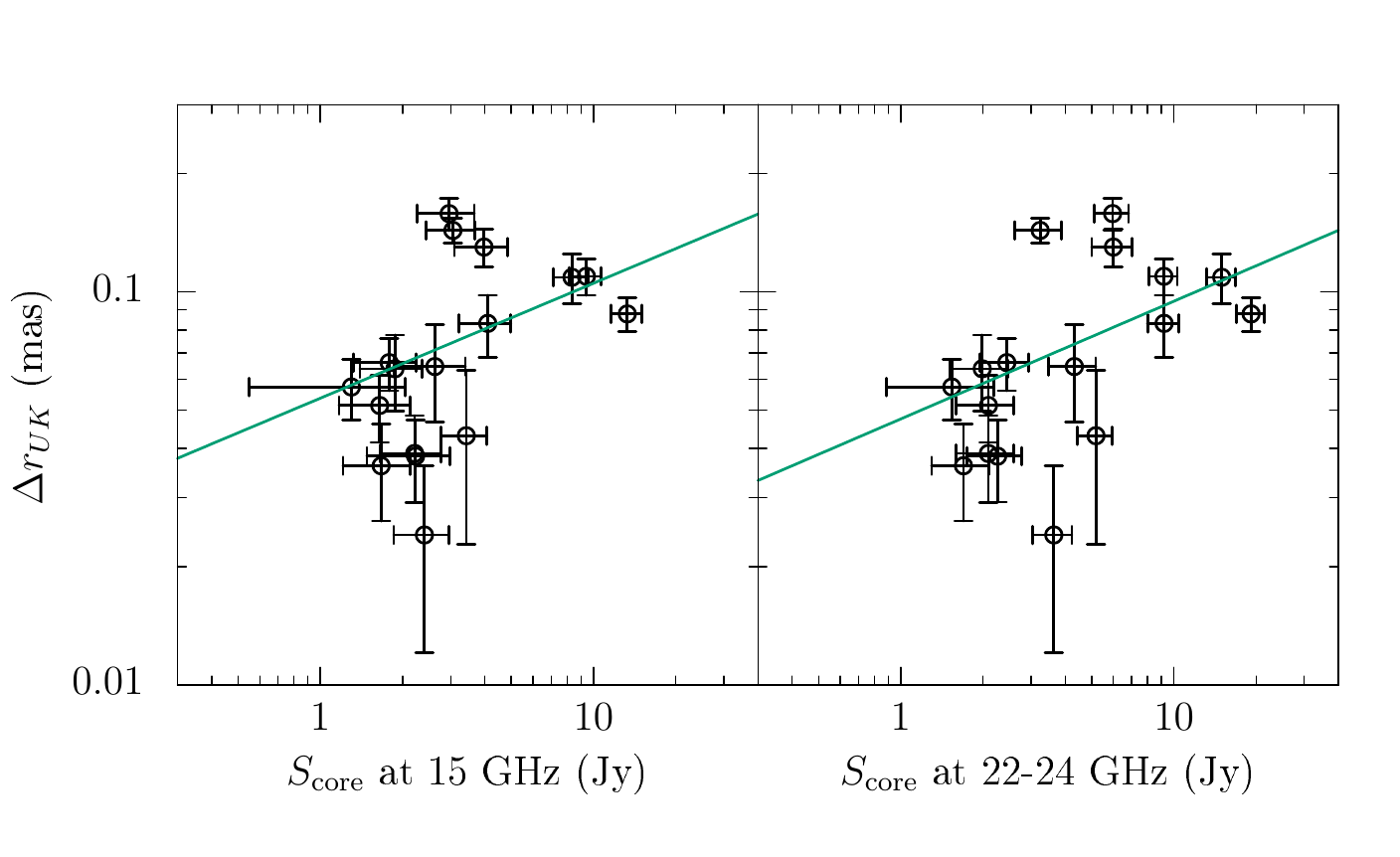}
    \caption{UK core-shifts versus core flux density at different frequencies. Positive correlations are found at 15 and 22$-$24\,GHz. The green line indicates a power-law fit, see Table~\ref{correlations} for further details.}
    \label{UKcs-flux}
\end{figure}

\begin{figure}[t]
\centering
\includegraphics[width=0.5\textwidth]{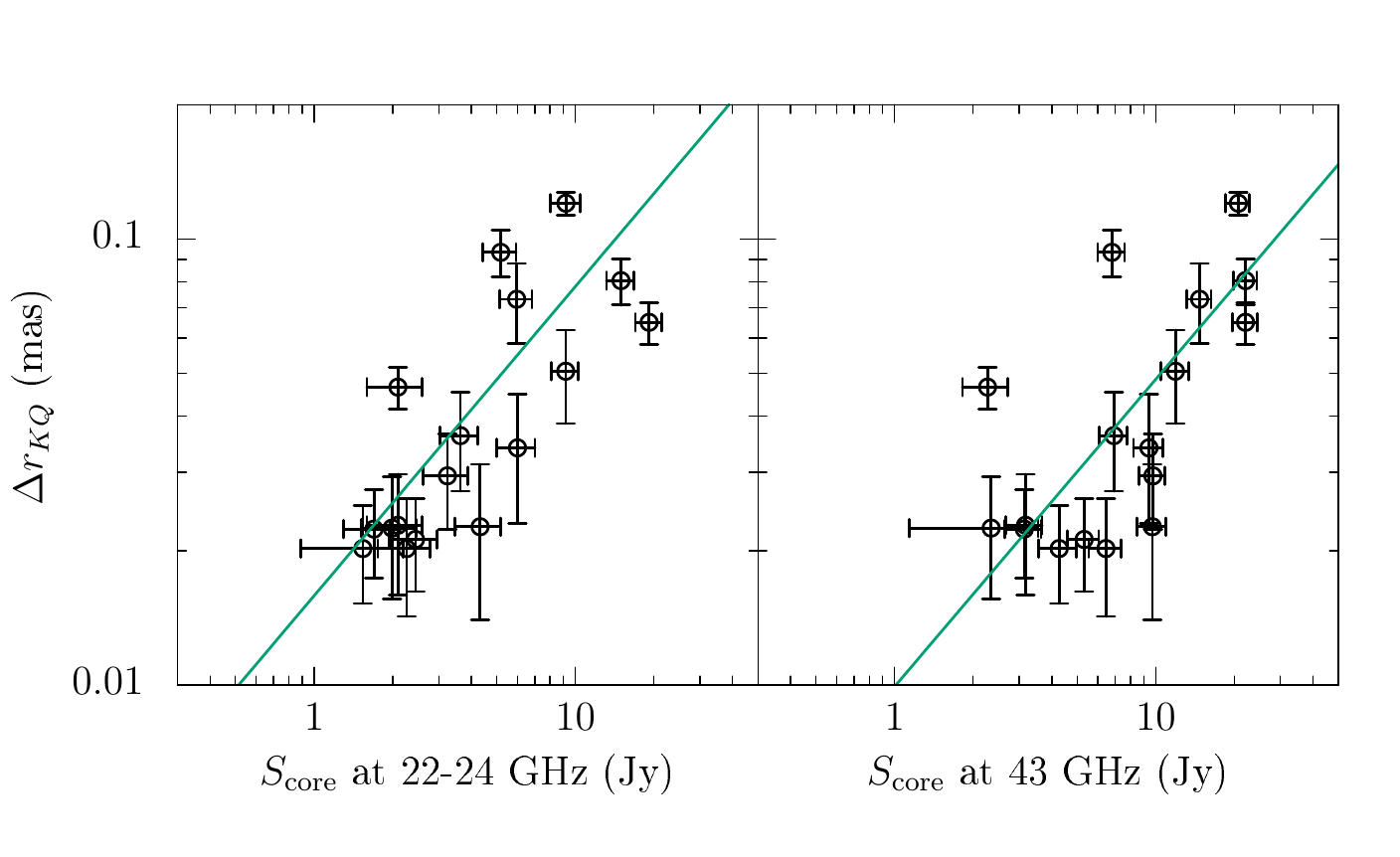}
    \caption{KQ core-shifts versus core flux density at different frequencies. Positive correlations are found at 22$-$24 and 43\,GHz. The green line indicates a power-law fit, see Table~\ref{correlations} for further details.}
    \label{KQcs-flux}
\end{figure}

\begin{figure}[t]
\centering
\includegraphics[width=0.5\textwidth]{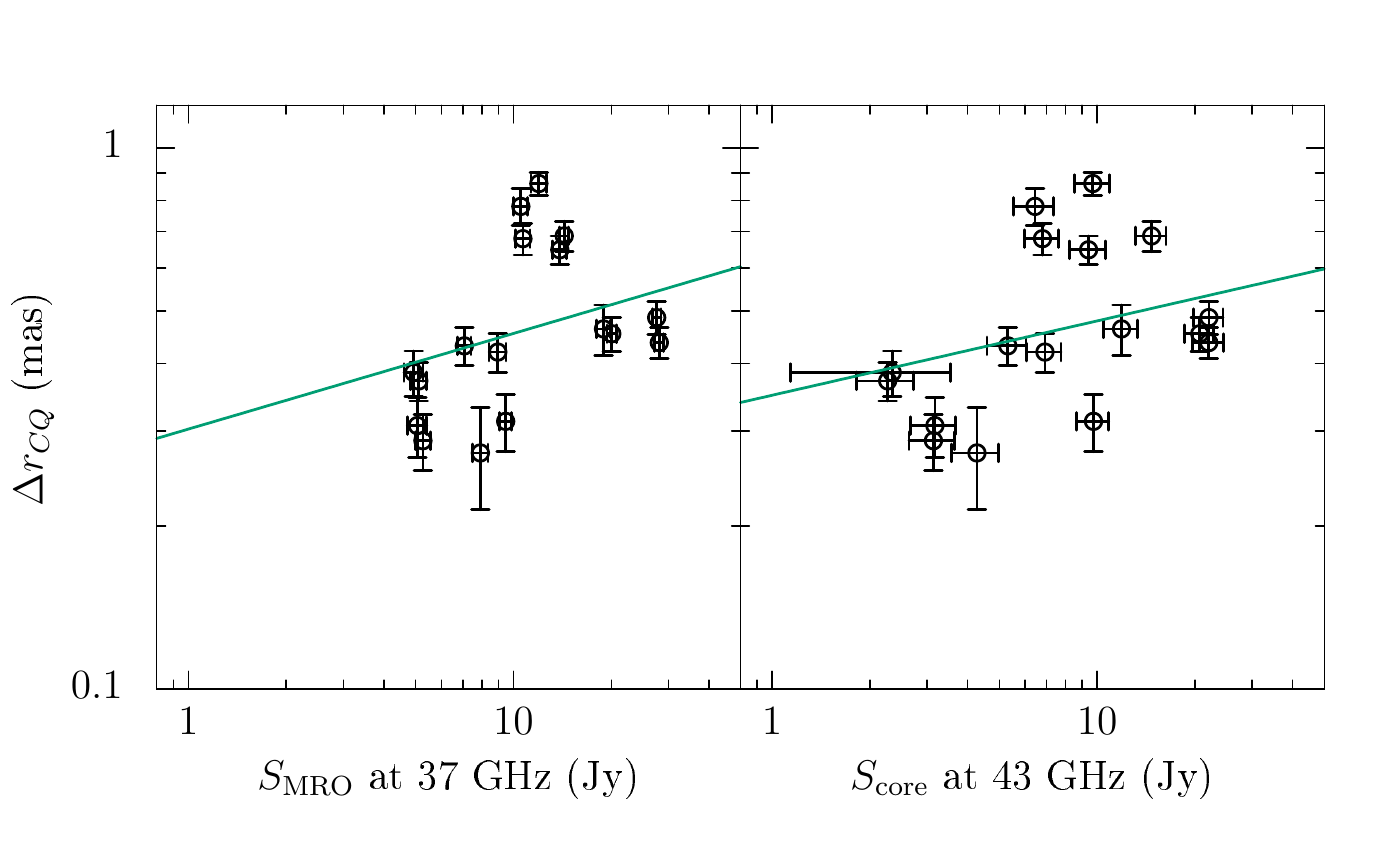}
    \caption{Full CQ core-shifts versus MRO 37\,GHz flux density and core flux density at 43\,GHz at different frequencies. Positive correlations are found at both frequencies. The green line indicates a power-law fit, see Table~\ref{correlations} for further details.}
    \label{CQcs-flux}
\end{figure}

%------––––––––––––– end  -----------------------------

%------ core-position versus core flux density figures below -----

\begin{figure*}[t]
\centering
\includegraphics[width=1\textwidth]{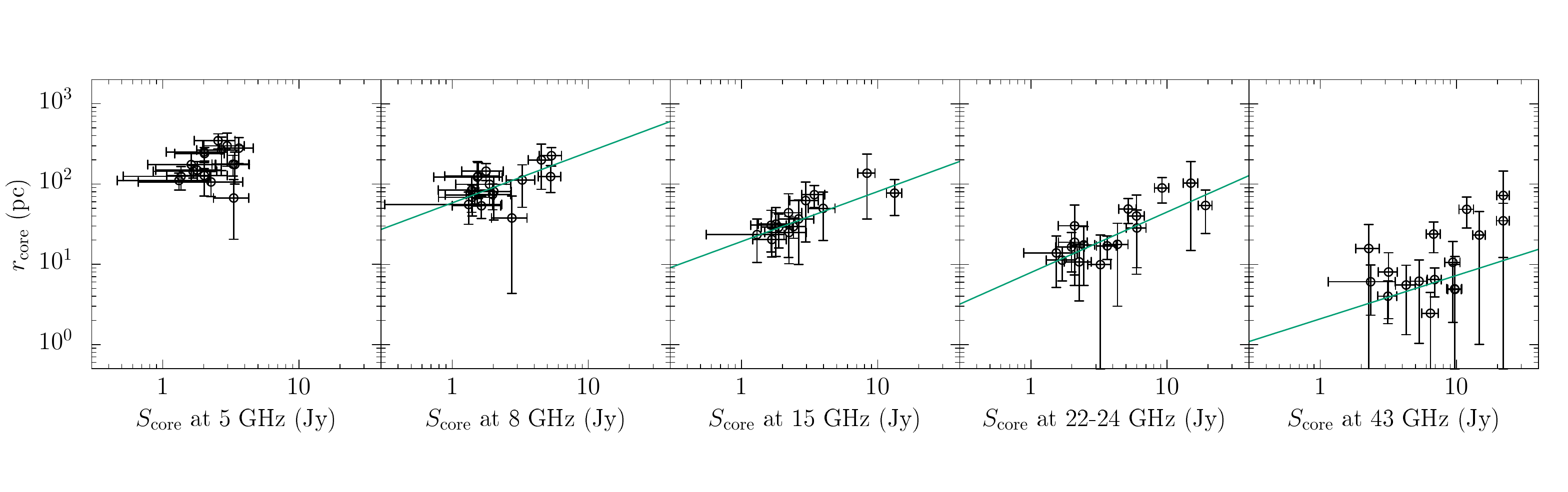}
    \caption{Core position versus core flux density at different frequencies. Positive correlations are found at 8, 15, 22$-$24, 43\,GHz. The power-law fit is shown in green color. Details of the fits are given in Table~\ref{correlations}.}
    \label{rcore-flux}
\end{figure*}

%---––––––––––––––––– end  -----------------------------

%------ k_r versus core flux density figures below -----

\begin{figure*}[t]
\centering
\includegraphics[width=0.7\textwidth]{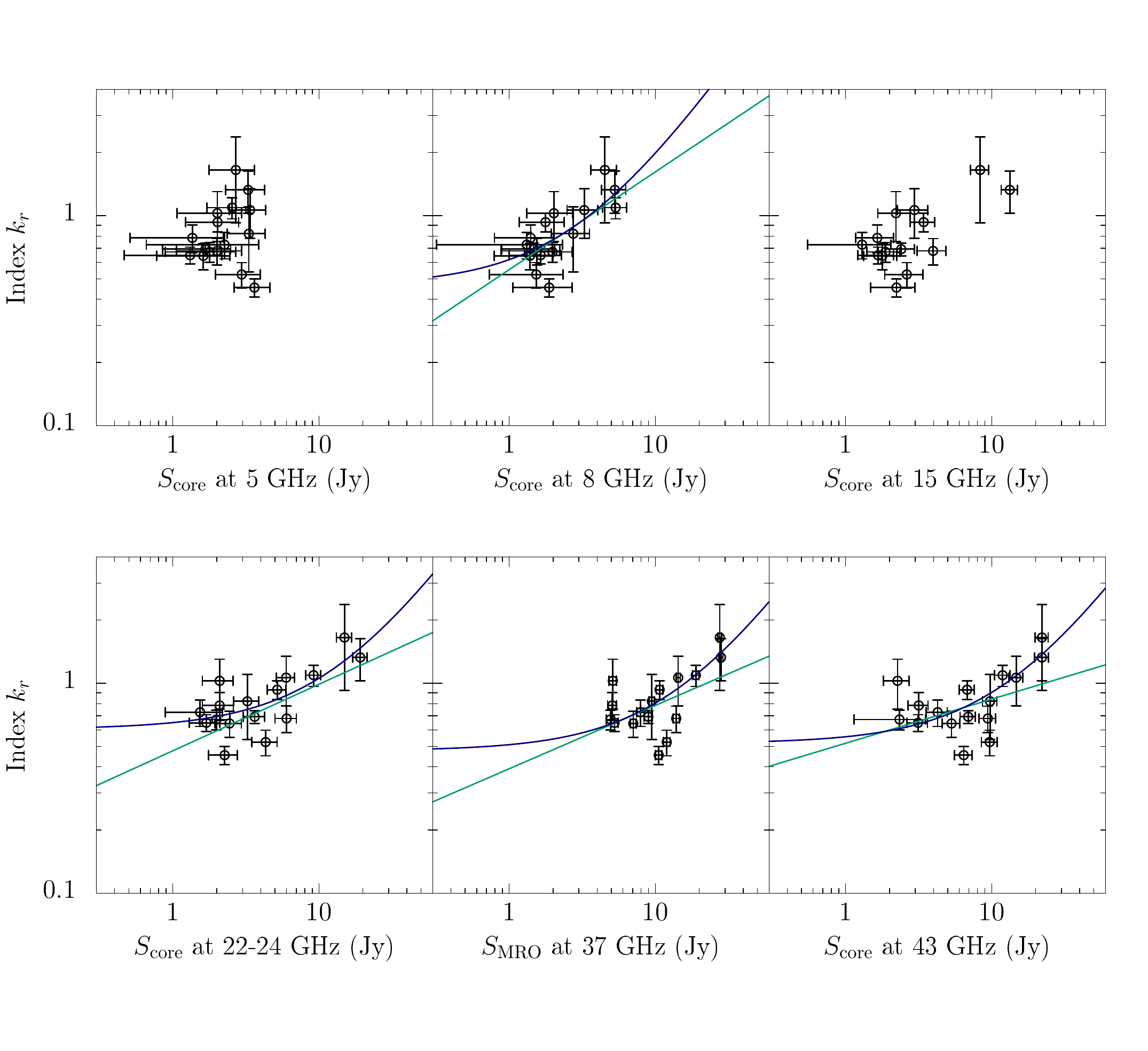}
 \caption{Core-shift index $k_r$ versus flux density at each frequency. No evident correlation is found at 5 and 15\,GHz.  Positive and moderate correlations were found at $8$, 22$-$24, $37$ and $43$\,GHz respectively. Power-law fits are shown with the green line and linear fits with a blue line. It appears that the $k_r$ index increases modestly when the core flux rises. All fitting parameters are displayed in Table~\ref{correlations}.}
    \label{kr-flux}
\end{figure*}

%---––––––––––––––––– end  -----------------------------

\subsection{Light curves and correlation analysis} \label{Sect:correlation}

The long term light curve of \object{3C\,454.3} at 37\,GHz obtained from the Mets\"ahovi Radio Observatory is displayed in Figure~\ref{2-plots}. In this period, three remarkable flux density peaks have been observed, and they are linked to major gamma-ray flare events as the one observed on 2009-12-02 \citep{Paccianietal2010}. 

The core flux density at each frequency as a function of time is shown in Figure~\ref{corelightcurves}. At 43\,GHz the core flux density from 2009 to 2010 corresponds to the sum of flux densities of the components 'Qa' and 'Qb'.
%\tscom{Explain that the 43\,GHz core flux density curve includes a sum of flux densities of 'fx' and 'fy' in 2009$-$2010.} 
Due to opacity effects, the 2009 flare starts at 43\,GHz and then progresses to lower frequencies; no flux density increase is seen at 5\,GHz yet. The first three VLBA observations in 2005 at 22 and 43\,GHz bands coincide well with the rising side of the flux at\,37 GHz, whereas our last observing epoch in 2005 coincides with the 'short-lived' plateau at 37\,GHz. All multi-frequency core flux density observations from mid-2006 to mid-2007 coincide with the low state at 37\,GHz. After that, the core flux density increased moderately until January 2008. The last observation in December 2008 coincides with the declining or post-flare phase at 37\,GHz. Finally, the last five observations at the 8, 15, 24, and 43\,GHz bands from September 2009 until February 2010 coincided with the sudden increase of flux density at 37\,GHz. The peak in the 43\,GHz core flux density matches well with the 37\,GHz peak in January 2010.

To understand the influence of flares on the core-shift effect, we explored the possible dependence of the core-shift, $\Delta r_{\nu_1\nu_2}$, the core position, $r_\mathrm{core}$ with the core flux density, $S_\mathrm{core}$ and with the total flux density ($S_\mathrm{MRO}$) at 37\,GHz. We calculated Spearman's rank correlation coefficient ($\rho$) and the $p$-value in $\tt R$. $p$-values below 0.05 are taken as possible correlations and those below 0.01 as correlations (all shown in bold face in Table~\ref{correlations}). 
We also include Pearson's correlation coefficient to evaluate the strength of the possible linear correlations. For the core-shift versus core flux density correlations, we used a generic fitting function of the form: $\Delta r_{\nu_1\nu_2} \propto S_\mathrm{core}^t$, where $t$ is the power-law index.

We present in Figures~\ref{CXcs-flux},\ref{XUcs-flux}, \ref{UKcs-flux}, \ref{KQcs-flux}  and \ref{CQcs-flux} scatter-plots of $\Delta r_{\nu_1\nu_2}$ -- $S_\mathrm{core}$ at each frequency. The scatter-plots are given for the following relationships: $\Delta r_\mathrm{CX}$ -- $S_\mathrm{core}$ at 5\,GHz and 8\,GHz, $\Delta r_\mathrm{XU}$ -- $S_\mathrm{core}$ at 8\,GHz and 15\,GHz, $\Delta r_\mathrm{UK}$ -- $S_\mathrm{core}$ at 15\,GHz and 22$-$24\,GHz and $\Delta r_\mathrm{KQ}$ -- $S_\mathrm{core}$ at 22$-$24\,GHz and 43\,GHz. With regard to $\Delta r_\mathrm{CX}$ -- $S_\mathrm{core}$ no correlations were found. A negative but weak correlation was obtained for $\Delta r_\mathrm{XU}$ -- $S_\mathrm{core}$ at 8\,GHz, but not at 15\,GHz. Positive and moderate correlations were found for $\Delta r_\mathrm{UK}$ -- $S_\mathrm{core}$ at 15\,GHz and 22$-$24\,GHz. A strong correlation was found at the higher frequencies for $\Delta r_\mathrm{KQ}$ -- $S_\mathrm{core}$ at 22$-$24\,GHz with a power-law index $t=0.7\pm 0.2$. A moderate correlation at 43\,GHz was also found, although the correlation is less robust than at 22$-$24\,GHz. Additionally, we searched for possible correlations of the full CQ core shift with the core flux at 43\,GHz and at 37\,GHz. 
The results suggest possible correlations, although there is not a clear monotonic relationship between the two variables as seen in Figure\ref{CQcs-flux}.

Figure~\ref{rcore-flux} displays the scatter-plots of the core positions as a function of the core flux at different frequencies, $r_\mathrm{core}$ -- $S_\mathrm{core}$. As is evident from the plots, there is no correlation between the variables at 5\,GHz, but there are moderate to strong positive correlations at all the other frequencies. Taking only the strongest correlations at 8, 15 and 22$-$24\,GHz, the fitted power-laws have indices from 0.6 to 0.8 with an average of 0.7. This indicates that the core position appears to follow approximately $r_\mathrm{core} \propto S^{2/3}$ as expected in the \citetalias{BlandfordKonigl} model if the flares are mainly due to increased particle density \citep{Lobanov1997, Kovalevetal2008}. We discuss this in Section~\ref{Sect:Discussion}.
 
We have also searched for correlations between the index $k_r$ and the core flux density, $k_r$ -- $S_\mathrm{core}$, at all the VLBA frequencies and at 37\,GHz as seen in Figure~\ref{kr-flux}. A significant correlation was found at 8\,GHz. Possible correlations were found at 22$-$24, 37 and 43\,GHz. In all the cases the fitted power-law indices are in the range of 0.2 to 0.5 (see Table~\ref{correlations}) and the Pearson correlation coefficients are high, therefore we also fitted the data with a linear function. This is an interesting new finding and it may be driven by the flaring behaviour as this is further discussed in Section~\ref{Sect:Discussion}. 

\begin{figure}[t]
\centering
\includegraphics[width=0.4\textwidth]{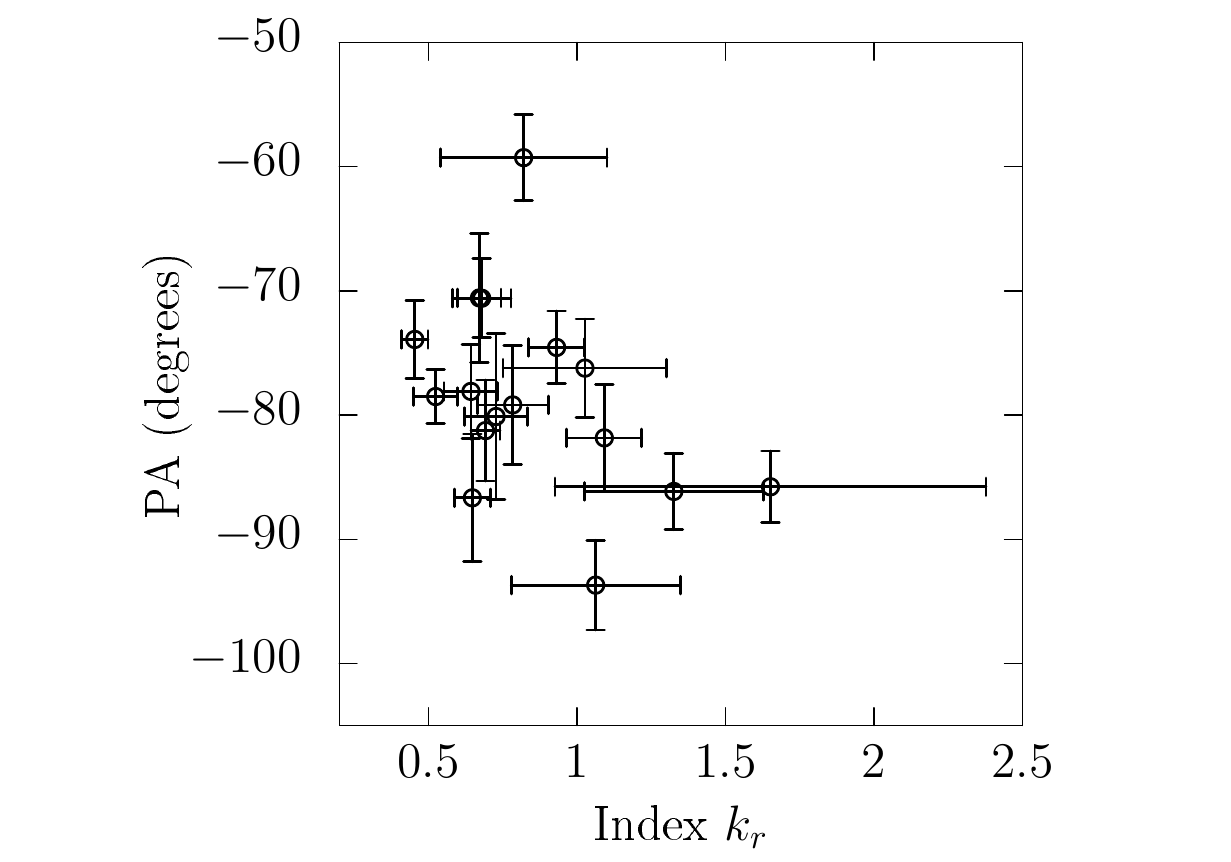}
   
    \caption{Jet position angle (PA) versus index $k_r$. The correlation between the variables is very weak, although the negative trend suggests that when $k_r>1$ the PA values decrease.}
    \label{PA-kr}
\end{figure}

In addition, correlations of index $k_r$ with jet PA have been also investigated. The scatter-plot is displayed in Figure~\ref{PA-kr}. Although the scatter plot appears to show hints of a negative trend for $k_r$ with the PA, there is no statistically significant correlation. Furthermore correlation studies of jet PA and flux density have also been considered, but we have not found evident correlations for PA -- $S_\mathrm{core}$, PA -- $S_\mathrm{MRO}$ at 37\,GHz and PA -- $\Delta r$ relationships. In general, the data appears scattered and disordered. A monotonic function cannot describe the relations. 

\begin{figure*}[t]
\centering
\includegraphics[width=0.7\textwidth]{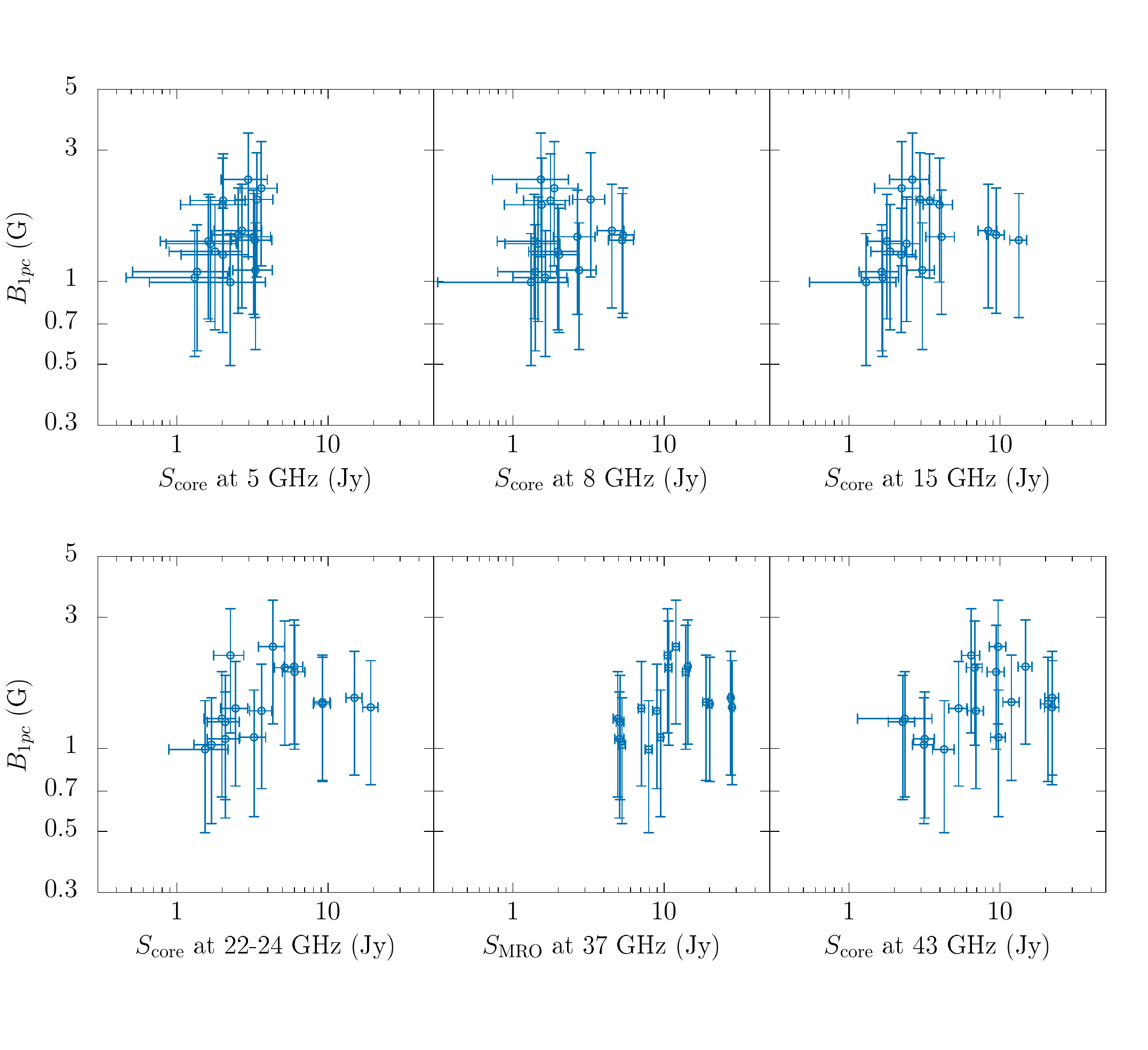}
    \caption{$B_\mathrm{1pc}$ as a function of the core flux density. The results are displayed assuming $k_r=1$ and using CQ core shifts.}
    \label{kr1B1CQ-flux}
\end{figure*}

Finally, for the sake of curiosity, we searched for possible correlations of $B_\mathrm{1pc}$ with the core flux density and MRO 37\,GHz flux density assuming entirely $k_r=1$. The results show positive Spearman correlation coefficients, although the data do not follow strong monotonic relationships, as seen in the scatter plots of Figure \ref{kr1B1CQ-flux}. It is evident from these plots that the $B_\mathrm{1pc}$ values are relatively stable at all frequencies. Since we used  $k_r=1$, and the same $\delta$, $\theta$ and $\theta_j$ at all the epochs, $B_\mathrm{1pc}$  depends only on $\Omega_{r\nu}$ which depends on $r_\mathrm{core}$. Hence, the $B_\mathrm{1pc} - S_\mathrm{core}$ correlation is a direct consequence of $r_\mathrm{core} - S_\mathrm{core}$ correlation.

\begin{table*}[t]
\centering
\begin{threeparttable}
\caption{Spearman's rank and Pearson correlation coefficients. Power-law fits are given only for the relations with $p$-values\,$< 0.05$. Additional linear fits are provided for the $k_r-S_\mathrm{core}$ relationship.}
\label{correlations}
%The tests are applied only to the relations resembling a monotonic association between the variables.}
\begin{tabular}{@{}lcccc@{}}
\hline
\multirow{2}{*}{Relations} & \multicolumn{2}{c}{Spearman} & Pearson & Fitting function \\ \cline{2-5} 
 & \begin{tabular}[c]{@{}c@{}}Correlation\\ coefficient\\ ($\rho$)\end{tabular} & p-value & \begin{tabular}[c]{@{}c@{}}Correlation\\ coefficient\\ ($r$)\end{tabular} & \begin{tabular}[c]{@{}c@{}} Power-law \end{tabular} \\ \hline\noalign{\smallskip}
$\Delta r_{CX}$ -- $S_\mathrm{core}$ at 5 GHz & 0.37 & 0.14 & & \\ \noalign{\smallskip}
$\Delta r_{CX}$ -- $S_\mathrm{core}$ at 8 GHz & 0.19 & 0.46   & & \\ \noalign{\smallskip}
$\Delta r_{XU}$ -- $S_\mathrm{core}$ at 8 GHz & -0.48 & \textbf{0.05}  & -0.46 & 
(0.24 $\pm$ 0.06)\,$S_\mathrm{core}^{-0.7 \pm 0.3}$\\\noalign{\smallskip}
$\Delta r_{XU}$ -- $S_\mathrm{core}$ at 15 GHz & -0.12 & 0.66   & & \\ \noalign{\smallskip}
$\Delta r_{UK}$ -- $S_\mathrm{core}$ at 15 GHz & 0.59 & \textbf{0.01} & 0.37 & (0.05 $\pm$ 0.01)\,$S_\mathrm{core}^{0.3 \pm 0.2}$\\\noalign{\smallskip}
$\Delta r_{UK}$ -- $S_\mathrm{core}$ at 22$-$24 GHz & 0.60 & \textbf{0.01} & 0.40 & (0.05 $\pm$ 0.01)\,$S_\mathrm{core}^{0.3 \pm 0.2}$\\\noalign{\smallskip}
$\Delta r_{KQ}$ -- $S_\mathrm{core}$ at 22$-$24 GHz & 0.77 & \textbf{0.0003}  & 0.60 & (0.02 $\pm$ 0.01)\,$S_\mathrm{core}^{0.7 \pm 0.2}$  \\\noalign{\smallskip}
$\Delta r_{KQ}$ -- $S_\mathrm{core}$ at 43 GHz  & 0.64 & \textbf{0.006}  & 0.71 & (0.010 $\pm$ 0.004)\,$S_\mathrm{core}^{0.7 \pm 0.2}$  \\\noalign{\smallskip}
$\Delta r_{CQ}$ -- $S_\mathrm{MRO}$ at 37 GHz & 0.63 & \textbf{0.008} & 0.22  & (0.3 $\pm$ 0.1)\,$S_\mathrm{core}^{0.2 \pm 0.2}$  \\\noalign{\smallskip} 
$\Delta r_{CQ}$ -- $S_\mathrm{core}$ at 43 GHz & 0.52 & \textbf{0.03} & 0.23 & (0.35 $\pm$ 0.09)\,$S_\mathrm{core}^{0.1 \pm 0.1}$  \\ 
\noalign{\smallskip}\hline \noalign{\smallskip}
$r_\mathrm{core}$ -- $S_\mathrm{core}$ at 5 GHz & 0.40 & 0.12  & &  \\\noalign{\smallskip}  
$r_\mathrm{core}$ -- $S_\mathrm{core}$ at 8 GHz & 0.61  & \textbf{0.02}\tnote{*} & 0.75 & (58.3 $\pm$ 9.3)\,$S_{core}^{0.6 \pm 0.2}$ \\\noalign{\smallskip}  
$r_\mathrm{core}$ -- $S_\mathrm{core}$ at 15 GHz& 0.84 & \textbf{0.0003} & 0.71 & (19.3 $\pm$ 2.8)\,$S_\mathrm{core}^{0.6 \pm 0.1}$\\ \noalign{\smallskip} 
$r_\mathrm{core}$ -- $S_\mathrm{core}$ at 22$-$24 GHz& 0.74 & \textbf{0.001} & 0.77 & (7.9 $\pm$ 1.7)\,$S_\mathrm{core}^{0.8 \pm 0.2}$\\ \noalign{\smallskip} 
$r_\mathrm{core}$ -- $S_\mathrm{core}$ at 43 GHz & 0.49 & \textbf{0.05}  & 0.77 & (2.1 $\pm$ 2.0)\,$S_\mathrm{core}^{0.5 \pm 0.5}$\\\noalign{\smallskip}\hline \noalign{\smallskip}
$k_r$ -- $S_\mathrm{core}$ at 5 GHz & 0.28  & 0.30  &  &    \\
$k_r$ -- $S_\mathrm{core}$ at 8 GHz & 0.67  & \textbf{0.006} & 0.80 & (0.55 $\pm$ 0.05)\,$S_\mathrm{core}^{0.5 \pm 0.1}$ \\\noalign{\smallskip}
 &   &  &  &   (0.5 $\pm$ 0.1)+(0.15 $\pm$ 0.03)\,$S_\mathrm{core}$ \\\noalign{\smallskip}
$k_r$ -- $S_\mathrm{core}$ at 15 GHz & 0.44 & 0.12 & 0.74 & \\
$k_r$ -- $S_\mathrm{core}$ at 22$-$24 GHz & 0.56 & \textbf{0.03} & 0.81 & (0.48 $\pm$ 0.06)\,$S_\mathrm{core}^{0.3 \pm 0.1}$  \\ \noalign{\smallskip}
 &   &  &  &   (0.6 $\pm$ 0.1)+(0.04 $\pm$ 0.01)\,$S_\mathrm{core}$ \\\noalign{\smallskip}
$k_r$ -- $S_\mathrm{MRO}$ at 37 GHz & 0.50 & \textbf{0.05} & 0.77 &  (0.39 $\pm$ 0.08)\,$S_\mathrm{core}^{0.30\pm 0.09}$ \tnote{$\dagger$} \\\noalign{\smallskip}
 &   &  &  &   (0.5 $\pm$ 0.1)+(0.03 $\pm$ 0.01)\,$S_\mathrm{core}$ \\
 $k_r$ -- $S_\mathrm{core}$ at 43 GHz & 0.52 & \textbf{0.04}  & 0.76  & (0.52 $\pm$ 0.09)\,$S_\mathrm{core}^{0.21 \pm 0.09}$ \tnote{$\dagger$}\\ \noalign{\smallskip}
 &   &  &  &   (0.5 $\pm$ 0.1)+(0.04 $\pm$ 0.01)\,$S_\mathrm{core}$ \\
\noalign{\smallskip}\hline \noalign{\smallskip}
PA -- $k_r$  & -0.42  & 0.11 & &  \\\noalign{\smallskip}\hline \noalign{\smallskip}
\multicolumn{5}{c}{Correlations with $B_\mathrm{1pc}$ values assuming $k_r$=1} \\  \noalign{\smallskip}\hline\noalign{\smallskip}
$B_\mathrm{CQ,1pc}$ -- $S_\mathrm{core}$ at 5 GHz & 0.54 & \textbf{0.03}  & 0.47 &   \\
$B_\mathrm{CQ,1pc}$ -- $S_\mathrm{core}$ at 8 GHz & 0.29 & 0.26 &  &\\
$B_\mathrm{CQ,1pc}$-- $S_\mathrm{core}$ at 15 GHz & 0.56 & \textbf{0.02} & 0.05 & \\
$B_\mathrm{CQ,1pc}$ -- $S_\mathrm{core}$ at 22$-$24 GHz & 0.61 & \textbf{0.01} &  0.12 &   \\
$B_\mathrm{CQ,1pc}$ -- $S_\mathrm{core}$ at 43 GHz & 0.52 & \textbf{0.03} & 0.23 &\\
$B_\mathrm{CQ,1pc}$ -- $S_\mathrm{MRO}$ at 37 GHz & 0.63 & \textbf{0.008}  & 0.24 &  \\\noalign{\smallskip}\hline\noalign{\smallskip}
\end{tabular}
\begin{tablenotes}
\item[*] Excluding the data point of 37\,pc.
\item[$\dagger$] The fits do not include the data at $k_r=0.45$ and $k_r=0.52$. These are taken as outliers in the scatter-plots. By including these points the power-law fits become much less steep with indices: 0.05 $\pm$ 0.18 at 37\,GHz and 0.04 $\pm$ 0.14 at 43\,GHz.

%\tscom{This does not sound well-motivated. What happens if you include these points?}
% \item[**] \textbf{As seen from Figure~\ref{kr1B1CQ-flux} the connection between the variables is not strictly monotonic, hence, we do not fit any function to the data.} \tscom{This does not make sense. Very few things in observational astronomy come out as strictly monotonic. I would recommend to drop this note.}
\end{tablenotes}
\end{threeparttable}
\end{table*}
\section{Discussion}

\subsection{Core-shift and flares}
\label{Sect:Discussion}

In our search for correlations between core-shift and core flux density, we found that $r_\mathrm{core}$ depends approximately on $S_\mathrm{core}^{0.7}$. This relationship matches well the theoretical prediction by \citetalias{BlandfordKonigl} that the observed radius '$r_\mathrm{max,ob}$', where the maximum brightness temperature is achieved, scales with the synchrothron luminosity, $L_\mathrm{s}$ as $r_\mathrm{max,ob} \propto L_\mathrm{s}^{2/3} \propto L_\mathrm{s}^{0.7}$ and consequently has the same dependence with flux density. 

The connection between the core position and core flux density during jet flares has been previously investigated by \citetalias{Plavin2019}. They use a phenomenological model of the form $r_\mathrm{core}=b S_\mathrm{core}^{K_\mathrm{rS}}$ and independent of any physical assumptions model the core-shift variability at 2.3/8.4\,GHz in a sample of 40 AGN. Using a general time-dependent model, they determine $K_\mathrm{rS} = 0.28 \pm 0.05$ for the whole sample. Our results show a mean $K_\mathrm{rS} = 0.7$ for $r_\mathrm{core} \propto S_\mathrm{core}^{K_\mathrm{rS}}$, which differs significantly from the results in \citetalias{Plavin2019}. On the other hand, our $\Delta r$ shows a weaker dependence on $S_\mathrm{core}$ with $K_\mathrm{rS} = 0.3 \pm 0.2$ at 15 and 22$-$24\,GHz and even a negative $K_\mathrm{rS}$ at 8\,GHz (see Table~\ref{correlations}). This discrepancy may be related to i) varying $k_r$ in our case, ii) lower observing frequencies used in \citetalias{Plavin2019}, or iii) 3C\,454.3 possibly having a behaviour that differs significantly from a typical AGN.

As pointed out by \citetalias[][with references therein]{Plavin2019}, the physical mechanisms producing variations of core flux density can be associated with variations in particle density, magnetic field strength or bulk Doppler factor as $S_\mathrm{core}\propto N_\mathrm{core}\,B_\mathrm{core}^{1-\alpha}\,\delta^{3-\alpha}$, where $\alpha$ is the optically thin spectral index. Thus, flares, and consequently, core position variations can be due to variations in any of these parameters. \citetalias{Plavin2019} list four basic scenarios for the flares: i) Flaring emission is in equipartition, i.e., $N_\mathrm{core} \propto B_\mathrm{core}^2$, while $\delta$ is constant; ii) Flares are mainly due to particle density variability; iii) Flares are mainly due to variations in magnetic field strength; and iv) Flares are due to variations in the jet Doppler factor. The fact that $k_r \neq 1$ in 3C\,454.3 for most of the observing epochs -- also outside of the flaring periods -- indicates that the source is unlikely to be in equipartition, making scenario (i) also unlikely. Scenario (ii) is perhaps the simplest case matching our observed relationship $r_\mathrm{core} \propto S_\mathrm{core}^{0.7}$, since it gives $r_\mathrm{core} \propto N_\mathrm{core}^{2/3} \propto S_\mathrm{core}^{2/3}$. Scenario (iii) cannot be ruled out. Scenario (iv) appears unlikely, since considering that in blazars in general \citep{Savolainen2002}, and in 3C\,454.3 in particular \citep{Jorstadetal2010,Jorstadetal2013}, the flares are connected to travelling disturbances in the VLBI jet, it is likely that changes in the jet Doppler factor alone do not produce them. Furthermore, $r_\mathrm{core} \propto \Gamma^{-4/3} \beta^{-2/3} \delta^{2/3} \sin^{-1/3}\theta\, S_\mathrm{core}^{2/3}$ in \citetalias{BlandfordKonigl}, i.e., one would expect $K_\mathrm{rS}$ to differ from $2/3$ in scenario (iv). In a general case both $N_\mathrm{core}$ and $B_\mathrm{core}$ likely vary simultaneously, but results from \citetalias{Plavin2019} and for example from \citet{Lobanov1999} suggest that variations in $N_\mathrm{core}$ dominate the flaring behaviour.

\subsection{Frequency dependency of the core-shift}

We found that for most of the time the frequency dependency of the core position does not follow $\nu_\mathrm{obs}^{-1}$ in \object{3C\,454.3}. Instead, the index $k_r$ was found to be a variable parameter in the period of our study. We have here demonstrated the variability of $k_r$ by direct measurements using multi-epoch, quasi-simultaneous multi-band VLBI data for the first time. Frequency dependency with $k_r < 1$ values were found both during flaring and quiescent states. This would mean that either energy equipartition or conical shape assumptions do not always hold or the jet cannot be described by the \citetalias{BlandfordKonigl} model. For a \citetalias{BlandfordKonigl} jet, one might assume that $k_r$ should be close to one and stable in quiescent states. During flaring episodes the parsec-scale jet would 'temporarily' become particle dominated and deviation from equipartition would be natural, implying a consequent deviation from $k_r=1$. This, however, does not seem to be the case in \object{3C\,454.3}. Instead, it appears that $k_r$ is \textit{typically} below one in this source. 

We note that, in general, core-shift and core position measurements can be positively biased due to the use of a Gaussian template, which is only a crude model for the actual jet brightness distribution, and consequently bias the $\nu^{-1/k_r}$ dependency when $k_r\neq1$ as shown by 
\cite{Pashchenko2020}. Their results show that this bias leads to observed $k_r$ values that are closer to one than the true $k_r$. Hence, the deviation of $k_r$ from one in \object{3C\,454.3} can be even larger than what is observed here.
%According to PS20, a core-shift bias could potentially result from geometrical estimates related to fitting the core with circular or elliptical Gaussian s. Their results show that sources with the smallest viewing angles have the most biased core shifts. Furthermore, in the present work, we study the core-shift variability on \object{3C\,454.3} and not independent specific core-shift values. The study of core-shift biasing in our data-set goes beyond the scope of this work which could be presented in future work.
% \textit{Wara: If the core-shift bias is aimed to be analysed for our data set, this must be done carefully and independently for each epoch by comparing the results with pure elliptical s to the core and using the formulas of Paschenko et al. to correct the bias. So this is very interesting but this was not the initial aim of this work. Given the time and pressure I am in, I leave out this study for the future. }

We stress that the large uncertainties of $k_r$ in 2009 to 2010 were produced by the strong flares that disrupted the core-shift effect. The ejection of a new moving feature at 43\,GHz led to ambivalence in measuring the $k_r$ parameter, judging by the results demonstrated in Figures from \ref{CSepoch15} to \ref{CSepoch19}. In these strong flaring epochs, the core-shift data are not always compatible with a single power-law dependence anymore, which supports the findings of \citet{Kutkinetal2019} and \citetalias{Plavin2019}.

We found a significant correlation of $k_r$ with the core flux density, $S_\mathrm{core}$ as shown in Figure~\ref{kr-flux}. It is seen that the index $k_r$ tends to become larger when the flux density is higher. This effect appears to be more steep and pronounced at 8\,GHz ($k_r-S_\mathrm{core}^{0.5}$) and less pronounced from 22$-$24 to 43\,GHz where the power-law indices range from 0.2 to 0.3.
This would mean that dependency of the core position on frequency ($\nu^{-1/k_r}$) is less  steep when flux density is higher. This could be explained by the fact that flares start at high frequencies and they affect the core-shift first at the high-frequency end of the spectrum and later at the low frequencies. Furthermore, if the KQ core-shift first increases due to a higher particle density while lower frequency core position still remains the same (since the flare has not yet propagated there) this could look like $k_r$ is larger. For example, Figure~\ref{CSepoch17} shows a situation in which the shift between 24\,GHz and 43\,GHz is suddenly very large during the rising part of a large flare. Such an effect agrees well with the flaring jet model presented by \citetalias{Plavin2019}. More observations are necessary for following closely the changes in the core position at different frequencies and consequently, $k_r$ during pre-flare, peak, and post-flare as well as extensive observations during quiescent states. Furthermore, it is important to note that the increase of $k_r$ with the flux density has also been observed in another source, \object{3C\,345} at 14.5\,GHz for a few observing epochs \citep{Kudryavtseva2011}.

\subsection{Consequences to the magnetic field estimation}

Significant deviations of $k_r$ from one imply that this parameter can also be variable in other radio sources. This should be taken into account when measuring magnetic field strengths using just a pair of frequencies and assuming \textit{a priori} $k_r=1$. In general, having three or more frequencies in core shift measurements in order to verify the $k_r=1$ condition is important if these measurements are used to calculate magnetic field strengths. Furthermore, making the measurements outside of flaring periods is likely to yield ``undisturbed'' core-shift measurement.

A key problem found in this study is the large differences in $B_\mathrm{1pc}$ with $k_r=1$ and $k_r \neq 1$. Using the same formula for both cases indicate that the current expression for $B_\mathrm{1pc}$ is not applicable and fails for high or low $k_r$ values. This is not surprising, since the deviations from $k_r=1$ indicate that one of the assumptions used to derive $B_\mathrm{1pc}$ does not hold. The expression for $B_\mathrm{1pc}$ in equipartition with a conical jet holds strictly as long as $k_r$ is near one. When $k_r=1$ epochs are selected, the magnetic field estimates are quite consistent even though the amount of core-shift varies significantly with time.

Since the equipartition expression for $B_\mathrm{1pc}$ does not hold when measured $k_r$ is different from one, one should consider measuring the magnetic field strength by incorporating the core flux density in the core shift measurements like done by \citet{Zdziarskietal2015} or by measuring the synchrotron self-absorption turnover frequency and turnover flux density of a resolved jet  \citep{Marscher1983, Savolainen2008}. In both cases, one can estimate the magnetic field strength without assuming equipartition. The known downside of these approaches is the high sensitivity of the derived magnetic field values to the accuracy of the measured quantities. In \object{3C\,454.3}, the core spectrum is inverted in nearly all observations which in most cases do not constrain well the turnover frequency (see the results in Appendix~\ref{appendix:core-shift-results}). 

\subsection{Consequences to astrometry and geodetic VLBI}

%\ykcom{Apparently, you will mention \citet{Porcas2009} which is broken in case if the effect is not $\propto\nu^{-1}$. Do not forget about importance of it for the VLBI-Gaia connection \citep[e.g.,][]{PetKov2017,PKP2019}.}

High-accuracy astrometric VLBI measurements of extragalactic radio sources are the basis of the International Celestial Reference Frame (ICRF). In its third realization  \citep[ICRF3,][]{Charlot2020} as well as in the independently produced Radio Fundamental Catalog (RFC\footnote{\url{http://astrogeo.org/rfc/}}), the absolute source positions are measured at sub-milliarcsecond accuracy. Together with other space geodetic techniques, VLBI is also essential for realizing the International Terrestrial Reference Frame (ITRF) and obtaining the full set of Earth Orientation Parameters that provide a link between the ITRF and the ICRF \citep[e.g.,][]{Petrov2009}. In astrometric and geodetic VLBI observations, core shift adds another frequency-dependent phase term \citep{Kovalevetal2008} that behaves like an extra path through a dispersive medium. \citet{Porcas2009} showed that if $k_r = 1$, the contribution of the core shift to the measured \textit{group delays} is, in fact, zero and the source coordinates derived from the group delays refer to the jet base, a fiducial point upstream of the core. If $k_r \ne 1$, this is, however, not the case and the core-shift-induced group delays are non-zero. On the other hand, the coordinates measured from the single-band \textit{phase delays} refer to the position of the radio emission at the given observing band and always depend on the core shift.

\citetalias{Plavin2019} points out that the core-shift \textit{variability} can affect the astrometric measurements using the group delays even if formally $k_r = 1$ and strong flares can disrupt any regular frequency dependency of $k_r$. Our results further show that in \object{3C\,454.3} $k_r \ne 1$ even during quiescent periods, i.e., the core shift can affect the measured group delays even outside of the flaring periods in this source. Therefore, unless \object{3C\,454.3} is an exceptional case, knowing the core shift -- including the actual $k_r$ -- is important for improving the accuracy of VLBI astrometry. The significant core-shift variability observed here and in \citetalias{Plavin2019} indicates that ideally core shift in e.g., the ICRF3 defining sources should be regularly monitored.

Several studies have established that systematic effects due to non-point-like structure of extragalactic radio sources is currently the major source of errors in geodetic VLBI measurements \citep[e.g.,][]{Xu2017,Anderson2018,Bolotin2019,Xu2021c}. With the advent of the new broadband (four 512\,MHz wide bands over 2$-$14\,GHz) observing system, VGOS \citep[VLBI Global Observing System;][]{Niell2018}, the source structure effects now dominate the thermal errors by about an order of magnitude and are of the same order or larger than the errors due to uncertainties in the atmospheric modeling \citep{Xu2021c}. VGOS has an ambitious goal of 1\,mm accuracy for the station positions on the ground, which translates to $\sim 30$\,$\mu$as accuracy of the source positions on the sky. Therefore, it is clear that accurate modeling of the time and frequency-dependent source structure is crucial for VGOS. This has led to active efforts to remove the source structure effects by imaging the geodetic VLBI data at the four bands of the VGOS \citep{Xu2021a}. In order to use these images to correct the visibility data, it is necessary to accurately align them, which amounts to determining the core shift effect. Our results presented in this paper directly demonstrate both  the variability of the magnitude of the core shift that was shown in \citetalias{Plavin2019} as well as the variability of $k_r$ which \citetalias{Plavin2019} deduced from a model analysis of two-frequency core-shift time series. The former means that core-shift should be regularly monitored also for the geodetic VLBI purposes and best if at more than two frequencies. The latter should be taken into account when attempting to e.g., fit the core-shift to the geodetic VLBI measurements -- an \textit{a priori} assumption of $k_r = 1$ may not always hold.

Finally, we comment on the offsets between optical and radio positions of AGN. The European Space Agency's astrometry mission \textit{Gaia} has recently provided sub-milliarcsecond positions in optical band \citep{2021A&A...649A...1G} for well over a billion celestial sources with a limiting magnitude of $G\approx21$. While a good overall agreement exists between the ICRF3 and RFC radio positions and the \textit{Gaia} optical positions, statistically significant offsets are seen for about 10\,\% of the matching AGN \citep[e.g.,][]{PetKov2017letter,PetKov2017,Gaia2018,Petrov2019}. \citet{Kovalev2017} found that the significant VLBI-\textit{Gaia} offsets are preferentially parallel to the direction of the radio jet. This result was later confirmed and studied by, e.g., \citet{PKP2019}, \citet{Kovalev2020gaia}, and \citet{Xu2021b}. The offsets therefore constitute a genuine astrophysical effect and core shift as well as the source structure can contribute to this effect at least in some cases \citep{PetKov2017}. If $k_r \ne 1$, like we report here for \object{3C\,454.3}, the radio source position derived from the group delays does not correspond to the jet base and one can have an offset from the nucleus-dominated optical position. On the other hand, if the optical emission is dominated by the jet, offsets are expected also for the $k_r = 1$ case, just in the opposite direction  \citep{Kovalev2017,PKP2019}. Knowing both the amount of the core-shift as well as its time and frequency dependency is useful for interpreting the offsets and underlines the importance of such measurements.

\section{Summary and conclusions}

The core-shift effect has been broadly used for a variety of studies which include jet geometry \citep[e.g][]{Pushkarev2018}, astrometry,
%(to be added: e.g Porcas, Abellan?, Rioja), 
and estimation of particle densities, magnetic field strengths and magnetic fluxes in a broad range of AGN \citep[e.g.,][]{Pushkarevetal2012, Zamaninasab2014, Plavin2019}. These have been possible by employing the VLBI technique that allows the highest angular resolution measurements. Identifying the surface where the opacity is near unity is essential for locating the core and measuring the core shift.

We studied the time variability of the core shift effect in the jet of \object{3C\,454.3} by analyzing multi-epoch and multi-frequency VLBA data from 2005 to 2010 (nineteen epochs). Core shift measurements were performed for the following adjacent frequency bands: CX (5\,GHz and 8\,GHz), XU (8\,GHz and 15\,GHz), UK (15\,GHz and {22$-$24\,GHz}), KQ (22$-$24\,GHz and 43\,GHz). These data allowed us to examine the time variability of both the core shift and core shift index, $k_r$, for the first time. The current study found significant variability of the core shift from 0.27\,mas up to 0.86\,mas for the frequencies between 5 and 43\,GHz.
% Additionally, the variation of the position angle spanned from $-60\degr$ to $-98\degr$, 
% suggesting a wobbling effect in the jet (\textbf{to relate this with Jorstad et al.TBA}). \tscom{In 4.1 it is said that the PA variability is not statistically significant. Perhaps leave the previous comment out?} 

Investigation of the variable nature of the core shift effect has been earlier carried out by \citetalias{Plavin2019} who found significant core shift variability in a sample of 40 sources (not including \object{3C\,454.3}), although only  between two frequencies, 2\,GHz and 8\,GHz. In this paper, we confirm the time variability of the core shift effect found by \citetalias{Plavin2019} and present for the first time direct evidence for the variability of the index $k_r$. We found significant deviations from the ideal Blandford \& Königl conical jet in equipartition ($k_r=1$). The full range of measured $k_r$ values goes from 0.45 to 1.7. The large values are, however, related to the flaring period in 2009$-$2010 and have large uncertainties. Typically $k_r$ was below one with a mean value of $k_r = 0.85\pm0.08$ for the entire study period.

The disorder of the core shifts in the observations from late 2009 to the beginning of 2010 are attributable to the strong outbursts which took place in that period \citep{Paccianietal2010}. Such a substantial variability of the core shift connected with nuclear flares has been also observed by \citetalias{Plavin2019}. In addition to this, the emergence of a new moving feature detected at 43\,GHz has also shifted the core position, hampering the accurate
localization of the core. Such disturbances introduce additional shifts at the high frequencies, hindering accurate estimations of the $k_r$ indices. The nuclear region model fitting at 43\,GHz revealed that the new feature was moving in the direction of the jet at an apparent speed of $\beta_\mathrm{app}$ = 5.0 $\pm$ 1.0 from 2009-09-22 to 2010-02-21.
%During these epochs, the source went through major outbursts, especially in December 2009 \citep{Paccianietal2010}. 

%How to compare our kr results with those of Mohan+2015? They obtained full kr for two different 'segments'. The first from 1966-2007, kr=0.97+-0.24. The second segment from 2007-2013, kr=1.22+-0.33. Note that they use time lags vs frequencies for kr measurements.

Using core-shift measurements and the expression for the magnetic field strength at one parsec in equipartition, we estimated $B_\mathrm{1pc}$ when $k_r=1$ and $k_r\neq1$. The results showed significant discrepancies which highlights the importance of first observationally validating the equipartition assumption before using the core shift measurements to obtain magnetic field strength estimates. Thus, $B_\mathrm{1pc}$ values were calculated only for the observations when $k_r$ was close to one, leading to $B_\mathrm{1pc}$ values that ranged from 1.3\,G to 2.0\,G. Subsequently, we estimated the magnetic flux of \object{3C\,454.3}. We found that the source was in the MAD state, which agrees with \citet{Zamaninasab2014}.

To identify a link between the core-shift effect and outbursts, we searched for correlations between the core shift and the core flux density and the 37\,GHz total flux density from single dish observations. We generally encountered a good correlation of the UK core-shift with the core flux density at 15 and 22$-$24\,GHz and the KQ core shift with the core flux density at 22$-$24 and $43$\,GHz. A good correlation was also found of the full CQ core shift with the single dish flux density at 37\,GHz and core flux density at 43\,GHz. The relationships follow a generic power-law as $\Delta r_{\nu_1\nu_2}\propto S^{t}_\mathrm{core}$, with $t$ ranging from 0.1 to 0.7. The large index of 0.7 has been found for the high frequency pair, KQ. Hence, there seems to be a tendency of increasing dependency between core-shift and flux density. The relationship becomes steeper for the higher frequency pair. The latter suggests that the core shift at high frequencies increases with strong flares. 

We did also find correlations between the core position (at a given frequency) and the core flux, $r_\mathrm{core}-S_\mathrm{core}$. At most frequencies expect at 5\,GHz, there are moderate to strong positive correlations between the variables. The data can be fitted well by a power-law with indices ranging from 0.6 to 0.8 (for 8, 15 and 22$-$24\,GHz) and with a mean of 0.7. This  matches very well the $r_\mathrm{core} \propto S_\mathrm{core}^{2/3}$ relationship predicted for a Blandford \& K\"onigl type jet when flares are due to changes in particle density (or magnetic field strength).

An interesting finding is the correlation of the $k_r$ index with core flux density. %Power-law \textbf{and linear} functions can fit well the data, showing that 
$k_r$ tends to increase as the source flares. This would mean that extreme and drastic outbursts in \object{3C\,454.3} tend to increase the core shift at the high frequencies. On the other hand, we did not find any evident correlation between the jet PA and $k_r$.

The present study suggests that full multi-frequency core-shift measurements should be carried out to measure $k_r$ before \textit{a priori} assuming anything about e.g., equipartition for calculation of $B_\mathrm{1pc}$, since deviations from $k_r=1$ indicate that the assumptions of Equation~\ref{eq:B1eq} may not hold and this can affect the jet's magnetic field strength calculations.

%In the large majority of the epochs we found deviations from conical shape with $\epsilon \neq1$. Deviations can go up down to 0.5 and up to 1.7. However, on average, the jet stayed far from a conical shape as $\epsilon=0.88\pm0.08$ for the study period.

%Nonetheless departures from $kr=1$ and $\epsilon=1$ were found which indicates that  Hence, core shift measurements should optimally be performed during the onset, peaking, and decline of flares to closely investigate both parameters' stability.

%and eventually to produce departures from the ideal conical jet in equipartition conditions. It would be compelling to test further the core shift versus flux with more observations in the future. 

\begin{acknowledgements}
We thank Ming~H.~Xu for useful discussions regarding the astrometric aspects of the core-shift effect.
This work was partly supported by the Academy of Finland under the project "Physics of Black Hole Powered Jets" (numbers 274477, 284495, and 312496) and the project "NT-VGOS" (number 315721). YYK is supported in the framework of the State project ``Science'' by the Ministry of Science and Higher Education of the Russian Federation under the contract 075-15-2020-778.
The Very Long Baseline Array and the National Radio Astronomy Observatory are facilities of the National Science Foundation operated under cooperative agreement by Associated Universities, Inc. This work made use of the Swinburne University of Technology software correlator \citep{2011PASP..123..275D}, developed as part of the Australian Major National Research Facilities Programme and operated under licence.
This paper has made use of data from the \mbox{MOJAVE} database that is maintained by the \mbox{MOJAVE} team \citep{Lister2018} and data obtained at Mets\"ahovi Radio Observatory, operated by Aalto University in Finland.
\end{acknowledgements}

% \ericom{Note that the journal MNRAS is spelled out while A\&A or ApJ use abbreviations. Unify.}

\bibliographystyle{aa}
\bibliography{references}

\begin{thebibliography}{95}
\expandafter\ifx\csname natexlab\endcsname\relax\def\natexlab#1{#1}\fi

\bibitem[{{Abdo} {et~al.}(2011){Abdo}, {Ackermann}, {Ajello}, {Allafort},
  {Baldini}, {Ballet}, {Barbiellini}, {Bastieri}, {Bellazzini}, {Berenji},
  {Blandford}, {Bloom}, {Bonamente}, {Borgland}, {Bouvier}, {Bregeon},
  {Brigida}, {Bruel}, {Buehler}, {Buson}, {Caliandro}, {Cameron}, {Caraveo},
  {Casandjian}, {Cavazzuti}, {Cecchi}, {Charles}, {Chekhtman}, {Cheung},
  {Chiang}, {Ciprini}, {Claus}, {Conrad}, {Cutini}, {D'Ammando}, {de Angelis},
  {de Palma}, {Dermer}, {Digel}, {Silva}, {Drell}, {Dubois}, {Dumora},
  {Escande}, {Favuzzi}, {Fegan}, {Ferrara}, {Fortin}, {Fukazawa}, {Fusco},
  {Gargano}, {Gasparrini}, {Gehrels}, {Germani}, {Giglietto}, {Giommi},
  {Giordano}, {Giroletti}, {Glanzman}, {Godfrey}, {Grenier}, {Grove},
  {Guiriec}, {Hadasch}, {Hayashida}, {Hays}, {Horan}, {Itoh},
  {J{\'o}hannesson}, {Johnson}, {Kamae}, {Katagiri}, {Kataoka},
  {Kn{\"o}dlseder}, {Kuss}, {Lande}, {Larsson}, {Latronico}, {Lee}, {Longo},
  {Loparco}, {Lott}, {Lovellette}, {Lubrano}, {Madejski}, {Makeev},
  {Mazziotta}, {McConville}, {McEnery}, {Michelson}, {Mitthumsiri}, {Mizuno},
  {Moiseev}, {Monte}, {Monzani}, {Morselli}, {Moskalenko}, {Murgia},
  {Naumann-Godo}, {Nishino}, {Nolan}, {Norris}, {Nuss}, {Ohsugi}, {Okumura},
  {Orlando}, {Ormes}, {Paneque}, {Pelassa}, {Pesce-Rollins}, {Pierbattista},
  {Piron}, {Porter}, {Rain{\`o}}, {Rando}, {Razzaque}, {Reimer}, {Reimer},
  {Ritz}, {Roth}, {Sadrozinski}, {Sanchez}, {Scargle}, {Schalk}, {Sgr{\`o}},
  {Siskind}, {Smith}, {Spandre}, {Spinelli}, {Strickman}, {Takahashi},
  {Takahashi}, {Tanaka}, {Tanaka}, {Thayer}, {Thayer}, {Thompson}, {Tibaldo},
  {Torres}, {Tosti}, {Tramacere}, {Troja}, {Vandenbroucke}, {Vasileiou},
  {Vianello}, {Vilchez}, {Vitale}, {Waite}, {Wang}, {Winer}, {Wood}, {Yang}, \&
  {Ziegler}}]{Abdo2011}
{Abdo}, A.~A., {Ackermann}, M., {Ajello}, M., {et~al.} 2011, \apjl, 733, L26

\bibitem[{{Abdo} {et~al.}(2009){Abdo}, {Ackermann}, {Ajello}, {Atwood},
  {Axelsson}, {Baldini}, {Ballet}, {Barbiellini}, {Bastieri}, {Battelino},
  {Baughman}, {Bechtol}, {Bellazzini}, {Berenji}, {Blandford}, {Bloom},
  {Bonamente}, {Borgland}, {Bouvier}, {Bregeon}, {Brez}, {Brigida}, {Bruel},
  {Burnett}, {Caliandro}, {Cameron}, {Caraveo}, {Casandjian}, {Cavazzuti},
  {Cecchi}, {Charles}, {Chaty}, {Chekhtman}, {Cheung}, {Chiang}, {Ciprini},
  {Claus}, {Cohen-Tanugi}, {Cominsky}, {Conrad}, {Costamante}, {Cutini},
  {Dermer}, {de Angelis}, {de Palma}, {Digel}, {Silva}, {Donato}, {Drell},
  {Dubois}, {Dumora}, {Farnier}, {Favuzzi}, {Focke}, {Foschini}, {Frailis},
  {Fuhrmann}, {Fukazawa}, {Funk}, {Fusco}, {Gargano}, {Gasparrini}, {Gehrels},
  {Germani}, {Giebels}, {Giglietto}, {Giommi}, {Giordano}, {Glanzman},
  {Godfrey}, {Grenier}, {Grondin}, {Grove}, {Guillemot}, {Guiriec}, {Hanabata},
  {Harding}, {Hartman}, {Hayashida}, {Hays}, {Hughes}, {J{\'o}hannesson},
  {Johnson}, {Johnson}, {Johnson}, {Kamae}, {Katagiri}, {Kataoka}, {Kawai},
  {Kerr}, {Kn{\"o}dlseder}, {Kocian}, {Kuehn}, {Kuss}, {Latronico}, {Lee},
  {Lemoine-Goumard}, {Longo}, {Loparco}, {Lott}, {Lovellette}, {Lubrano},
  {Madejski}, {Makeev}, {Massaro}, {Mazziotta}, {McEnery}, {McGlynn}, {Meurer},
  {Michelson}, {Mitthumsiri}, {Mizuno}, {Moiseev}, {Monte}, {Monzani},
  {Morselli}, {Moskalenko}, {Murgia}, {Nolan}, {Norris}, {Nuss}, {Ohsugi},
  {Omodei}, {Orlando}, {Ormes}, {Paneque}, {Panetta}, {Parent}, {Pelassa},
  {Pepe}, {Pesce-Rollins}, {Piron}, {Porter}, {Rain{\`o}}, {Rando}, {Razzano},
  {Reimer}, {Reimer}, {Reposeur}, {Reyes}, {Ritz}, {Rochester}, {Rodriguez},
  {Rahoui}, {Ryde}, {Sadrozinski}, {Sambruna}, {Sanchez}, {Sander},
  {Parkinson}, {Sgr{\`o}}, {Shaw}, {Smith}, {Smith}, {Spandre}, {Spinelli},
  {Starck}, {Strickman}, {Suson}, {Tajima}, {Takahashi}, {Takahashi}, {Tanaka},
  {Thayer}, {Thayer}, {Thompson}, {Tibaldo}, {Torres}, {Tosti}, {Tramacere},
  {Uchiyama}, {Usher}, {Vilchez}, {Villata}, {Vitale}, {Waite}, {Winer},
  {Wood}, {Ylinen}, {Zensus}, \& {Ziegler}}]{Abdo2009}
{Abdo}, A.~A., {Ackermann}, M., {Ajello}, M., {et~al.} 2009, \apj, 699, 817

\bibitem[{{Ackermann} {et~al.}(2010){Ackermann}, {Ajello}, {Baldini}, {Ballet},
  {Barbiellini}, {Bastieri}, {Bechtol}, {Bellazzini}, {Berenji}, {Blandford},
  {Bonamente}, {Borgland}, {Bregeon}, {Brigida}, {Bruel}, {Buehler}, {Burnett},
  {Buson}, {Caliandro}, {Cameron}, {Caraveo}, {Carrigan}, {Casandjian},
  {Cavazzuti}, {Cecchi}, {{\c{C}}elik}, {Chekhtman}, {Cheung}, {Chiang},
  {Ciprini}, {Claus}, {Cohen-Tanugi}, {Corbel}, {Cutini}, {D'Ammando},
  {Dermer}, {de Angelis}, {de Palma}, {Digel}, {Silva}, {Drell}, {Dubois},
  {Dumora}, {Escande}, {Favuzzi}, {Fegan}, {Ferrara}, {Fuhrmann}, {Fukazawa},
  {Fusco}, {Gargano}, {Gasparrini}, {Gehrels}, {Germani}, {Giebels},
  {Giglietto}, {Giommi}, {Giordano}, {Giroletti}, {Glanzman}, {Godfrey},
  {Grenier}, {Grove}, {Guiriec}, {Hadasch}, {Hayashida}, {Hays},
  {J{\'o}hannesson}, {Johnson}, {Johnson}, {Kamae}, {Katagiri}, {Kataoka},
  {Kn{\"o}dlseder}, {Kuss}, {Lande}, {Larsson}, {Latronico}, {Lee}, {Llena
  Garde}, {Longo}, {Loparco}, {Lott}, {Lubrano}, {Madejski}, {Makeev},
  {Marchili}, {Mazziotta}, {McEnery}, {Mehault}, {Michelson}, {Mizuno},
  {Monte}, {Monzani}, {Morselli}, {Moskalenko}, {Murgia}, {Nakamori},
  {Nalewajko}, {Naumann-Godo}, {Nolan}, {Norris}, {Nuss}, {Ohsugi}, {Okumura},
  {Omodei}, {Orlando}, {Ormes}, {Pelassa}, {Pepe}, {Pesce-Rollins}, {Piron},
  {Porter}, {Rain{\`o}}, {Rando}, {Razzano}, {Reimer}, {Reimer}, {Reyes},
  {Ripken}, {Ritz}, {Roth}, {Sadrozinski}, {Sanchez}, {Sander}, {Scargle},
  {Sgr{\`o}}, {Sikora}, {Siskind}, {Spandre}, {Spinelli}, {Strickman}, {Suson},
  {Takahashi}, {Takahashi}, {Tanaka}, {Tanaka}, {Thayer}, {Thayer}, {Thompson},
  {Tibaldo}, {Torres}, {Tosti}, {Tramacere}, {Usher}, {Vandenbroucke},
  {Vilchez}, {Vitale}, {Waite}, {Wang}, {Wehrle}, {Winer}, {Yang}, {Ylinen}, \&
  {Ziegler}}]{Ackermann2010}
{Ackermann}, M., {Ajello}, M., {Baldini}, L., {et~al.} 2010, \apj, 721, 1383

\bibitem[{{Algaba} {et~al.}(2019){Algaba}, {Hodgson}, {Kang}, {Kim}, {Kim},
  {Lee}, {Lee}, \& {Trippe}}]{Algaba2019}
{Algaba}, J.-C., {Hodgson}, J., {Kang}, S.-C., {et~al.} 2019, Journal of Korean
  Astronomical Society, 52, 31

\bibitem[{{Anderson} \& {Xu}(2018)}]{Anderson2018}
{Anderson}, J.~M. \& {Xu}, M.~H. 2018, Journal of Geophysical Research (Solid
  Earth), 123, 10,162

\bibitem[{{Blandford} \& {K{\"o}nigl}(1979)}]{BlandfordKonigl}
{Blandford}, R.~D. \& {K{\"o}nigl}, A. 1979, \apj, 232, 34

\bibitem[{{Bolotin} {et~al.}(2019){Bolotin}, {Baver}, {Bolotina}, {Gipson},
  {Gordon}, {Le Bail}, \& {MacMillan}}]{Bolotin2019}
{Bolotin}, S., {Baver}, K., {Bolotina}, O., {et~al.} 2019, in Proceedings of
  the 24th European VLBI Group for Geodesy and Astrometry Working Meeting, ed.
  R.~{Haas}, S.~{Garcia-Espada}, \& J.~A. {L{\'o}pez Fern{\'a}ndez}, Vol.~24,
  224--228

\bibitem[{{Bonnoli} {et~al.}(2011){Bonnoli}, {Ghisellini}, {Foschini},
  {Tavecchio}, \& {Ghirlanda}}]{Bonnoli2011}
{Bonnoli}, G., {Ghisellini}, G., {Foschini}, L., {Tavecchio}, F., \&
  {Ghirlanda}, G. 2011, \mnras, 410, 368

\bibitem[{{Britzen} {et~al.}(2013){Britzen}, {Qian}, {Witzel}, {Krichbaum},
  {Aller}, {Aller}, {Kurtanidze}, {Vercellone}, \& {Richter}}]{Britzen2013}
{Britzen}, S., {Qian}, S.-J., {Witzel}, A., {et~al.} 2013, \aap, 557, A37

\bibitem[{Chamani {et~al.}(2021)Chamani, Savolainen, Hada, \& Xu}]{Chamani2021}
Chamani, W., Savolainen, T., Hada, K., \& Xu, M.~H. 2021, A\&A, 652, A14

\bibitem[{{Charlot} {et~al.}(2020){Charlot}, {Jacobs}, {Gordon}, {Lambert}, {de
  Witt}, {B{\"o}hm}, {Fey}, {Heinkelmann}, {Skurikhina}, {Titov}, {Arias},
  {Bolotin}, {Bourda}, {Ma}, {Malkin}, {Nothnagel}, {Mayer}, {MacMillan},
  {Nilsson}, \& {Gaume}}]{Charlot2020}
{Charlot}, P., {Jacobs}, C.~S., {Gordon}, D., {et~al.} 2020, \aap, 644, A159

\bibitem[{Croke \& Gabuzda(2008)}]{CrokeGabuzda2008}
Croke, S.~M. \& Gabuzda, D.~C. 2008, \mnras, 386, 619

\bibitem[{{Deller} {et~al.}(2011){Deller}, {Brisken}, {Phillips}, {Morgan},
  {Alef}, {Cappallo}, {Middelberg}, {Romney}, {Rottmann}, {Tingay}, \&
  {Wayth}}]{2011PASP..123..275D}
{Deller}, A.~T., {Brisken}, W.~F., {Phillips}, C.~J., {et~al.} 2011, \pasp,
  123, 275

\bibitem[{{Donnarumma} {et~al.}(2009){Donnarumma}, {Pucella}, {Vittorini},
  {D'Ammando}, {Vercellone}, {Raiteri}, {Villata}, {Perri}, {Chen}, {Smart},
  {Kataoka}, {Kawai}, {Mori}, {Tosti}, {Impiombato}, {Takahashi}, {Sato},
  {Tavani}, {Bulgarelli}, {Chen}, {Giuliani}, {Longo}, {Pacciani}, {Argan},
  {Barbiellini}, {Boffelli}, {Caraveo}, {Cattaneo}, {Cocco}, {Contessi},
  {Costa}, {Del Monte}, {De Paris}, {Di Cocco}, {Evangelista}, {Feroci},
  {Ferrari}, {Fiorini}, {Froysland}, {Frutti}, {Fuschino}, {Galli}, {Gianotti},
  {Labanti}, {Lapshov}, {Lazzarotto}, {Lipari}, {Marisaldi}, {Mastropietro},
  {Mereghetti}, {Morelli}, {Moretti}, {Morselli}, {Pellizzoni}, {Perotti},
  {Piano}, {Picozza}, {Pilia}, {Porrovecchio}, {Prest}, {Rapisarda},
  {Rappoldi}, {Rubini}, {Sabatini}, {Scalise}, {Soffitta}, {Striani},
  {Trifoglio}, {Trois}, {Vallazza}, {Zambra}, {Zanello}, {Pittori},
  {Santolamazza}, {Verrecchia}, {Giommi}, {Antonelli}, {Colafrancesco}, \&
  {Salotti}}]{Donnarumaeral2009}
{Donnarumma}, I., {Pucella}, G., {Vittorini}, V., {et~al.} 2009, \apj, 707,
  1115

\bibitem[{{Fromm} {et~al.}(2015){Fromm}, {Perucho}, {Ros}, {Savolainen}, \&
  {Zensus}}]{Fromm2015}
{Fromm}, C.~M., {Perucho}, M., {Ros}, E., {Savolainen}, T., \& {Zensus}, J.~A.
  2015, \aap, 576, A43

\bibitem[{{Fromm} {et~al.}(2013{\natexlab{a}}){Fromm}, {Ros}, {Perucho},
  {Savolainen}, {Mimica}, {Kadler}, {Lobanov}, {Lister}, {Kovalev}, \&
  {Zensus}}]{Fromm2013a}
{Fromm}, C.~M., {Ros}, E., {Perucho}, M., {et~al.} 2013{\natexlab{a}}, \aap,
  551, A32

\bibitem[{{Fromm} {et~al.}(2013{\natexlab{b}}){Fromm}, {Ros}, {Perucho},
  {Savolainen}, {Mimica}, {Kadler}, {Lobanov}, \& {Zensus}}]{Fromm2013b}
{Fromm}, C.~M., {Ros}, E., {Perucho}, M., {et~al.} 2013{\natexlab{b}}, \aap,
  557, A105

\bibitem[{{Fromm} {et~al.}(2010){Fromm}, {Ros}, {Savolainen}, {Lobanov},
  {Perucho}, {Zensus}, {Aller}, {Aller}, {Gurwell}, \&
  {Laehteenmaeki}}]{Fromm2010}
{Fromm}, C.~M., {Ros}, E., {Savolainen}, T., {et~al.} 2010, in Fermi meets
  Jansky - AGN in Radio and Gamma Rays. Proceedings of a Workshop held 21-23
  June 2010 at the Max-Planck-Institut f. Radioastronomie, Bonn, Germany.
  Edited by Tuomas Savolainen, Eduardo Ros, Richard W. Porcas and J. Anton
  Zensus. Bonn: Max-Planck-Institut f. Radioastronomie.

\bibitem[{{Gaia Collaboration} {et~al.}(2021){Gaia Collaboration}, {Brown},
  {Vallenari}, {Prusti}, {de Bruijne}, {Babusiaux}, {Biermann}, {Creevey},
  {Evans}, {Eyer}, {Hutton}, {Jansen}, {Jordi}, {Klioner}, {Lammers},
  {Lindegren}, {Luri}, {Mignard}, {Panem}, {Pourbaix}, {Randich}, {Sartoretti},
  {Soubiran}, {Walton}, {Arenou}, {Bailer-Jones}, {Bastian}, {Cropper},
  {Drimmel}, {Katz}, {Lattanzi}, {van Leeuwen}, {Bakker}, {Cacciari},
  {Casta{\~n}eda}, {De Angeli}, {Ducourant}, {Fabricius}, {Fouesneau},
  {Fr{\'e}mat}, {Guerra}, {Guerrier}, {Guiraud}, {Jean-Antoine Piccolo},
  {Masana}, {Messineo}, {Mowlavi}, {Nicolas}, {Nienartowicz}, {Pailler},
  {Panuzzo}, {Riclet}, {Roux}, {Seabroke}, {Sordo}, {Tanga}, {Th{\'e}venin},
  {Gracia-Abril}, {Portell}, {Teyssier}, {Altmann}, {Andrae}, {Bellas-Velidis},
  {Benson}, {Berthier}, {Blomme}, {Brugaletta}, {Burgess}, {Busso}, {Carry},
  {Cellino}, {Cheek}, {Clementini}, {Damerdji}, {Davidson}, {Delchambre},
  {Dell'Oro}, {Fern{\'a}ndez-Hern{\'a}ndez}, {Galluccio}, {Garc{\'\i}a-Lario},
  {Garcia-Reinaldos}, {Gonz{\'a}lez-N{\'u}{\~n}ez}, {Gosset}, {Haigron},
  {Halbwachs}, {Hambly}, {Harrison}, {Hatzidimitriou}, {Heiter},
  {Hern{\'a}ndez}, {Hestroffer}, {Hodgkin}, {Holl}, {Jan{\ss}en}, {Jevardat de
  Fombelle}, {Jordan}, {Krone-Martins}, {Lanzafame}, {L{\"o}ffler}, {Lorca},
  {Manteiga}, {Marchal}, {Marrese}, {Moitinho}, {Mora}, {Muinonen}, {Osborne},
  {Pancino}, {Pauwels}, {Petit}, {Recio-Blanco}, {Richards}, {Riello},
  {Rimoldini}, {Robin}, {Roegiers}, {Rybizki}, {Sarro}, {Siopis}, {Smith},
  {Sozzetti}, {Ulla}, {Utrilla}, {van Leeuwen}, {van Reeven}, {Abbas}, {Abreu
  Aramburu}, {Accart}, {Aerts}, {Aguado}, {Ajaj}, {Altavilla}, {{\'A}lvarez},
  {{\'A}lvarez Cid-Fuentes}, {Alves}, {Anderson}, {Anglada Varela}, {Antoja},
  {Audard}, {Baines}, {Baker}, {Balaguer-N{\'u}{\~n}ez}, {Balbinot}, {Balog},
  {Barache}, {Barbato}, {Barros}, {Barstow}, {Bartolom{\'e}}, {Bassilana},
  {Bauchet}, {Baudesson-Stella}, {Becciani}, {Bellazzini}, {Bernet}, {Bertone},
  {Bianchi}, {Blanco-Cuaresma}, {Boch}, {Bombrun}, {Bossini}, {Bouquillon},
  {Bragaglia}, {Bramante}, {Breedt}, {Bressan}, {Brouillet}, {Bucciarelli},
  {Burlacu}, {Busonero}, {Butkevich}, {Buzzi}, {Caffau}, {Cancelliere},
  {C{\'a}novas}, {Cantat-Gaudin}, {Carballo}, {Carlucci}, {Carnerero},
  {Carrasco}, {Casamiquela}, {Castellani}, {Castro-Ginard}, {Castro Sampol},
  {Chaoul}, {Charlot}, {Chemin}, {Chiavassa}, {Cioni}, {Comoretto}, {Cooper},
  {Cornez}, {Cowell}, {Crifo}, {Crosta}, {Crowley}, {Dafonte}, {Dapergolas},
  {David}, {David}, {de Laverny}, {De Luise}, {De March}, {De Ridder}, {de
  Souza}, {de Teodoro}, {de Torres}, {del Peloso}, {del Pozo}, {Delbo},
  {Delgado}, {Delgado}, {Delisle}, {Di Matteo}, {Diakite}, {Diener},
  {Distefano}, {Dolding}, {Eappachen}, {Edvardsson}, {Enke}, {Esquej}, {Fabre},
  {Fabrizio}, {Faigler}, {Fedorets}, {Fernique}, {Fienga}, {Figueras},
  {Fouron}, {Fragkoudi}, {Fraile}, {Franke}, {Gai}, {Garabato},
  {Garcia-Gutierrez}, {Garc{\'\i}a-Torres}, {Garofalo}, {Gavras}, {Gerlach},
  {Geyer}, {Giacobbe}, {Gilmore}, {Girona}, {Giuffrida}, {Gomel}, {Gomez},
  {Gonzalez-Santamaria}, {Gonz{\'a}lez-Vidal}, {Granvik},
  {Guti{\'e}rrez-S{\'a}nchez}, {Guy}, {Hauser}, {Haywood}, {Helmi}, {Hidalgo},
  {Hilger}, {H{\l}adczuk}, {Hobbs}, {Holland}, {Huckle}, {Jasniewicz},
  {Jonker}, {Juaristi Campillo}, {Julbe}, {Karbevska}, {Kervella}, {Khanna},
  {Kochoska}, {Kontizas}, {Kordopatis}, {Korn}, {Kostrzewa-Rutkowska},
  {Kruszy{\'n}ska}, {Lambert}, {Lanza}, {Lasne}, {Le Campion}, {Le Fustec},
  {Lebreton}, {Lebzelter}, {Leccia}, {Leclerc}, {Lecoeur-Taibi}, {Liao},
  {Licata}, {Lindstr{\o}m}, {Lister}, {Livanou}, {Lobel}, {Madrero Pardo},
  {Managau}, {Mann}, {Marchant}, {Marconi}, {Marcos Santos}, {Marinoni},
  {Marocco}, {Marshall}, {Martin Polo}, {Mart{\'\i}n-Fleitas}, {Masip},
  {Massari}, {Mastrobuono-Battisti}, {Mazeh}, {McMillan}, {Messina},
  {Michalik}, {Millar}, {Mints}, {Molina}, {Molinaro}, {Moln{\'a}r},
  {Montegriffo}, {Mor}, {Morbidelli}, {Morel}, {Morris}, {Mulone}, {Munoz},
  {Muraveva}, {Murphy}, {Musella}, {Noval}, {Ord{\'e}novic}, {Orr{\`u}},
  {Osinde}, {Pagani}, {Pagano}, {Palaversa}, {Palicio}, {Panahi}, {Pawlak},
  {Pe{\~n}alosa Esteller}, {Penttil{\"a}}, {Piersimoni}, {Pineau}, {Plachy},
  {Plum}, {Poggio}, {Poretti}, {Poujoulet}, {Pr{\v{s}}a}, {Pulone}, {Racero},
  {Ragaini}, {Rainer}, {Raiteri}, {Rambaux}, {Ramos}, {Ramos-Lerate}, {Re
  Fiorentin}, {Regibo}, {Reyl{\'e}}, {Ripepi}, {Riva}, {Rixon}, {Robichon},
  {Robin}, {Roelens}, {Rohrbasser}, {Romero-G{\'o}mez}, {Rowell}, {Royer},
  {Rybicki}, {Sadowski}, {Sagrist{\`a} Sell{\'e}s}, {Sahlmann}, {Salgado},
  {Salguero}, {Samaras}, {Sanchez Gimenez}, {Sanna}, {Santove{\~n}a},
  {Sarasso}, {Schultheis}, {Sciacca}, {Segol}, {Segovia}, {S{\'e}gransan},
  {Semeux}, {Shahaf}, {Siddiqui}, {Siebert}, {Siltala}, {Slezak}, {Smart},
  {Solano}, {Solitro}, {Souami}, {Souchay}, {Spagna}, {Spoto}, {Steele},
  {Steidelm{\"u}ller}, {Stephenson}, {S{\"u}veges}, {Szabados}, {Szegedi-Elek},
  {Taris}, {Tauran}, {Taylor}, {Teixeira}, {Thuillot}, {Tonello}, {Torra},
  {Torra}, {Turon}, {Unger}, {Vaillant}, {van Dillen}, {Vanel}, {Vecchiato},
  {Viala}, {Vicente}, {Voutsinas}, {Weiler}, {Wevers}, {Wyrzykowski}, {Yoldas},
  {Yvard}, {Zhao}, {Zorec}, {Zucker}, {Zurbach}, \&
  {Zwitter}}]{2021A&A...649A...1G}
{Gaia Collaboration}, {Brown}, A.~G.~A., {Vallenari}, A., {et~al.} 2021, \aap,
  649, A1

\bibitem[{{Gaia Collaboration} {et~al.}(2018){Gaia Collaboration}, {Mignard},
  {Klioner}, {Lindegren}, {Hern{\'a}ndez}, {Bastian}, {Bombrun}, {Hobbs},
  {Lammers}, {Michalik}, {Ramos-Lerate}, {Biermann},
  {Fern{\'a}ndez-Hern{\'a}ndez}, {Geyer}, {Hilger}, {Siddiqui},
  {Steidelm{\"u}ller}, {Babusiaux}, {Barache}, {Lambert}, {Andrei}, {Bourda},
  {Charlot}, {Brown}, {Vallenari}, {Prusti}, {de Bruijne}, {Bailer-Jones},
  {Evans}, {Eyer}, {Jansen}, {Jordi}, {Luri}, {Panem}, {Pourbaix}, {Randich},
  {Sartoretti}, {Soubiran}, {van Leeuwen}, {Walton}, {Arenou}, {Cropper},
  {Drimmel}, {Katz}, {Lattanzi}, {Bakker}, {Cacciari}, {Casta{\~n}eda},
  {Chaoul}, {Cheek}, {De Angeli}, {Fabricius}, {Guerra}, {Holl}, {Masana},
  {Messineo}, {Mowlavi}, {Nienartowicz}, {Panuzzo}, {Portell}, {Riello},
  {Seabroke}, {Tanga}, {Th{\'e}venin}, {Gracia-Abril}, {Comoretto},
  {Garcia-Reinaldos}, {Teyssier}, {Altmann}, {Andrae}, {Audard},
  {Bellas-Velidis}, {Benson}, {Berthier}, {Blomme}, {Burgess}, {Busso},
  {Carry}, {Cellino}, {Clementini}, {Clotet}, {Creevey}, {Davidson}, {De
  Ridder}, {Delchambre}, {Dell'Oro}, {Ducourant}, {Fouesneau}, {Fr{\'e}mat},
  {Galluccio}, {Garc{\'\i}a-Torres}, {Gonz{\'a}lez-N{\'u}{\~n}ez},
  {Gonz{\'a}lez-Vidal}, {Gosset}, {Guy}, {Halbwachs}, {Hambly}, {Harrison},
  {Hestroffer}, {Hodgkin}, {Hutton}, {Jasniewicz}, {Jean-Antoine-Piccolo},
  {Jordan}, {Korn}, {Krone-Martins}, {Lanzafame}, {Lebzelter}, {L{\"o}ffler},
  {Manteiga}, {Marrese}, {Mart{\'\i}n-Fleitas}, {Moitinho}, {Mora}, {Muinonen},
  {Osinde}, {Pancino}, {Pauwels}, {Petit}, {Recio-Blanco}, {Richards},
  {Rimoldini}, {Robin}, {Sarro}, {Siopis}, {Smith}, {Sozzetti}, {S{\"u}veges},
  {Torra}, {van Reeven}, {Abbas}, {Abreu Aramburu}, {Accart}, {Aerts},
  {Altavilla}, {{\'A}lvarez}, {Alvarez}, {Alves}, {Anderson}, {Anglada Varela},
  {Antiche}, {Antoja}, {Arcay}, {Astraatmadja}, {Bach}, {Baker},
  {Balaguer-N{\'u}{\~n}ez}, {Balm}, {Barata}, {Barbato}, {Barblan}, {Barklem},
  {Barrado}, {Barros}, {Barstow}, {Bartholom{\'e} Mu{\~n}oz}, {Bassilana},
  {Becciani}, {Bellazzini}, {Berihuete}, {Bertone}, {Bianchi}, {Bienaym{\'e}},
  {Blanco-Cuaresma}, {Boch}, {Boeche}, {Borrachero}, {Bossini}, {Bouquillon},
  {Bragaglia}, {Bramante}, {Breddels}, {Bressan}, {Brouillet},
  {Br{\"u}semeister}, {Brugaletta}, {Bucciarelli}, {Burlacu}, {Busonero},
  {Butkevich}, {Buzzi}, {Caffau}, {Cancelliere}, {Cannizzaro}, {Cantat-Gaudin},
  {Carballo}, {Carlucci}, {Carrasco}, {Casamiquela}, {Castellani},
  {Castro-Ginard}, {Chemin}, {Chiavassa}, {Cocozza}, {Costigan}, {Cowell},
  {Crifo}, {Crosta}, {Crowley}, {Cuypers}, {Dafonte}, {Damerdji}, {Dapergolas},
  {David}, {David}, {de Laverny}, {De Luise}, {De March}, {de Souza}, {de
  Torres}, {Debosscher}, {del Pozo}, {Delbo}, {Delgado}, {Delgado}, {Diakite},
  {Diener}, {Distefano}, {Dolding}, {Drazinos}, {Dur{\'a}n}, {Edvardsson},
  {Enke}, {Eriksson}, {Esquej}, {Eynard Bontemps}, {Fabre}, {Fabrizio},
  {Faigler}, {Falc{\~a}o}, {Farr{\`a}s Casas}, {Federici}, {Fedorets},
  {Fernique}, {Figueras}, {Filippi}, {Findeisen}, {Fonti}, {Fraile}, {Fraser},
  {Fr{\'e}zouls}, {Gai}, {Galleti}, {Garabato}, {Garc{\'\i}a-Sedano},
  {Garofalo}, {Garralda}, {Gavel}, {Gavras}, {Gerssen}, {Giacobbe}, {Gilmore},
  {Girona}, {Giuffrida}, {Glass}, {Gomes}, {Granvik}, {Gueguen}, {Guerrier},
  {Guiraud}, {Guti{\'e}}, {Haigron}, {Hatzidimitriou}, {Hauser}, {Haywood},
  {Heiter}, {Helmi}, {Heu}, {Hofmann}, {Holland}, {Huckle}, {Hypki}, {Icardi},
  {Jan{\ss}en}, {Jevardat de Fombelle}, {Jonker}, {Juh{\'a}sz}, {Julbe},
  {Karampelas}, {Kewley}, {Klar}, {Kochoska}, {Kohley}, {Kolenberg},
  {Kontizas}, {Kontizas}, {Koposov}, {Kordopatis}, {Kostrzewa-Rutkowska},
  {Koubsky}, {Lanza}, {Lasne}, {Lavigne}, {Le Fustec}, {Le Poncin-Lafitte},
  {Lebreton}, {Leccia}, {Leclerc}, {Lecoeur-Taibi}, {Lenhardt}, {Leroux},
  {Liao}, {Licata}, {Lindstr{\o}m}, {Lister}, {Livanou}, {Lobel}, {L{\'o}pez},
  {Managau}, {Mann}, {Mantelet}, {Marchal}, {Marchant}, {Marconi}, {Marinoni},
  {Marschalk{\'o}}, {Marshall}, {Martino}, {Marton}, {Mary}, {Massari},
  {Matijevi{\v{c}}}, {Mazeh}, {McMillan}, {Messina}, {Millar}, {Molina},
  {Molinaro}, {Moln{\'a}r}, {Montegriffo}, {Mor}, {Morbidelli}, {Morel},
  {Morris}, {Mulone}, {Muraveva}, {Musella}, {Nelemans}, {Nicastro}, {Noval},
  {O'Mullane}, {Ord{\'e}novic}, {Ord{\'o}{\~n}ez-Blanco}, {Osborne}, {Pagani},
  {Pagano}, {Pailler}, {Palacin}, {Palaversa}, {Panahi}, {Pawlak},
  {Piersimoni}, {Pineau}, {Plachy}, {Plum}, {Poggio}, {Poujoulet},
  {Pr{\v{s}}a}, {Pulone}, {Racero}, {Ragaini}, {Rambaux}, {Regibo},
  {Reyl{\'e}}, {Riclet}, {Ripepi}, {Riva}, {Rivard}, {Rixon}, {Roegiers},
  {Roelens}, {Romero-G{\'o}mez}, {Rowell}, {Royer}, {Ruiz-Dern}, {Sadowski},
  {Sagrist{\`a} Sell{\'e}s}, {Sahlmann}, {Salgado}, {Salguero}, {Sanna},
  {Santana-Ros}, {Sarasso}, {Savietto}, {Schultheis}, {Sciacca}, {Segol},
  {Segovia}, {S{\'e}gransan}, {Shih}, {Siltala}, {Silva}, {Smart}, {Smith},
  {Solano}, {Solitro}, {Sordo}, {Soria Nieto}, {Souchay}, {Spagna}, {Spoto},
  {Stampa}, {Steele}, {Stephenson}, {Stoev}, {Suess}, {Surdej}, {Szabados},
  {Szegedi-Elek}, {Tapiador}, {Taris}, {Tauran}, {Taylor}, {Teixeira},
  {Terrett}, {Teyssandier}, {Thuillot}, {Titarenko}, {Torra Clotet}, {Turon},
  {Ulla}, {Utrilla}, {Uzzi}, {Vaillant}, {Valentini}, {Valette}, {van Elteren},
  {Van Hemelryck}, {van Leeuwen}, {Vaschetto}, {Vecchiato}, {Veljanoski},
  {Viala}, {Vicente}, {Vogt}, {von Essen}, {Voss}, {Votruba}, {Voutsinas},
  {Walmsley}, {Weiler}, {Wertz}, {Wevers}, {Wyrzykowski}, {Yoldas},
  {{\v{Z}}erjal}, {Ziaeepour}, {Zorec}, {Zschocke}, {Zucker}, {Zurbach}, \&
  {Zwitter}}]{Gaia2018}
{Gaia Collaboration}, {Mignard}, F., {Klioner}, S.~A., {et~al.} 2018, \aap,
  616, A14

\bibitem[{{Ghisellini} {et~al.}(2007){Ghisellini}, {Foschini}, {Tavecchio}, \&
  {Pian}}]{Ghisellini2007}
{Ghisellini}, G., {Foschini}, L., {Tavecchio}, F., \& {Pian}, E. 2007, \mnras,
  382, L82

\bibitem[{{Giommi} {et~al.}(2006){Giommi}, {Blustin}, {Capalbi},
  {Colafrancesco}, {Cucchiara}, {Fuhrmann}, {Krimm}, {Marchili}, {Massaro},
  {Perri}, {Tagliaferri}, {Tosti}, {Tramacere}, {Burrows}, {Chincarini},
  {Falcone}, {Gehrels}, {Kennea}, \& {Sambruna}}]{Giommi2006}
{Giommi}, P., {Blustin}, A.~J., {Capalbi}, M., {et~al.} 2006, \aap, 456, 911

\bibitem[{{Greisen}(2003)}]{Greisen2003}
{Greisen}, E.~W. 2003, in Astrophysics and Space Science Library, Vol. 285,
  Information Handling in Astronomy - Historical Vistas, ed. A.~{Heck}, 109

\bibitem[{{Hada} {et~al.}(2011){Hada}, {Doi}, {Kino}, {Nagai}, {Hagiwara}, \&
  {Kawaguchi}}]{Hada2011}
{Hada}, K., {Doi}, A., {Kino}, M., {et~al.} 2011, \nat, 477, 185

\bibitem[{{Hirotani}(2005)}]{Hirotani2005}
{Hirotani}, K. 2005, \apj, 619, 73

\bibitem[{Homan {et~al.}(2021)Homan, Cohen, Hovatta, Kellermann, Kovalev,
  Lister, Popkov, Pushkarev, Ros, \& Savolainen}]{Homan2021}
Homan, D.~C., Cohen, M.~H., Hovatta, T., {et~al.} 2021, The Astrophysical
  Journal, 923, 67

\bibitem[{{Hovatta} {et~al.}(2012){Hovatta}, {Lister}, {Aller}, {Aller},
  {Homan}, {Kovalev}, {Pushkarev}, \& {Savolainen}}]{Hovattaetal2012}
{Hovatta}, T., {Lister}, M.~L., {Aller}, M.~F., {et~al.} 2012, \aj, 144, 105

\bibitem[{{Hu} {et~al.}(2021){Hu}, {Yan}, \& {Hu}}]{Hu2021}
{Hu}, W., {Yan}, D., \& {Hu}, Q. 2021, \mnras, 503, 2523

\bibitem[{{Jackson} \& {Browne}(1991)}]{Jackson1991}
{Jackson}, N. \& {Browne}, I.~W.~A. 1991, \mnras, 250, 414

\bibitem[{Jorstad {et~al.}(2010)Jorstad, Marscher, Larionov, Agudo, Smith,
  Gurwell, Lähteenmäki, Tornikoski, Markowitz, Arkharov, Blinov, Chatterjee,
  D'Arcangelo, Falcone, Gómez, Hagen-Thorn, Jordan, Kimeridze, Konstantinova,
  Kopatskaya, Kurtanidze, Larionova, Larionova, McHardy, Melnichuk,
  Roca-Sogorb, Schmidt, Skiff, Taylor, Thum, Troitsky, \&
  Wiesemeyer}]{Jorstadetal2010}
Jorstad, S.~G., Marscher, A.~P., Larionov, V.~M., {et~al.} 2010, The
  Astrophysical Journal, 715, 362

\bibitem[{{Jorstad} {et~al.}(2005){Jorstad}, {Marscher}, {Lister}, {Stirling},
  {Cawthorne}, {Gear}, {G{\'o}mez}, {Stevens}, {Smith}, {Forster}, \&
  {Robson}}]{Jorstad2005}
{Jorstad}, S.~G., {Marscher}, A.~P., {Lister}, M.~L., {et~al.} 2005, \aj, 130,
  1418

\bibitem[{Jorstad {et~al.}(2013)Jorstad, Marscher, Smith, Larionov, Agudo,
  Gurwell, Wehrle, Lähteenmäki, Nikolashvili, Schmidt, Arkharov, Blinov,
  Blumenthal, Casadio, Chigladze, Efimova, Eggen, Gómez, Grupe, Hagen-Thorn,
  Joshi, Kimeridze, Konstantinova, Kopatskaya, Kurtanidze, Kurtanidze,
  Larionova, Larionova, Sigua, MacDonald, Maune, McHardy, Miller, Molina,
  Morozova, Scott, Taylor, Tornikoski, Troitsky, Thum, Walker, Williamson,
  Sallum, Consiglio, \& Strelnitski}]{Jorstadetal2013}
Jorstad, S.~G., Marscher, A.~P., Smith, P.~S., {et~al.} 2013, The Astrophysical
  Journal, 773, 147

\bibitem[{{Komatsu} {et~al.}(2009){Komatsu}, {Dunkley}, {Nolta}, {Bennett},
  {Gold}, {Hinshaw}, {Jarosik}, {Larson}, {Limon}, {Page}, {Spergel},
  {Halpern}, {Hill}, {Kogut}, {Meyer}, {Tucker}, {Weiland}, {Wollack}, \&
  {Wright}}]{Komatsu2009}
{Komatsu}, E., {Dunkley}, J., {Nolta}, M.~R., {et~al.} 2009, \apjs, 180, 330

\bibitem[{{K\"onigl}(1981)}]{Konigl1981}
{K\"onigl}, A. 1981, \apj, 243, 700

\bibitem[{{Kovalev} {et~al.}(2008){Kovalev}, {Lobanov}, {Pushkarev}, \&
  {Zensus}}]{Kovalevetal2008}
{Kovalev}, Y.~Y., {Lobanov}, A.~P., {Pushkarev}, A.~B., \& {Zensus}, J.~A.
  2008, A\&A, 483, 759

\bibitem[{{Kovalev} {et~al.}(2017){Kovalev}, {Petrov}, \&
  {Plavin}}]{Kovalev2017}
{Kovalev}, Y.~Y., {Petrov}, L., \& {Plavin}, A.~V. 2017, \aap, 598, L1

\bibitem[{{Kovalev} {et~al.}(2020){Kovalev}, {Zobnina}, {Plavin}, \&
  {Blinov}}]{Kovalev2020gaia}
{Kovalev}, Y.~Y., {Zobnina}, D.~I., {Plavin}, A.~V., \& {Blinov}, D. 2020,
  \mnras, 493, L54

\bibitem[{{Kravchenko} {et~al.}(2016){Kravchenko}, {Kovalev}, {Hovatta}, \&
  {Ramakrishnan}}]{Kravchenko2016}
{Kravchenko}, E.~V., {Kovalev}, Y.~Y., {Hovatta}, T., \& {Ramakrishnan}, V.
  2016, \mnras, 462, 2747

\bibitem[{{Kudryavtseva} {et~al.}(2011){Kudryavtseva}, {Gabuzda}, {Aller}, \&
  {Aller}}]{Kudryavtseva2011}
{Kudryavtseva}, N.~A., {Gabuzda}, D.~C., {Aller}, M.~F., \& {Aller}, H.~D.
  2011, \mnras, 415, 1631

\bibitem[{{Kutkin} {et~al.}(2019){Kutkin}, {Pashchenko}, {Sokolovsky},
  {Kovalev}, {Aller}, \& {Aller}}]{Kutkinetal2019}
{Kutkin}, A.~M., {Pashchenko}, I.~N., {Sokolovsky}, K.~V., {et~al.} 2019,
  \mnras, 486, 430

\bibitem[{Kutkin {et~al.}(2014)Kutkin, Sokolovsky, Lisakov, Kovalev,
  Savolainen, Voytsik, Lobanov, Aller, Aller, Lahteenmaki, Tornikoski, Volvach,
  \& Volvach}]{Kutkinetal2014}
Kutkin, A.~M., Sokolovsky, K.~V., Lisakov, M.~M., {et~al.} 2014, \mnras, 437,
  3396

\bibitem[{{Lister} {et~al.}(2009){Lister}, {Aller}, {Aller}, {Cohen}, {Homan},
  {Kadler}, {Kellermann}, {Kovalev}, {Ros}, {Savolainen}, {Zensus}, \&
  {Vermeulen}}]{Lister2009a}
{Lister}, M.~L., {Aller}, H.~D., {Aller}, M.~F., {et~al.} 2009, \aj, 137, 3718

\bibitem[{{Lister} {et~al.}(2018){Lister}, {Aller}, {Aller}, {Hodge}, {Homan},
  {Kovalev}, {Pushkarev}, \& {Savolainen}}]{Lister2018}
{Lister}, M.~L., {Aller}, M.~F., {Aller}, H.~D., {et~al.} 2018, \apjs, 234, 12

\bibitem[{{Lobanov}(1998)}]{Lobanov1997}
{Lobanov}, A.~P. 1998, \aap, 330, 79

\bibitem[{{Lobanov} \& {Zensus}(1999)}]{Lobanov1999}
{Lobanov}, A.~P. \& {Zensus}, J.~A. 1999, \apj, 521, 509

\bibitem[{{Marcaide} \& {Shapiro}(1984)}]{MarcaideShapiro1984}
{Marcaide}, J.~M. \& {Shapiro}, I.~I. 1984, \apj, 276, 56

\bibitem[{Marscher(2010)}]{Marscher2010}
Marscher, A. 2010, in The Jet Paradigm: From Microquasars to Quasars, ed.
  T.~Belloni (Berlin, Heidelberg: Springer Berlin Heidelberg), 173--201

\bibitem[{{Marscher}(1983)}]{Marscher1983}
{Marscher}, A.~P. 1983, \apj, 264, 296

\bibitem[{McKinney {et~al.}(2012)McKinney, Tchekhovskoy, \&
  Blandford}]{McKinney2012}
McKinney, J.~C., Tchekhovskoy, A., \& Blandford, R.~D. 2012, \mnras, 423, 3083

\bibitem[{{Mohan} {et~al.}(2015){Mohan}, {Agarwal}, {Mangalam}, {Gupta},
  {Wiita}, {Volvach}, {Aller}, {Aller}, {Gu}, {L{\"a}hteenm{\"a}ki},
  {Tornikoski}, \& {Volvach}}]{Mohan2015}
{Mohan}, P., {Agarwal}, A., {Mangalam}, A., {et~al.} 2015, \mnras, 452, 2004

\bibitem[{{Narayan} {et~al.}(2003){Narayan}, {Igumenshchev}, \&
  {Abramowicz}}]{Narayan2003}
{Narayan}, R., {Igumenshchev}, I.~V., \& {Abramowicz}, M.~A. 2003, \pasj, 55,
  L69

\bibitem[{{Niell} {et~al.}(2018){Niell}, {Barrett}, {Burns}, {Cappallo},
  {Corey}, {Derome}, {Eckert}, {Elosegui}, {McWhirter}, {Poirier},
  {Rajagopalan}, {Rogers}, {Ruszczyk}, {SooHoo}, {Titus}, {Whitney}, {Behrend},
  {Bolotin}, {Gipson}, {Gordon}, {Himwich}, \& {Petrachenko}}]{Niell2018}
{Niell}, A., {Barrett}, J., {Burns}, A., {et~al.} 2018, Radio Science, 53, 1269

\bibitem[{O'Sullivan \& Gabuzda(2009)}]{OSullivanGabuzda2009}
O'Sullivan, S.~P. \& Gabuzda, D.~C. 2009, \mnras, 400, 26

\bibitem[{{Pacciani} {et~al.}(2010){Pacciani}, {Vittorini}, {Tavani},
  {Fiocchi}, {Vercellone}, {D'Ammando}, {Sakamoto}, {Pian}, {Raiteri},
  {Villata}, {Sasada}, {Itoh}, {Yamanaka}, {Uemura}, {Striani}, {Fugazza},
  {Tiengo}, {Krimm}, {Stroh}, {Falcone}, {Curran}, {Sadun}, {Lahteenmaki},
  {Tornikoski}, {Aller}, {Aller}, {Lin}, {Larionov}, {Leto}, {Takalo},
  {Berdyugin}, {Gurwell}, {Bulgarelli}, {Chen}, {Donnarumma}, {Giuliani},
  {Longo}, {Pucella}, {Argan}, {Barbiellini}, {Caraveo}, {Cattaneo}, {Costa},
  {De Paris}, {Del Monte}, {Di Cocco}, {Evangelista}, {Ferrari}, {Feroci},
  {Fiorini}, {Fuschino}, {Galli}, {Gianotti}, {Labanti}, {Lapshov},
  {Lazzarotto}, {Lipari}, {Marisaldi}, {Mereghetti}, {Morelli}, {Moretti},
  {Morselli}, {Pellizzoni}, {Perotti}, {Piano}, {Picozza}, {Pilia}, {Prest},
  {Rapisarda}, {Rappoldi}, {Rubini}, {Sabatini}, {Soffitta}, {Trifoglio},
  {Trois}, {Vallazza}, {Zanello}, {Colafrancesco}, {Pittori}, {Verrecchia},
  {Santolamazza}, {Lucarelli}, {Giommi}, \& {Salotti}}]{Paccianietal2010}
{Pacciani}, L., {Vittorini}, V., {Tavani}, M., {et~al.} 2010, \apjl, 716, L170

\bibitem[{{Pashchenko} {et~al.}(2020){Pashchenko}, {Plavin}, {Kutkin}, \&
  {Kovalev}}]{Pashchenko2020}
{Pashchenko}, I.~N., {Plavin}, A.~V., {Kutkin}, A.~M., \& {Kovalev}, Y.~Y.
  2020, \mnras, 499, 4515

\bibitem[{{Pauliny-Toth} {et~al.}(1987){Pauliny-Toth}, {Porcas}, {Zensus},
  {Kellermann}, {Wu}, {Nicholson}, \& {Mantovani}}]{Pauliny-Toth1987}
{Pauliny-Toth}, I.~I.~K., {Porcas}, R.~W., {Zensus}, J.~A., {et~al.} 1987,
  \nat, 328, 778

\bibitem[{{Petrov} {et~al.}(2009){Petrov}, {Gordon}, {Gipson}, {MacMillan},
  {Ma}, {Fomalont}, {Walker}, \& {Carabajal}}]{Petrov2009}
{Petrov}, L., {Gordon}, D., {Gipson}, J., {et~al.} 2009, Journal of Geodesy,
  83, 859

\bibitem[{{Petrov} \& {Kovalev}(2017{\natexlab{a}})}]{PetKov2017}
{Petrov}, L. \& {Kovalev}, Y.~Y. 2017{\natexlab{a}}, \mnras, 471, 3775

\bibitem[{{Petrov} \& {Kovalev}(2017{\natexlab{b}})}]{PetKov2017letter}
{Petrov}, L. \& {Kovalev}, Y.~Y. 2017{\natexlab{b}}, \mnras, 467, L71

\bibitem[{{Petrov} {et~al.}(2019){Petrov}, {Kovalev}, \& {Plavin}}]{Petrov2019}
{Petrov}, L., {Kovalev}, Y.~Y., \& {Plavin}, A.~V. 2019, \mnras, 482, 3023

\bibitem[{{Pian} {et~al.}(2006){Pian}, {Foschini}, {Beckmann}, {Soldi},
  {T{\"u}rler}, {Gehrels}, {Ghisellini}, {Giommi}, {Maraschi}, {Pursimo},
  {Raiteri}, {Tagliaferri}, {Tornikoski}, {Tosti}, {Treves}, {Villata}, {Barr},
  {Courvoisier}, {Di Cocco}, {Hudec}, {Fuhrmann}, {Malaguti}, {Persic},
  {Tavecchio}, \& {Walter}}]{Pian2006}
{Pian}, E., {Foschini}, L., {Beckmann}, V., {et~al.} 2006, \aap, 449, L21

\bibitem[{{Plavin} {et~al.}(2019{\natexlab{a}}){Plavin}, {Kovalev}, \&
  {Petrov}}]{PKP2019}
{Plavin}, A.~V., {Kovalev}, Y.~Y., \& {Petrov}, L.~Y. 2019{\natexlab{a}}, \apj,
  871, 143

\bibitem[{{Plavin} {et~al.}(2019{\natexlab{b}}){Plavin}, {Kovalev},
  {Pushkarev}, \& {Lobanov}}]{Plavin2019}
{Plavin}, A.~V., {Kovalev}, Y.~Y., {Pushkarev}, A.~B., \& {Lobanov}, A.~P.
  2019{\natexlab{b}}, \mnras, 485, 1822

\bibitem[{{Porcas}(2009)}]{Porcas2009}
{Porcas}, R.~W. 2009, A\&A, 505, L1

\bibitem[{Pushkarev {et~al.}(2018)Pushkarev, Butuzova, Kovalev, \&
  Hovatta}]{Pushkarev2018}
Pushkarev, A.~B., Butuzova, M.~S., Kovalev, Y.~Y., \& Hovatta, T. 2018, \mnras,
  482, 2336

\bibitem[{{Pushkarev} {et~al.}(2012){Pushkarev}, {Hovatta}, {Kovalev},
  {Lister}, {Lobanov}, {Savolainen}, \& {Zensus}}]{Pushkarevetal2012}
{Pushkarev}, A.~B., {Hovatta}, T., {Kovalev}, Y.~Y., {et~al.} 2012, A\&A, 545,
  A113

\bibitem[{{Qian} {et~al.}(2021){Qian}, {Britzen}, {Krichbaum}, \&
  {Witzel}}]{Qian2021}
{Qian}, S.~J., {Britzen}, S., {Krichbaum}, T.~P., \& {Witzel}, A. 2021, \aap,
  653, A7

\bibitem[{{Raiteri} {et~al.}(2011){Raiteri}, {Villata}, {Aller}, {Gurwell},
  {Kurtanidze}, {L{\"a}hteenm{\"a}ki}, {Larionov}, {Romano}, {Vercellone},
  {Agudo}, {Aller}, {Arkharov}, {Bach}, {Ben{\'\i}tez}, {Berdyugin}, {Blinov},
  {Borisova}, {B{\"o}ttcher}, {Bravo Calle}, {Buemi}, {Calcidese}, {Carosati},
  {Casas}, {Chen}, {Efimova}, {G{\'o}mez}, {Gusbar}, {Hawkins}, {Heidt},
  {Hiriart}, {Hsiao}, {Jordan}, {Jorstad}, {Joshi}, {Kimeridze}, {Koptelova},
  {Konstantinova}, {Kopatskaya}, {Kurtanidze}, {Larionova}, {Larionova},
  {Leto}, {Li}, {Ligustri}, {Lindfors}, {Lister}, {Marscher}, {Molina},
  {Morozova}, {Nieppola}, {Nikolashvili}, {Nilsson}, {Palma}, {Pasanen},
  {Reinthal}, {Roberts}, {Ros}, {Roustazadeh}, {Sadun}, {Sakamoto}, {Schwartz},
  {Sigua}, {Sillanp{\"a}{\"a}}, {Takalo}, {Tammi}, {Taylor}, {Tornikoski},
  {Trigilio}, {Troitsky}, {Umana}, {Volvach}, \& {Yuldasheva}}]{Raiteri2011}
{Raiteri}, C.~M., {Villata}, M., {Aller}, M.~F., {et~al.} 2011, \aap, 534, A87

\bibitem[{{Raiteri} {et~al.}(2008){Raiteri}, {Villata}, {Larionov}, {Gurwell},
  {Chen}, {Kurtanidze}, {Aller}, {B{\"o}ttcher}, {Calcidese}, {Hroch},
  {L{\"a}hteenm{\"a}ki}, {Lee}, {Nilsson}, {Ohlert}, {Papadakis}, {Agudo},
  {Aller}, {Angelakis}, {Arkharov}, {Bach}, {Bachev}, {Berdyugin}, {Buemi},
  {Carosati}, {Charlot}, {Chatzopoulos}, {Forn{\'e}}, {Frasca}, {Fuhrmann},
  {G{\'o}mez}, {Gupta}, {Hagen-Thorn}, {Hsiao}, {Jordan}, {Jorstad},
  {Konstantinova}, {Kopatskaya}, {Krichbaum}, {Lanteri}, {Larionova}, {Latev},
  {Le Campion}, {Leto}, {Lin}, {Marchili}, {Marilli}, {Marscher}, {McBreen},
  {Mihov}, {Nesci}, {Nicastro}, {Nikolashvili}, {Novak}, {Ovcharov}, {Pian},
  {Principe}, {Pursimo}, {Ragozzine}, {Ros}, {Sadun}, {Sagar}, {Semkov},
  {Smart}, {Smith}, {Strigachev}, {Takalo}, {Tavani}, {Tornikoski}, {Trigilio},
  {Uckert}, {Umana}, {Valcheva}, {Vercellone}, {Volvach}, \&
  {Wiesemeyer}}]{Raiteri2008}
{Raiteri}, C.~M., {Villata}, M., {Larionov}, V.~M., {et~al.} 2008, \aap, 491,
  755

\bibitem[{{Remillard}(2005)}]{Remillard2005}
{Remillard}, R. 2005, The Astronomer's Telegram, 484, 1

\bibitem[{{Sargent} {et~al.}(1988){Sargent}, {Steidel}, \&
  {Boksenberg}}]{Sargent1988}
{Sargent}, W. L.~W., {Steidel}, C.~C., \& {Boksenberg}, A. 1988, \apj, 334, 22

\bibitem[{{Sarkar} {et~al.}(2019){Sarkar}, {Chitnis}, {Gupta}, {Gaur}, {Patel},
  {Wiita}, {Volvach}, {Tornikoski}, {Chamani}, {Enestam},
  {L{\"a}hteenm{\"a}ki}, {Tammi}, {Vera}, \& {Volvach}}]{Sarkar2019}
{Sarkar}, A., {Chitnis}, V.~R., {Gupta}, A.~C., {et~al.} 2019, \apj, 887, 185

\bibitem[{{Savolainen} {et~al.}(2002){Savolainen}, {Wiik}, {Valtaoja},
  {Jorstad}, \& {Marscher}}]{Savolainen2002}
{Savolainen}, T., {Wiik}, K., {Valtaoja}, E., {Jorstad}, S.~G., \& {Marscher},
  A.~P. 2002, \aap, 394, 851

\bibitem[{{Savolainen} {et~al.}(2008{\natexlab{a}}){Savolainen}, {Wiik},
  {Valtaoja}, \& {Tornikoski}}]{Savolainen2008}
{Savolainen}, T., {Wiik}, K., {Valtaoja}, E., \& {Tornikoski}, M.
  2008{\natexlab{a}}, in Astronomical Society of the Pacific Conference Series,
  Vol. 386, Extragalactic Jets: Theory and Observation from Radio to Gamma Ray,
  ed. T.~A. {Rector} \& D.~S. {De Young}, 451

\bibitem[{{Savolainen} {et~al.}(2008{\natexlab{b}}){Savolainen}, {Wiik},
  {Valtaoja}, \& {Tornikoski}}]{Savolainen2008b}
{Savolainen}, T., {Wiik}, K., {Valtaoja}, E., \& {Tornikoski}, M.
  2008{\natexlab{b}}, in The role of VLBI in the Golden Age for Radio
  Astronomy, Vol.~9, 9

\bibitem[{{Sharma} {et~al.}(2022){Sharma}, {Massi}, \& {Torricelli-Ciamponi
  }}]{Sharma2022}
{Sharma}, R., {Massi}, H., \& {Torricelli-Ciamponi }, G. 2022, \aap, in press

\bibitem[{{Shepherd}(1997)}]{Shepherd1997}
{Shepherd}, M.~C. 1997, in Astronomical Society of the Pacific Conference
  Series, Vol. 125, Astronomical Data Analysis Software and Systems VI, ed.
  G.~{Hunt} \& H.~{Payne}, 77

\bibitem[{{Sokolovsky} {et~al.}(2011){Sokolovsky}, {Kovalev}, {Pushkarev}, \&
  {Lobanov}}]{Sokolovskyetal2011}
{Sokolovsky}, K.~V., {Kovalev}, Y.~Y., {Pushkarev}, A.~B., \& {Lobanov}, A.~P.
  2011, \aap, 532, A38

\bibitem[{{Tchekhovskoy}(2015)}]{Tchekhovskoy2015}
{Tchekhovskoy}, A. 2015, in Astrophysics and Space Science Library, Vol. 414,
  The Formation and Disruption of Black Hole Jets, ed. I.~{Contopoulos},
  D.~{Gabuzda}, \& N.~{Kylafis}, 45

\bibitem[{{Tchekhovskoy} {et~al.}(2011){Tchekhovskoy}, {Narayan}, \&
  {McKinney}}]{Tchekhovskoy2011}
{Tchekhovskoy}, A., {Narayan}, R., \& {McKinney}, J.~C. 2011, \mnras, 418, L79

\bibitem[{{Ter\"asranta} {et~al.}(1998){Ter\"asranta}, {Tornikoski}, {Mujunen},
  {Karlamaa}, {Valtonen}, {Henelius}, {Urpo}, {Lainela}, {Pursimo}, {Nilsson},
  {Wiren}, {Laehteenmaeki}, {Korpi}, {Rekola}, {Heinaemaeki}, {Hanski},
  {Nurmi}, {Kokkonen}, {Keinaenen}, {Joutsamo}, {Oksanen}, {Pietilae},
  {Valtaoja}, {Valtonen}, \& {Koenoenen}}]{Teraesrantaetal}
{Ter\"asranta}, H., {Tornikoski}, M., {Mujunen}, A., {et~al.} 1998, \aaps, 132,
  305

\bibitem[{{Vercellone} {et~al.}(2009){Vercellone}, {Chen}, {Vittorini},
  {Giuliani}, {D'Ammando}, {Tavani}, {Donnarumma}, {Pucella}, {Raiteri},
  {Villata}, {Chen}, {Tosti}, {Impiombato}, {Romano}, {Belfiore}, {De Luca},
  {Novara}, {Senziani}, {Bazzano}, {Fiocchi}, {Ubertini}, {Ferrari}, {Argan},
  {Barbiellini}, {Boffelli}, {Bulgarelli}, {Caraveo}, {Cattaneo}, {Cocco},
  {Costa}, {del Monte}, {DeParis}, {Di Cocco}, {Evangelista}, {Feroci},
  {Fiorini}, {Fornari}, {Froysland}, {Fuschino}, {Galli}, {Gianotti},
  {Labanti}, {Lapshov}, {Lazzarotto}, {Lipari}, {Longo}, {Marisaldi},
  {Mereghetti}, {Morselli}, {Pellizzoni}, {Pacciani}, {Perotti}, {Picozza},
  {Prest}, {Rapisarda}, {Rappoldi}, {Soffitta}, {Trifoglio}, {Trois},
  {Vallazza}, {Zambra}, {Zanello}, {Pittori}, {Verrecchia}, {Santolamazza},
  {Preger}, {Gasparrini}, {Cutini}, {Giommi}, {Colafrancesco}, \&
  {Salotti}}]{Vercelloneetal2009}
{Vercellone}, S., {Chen}, A.~W., {Vittorini}, V., {et~al.} 2009, \apj, 690,
  1018

\bibitem[{{Vercellone} {et~al.}(2010){Vercellone}, {D'Ammando}, {Vittorini},
  {Donnarumma}, {Pucella}, {Tavani}, {Ferrari}, {Raiteri}, {Villata}, {Romano},
  {Krimm}, {Tiengo}, {Chen}, {Giovannini}, {Venturi}, {Giroletti}, {Kovalev},
  {Sokolovsky}, {Pushkarev}, {Lister}, {Argan}, {Barbiellini}, {Bulgarelli},
  {Caraveo}, {Cattaneo}, {Cocco}, {Costa}, {Del Monte}, {De Paris}, {Di Cocco},
  {Evangelista}, {Feroci}, {Fiorini}, {Fornari}, {Froysland}, {Fuschino},
  {Galli}, {Gianotti}, {Labanti}, {Lapshov}, {Lazzarotto}, {Lipari}, {Longo},
  {Giuliani}, {Marisaldi}, {Mereghetti}, {Morselli}, {Pellizzoni}, {Pacciani},
  {Perotti}, {Piano}, {Picozza}, {Pilia}, {Prest}, {Rapisarda}, {Rappoldi},
  {Sabatini}, {Soffitta}, {Striani}, {Trifoglio}, {Trois}, {Vallazza},
  {Zambra}, {Zanello}, {Pittori}, {Verrecchia}, {Santolamazza}, {Giommi},
  {Colafrancesco}, {Salotti}, {Agudo}, {Aller}, {Aller}, {Arkharov}, {Bach},
  {Bachev}, {Beltrame}, {Ben{\'{\i}}tez}, {B{\"o}ttcher}, {Buemi}, {Calcidese},
  {Capezzali}, {Carosati}, {Chen}, {Da Rio}, {Di Paola}, {Dolci}, {Dultzin},
  {Forn{\'e}}, {G{\'o}mez}, {Gurwell}, {Hagen-Thorn}, {Halkola}, {Heidt},
  {Hiriart}, {Hovatta}, {Hsiao}, {Jorstad}, {Kimeridze}, {Konstantinova},
  {Kopatskaya}, {Koptelova}, {Kurtanidze}, {L{\"a}hteenm{\"a}ki}, {Larionov},
  {Leto}, {Ligustri}, {Lindfors}, {Lopez}, {Marscher}, {Mujica},
  {Nikolashvili}, {Nilsson}, {Mommert}, {Palma}, {Pasanen}, {Roca-Sogorb},
  {Ros}, {Roustazadeh}, {Sadun}, {Saino}, {Sigua}, {Sorcia}, {Takalo},
  {Tornikoski}, {Trigilio}, {Turchetti}, \& {Umana}}]{Vercelloneeatlal2010}
{Vercellone}, S., {D'Ammando}, F., {Vittorini}, V., {et~al.} 2010, \apj, 712,
  405

\bibitem[{{Villata} {et~al.}(2006){Villata}, {Raiteri}, {Balonek}, {Aller},
  {Jorstad}, {Kurtanidze}, {Nicastro}, {Nilsson}, {Aller}, {Arai}, {Arkharov},
  {Bach}, {Ben{\'\i}tez}, {Berdyugin}, {Buemi}, {B{\"o}ttcher}, {Carosati},
  {Casas}, {Caulet}, {Chen}, {Chiang}, {Chou}, {Ciprini}, {Coloma}, {di Rico},
  {D{\'\i}az}, {Efimova}, {Forsyth}, {Frasca}, {Fuhrmann}, {Gadway}, {Gupta},
  {Hagen-Thorn}, {Harvey}, {Heidt}, {Hernandez-Toledo}, {Hroch}, {Hu}, {Hudec},
  {Ibrahimov}, {Imada}, {Kamata}, {Kato}, {Katsuura}, {Konstantinova},
  {Kopatskaya}, {Kotaka}, {Kovalev}, {Kovalev}, {Krichbaum}, {Kubota},
  {Kurosaki}, {Lanteri}, {Larionov}, {Larionova}, {Laurikainen}, {Lee}, {Leto},
  {L{\"a}hteenm{\"a}ki}, {L{\'o}pez-Cruz}, {Marilli}, {Marscher}, {McHardy},
  {Mondal}, {Mullan}, {Napoleone}, {Nikolashvili}, {Ohlert}, {Postnikov},
  {Pursimo}, {Ragni}, {Ros}, {Sadakane}, {Sadun}, {Savolainen}, {Sergeeva},
  {Sigua}, {Sillanp{\"a}{\"a}}, {Sixtova}, {Sumitomo}, {Takalo},
  {Ter{\"a}sranta}, {Tornikoski}, {Trigilio}, {Umana}, {Volvach}, {Voss}, \&
  {Wortel}}]{Villata2006}
{Villata}, M., {Raiteri}, C.~M., {Balonek}, T.~J., {et~al.} 2006, \aap, 453,
  817

\bibitem[{{Voitsik} {et~al.}(2018){Voitsik}, {Pushkarev}, {Kovalev}, {Plavin},
  {Lobanov}, \& {Ipatov}}]{Voitsik2018}
{Voitsik}, P.~A., {Pushkarev}, A.~B., {Kovalev}, Y.~Y., {et~al.} 2018,
  Astronomy Reports, 62, 787

\bibitem[{{Volvach} {et~al.}(2021){Volvach}, {Volvach}, \&
  {Larionov}}]{Volvach2021}
{Volvach}, A.~E., {Volvach}, L.~N., \& {Larionov}, M.~G. 2021, \aap, 648, A27

\bibitem[{Walker {et~al.}(2000)Walker, Dhawan, Romney, Kellermann, \&
  Vermeulen}]{Walkeretal2000}
Walker, R.~C., Dhawan, V., Romney, J.~D., Kellermann, K.~I., \& Vermeulen,
  R.~C. 2000, The Astrophysical Journal, 530, 233

\bibitem[{{Wright}(2006)}]{Wright2006}
{Wright}, E.~L. 2006, \pasp, 118, 1711

\bibitem[{{Xu} {et~al.}(2021{\natexlab{a}}){Xu}, {Anderson}, {Heinkelmann},
  {Lunz}, {Schuh}, \& {Wang}}]{Xu2021c}
{Xu}, M.~H., {Anderson}, J.~M., {Heinkelmann}, R., {et~al.} 2021{\natexlab{a}},
  Journal of Geodesy, 95, 51

\bibitem[{{Xu} {et~al.}(2017){Xu}, {Heinkelmann}, {Anderson}, {Mora-Diaz},
  {Karbon}, {Schuh}, \& {Wang}}]{Xu2017}
{Xu}, M.~H., {Heinkelmann}, R., {Anderson}, J.~M., {et~al.} 2017, Journal of
  Geodesy, 91, 767

\bibitem[{{Xu} {et~al.}(2021{\natexlab{b}}){Xu}, {Lunz}, {Anderson},
  {Savolainen}, {Zubko}, \& {Schuh}}]{Xu2021b}
{Xu}, M.~H., {Lunz}, S., {Anderson}, J.~M., {et~al.} 2021{\natexlab{b}}, \aap,
  647, A189

\bibitem[{{Xu} {et~al.}(2021{\natexlab{c}}){Xu}, {Savolainen}, {Zubko},
  {Poutanen}, {Lunz}, {Schuh}, \& {Wang}}]{Xu2021a}
{Xu}, M.~H., {Savolainen}, T., {Zubko}, N., {et~al.} 2021{\natexlab{c}},
  Journal of Geophysical Research (Solid Earth), 126, e21238

\bibitem[{{Zamaninasab} {et~al.}(2014){Zamaninasab}, {Clausen-Brown},
  {Savolainen}, \& {Tchekhovskoy}}]{Zamaninasab2014}
{Zamaninasab}, M., {Clausen-Brown}, E., {Savolainen}, T., \& {Tchekhovskoy}, A.
  2014, \nat, 510, 126

\bibitem[{{Zamaninasab} {et~al.}(2013){Zamaninasab}, {Savolainen},
  {Clausen-Brown}, {Hovatta}, {Lister}, {Krichbaum}, {Kovalev}, \&
  {Pushkarev}}]{Zamaninasab2013}
{Zamaninasab}, M., {Savolainen}, T., {Clausen-Brown}, E., {et~al.} 2013,
  \mnras, 436, 3341

\bibitem[{Zdziarski {et~al.}(2015)Zdziarski, Sikora, Pjanka, \&
  Tchekhovskoy}]{Zdziarskietal2015}
Zdziarski, A.~A., Sikora, M., Pjanka, P., \& Tchekhovskoy, A. 2015, \mnras,
  451, 927

\end{thebibliography}

\begin{appendix}

\onecolumn
\section{Core-shift magnitudes}
\label{appendix:core-shift-values}
\begin{longtable}{cccccc}
 \caption{\label{coreshift-table}Core-shift values per epoch and frequency pair. }
% \ericom{See comment on significant digits in the error bar when starting with 1.  Use mathematical minus when this happens. }
\\
\hline\hline
Date & \begin{tabular}[c]{@{}c@{}}Frequency\\ pair\end{tabular} & \begin{tabular}[c]{@{}c@{}}R.A.\\ (mas)\end{tabular} & \begin{tabular}[c]{@{}c@{}}Dec\\ (mas)\end{tabular} & \begin{tabular}[c]{@{}c@{}}Absolute values\\ (mas)\end{tabular} & \begin{tabular}[c]{@{}c@{}}Projected absolute\\  values (mas)\end{tabular} \\
\hline 
\endfirsthead
\caption{continued.}\\
\hline\hline 
Date & \begin{tabular}[c]{@{}c@{}}Frequency\\ pair\end{tabular} & \begin{tabular}[c]{@{}c@{}}R.A.\\ (mas)\end{tabular} & \begin{tabular}[c]{@{}c@{}}Dec\\ (mas)\end{tabular} & \begin{tabular}[c]{@{}c@{}}Absolute values\\ (mas)\end{tabular} & \begin{tabular}[c]{@{}c@{}}Projected absolute\\  values (mas)\end{tabular} \\
\hline 
\endhead
\hline
\endfoot
\multirow{4}{*}{2005-05-19\tablefootmark{a}} 
& CX &	$-$0.15 $\pm$ 0.06 & 0.03 $\pm$ 0.07 & 0.15 $\pm$ 0.06 & 0.15 $\pm$ 0.06 \\
& XU &	$-$0.06 $\pm$ 0.02 & $-$0.04 $\pm$ 0.02 & 0.07 $\pm$ 0.02  & 0.06 $\pm$ 0.03 \\
& UK &	$-$0.04 $\pm$ 0.01 & 0.04 $\pm$ 0.01 & 0.06 $\pm$ 0.01  & 0.05 $\pm$ 0.02 \\  
& KQ &  $-$0.020 $\pm$ 0.005 & $-$0.003 $\pm$ 0.005  & 0.020 $\pm$ 0.005  & 0.020 $\pm$ 0.005 \\
& CQ & $-$0.27 $\pm$ 0.06  & 0.03 $\pm$ 0.07 & 0.27 $\pm$ 0.06   & 0.27  $\pm$ 0.06  \\ \hline \noalign{\smallskip}
\multirow{4}{*}{2005-07-14\tablefootmark{a}} 
& CX &  $-$0.42 $\pm$ 0.06 & 0.04 $\pm$ 0.07 & 0.42 $\pm$ 0.06 & 0.42 $\pm$ 0.06 \\
& XU &	$-$0.29 $\pm$ 0.03 & 0.10 $\pm$ 0.02 & 0.31 $\pm$ 0.03 & 0.30 $\pm$ 0.03 \\
& UK &	$-$0.04 $\pm$ 0.01 & 0.004 $\pm$ 0.009 & 0.04 $\pm$ 0.01 & 0.04 $\pm$ 0.01  \\
& KQ &	$-$0.020 $\pm$ 0.006 & 0.003 $\pm$ 0.005 & 0.020 $\pm$ 0.006 & 0.020 $\pm$ 0.006\\
& CQ & $-$0.77 $\pm$ 0.06  & 0.15 $\pm$ 0.07 & 0.78 $\pm$ 0.06   & 0.78 $\pm$ 0.06  \\\hline \noalign{\smallskip}
\multirow{4}{*}{2005-09-01\tablefootmark{a}} 
& CX &	$-$0.41 $\pm$ 0.03 & 0.10 $\pm$ 0.03 & 0.42 $\pm$ 0.03 & 0.42 $\pm$ 0.03 \\
& XU &	$-$0.35 $\pm$ 0.02 & 0.06 $\pm$ 0.02 & 0.35 $\pm$ 0.02 & 0.35 $\pm$ 0.02 \\
& UK &	$-$0.06 $\pm$ 0.02 & 0.01 $\pm$ 0.01 & 0.06 $\pm$ 0.02 & 0.06 $\pm$ 0.02 \\
& KQ &	$-$0.02 $\pm$ 0.01 & 0.02 $\pm$ 0.01 & 0.02 $\pm$ 0.01 & 0.02 $\pm$ 0.01 \\
& CQ &  $-$0.84 $\pm$ 0.04 & 0.18 $\pm$ 0.04 & 0.86 $\pm$ 0.04 & 0.86 $\pm$ 0.04  \\\hline \noalign{\smallskip}
\multirow{4}{*}{2005-12-04\tablefootmark{b}} 
& CX &	$-$0.34 $\pm$ 0.03 & 0.07 $\pm$ 0.03 & 0.35 $\pm$ 0.03  & 0.35 $\pm$ 0.03  \\
& XU &	$-$0.16 $\pm$ 0.02 & 0.01 $\pm$ 0.02 & 0.16 $\pm$ 0.02  & 0.16 $\pm$ 0.02   \\
& UK &	$-$0.09 $\pm$ 0.02 & 0.09 $\pm$ 0.01 & 0.13 $\pm$ 0.02  & 0.12 $\pm$ 0.02 \\
& KQ &	$-$0.02 $\pm$ 0.02 & 0.03 $\pm$ 0.01 & 0.03 $\pm$ 0.01  & 0.02 $\pm$ 0.02  \\
& CQ &  $-$0.62 $\pm$ 0.04 & 0.20 $\pm$ 0.04 & 0.65 $\pm$ 0.04  & 0.65 $\pm$ 0.04  \\\hline \noalign{\smallskip}
\multirow{4}{*}{2006-10-03\tablefootmark{b}} 
& CX &	$-$0.23 $\pm$ 0.03 & 0.07 $\pm$ 0.03 & 0.24 $\pm$ 0.03  & 0.24 $\pm$ 0.03  \\
& XU &	$-$0.06 $\pm$ 0.02 & $-$0.001 $\pm$	0.017 & 0.06 $\pm$ 0.02 & 0.05 $\pm$ 0.02   \\
& UK &	$-$0.03 $\pm$ 0.01 & 0.02 $\pm$ 0.01 & 0.04 $\pm$ 0.01 & 0.04 $\pm$ 0.01   \\
& KQ &	$-$0.046 $\pm$ 0.005 & 0.007 $\pm$ 0.005 & 0.046 $\pm$ 0.005 & 0.046 $\pm$ 0.005  \\
& CQ &  $-$0.36  $\pm$ 0.03  & 0.09 $\pm$ 0.03 & 0.37 $\pm$ 0.03 & 0.37 $\pm$ 0.03  \\\hline \noalign{\smallskip}
\multirow{4}{*}{2006-12-04}
& CX &	$-$0.19 $\pm$ 0.03 & 0.09 $\pm$ 0.03  & 0.21 $\pm$ 0.03 & 0.21 $\pm$ 0.03  \\ 
& XU &	$-$0.11 $\pm$ 0.02 & $-$0.01 $\pm$ 0.02  & 0.11 $\pm$ 0.02 & 0.10 $\pm$ 0.02  \\ 
& UK &	$-$0.05 $\pm$ 0.02 & 0.04 $\pm$ 0.02  & 0.06 $\pm$ 0.02  & 0.06 $\pm$ 0.02  \\ 
& KQ &	$-$0.021 $\pm$ 0.007 & 0.008 $\pm$ 0.006  & 0.02 $\pm$ 0.01 & 0.02 $\pm$ 0.01  \\
& CQ &  $-$0.36 $\pm$ 0.04  & 0.13 $\pm$ 0.04  & 0.38 $\pm$ 0.04 & 0.38 $\pm$ 0.04  \\\hline  \noalign{\smallskip}
\multirow{4}{*}{2007-01-26\tablefootmark{b}} 
& CX &	$-$0.11 $\pm$ 0.03 & 0.11 $\pm$ 0.03  & 0.15 $\pm$ 0.03  & 0.13 $\pm$ 0.03  \\
& XU &	$-$0.11 $\pm$ 0.02 & $-$0.02 $\pm$ 0.02  & 0.12 $\pm$ 0.02 & 0.10 $\pm$ 0.02 \\
& UK &	$-$0.04 $\pm$ 0.01 &  0.01 $\pm$ 0.01  & 0.04 $\pm$ 0.01 & 0.04 $\pm$ 0.01   \\
& KQ &	$-$0.020 $\pm$ 0.005 & $-$0.010 $\pm$ 0.005  & 0.022 $\pm$ 0.005 & 0.016 $\pm$ 0.005  \\ 
& CQ &  $-$0.28 $\pm$ 0.04  & 0.08 $\pm$ 0.03 & 0.29 $\pm$ 0.04 & 0.29 $\pm$ 0.04  \\\hline  \noalign{\smallskip}
\multirow{4}{*}{2007-04-26\tablefootmark{b}} 
& CX &	$-$0.09 $\pm$ 0.03 & 0.08 $\pm$ 0.03 & 0.12 $\pm$ 0.03  & 0.11 $\pm$ 0.04 \\
& XU &	$-$0.13 $\pm$ 0.02 & $-$0.001 $\pm$ 0.014 & 0.13 $\pm$ 0.02 & 0.12 $\pm$ 0.02  \\
& UK &	$-$0.05 $\pm$ 0.01 & 0.02 $\pm$ 0.01 & 0.05 $\pm$ 0.01  & 0.05 $\pm$ 0.01   \\ 
& KQ &	$-$0.022 $\pm$ 0.007 & 0.006 $\pm$ 0.005  & 0.023 $\pm$ 0.007  & 0.023 $\pm$ 0.007  \\  
& CQ & $-$0.29 $\pm$ 0.04  & 0.10  $\pm$ 0.04 & 0.31 $\pm$ 0.04 & 0.31 $\pm$ 0.04 \\\hline  \noalign{\smallskip}
\multirow{4}{*}{2007-06-16} 
& CX &	$-$0.18 $\pm$ 0.03 & 0.08 $\pm$ 0.03 & 0.19 $\pm$ 0.03  & 0.19 $\pm$ 0.03 \\
& XU &	$-$0.16 $\pm$ 0.02 & 0.006 $\pm$ 0.015 & 0.16 $\pm$ 0.02  & 0.16 $\pm$ 0.02  \\
& UK &	$-$0.06 $\pm$ 0.01 & 0.02 $\pm$ 0.01 & 0.07 $\pm$ 0.01 & 0.07 $\pm$ 0.01 \\
& KQ &	$-$0.020 $\pm$ 0.005 & 0.007 $\pm$ 0.005 & 0.021 $\pm$ 0.005  & 0.021 $\pm$ 0.005 \\ 
& CQ & $-$0.42 $\pm$ 0.04 & 0.11 $\pm$ 0.03  & 0.43 $\pm$ 0.04  & 0.43 $\pm$ 0.04  \\\hline \noalign{\smallskip}
\multirow{4}{*}{2007-07-25\tablefootmark{b}} 
& CX &	$-$0.20 $\pm$ 0.03 & 0.03 $\pm$ 0.03 & 0.20 $\pm$ 0.03  & 0.20 $\pm$ 0.03  \\
& XU &	$-$0.16 $\pm$ 0.02 & 0.03 $\pm$ 0.02 & 0.16 $\pm$ 0.02  & 0.16 $\pm$ 0.02  \\
& UK &	$-$0.02 $\pm$ 0.01 & $-$0.002 $\pm$ 0.008 & 0.02 $\pm$ 0.01  & 0.02 $\pm$ 0.01  \\
& KQ &	$-$0.04 $\pm$ 0.01 & 0.004 $\pm$ 0.011 & 0.04 $\pm$ 0.01  & 0.04 $\pm$ 0.01  \\ 
& CQ & $-$0.42 $\pm$ 0.04 & 0.06 $\pm$ 0.03 & 0.42 $\pm$ 0.04 & 0.42 $\pm$ 0.04  \\\hline \noalign{\smallskip}
\multirow{4}{*}{2007-09-13\tablefootmark{b}} 
& CX &	$-$0.31 $\pm$ 0.04 & 0.06 $\pm$ 0.04 & 0.32 $\pm$ 0.04 & 0.32 $\pm$ 0.04    \\
& XU &	$-$0.23 $\pm$ 0.02 & 0.04 $\pm$ 0.02 & 0.23 $\pm$ 0.02 & 0.23 $\pm$ 0.02   \\
& UK &	$-$0.04 $\pm$ 0.02 & 0.02 $\pm$ 0.02 & 0.04 $\pm$ 0.02 & 0.04 $\pm$ 0.02   \\
& KQ &	$-$0.08 $\pm$ 0.01 & 0.05 $\pm$ 0.01 & 0.09 $\pm$ 0.01 & 0.09 $\pm$ 0.01  \\ 
& CQ &  $-$0.66 $\pm$ 0.05 & 0.16 $\pm$ 0.04 & 0.68 $\pm$ 0.05 & 0.68 $\pm$ 0.05  \\\hline
\multirow{4}{*}{2008-01-03\tablefootmark{b}} 
& CX &  $-$0.92 $\pm$ 0.03 & 0.21 $\pm$ 0.04  & 0.94 $\pm$ 0.03  & 0.92 $\pm$ 0.04 \\
& XU &	$-$0.26 $\pm$ 0.02 & 0.18 $\pm$ 0.03  & 0.32 $\pm$ 0.02  & 0.32 $\pm$ 0.03  \\
& UK &	$-$0.13 $\pm$ 0.01 & 0.14 $\pm$ 0.02  & 0.19 $\pm$ 0.02  & 0.18 $\pm$ 0.01 \\
& KQ &	$-$0.15 $\pm$ 0.01 & 0.21 $\pm$ 0.02  & 0.26 $\pm$ 0.02  & 0.23 $\pm$ 0.01  \\ 
& CQ &  $-$1.47 $\pm$ 0.04 & 0.74 $\pm$ 0.05  & 1.64 $\pm$ 0.04  & 1.64 $\pm$ 0.05  \\\hline \noalign{\smallskip}
\multirow{4}{*}{2008-12-07} 
& CXl  & $-$0.22 $\pm$ 0.03 &	$-$0.01 $\pm$ 0.03  & 0.22 $\pm$ 0.03  & 0.22 $\pm$ 0.03 \\
& XlXh & $-$0.05 $\pm$ 0.03 &	0.007 $\pm$	0.026 & 0.05 $\pm$ 0.03  & 0.05 $\pm$ 0.03  \\
& XhU  & $-$0.02 $\pm$ 0.03 & $-$0.05 $\pm$	0.02  & 0.05 $\pm$ 0.02  & 0.02 $\pm$ 0.03  \\
& UKl  & $-$0.11 $\pm$ 0.01 &	0.05 $\pm$ 0.02   & 0.12 $\pm$ 0.01  & 0.10 $\pm$ 0.01 \\
& KlKh & $-$0.02 $\pm$ 0.01 &	0.001 $\pm$ 0.010 & 0.02 $\pm$ 0.01  & 0.02 $\pm$ 0.01  \\
& KhQ  & $-$0.05 $\pm$ 0.01 &	0.007 $\pm$	0.008 & 0.05 $\pm$ 0.01  & 0.05 $\pm$ 0.01  \\ 
& CQ   & $-$0.46 $\pm$ 0.05 & 0.002 $\pm$ 0.046 & 0.46 $\pm$ 0.05  & 0.46 $\pm$ 0.05  \\\hline \noalign{\smallskip}
 \multirow{4}{*}{2009-09-22} 
& CX &	$-$0.10 $\pm$ 0.03   & $-$0.03 $\pm$ 0.05   & 0.10 $\pm$ 0.04  & 0.09 $\pm$ 0.03  \\
& XU &	$-$0.09 $\pm$ 0.02   & $-$0.02 $\pm$ 0.02   & 0.09 $\pm$ 0.02  & 0.08 $\pm$ 0.02  \\
& UK &	$-$0.11 $\pm$ 0.01   &  0.09 $\pm$ 0.01   & 0.14 $\pm$ 0.01  & 0.13 $\pm$ 0.02 \\
& KQ &	$-$0.005 $\pm$ 0.011 &  0.029 $\pm$ 0.007 & 0.029 $\pm$ 0.007  & 0.01 $\pm$ 0.01 \\ 
& CQ & $-$0.30 $\pm$ 0.04 & 0.07 $\pm$ 0.05 & 0.31 $\pm$ 0.04 & 0.31 $\pm$ 0.04  \\\hline \noalign{\smallskip}
\multirow{4}{*}{2009-10-22} 
& CX &	$-$0.34 $\pm$ 0.04 &  0.13 $\pm$ 0.06 & 0.37 $\pm$ 0.04 & 0.35 $\pm$ 0.04  \\
& XU &	$-$0.13 $\pm$ 0.02 & $-$0.05 $\pm$ 0.02 & 0.13 $\pm$ 0.02 & 0.12 $\pm$ 0.02 \\
& UK &	$-$0.16 $\pm$ 0.02 &  0.01 $\pm$ 0.02 & 0.16 $\pm$ 0.02 & 0.16 $\pm$ 0.02 \\
& KQ &	$-$0.06 $\pm$ 0.02 & $-$0.04 $\pm$ 0.01 & 0.07 $\pm$ 0.02 & 0.06 $\pm$ 0.02  \\ 
& CQ &  $-$0.69 $\pm$ 0.05 & 0.05  $\pm$ 0.06 & 0.69 $\pm$ 0.05 & 0.69 $\pm$ 0.05  \\\hline \noalign{\smallskip}
 \multirow{4}{*}{2009-12-03} 
& CX &	$-$0.13 $\pm$ 0.03 & $-$0.05 $\pm$ 0.03 & 0.14 $\pm$ 0.03  & 0.14 $\pm$ 0.03  \\
& XU &	$-$0.13 $\pm$ 0.02 & $-$0.02 $\pm$ 0.02 & 0.13 $\pm$ 0.02  & 0.13 $\pm$ 0.02 \\
& UK &	$-$0.07 $\pm$ 0.02 &  0.05 $\pm$ 0.01 & 0.08 $\pm$ 0.02  & 0.07 $\pm$ 0.02  \\
& KQ &	$-$0.120 $\pm$ 0.007 & $-$0.007 $\pm$ 0.018 & 0.120 $\pm$ 0.007 & 0.120 $\pm$ 0.007 \\ 
& CQ &  $-$0.45 $\pm$ 0.03 & $-$0.04 $\pm$ 0.04 & 0.45 $\pm$ 0.03 & 0.45 $\pm$ 0.04  \\\hline \noalign{\smallskip}
 \multirow{4}{*}{2010-01-18\tablefootmark{b}} 
& CX &	$-$0.25 $\pm$ 0.03 & $-$0.08 $\pm$ 0.03 & 0.26 $\pm$ 0.03  & 0.25 $\pm$ 0.03  \\
& XU &	$-$0.06 $\pm$ 0.02 & $-$0.02 $\pm$ 0.02 & 0.06 $\pm$ 0.02  & 0.06 $\pm$ 0.02 \\
& UK &	$-$0.11 $\pm$ 0.02 &  0.021 $\pm$ 0.009 & 0.11 $\pm$ 0.02 & 0.10 $\pm$ 0.02  \\
& KQ &	$-$0.07 $\pm$ 0.01 &  0.032 $\pm$ 0.007 & 0.08 $\pm$ 0.01 & 0.07 $\pm$ 0.01 \\
& CQ &  $-$0.48 $\pm$ 0.04 & $-$0.04 $\pm$ 0.03 & 0.49 $\pm$ 0.04 & 0.49 $\pm$ 0.04  \\\hline \noalign{\smallskip}
 \multirow{4}{*}{2010-02-21\tablefootmark{b}} 
& CX &	$-$0.21 $\pm$ 0.03   & $-$0.09 $\pm$ 0.03  & 0.23 $\pm$ 0.03  & 0.22 $\pm$ 0.03  \\
& XU &	$-$0.09 $\pm$ 0.02   & $-$0.02 $\pm$ 0.01  & 0.09 $\pm$ 0.02  & 0.09 $\pm$ 0.02  \\
& UK &	$-$0.075 $\pm$ 0.008 &  0.05 $\pm$ 0.01  & 0.09 $\pm$ 0.01  & 0.07 $\pm$ 0.01  \\
& KQ &	$-$0.060 $\pm$ 0.007 &  0.025 $\pm$ 0.007 & 0.065 $\pm$ 0.007  & 0.058 $\pm$ 0.007  \\ 
& CQ & $-$0.44 $\pm$ 0.03 & $-$0.04 $\pm$ 0.03 & 0.44 $\pm$ 0.03 & 0.44 $\pm$ 0.03  \\\hline 
\end{longtable}
\tablefoot{
the nuclear region at the Q band can be model fitted by two components where a bright moving feature has been identified at all epochs in the period 2009-2010.\\
\tablefootmark{a}{The nuclear region at both C and X bands can be fitted with two components.}\\
\tablefootmark{b}{Only the C band can be fitted with two components.}
}

\newpage
\twocolumn
\section{Core-shift pair vector choice}
\label{appendix:choices}

\begin{figure}[h]
\centering
 \subfigure[]
    {
        \includegraphics[width=0.47\textwidth]{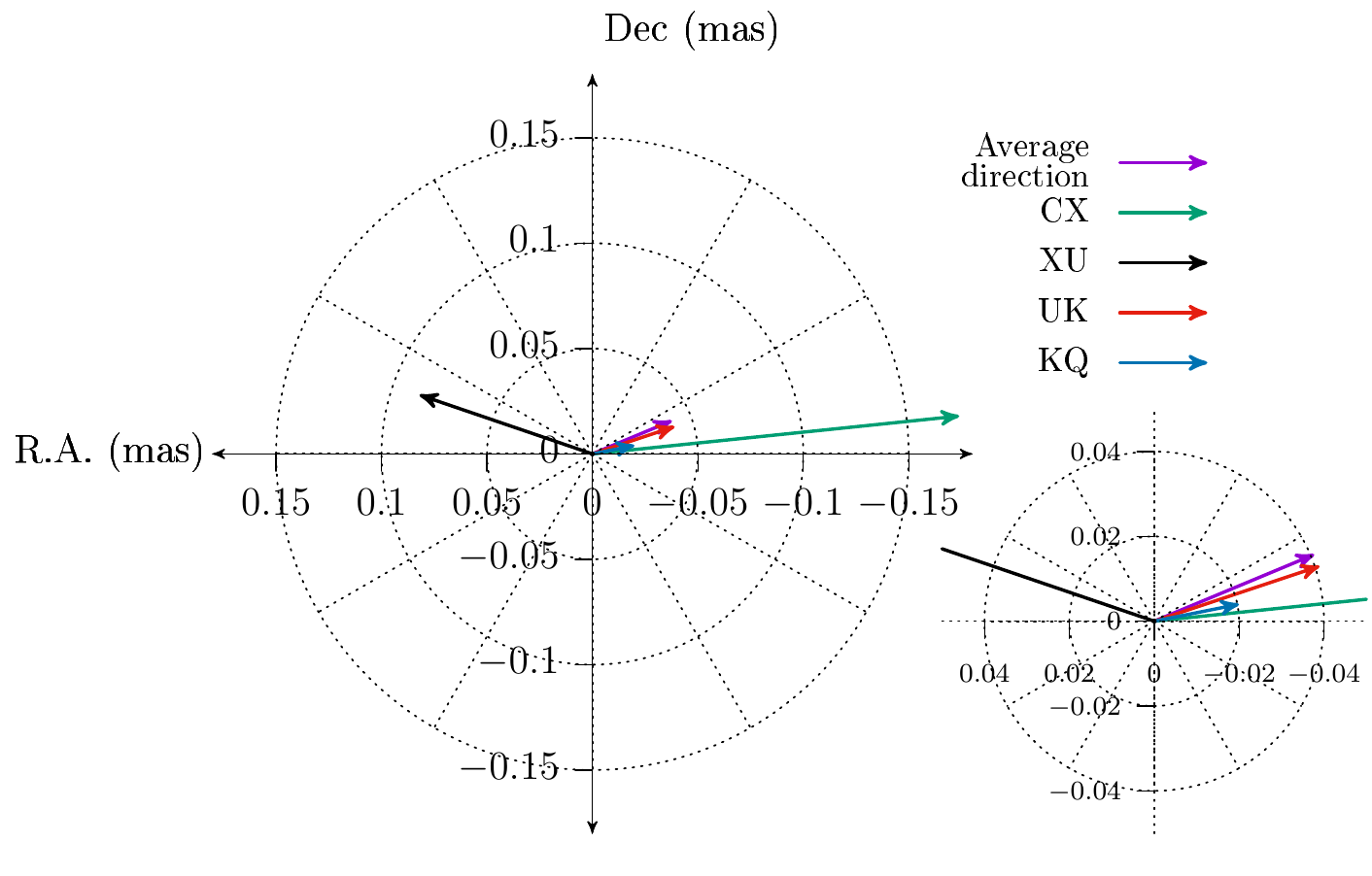}
    }
 \subfigure[]
    {
        \includegraphics[width=0.45\textwidth]{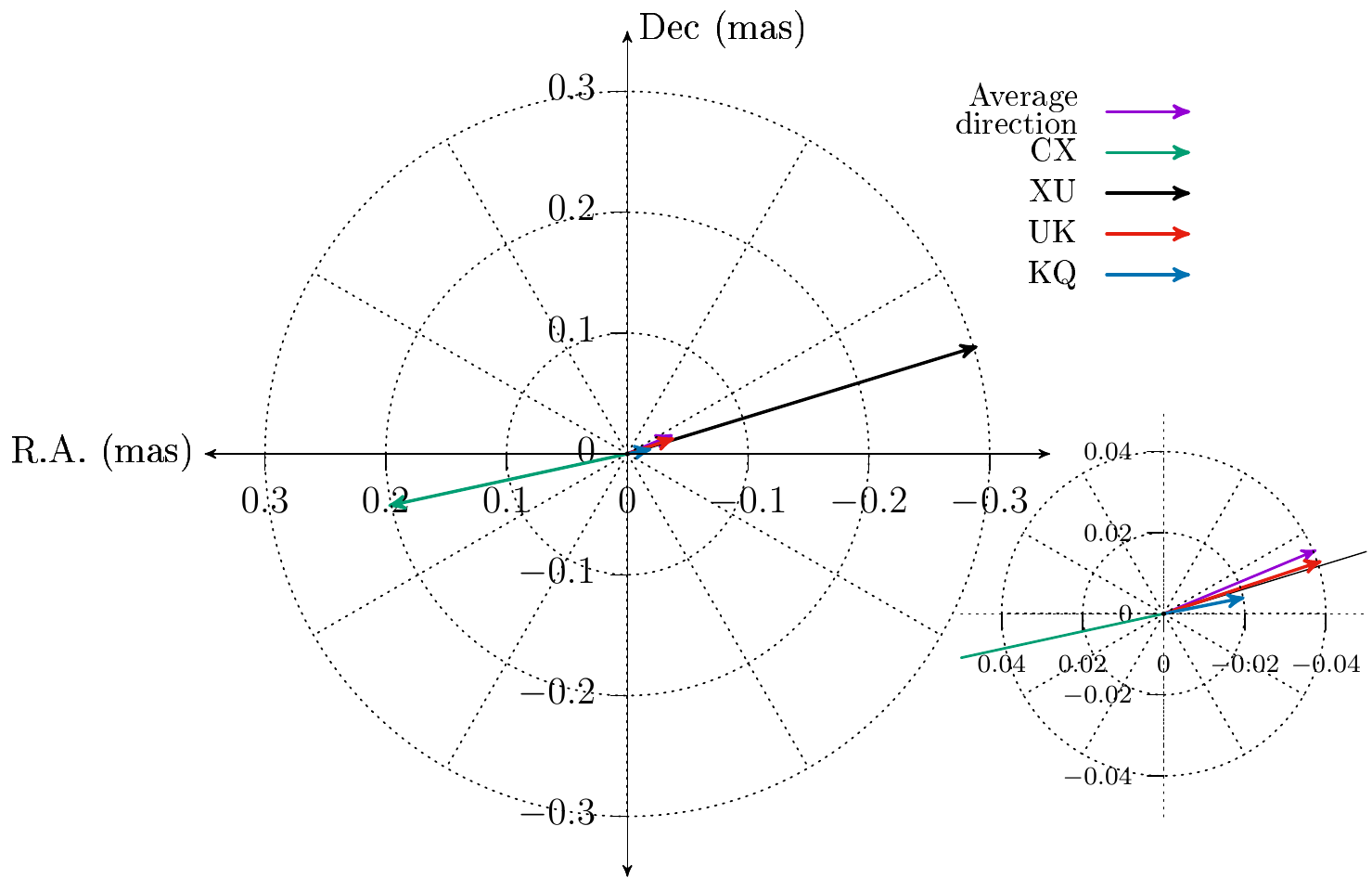}
    }
   \subfigure[]
    {
        \includegraphics[width=0.45\textwidth]{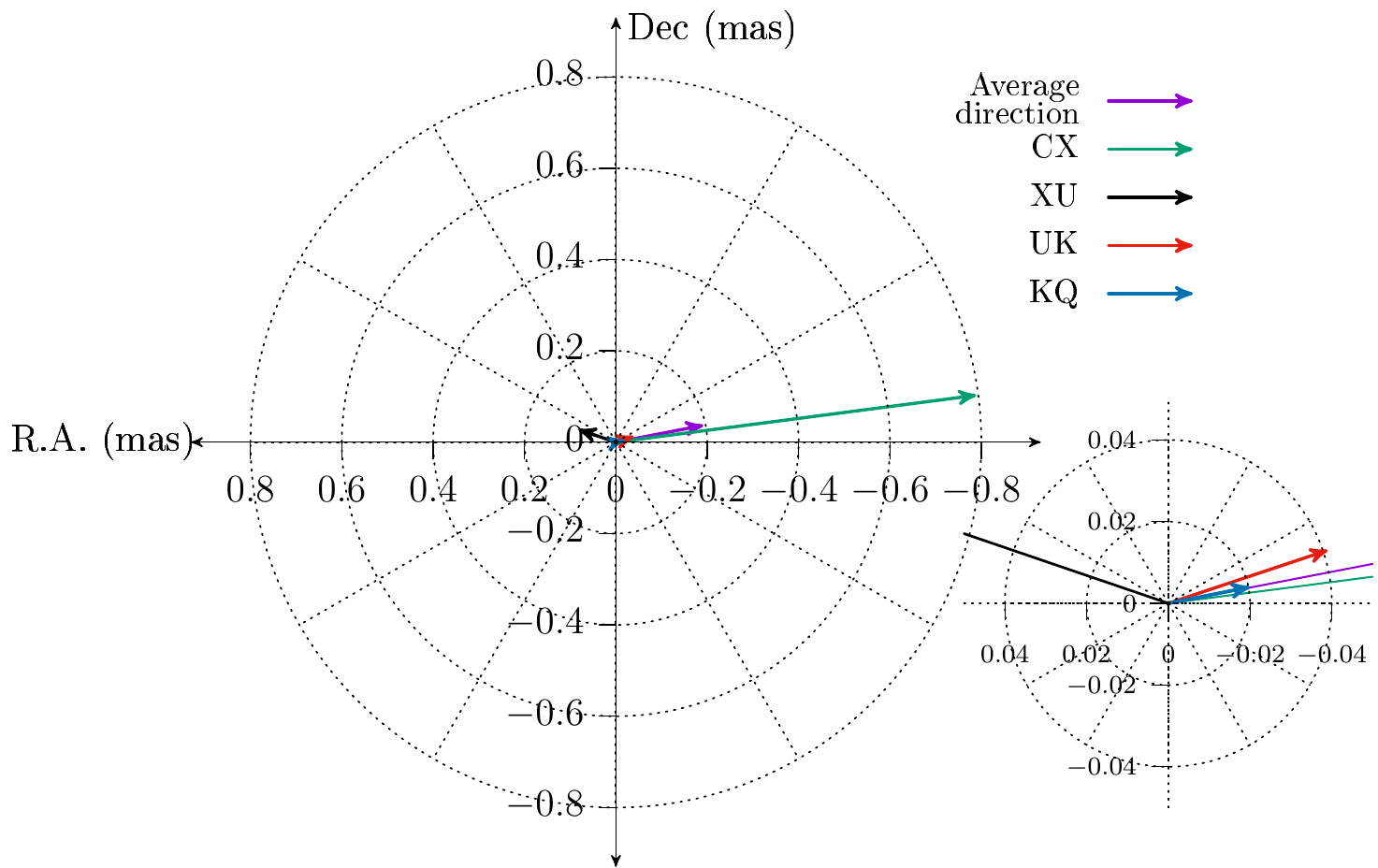}
    }
    \caption{Core-shift vector directions by combining the alternative options for the core at low frequencies, C/X bands. These are a) Ca/Xa, b) Ca/Xb, and c) Cb/Xa. In all options, at least one vector points in the opposite direction contrary to the jet direction that is towards West.}
    \label{other-vectors}
\end{figure}

As described in Figure~\ref{spectrum-csvectors}, here we show in Figure~\ref{other-vectors} a comparison of the core-shift vectors produced by the other choices. The optimal core choice at the C and X bands is the pair: Cb/Xb. This leads both CX and XU core-shift vectors pointing in the jet direction (towards West), as already demonstrated in Figure~\ref{spectrum-csvectors}b.

\section{Parameter's uncertainties}
\label{appendix:error-prop}

\subsection{Uncertainty of $\Omega_{r\nu}$}

The uncertainty of $\Omega_{r\nu}$, denoted by $\Delta \Omega_{r\nu}$ is obtained by propagating uncertainty in Equation~\ref{eq:omega} with both $\Delta r_{\nu_1\nu_2}$ and $k_r$ as variable parameters. For simplification purposes, we set here  $\Delta r_{\nu_1\nu_2} = l$ and the uncertainty of $\Delta r_{\nu_1\nu_2} = \Delta l$. The uncertainty of index $k_r$ is $\Delta k_r$. The expression for $\Delta \Omega_{r\nu}$ is given as
\begin{equation}
\Delta \Omega_{r\nu} = \sqrt{\left (  \frac{\partial \Omega }{\partial l}\cdot \Delta l\right )^2 + \left ( \frac{\partial \Omega }{\partial k_r}\cdot \Delta k_r \right )^2 },
\end{equation}
where
\begin{equation}
      \frac{\partial \Omega_{r\nu} }{\partial l} = \frac{\Omega_{r\nu}}{l},
\end{equation}
and
\begin{equation}
     \frac{\partial \Omega_{r\nu} }{\partial k_r} = \Omega_{r\nu} \frac{\left (\nu_1^{1/k_r} \ln(\nu_2) - \nu_2^{1/k_r} \ln(\nu_1) \right )}{k_r^2(\nu_2^{1/k_r}-\nu_1^{1/k_r})}.
\end{equation}
Plugging C.2 and C.3 into C.1, the uncertainty of $\Omega_{r\nu}$ is given by
\begin{equation}
\label{eq:error-omega}
     \Delta \Omega_{r\nu} = \Omega_{r\nu} \sqrt{\left ( \frac{\Delta l}{l} \right )^2 + \left ( \frac{\nu_1^{1/k_r} \ln(\nu_2) - \nu_2^{1/k_r} \ln(\nu_1) }{k_r^2(\nu_2^{1/k_r}-\nu_1^{1/k_r})} \cdot \Delta k_r \right )^2}.
\end{equation}
Equation C.4 is a generic formula to estimate the uncertainties in $\Omega_{r\nu}$ for each frequency pair: $\Delta \Omega_{CX}$, $\Delta \Omega_{XU}$, $\Delta \Omega_{UK}$ and $\Delta \Omega_{KQ}$. 
 
\subsection{Uncertainty of $r_\mathrm{core}$}
 
The uncertainty of $r_\mathrm{core}$ is denoted by $\Delta r_\mathrm{core}$. The measured uncertainties are propagated in Equation~\ref{eq:rcore} having $\Omega_{r\nu}$ and $k_r$ as variable parameters. Note that $\Omega_{r\nu}$ is an average-mean value per epoch, therefore the expression for $\Delta r_\mathrm{core}$ is
\begin{equation}
     \Delta r_\mathrm{core}=\sqrt{\left (  \frac{\partial r_\mathrm{core} }{\partial \Omega_{r\nu}}\cdot \Delta \Omega_{r\nu}\right )^2 + \left ( \frac{\partial r_\mathrm{core} }{\partial k_r}\cdot \Delta k_r \right )^2 },
\end{equation}
with
\begin{equation}
    \frac{\partial r_\mathrm{core} }{\partial \Omega_{r\nu}}=\frac{r_\mathrm{core}}{\Omega_{r\nu}},
\end{equation}
and
\begin{equation}
    \frac{\partial r_\mathrm{core} }{\partial k_r} = r_\mathrm{core}\frac{\ln(\nu)}{k_r^2}.
\end{equation}
Plugging C.6 and C.7 into C.5, the uncertainty of $r_\mathrm{core}$ is given by
\begin{equation}
     \Delta r_\mathrm{core}=r_\mathrm{core}\sqrt{\left(  \frac{\Delta \Omega_{r\nu}}{\Omega_{r\nu}} \right)^2 + \left( \frac{\ln(\nu)}{k_r^2} \Delta k_r \right)^2 }.
\end{equation}

\subsection{Uncertainty of $B_\mathrm{1pc}$}

The uncertainty of $B_\mathrm{1pc}$ is denoted by $\Delta B_\mathrm{1pc}$. The measured uncertainties are propagated in Equation~\ref{eq:B1eq} with respect to the CQ core-shift, $\Delta r_{CQ}$, $\delta$, $\theta_\mathrm{j}$, $\theta$ and $k_r$. Below $M=4.85\cdot10^{-9}\,D_L/(1+z)^2$. Inserting equation~\ref{eq:omega} into equation~\ref{eq:B1eq}, the formula for $B_\mathrm{1pc}$ transforms to
\begin{equation}
    B_\mathrm{1pc} \approx 0.025 \left [\frac{\sigma_\mathrm{rel}\,(1+z)^3}
    {\delta^2\, \theta_\mathrm{j}\, \mathrm{sin}^{3k_{r}-1}\theta}\, \nu_1^3\nu_2^3 
    \left ( \frac{M\Delta r_{CQ}}{\nu_2^{1/k_r}-\nu_1^{1/k_r}} \right )^{3k_r}
    \right ]^{\frac{1}{4}}[\mathrm{G}].
\end{equation}
For simplification purposes, we set here  $\Delta r_{CQ} = A$ and the uncertainty of the CQ core-shift as $\Delta A$. The error of $B_\mathrm{1pc}$ is denoted by $\Delta B_\mathrm{1pc}$ and obtained as
\begin{align}
    \Delta B_\mathrm{1pc}  = & \Bigg[ \left (  \frac{\partial B_\mathrm{1pc} }{\partial A}\cdot \Delta A\right )^2 +
    \left (  \frac{\partial B_\mathrm{1pc} }{\partial \delta}\cdot \Delta \delta \right )^2 + \left ( \frac{\partial B_\mathrm{1pc} }{\partial \theta_\mathrm{j}}\cdot \Delta \theta_\mathrm{j} \right )^2 \notag \\ 
    & + \left ( \frac{\partial B_\mathrm{1pc} }{\partial \theta}\cdot \Delta \theta \right )^2 +
    \left ( \frac{\partial B_\mathrm{1pc} }{\partial k_r}\cdot \Delta k_r \right )^2 \Bigg]^{1/2},
\end{align}
where 
\begin{equation}
      \frac{\partial B_\mathrm{1pc} }{\partial A} = \frac{3}{4}k_r \frac{B_\mathrm{1pc}}{\Delta A}, 
\end{equation}
\begin{equation}
     \frac{\partial B_\mathrm{1pc} }{\partial \delta} = -\frac{1}{2}\frac{B_\mathrm{1pc}}{\delta}, 
\end{equation}
\begin{equation}
     \frac{\partial B_\mathrm{1pc} }{\partial \theta_\mathrm{j}} = -\frac{1}{4}\frac{B_\mathrm{1pc}}{\theta_\mathrm{j}}, 
\end{equation}
\begin{equation}
     \frac{\partial B_\mathrm{1pc} }{\partial \theta} = -\frac{1}{4}(3k_r-1)\frac{\cos\theta}{\sin\theta}B_\mathrm{1pc}, 
\end{equation}
\begin{align}
    \frac{\partial B_\mathrm{1pc} }{\partial k_r} = & \frac{3}{4} \frac{1}{k_r} B_\mathrm{1pc} \frac{1}{\nu_2^{1/k_r} - \nu_1^{1/k_r}} \notag \\
    &  \Bigg[ k_r (\nu_2^{1/k_r}-\nu_1^{1/k_r})\,\Bigg( \ln \Bigg( \frac{MA}{\nu_2^{1/kr}-\nu_1^{1/k_r}} \Bigg) - \ln(\sin\theta) \Bigg) \notag \\
    & + \nu_2^{1/k_r}\ln(\nu_2) - \nu_1^{1/k_r}\ln(\nu_1) \Bigg]
\end{align}
Finally, plugging all the partial derivatives into C.10, the error of $B_\mathrm{1pc}$ is readily obtained.

\section{Core spectrum, core-shift vectors and power-law fits}
\label{appendix:core-shift-results}
\begin{figure}[h]
\centering
 \subfigure[]
    {
        \includegraphics[width=0.45\textwidth]{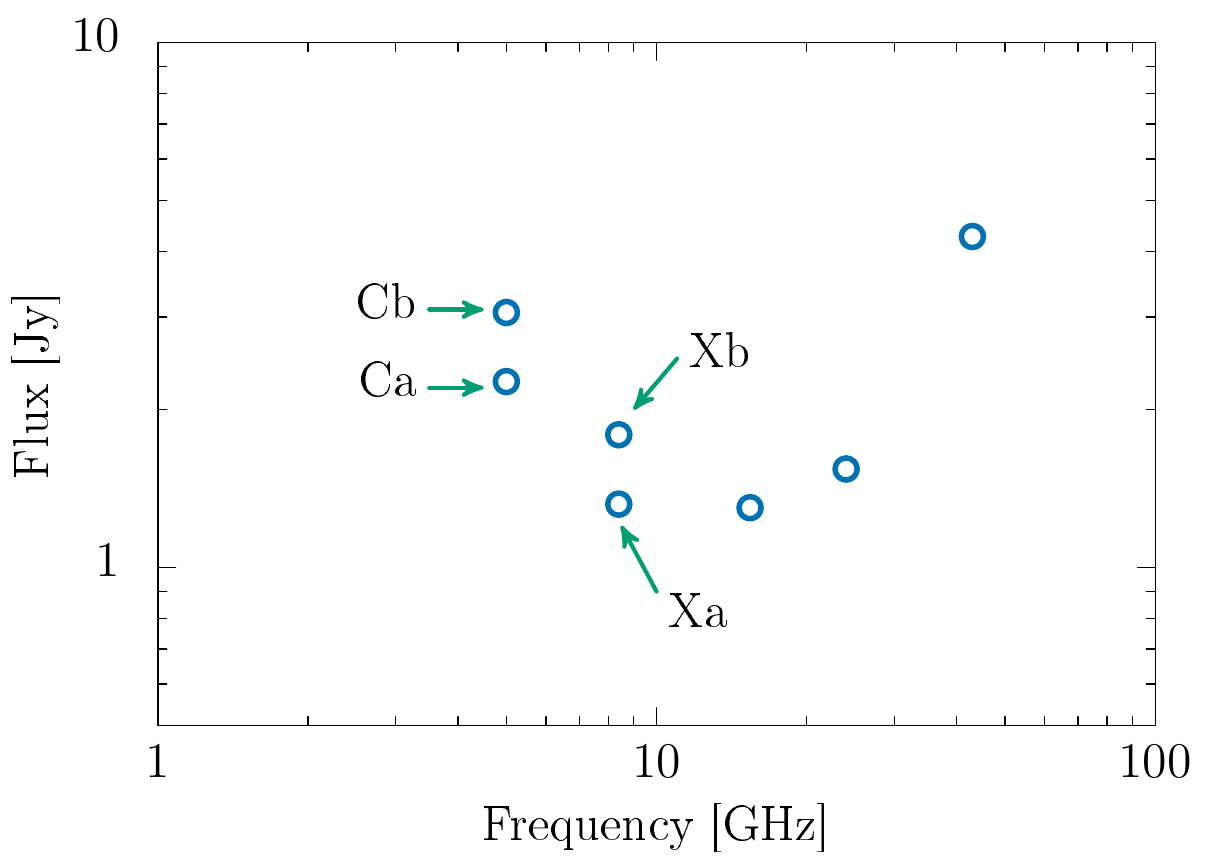}
    }
    \subfigure[]
    {
        \includegraphics[width=0.45\textwidth]{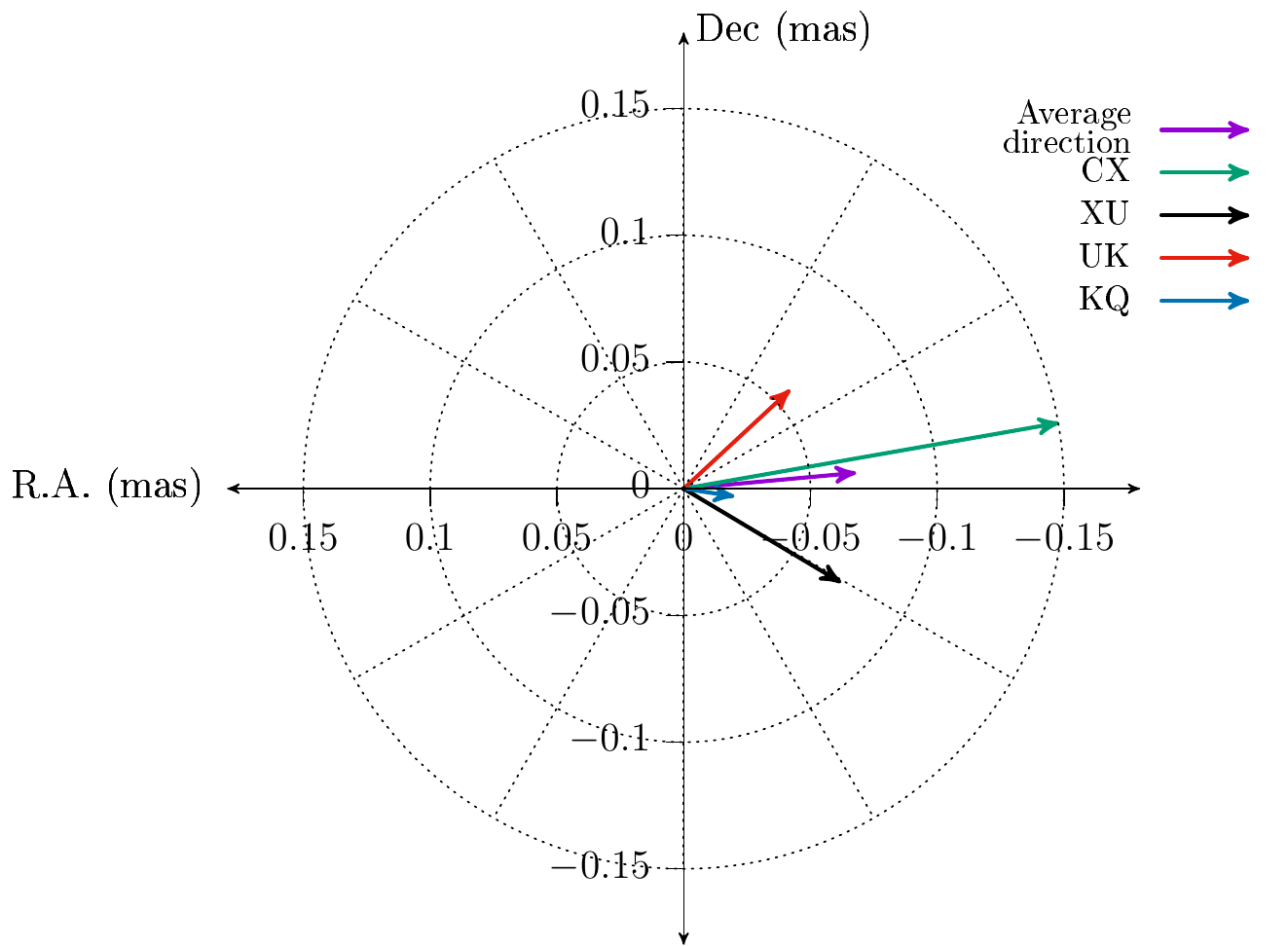}
    }
     \subfigure[]
    {
        \includegraphics[width=0.5\textwidth]{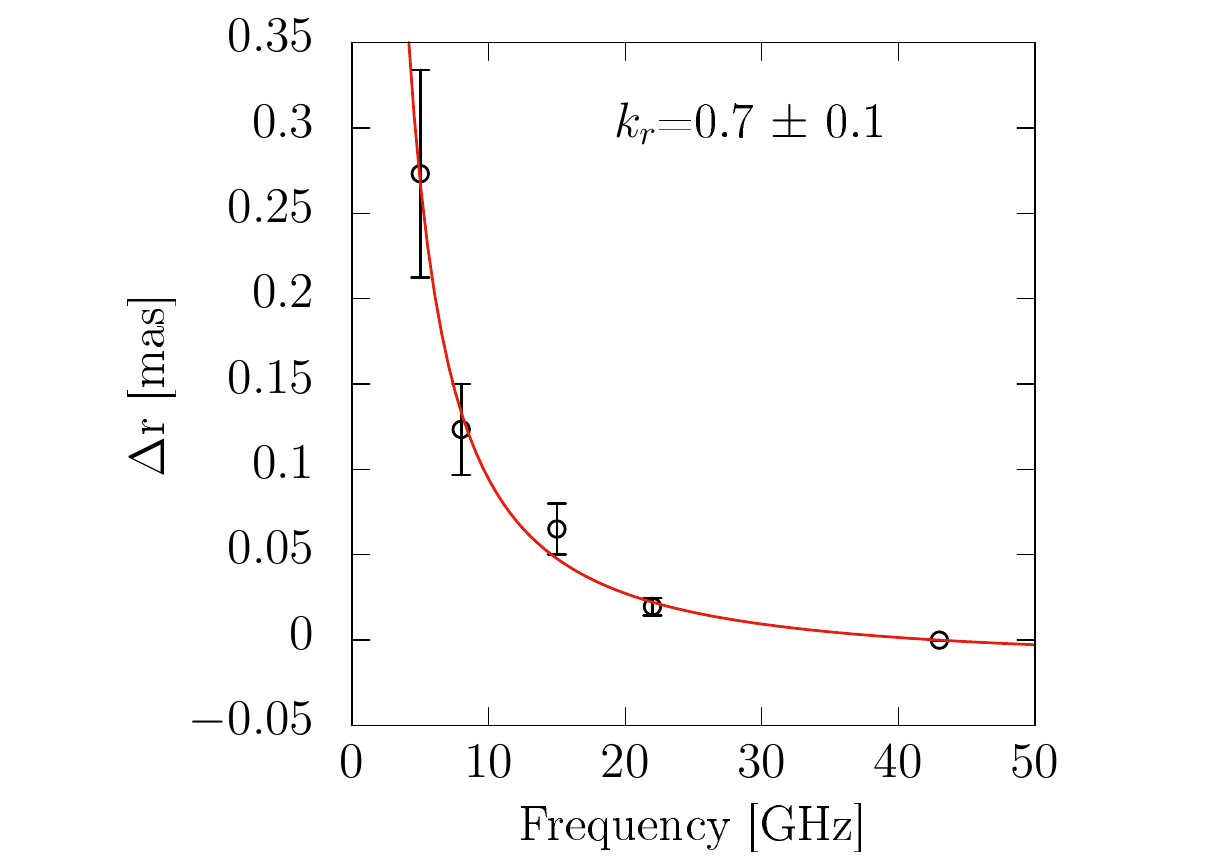}
    }
    \caption{Epoch 1, 2005-05-19. (a) Core spectrum, Ca/Xa represent the core. (b) Core-shift vectors of all frequency pairs. The choice of Ca/Xa cores lead to the correct direction of vectors CX and XU. (c) Power-law fit is shown with the red curve.}
    \label{CSepoch1}
\end{figure}

\begin{figure}[!h]
\centering
   \subfigure[]
    {
        \includegraphics[width=0.45\textwidth]{Epoch2-CS14-07-2005/14-07-2005-fv.pdf}
    }
    \subfigure[]
    {
        \includegraphics[width=0.47\textwidth]{Epoch2-CS14-07-2005/vectors_polar.pdf}
    }
     \subfigure[]
    {
        \includegraphics[width=0.5\textwidth]{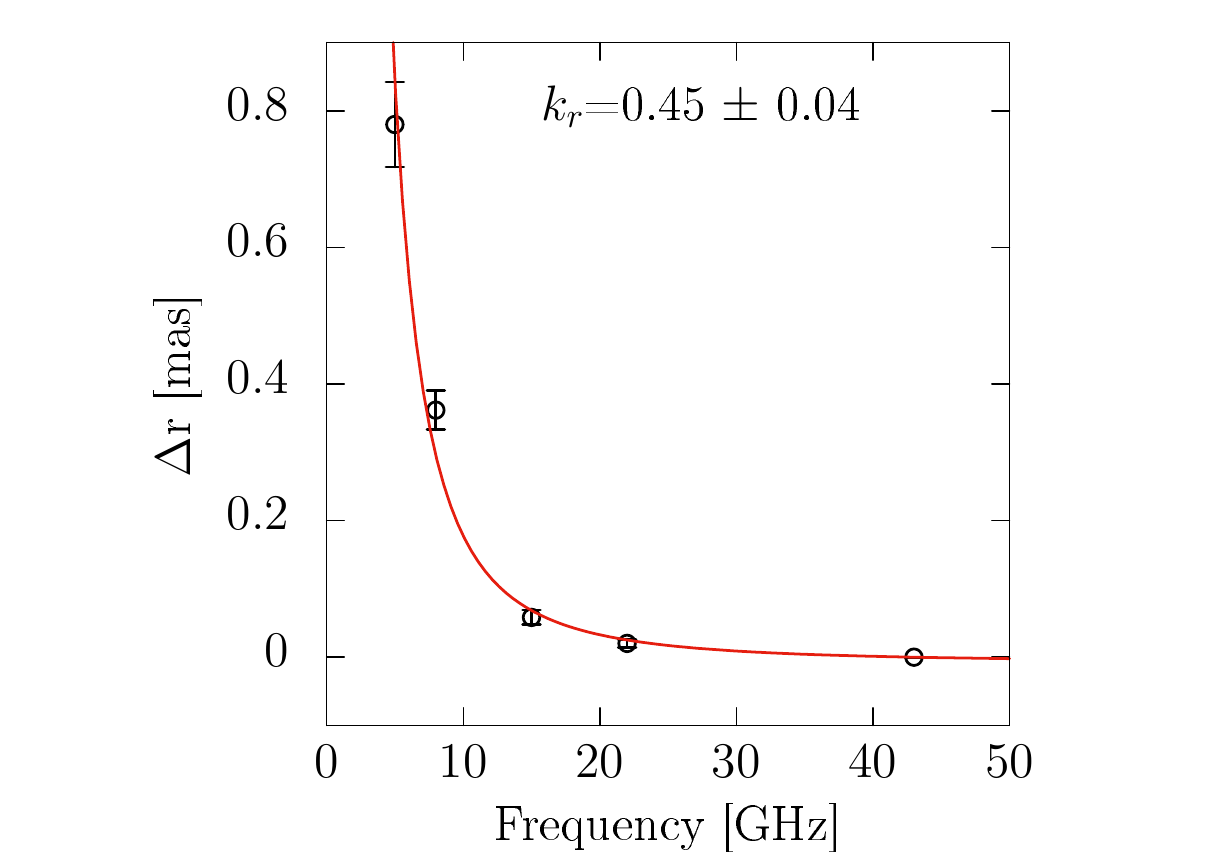}
    }
    \caption{Epoch 2, 2005-07-14: (a) Core spectrum, Cb/Xb represent the core. (b) Core-shift vectors of all frequency pairs. The choice of Cb/Xb cores lead to the correct direction of vectors CX and XU. (c) Power-law fit is shown with the red curve.}
    \label{CSepoch2}
\end{figure}

\begin{figure}[!h]
\centering
   \subfigure[]
    {
        \includegraphics[width=0.45\textwidth]{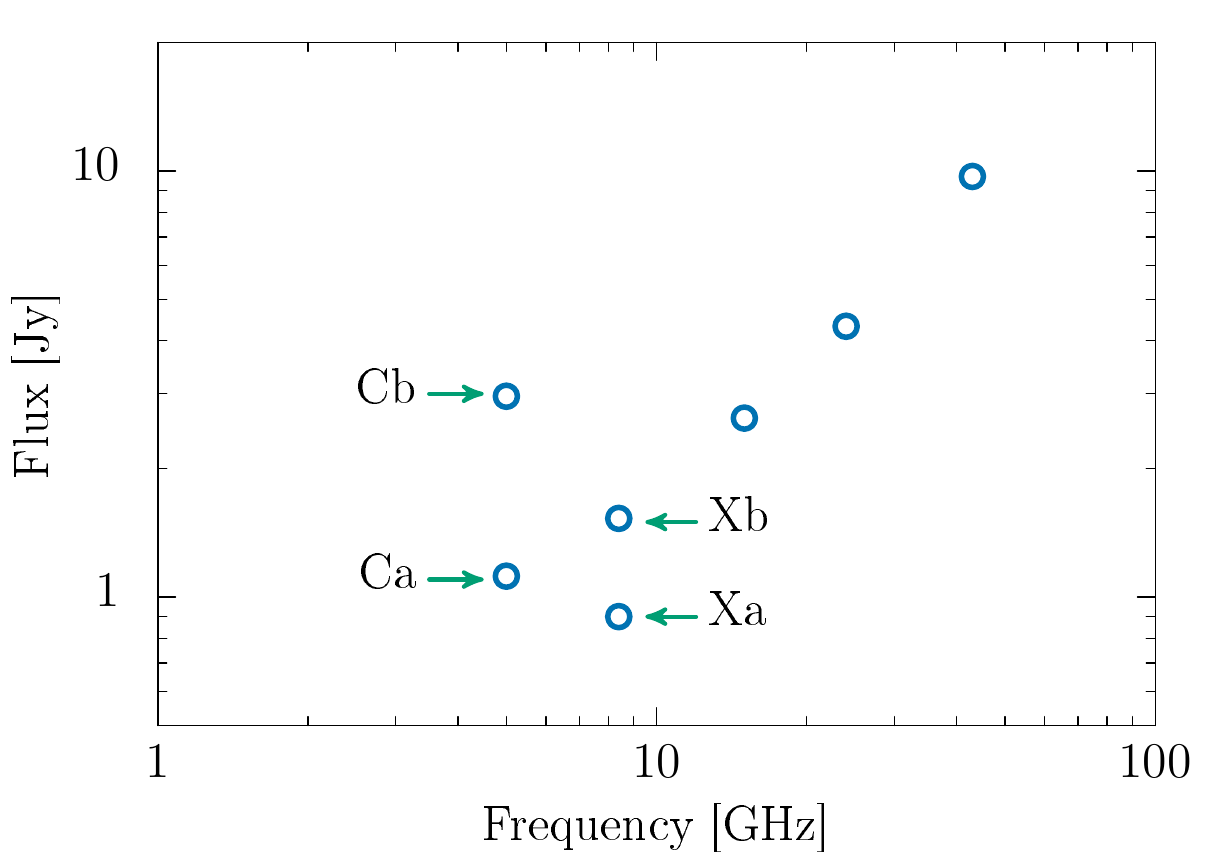}
    }
    \subfigure[]
    {
        \includegraphics[width=0.47\textwidth]{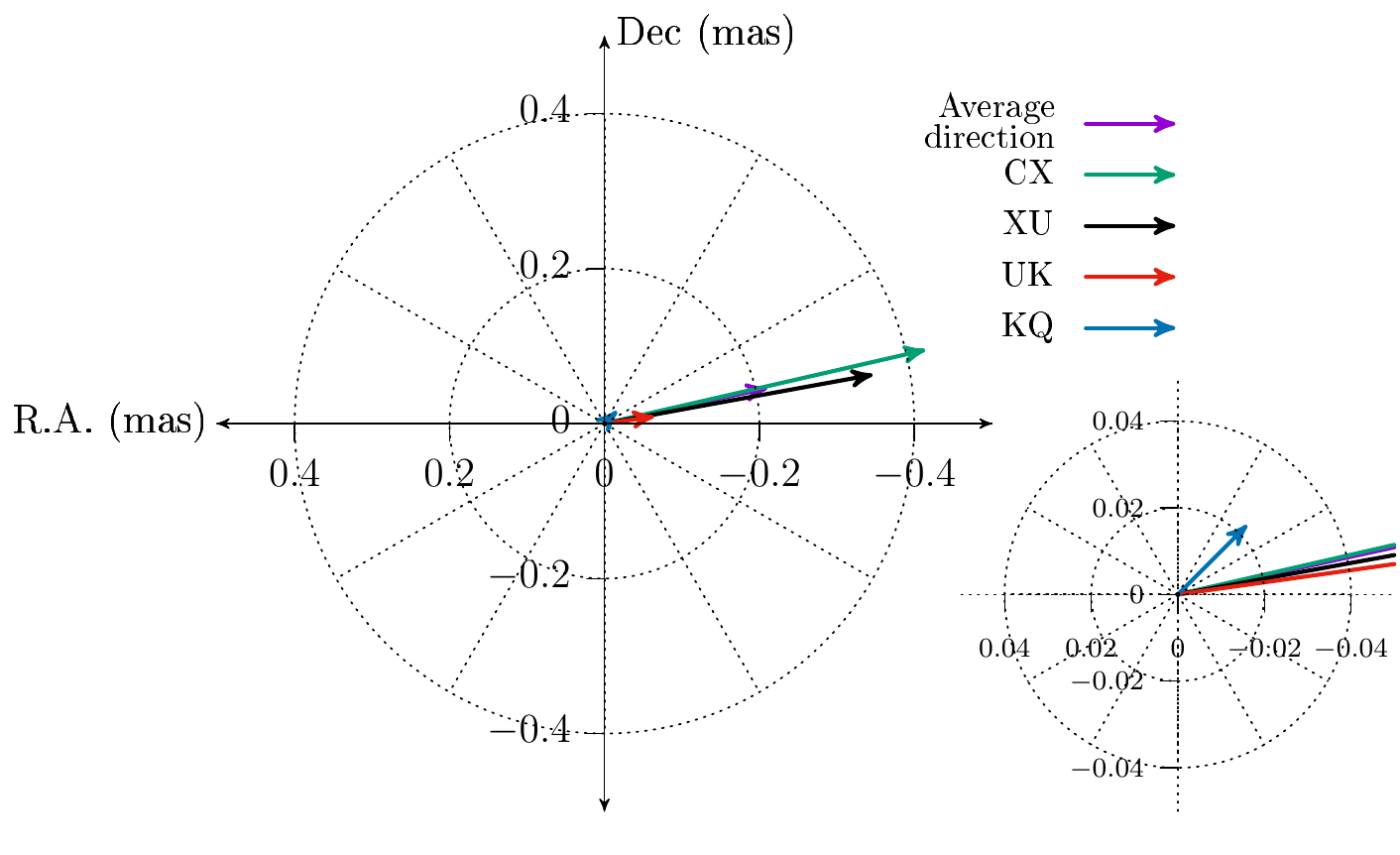}
    }
     \subfigure[]
    {
        \includegraphics[width=0.5\textwidth]{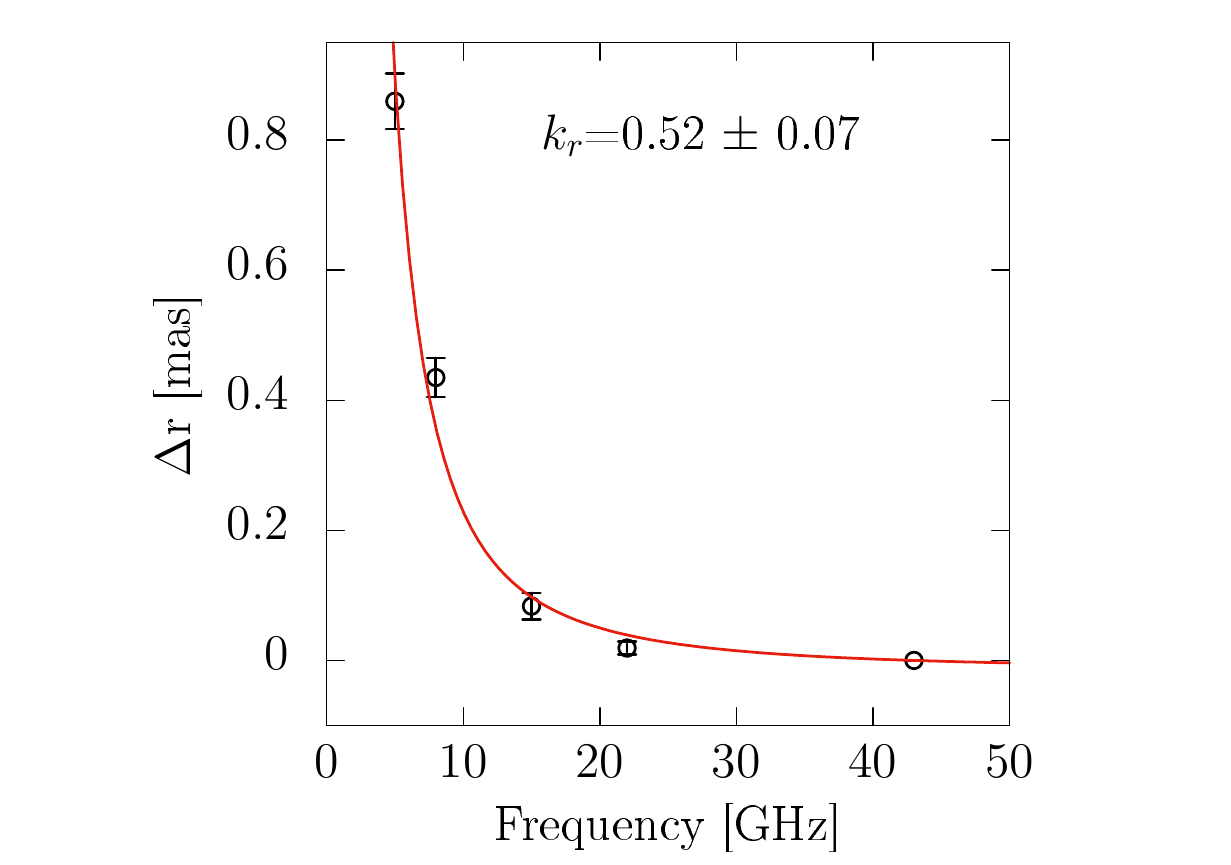}
    }
    \caption{Epoch 3, 2005-09-01: (a) Core spectrum, Cb/Xb represent the core. (b) Core-shift vectors of all frequency pairs. The choice of Cb/Xb cores lead to the correct direction of vectors CX and XU. (c) Power-law fit is shown with the red curve.}
    \label{CSepoch3}
\end{figure}

\begin{figure}[!h]
\centering
   \subfigure[]
    {
        \includegraphics[width=0.45\textwidth]{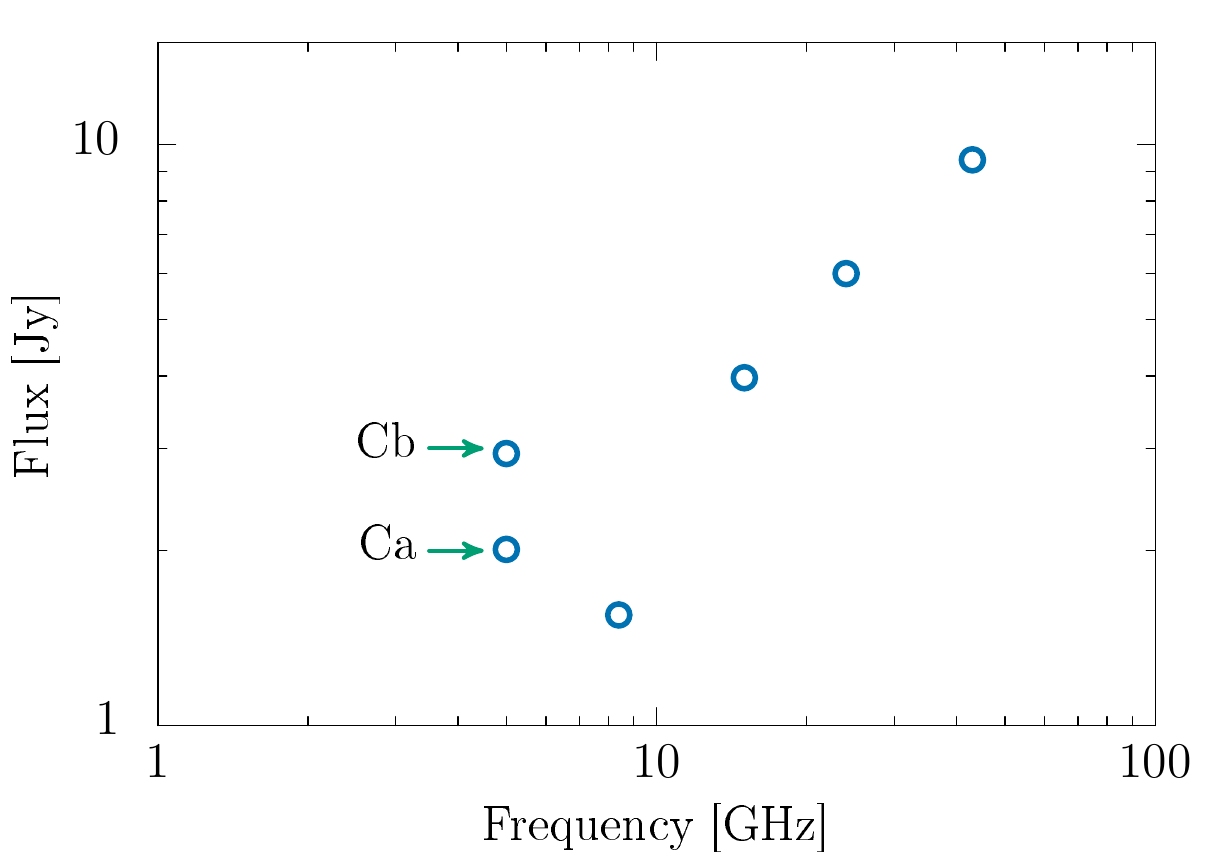}
    }
    \subfigure[]
    {
        \includegraphics[width=0.47\textwidth]{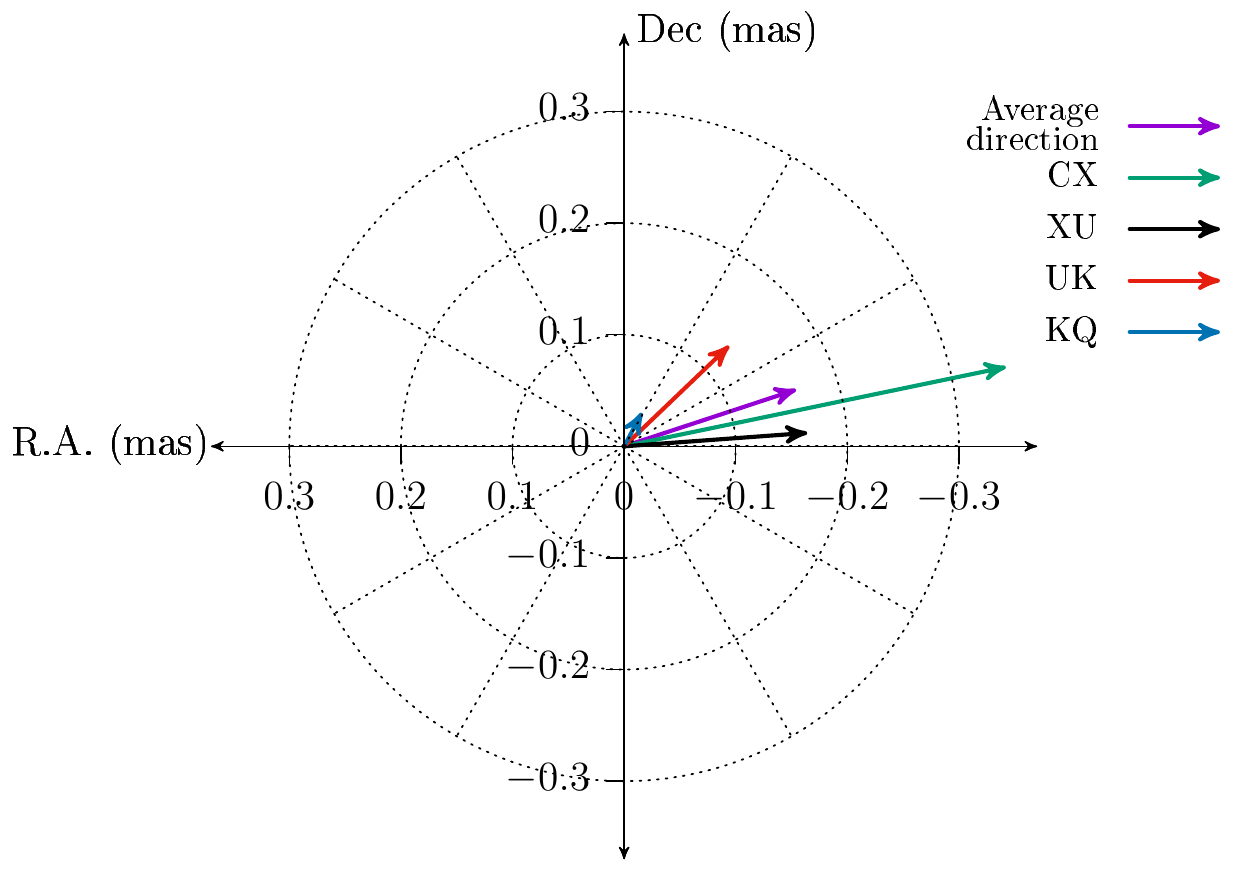}
    }
     \subfigure[]
    {
        \includegraphics[width=0.5\textwidth]{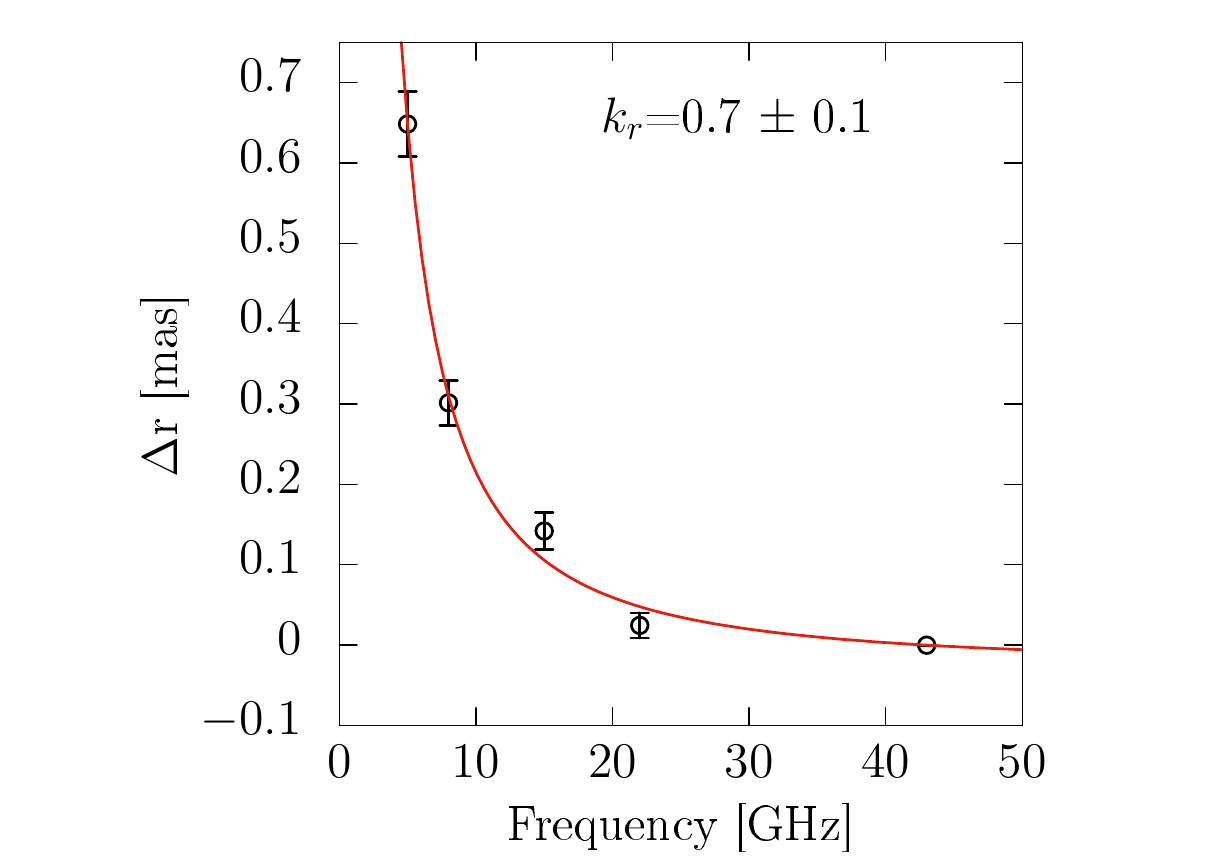}
    }
    \caption{Epoch 4, 2005-12-04. (a) Core spectrum, Ca represents the core. (b) Core-shift vectors of all frequency pairs. The choice of Ca core leads to the correct direction of the CX vector. (c) Power-law fit is shown with the red curve.}
    \label{CSepoch4}
\end{figure}

\begin{figure}[!h]
\centering
   \subfigure[]
    {
        \includegraphics[width=0.45\textwidth]{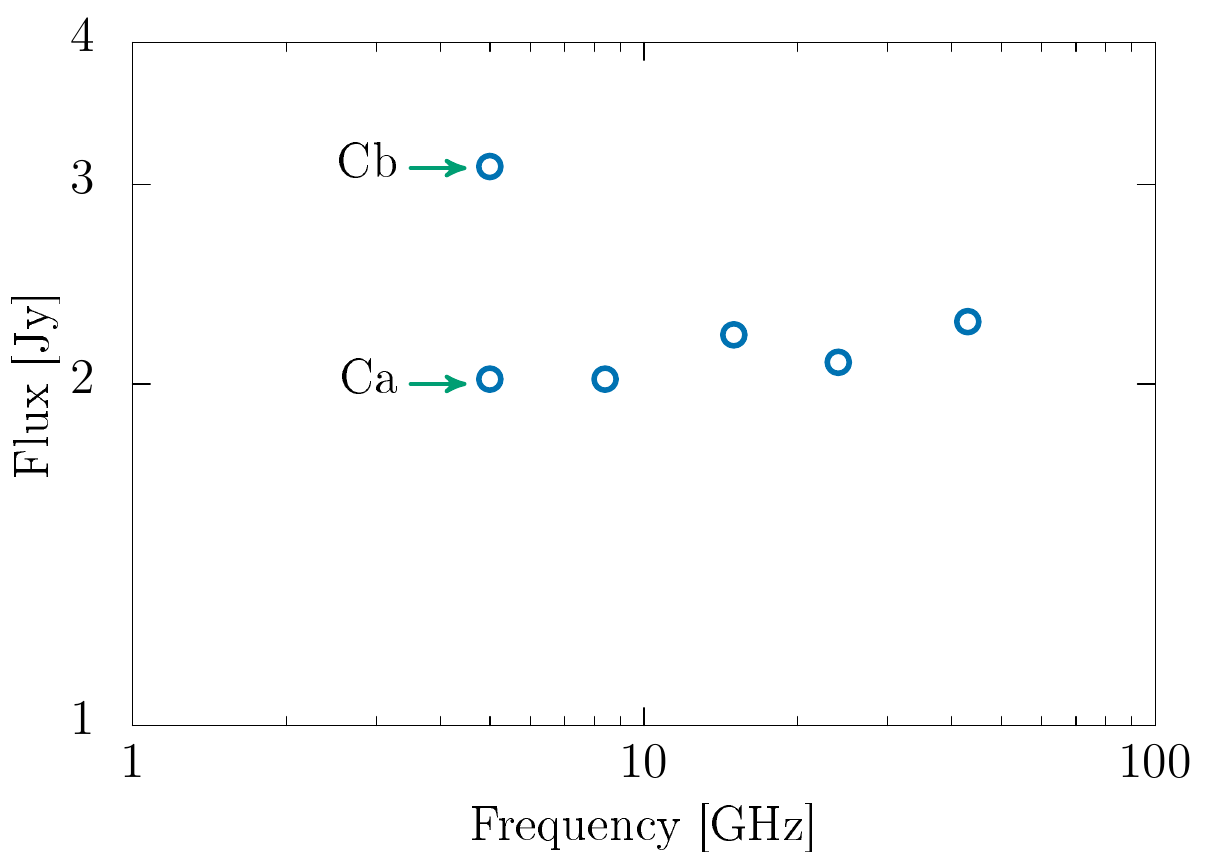}
    }
    \subfigure[]
    {
        \includegraphics[width=0.47\textwidth]{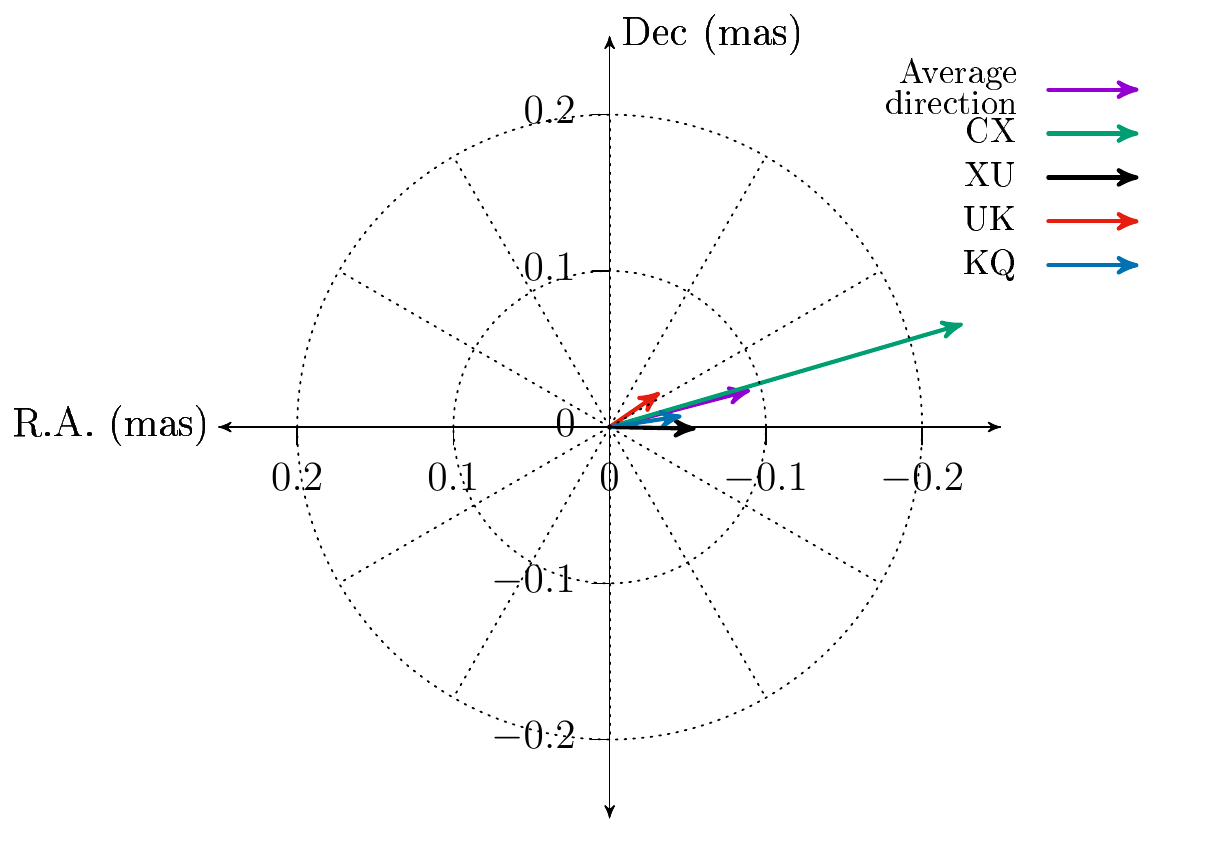}
    }
     \subfigure[]
    {
        \includegraphics[width=0.5\textwidth]{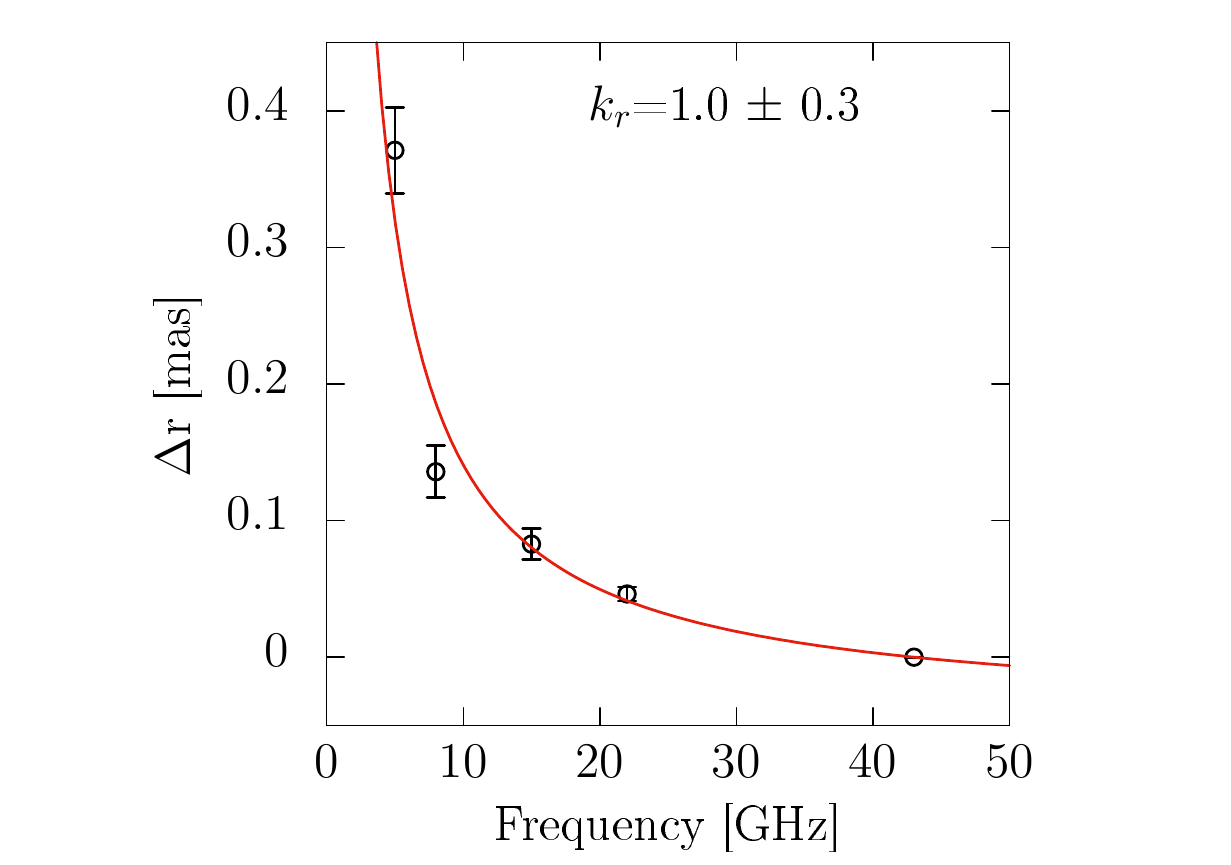}
    }
    \caption{Epoch 6, 2006-10-03. (a) Core spectrum, Ca represents the core. (b) Core-shift vectors of all frequency pairs. The choice of Ca core leads to the correct direction of the CX vector. (c) Power-law fit is shown with the red curve.}
    \label{CSepoch6}
\end{figure}

\begin{figure}[!h]
\centering
  \subfigure[]
    {
        \includegraphics[width=0.45\textwidth]{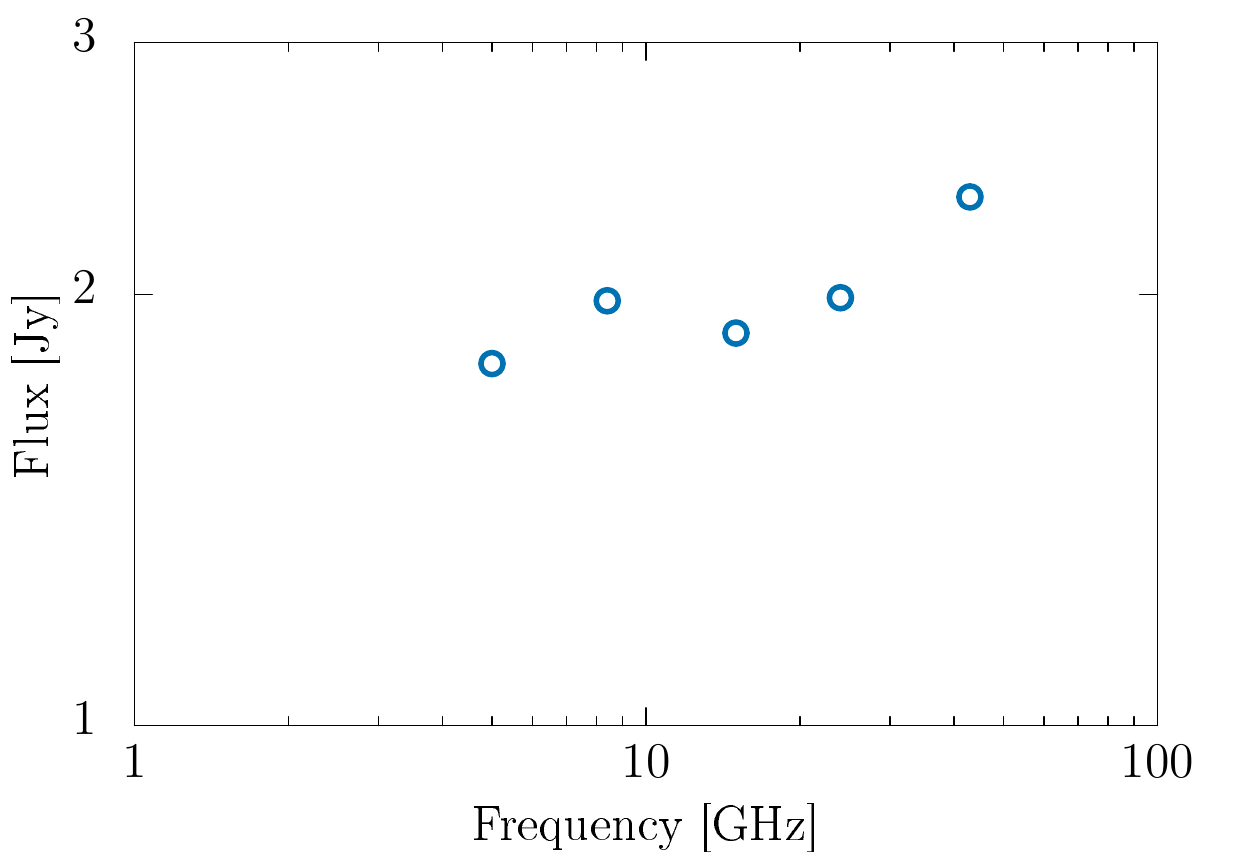}
    }
    \subfigure[]
    {
        \includegraphics[width=0.47\textwidth]{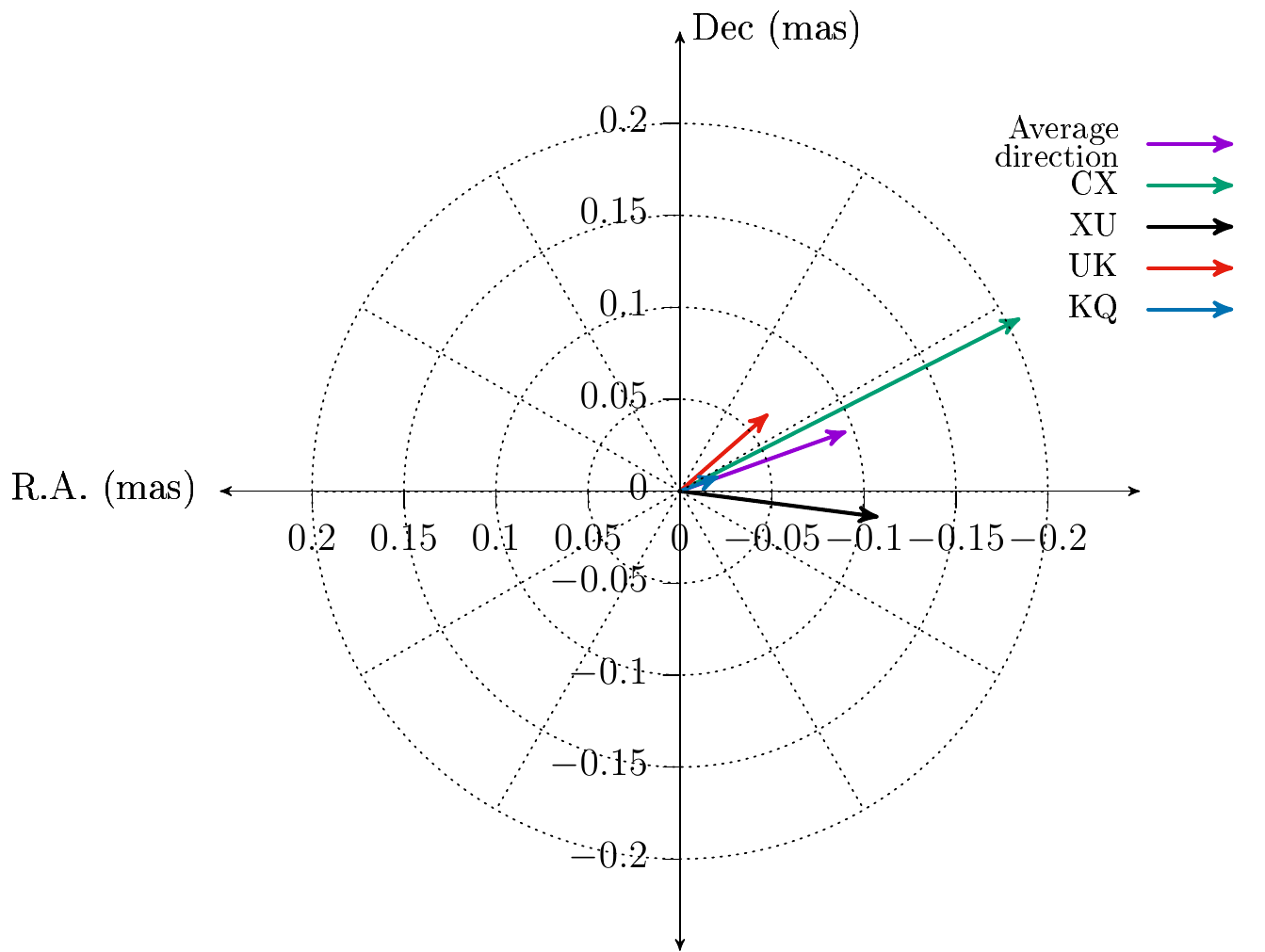}
    }
     \subfigure[]
    {
        \includegraphics[width=0.5\textwidth]{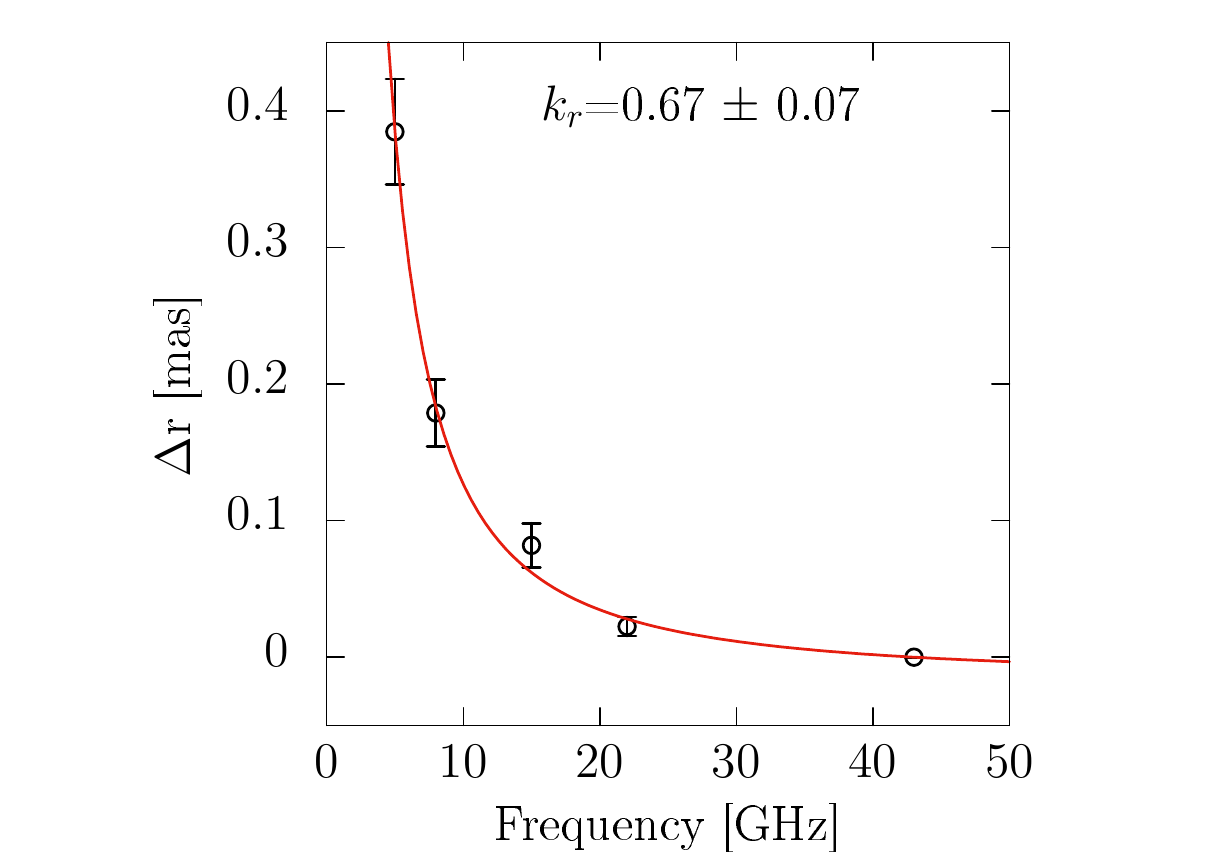}
    }
    \caption{Epoch 7, 2006-12-04. (a) Core spectrum. (b) Core-shift vectors of all frequency pairs. (c) Power-law fit is shown with the red curve.}
    \label{CSepoch7}
\end{figure}

\begin{figure}[!h]
\centering
   \subfigure[]
    {
        \includegraphics[width=0.45\textwidth]{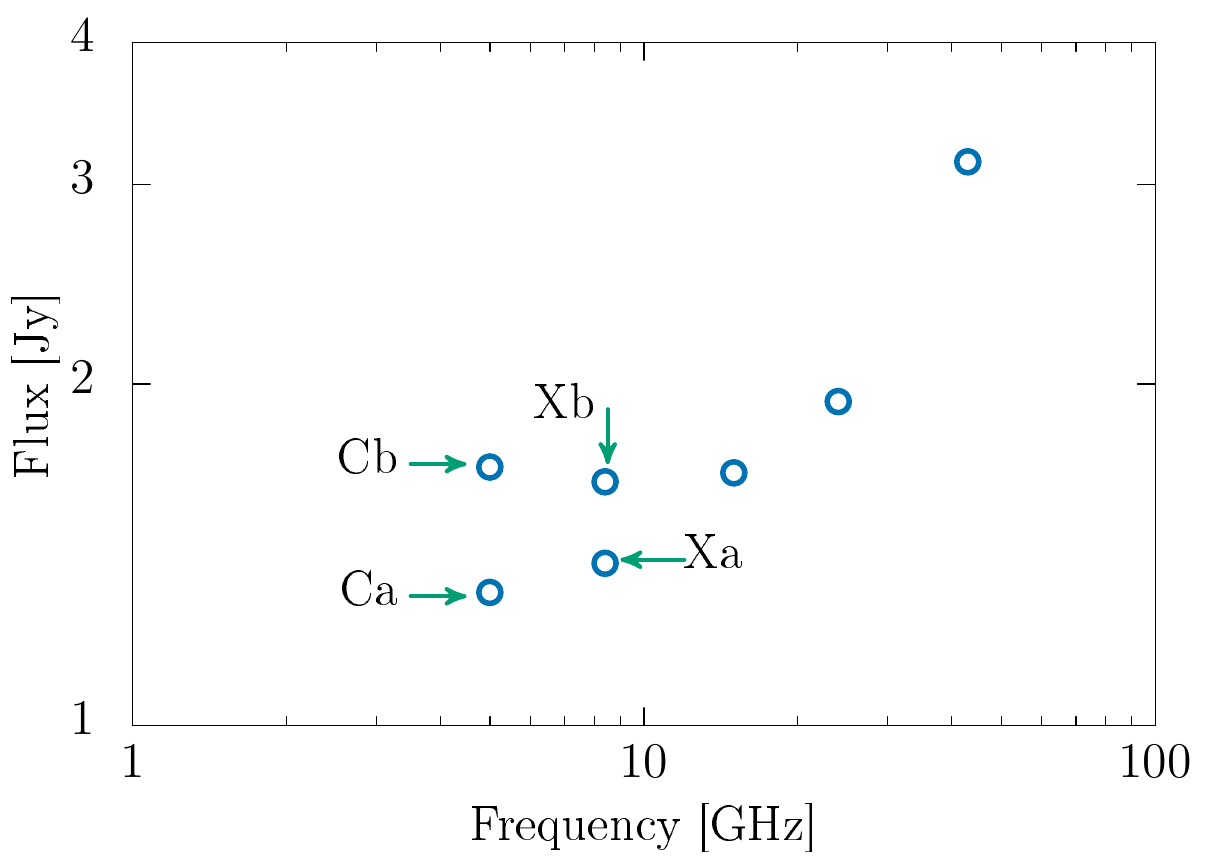}
    }
    \subfigure[]
    {
        \includegraphics[width=0.47\textwidth]{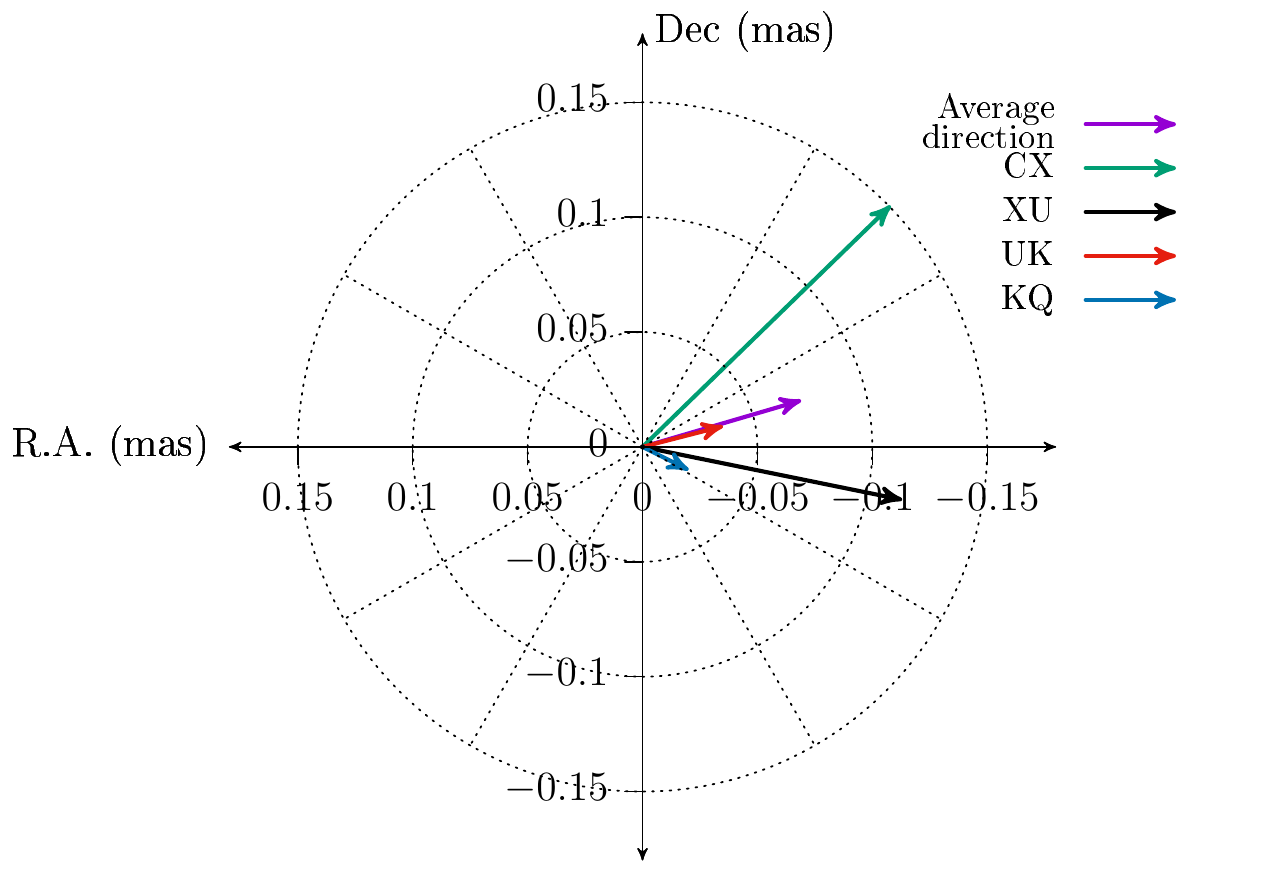}
    }
     \subfigure[]
    {
        \includegraphics[width=0.5\textwidth]{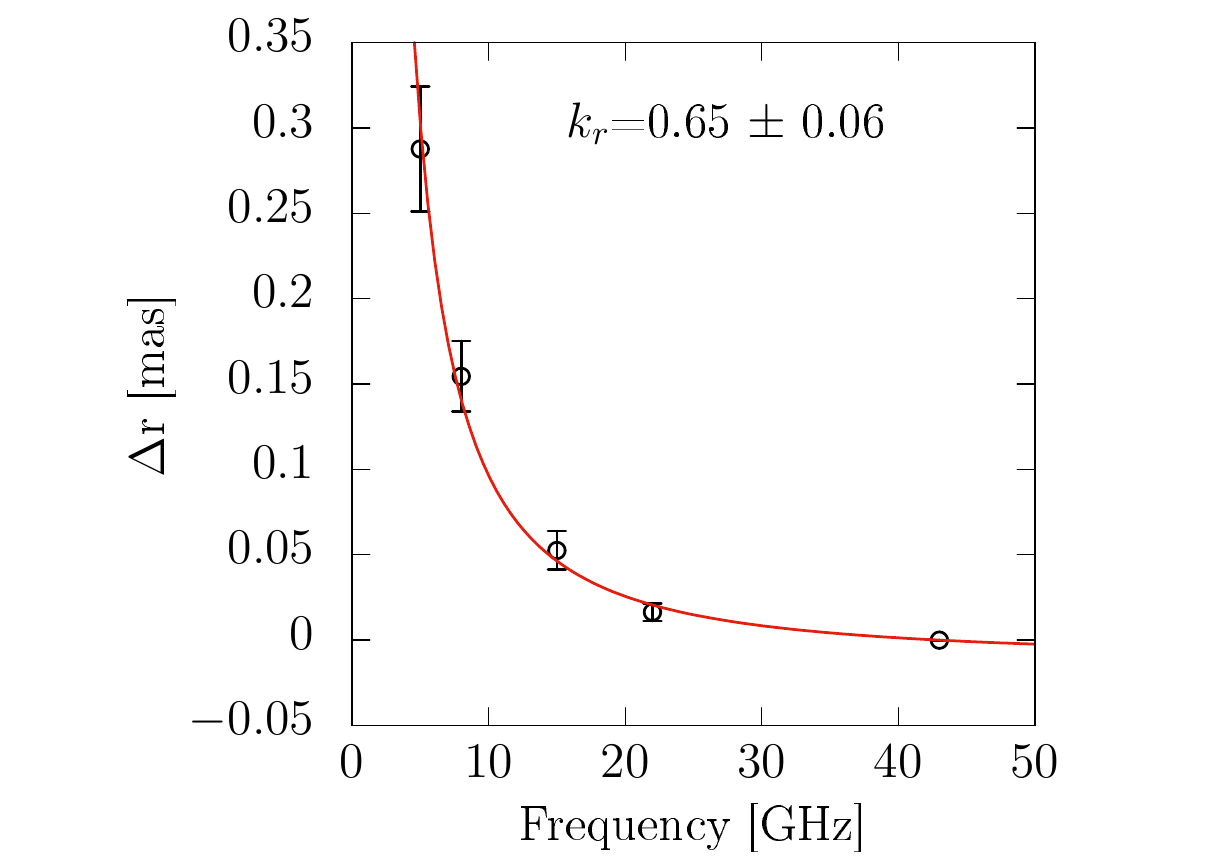}
    }
    \caption{Epoch 8, 2007-01-26. (a) Core spectrum, Ca/Xb represent the core. (b) Core-shift vectors of all frequency pairs. The choice of Ca/Xb cores lead to the correct direction of vectors CX and XU. (c) Power-law fit is shown with the red curve.}
    \label{CSepoch8}
\end{figure}

\begin{figure}[!h]
\centering
   \subfigure[]
    {
        \includegraphics[width=0.45\textwidth]{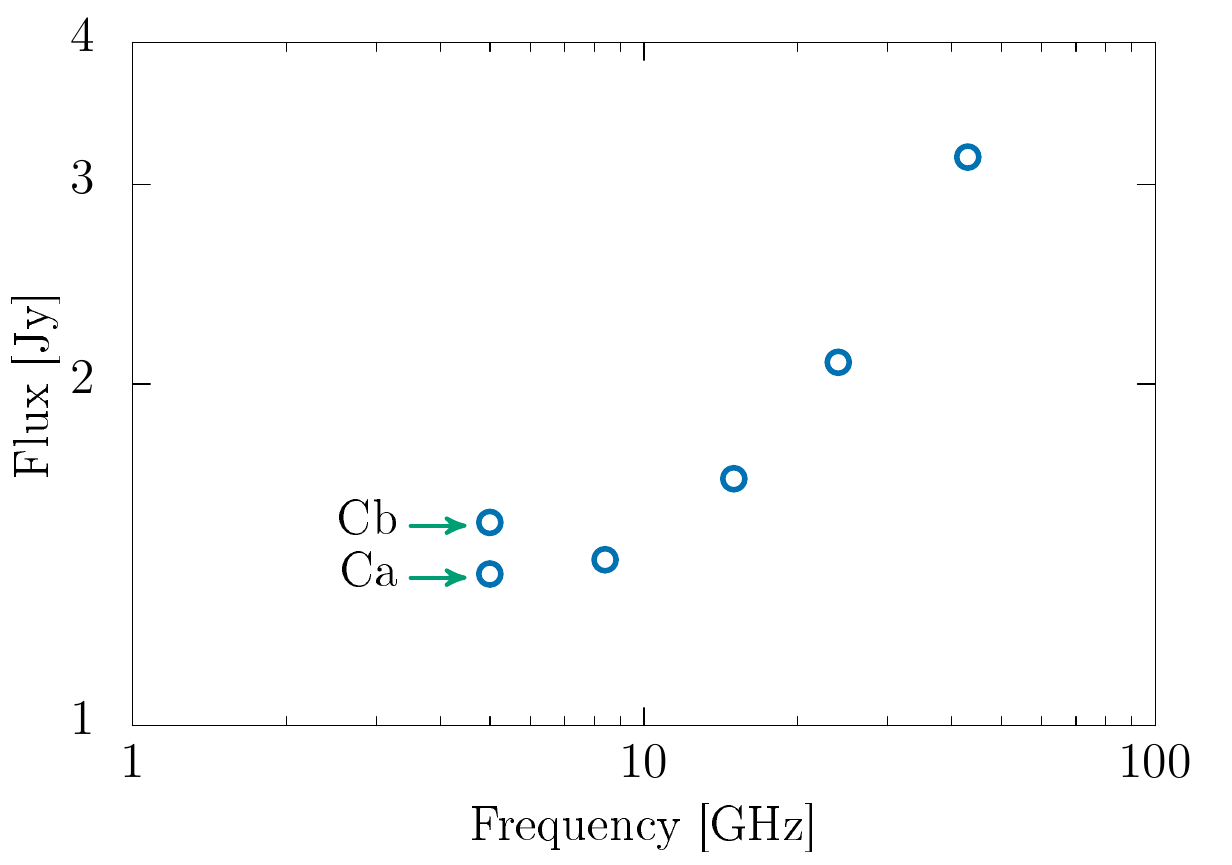}
    }
    \subfigure[]
    {
        \includegraphics[width=0.47\textwidth]{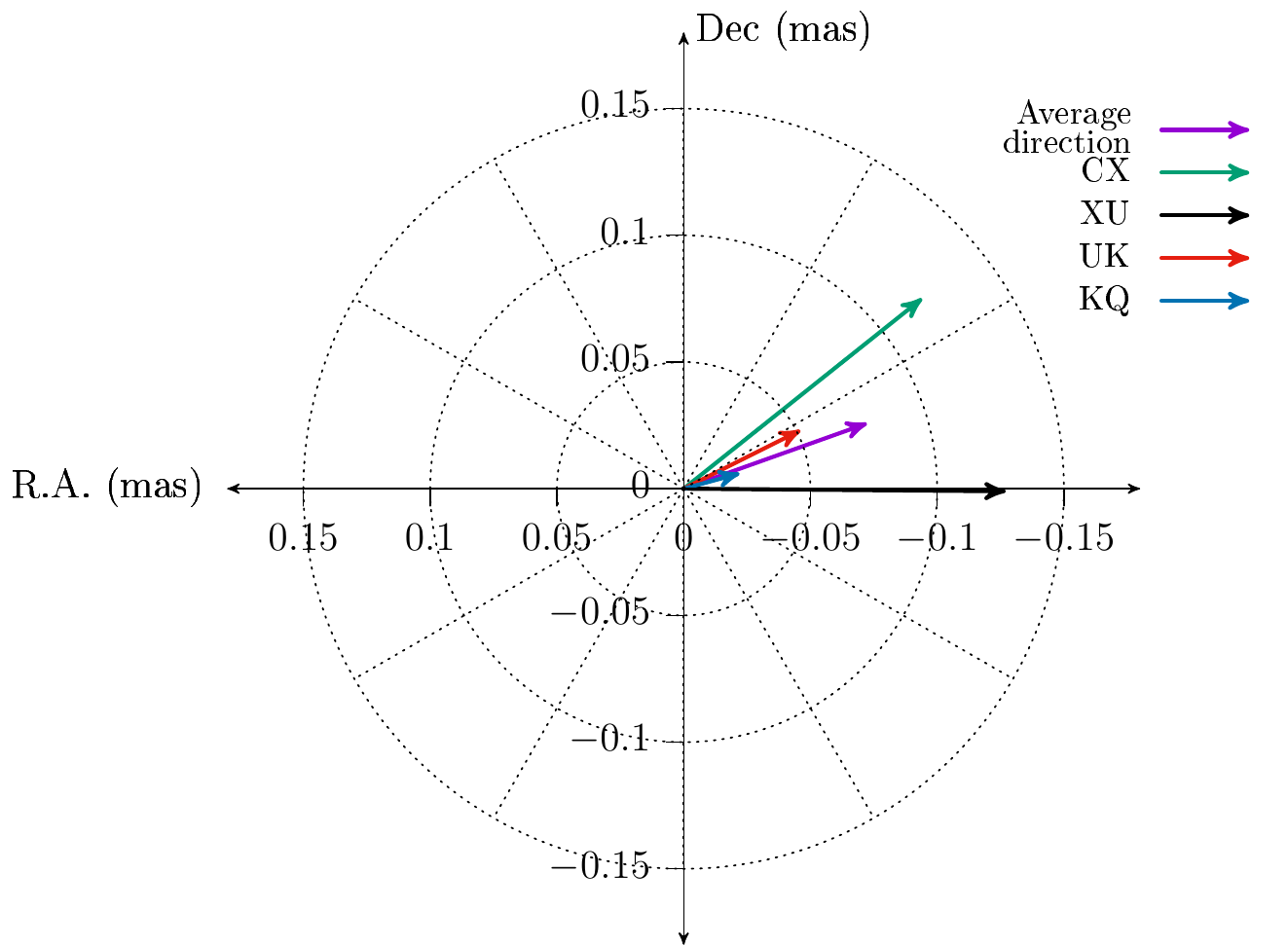}
    }
     \subfigure[]
    {
        \includegraphics[width=0.5\textwidth]{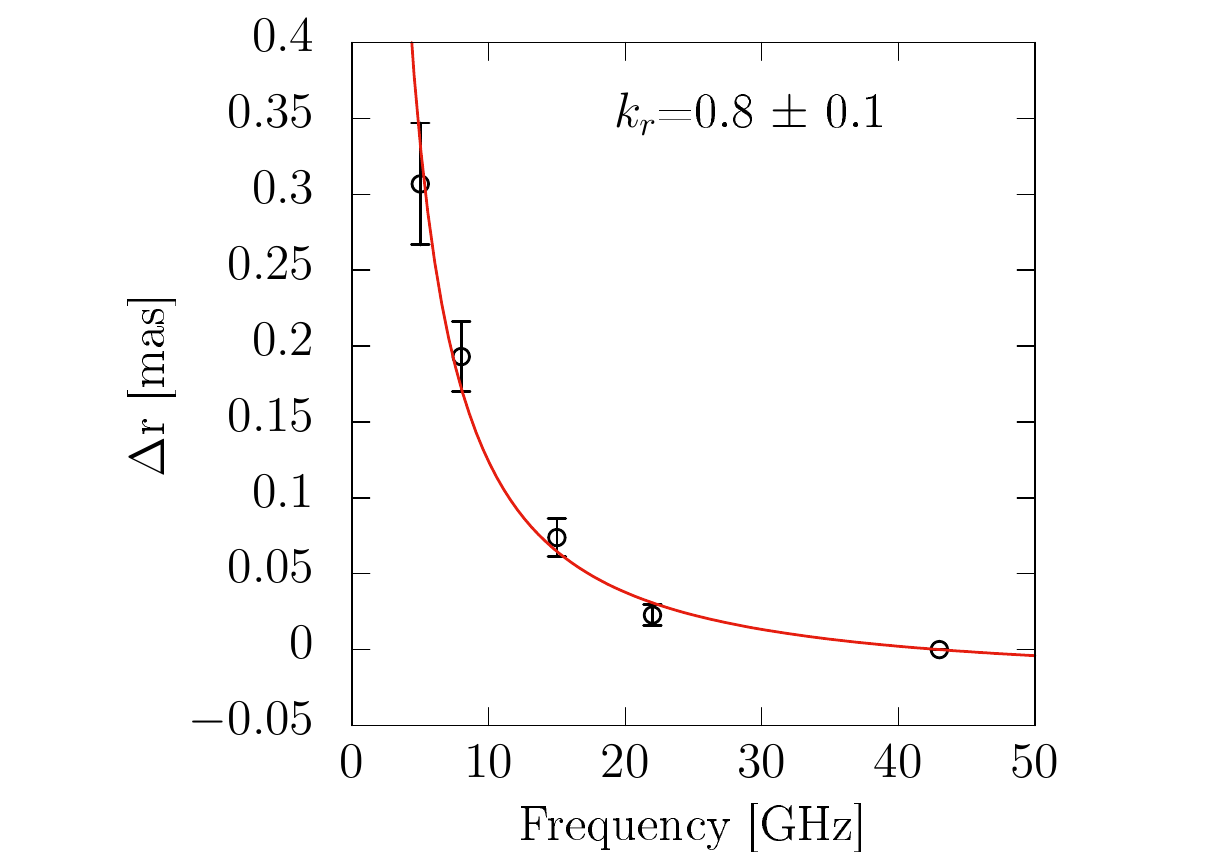}
    }
    \caption{Epoch 9, 2007-04-26. (a) Core spectrum, Ca represents the core. (b) Core-shift vectors of all frequency pairs. The choice of Ca core leads to the correct direction of the CX vector. (c) Power-law fit is shown with the red curve.}
    \label{CSepoch9}
\end{figure}

\begin{figure}[!h]
\centering
   \subfigure[]
    {
        \includegraphics[width=0.45\textwidth]{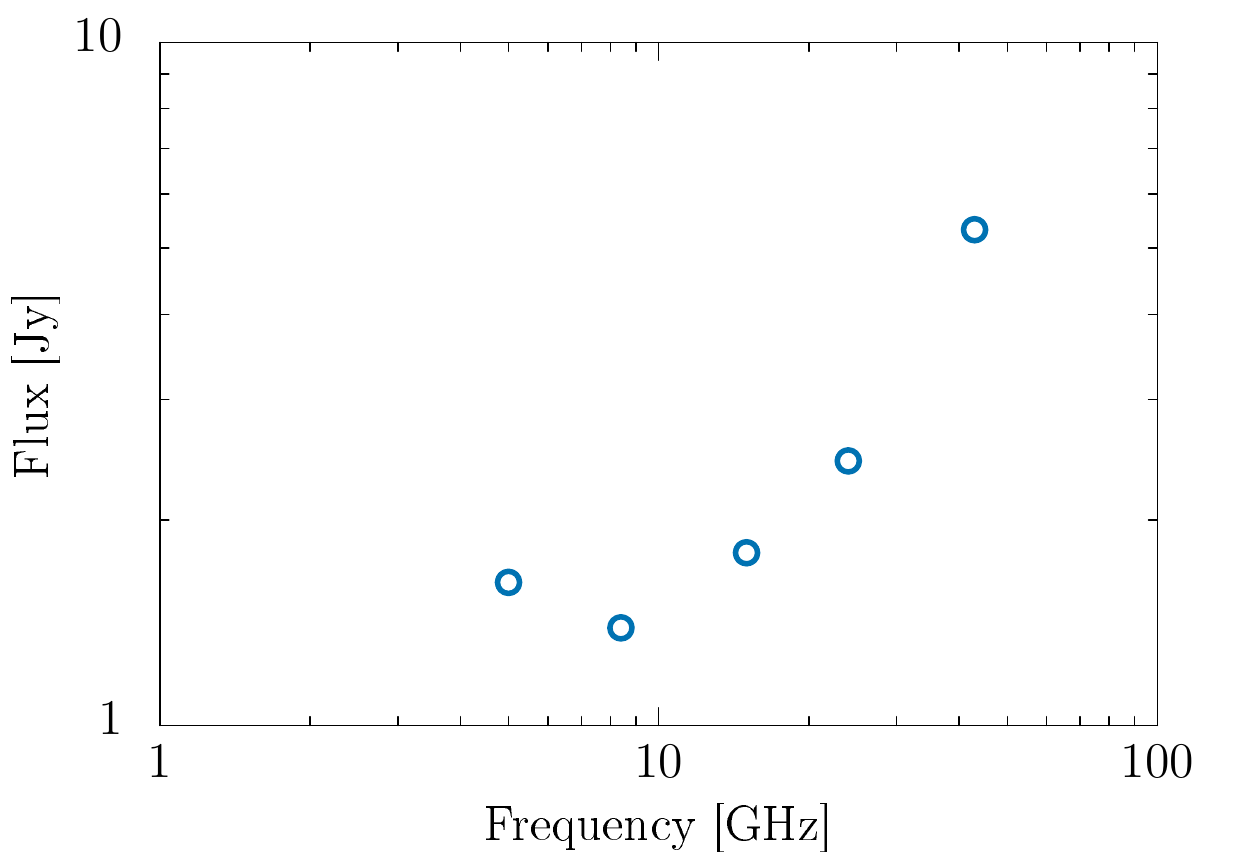}
    }
    \subfigure[]
    {
        \includegraphics[width=0.47\textwidth]{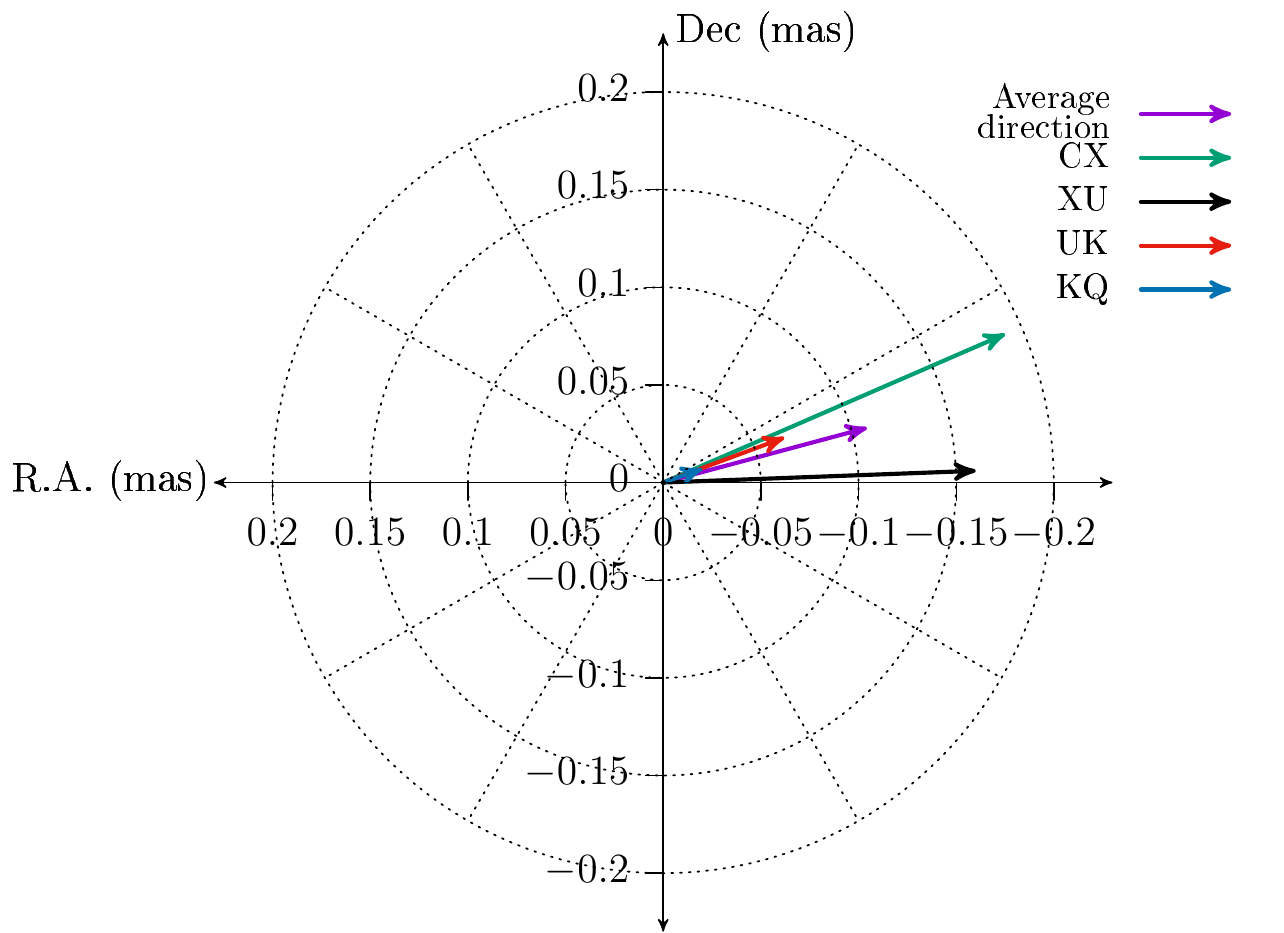}
    }
     \subfigure[]
    {
        \includegraphics[width=0.5\textwidth]{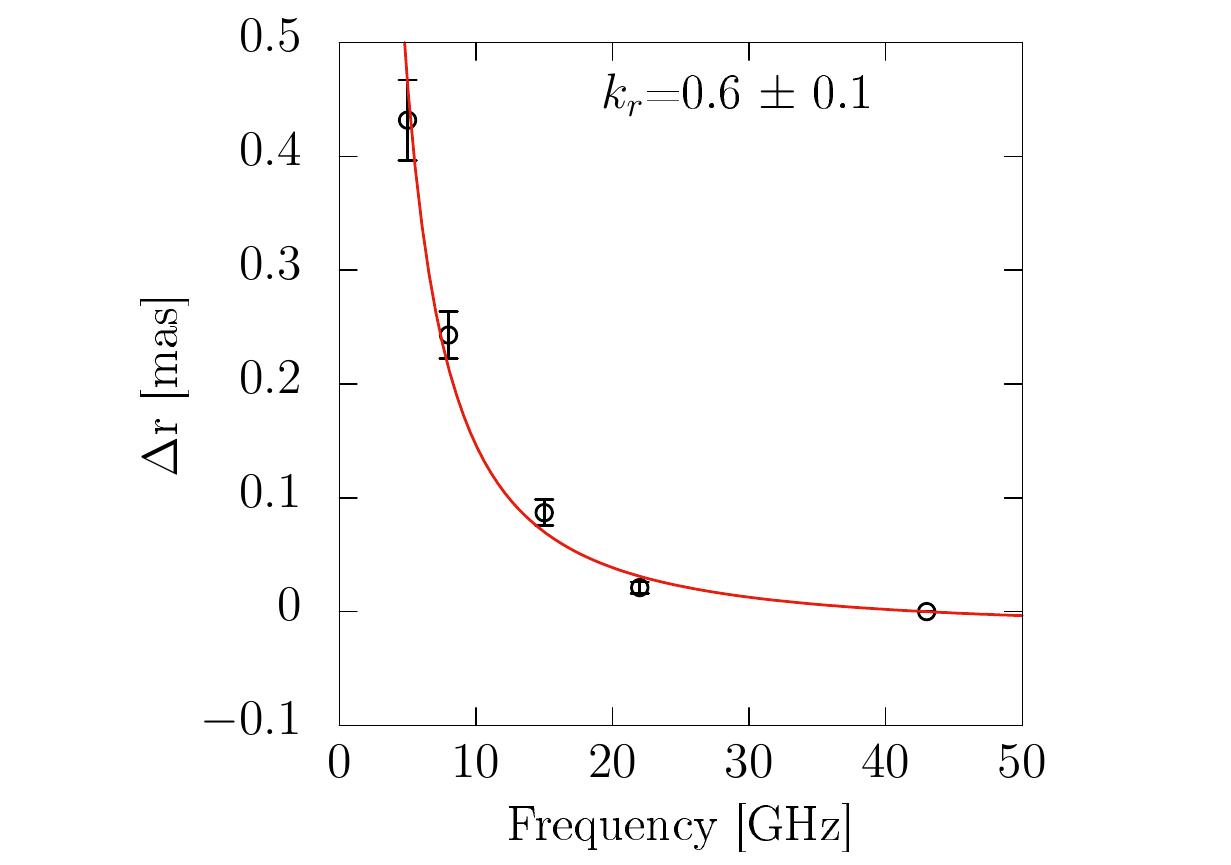}
    }
    \caption{Epoch 10, 2007-06-16. (a) Core spectrum. (b) Core-shift vectors of all frequency pairs. (c) Power-law fit is shown with the red curve.}
    \label{CSepoch10}
\end{figure}

\begin{figure}[!h]
\centering
   \subfigure[]
    {
        \includegraphics[width=0.45\textwidth]{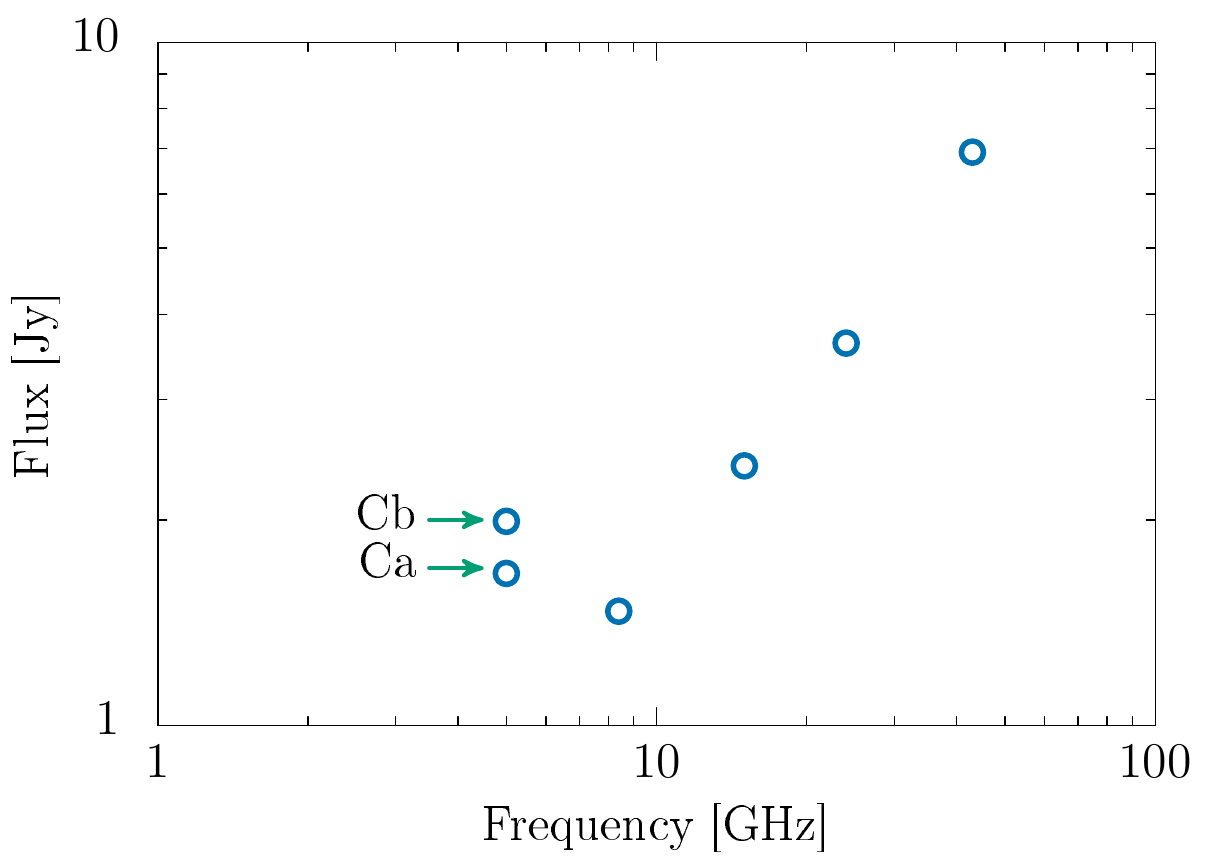}
    }
    \subfigure[]
    {
        \includegraphics[width=0.47\textwidth]{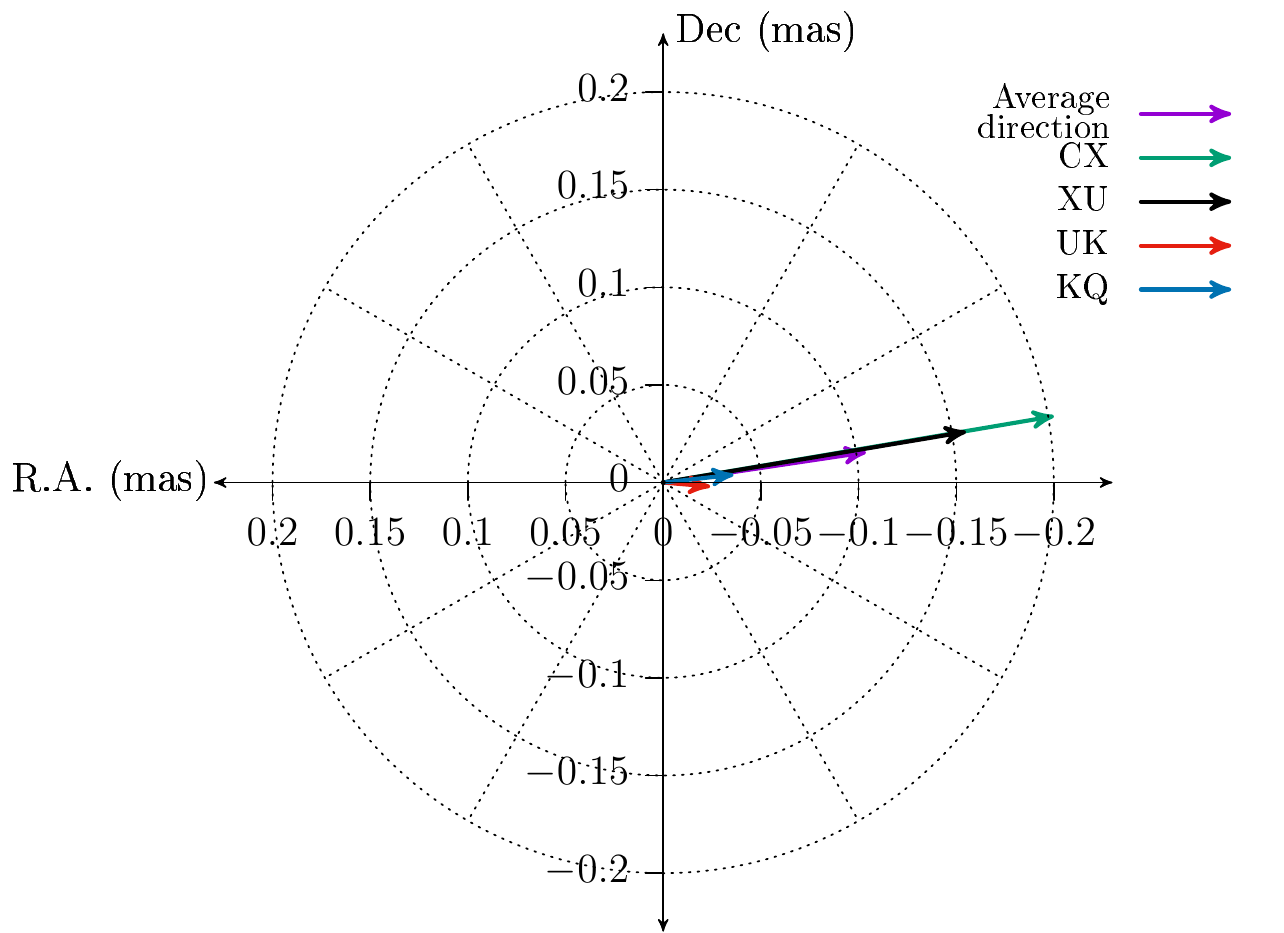}
    }
     \subfigure[]
    {
        \includegraphics[width=0.5\textwidth]{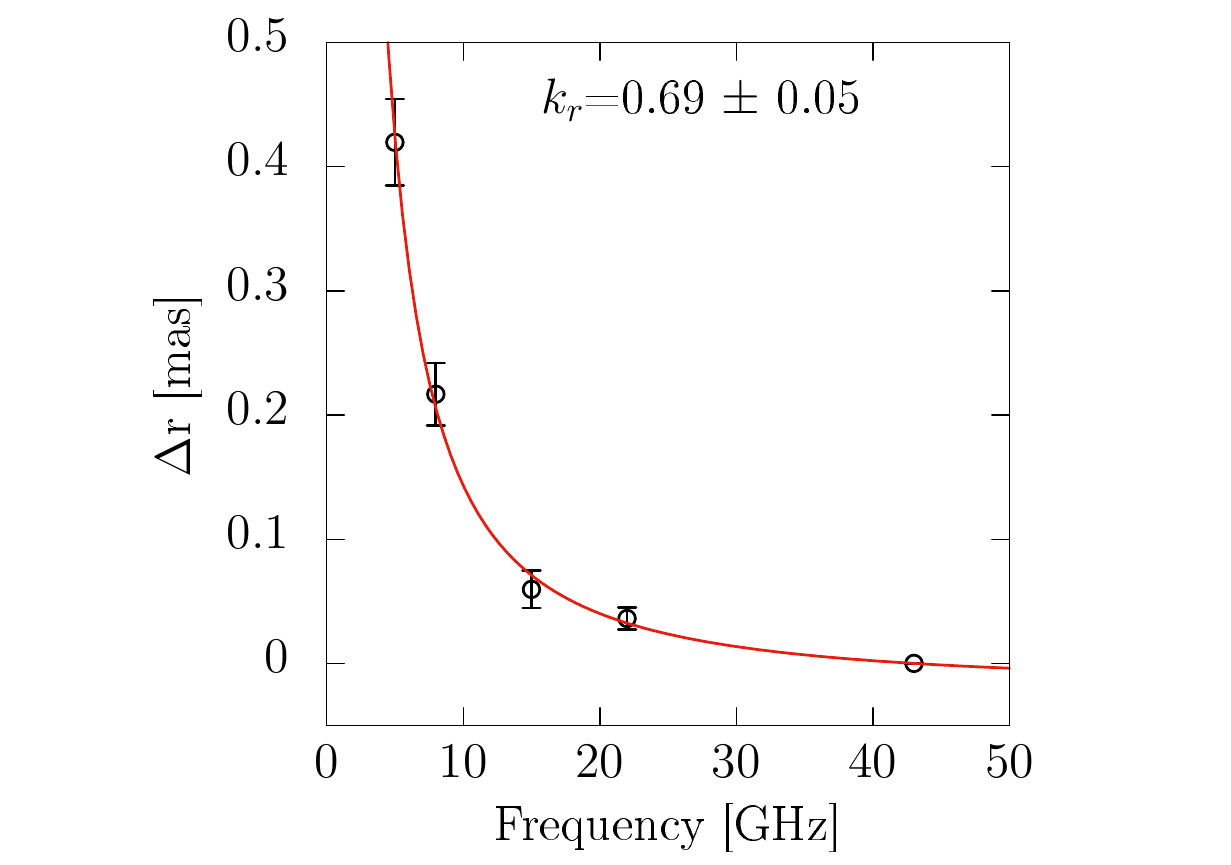}
    }
    \caption{Epoch 11, 2007-07-25.
    (a) Core spectrum, Ca represents the core. (b) Core-shift vectors of all frequency pairs. The choice of Ca core leads to the correct direction of the CX vector. (c) Power-law fit is shown with the red curve.}
    \label{CSepoch11}
\end{figure}

\begin{figure}[!h]
\centering
   \subfigure[]
    {
        \includegraphics[width=0.45\textwidth]{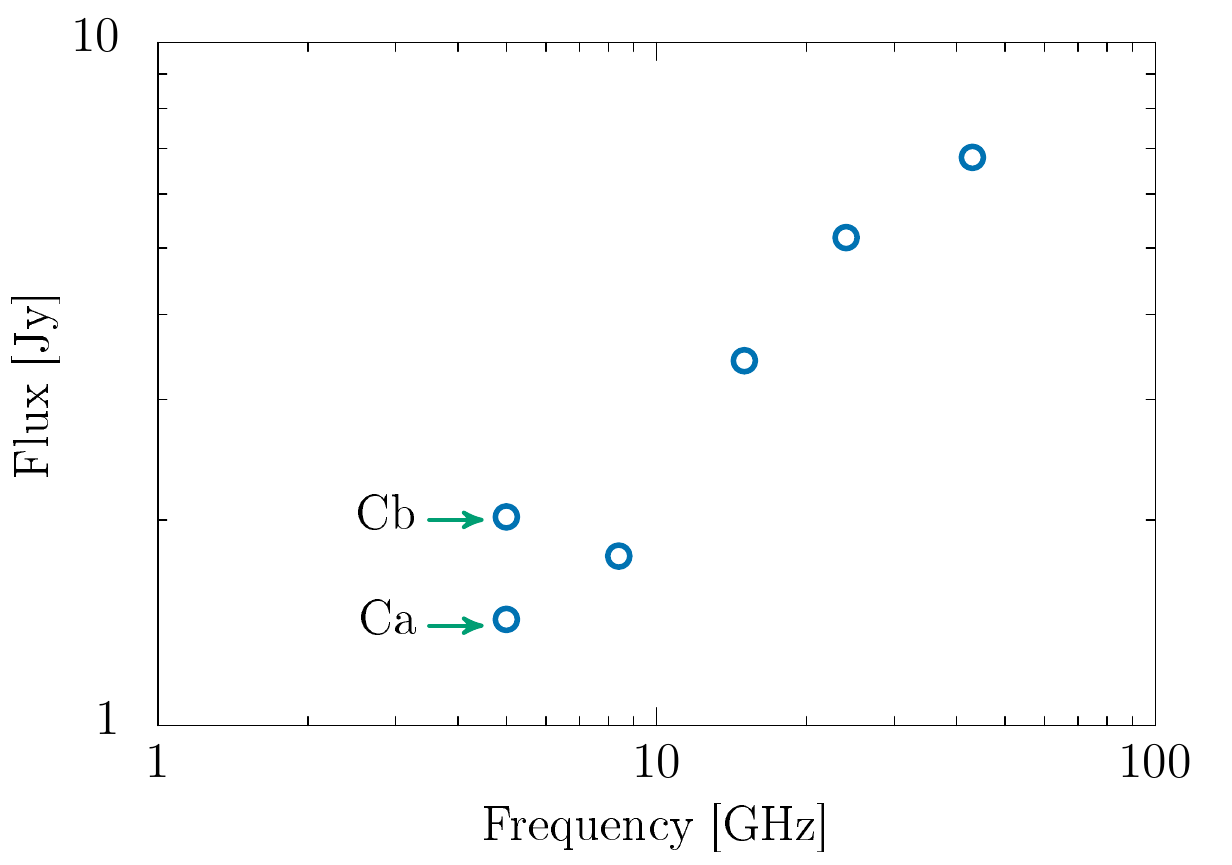}
    }
    \subfigure[]
    {
        \includegraphics[width=0.47\textwidth]{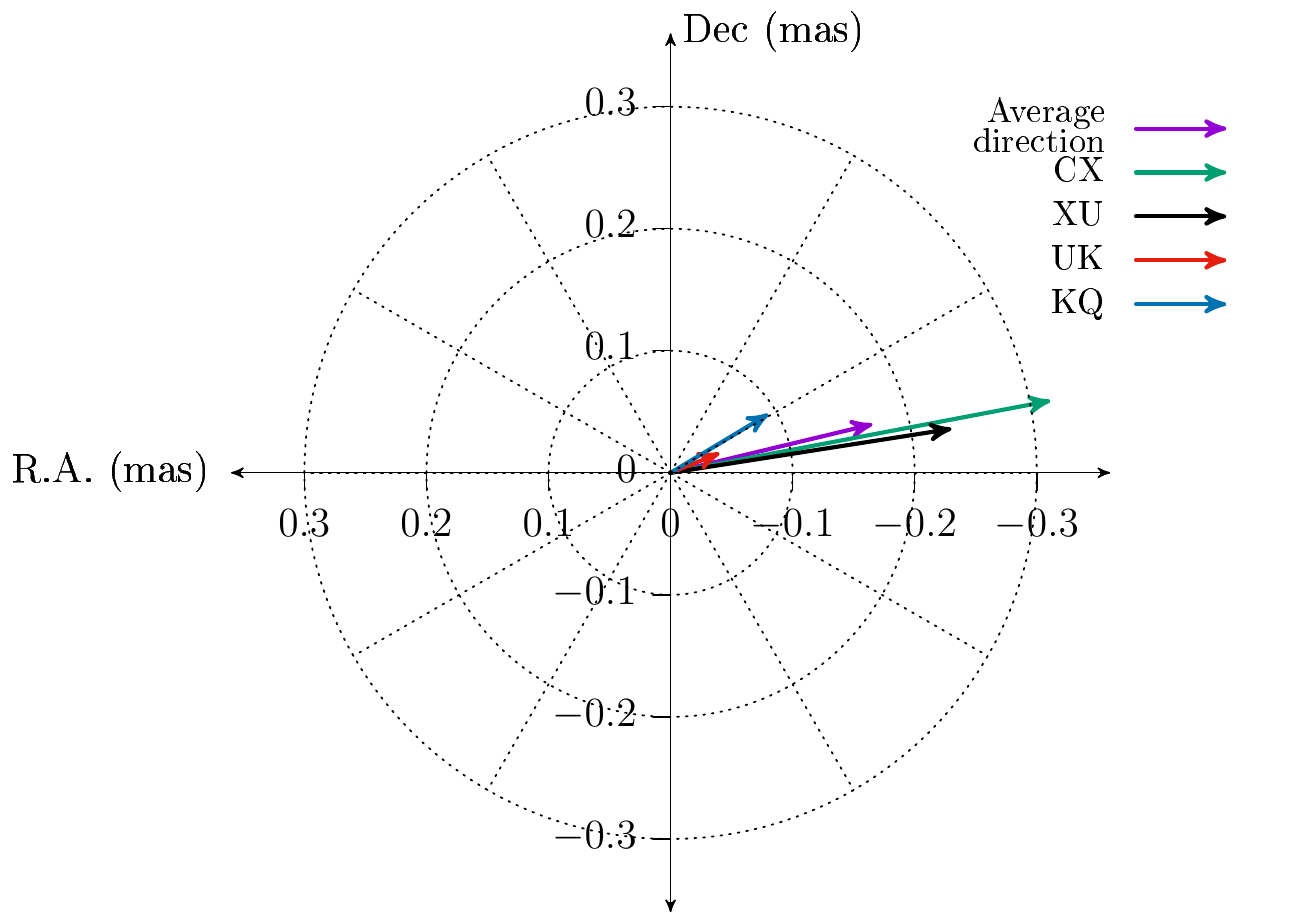}
    }
     \subfigure[]
    {
        \includegraphics[width=0.5\textwidth]{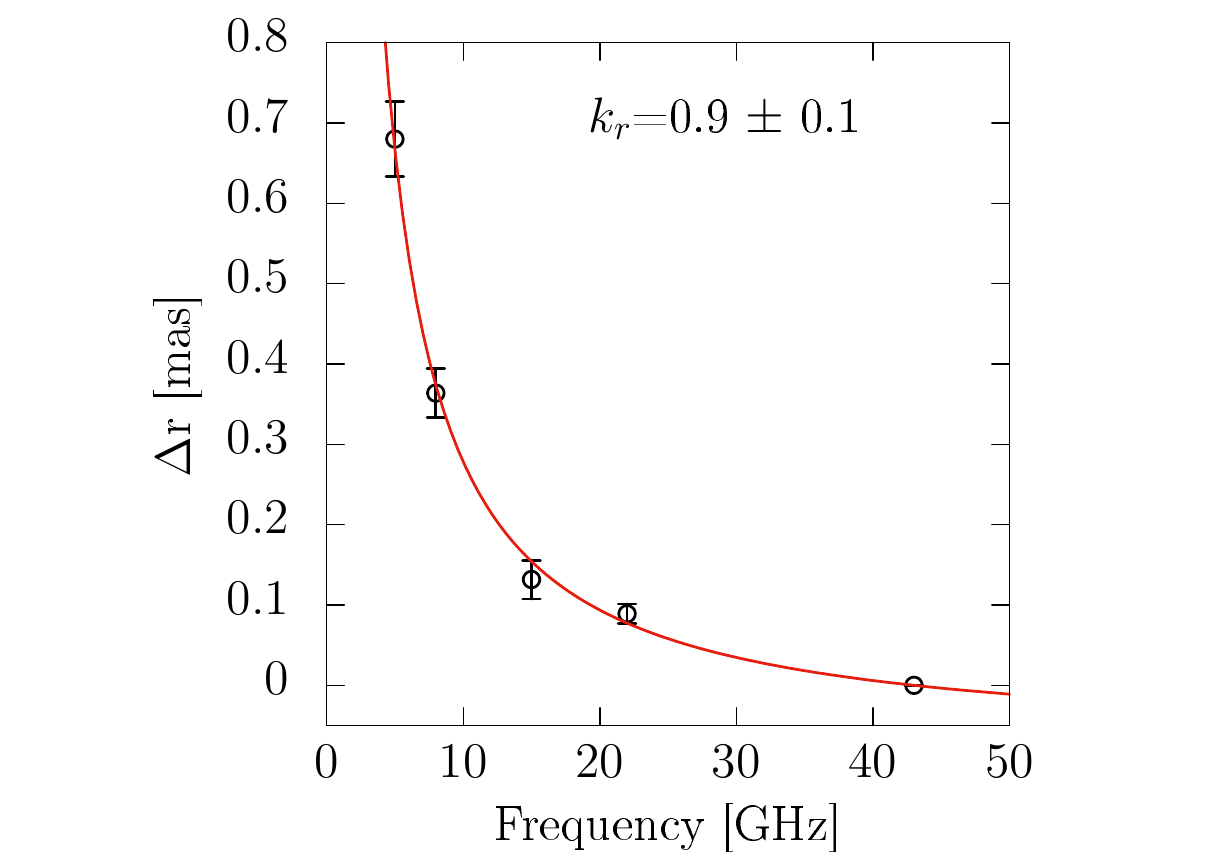}
    }
    \caption{Epoch 12, 2007-09-13. (a) Core spectrum, Cb represents the core. (b) Core-shift vectors of all frequency pairs. The choice of Cb core leads to the correct direction of the CX vector. (c) Power-law fit is shown with the red curve.}
    \label{CSepoch12}
\end{figure}

\begin{figure}[!h]
\centering
   \subfigure[]
    {
        \includegraphics[width=0.45\textwidth]{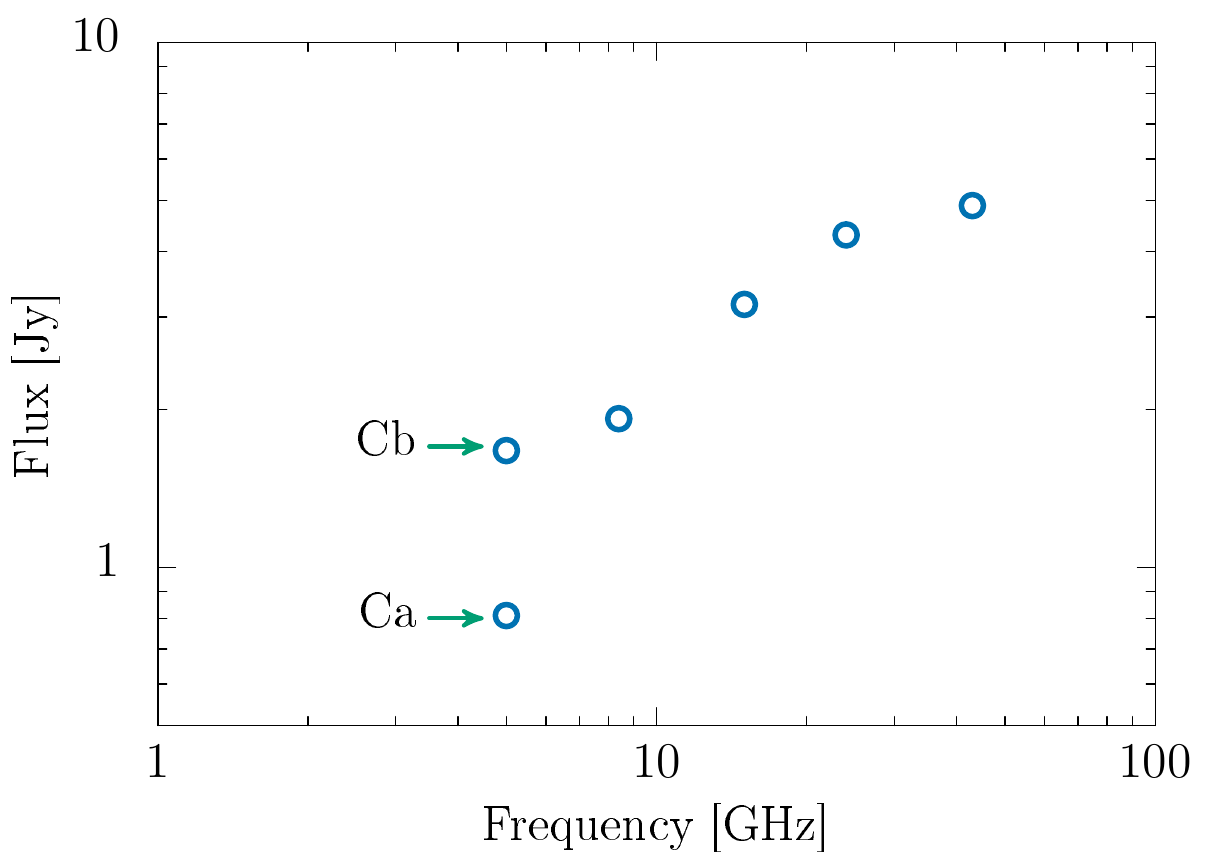}
    }
     \subfigure[]
    {
        \includegraphics[width=0.47\textwidth]{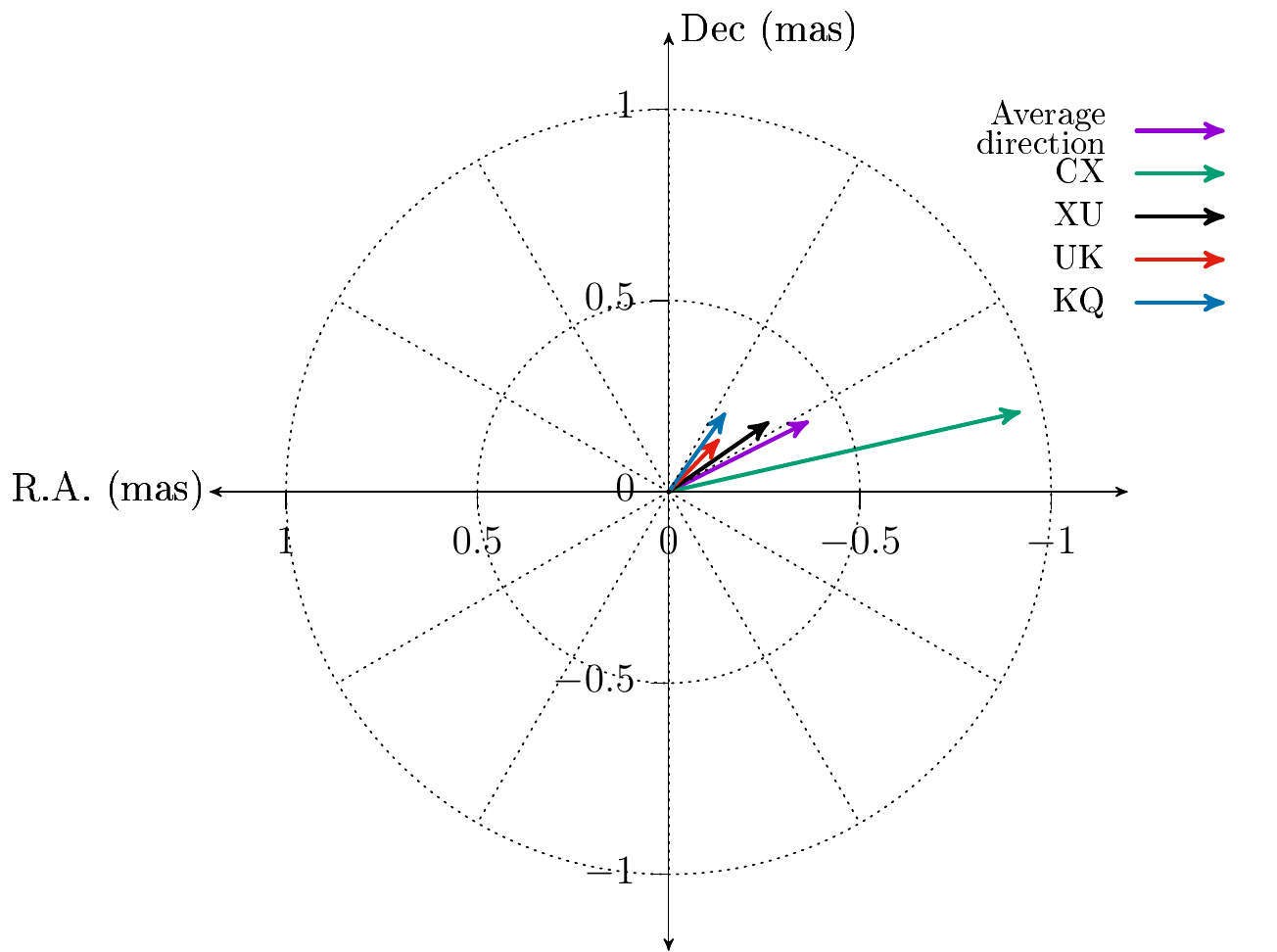}
    }
     \subfigure[]
    {
        \includegraphics[width=0.5\textwidth]{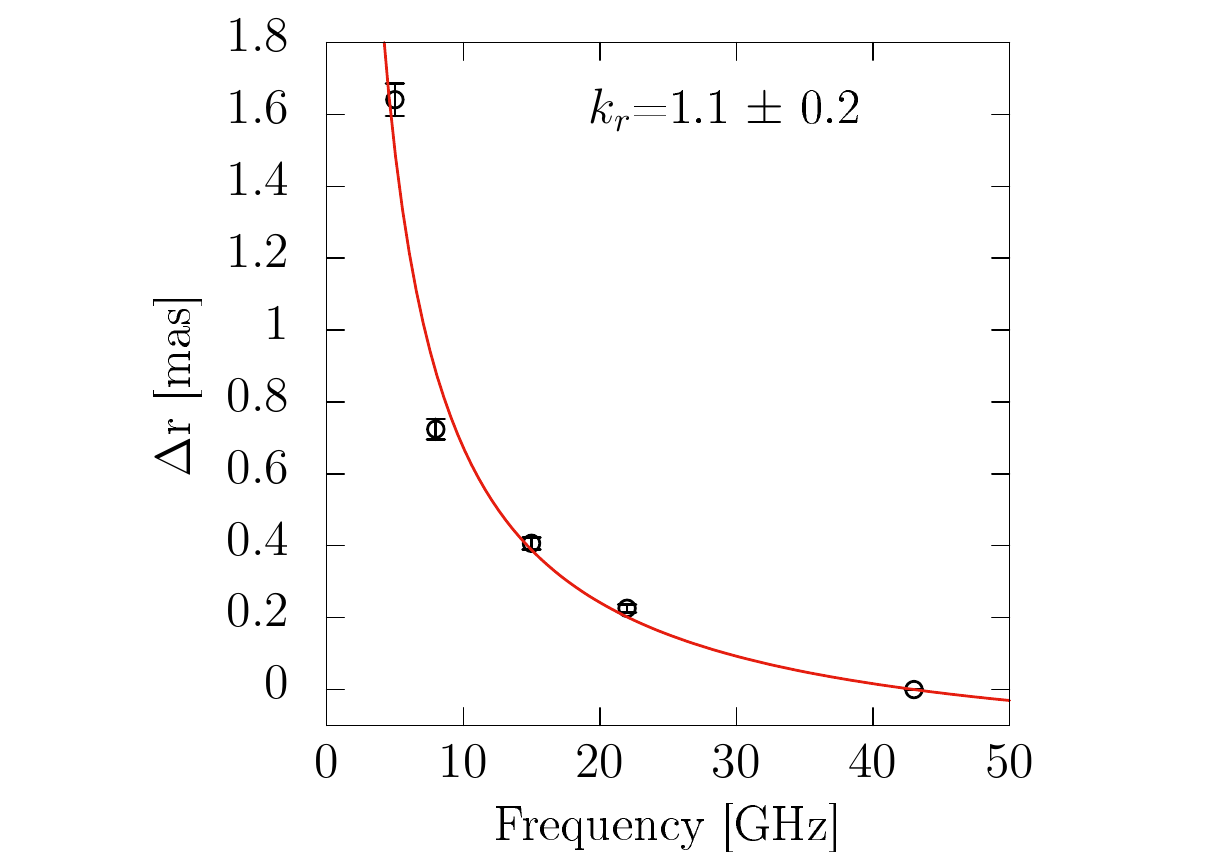}
    }
    \caption{Epoch 13, 2008-01-03. (a) Core spectrum, Cb possibly represents the core. (b) Core-shift vectors of all frequency pairs. (c) Power-law fit is shown with the red curve. Due to the poor $(u,v)$ coverage, the core at the C band cannot be well resolved, producing ambiguities on its location. As a result very large core-shift values above 1\,mas were measured. This observation was not included for the variability analysis.}
    \label{CSepoch13}
\end{figure}

\begin{figure*}[!h]
\centering
   \subfigure[]
    {
        \includegraphics[width=0.45\textwidth]{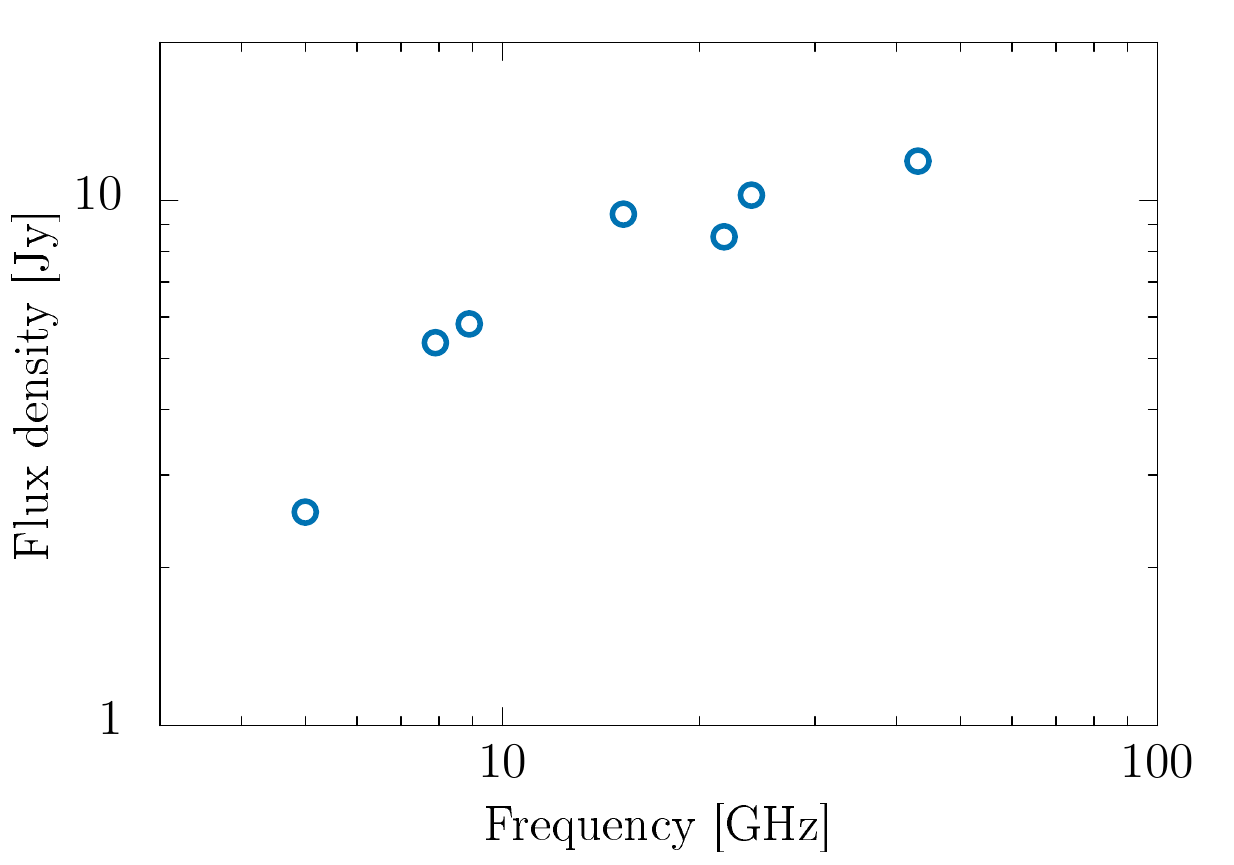}
    }
    \hspace{2cm}
    \subfigure[]
    {
        \includegraphics[width=0.52\textwidth]{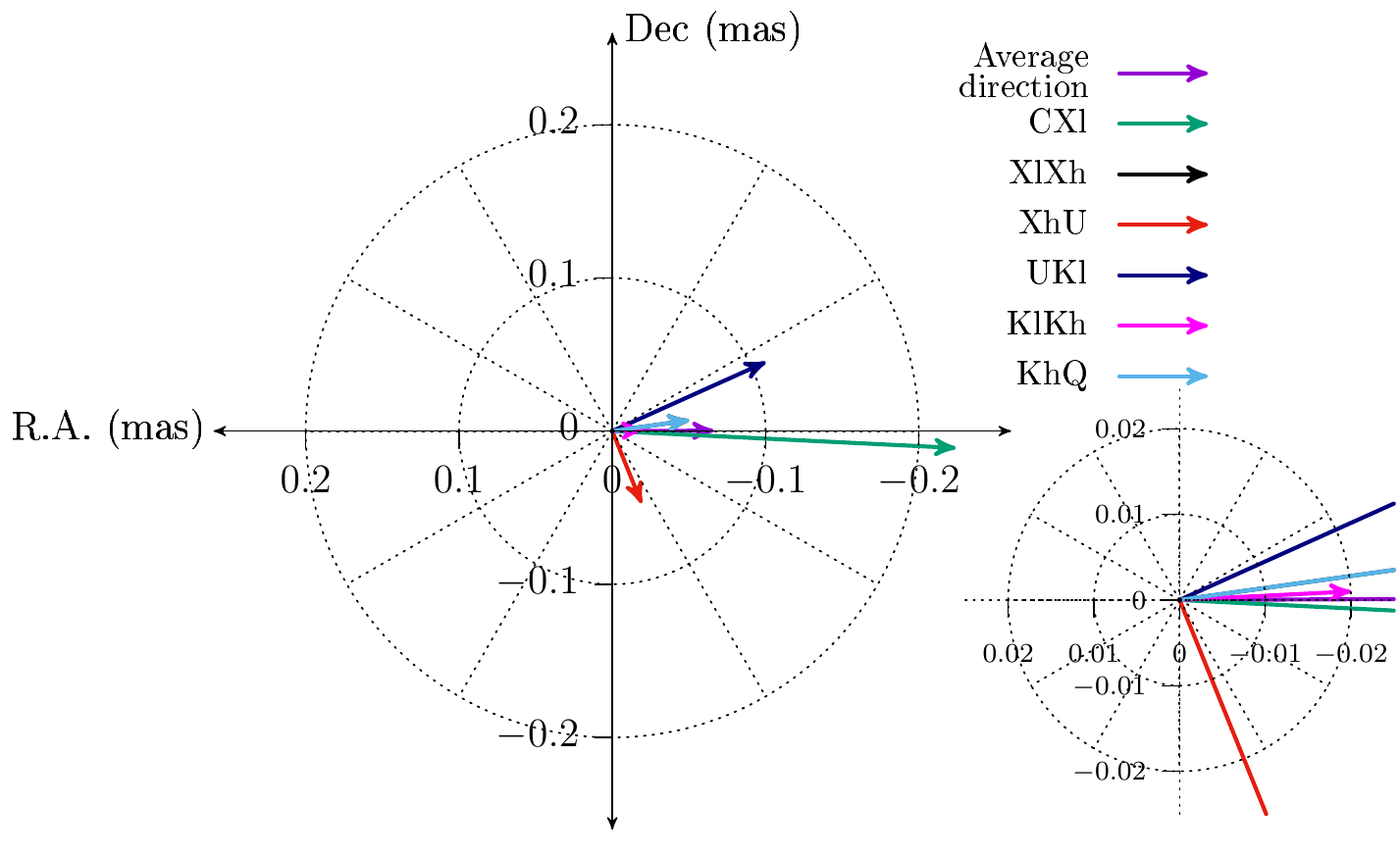}
    }
     \subfigure[]
    {
        \includegraphics[width=0.45\textwidth]{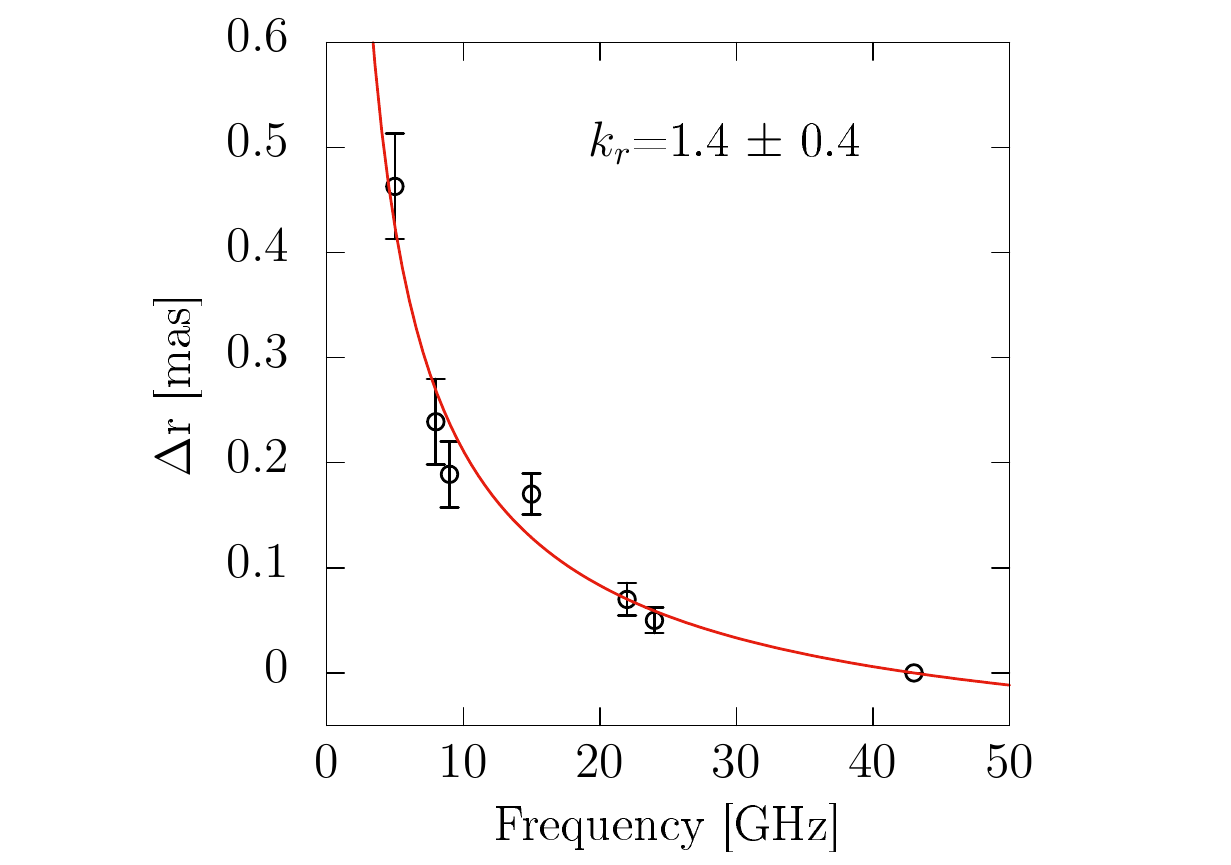}
    }    
    \subfigure[]
    {
        \includegraphics[width=0.52\textwidth]{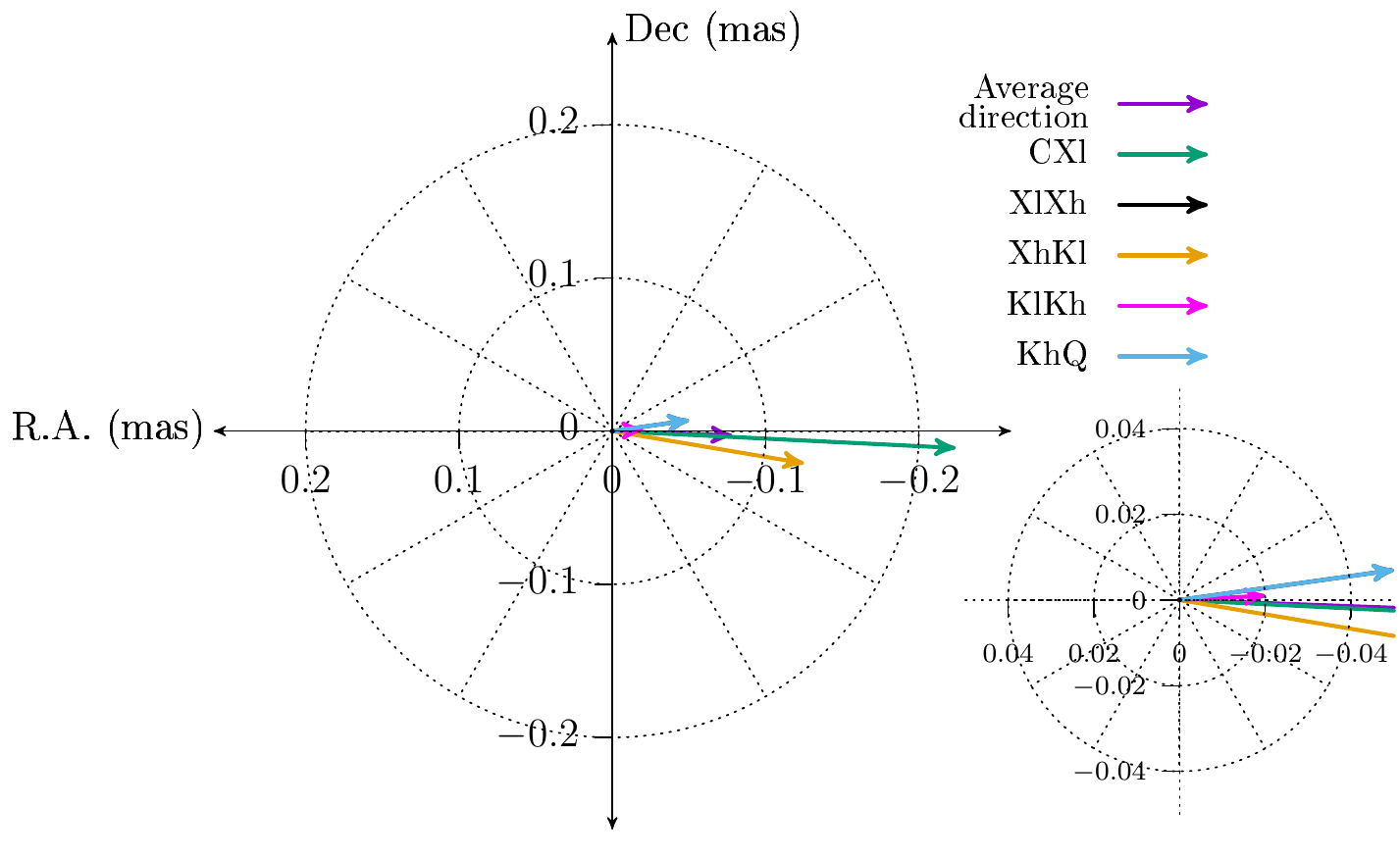}
    }
       \subfigure[]
    {
        \includegraphics[width=0.45\textwidth]{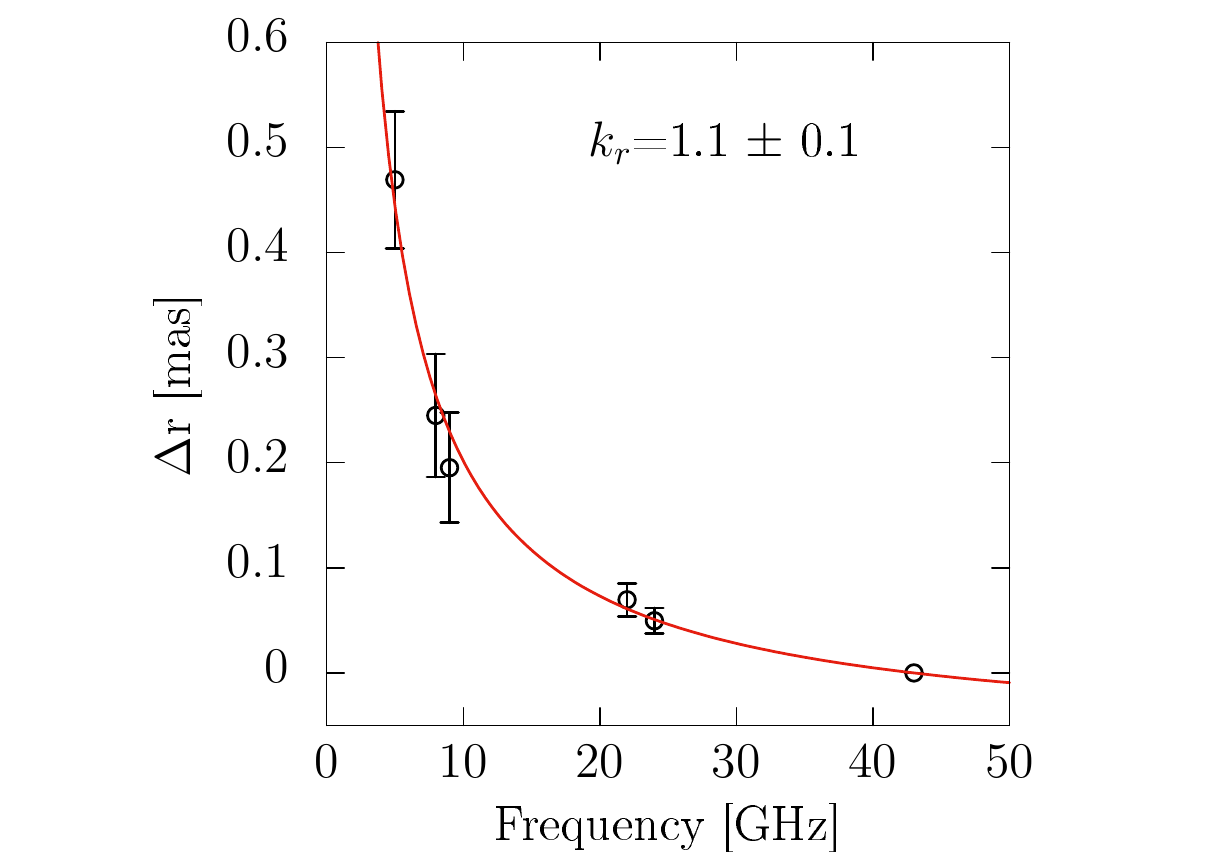}
    }
    \caption{Epoch 14, 2008-12-07. (a) Core spectrum. (b) Core-shift vectors of all frequency pairs including intermediate frequencies denoted by Xl (7.9\,GHz), Xh (8.9\,GHz), Kl (21.8\,GHz) and  Kh (24\,GHz). (c) Power-law fit (red curve) using all bands. (d) Core-shift vectors without the U-band, (e) Power-law fit (red curve) using all bands except the U-band. This approach leads to a better fit.}
    \label{CSepoch14}
\end{figure*}

\begin{figure*}[!h]
\centering
   \subfigure[]
    {
        \includegraphics[width=0.45\textwidth]{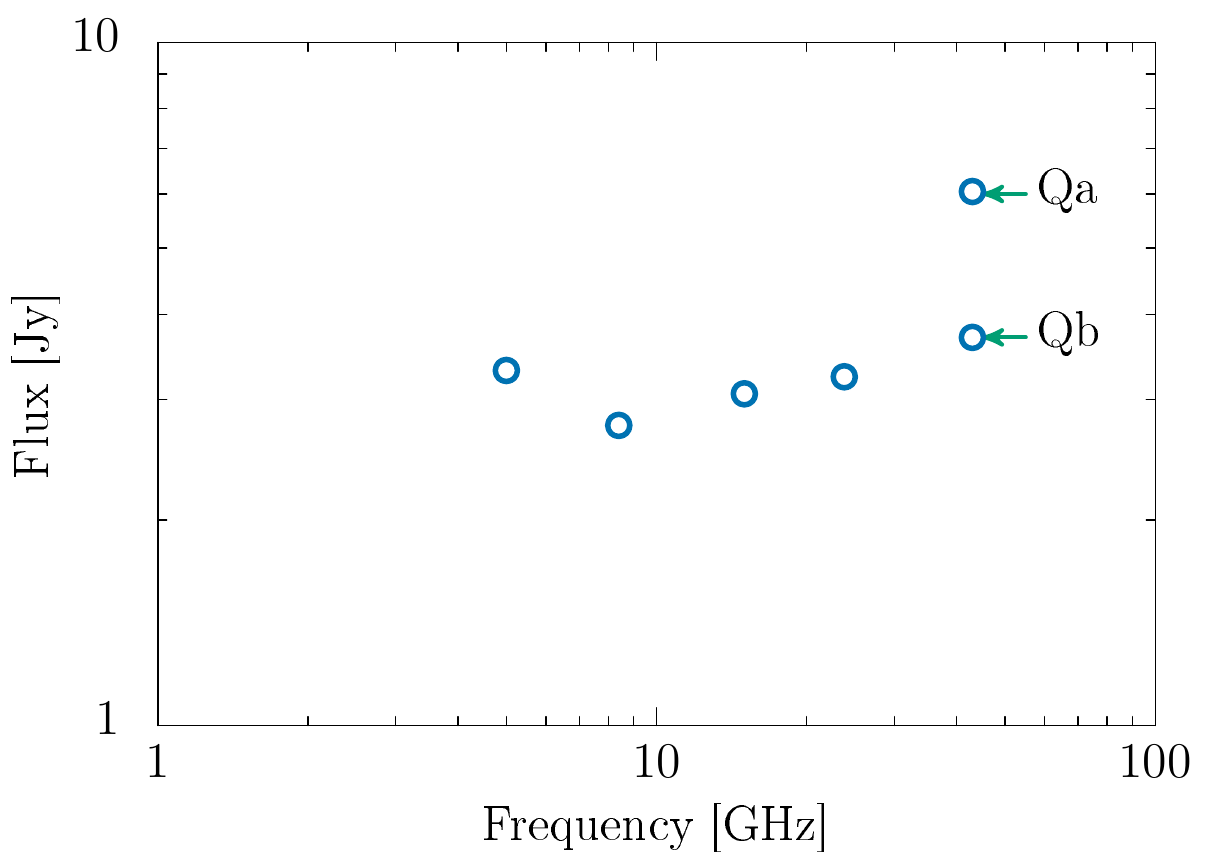}
    }
    \subfigure[]
    {
        \includegraphics[width=0.45\textwidth]{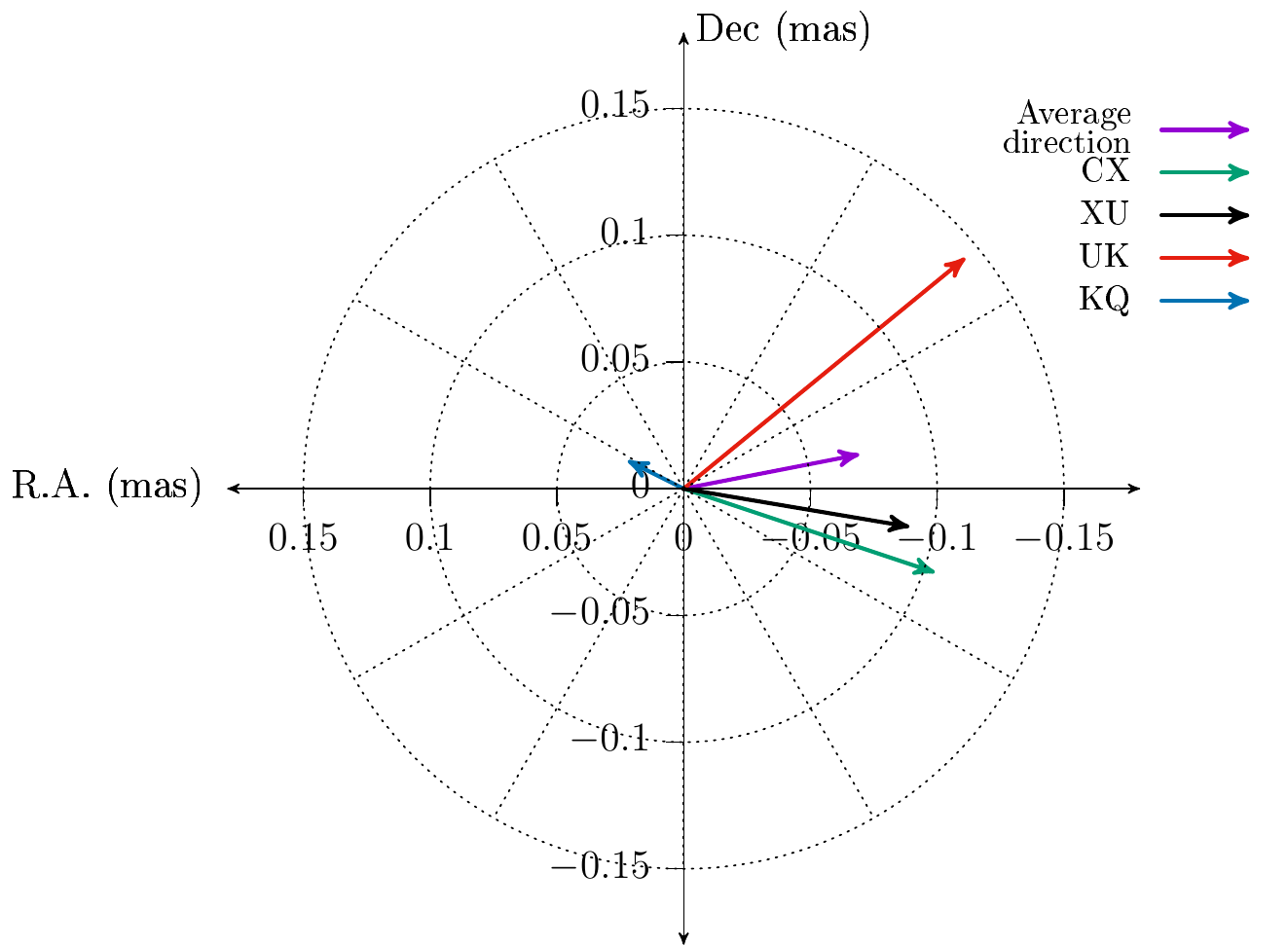}
    }
    \subfigure[]
    {
        \includegraphics[width=0.45\textwidth]{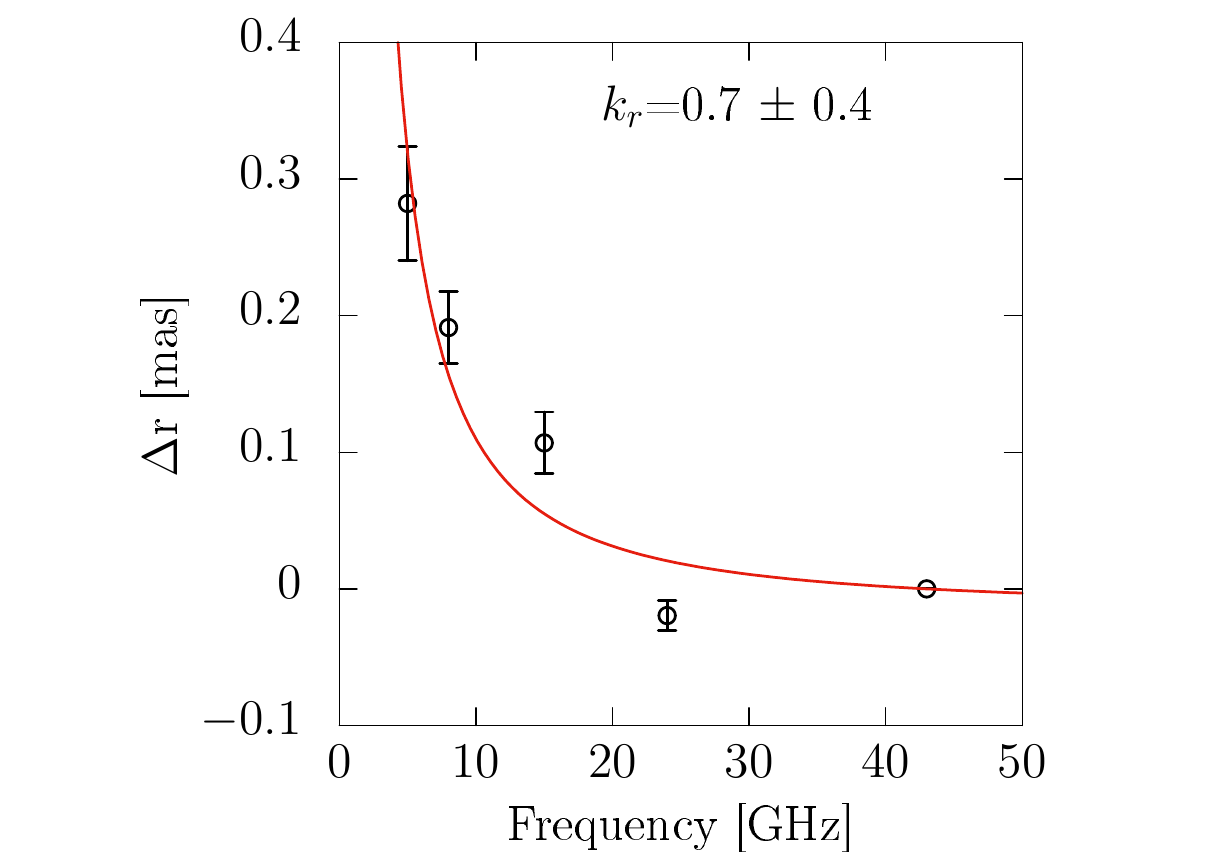}
    }
     \subfigure[]
    {
        \includegraphics[width=0.45\textwidth]{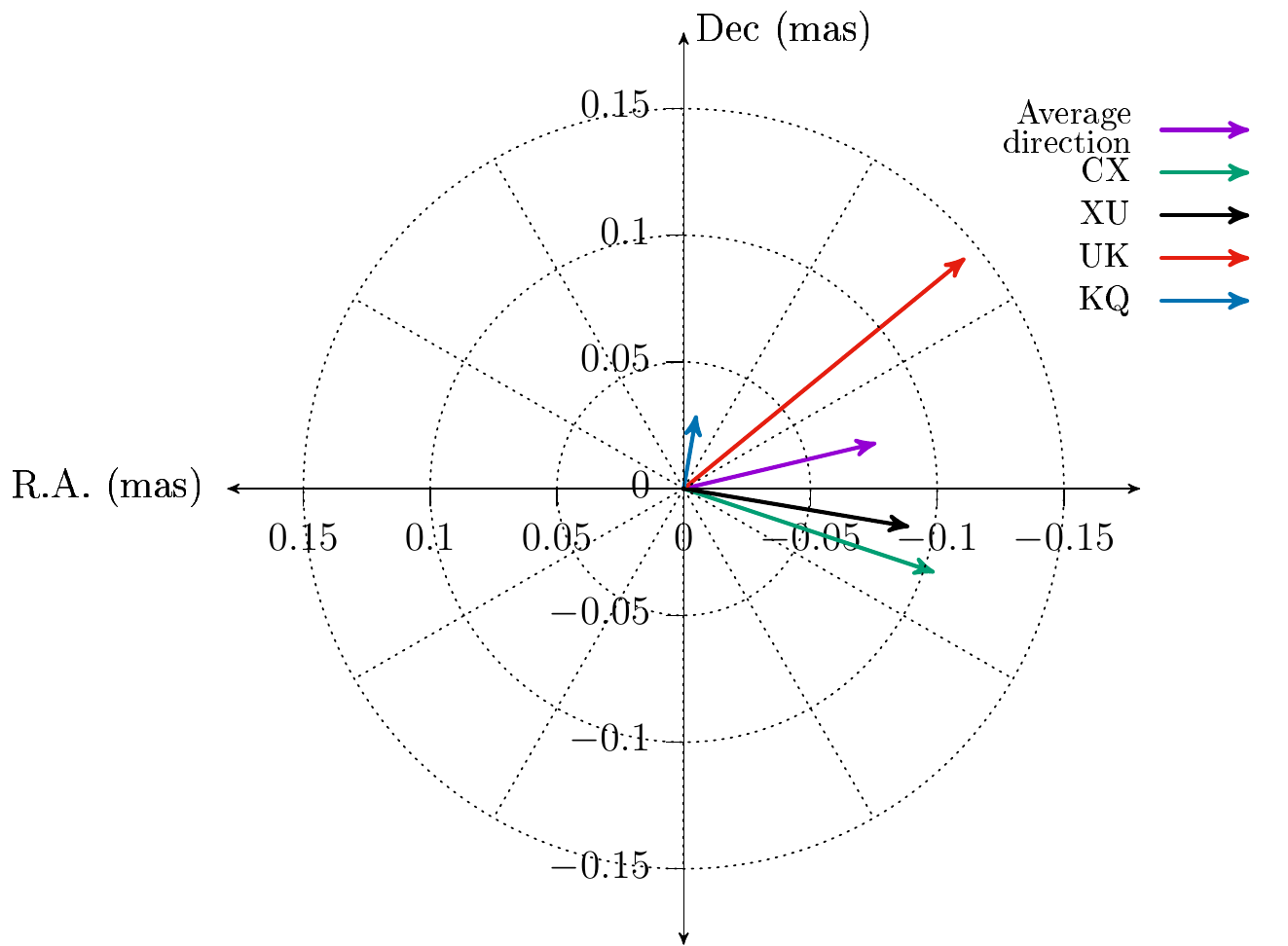}
    } 
     \subfigure[]
    {
        \includegraphics[width=0.45\textwidth]{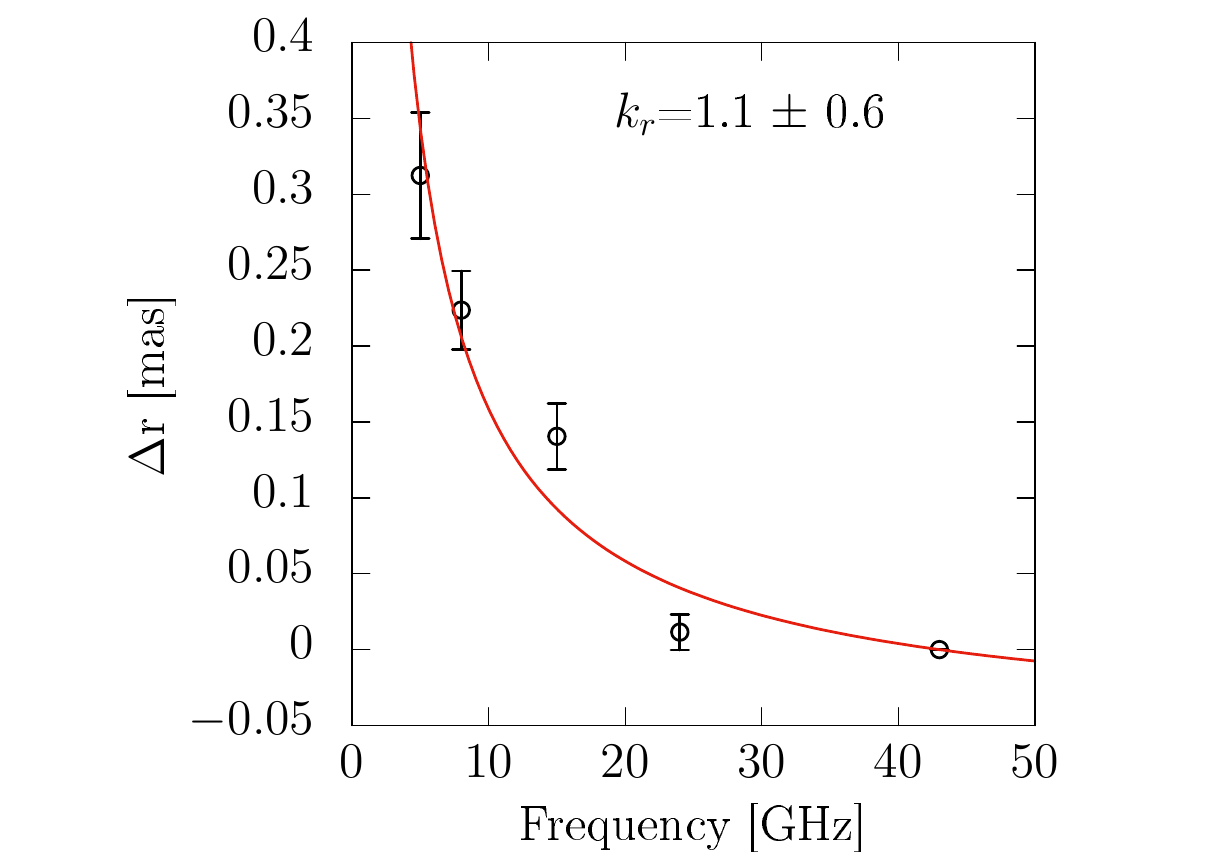}
    }
     \subfigure[]
    {
        \includegraphics[width=0.45\textwidth]{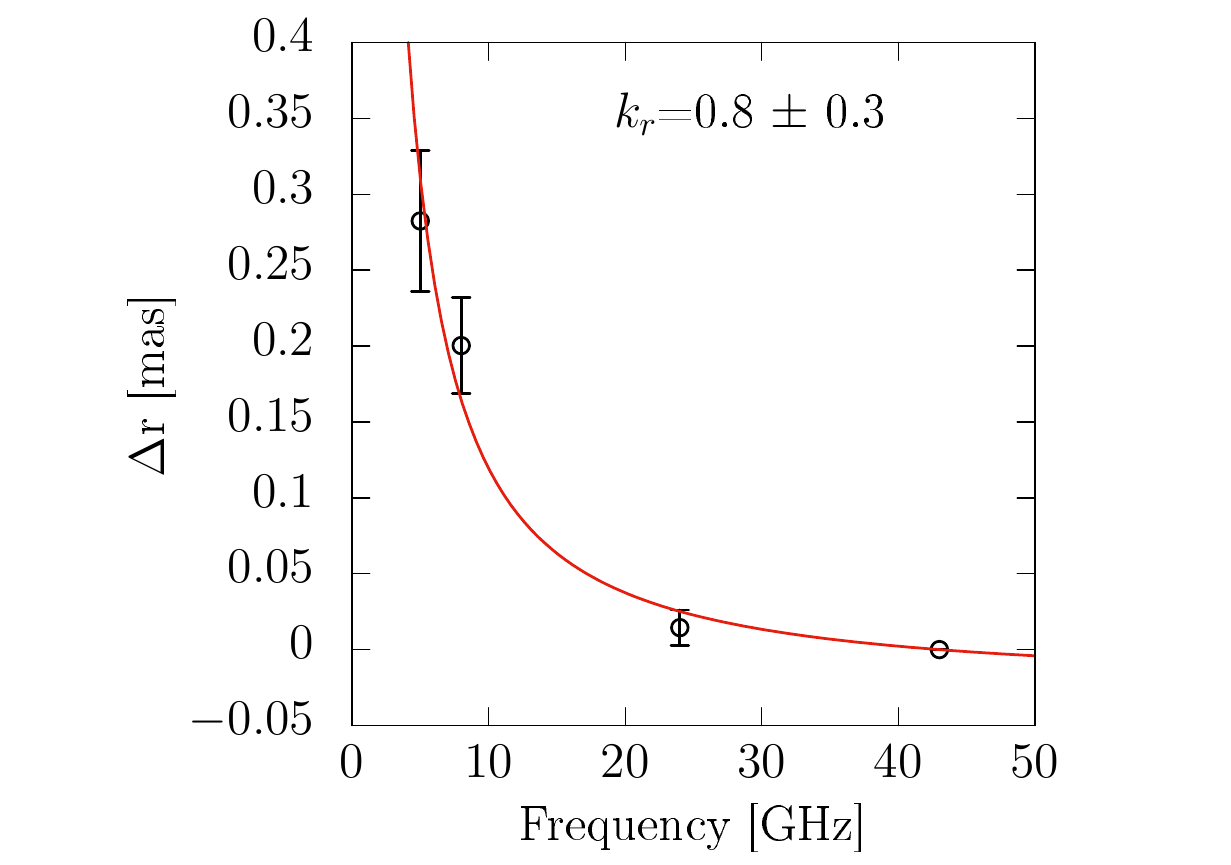}
    }
     \caption{Epoch 15, 2009-09-22. a) Core spectrum, where Qa represents the core and Qb the feature moving downstream. For comparisons see Figure~\ref{43GHzcore}a. (b) Core-shift vectors of all frequency pairs. Using component Qb makes the KQ core-shift vector to point in the opposite direction. c) Power law fit (red curve) using the Qb component. d) Core-shift vectors of all frequency pairs. Using component Qa, labeled as the core gives a reasonable direction of the KQ core shift vector. e) Power law fit (red curve) using Qa as the core but a better fit is obtained f) when the U band is excluded.}
    \label{CSepoch15}
\end{figure*}

\begin{figure*}[!h]
\centering
   \subfigure[.]
    {
        \includegraphics[width=0.45\textwidth]{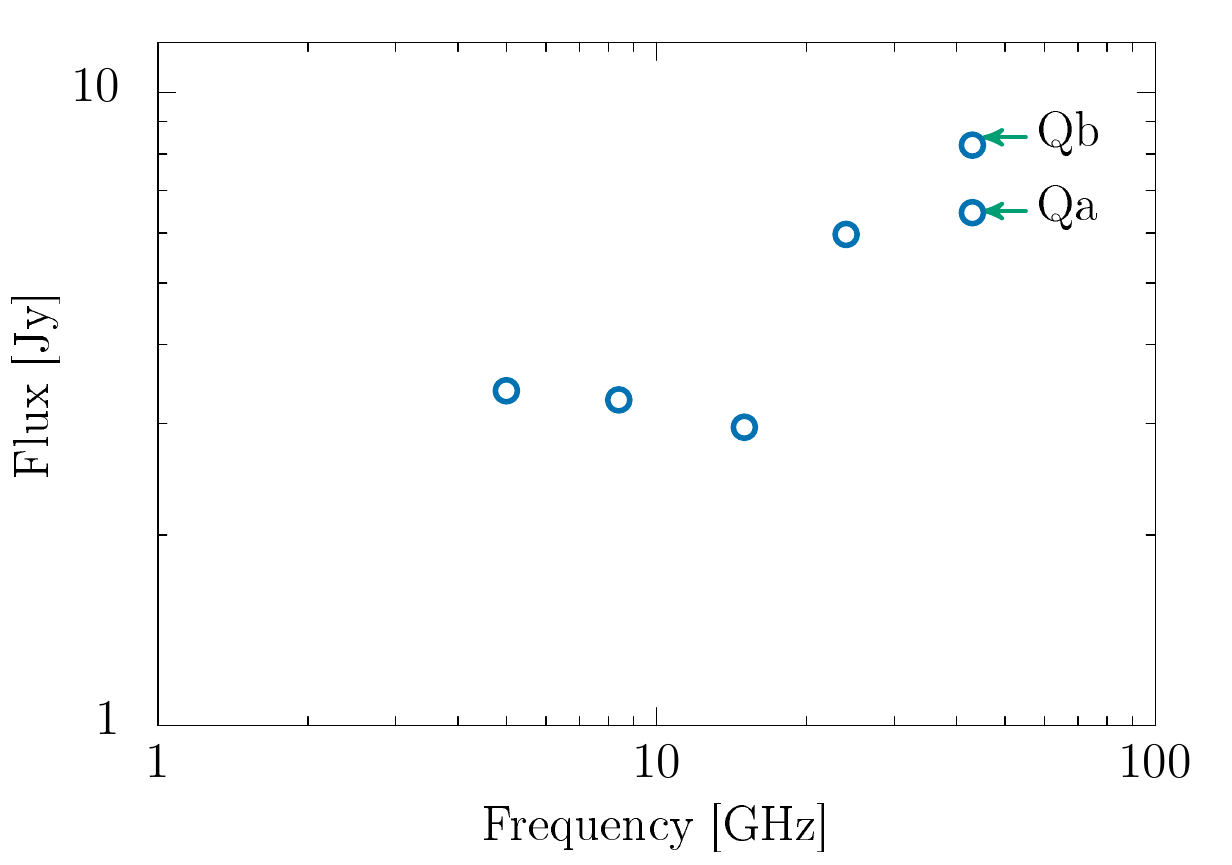}
    }
    \hspace{2cm}
    \subfigure[]
    {
        \includegraphics[width=0.45\textwidth]{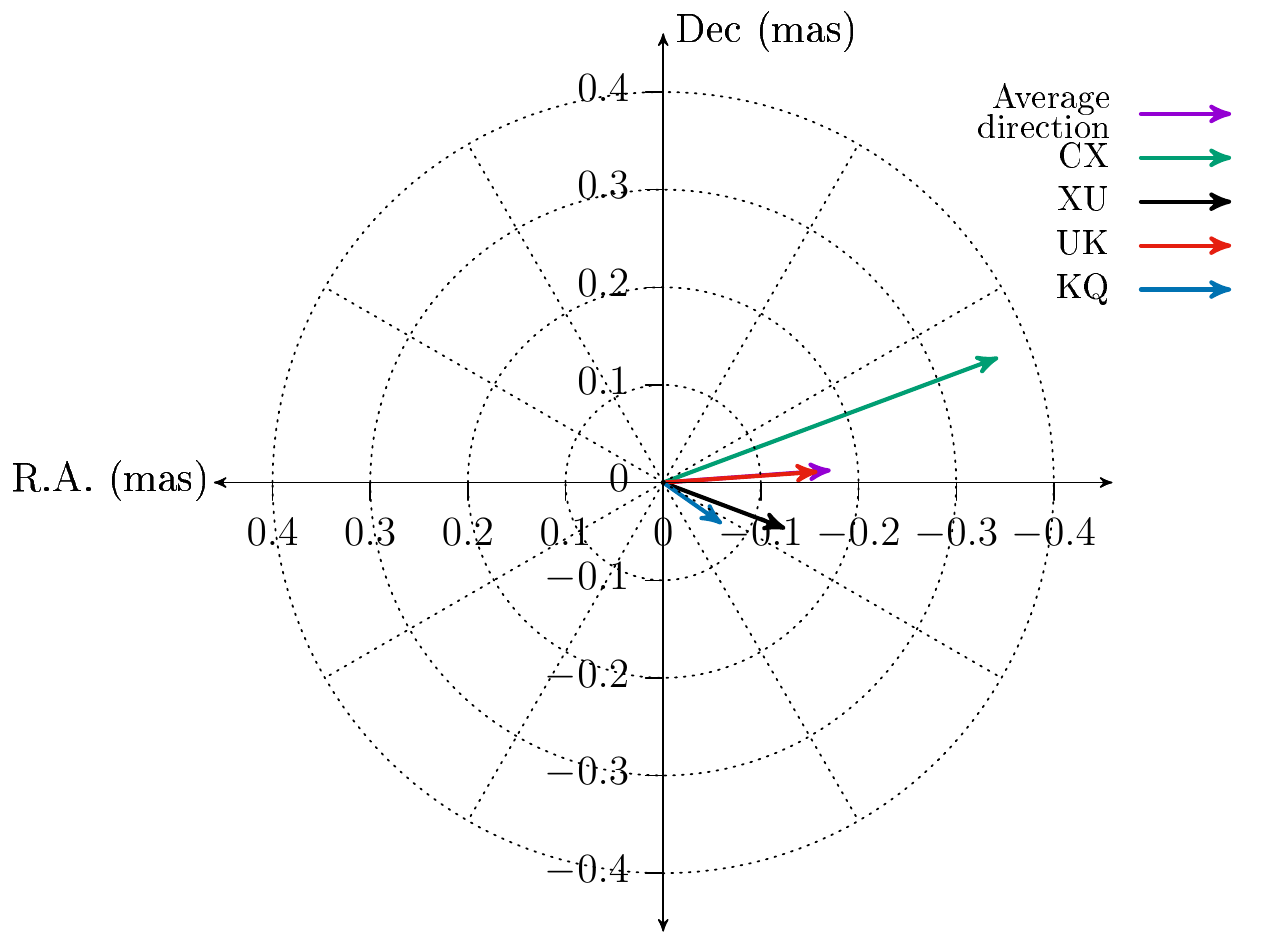}
    }
         \subfigure[]
    {
        \includegraphics[width=0.45\textwidth]{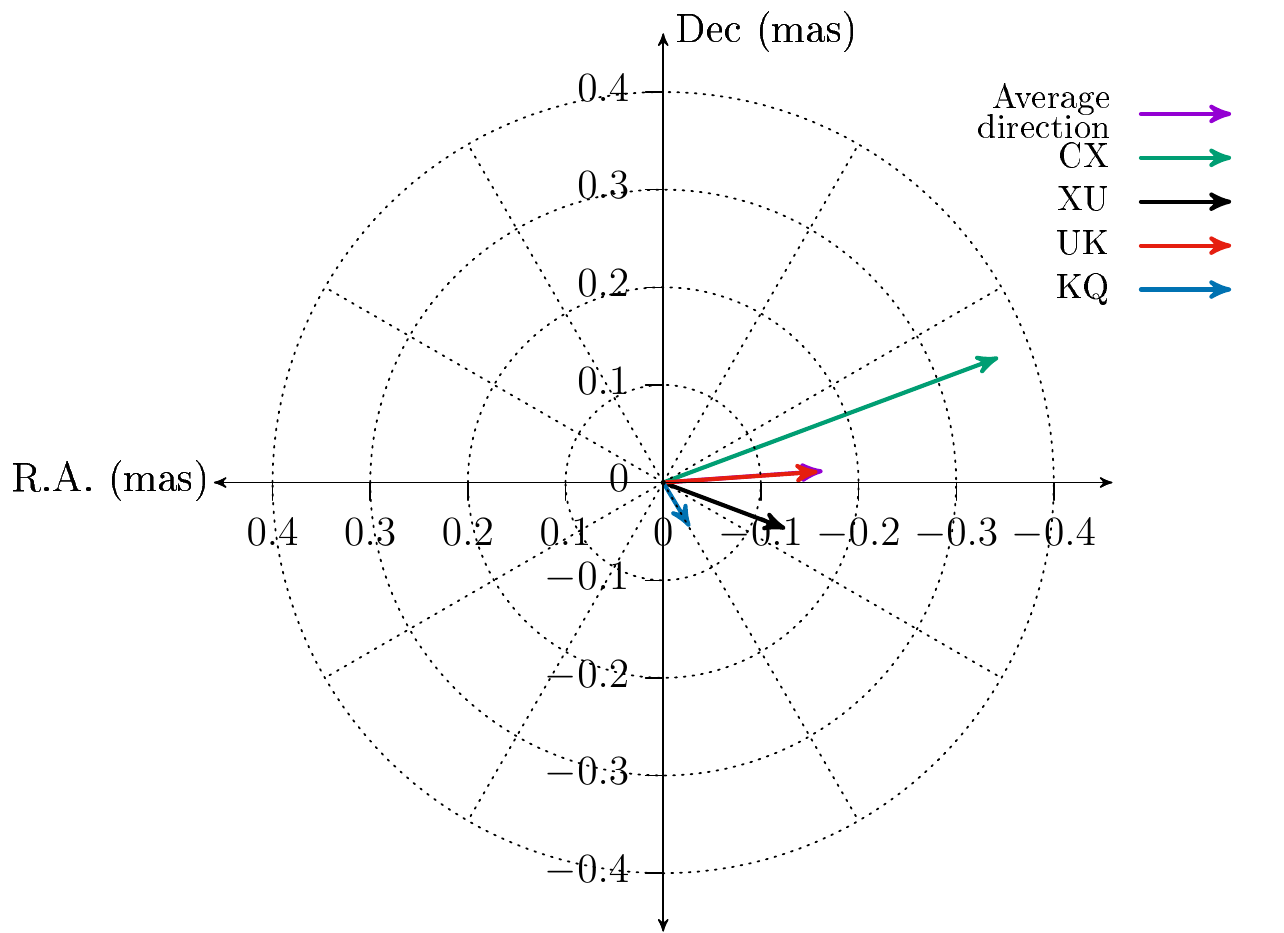}
    }
     \subfigure[]
    {
        \includegraphics[width=0.45\textwidth]{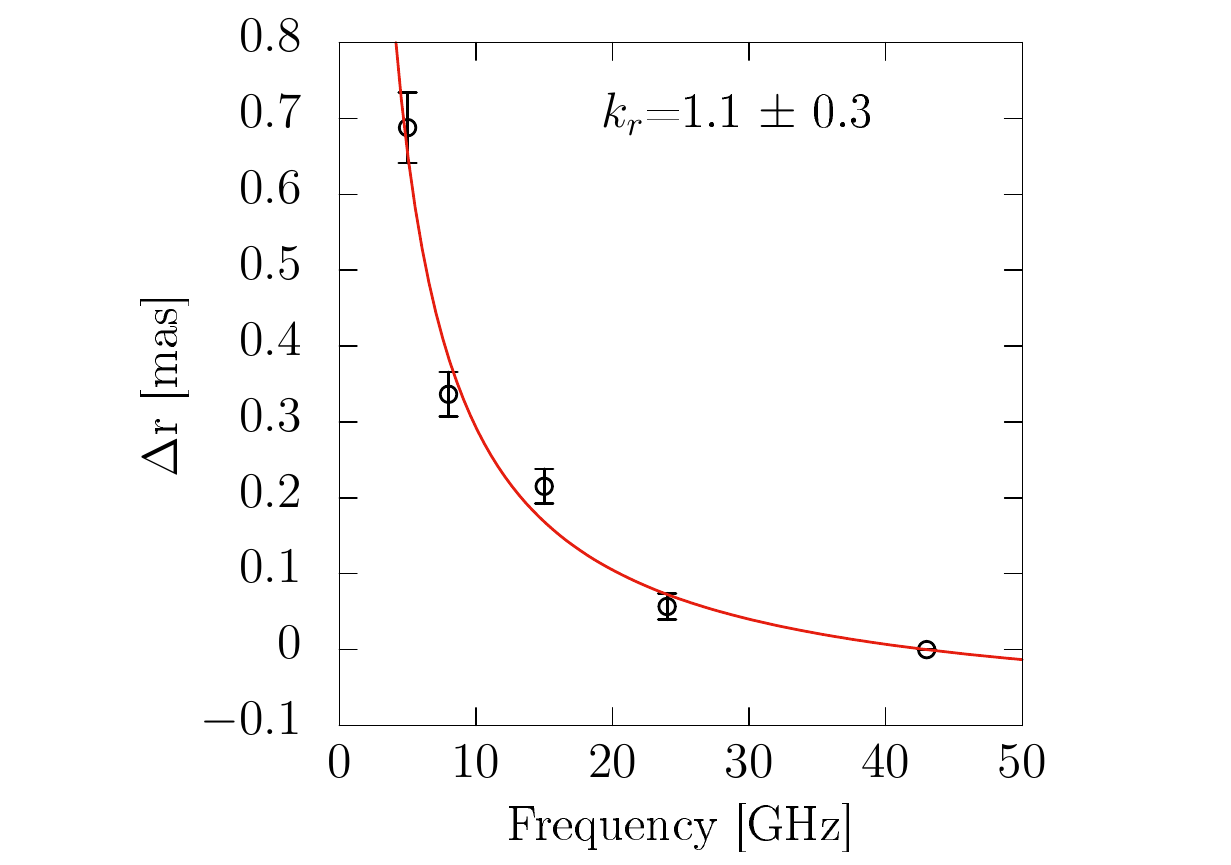}
    } 
     \subfigure[]
    {
        \includegraphics[width=0.45\textwidth]{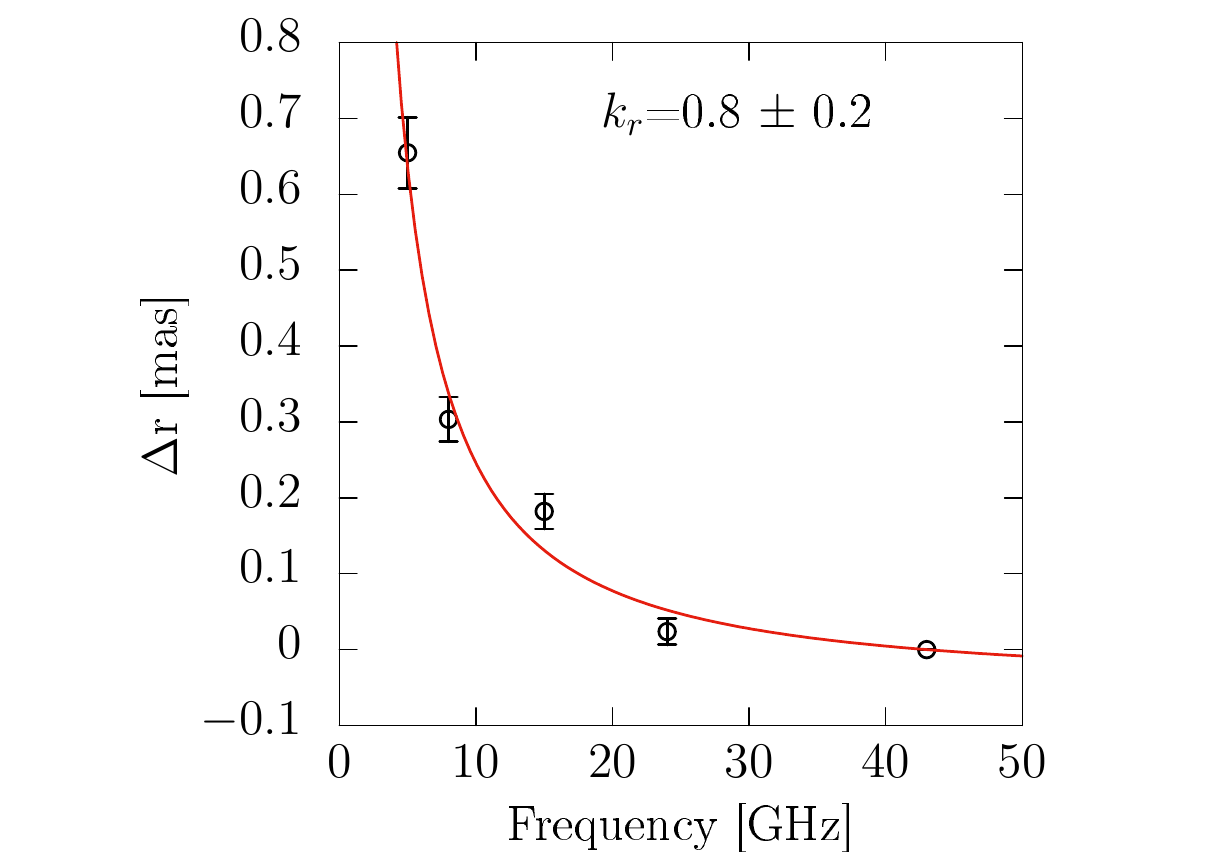}
    } 
    \caption{Epoch 16, 2009-10-22. a) Core spectrum, where Qa represents the core and Qb the feature moving downstream. For comparisons see Figure~\ref{43GHzcore}b. Core-shift vectors of all frequency pairs using (b) Qa and c) Qb. In this case, both components have similar impact on the direction of KQ core-shift vector. Power law fits (red curve) using d) Qa and e) Qb. In this epoch the flare appears to hinder the correct location of the core at the Q-band (43\,GHz).}
    \label{CSepoch16}
\end{figure*}

\begin{figure*}[!h]
\centering
   \subfigure[]
    {
        \includegraphics[width=0.45\textwidth]{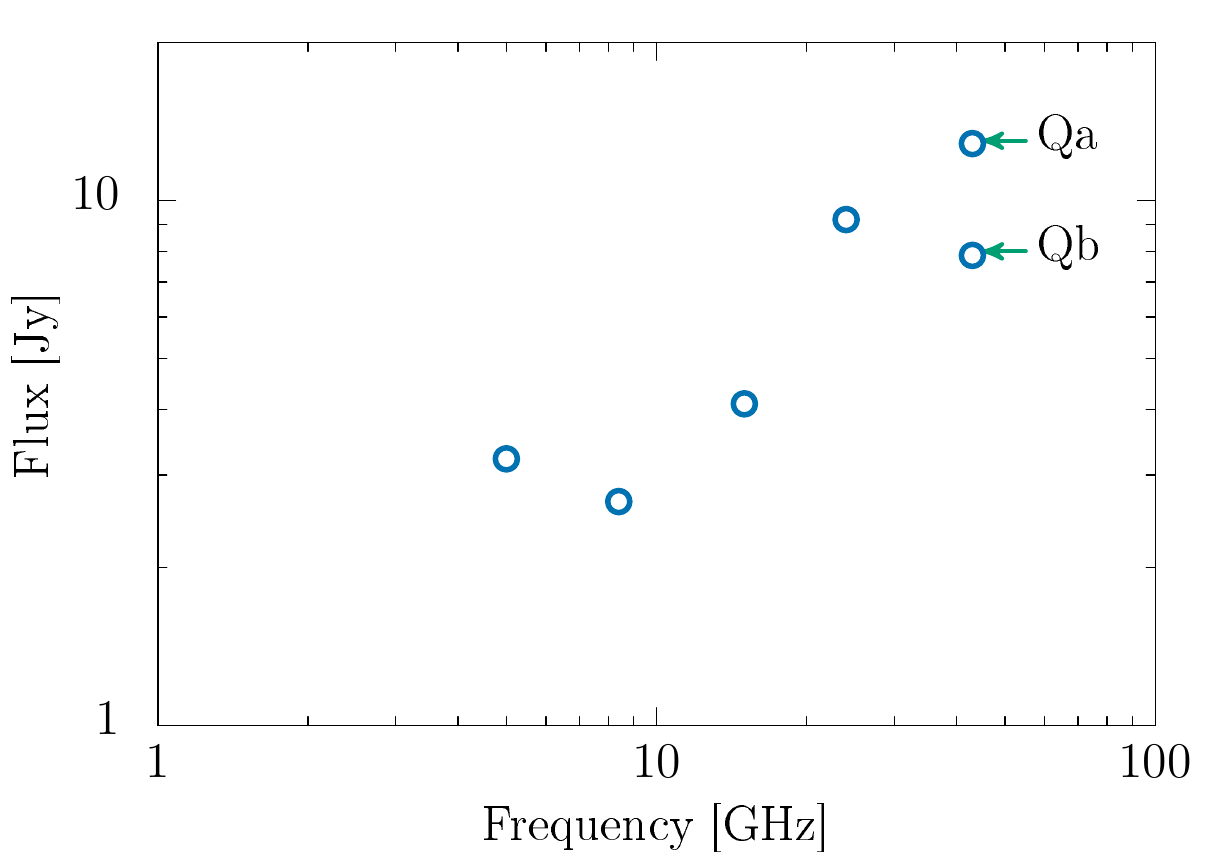}
    }
    \hspace{2cm}
     \subfigure[]
    {
        \includegraphics[width=0.45\textwidth]{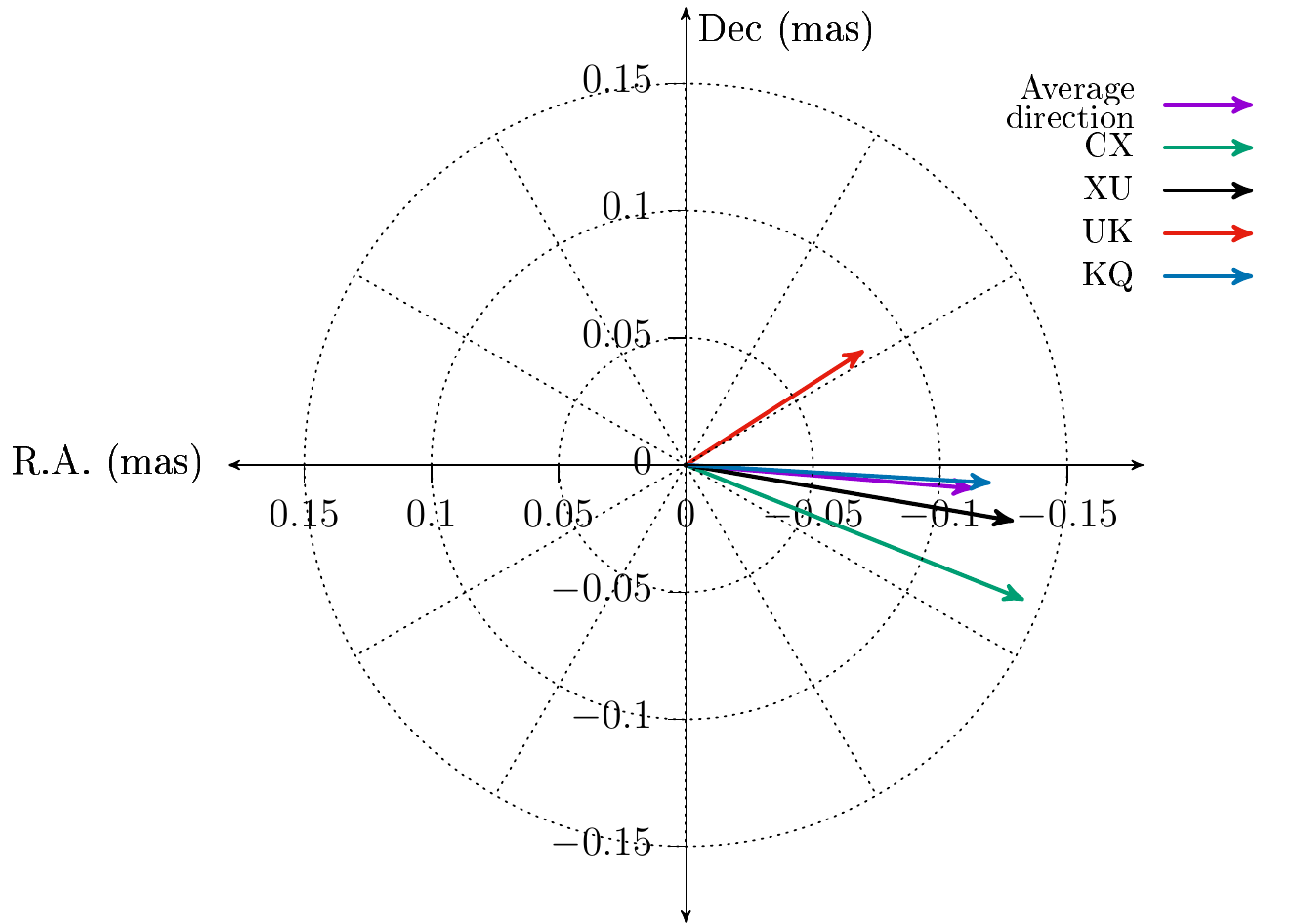}
    }
    \subfigure[]
    {
        \includegraphics[width=0.45\textwidth]{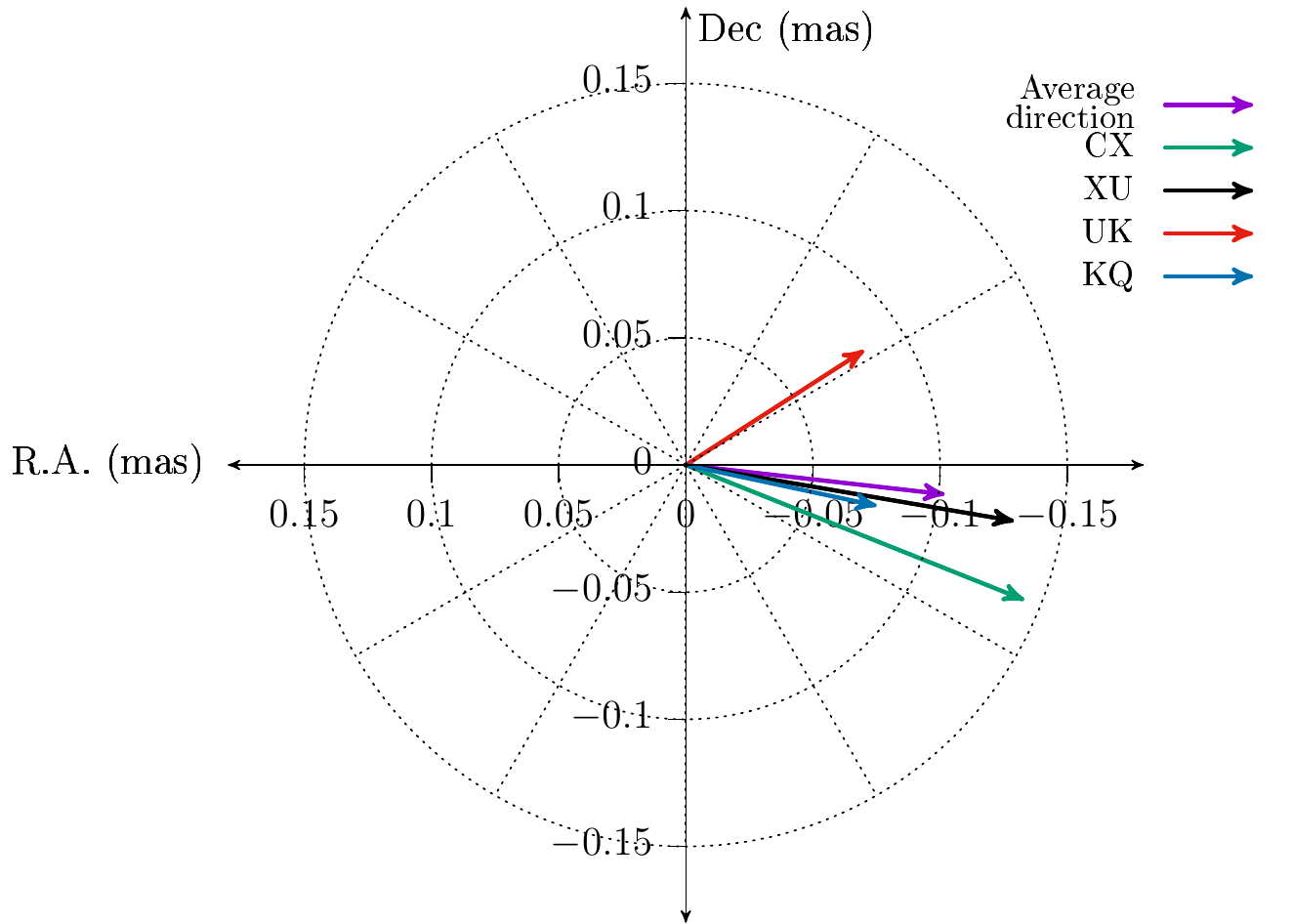}
    }
     \subfigure[]
    {
        \includegraphics[width=0.45\textwidth]{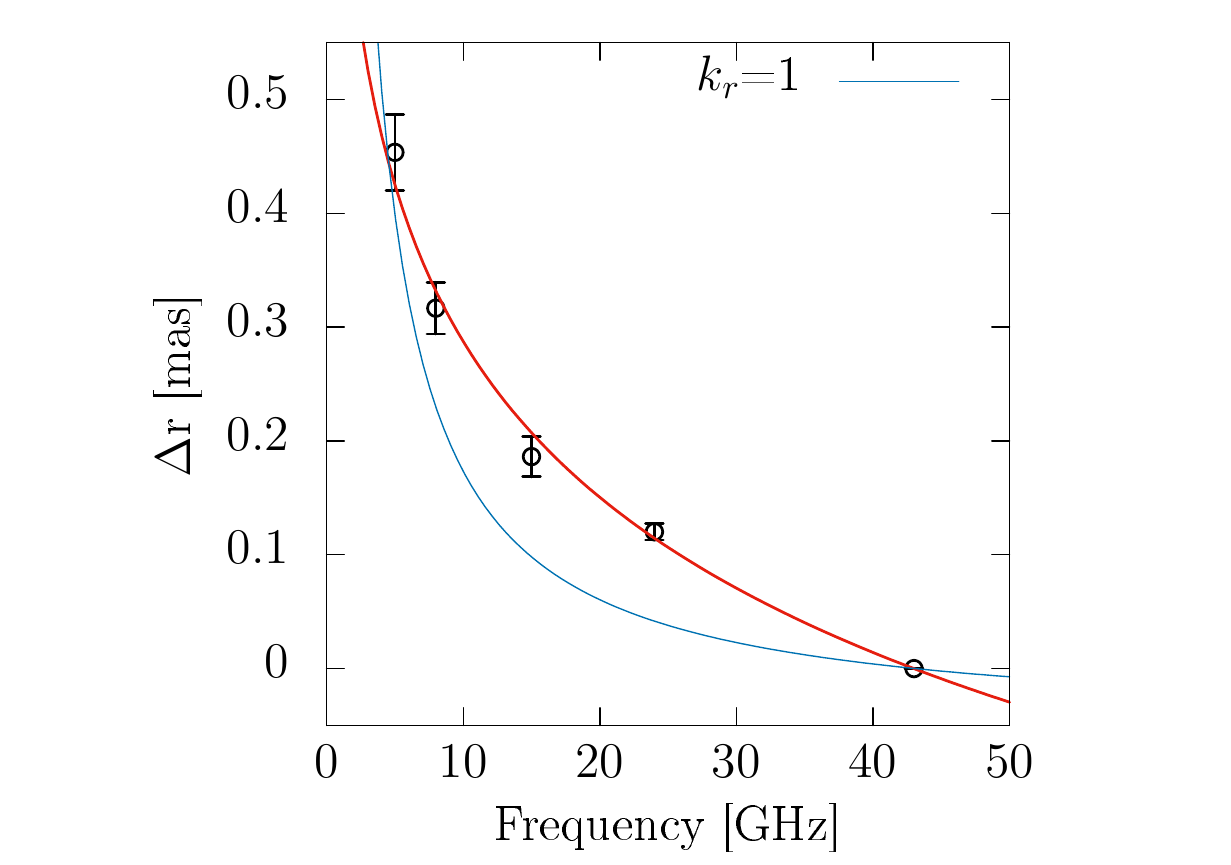}
    } 
     \subfigure[]
    {
        \includegraphics[width=0.45\textwidth]{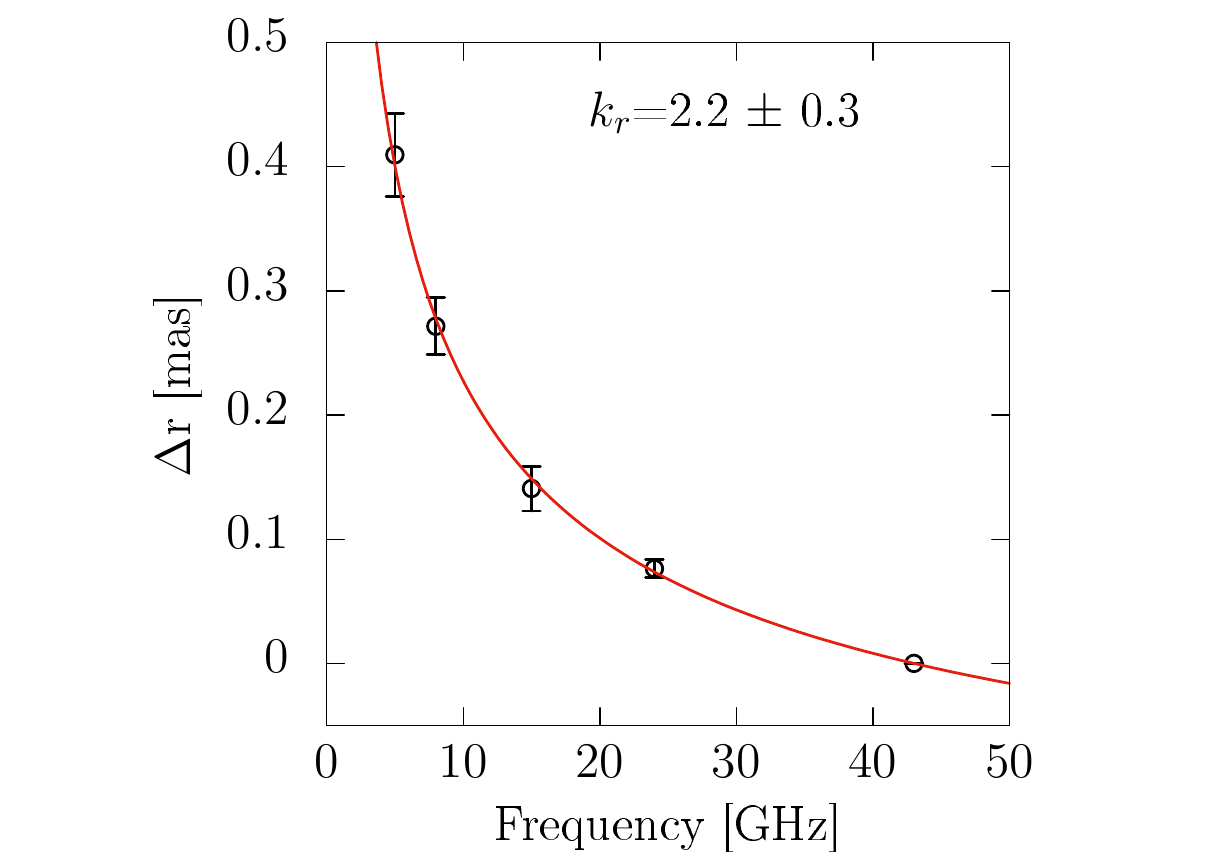}
    }
    \caption{Epoch 17, 2009-12-03. a) Core spectrum, where Qa represents the core and Qb the feature moving downstream. For comparisons see Figure~\ref{43GHzcore}c. Core-shift vectors of all frequency pairs using (b) Qa and c) Qb. Similarly as in the previous epoch, both components have similar impact on the direction of KQ core-shift vector. Power law fits (red curve) using d) Qa and e) Qb. In this epoch the flare appears to hinder the correct location of the core at the Q-band (43\,GHz). This ultimately disrupts the core-shift effect by increasing the core-shift values at the high frequencies as seen in d). Therefore, this observation it is not included in the variability analysis of index $k_r$.}
    \label{CSepoch17}
\end{figure*}

\begin{figure*}[!h]
\centering
   \subfigure[]
    {
        \includegraphics[width=0.45\textwidth]{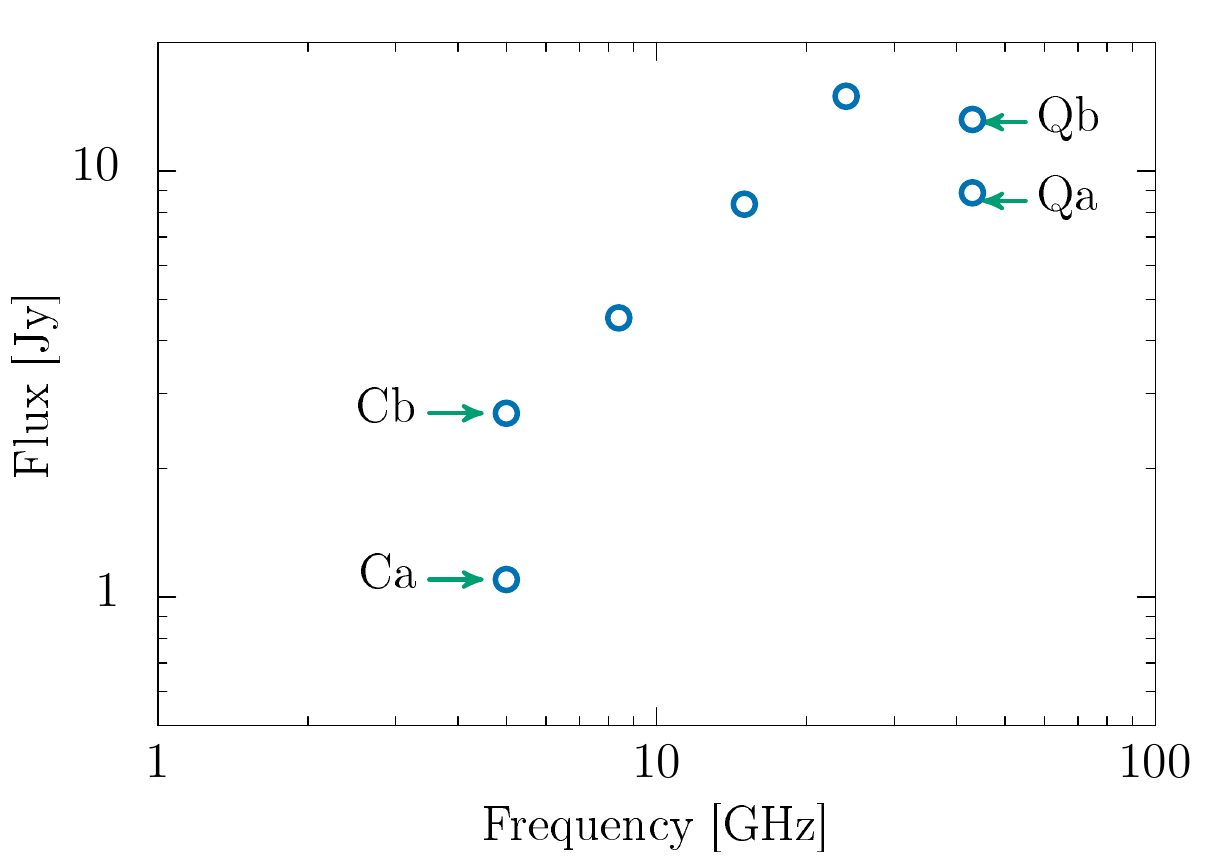}
    }
    \hspace{2cm}
        \subfigure[]
    {
        \includegraphics[width=0.45\textwidth]{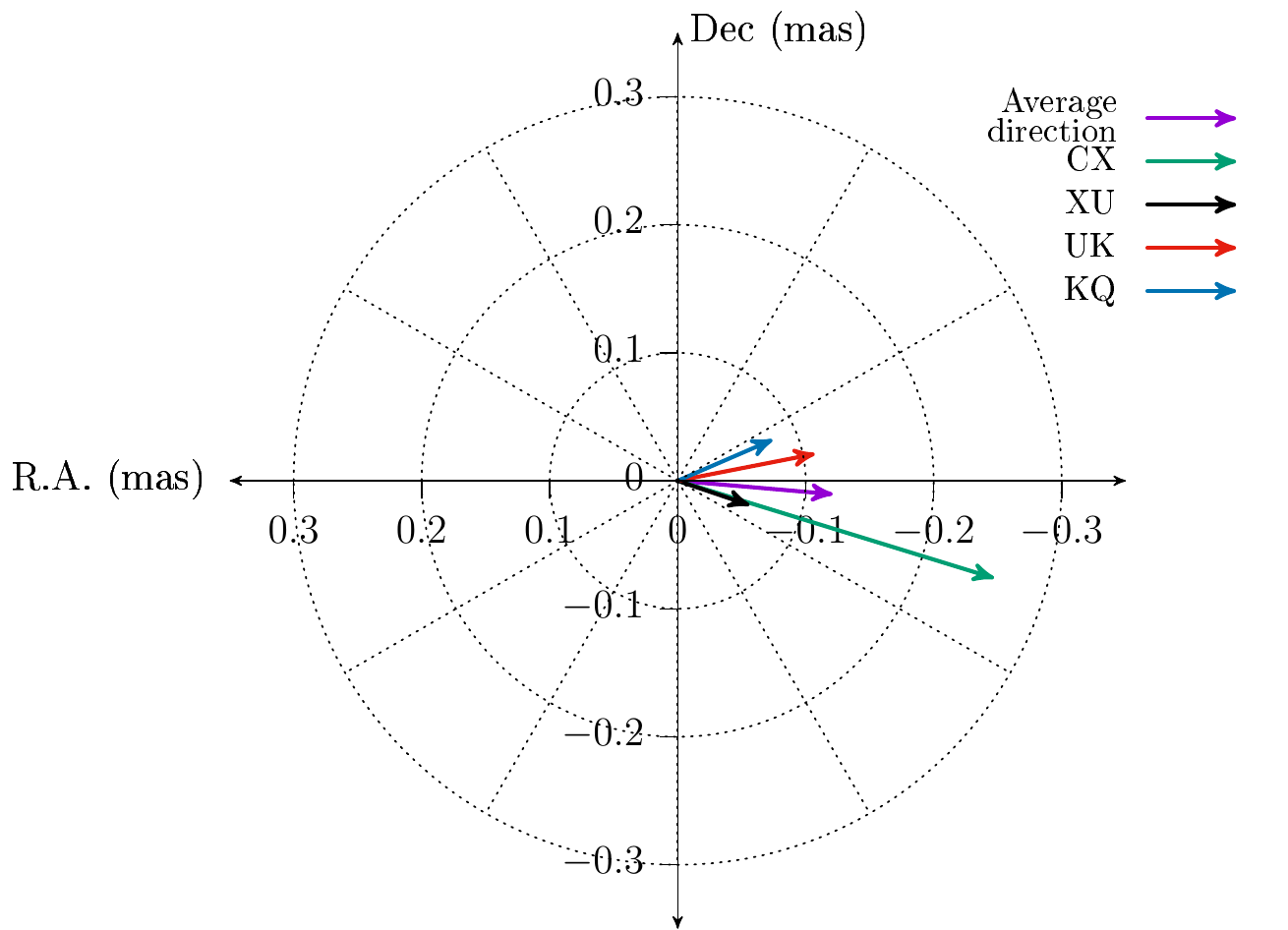}
    }
      \subfigure[]
    {
        \includegraphics[width=0.45\textwidth]{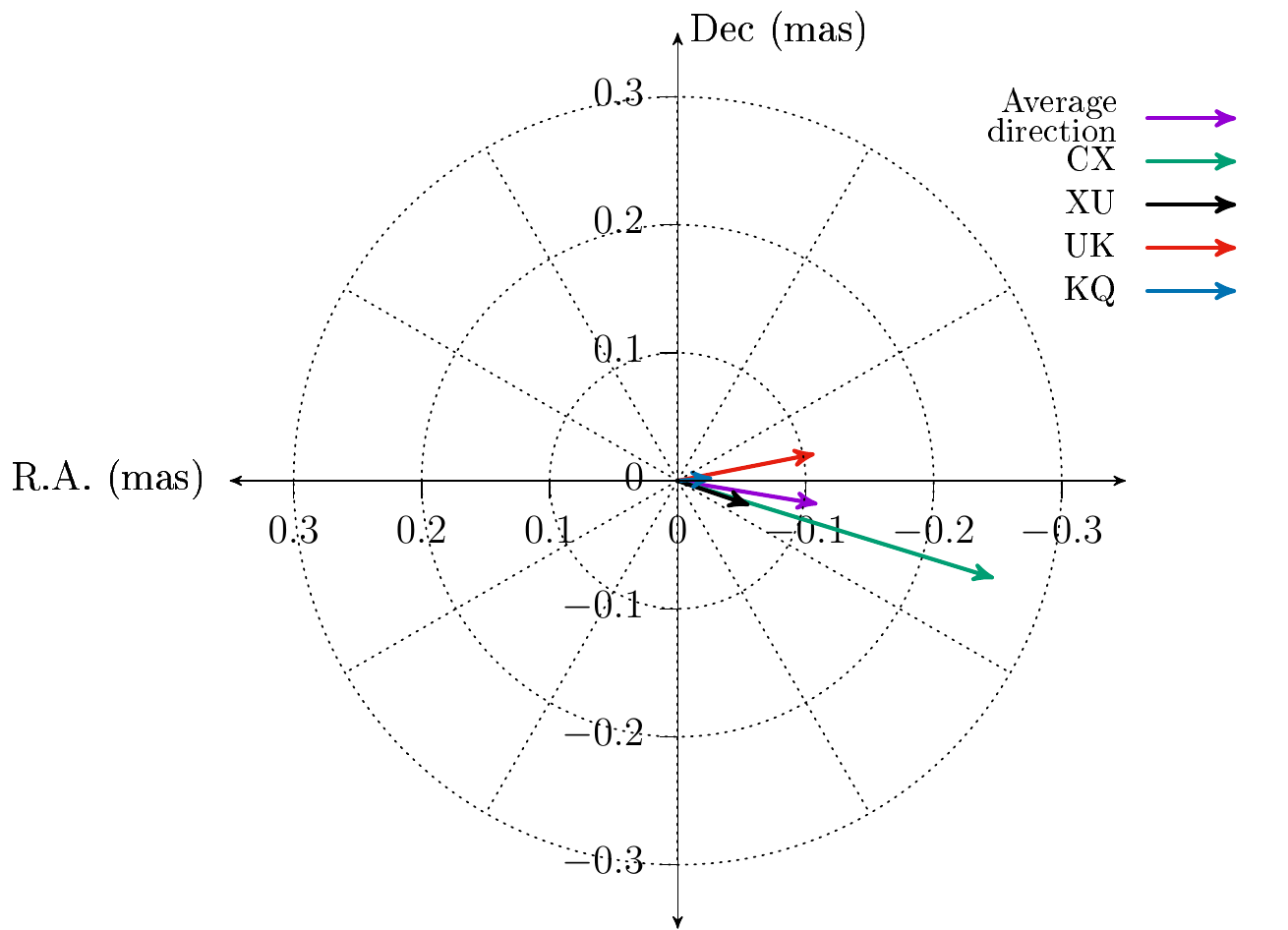}
    } 
     \subfigure[]
    {
        \includegraphics[width=0.45\textwidth]{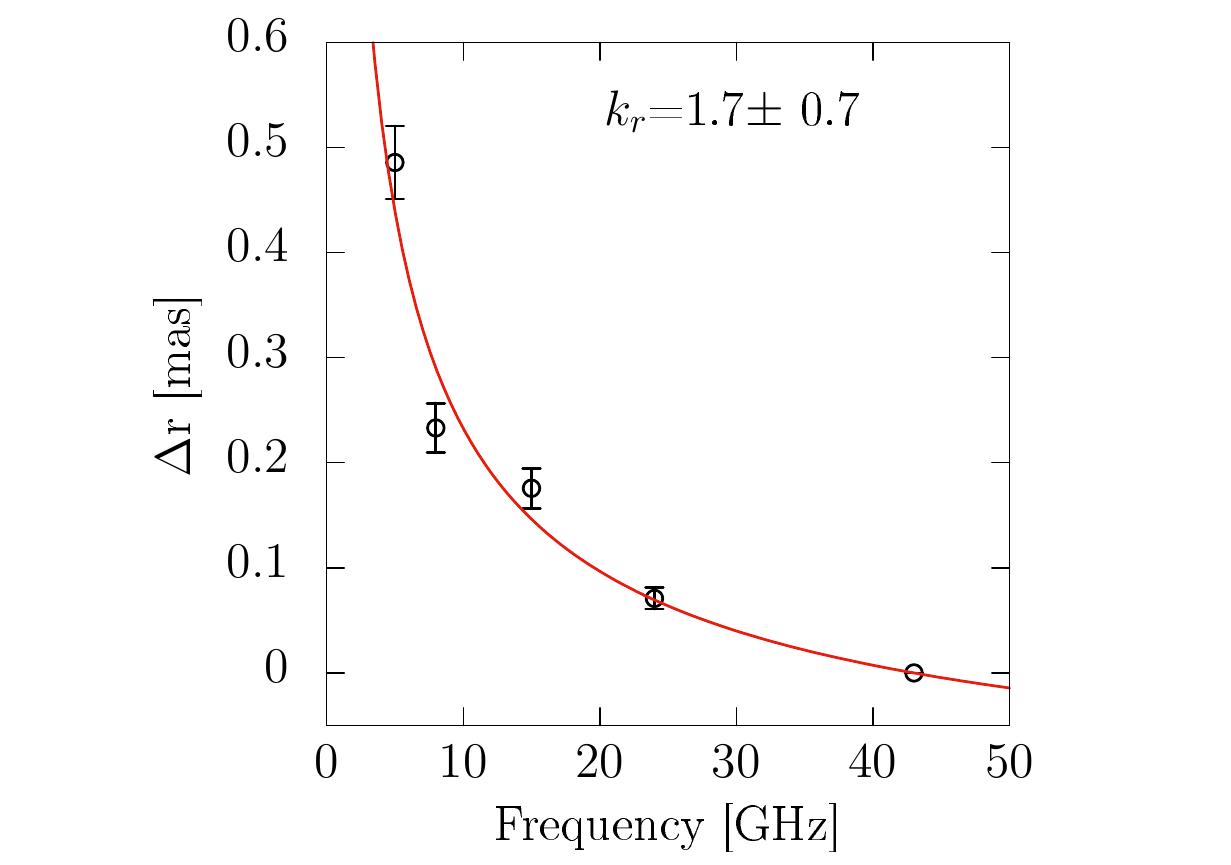}
    }
    \subfigure[]
    {
        \includegraphics[width=0.45\textwidth]{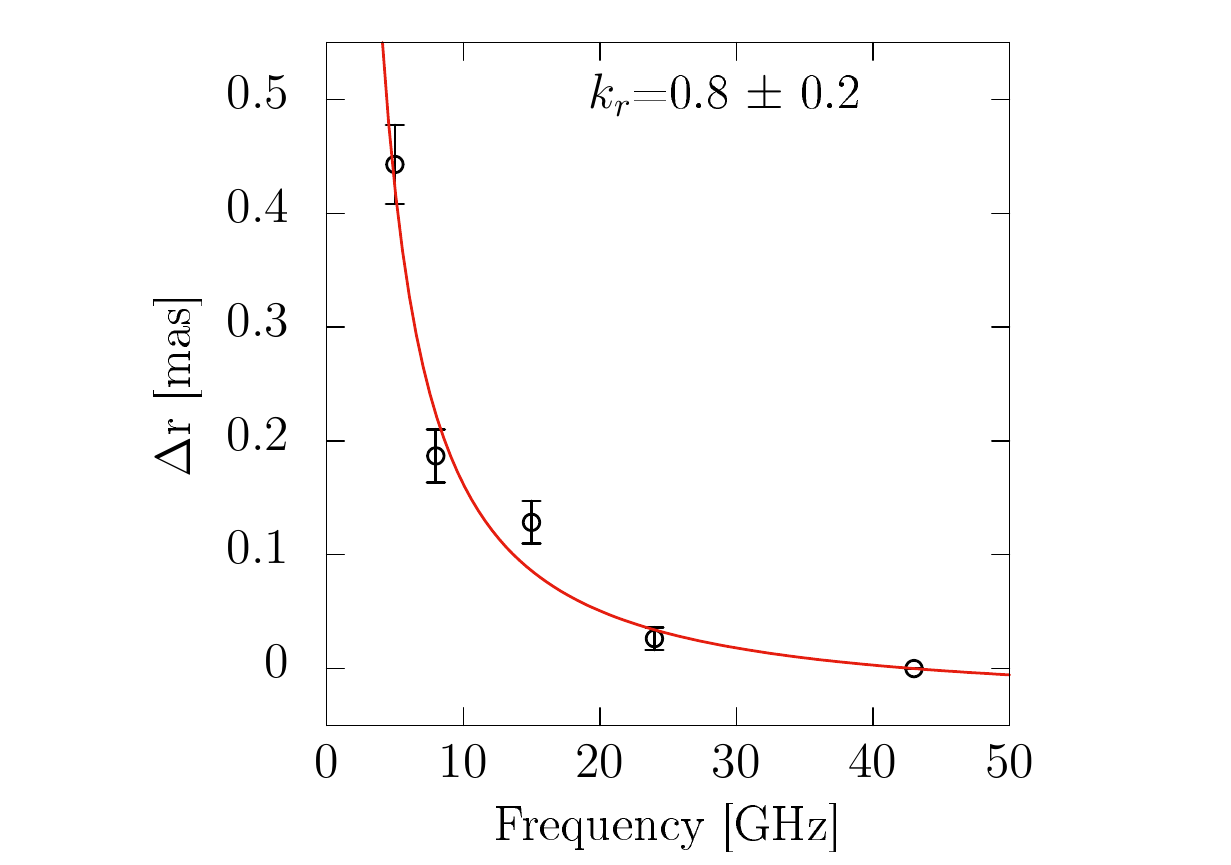}
    }
    \caption{Epoch 18, 2010-01-18. a) Core spectrum, where Cb and Qa represents the core at the C and Q-bands. Qb is the feature moving downstream. For comparisons see Figure~\ref{43GHzcore}d.  Core-shift vectors of all frequency pairs using (b) Qa and c) Qb. Similarly as in the previous epoch, both components have similar impact on the direction of KQ core-shift vector. Power law fits (red curve) using d) Qa and e) Qb. Again, the flare appears to hinder the correct location of the core at the Q-band (43\,GHz). The effect is less pronounced as in the previous epoch but it disordered the core-shift effect by increasing the core-shift values at the high frequencies as seen in d).}
    \label{CSepoch18}
\end{figure*}

\begin{figure*}[!h]
\centering
   \subfigure[]
    {
    \includegraphics[width=0.45\textwidth]{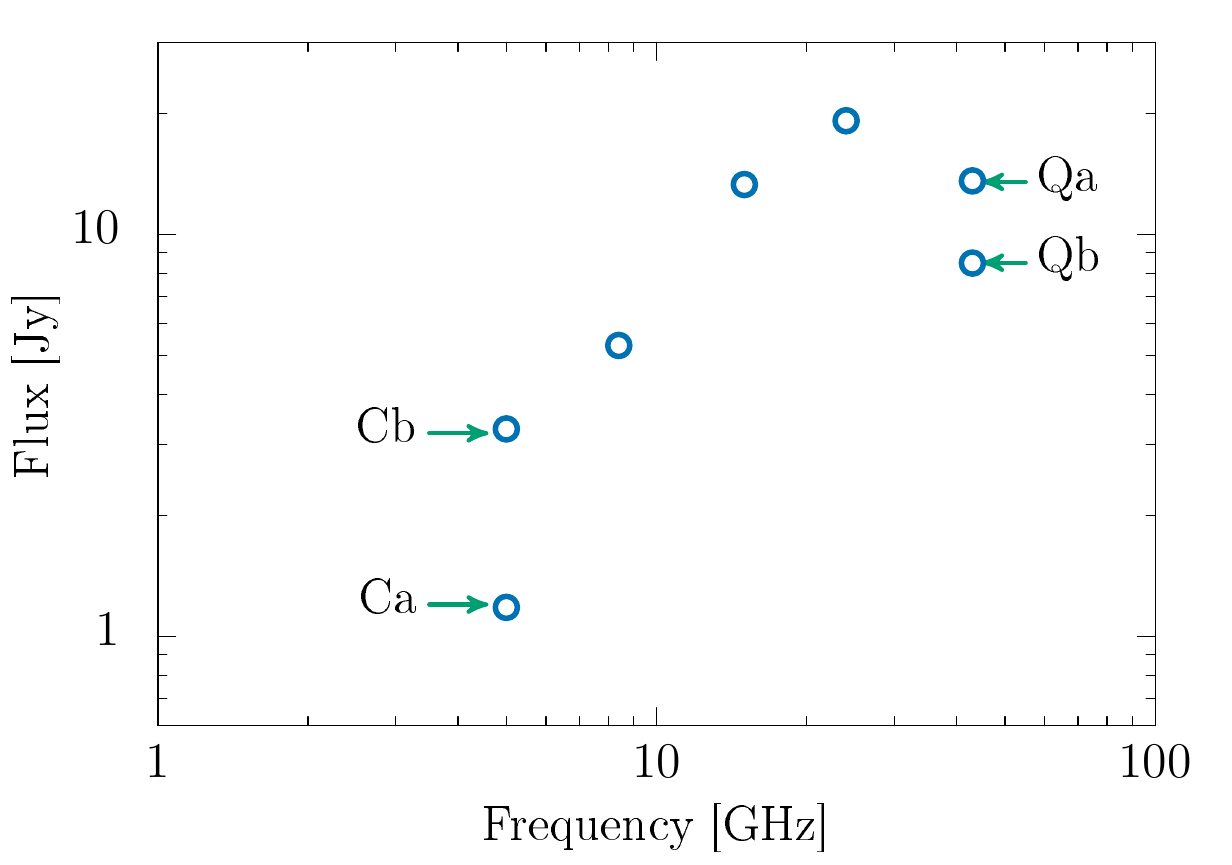}
    }
    \hspace{2cm}
    \subfigure[]
    {
    \includegraphics[width=0.44\textwidth]{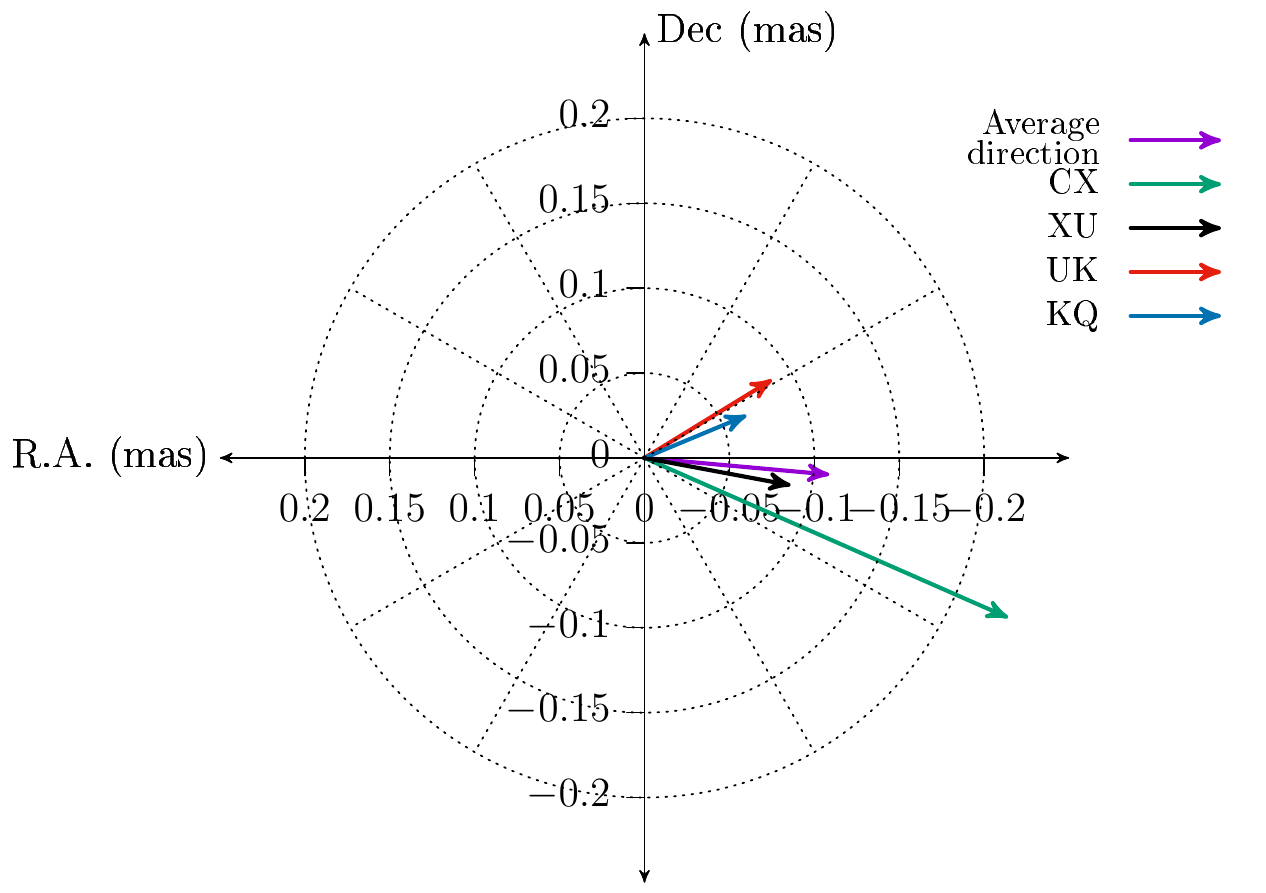}
    }
    \subfigure[]
    {
    \includegraphics[width=0.52\textwidth]{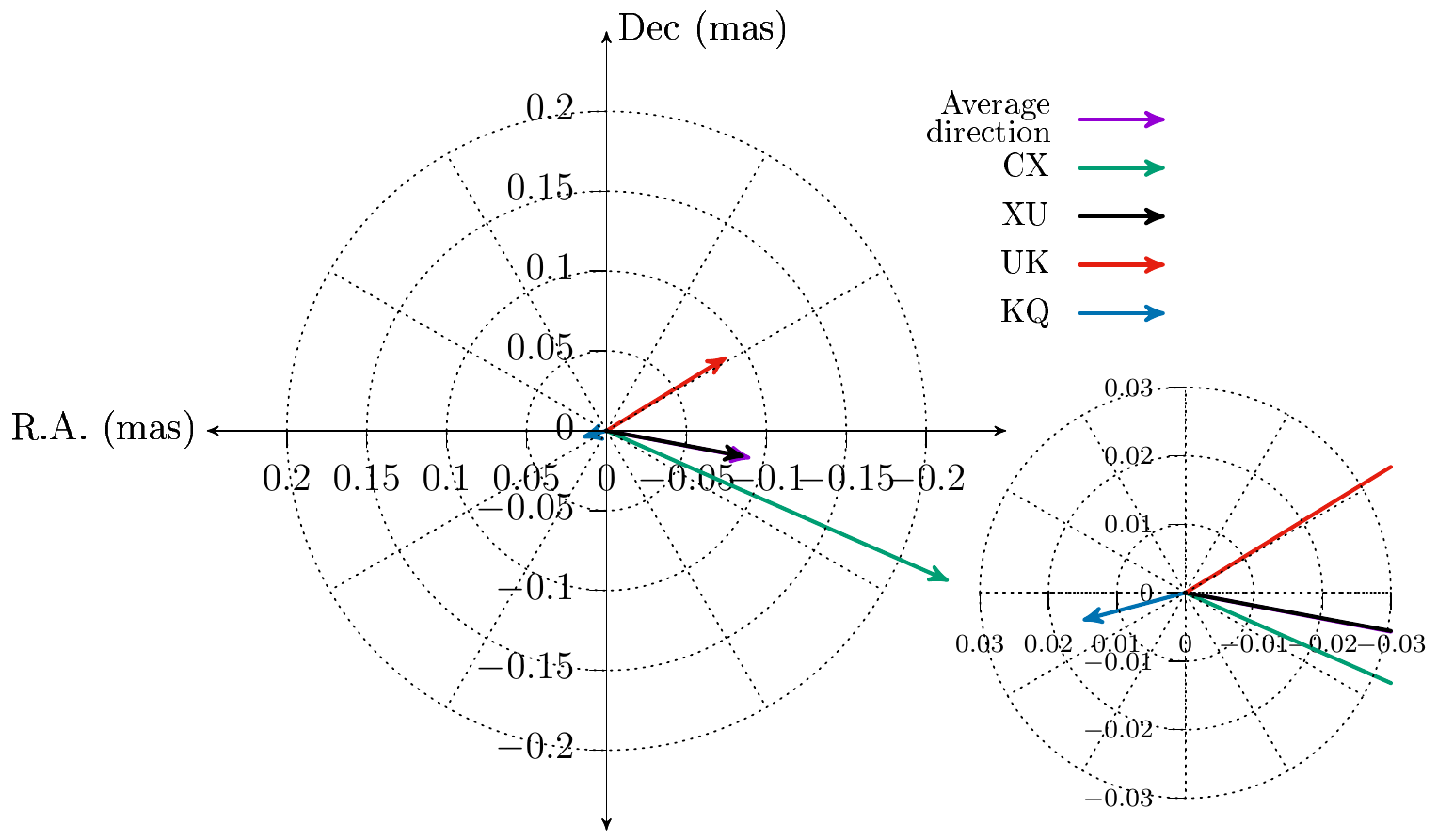}
    }
    \subfigure[]
    {
    \includegraphics[width=0.45\textwidth]{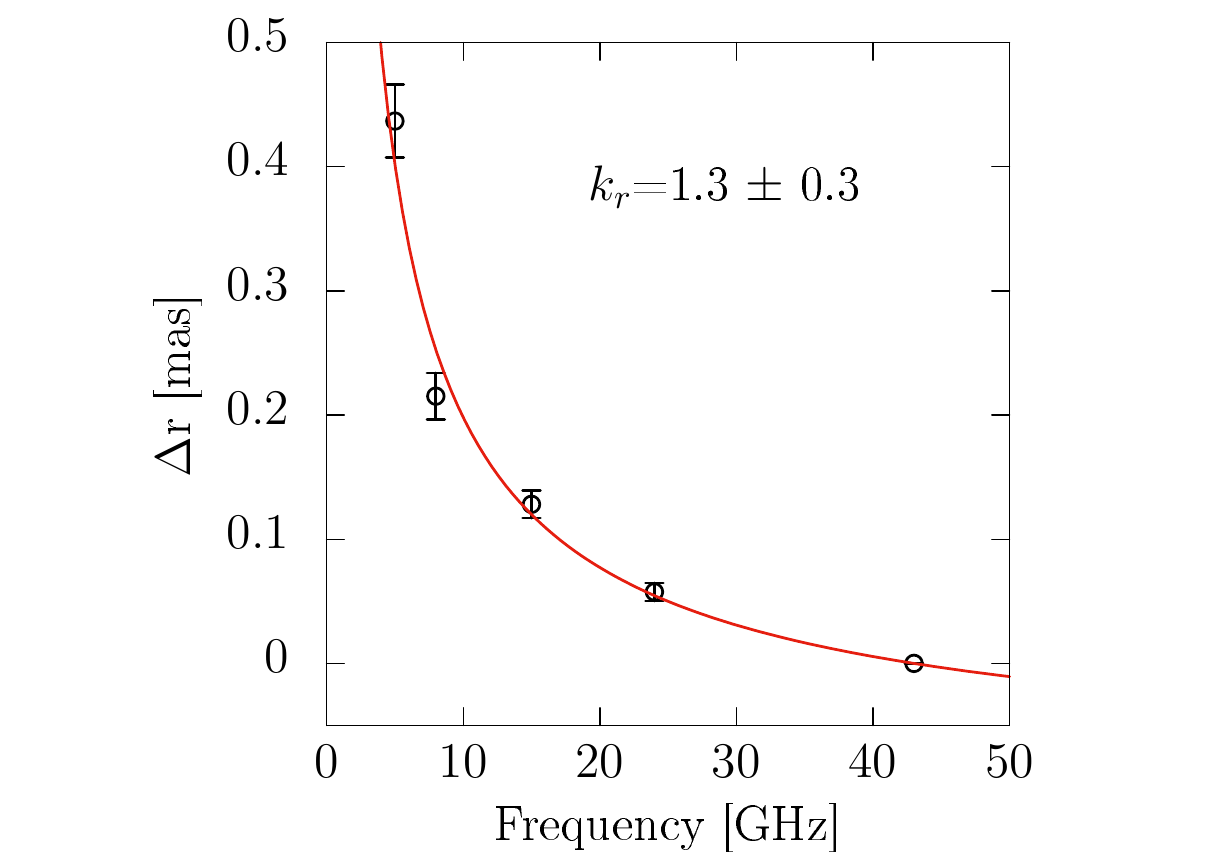}
    }
    \subfigure[]
    {
    \includegraphics[width=0.45\textwidth]{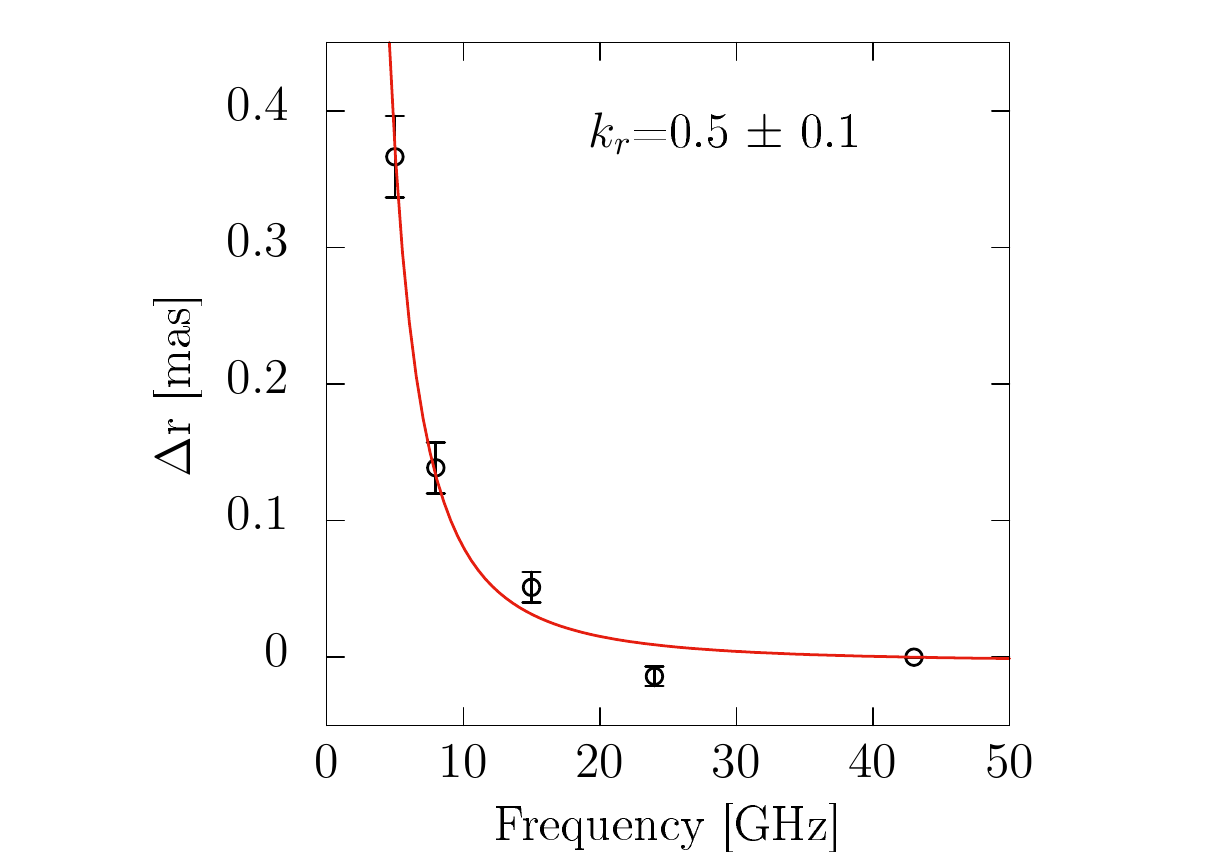}
    }
      \caption{Epoch 19, 2010-02-21. a) Core spectrum, where Cb and Qa represents the core at the C and Q-bands. Qb is the feature moving downstream. For comparisons see Figure~\ref{43GHzcore}e.  Core-shift vectors of all frequency pairs using (b) Qa and c) Qb. In this epoch, it is clear that Qa and Qb have a different effect on the KQ core-shift vector direction. The use of Qb leads to the wrong core-shift direction. Power law fits (red curve) using d) Qa and e) Qb. Similarly as in previous epochs, the flare appears to affect the location of the core at the Q-band (43\,GHz). In this observation, the flare effect is less pronounced but increased the core-shift values at the high frequencies as seen in d).}
    \label{CSepoch19}
\end{figure*}

Main results of the core-shift analysis for individual epochs (listed in Table~\ref{t1}) are shown in Figures~\ref{CSepoch1}$-$\ref{CSepoch19} here. The results for each epoch include the core spectrum, the core-shift vectors in polar coordinates and the core-shift power-law fit. Projected values of the core-shifts onto the average direction are used. All fitting parameters are summarized in Table~\ref{fits-parameters}.

We note that we still present the core-shift analysis for the observation on 2008-01-03 (epoch 13, Figure~\ref{CSepoch13}) as a demonstration of the case when large core-shifts are measured due to lack of enough resolution. Furthermore, as stressed in the main text, this observation was not included for the variability analysis.

% Epoch 1 below

% Epoch 2 below

% Epoch 3 below

% Epoch 4 below

% EPOCH 5: 03.08.2006--BAD EPOCH
% The Mauna Kea antenna is missing, not enough resolution to resolve the core at the higher frequencies. We will present images for this epoch only.
% \begin{figure*}[!h]
% \centering
%     \subfigure[]
%     {
%         \includegraphics[width=0.45\textwidth]{Epoch5-CS03-08-2006/03-08-2006-fv.pdf}
%         
%     }
%     \subfigure[]
%     {
%         \includegraphics[width=0.45\textwidth]{Epoch5-CS03-08-2006/CSvectors.pdf}
%         
%     }
%      \subfigure[]
%     {
%         \includegraphics[width=0.45\textwidth]{Epoch5-CS03-08-2006/Cshifts.pdf}
%         
%     }
%     \caption{03.08.2006. }
%     \label{fig:sample_subfigures}
% \end{figure*}

% Epoch 6 below

% Epoch 7 below

% Epoch 8 below

% Epoch 9 below

% Epoch 10 below

% Epoch 11 below

% Epoch 12 below

% Epoch 13 below

\newpage
% Epoch 14 below

% Epoch 15 below

% Epoch 16 below

% Epoch 17 below

% Epoch 18 below

% Epoch 19 below

\section{VLBA multi-epoch and multi-frequency images}
\label{fullVLBAimages}
\begin{figure*}[h]
\centering
   \subfigure[]
    {
        \includegraphics[width=0.3\textwidth]{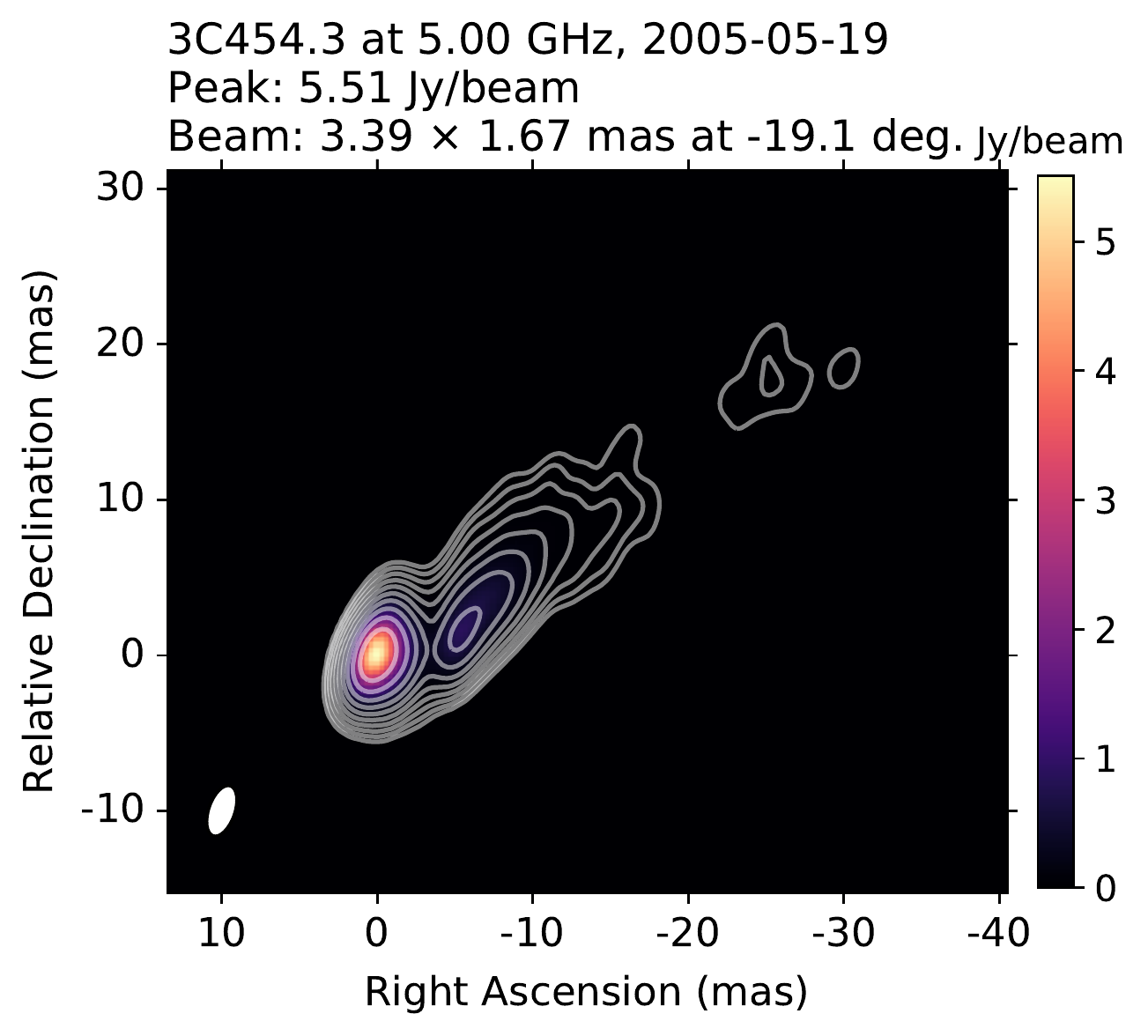}
    }
       \subfigure[]
    {
        \includegraphics[width=0.3\textwidth]{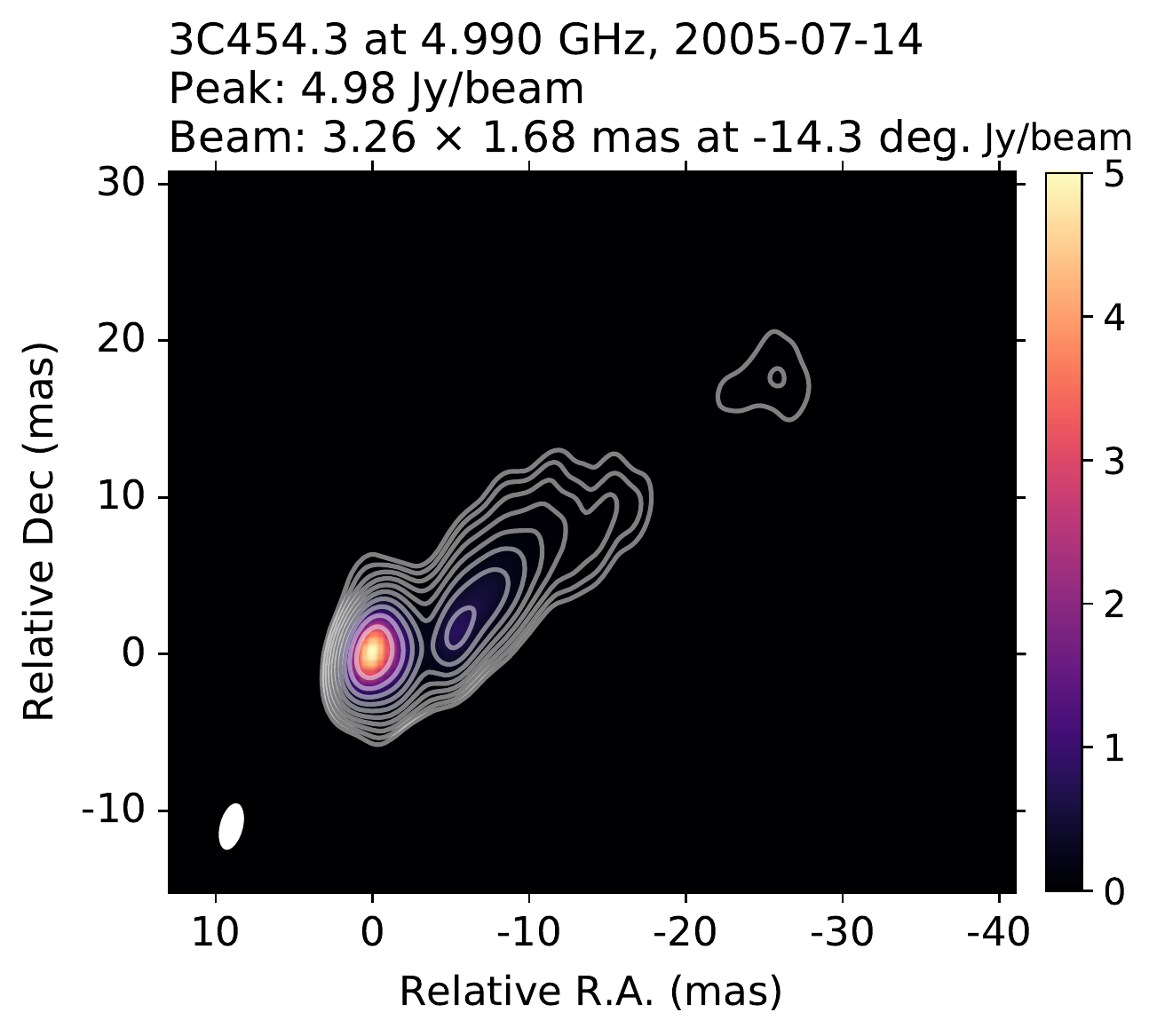}
    }
           \subfigure[]
    {
        \includegraphics[width=0.3\textwidth]{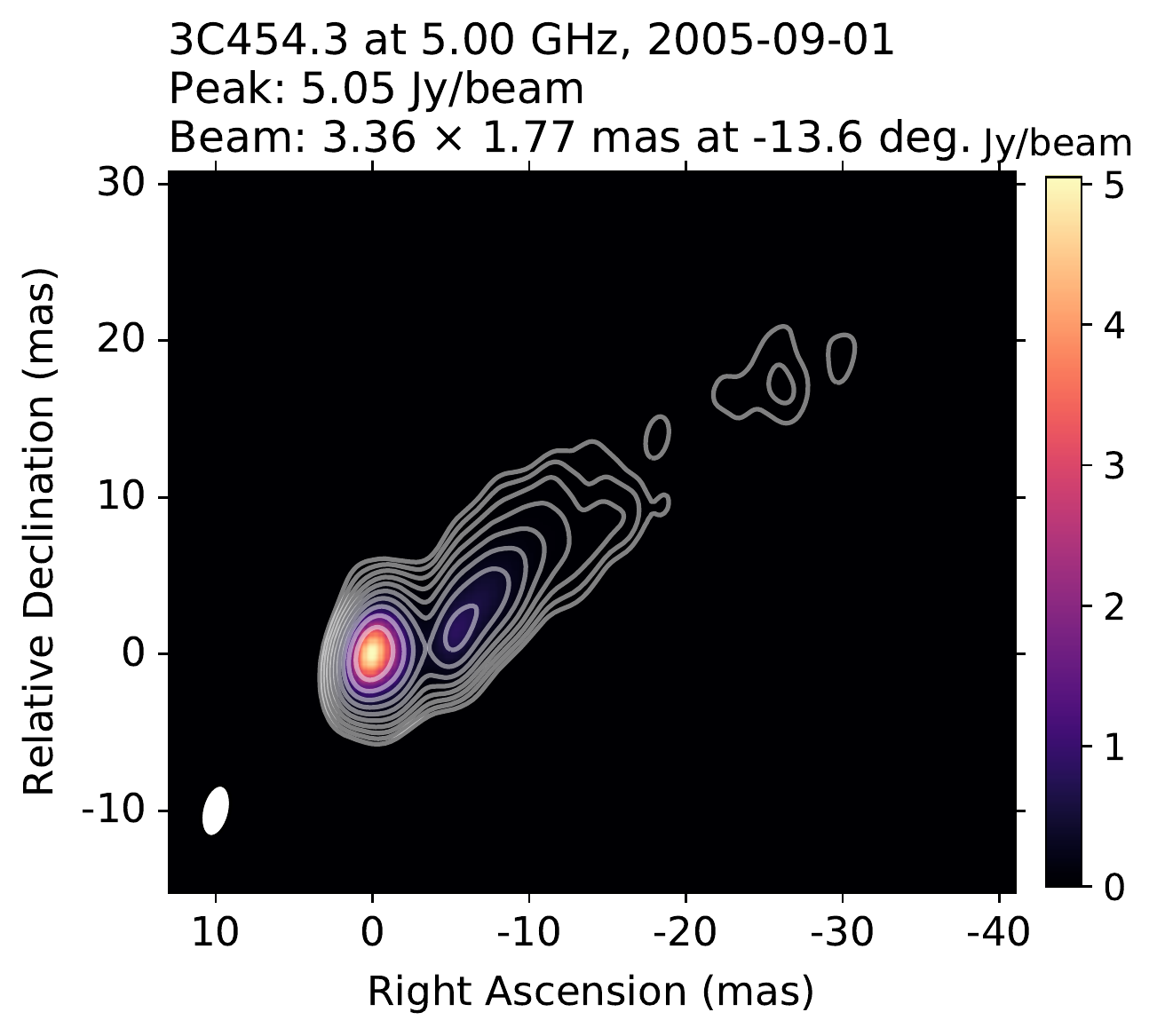}
    }
     \subfigure[]
    {
         \includegraphics[width=0.3\textwidth]{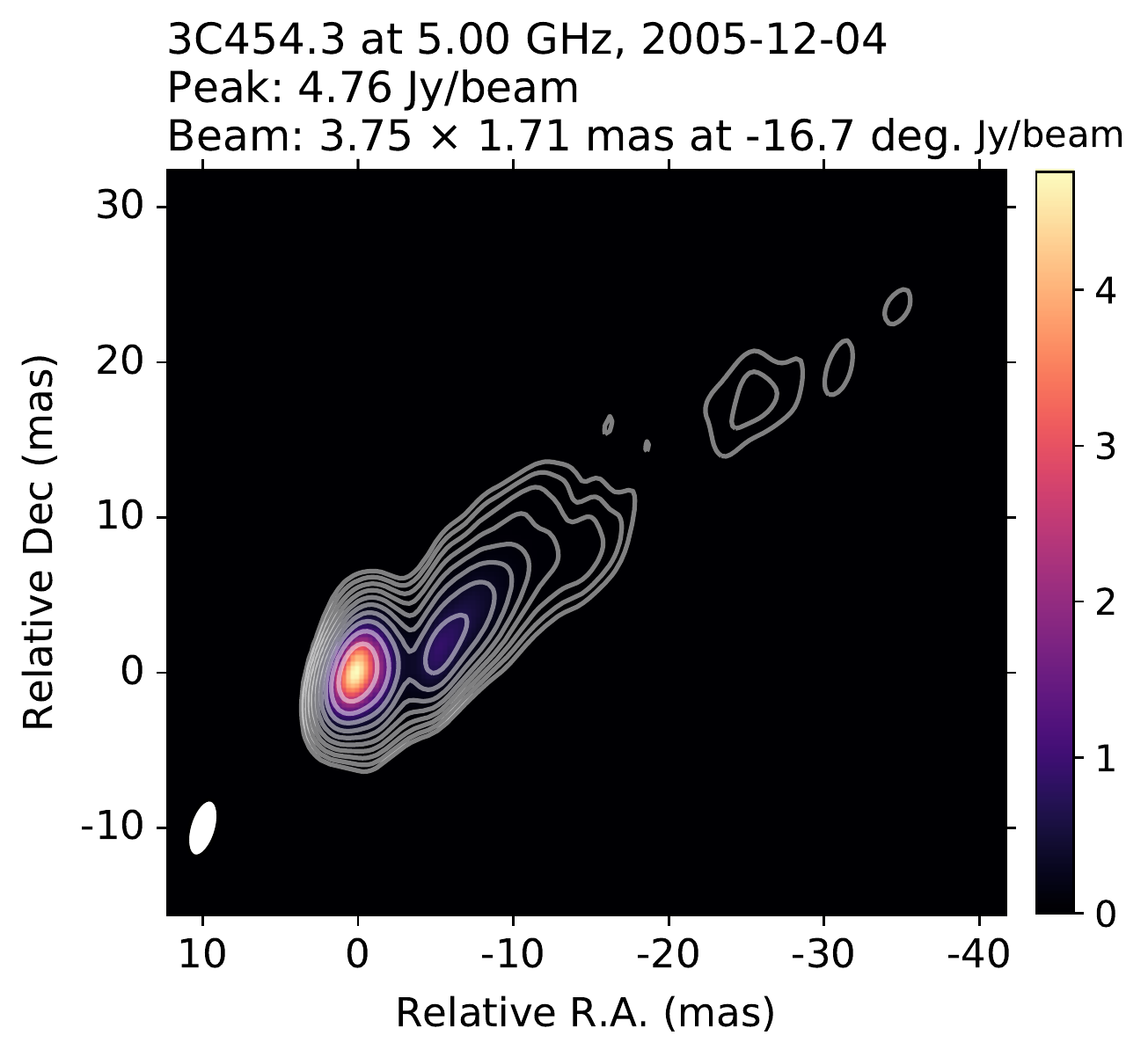}
    }
     \subfigure[]
    {
        \includegraphics[width=0.3\textwidth]{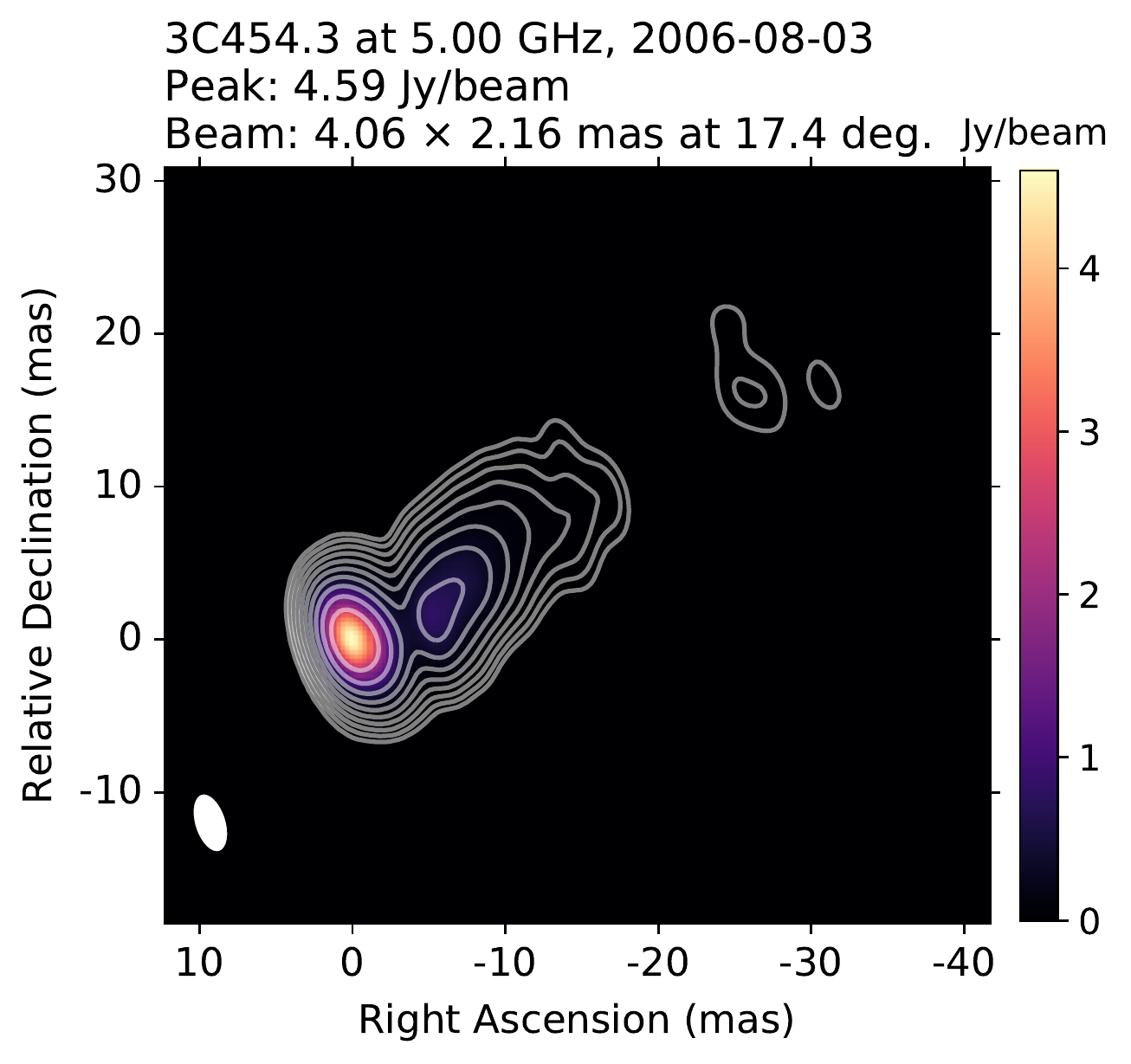}
        
    }   
       \subfigure[]
    {
         \includegraphics[width=0.3\textwidth]{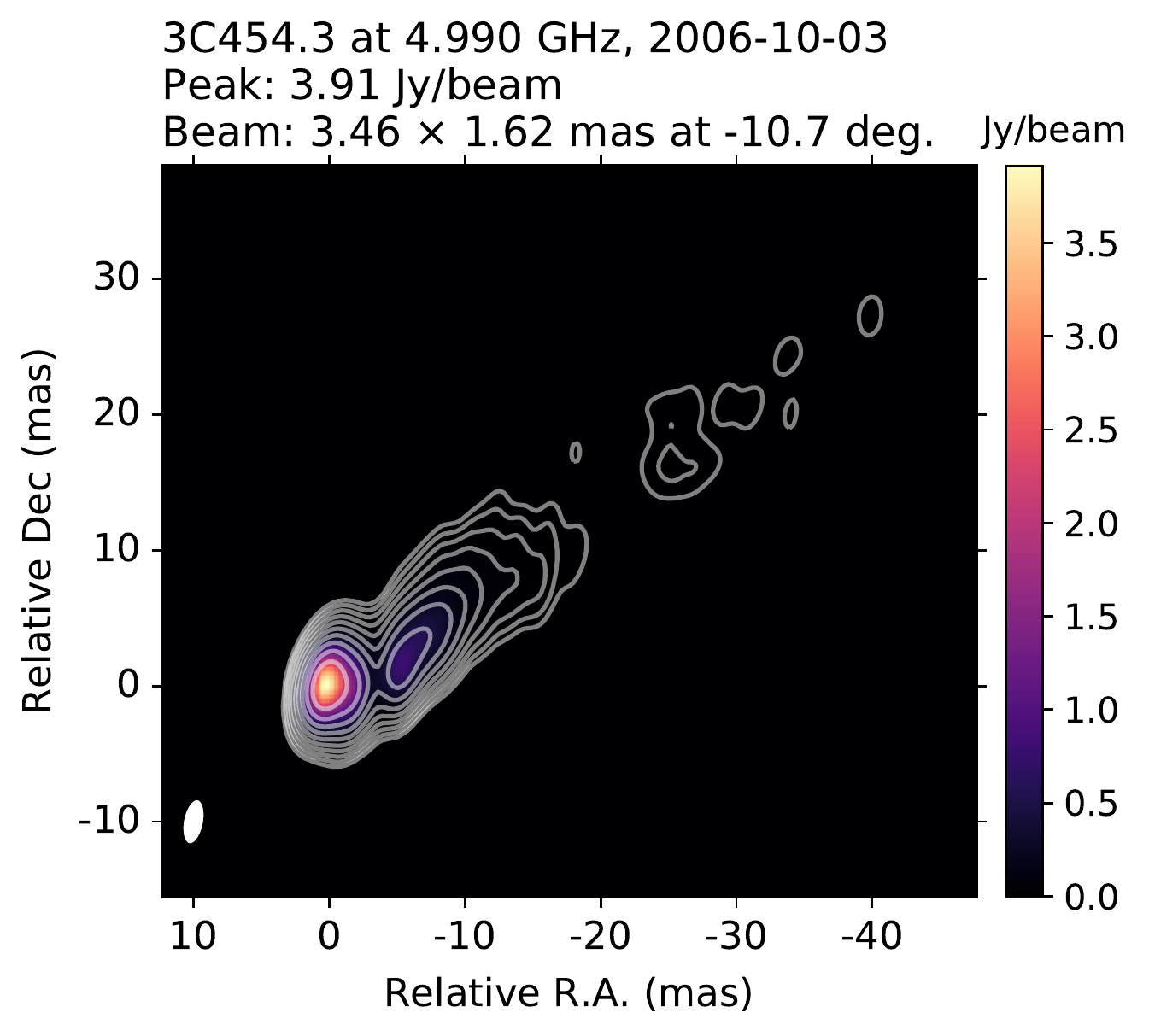}
    }
       \subfigure[]
    {
         \includegraphics[width=0.3\textwidth]{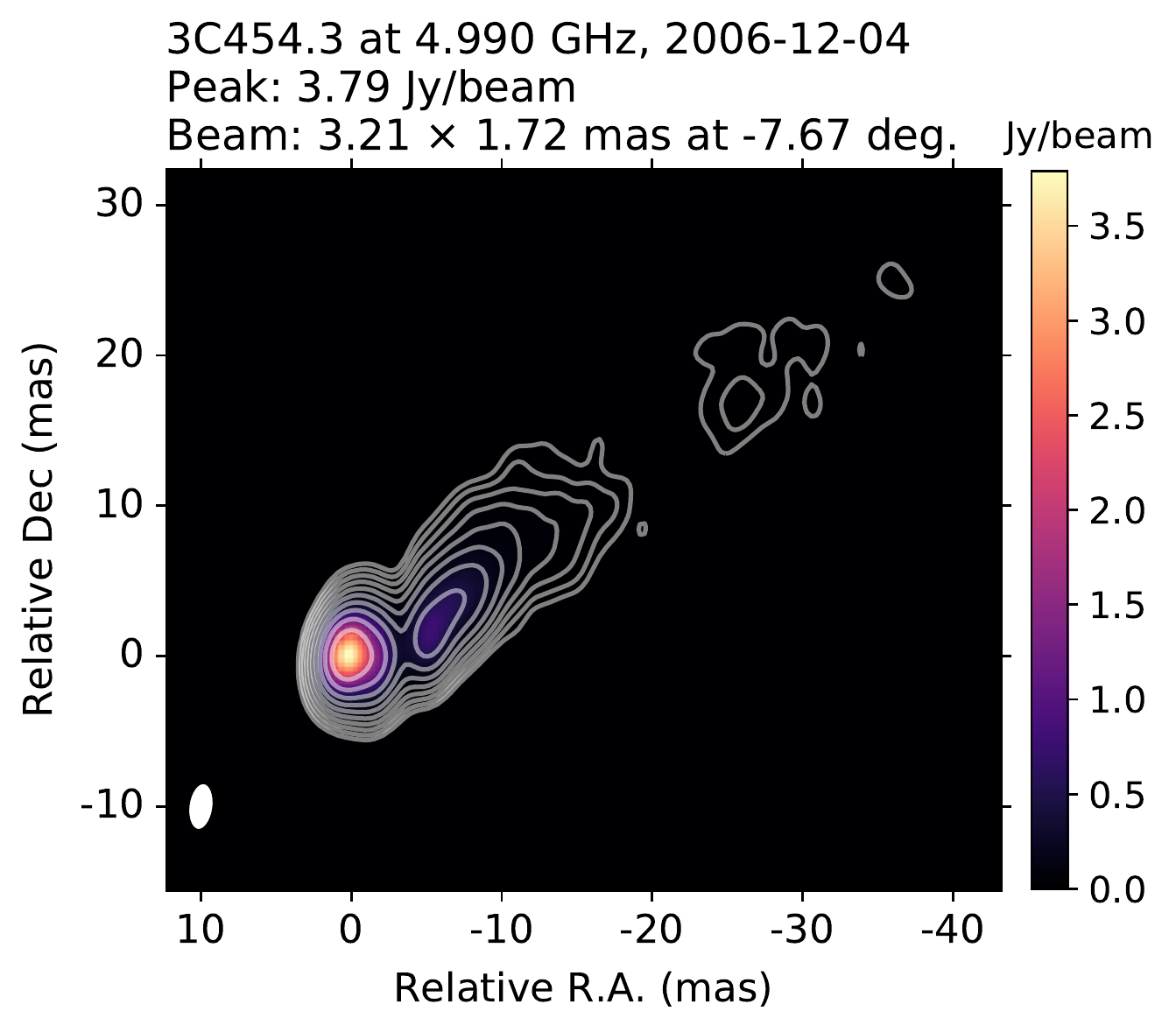}
    }
    \subfigure[]
    {
        \includegraphics[width=0.3\textwidth]{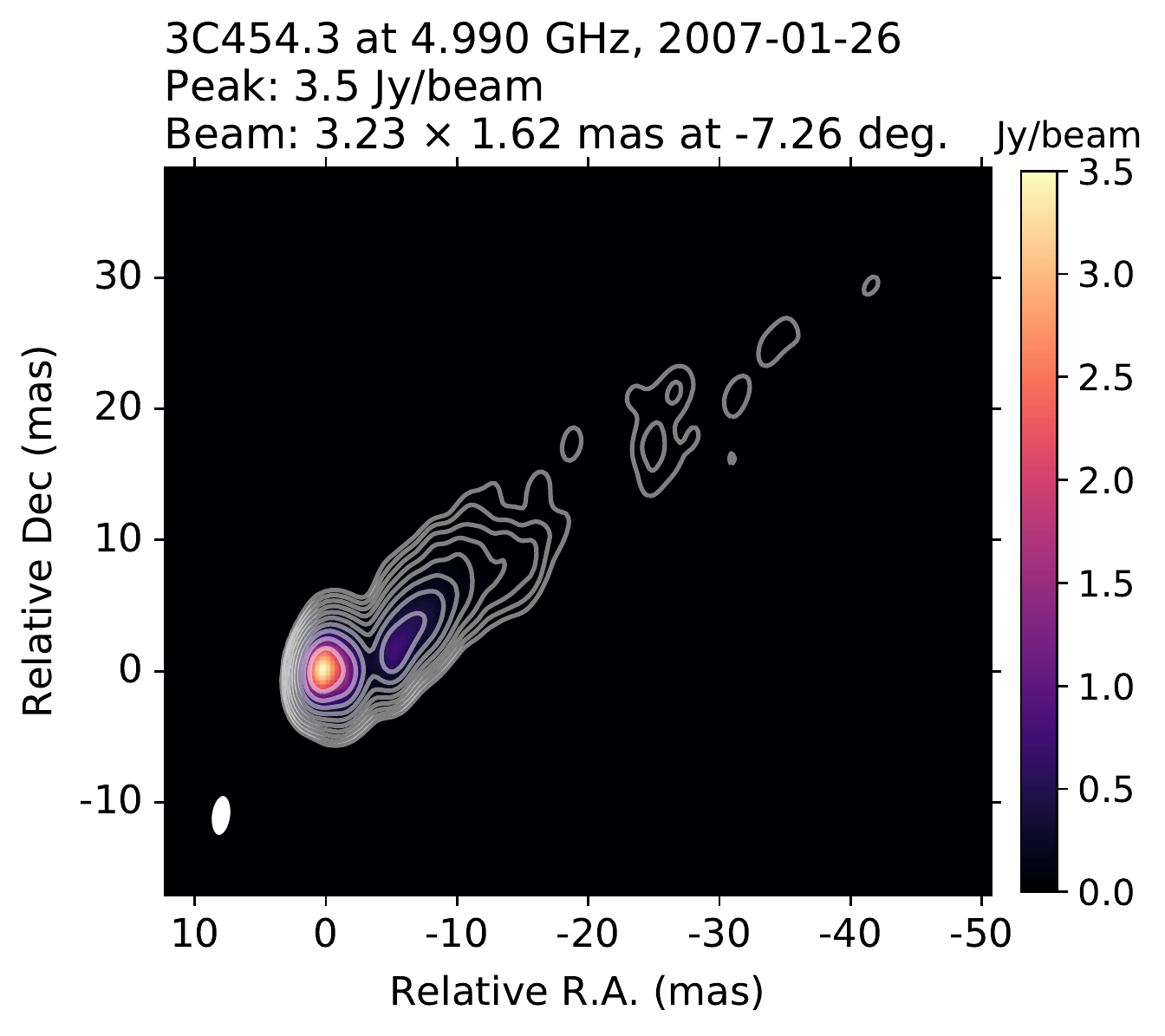}
    }
    \subfigure[]
    {
         \includegraphics[width=0.3\textwidth]{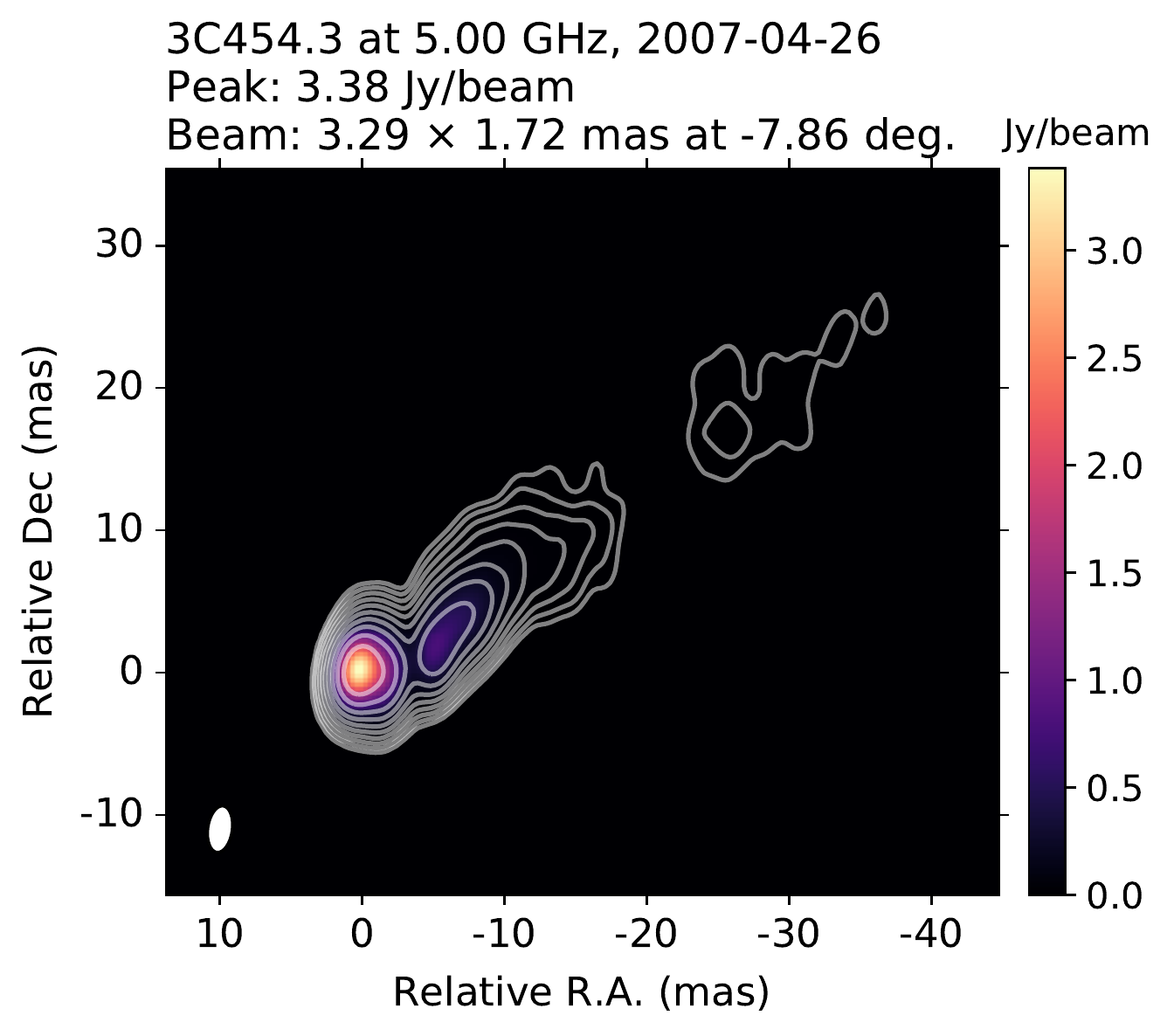}
    }
        \subfigure[]
    {
         \includegraphics[width=0.3\textwidth]{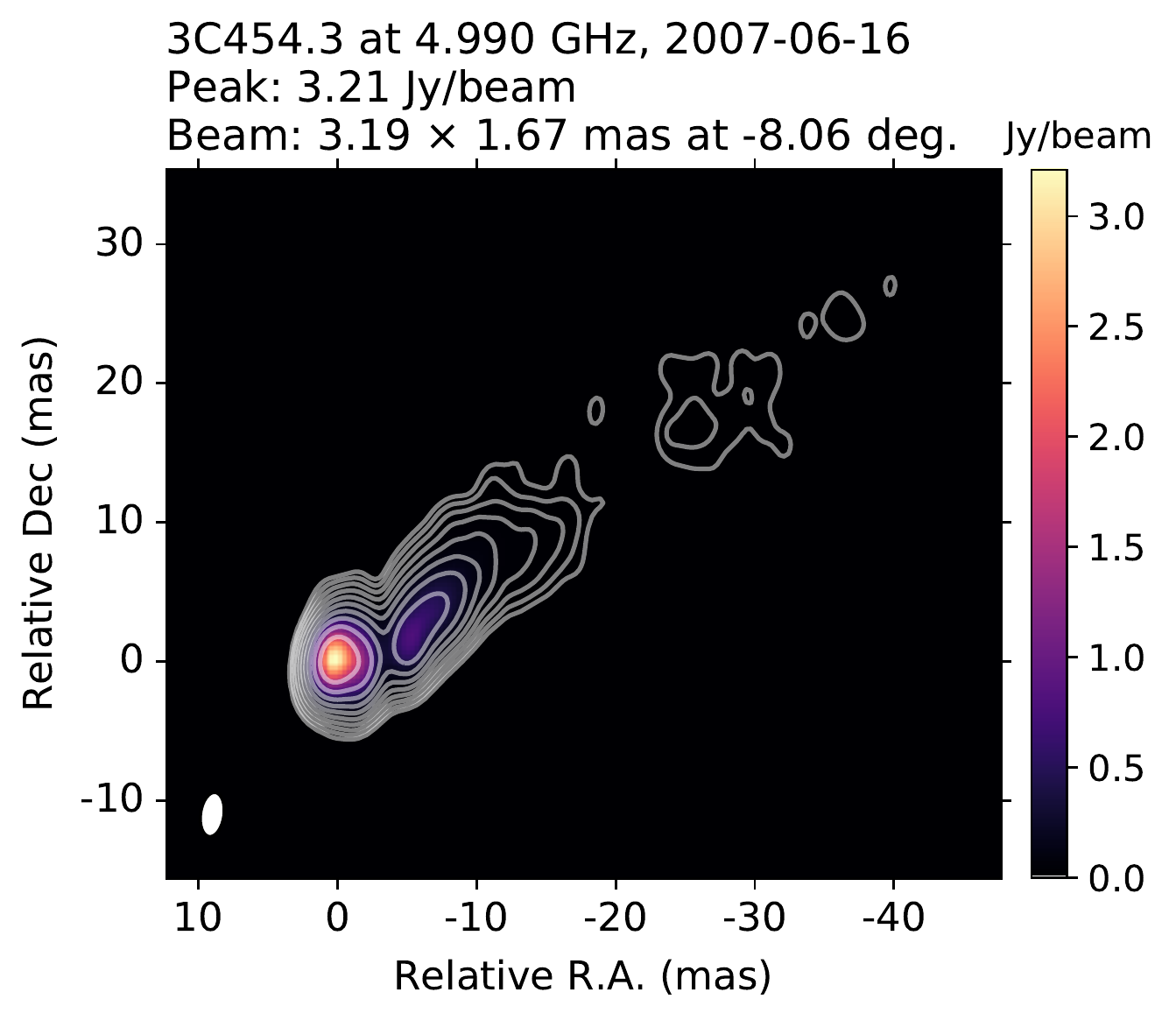}
    }
     \subfigure[]
    {
         \includegraphics[width=0.3\textwidth]{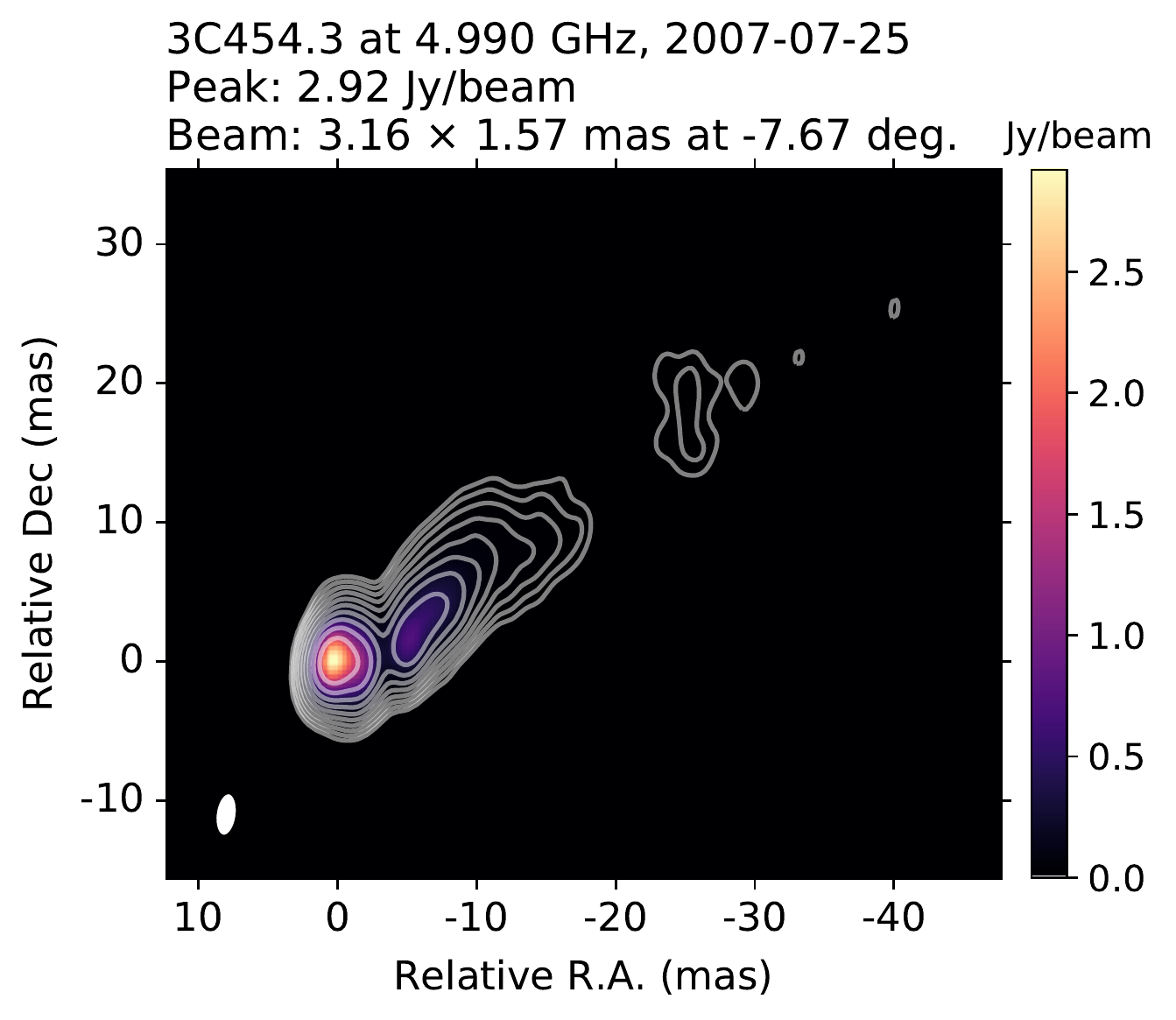}
    }    
         \subfigure[]
    {
        \includegraphics[width=0.3\textwidth]{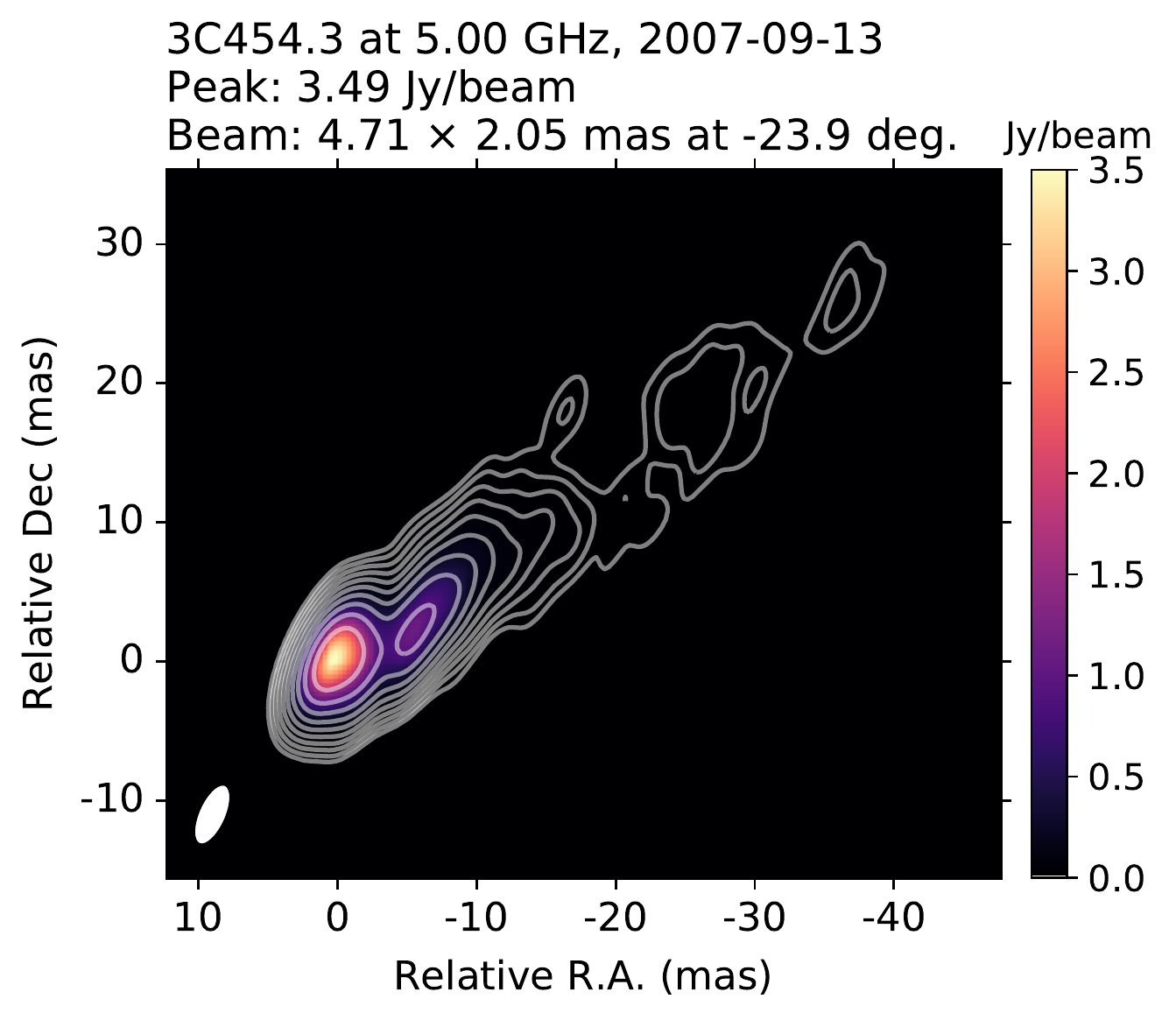}
    }   
     \caption{C-band (5\,GHz) CLEAN images of 3C454.3 from 2005-05-19 until 2007-09-13 with contours at -0.1\%, 0.1\%, 0.2\%, 0.4\%, 0.8\%, 1.6\%, 3.2\%, 6.4\%, 12.8\%, 25.6\%, and 51.2\% of the peak intensity at each image. }
    \label{Cbandimagesp1}
\end{figure*}

\begin{figure*}[]
\centering
    \subfigure[]
    {
         \includegraphics[width=0.3\textwidth]{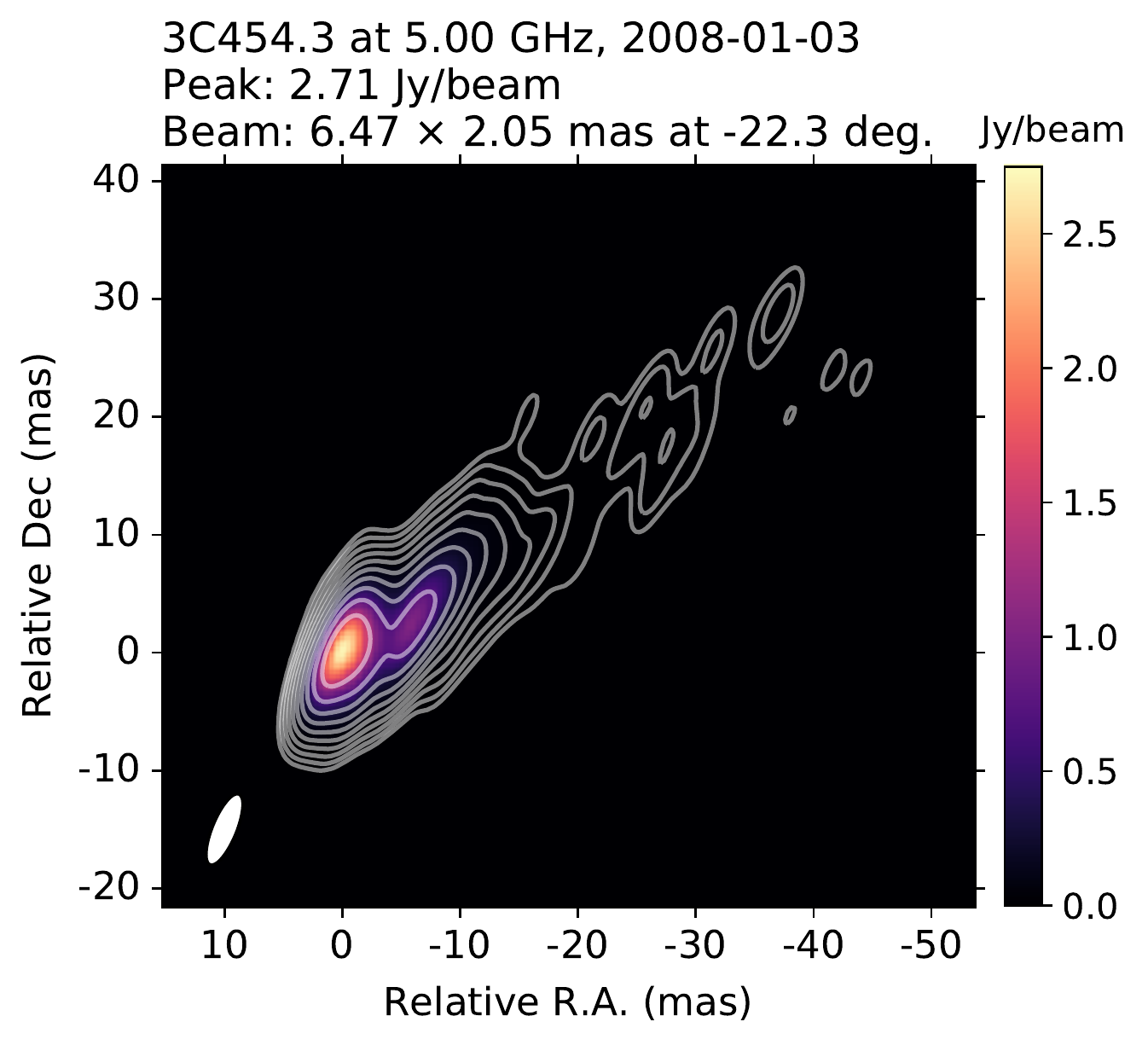}
    }
    \subfigure[]
    {
        \includegraphics[width=0.3\textwidth]{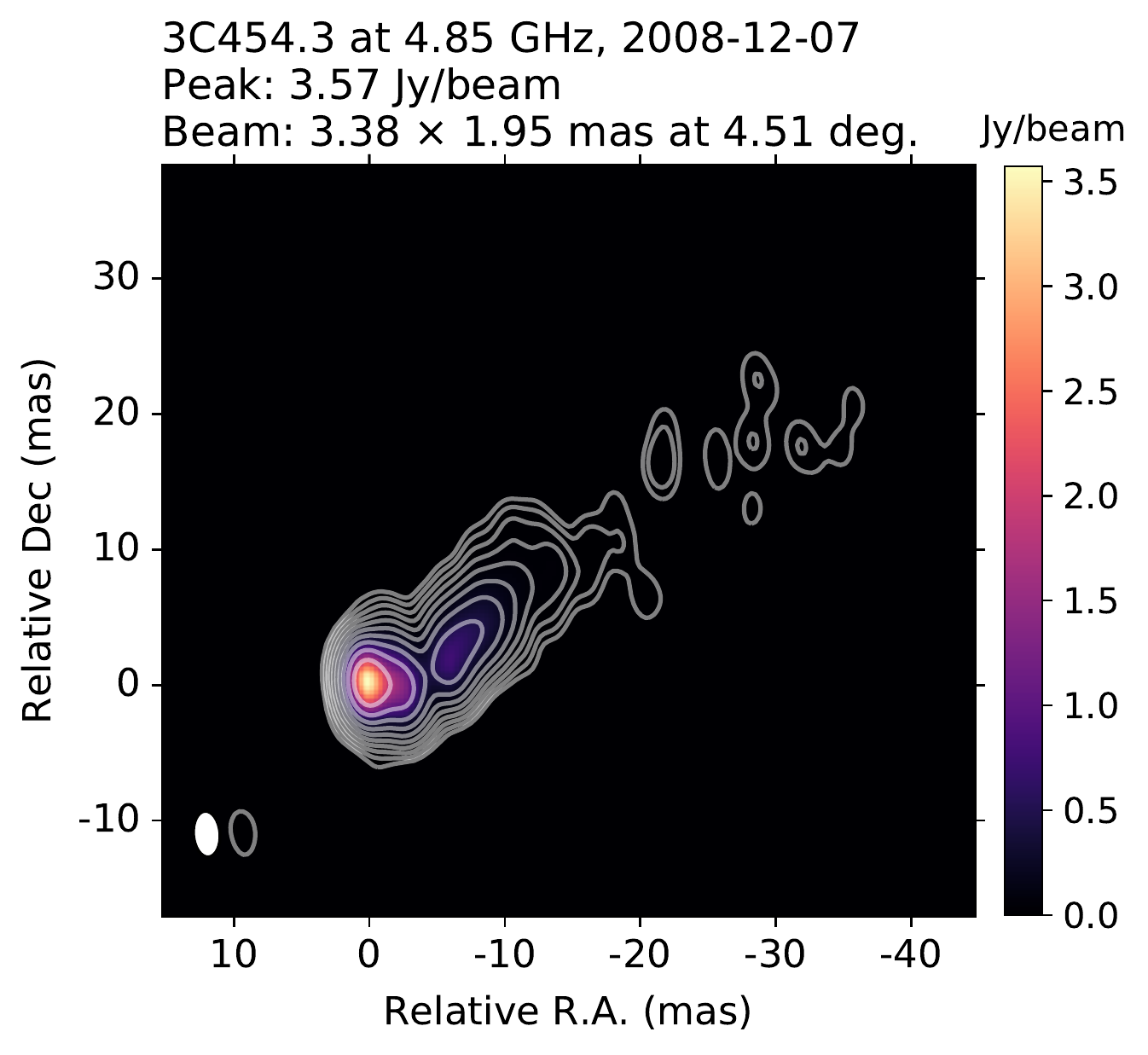}
    }
    \subfigure[]
    {
         \includegraphics[width=0.3\textwidth]{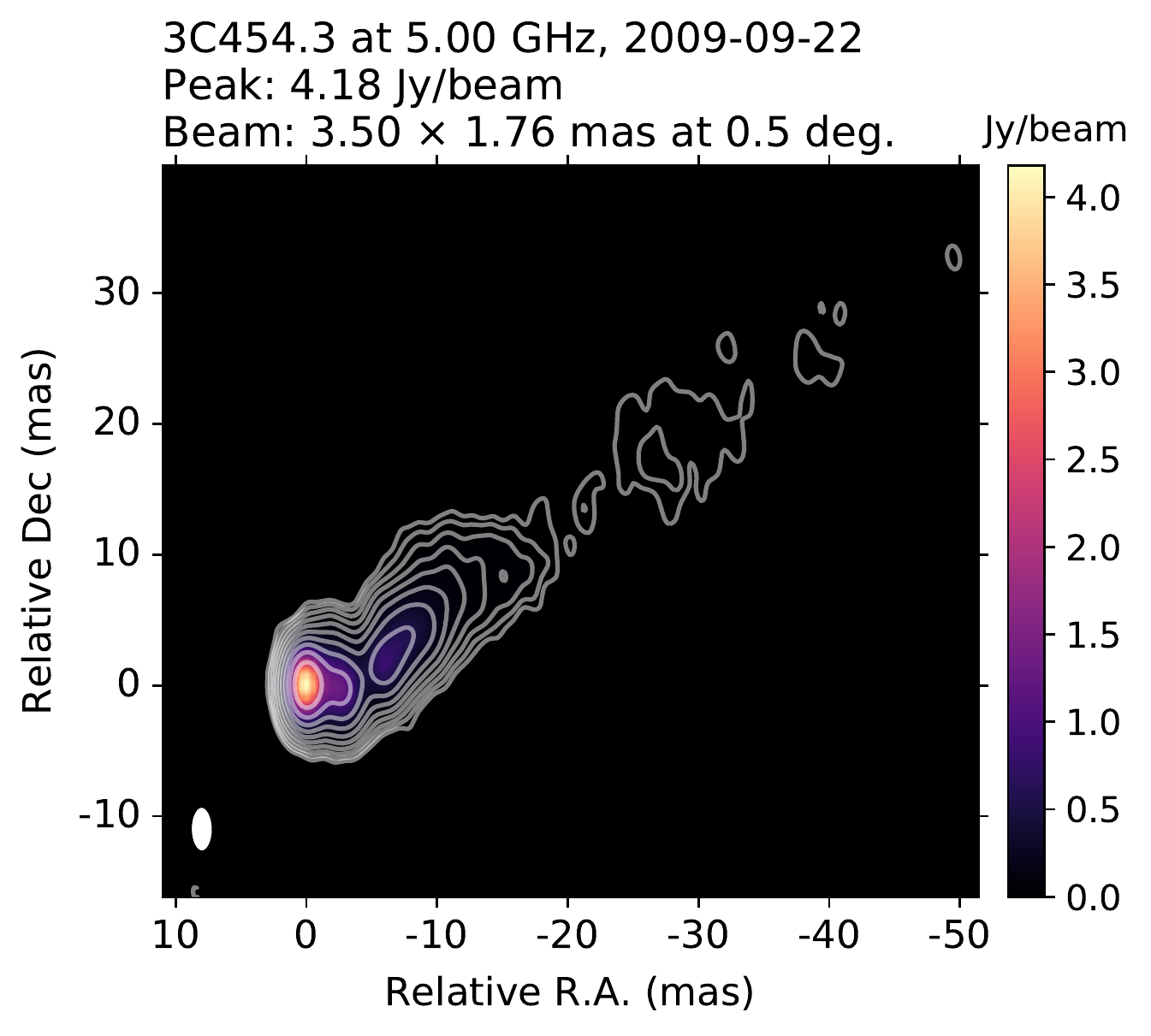}
    }
        \subfigure[]
    {
         \includegraphics[width=0.3\textwidth]{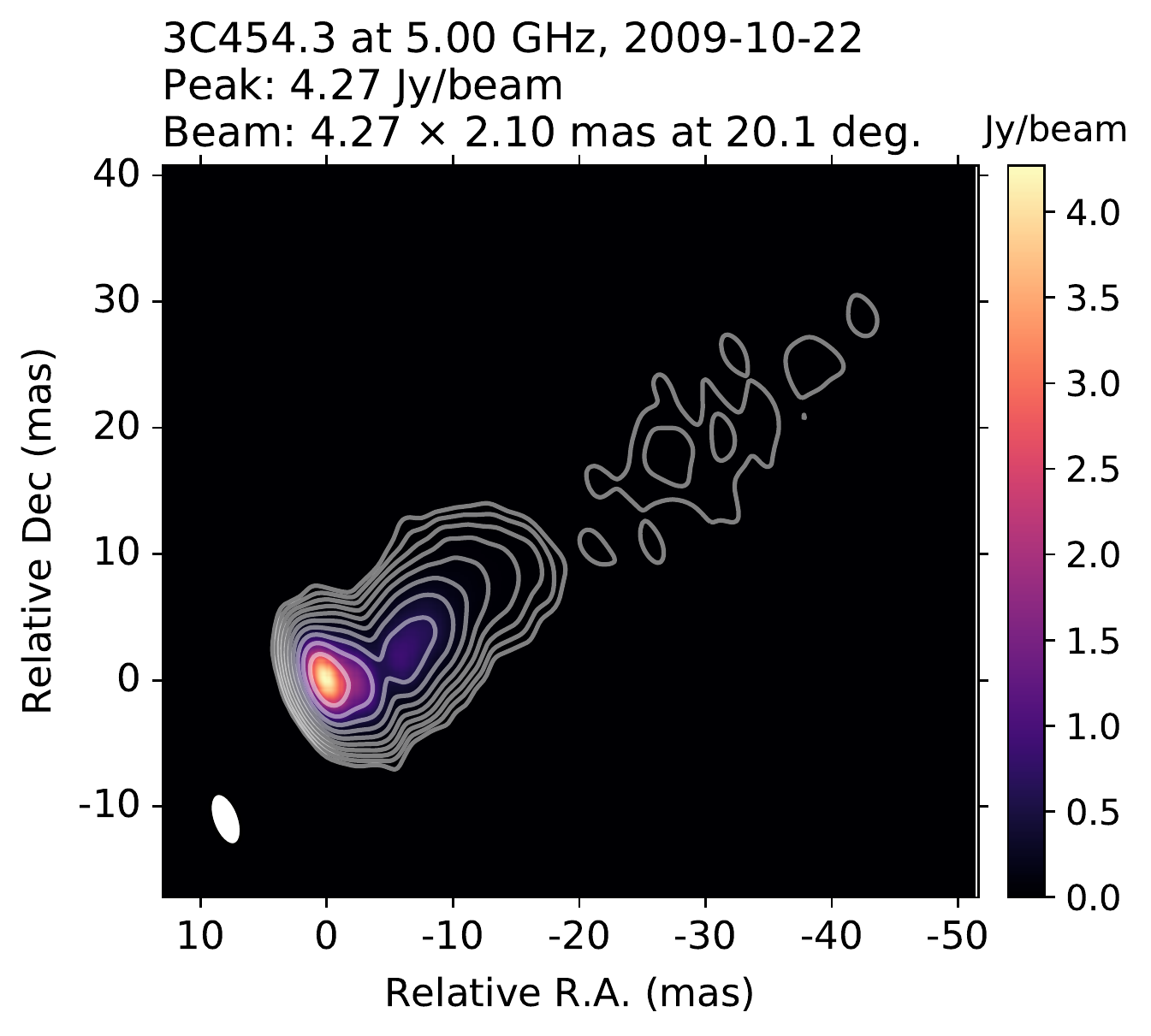}
    }
     \subfigure[]
    {
         \includegraphics[width=0.3\textwidth]{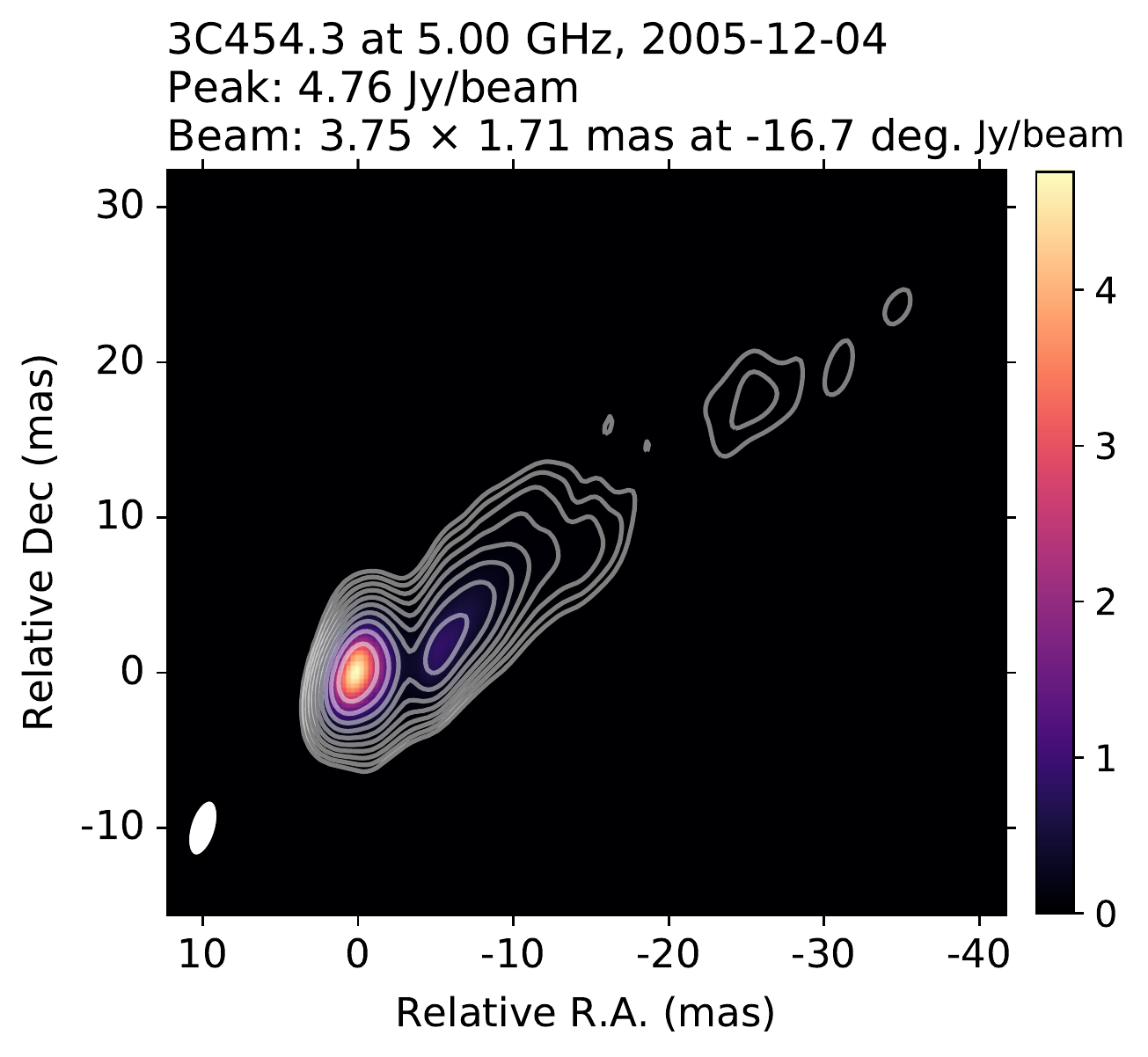}
    }    
         \subfigure[]
    {
        \includegraphics[width=0.3\textwidth]{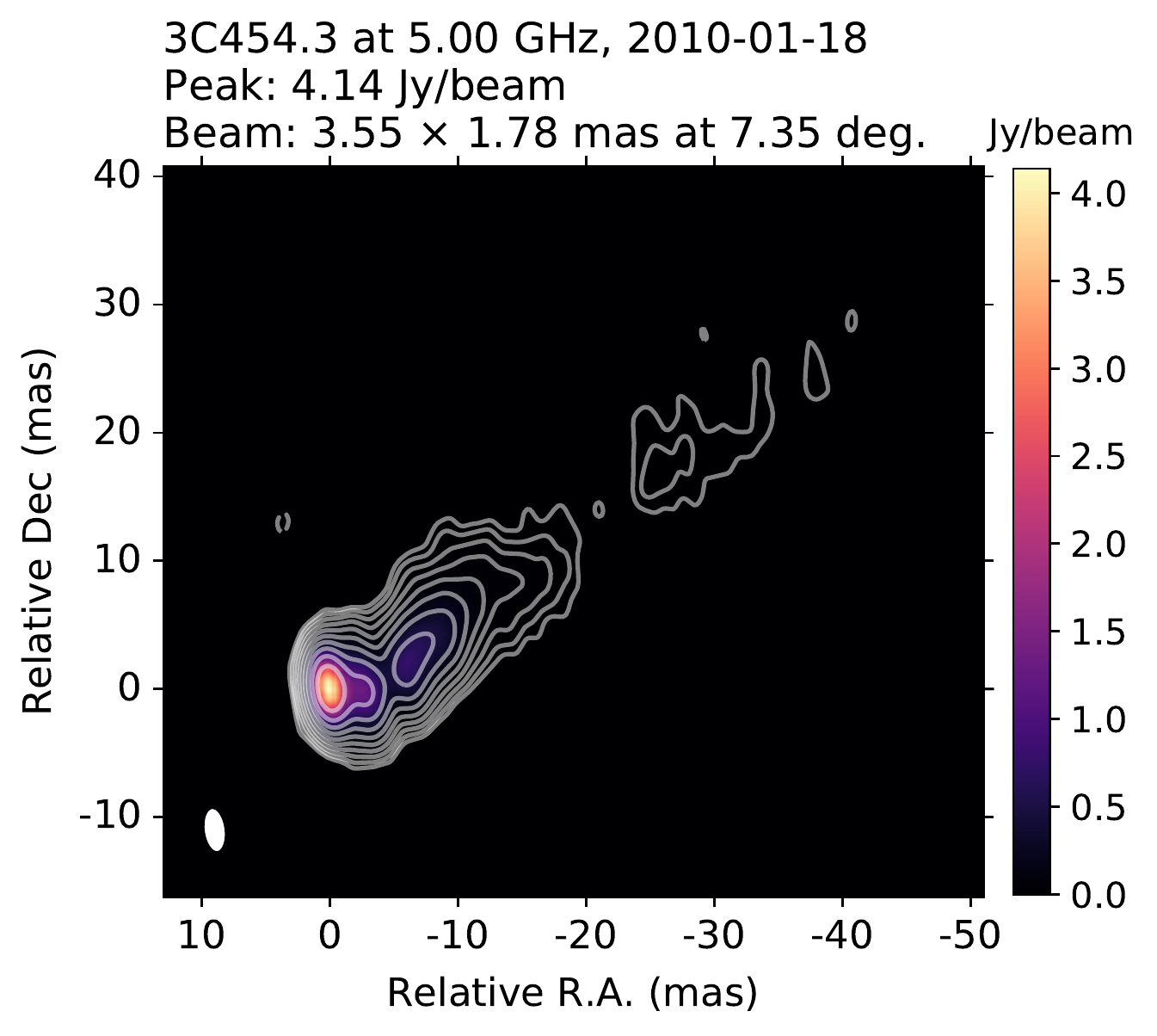}
    }   
         \subfigure[]
    {
        \includegraphics[width=0.3\textwidth]{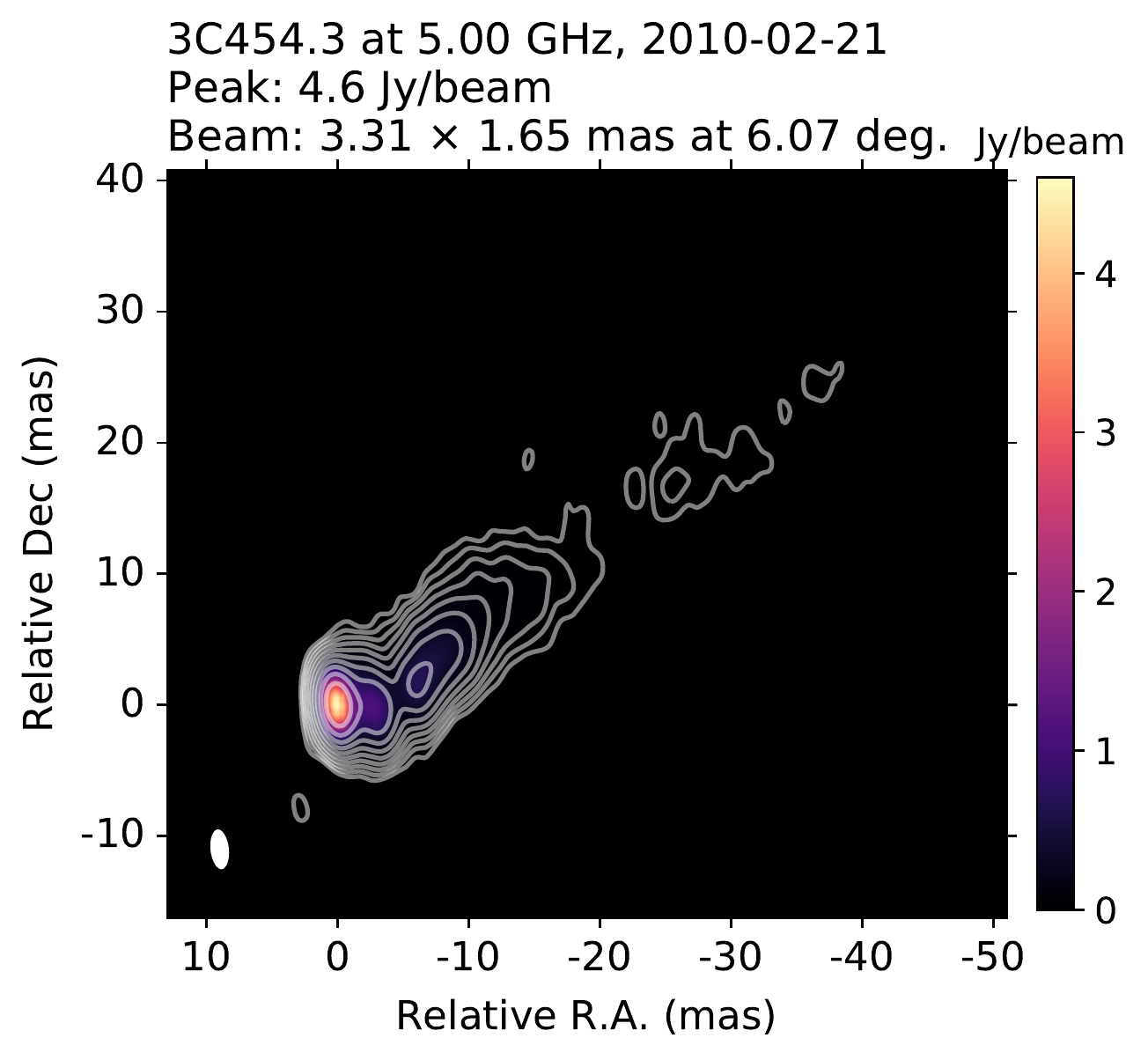}
    }   
     \caption{C-band (5\,GHz) CLEAN images of 3C454.3 from 2008-01-03 until 2010-02-21 with contours at -0.1\%, 0.1\%, 0.2\%, 0.4\%, 0.8\%, 1.6\%, 3.2\%, 6.4\%, 12.8\%, 25.6\%, and 51.2\% of the peak intensity at each image. }
    \label{Cbandimagesp2}
\end{figure*}

\begin{figure*}[]
\centering
   \subfigure[]
    {
        \includegraphics[width=0.3\textwidth]{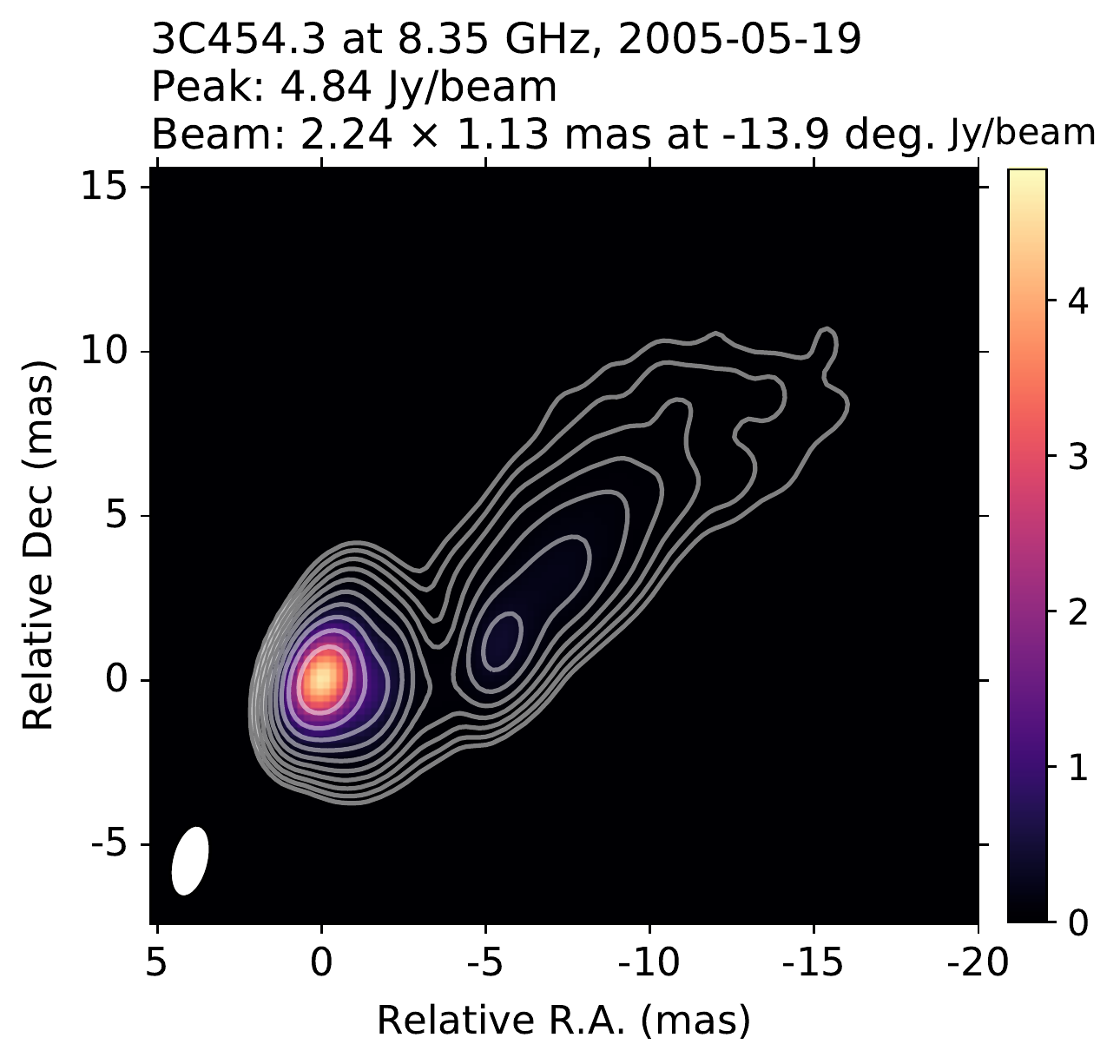}
    }
       \subfigure[]
    {
        \includegraphics[width=0.3\textwidth]{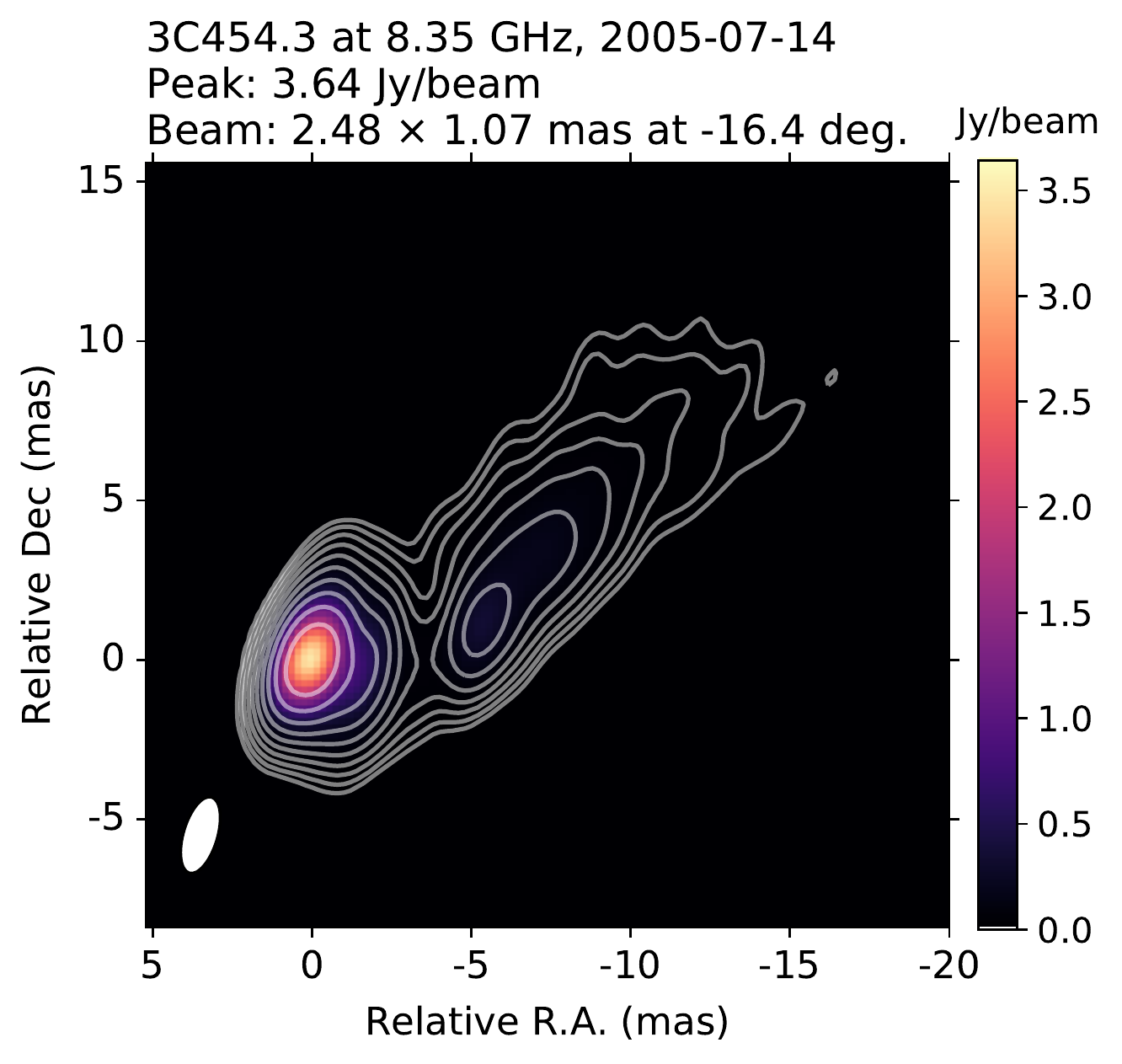}
    }
       \subfigure[]
    {
        \includegraphics[width=0.3\textwidth]{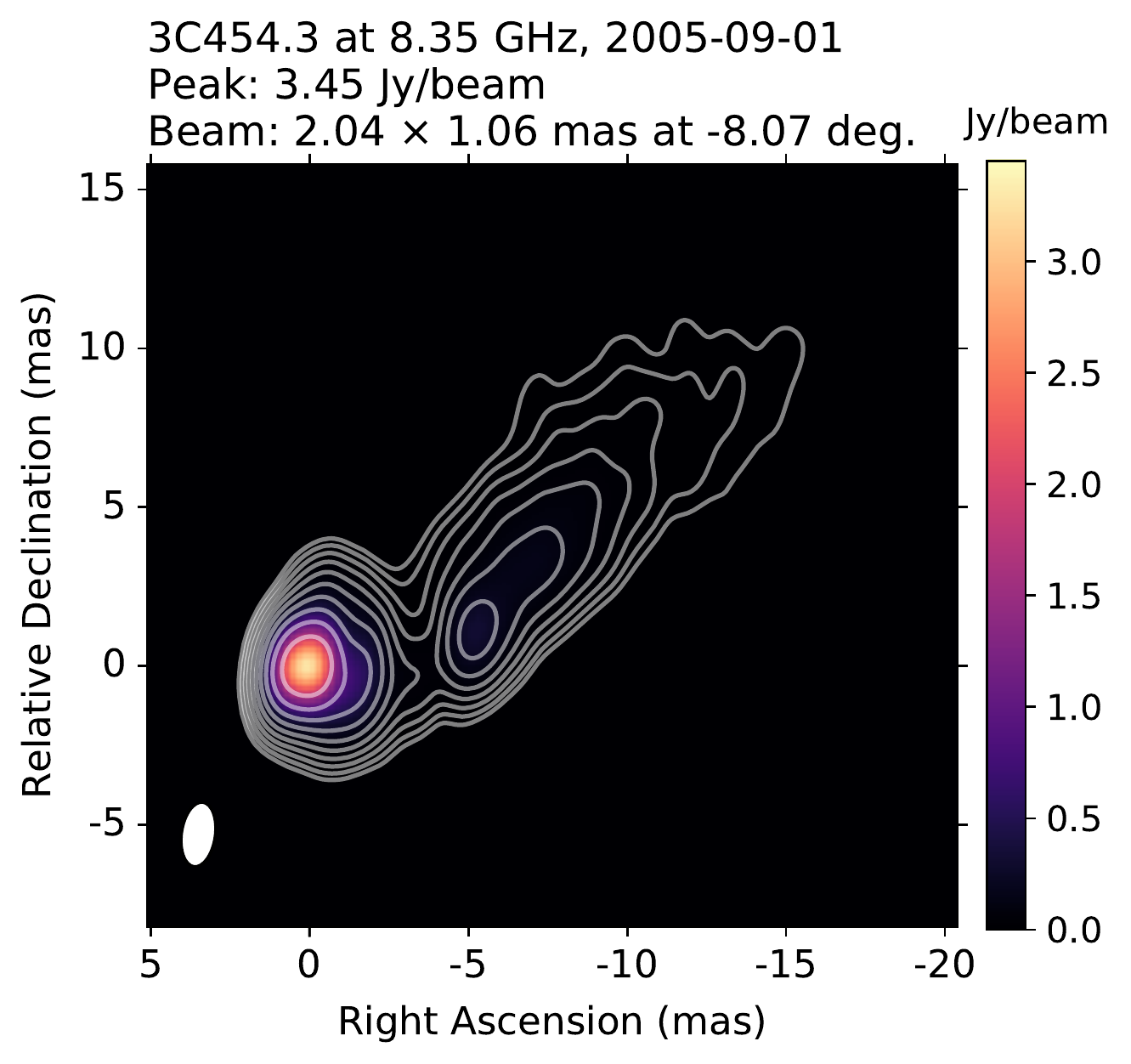}
    }
     \subfigure[]
    {
         \includegraphics[width=0.3\textwidth]{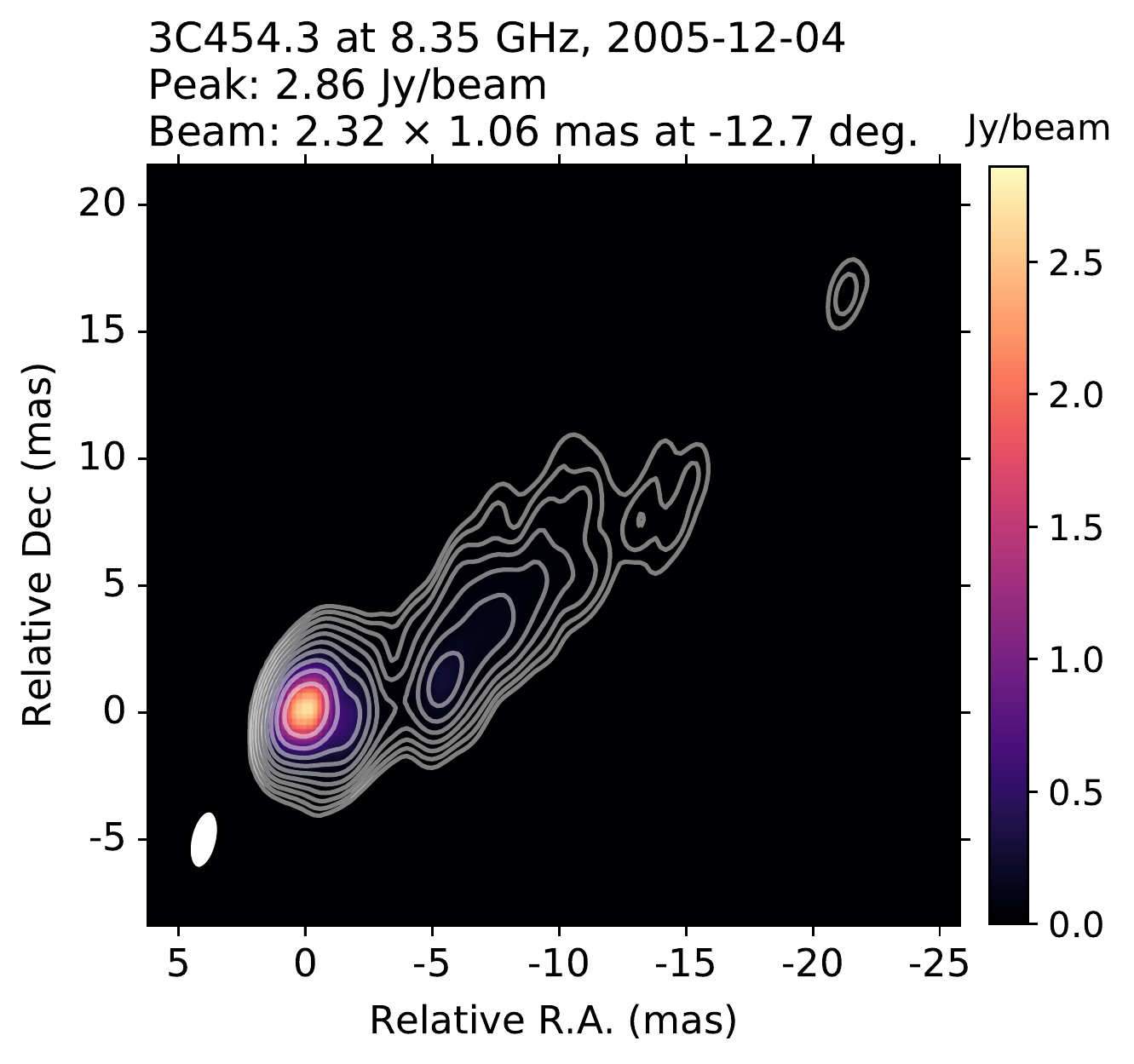}
    }
       \subfigure[]
    {
         \includegraphics[width=0.3\textwidth]{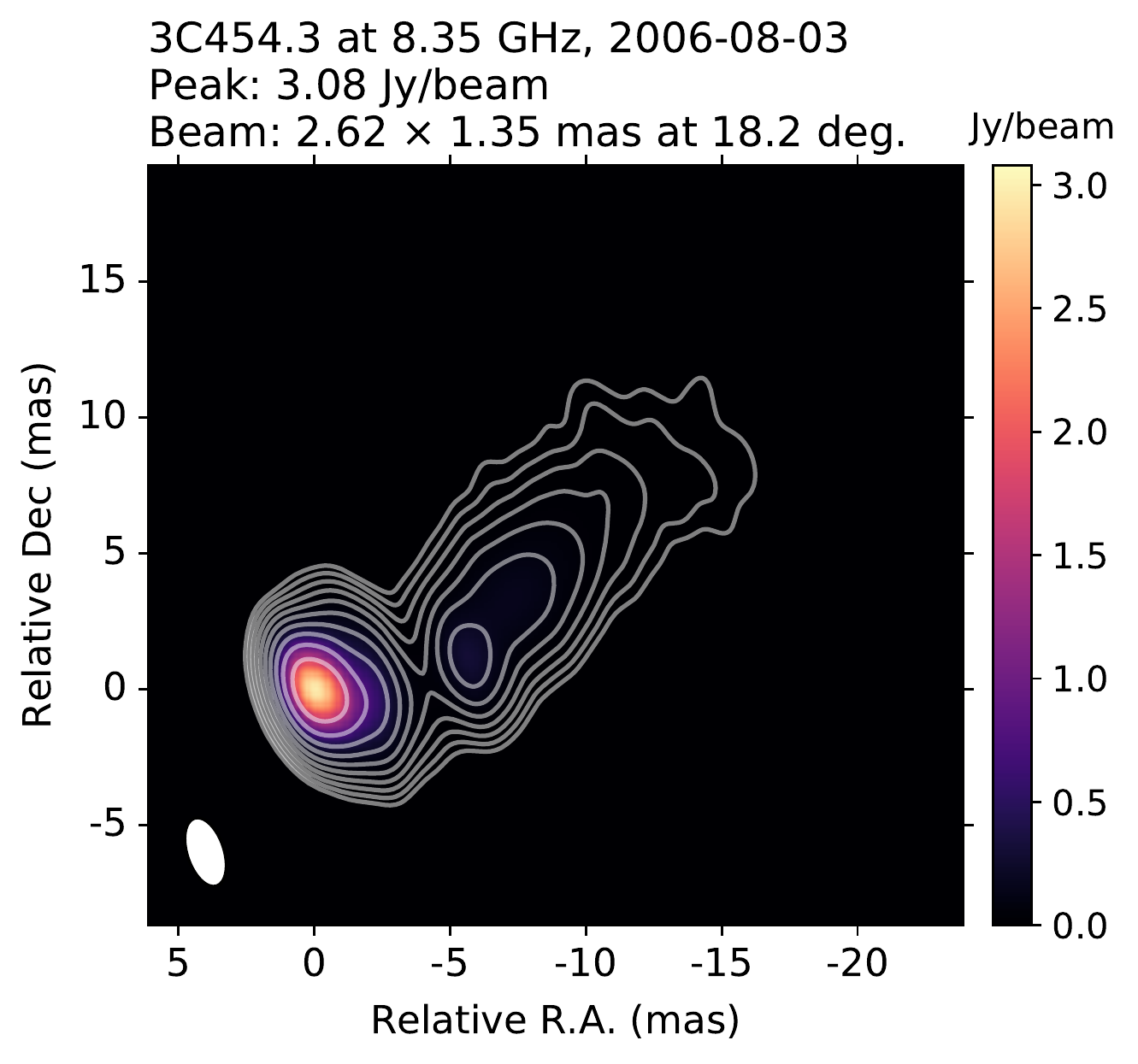}
    }
       \subfigure[]
    {
         \includegraphics[width=0.3\textwidth]{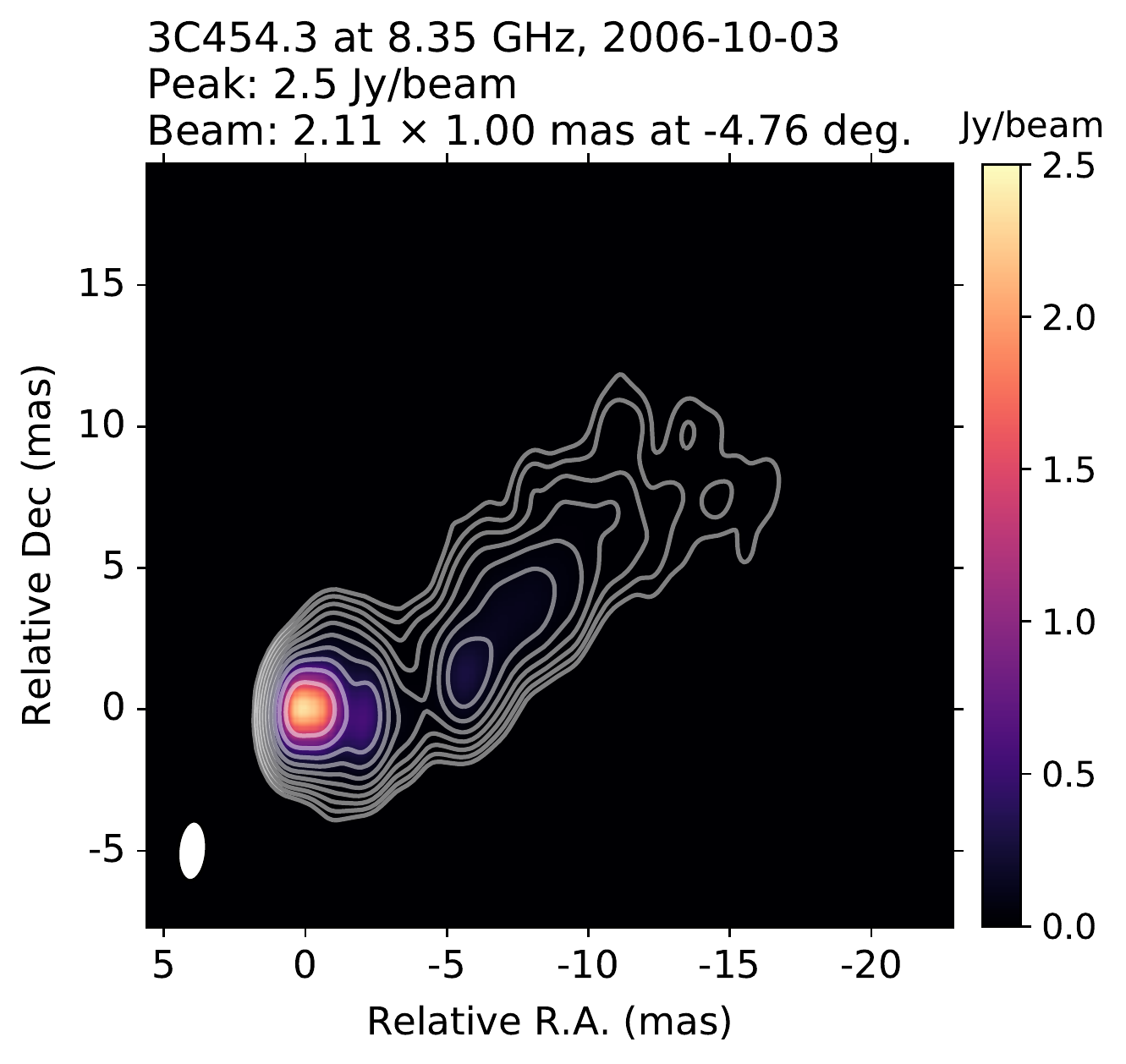}
    }
    \subfigure[]
    {
        \includegraphics[width=0.3\textwidth]{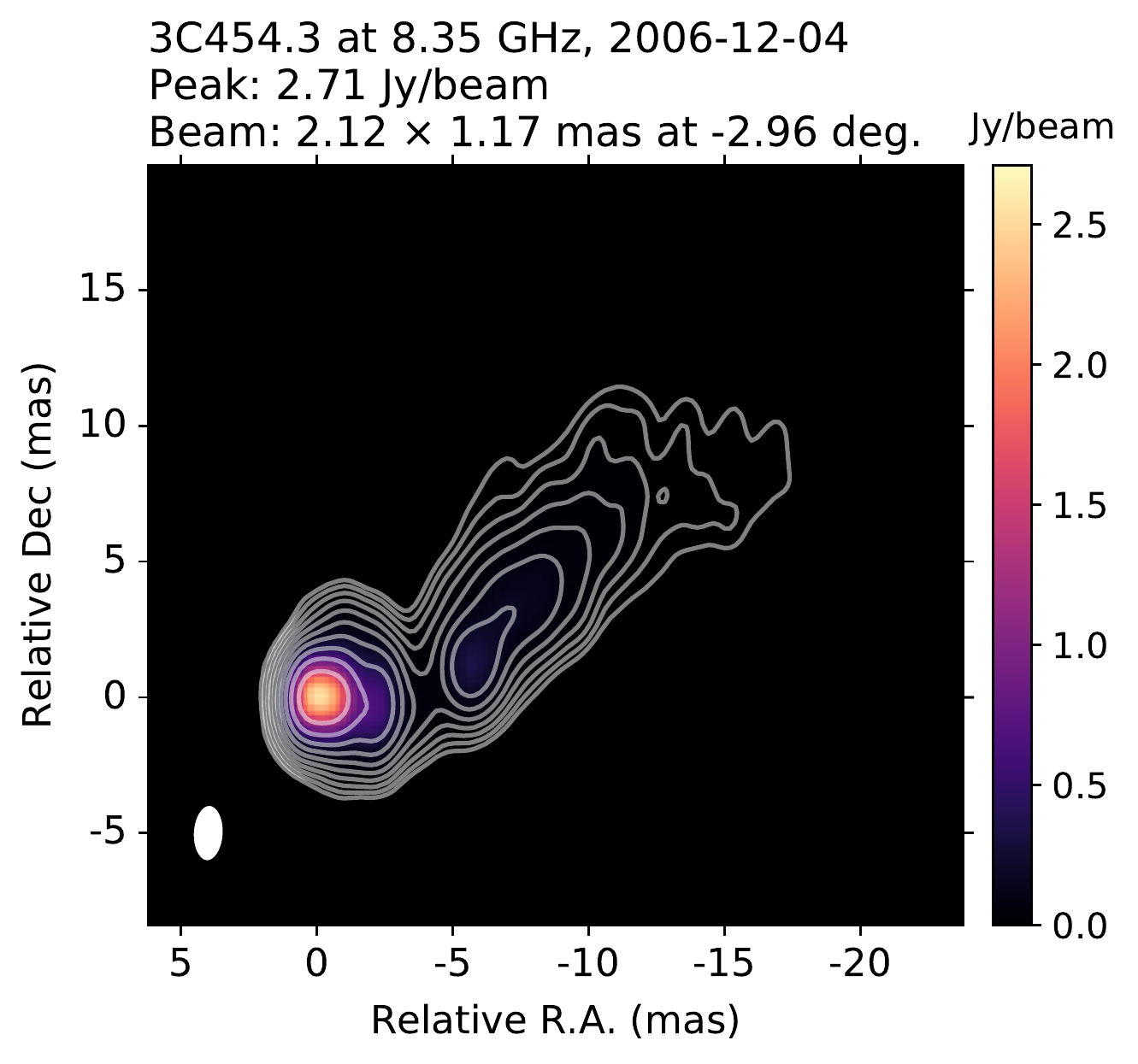}
    }
    \subfigure[]
    {
         \includegraphics[width=0.3\textwidth]{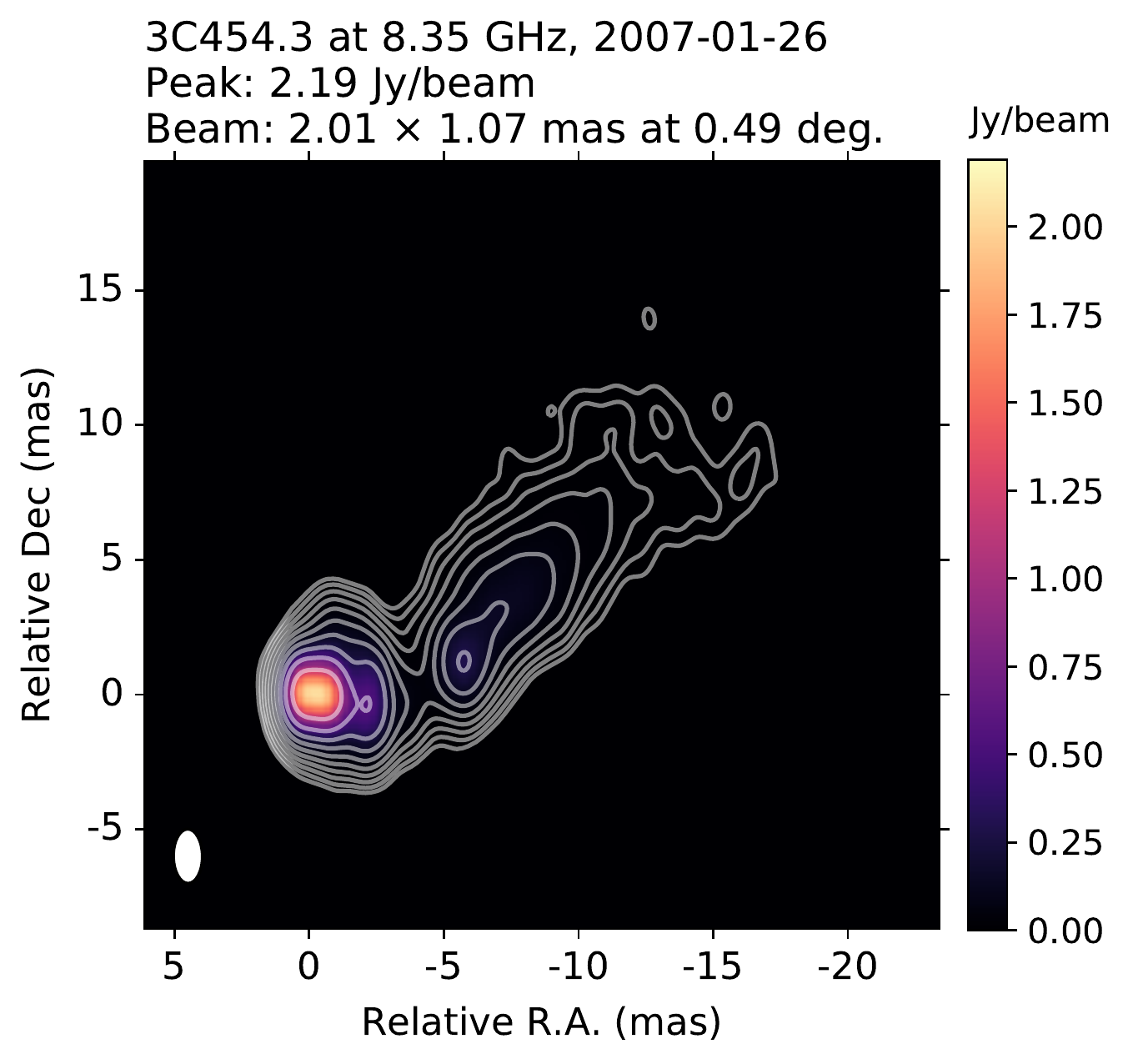}
    }
        \subfigure[]
    {
         \includegraphics[width=0.3\textwidth]{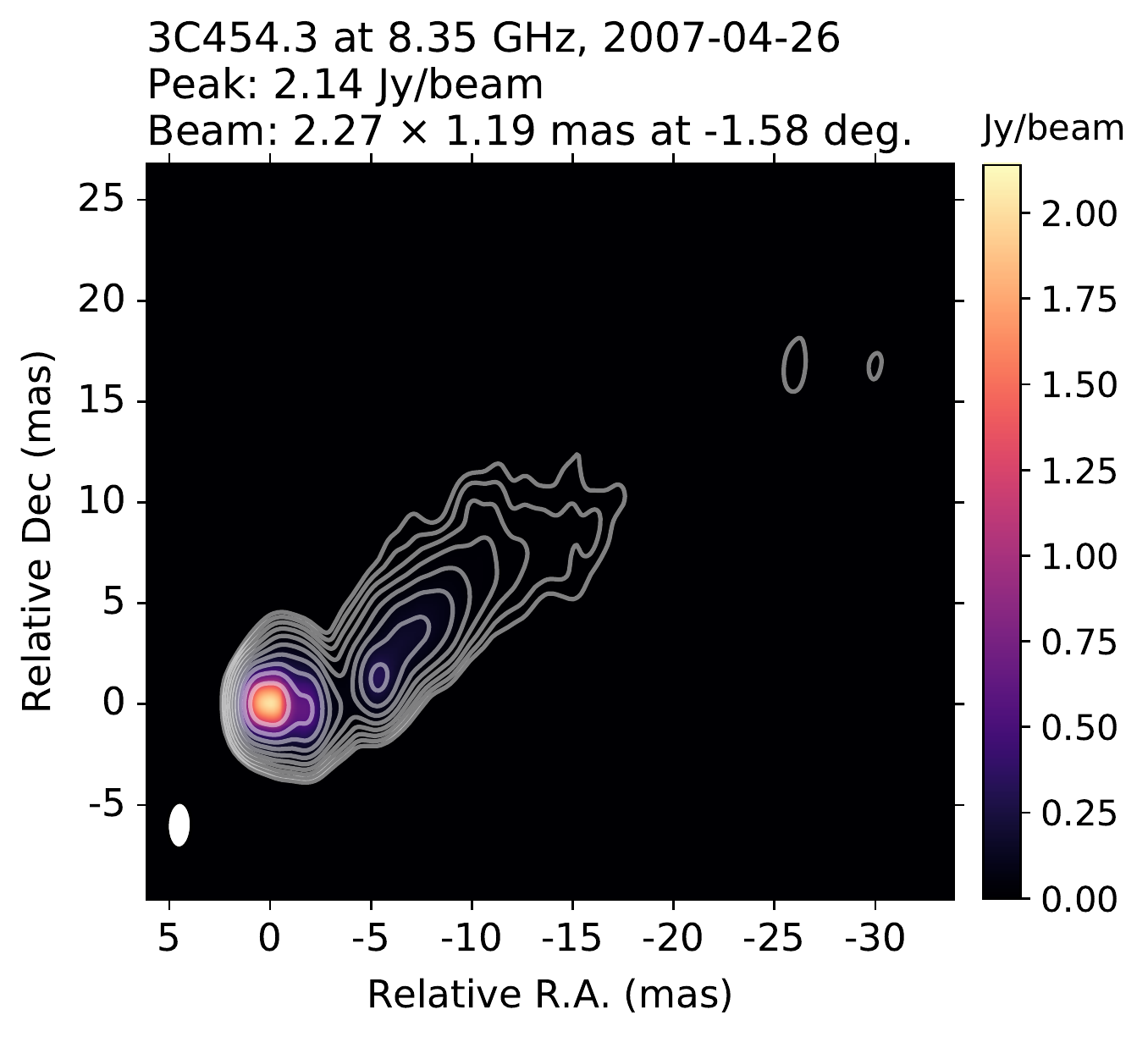}
    }
     \subfigure[]
    {
         \includegraphics[width=0.3\textwidth]{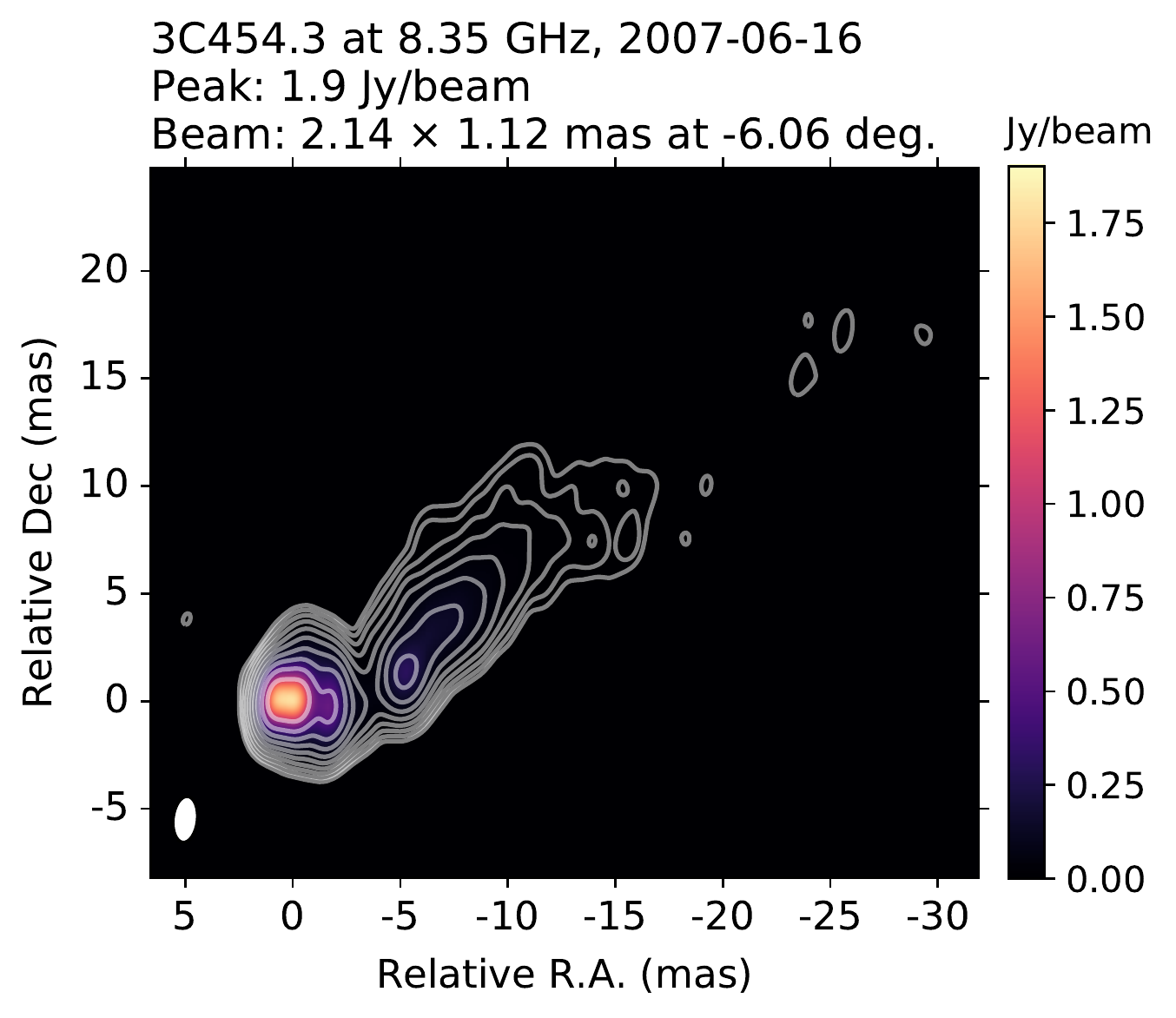}
    }    
         \subfigure[]
    {
        \includegraphics[width=0.3\textwidth]{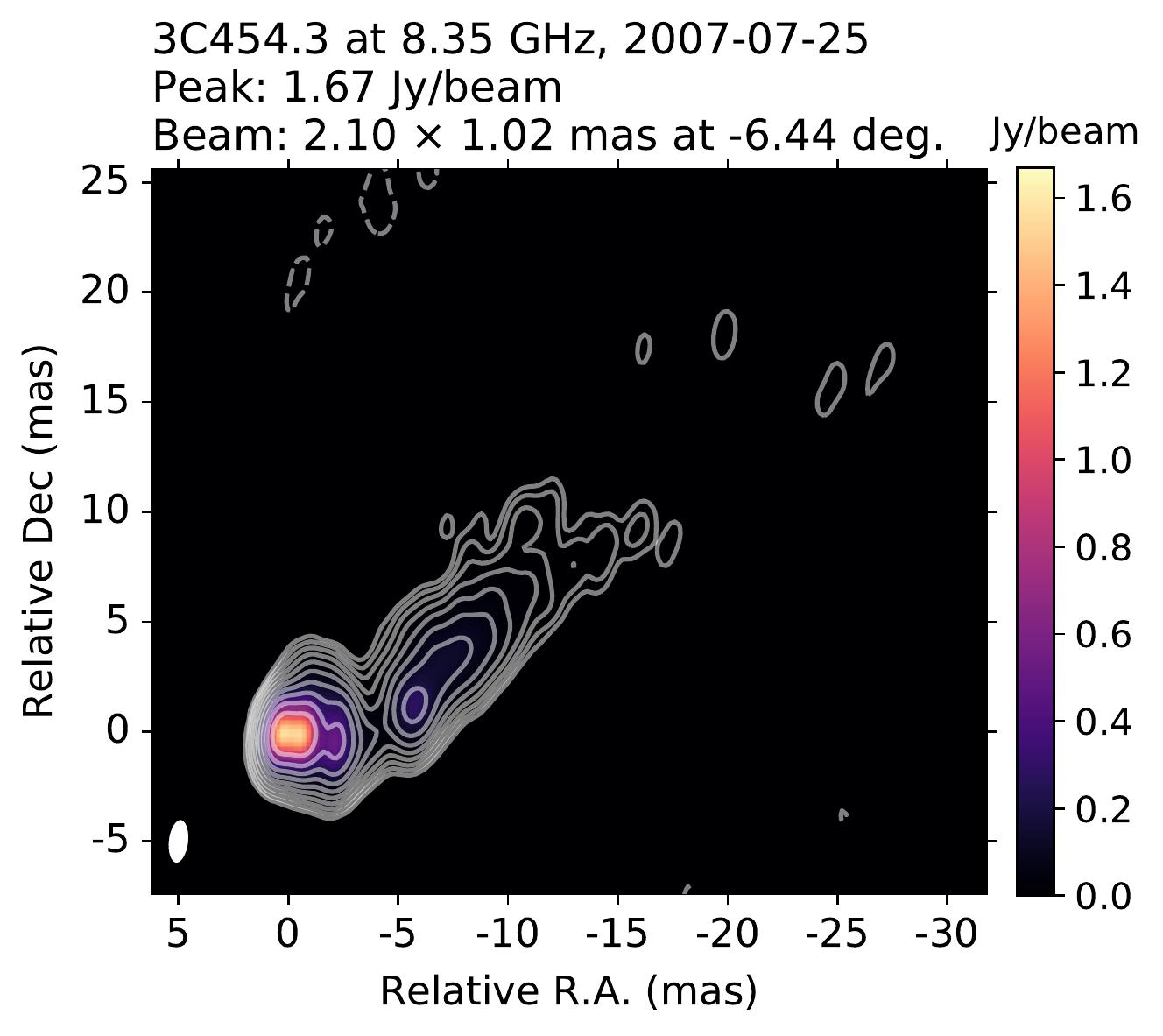}
    }   
         \subfigure[]
    {
        \includegraphics[width=0.3\textwidth]{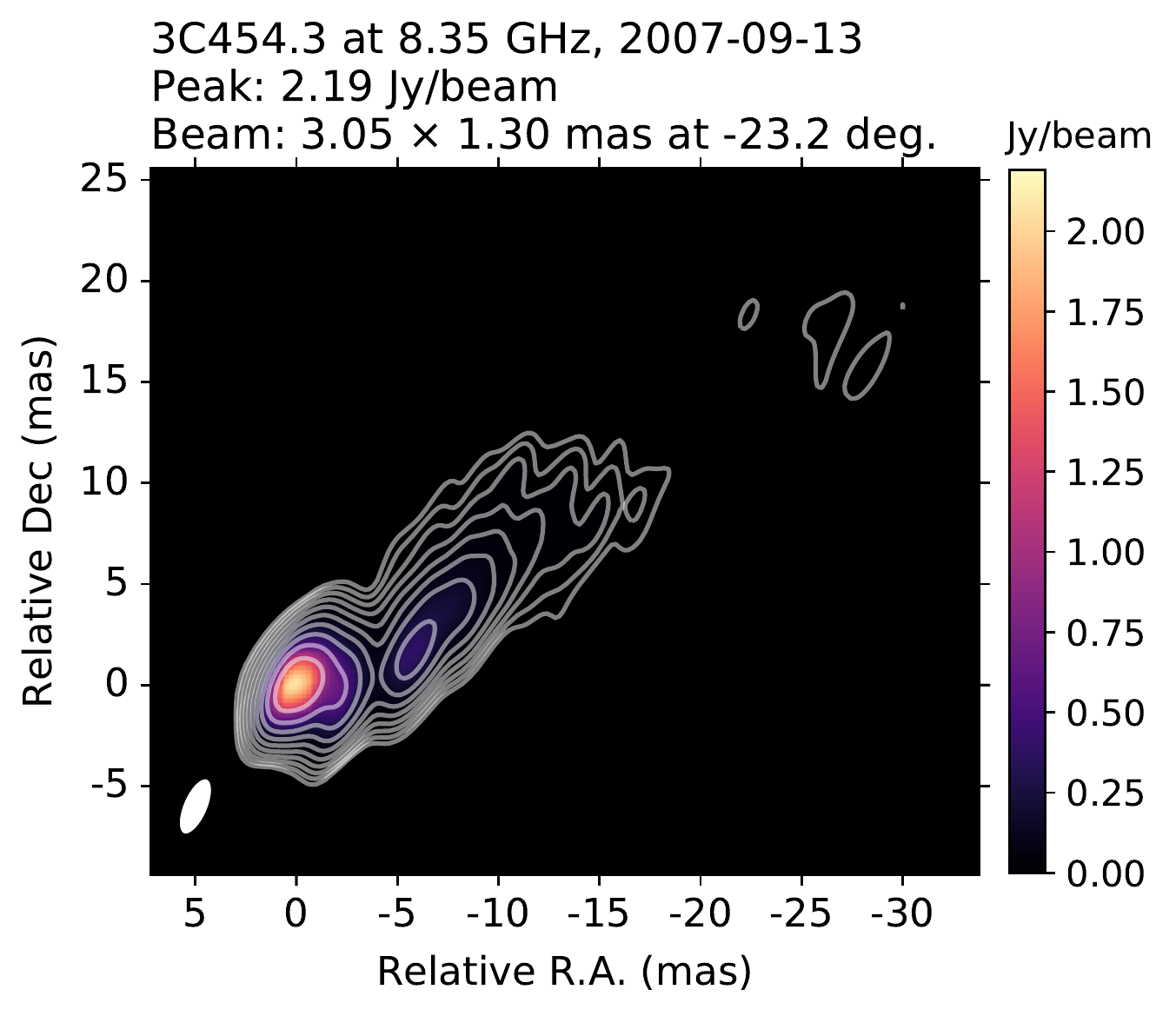}
    }   
     \caption{X-band (8\,GHz) CLEAN images of 3C454.3 from 2005-05-19 until 2007-09-13 with contours at -0.1\%, 0.1\%, 0.2\%, 0.4\%, 0.8\%, 1.6\%, 3.2\%, 6.4\%, 12.8\%, 25.6\%, and 51.2\% of the peak intensity at each image. }
    \label{Xbandimagesp1}
\end{figure*}

\begin{figure*}[]
\centering
    \subfigure[]
    {
         \includegraphics[width=0.3\textwidth]{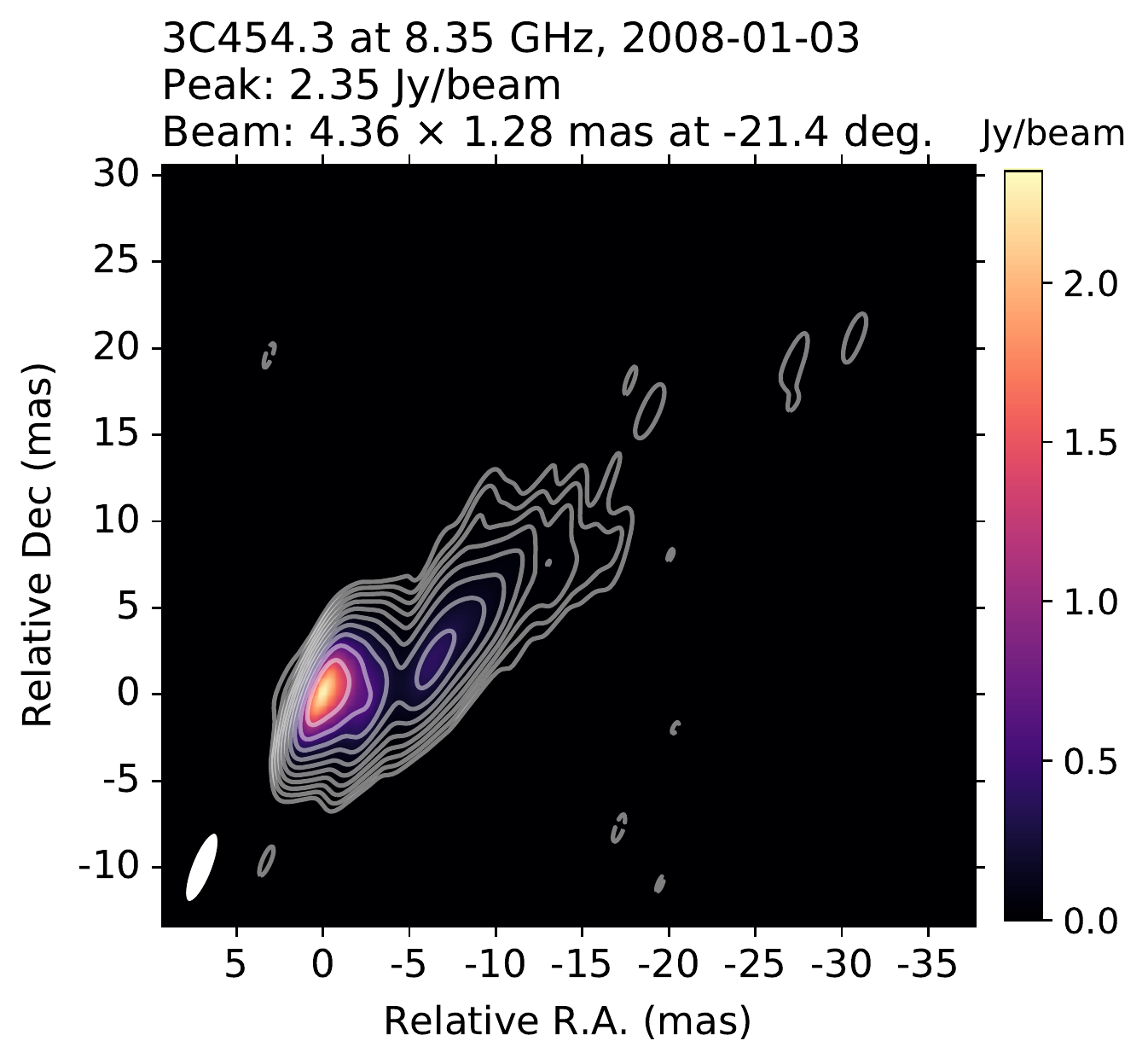}
    }
    \subfigure[]
    {
        \includegraphics[width=0.3\textwidth]{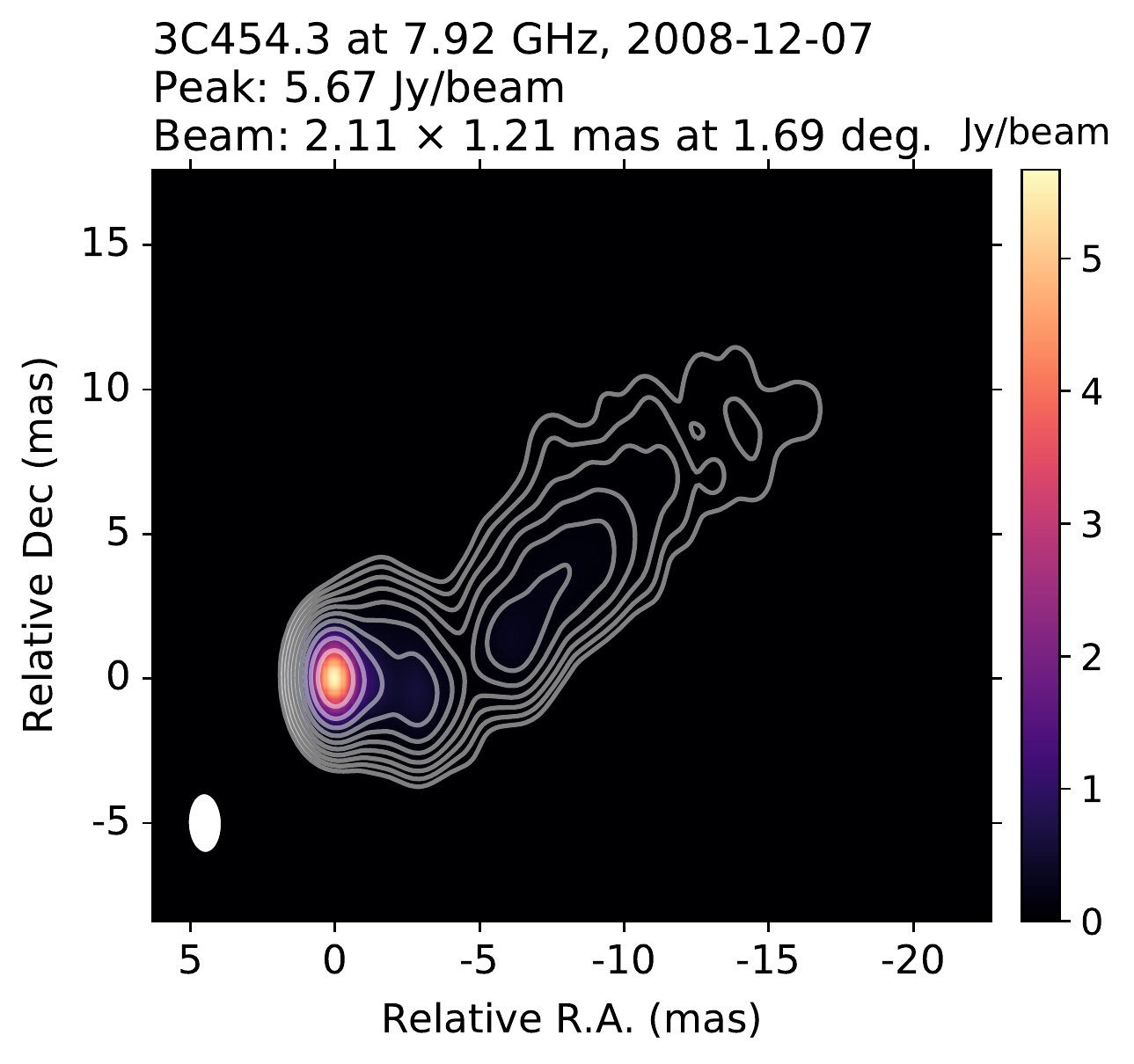}
    }
    \subfigure[]
    {
         \includegraphics[width=0.3\textwidth]{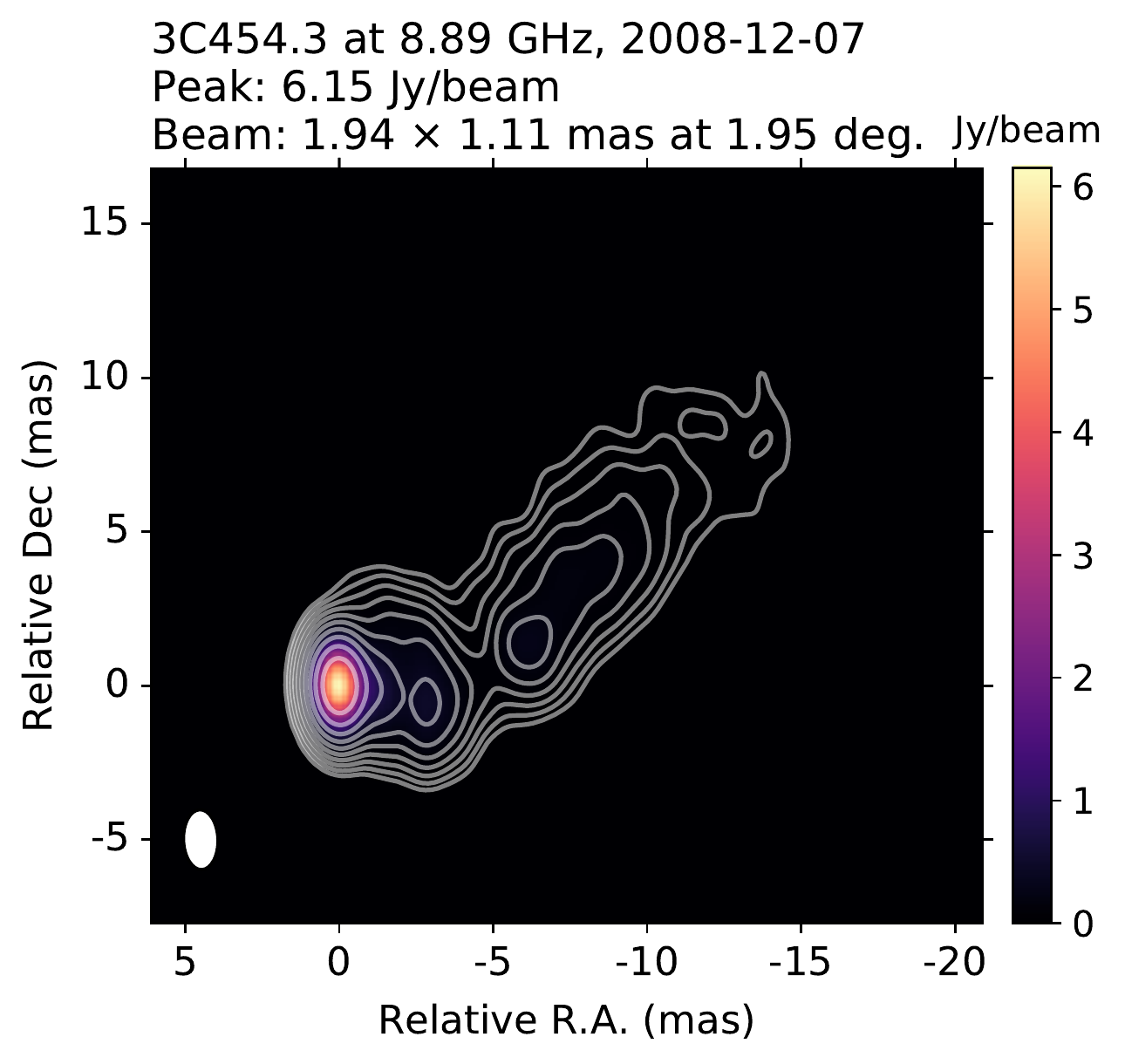}
    }
        \subfigure[]
    {
         \includegraphics[width=0.3\textwidth]{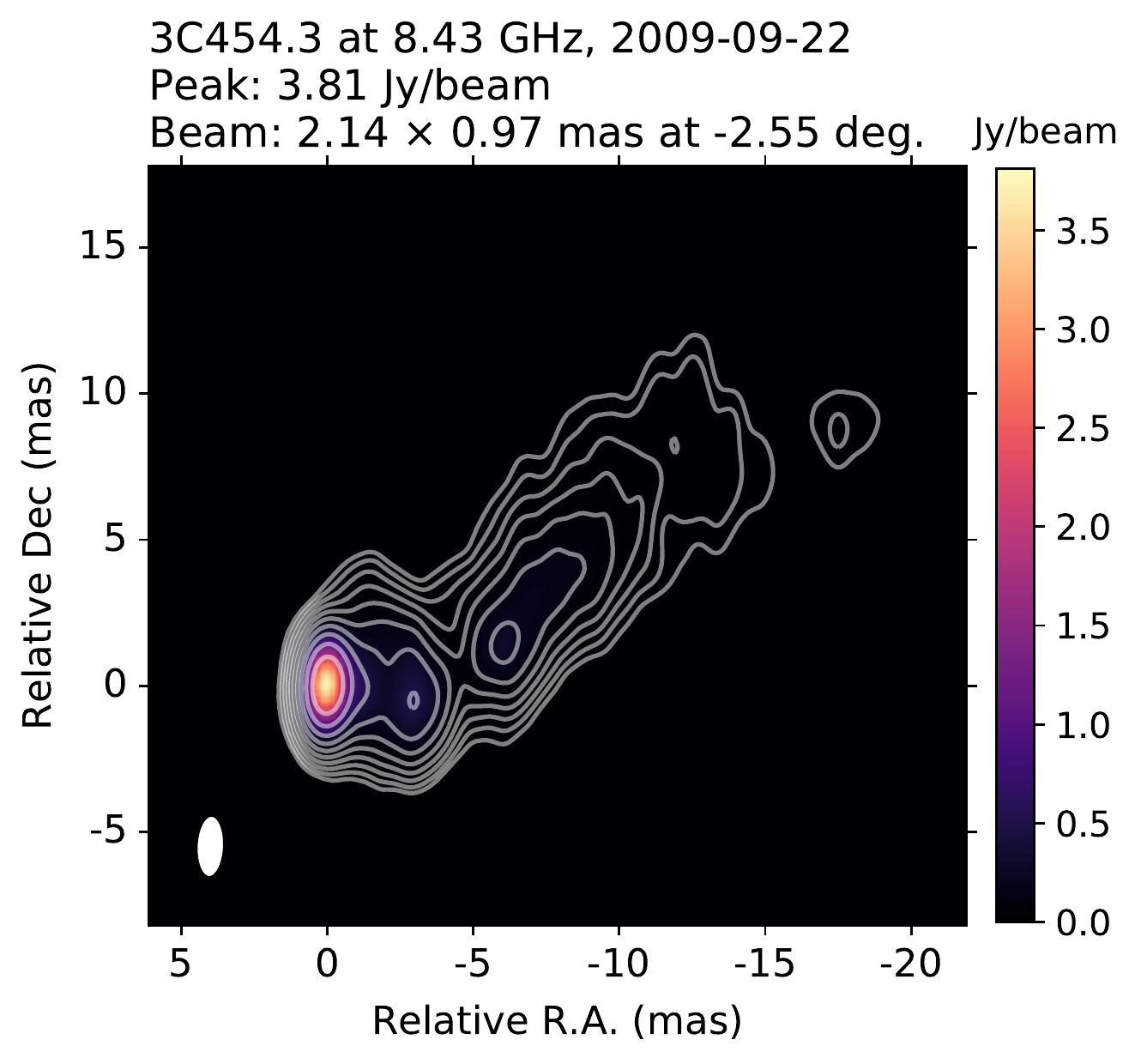}
    }
     \subfigure[]
    {
         \includegraphics[width=0.3\textwidth]{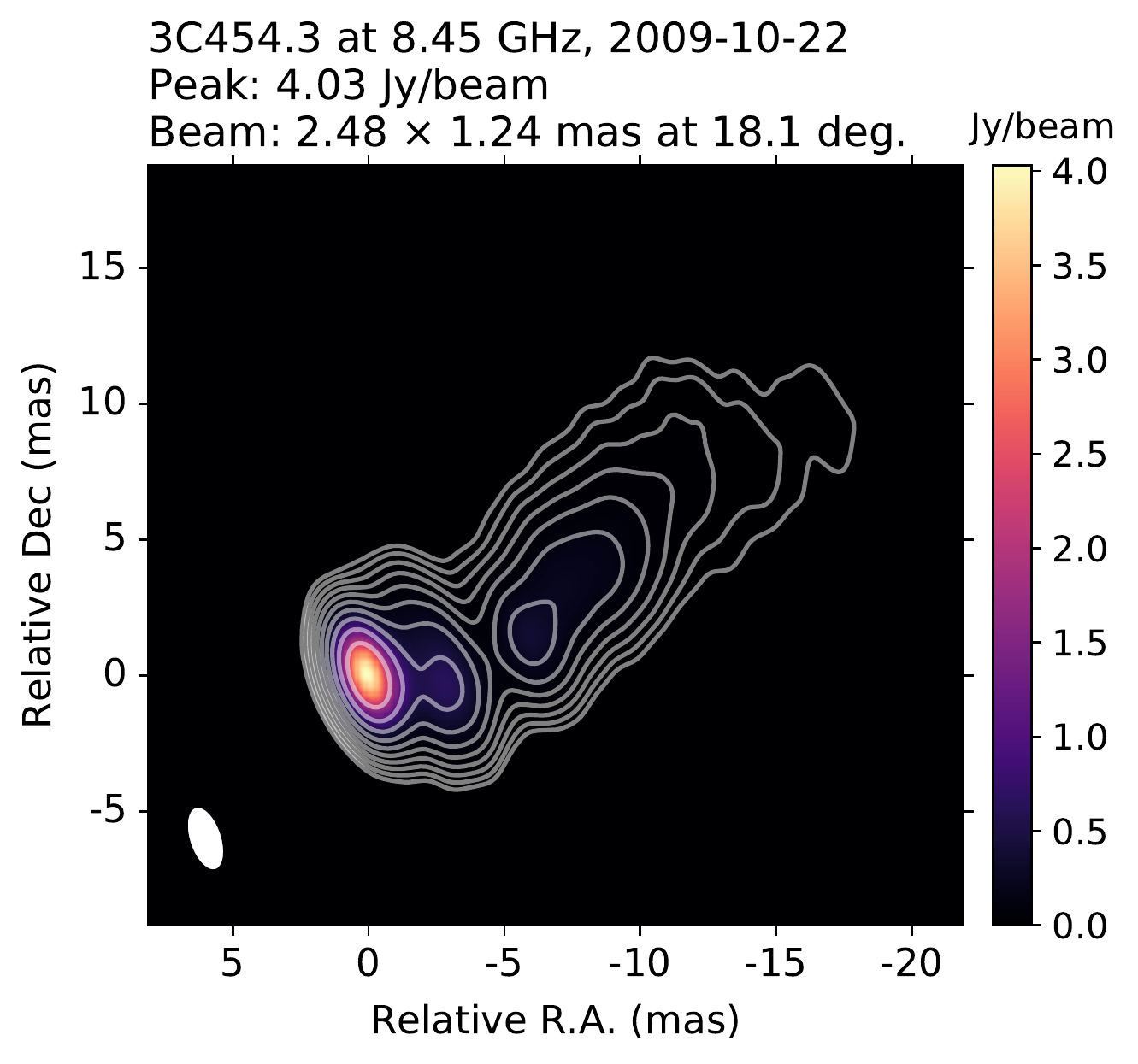}
    }    
         \subfigure[]
    {
        \includegraphics[width=0.3\textwidth]{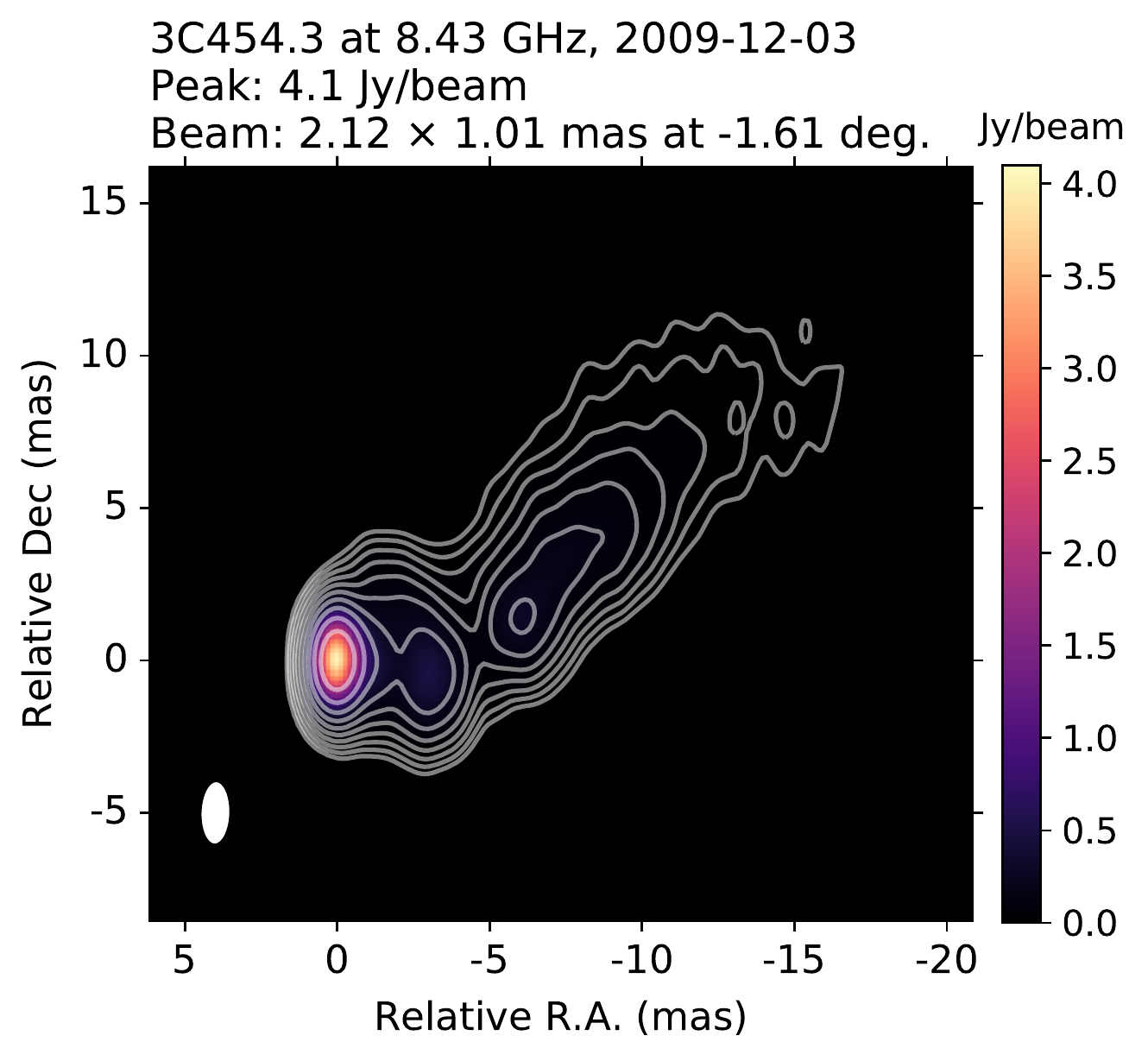}
    }   
         \subfigure[]
    {
        \includegraphics[width=0.3\textwidth]{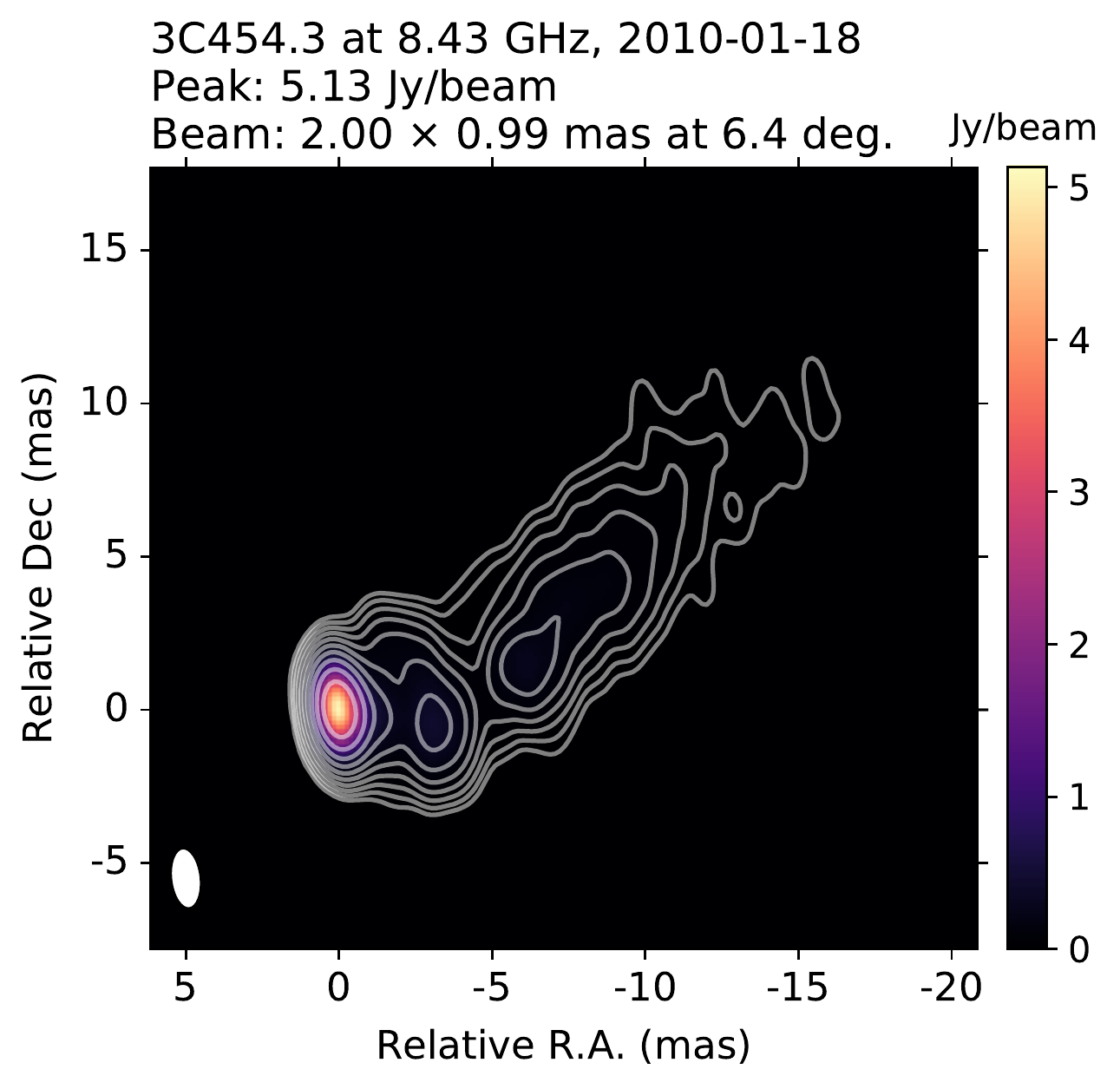}
    }   
       \subfigure[]
    {
        \includegraphics[width=0.3\textwidth]{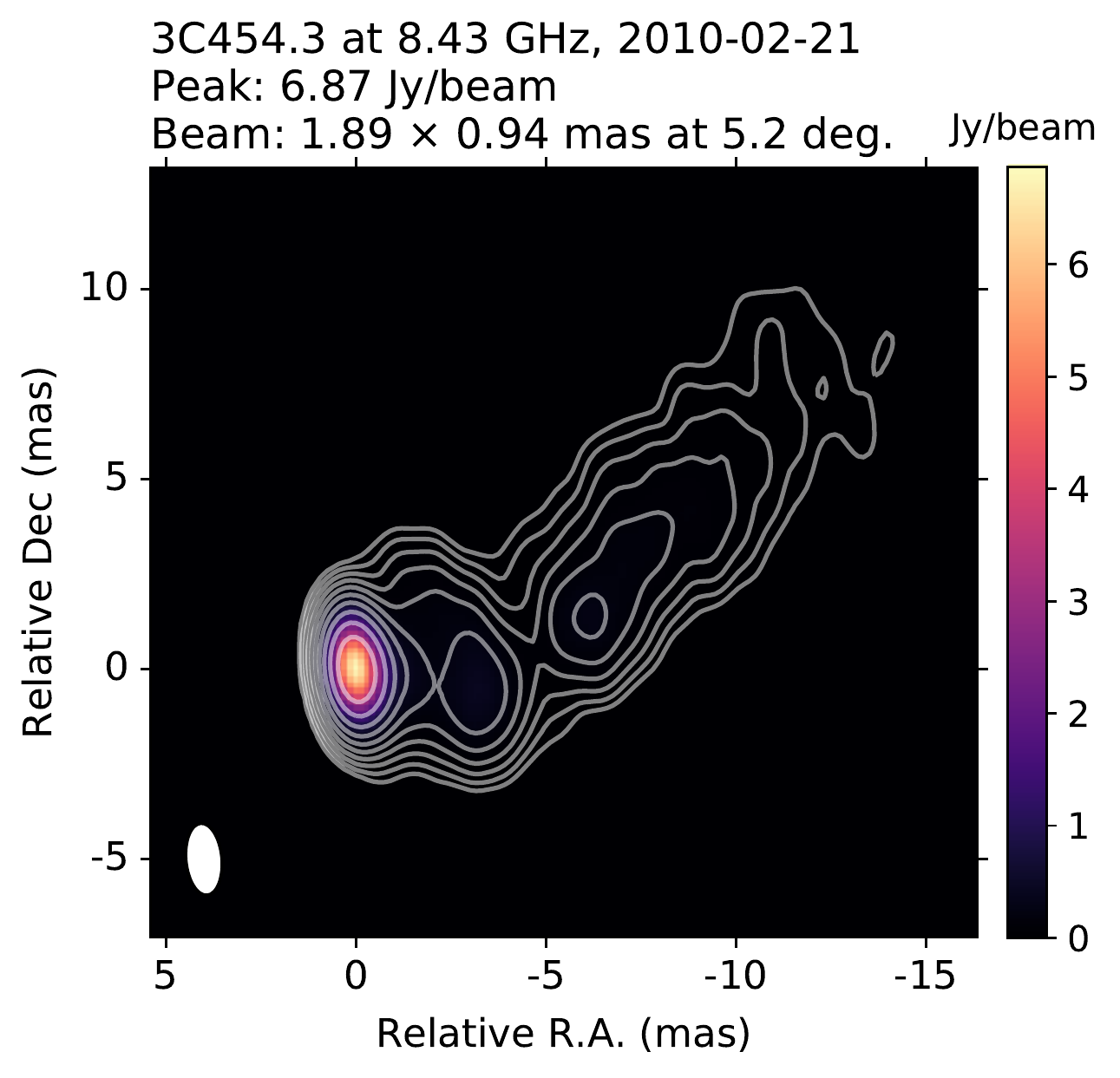}
    }   
     \caption{X-band (8\,GHz) CLEAN images of 3C454.3 from 2008-01-03 until 2010-02-21 with contours at -0.1\%, 0.1\%, 0.2\%, 0.4\%, 0.8\%, 1.6\%, 3.2\%, 6.4\%, 12.8\%, 25.6\%, and 51.2\% of the peak intensity at each image. }
    \label{Xbandimagesp2}
\end{figure*}

\begin{figure*}[]
\centering
    \subfigure[]
    {
         \includegraphics[width=0.3\textwidth]{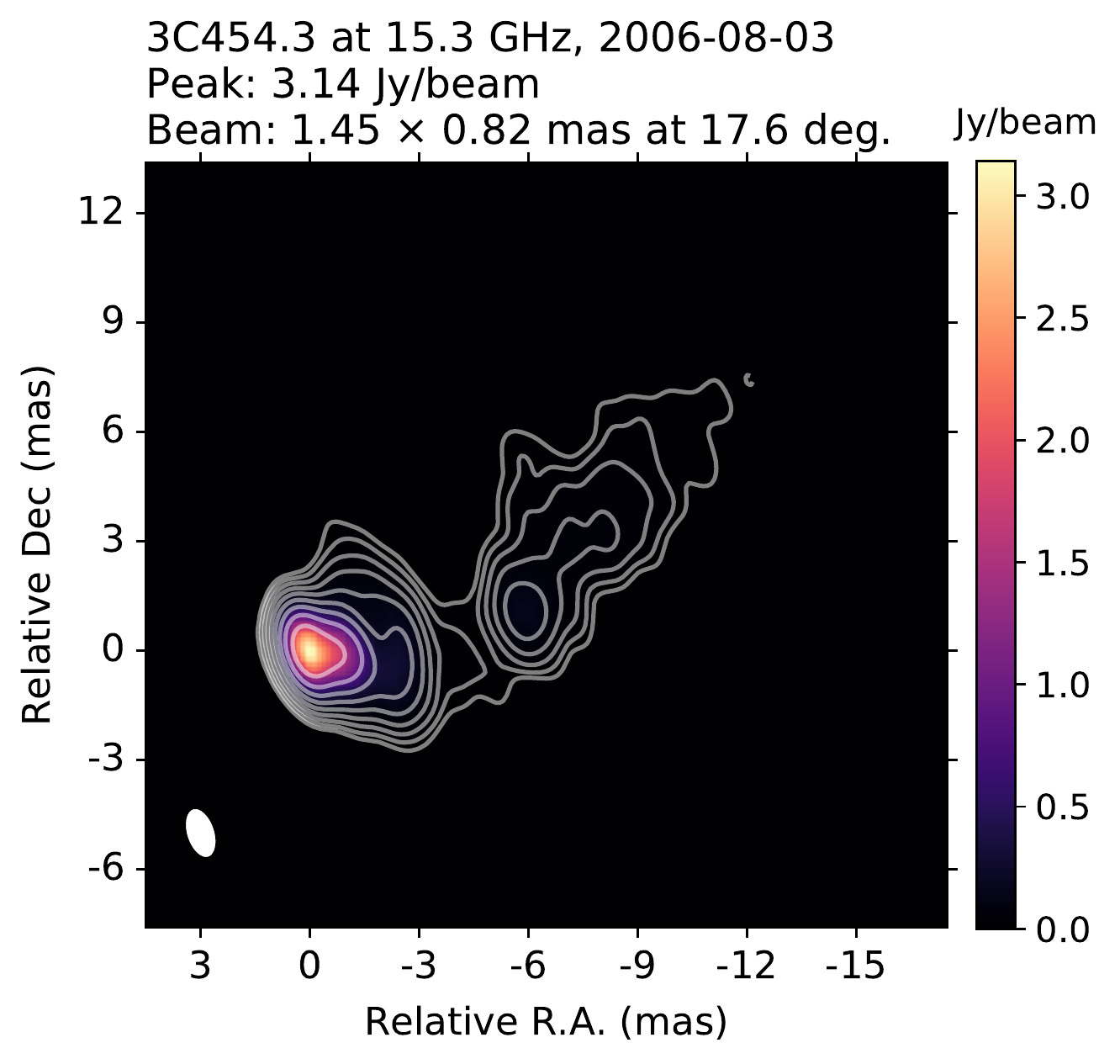}
    }
    \subfigure[]
    {
        \includegraphics[width=0.3\textwidth]{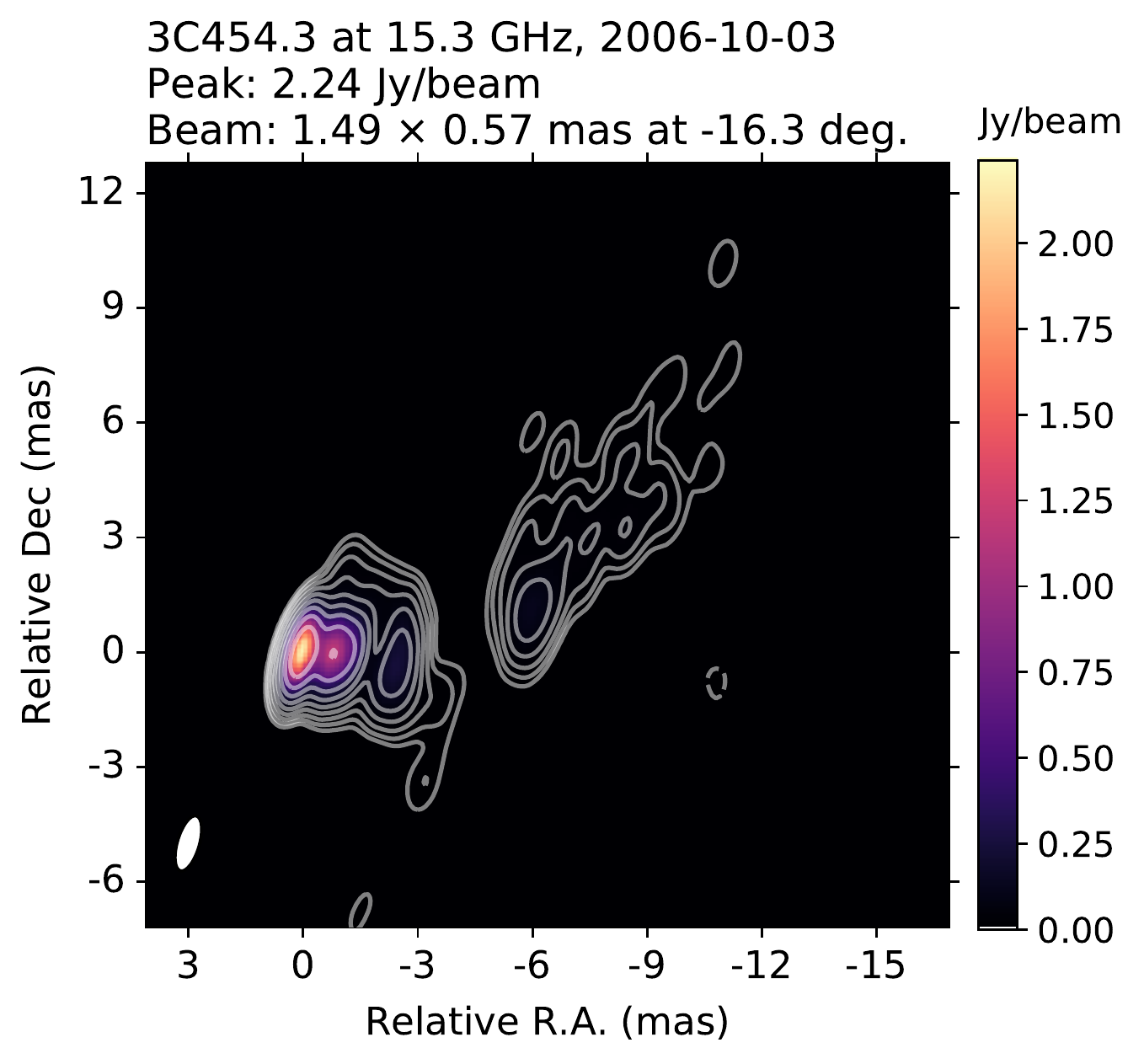}
    }
    \subfigure[]
    {
         \includegraphics[width=0.3\textwidth]{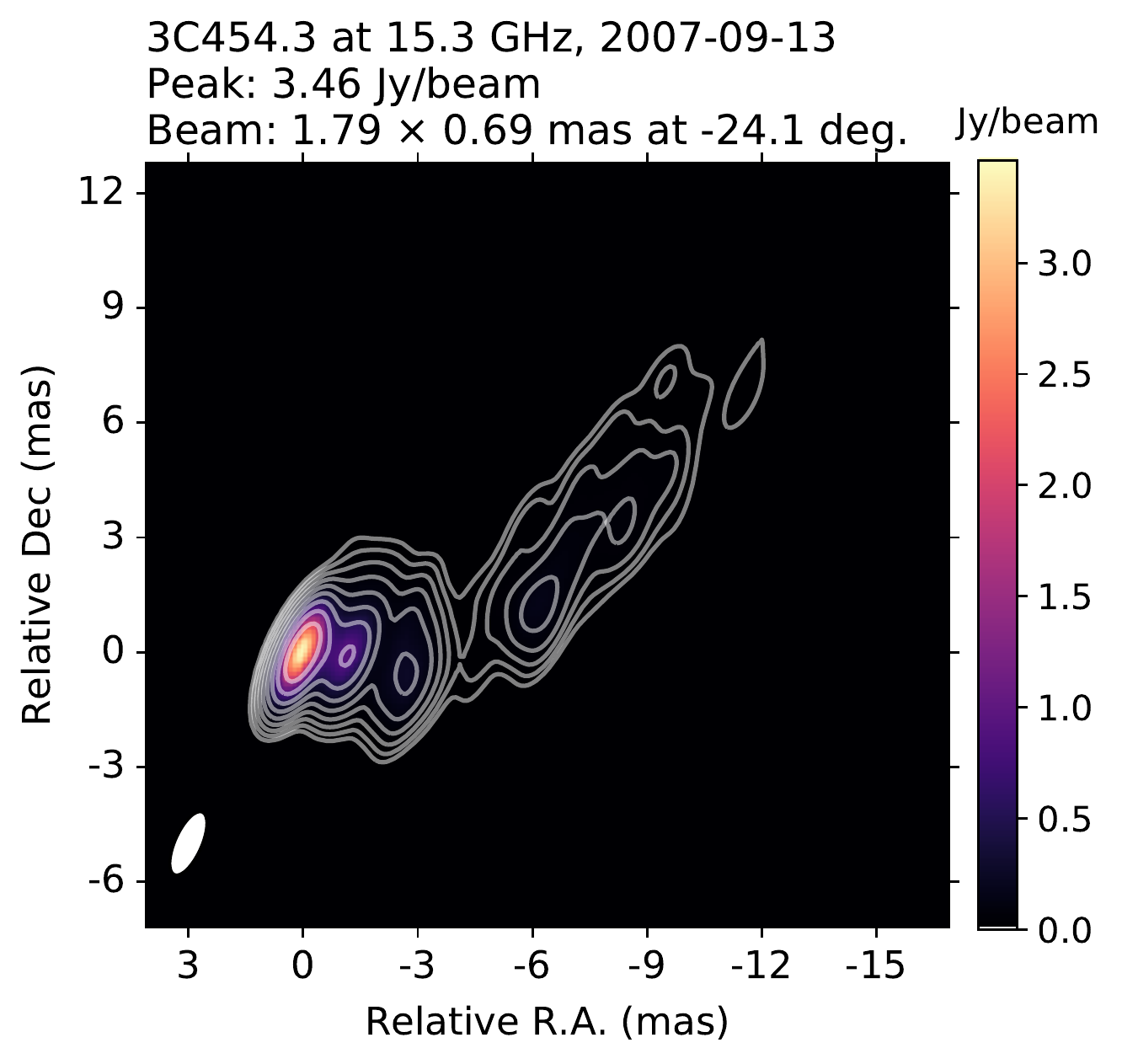}
    }
        \subfigure[]
    {
         \includegraphics[width=0.3\textwidth]{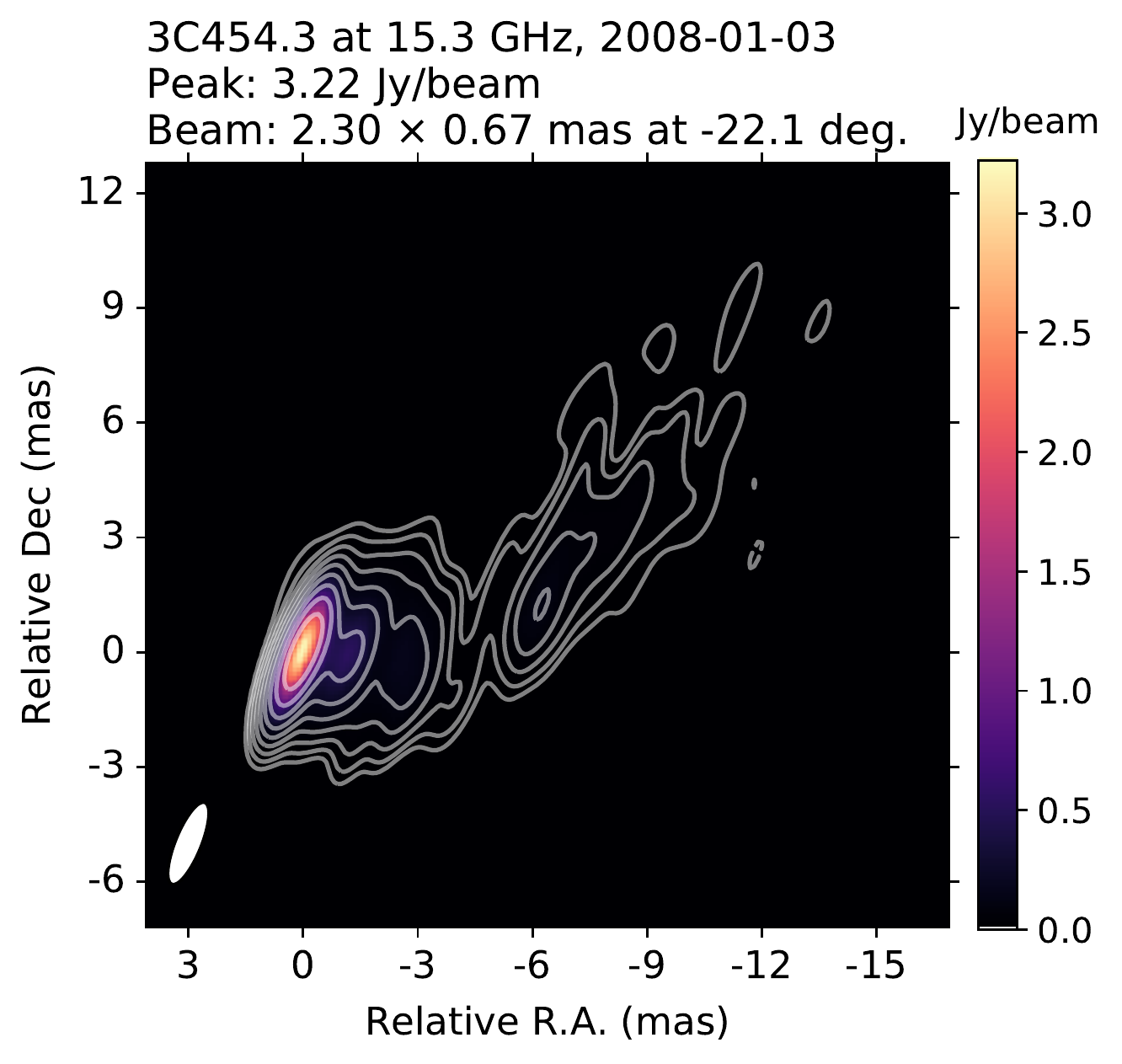}
     }
     \subfigure[]
    {
         \includegraphics[width=0.3\textwidth]{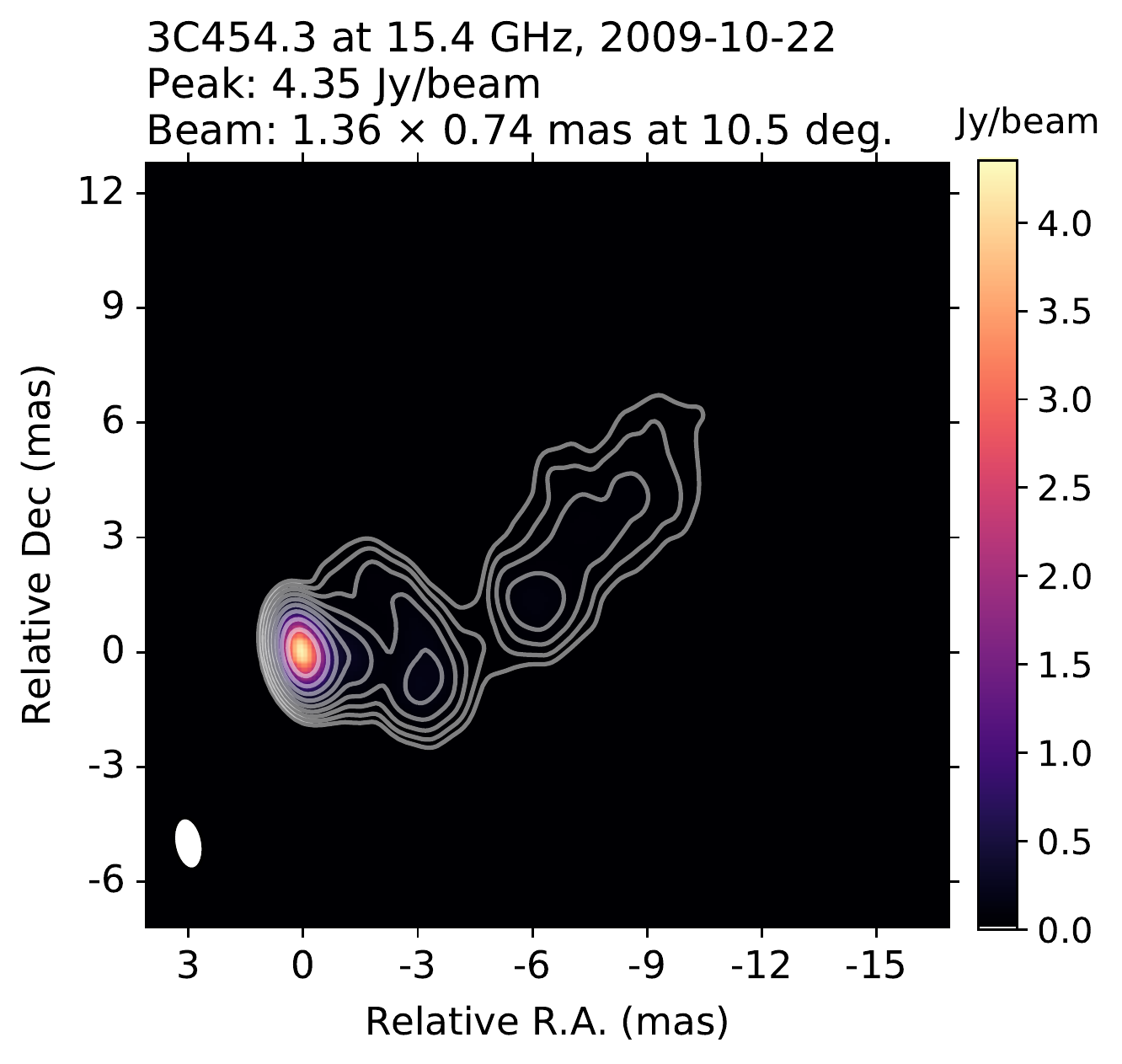}
    }    
    \caption{U-band (15\,GHz) CLEAN images of 3C454.3 from 2006-08-03 until 2010-02-21 with contours at -0.2\%, 0.2\%, 0.4\%, 0.8\%, 1.6\%, 3.2\%, 6.4\%, 12.8\%, 25.6\%, and 51.2\% of the peak intensity at each image. The missing images for the other epochs are already published by the MOJAVE team.}
    \label{Ubandimages}
\end{figure*}

\begin{figure*}[]
\centering
   \subfigure[]
    {
        \includegraphics[width=0.3\textwidth]{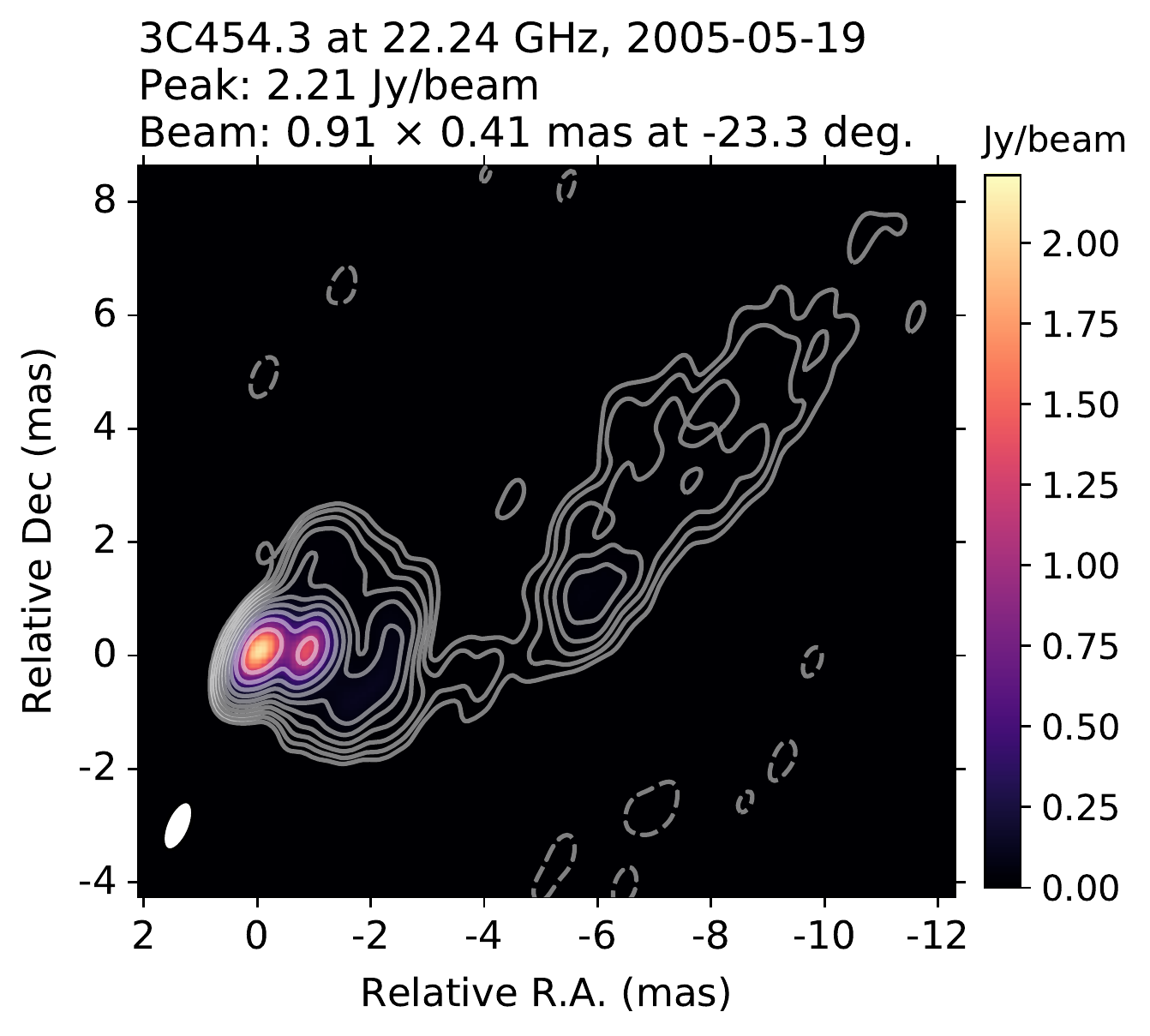}
    }
       \subfigure[]
    {
        \includegraphics[width=0.3\textwidth]{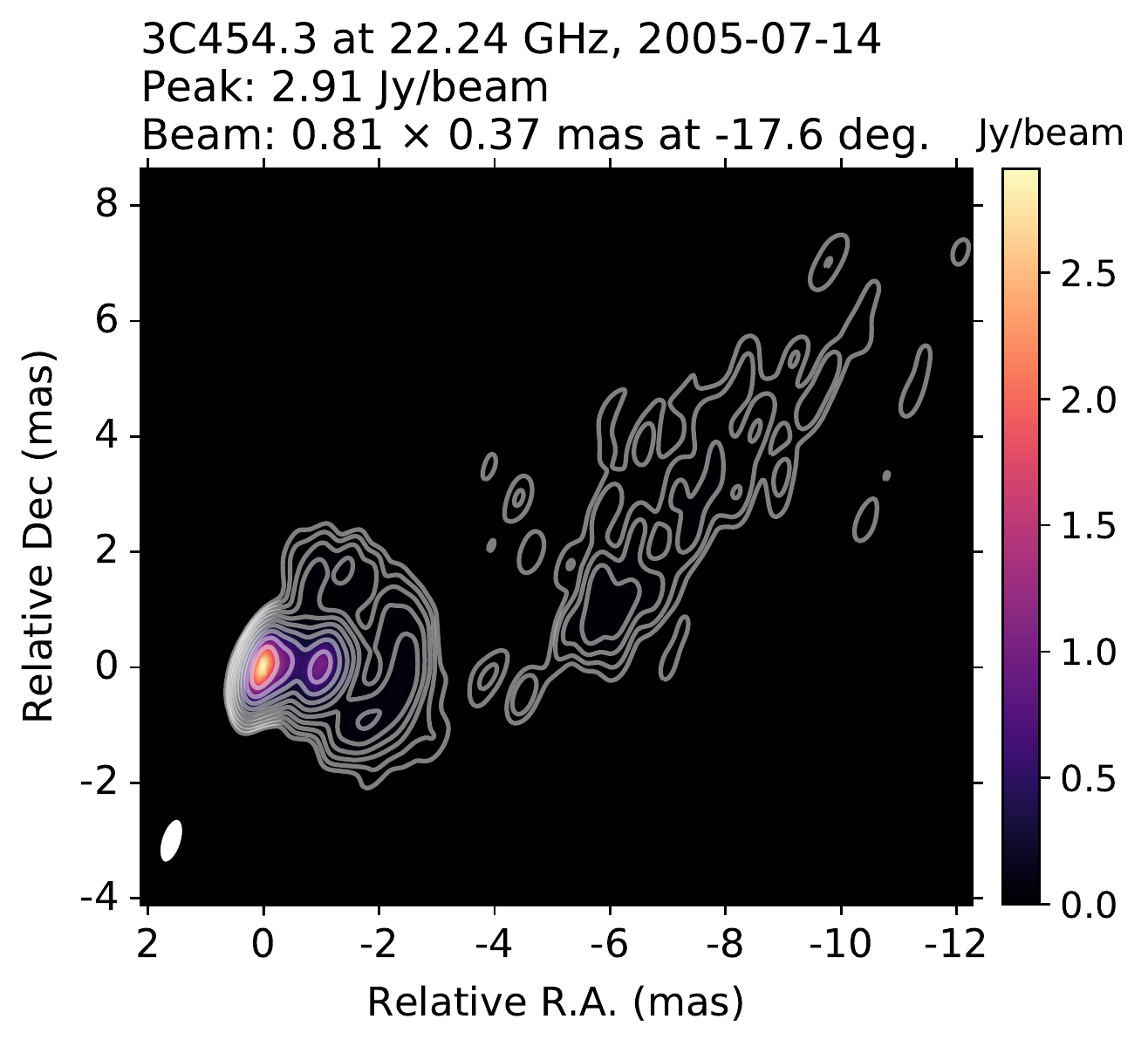}
    }
           \subfigure[]
    {
        \includegraphics[width=0.3\textwidth]{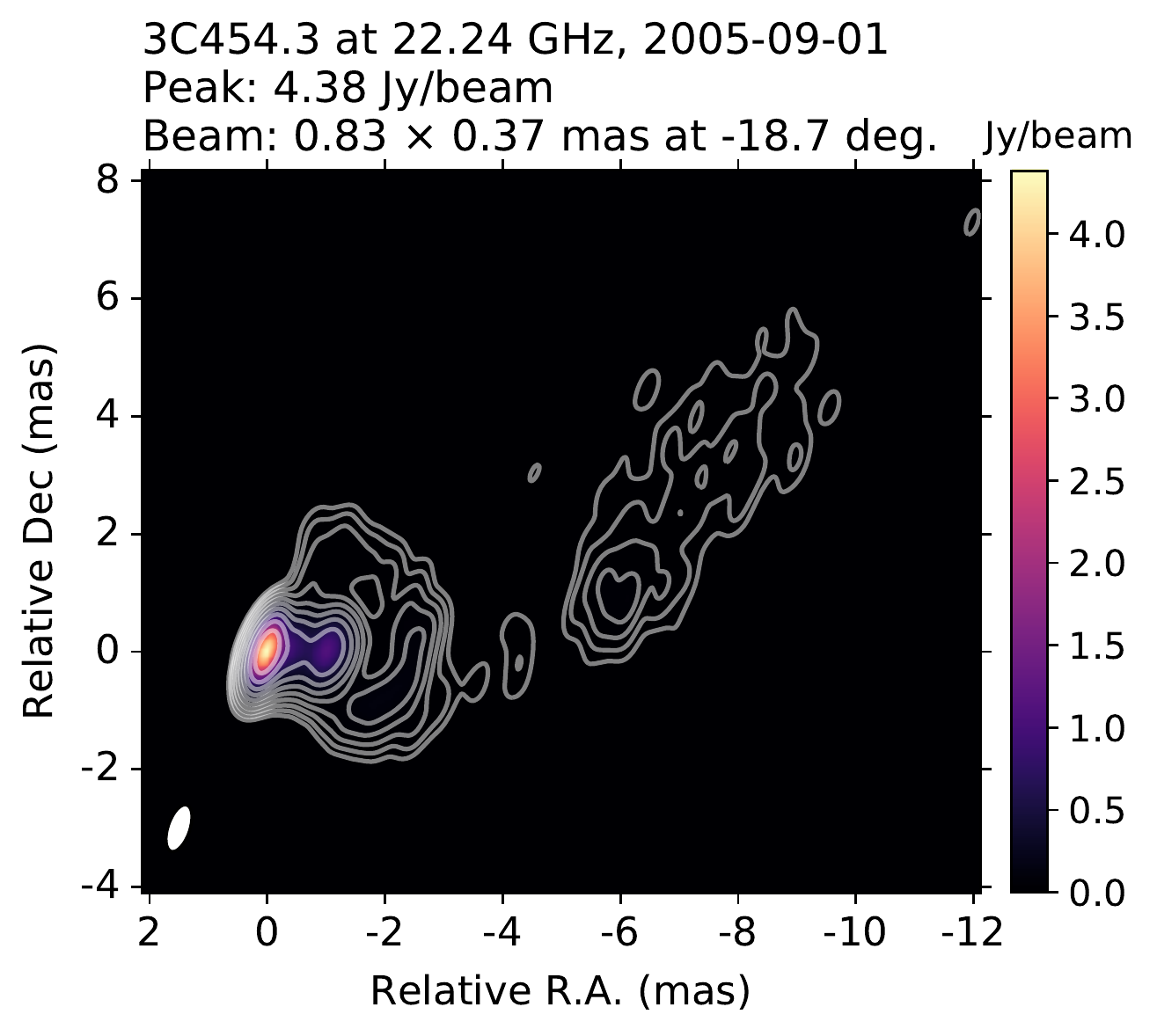}
    }
     \subfigure[]
    {
         \includegraphics[width=0.3\textwidth]{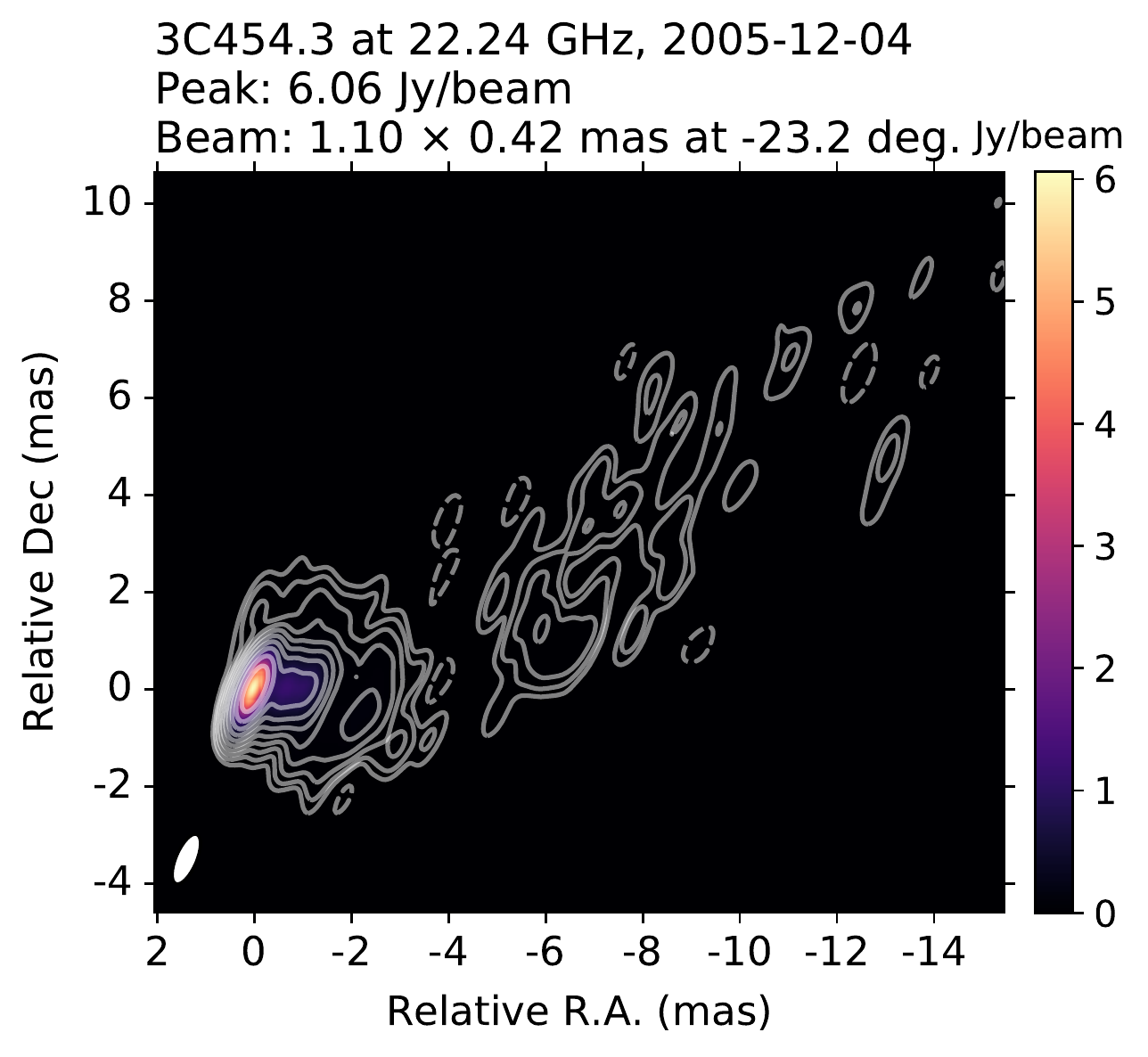}
    }
     \subfigure[]
    {
        \includegraphics[width=0.3\textwidth]{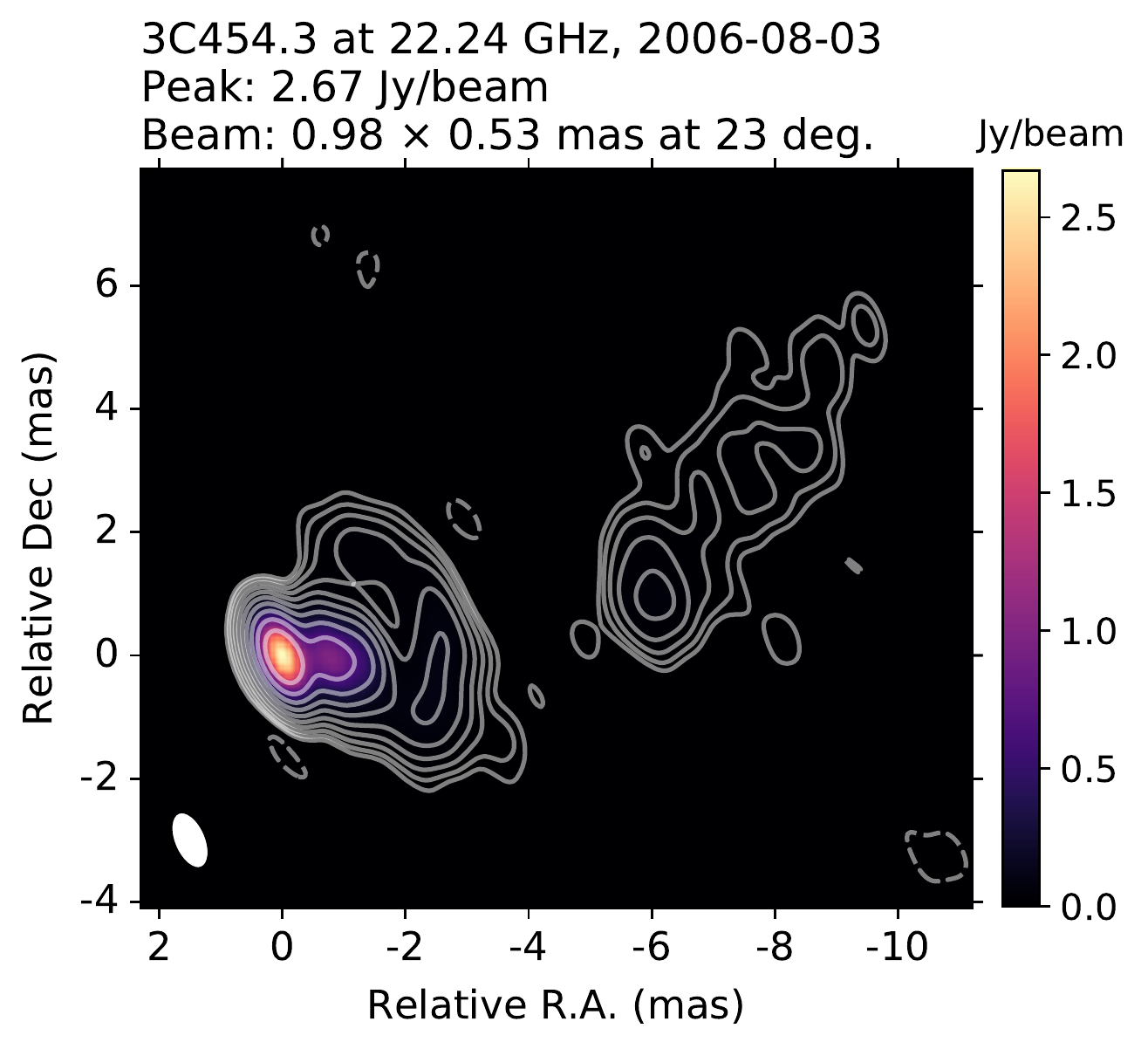}
    }   
       \subfigure[]
    {
         \includegraphics[width=0.3\textwidth]{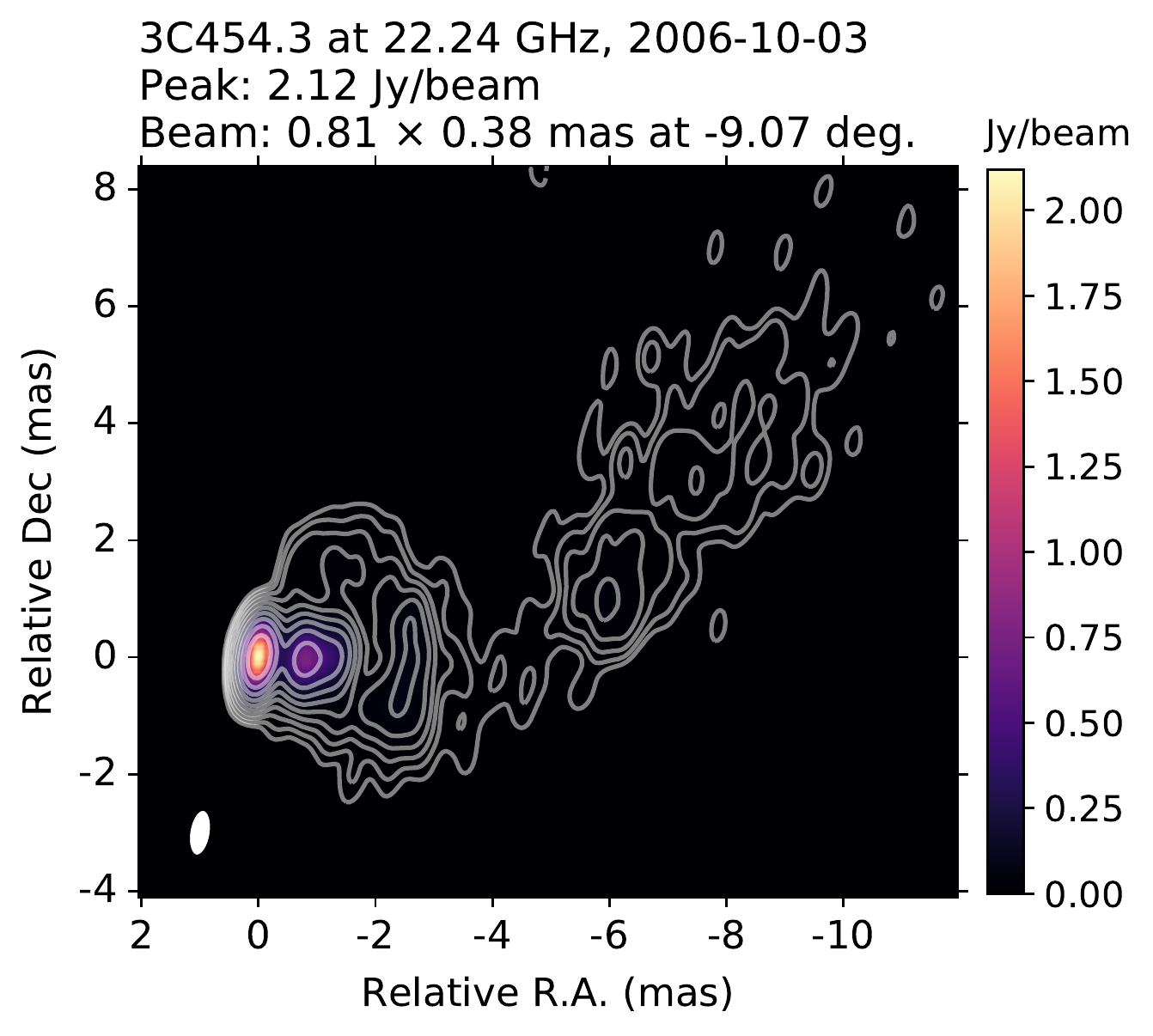}
        
    }
       \subfigure[]
    {
         \includegraphics[width=0.3\textwidth]{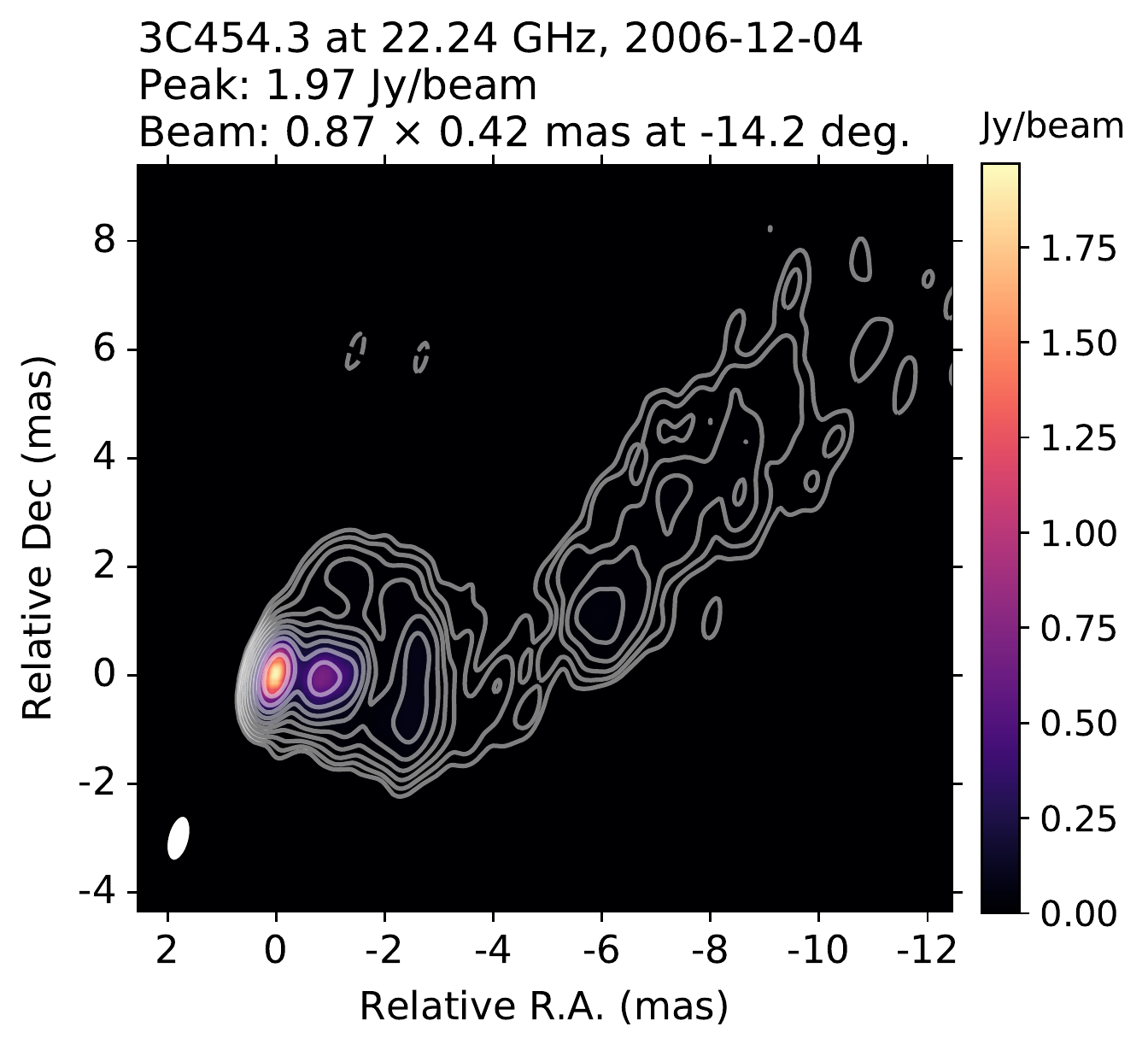}
    }
    \subfigure[]
    {
        \includegraphics[width=0.3\textwidth]{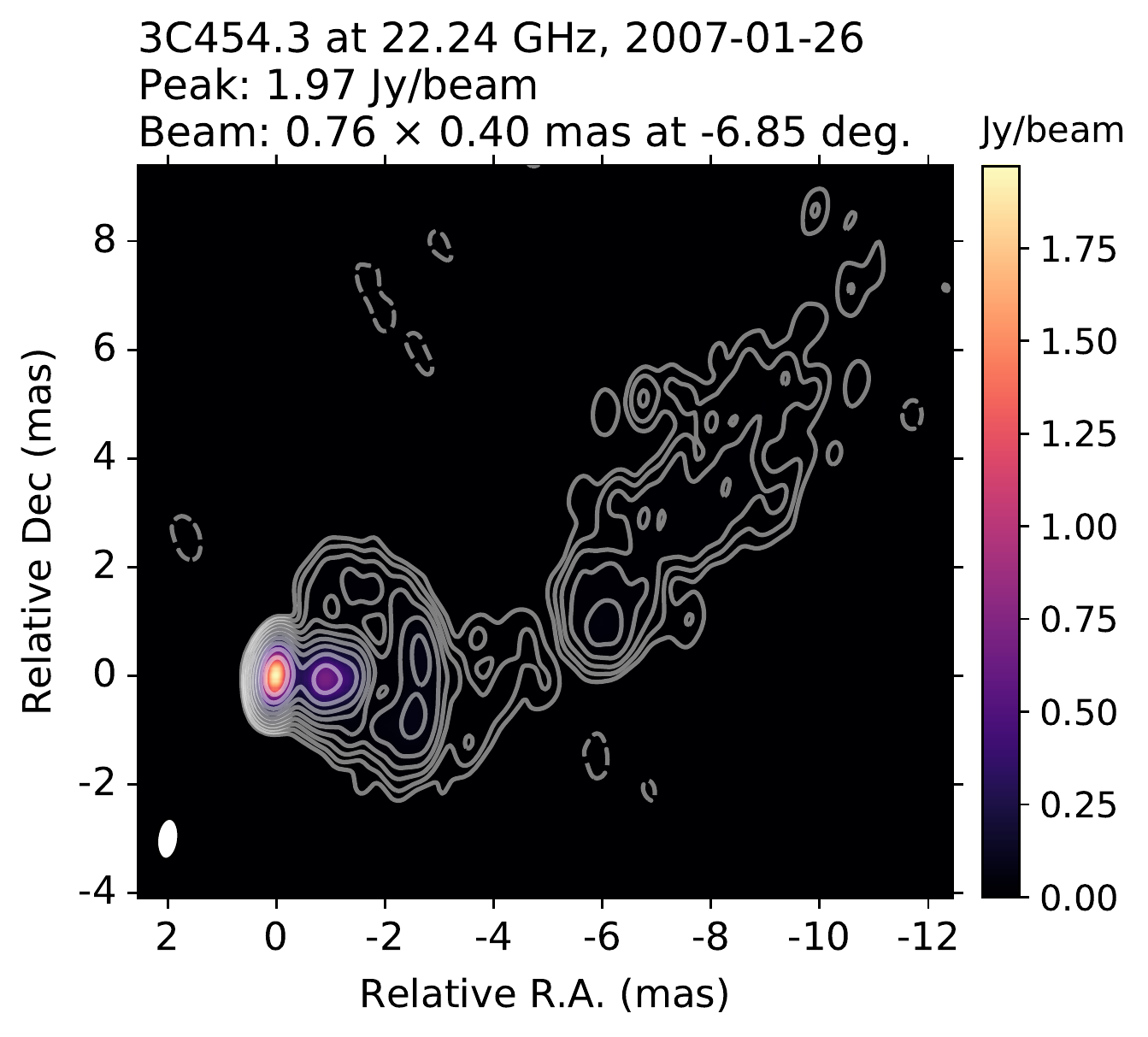}
    }
    \subfigure[]
    {
         \includegraphics[width=0.3\textwidth]{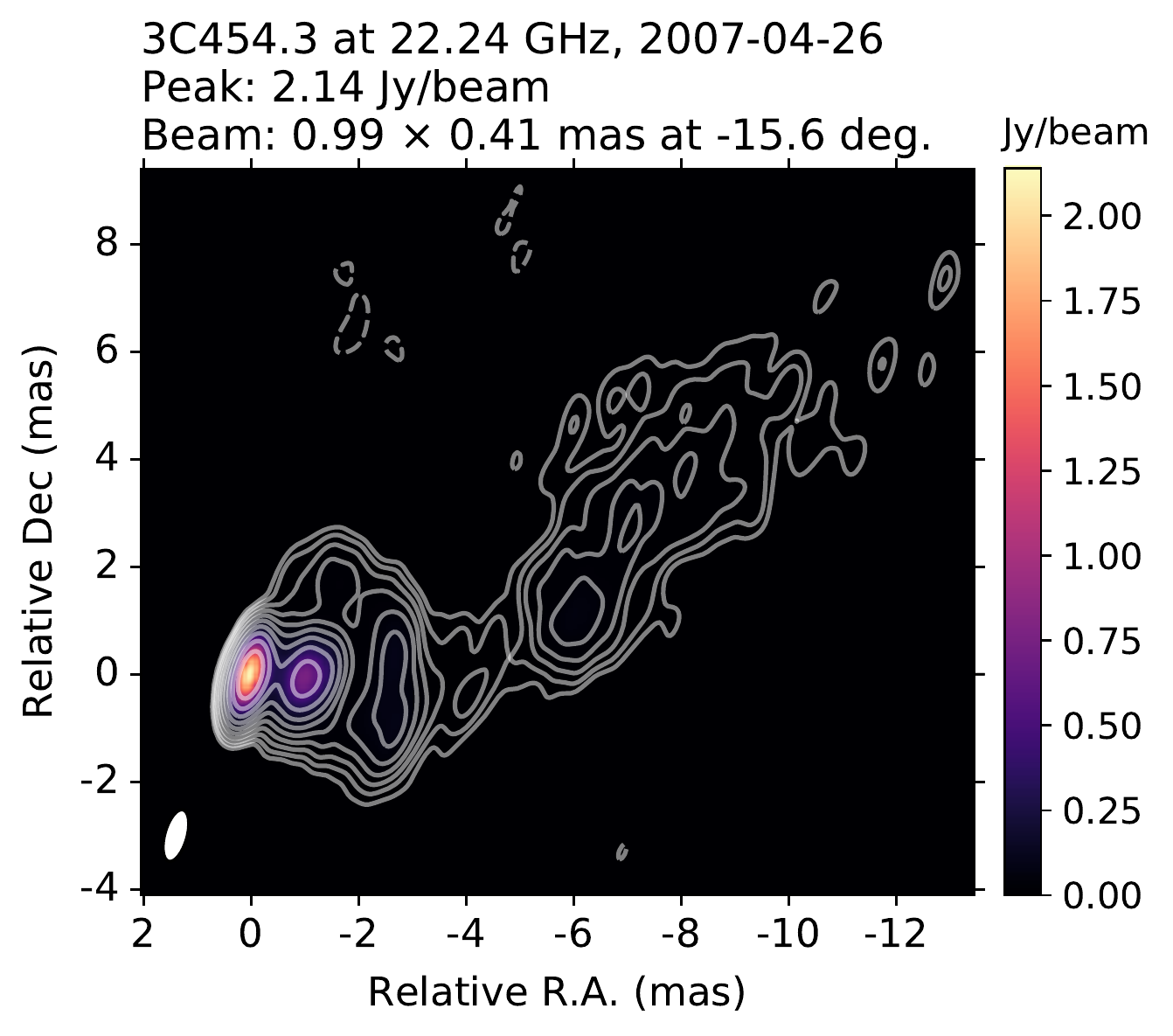}
    }
        \subfigure[]
    {
         \includegraphics[width=0.3\textwidth]{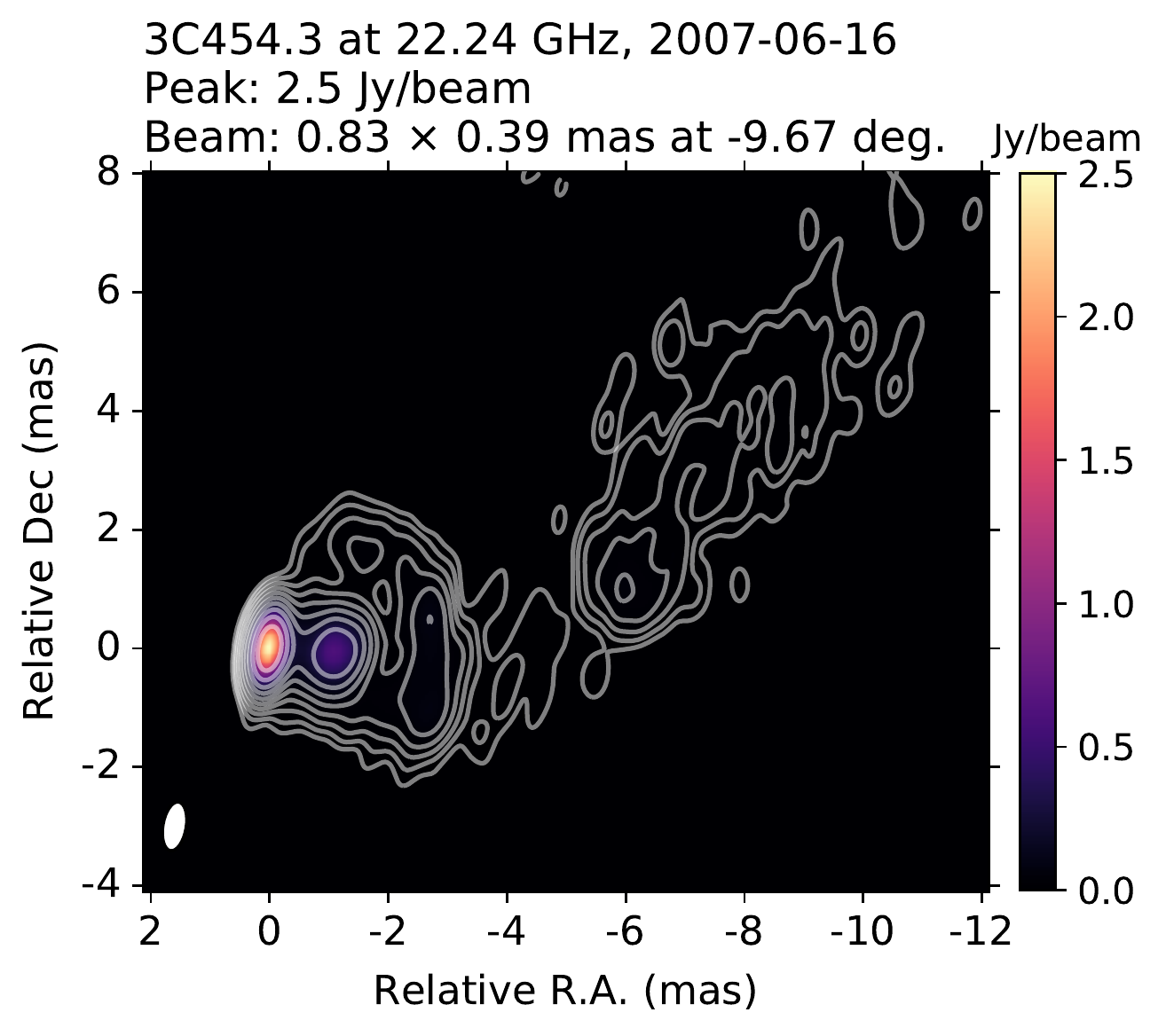}
    }
     \subfigure[]
    {
         \includegraphics[width=0.3\textwidth]{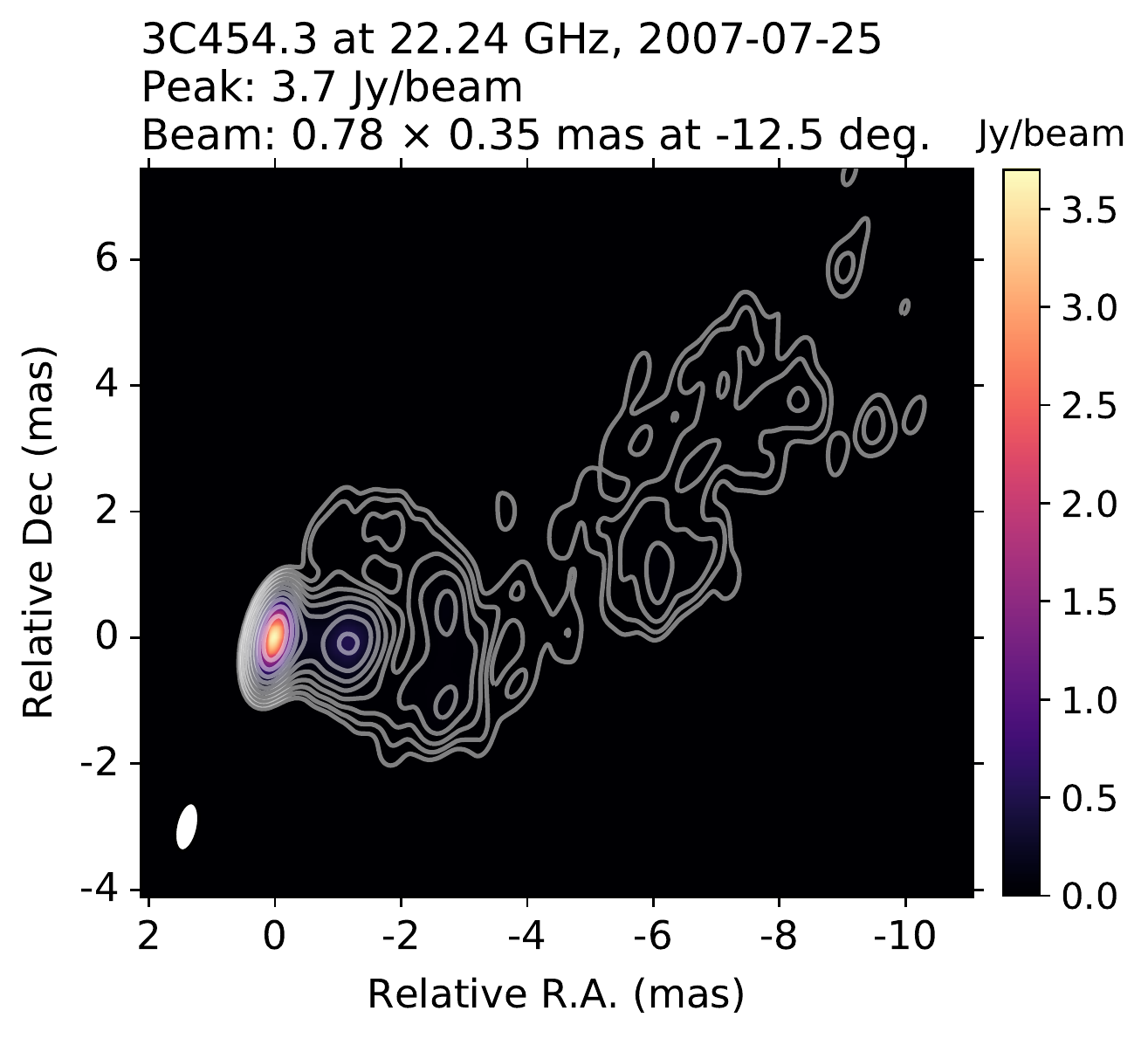}
    }    
         \subfigure[]
    {
        \includegraphics[width=0.3\textwidth]{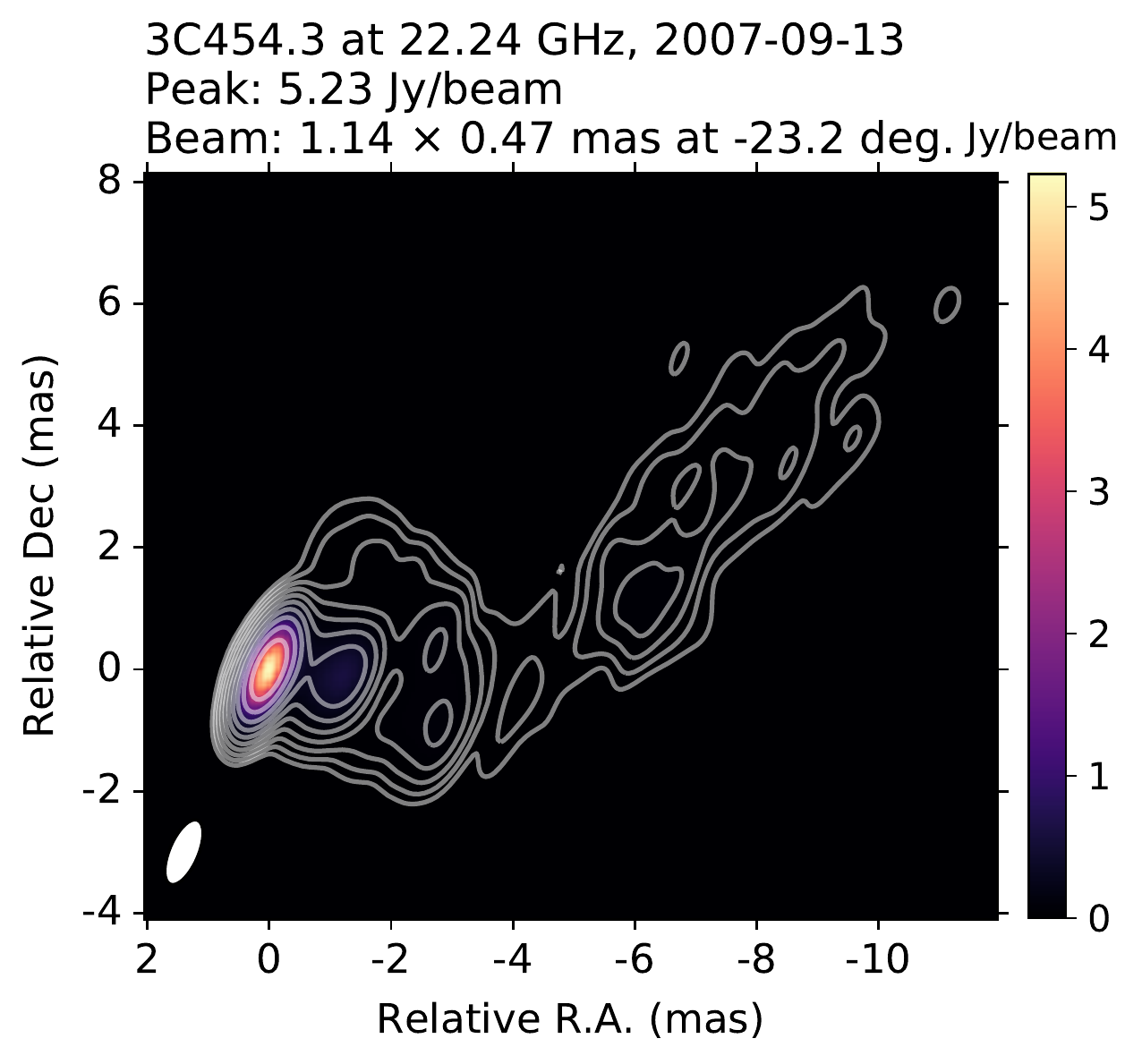}
    }   
     \caption{K-band (22$-$24\,GHz) CLEAN images of 3C454.3 from 2005-05-19 until 2007-09-13 with contours at -0.1\%,0.1\%, 0.2\%, 0.4\%, 0.8\%, 1.6\%, 3.2\%, 6.4\%, 12.8\%, 25.6\%, and 51.2\% of the peak intensity at each image. }
    \label{Kbandimagesp1}
\end{figure*}

\begin{figure*}[]
\centering
    \subfigure[]
    {
         \includegraphics[width=0.3\textwidth]{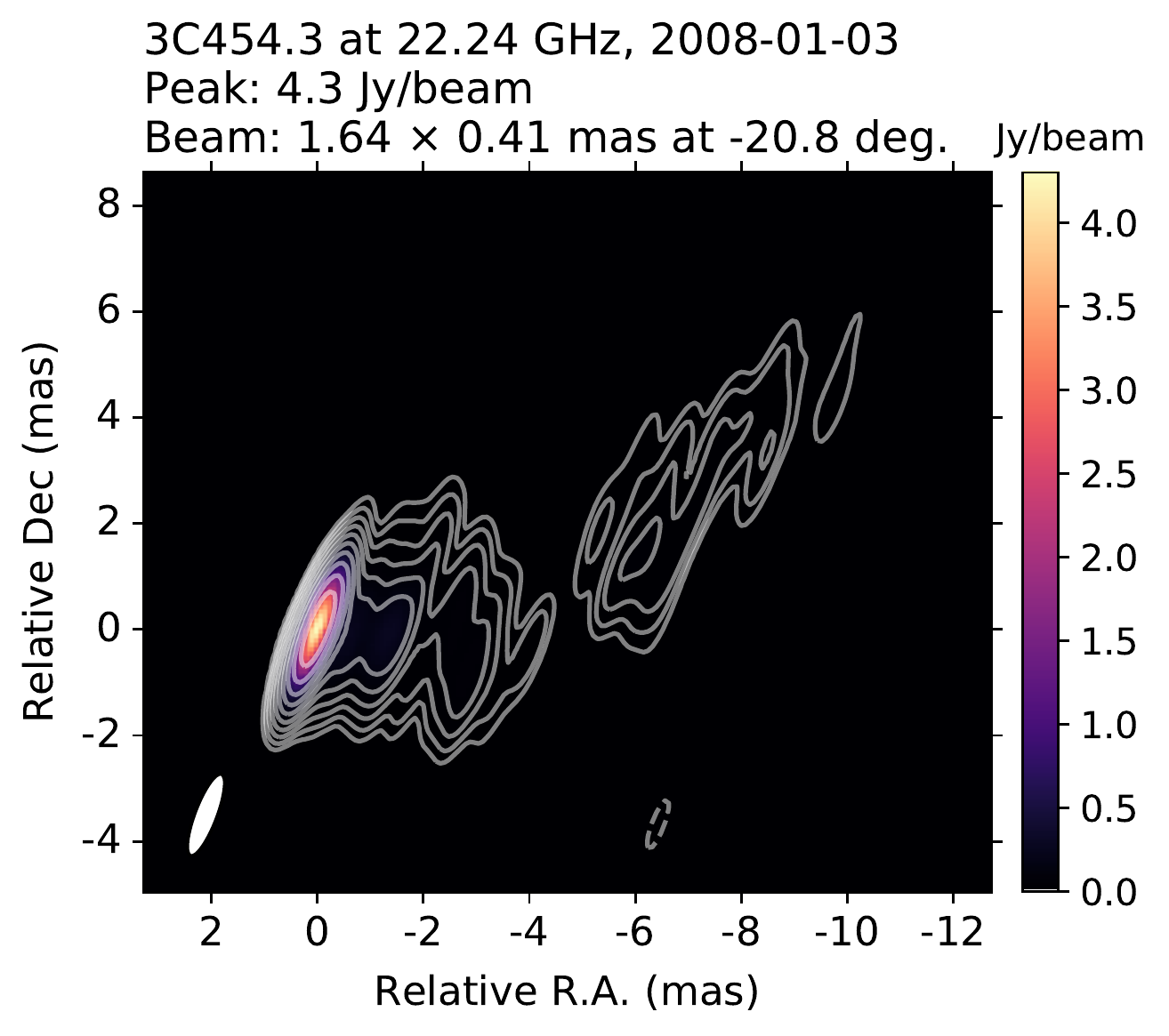}
     }
    \subfigure[]
    {
        \includegraphics[width=0.3\textwidth]{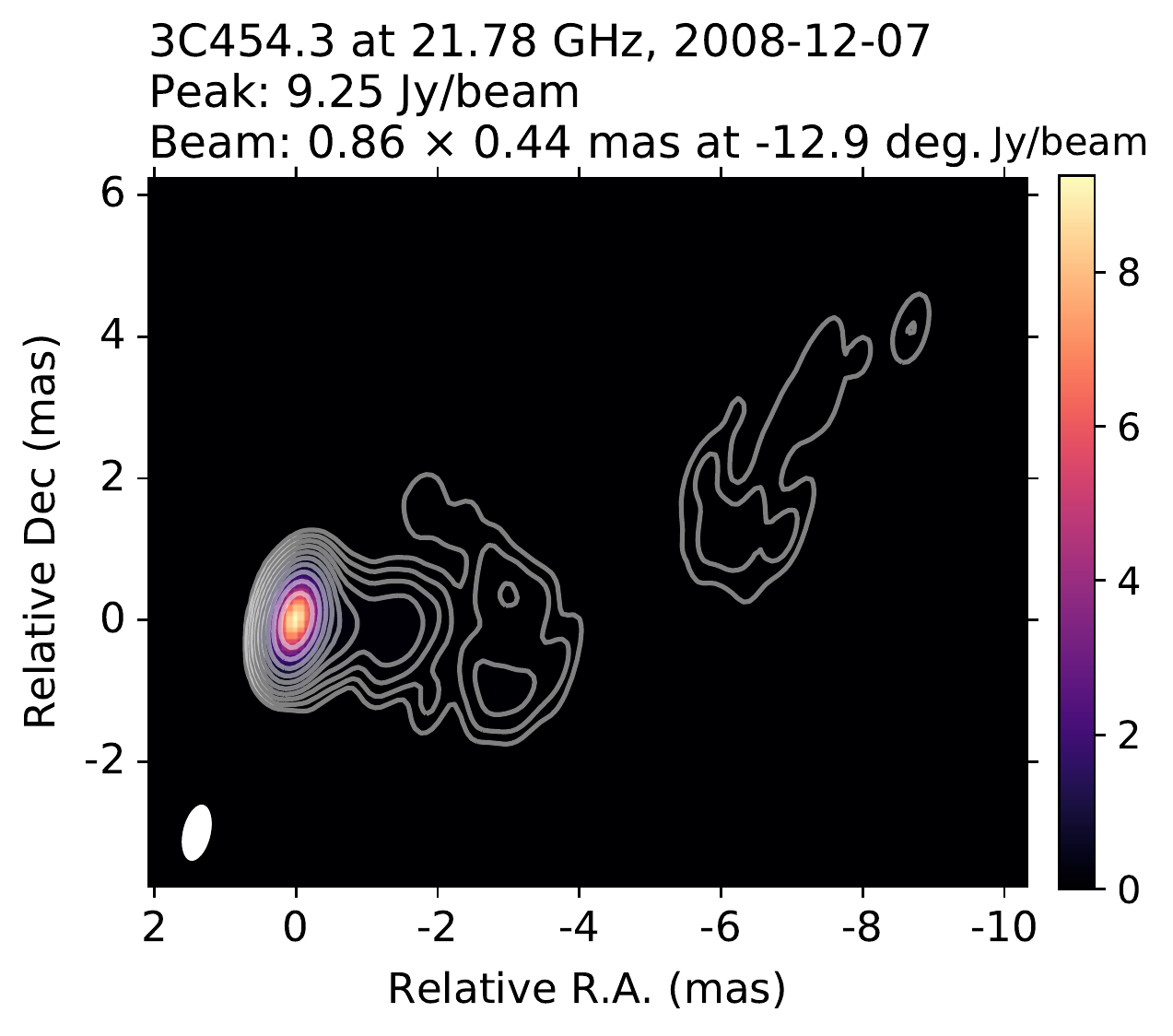}
    }
    \subfigure[]
    {
         \includegraphics[width=0.3\textwidth]{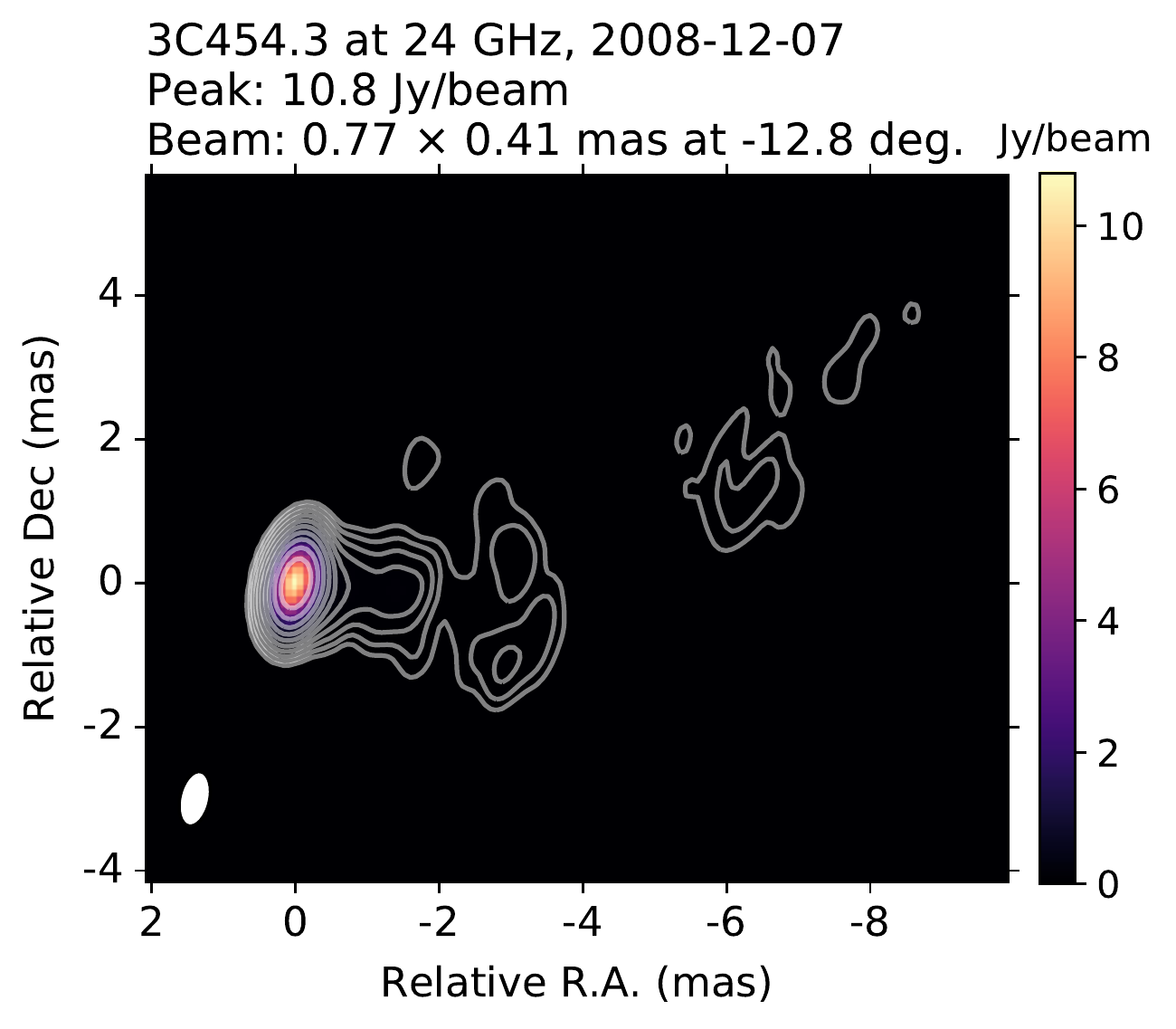}
    }
        \subfigure[]
    {
         \includegraphics[width=0.3\textwidth]{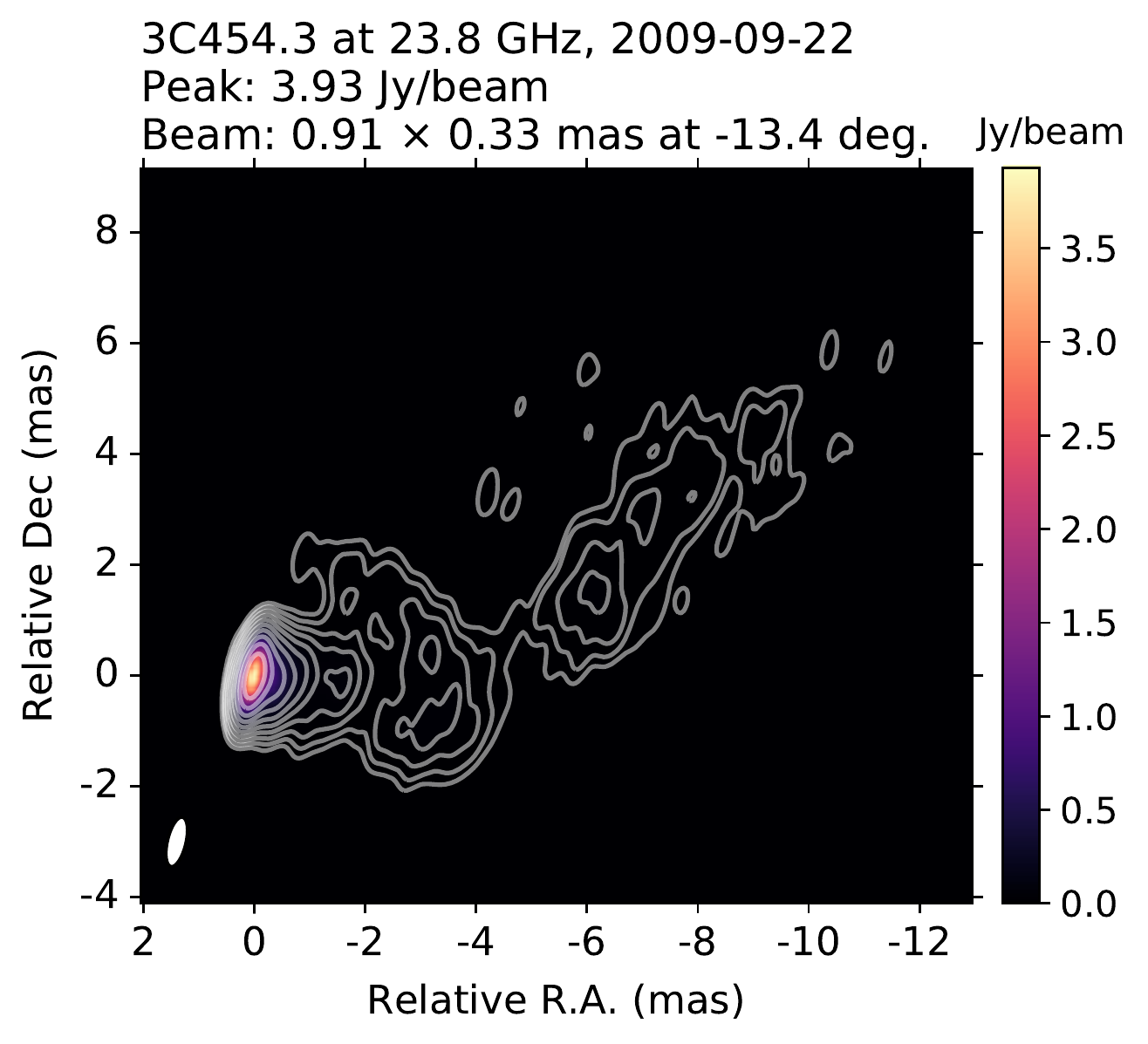}
    }
     \subfigure[]
    {
         \includegraphics[width=0.3\textwidth]{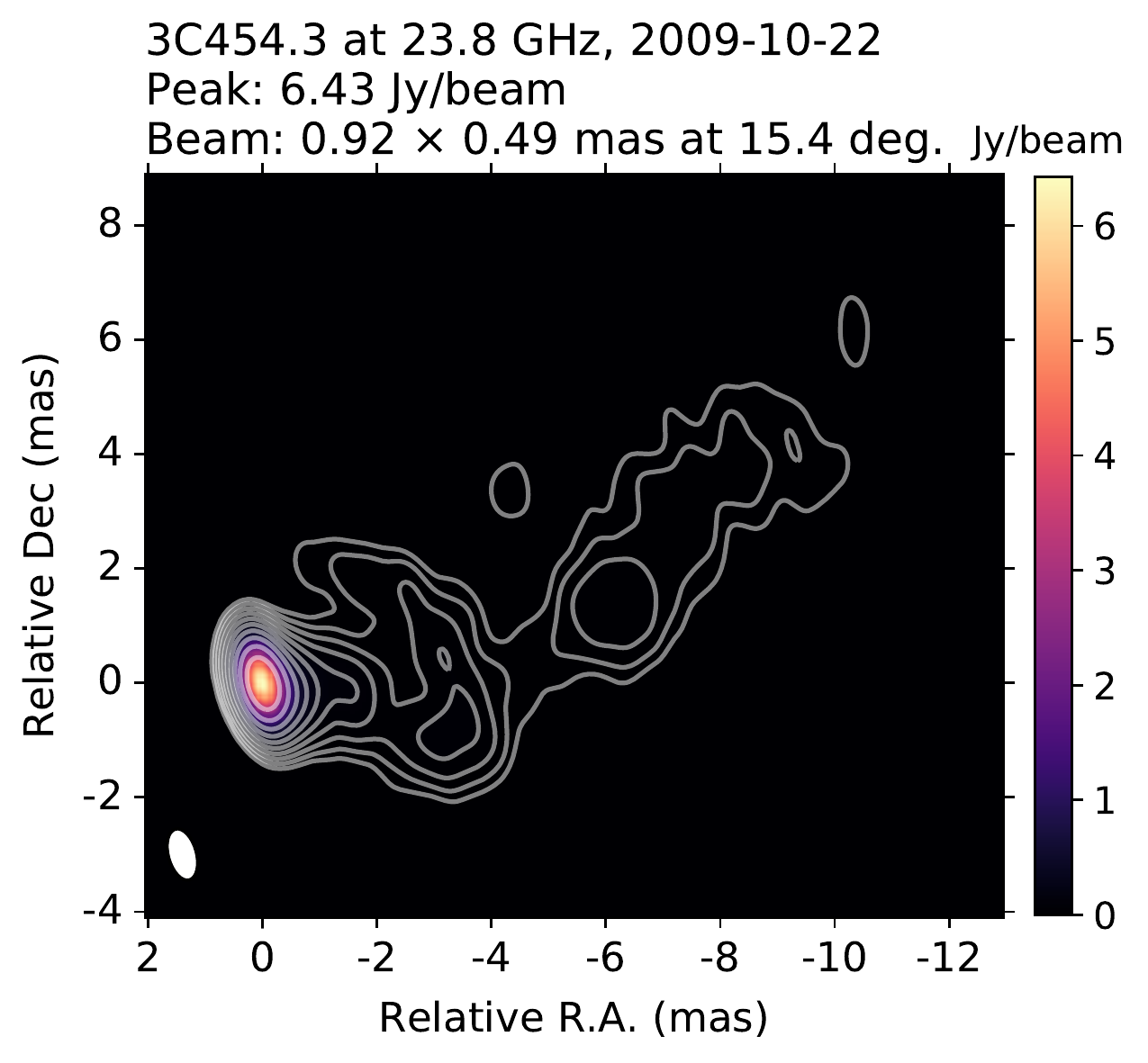}
    }    
         \subfigure[]
    {
        \includegraphics[width=0.3\textwidth]{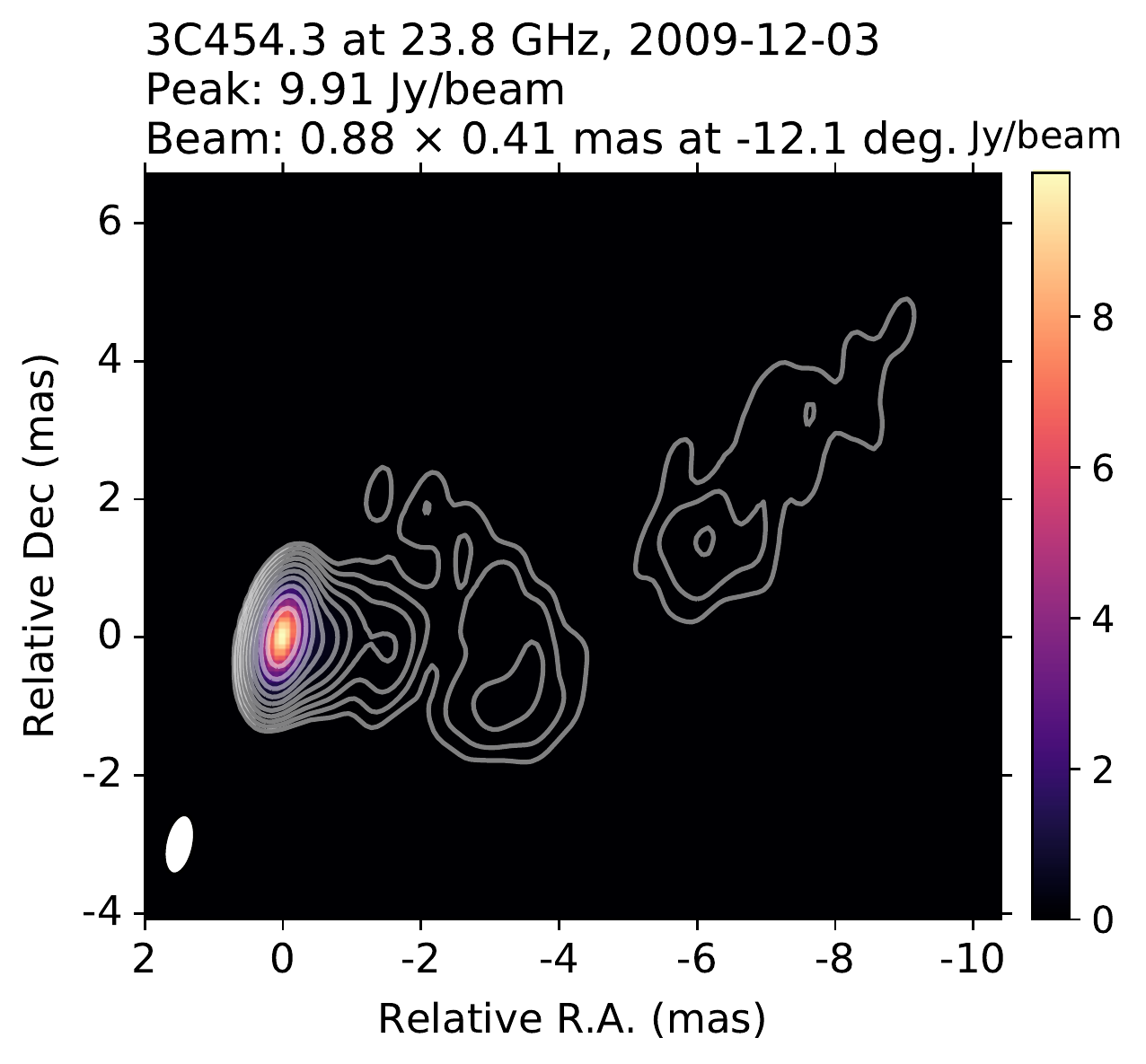}
    }   
         \subfigure[]
    {
        \includegraphics[width=0.3\textwidth]{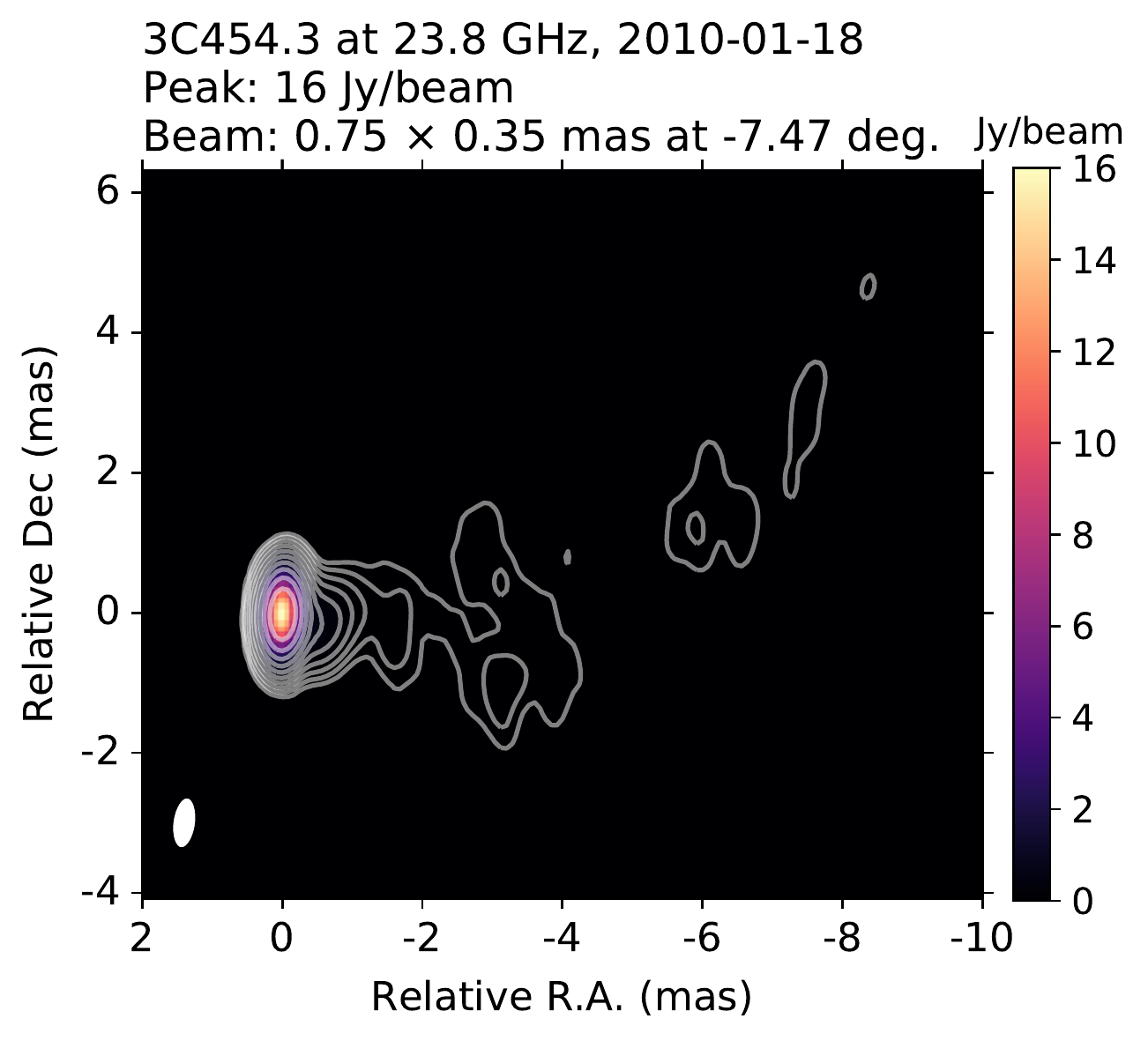}
    }   
       \subfigure[]
    {
        \includegraphics[width=0.3\textwidth]{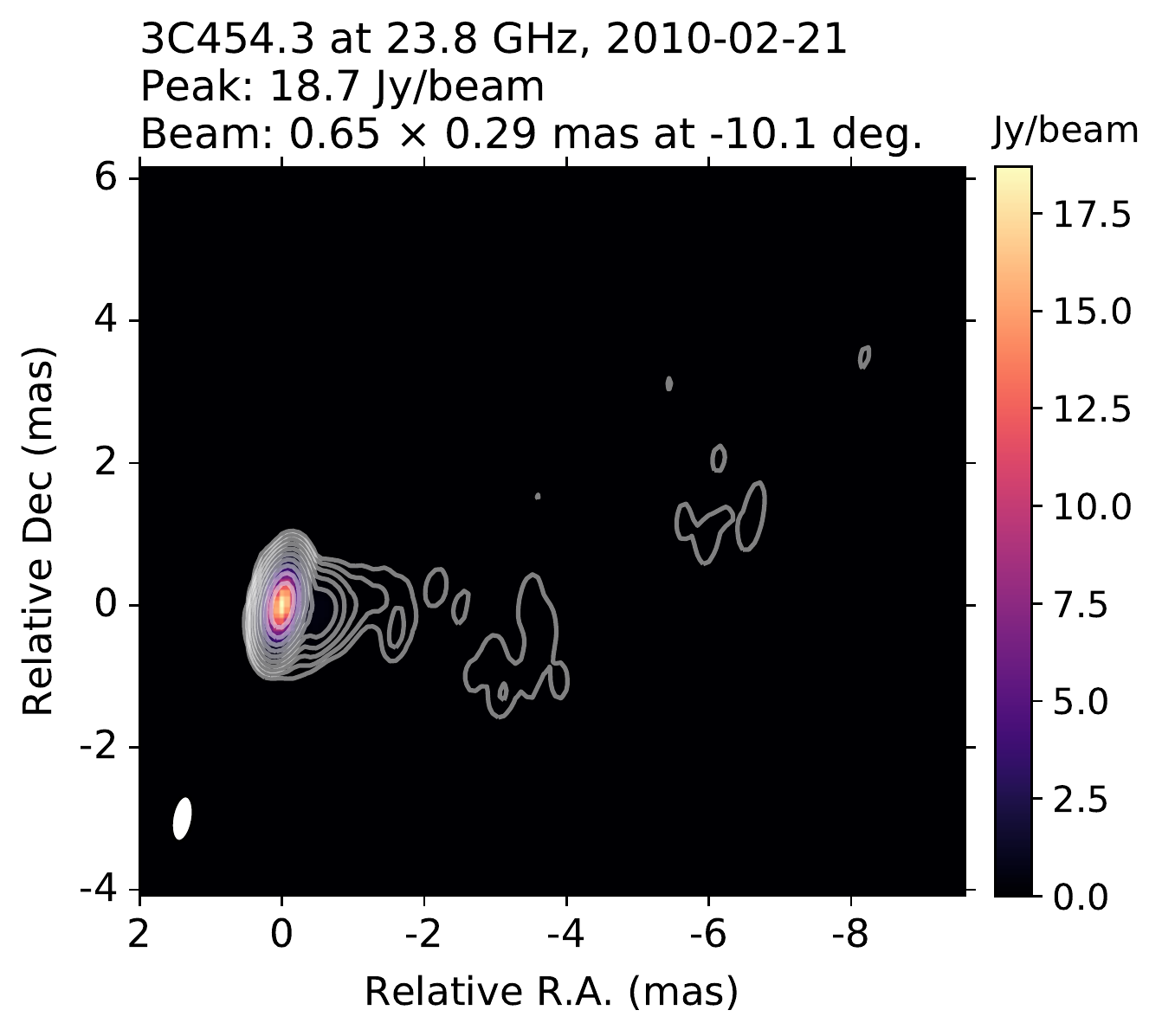}
    }   
     \caption{K-band (22$-$24\,GHz) CLEAN images of 3C454.3 from 2008-01-03 until 2010-02-21 with contours at -0.1\%, 0.1\%, 0.2\%, 0.4\%, 0.8\%, 1.6\%, 3.2\%, 6.4\%, 12.8\%, 25.6\%, and 51.2\% of the peak intensity at each image. }
    \label{Kbandimagesp2}
\end{figure*}

\begin{figure*}[]
\centering
   \subfigure[]
    {
        \includegraphics[width=0.3\textwidth]{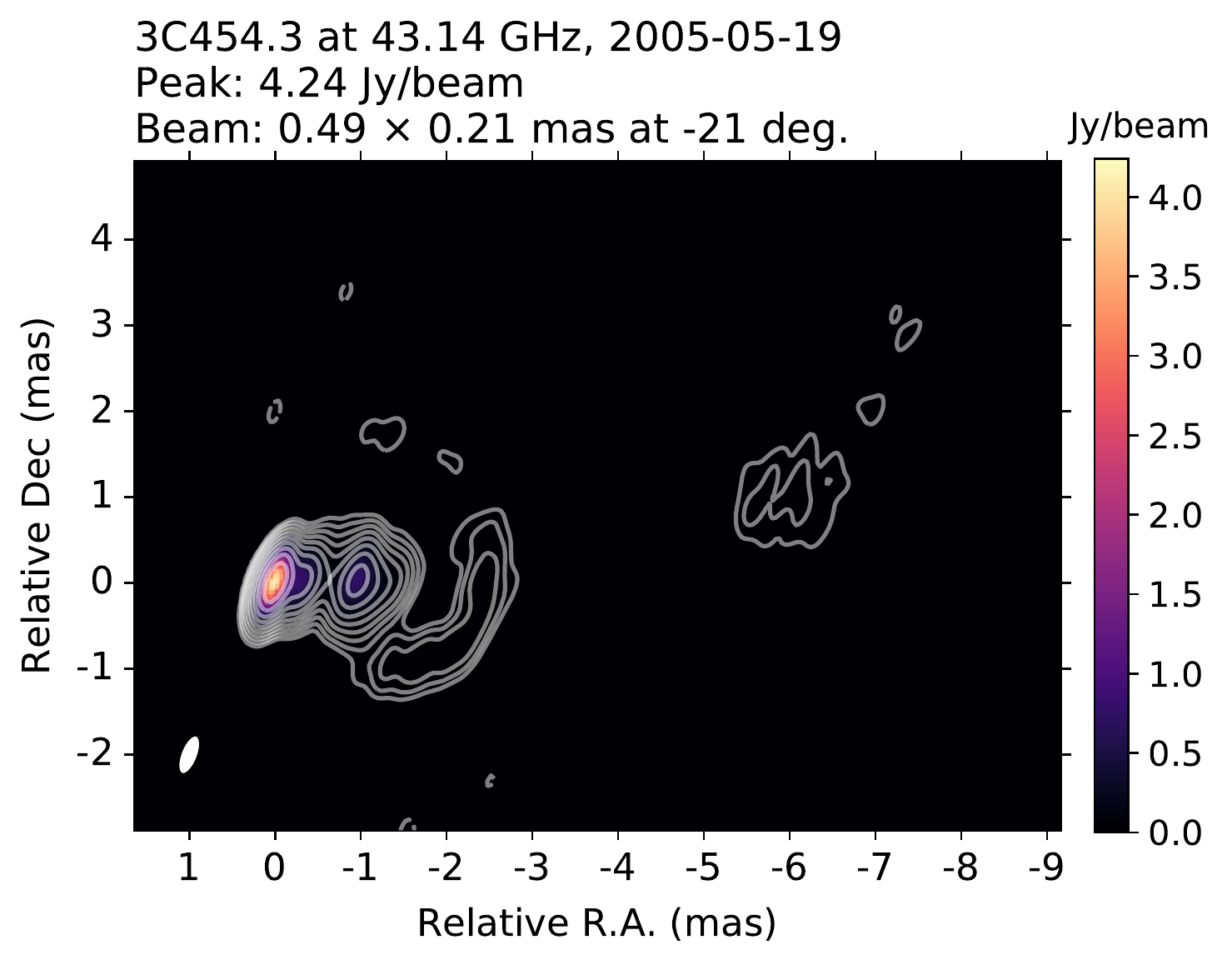}
    }
       \subfigure[]
    {
        \includegraphics[width=0.3\textwidth]{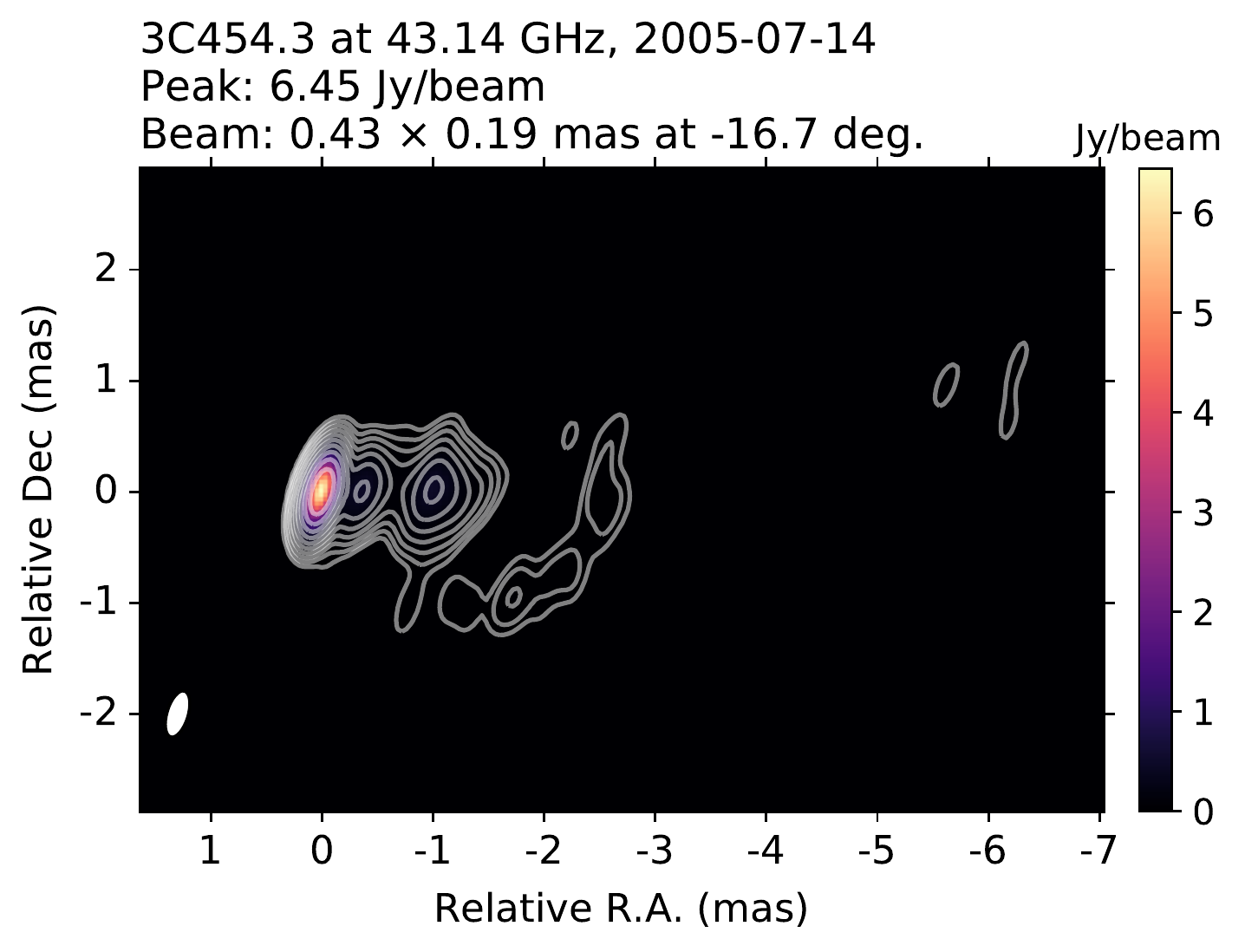}
    }
           \subfigure[]
    {
        \includegraphics[width=0.3\textwidth]{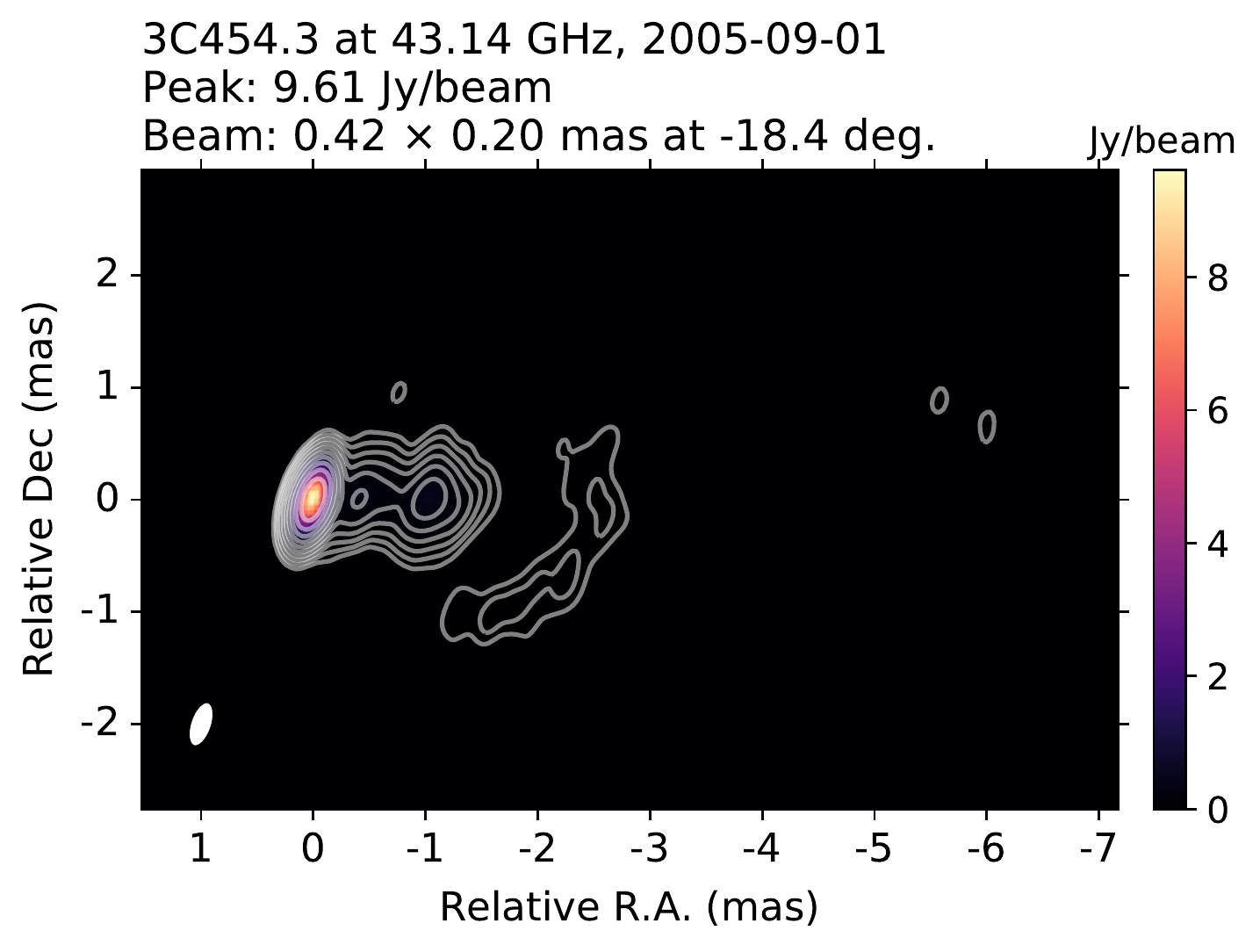}
    }
     \subfigure[]
    {
         \includegraphics[width=0.3\textwidth]{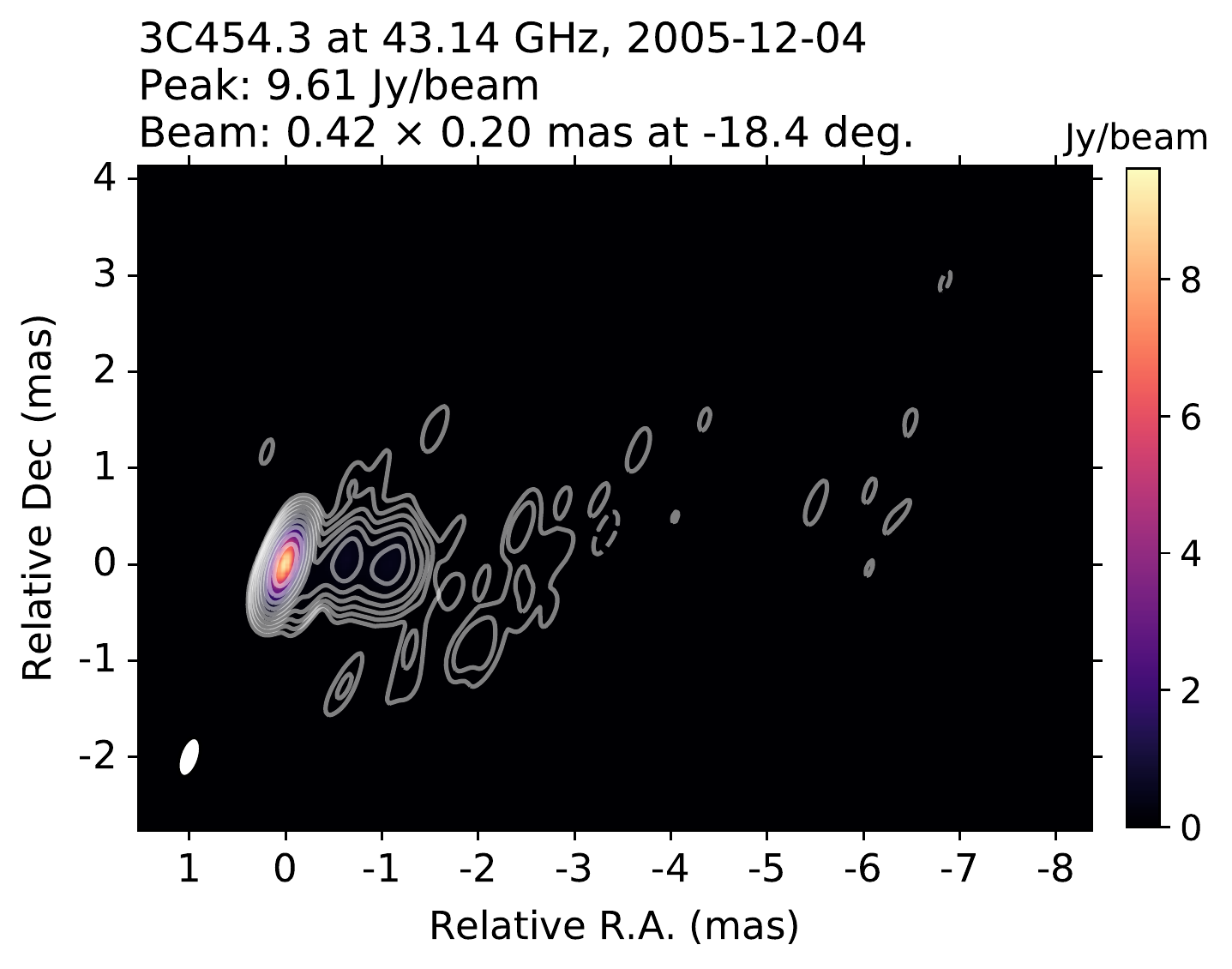}
    }
     \subfigure[]
    {
        \includegraphics[width=0.3\textwidth]{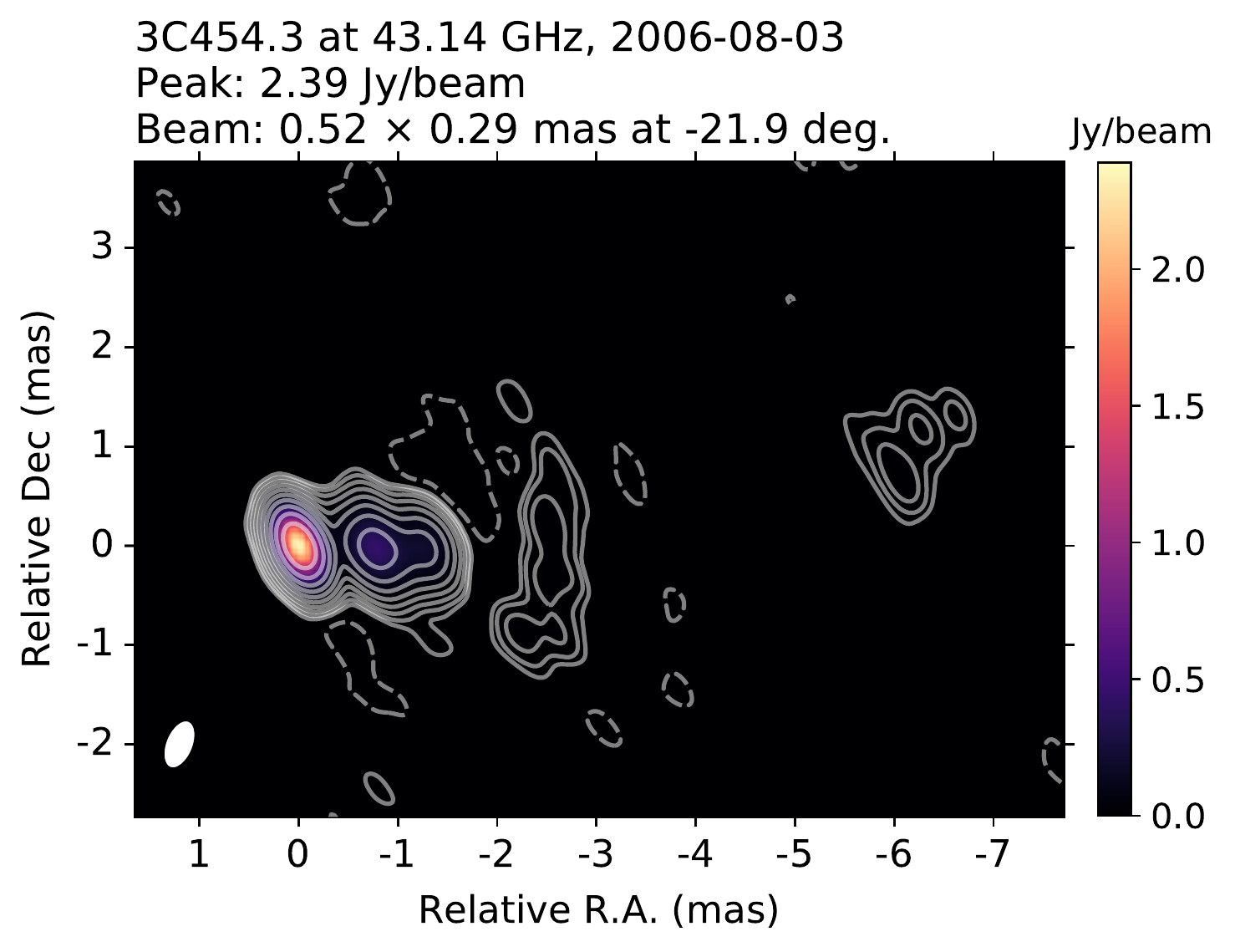}
    }   
       \subfigure[]
    {
         \includegraphics[width=0.3\textwidth]{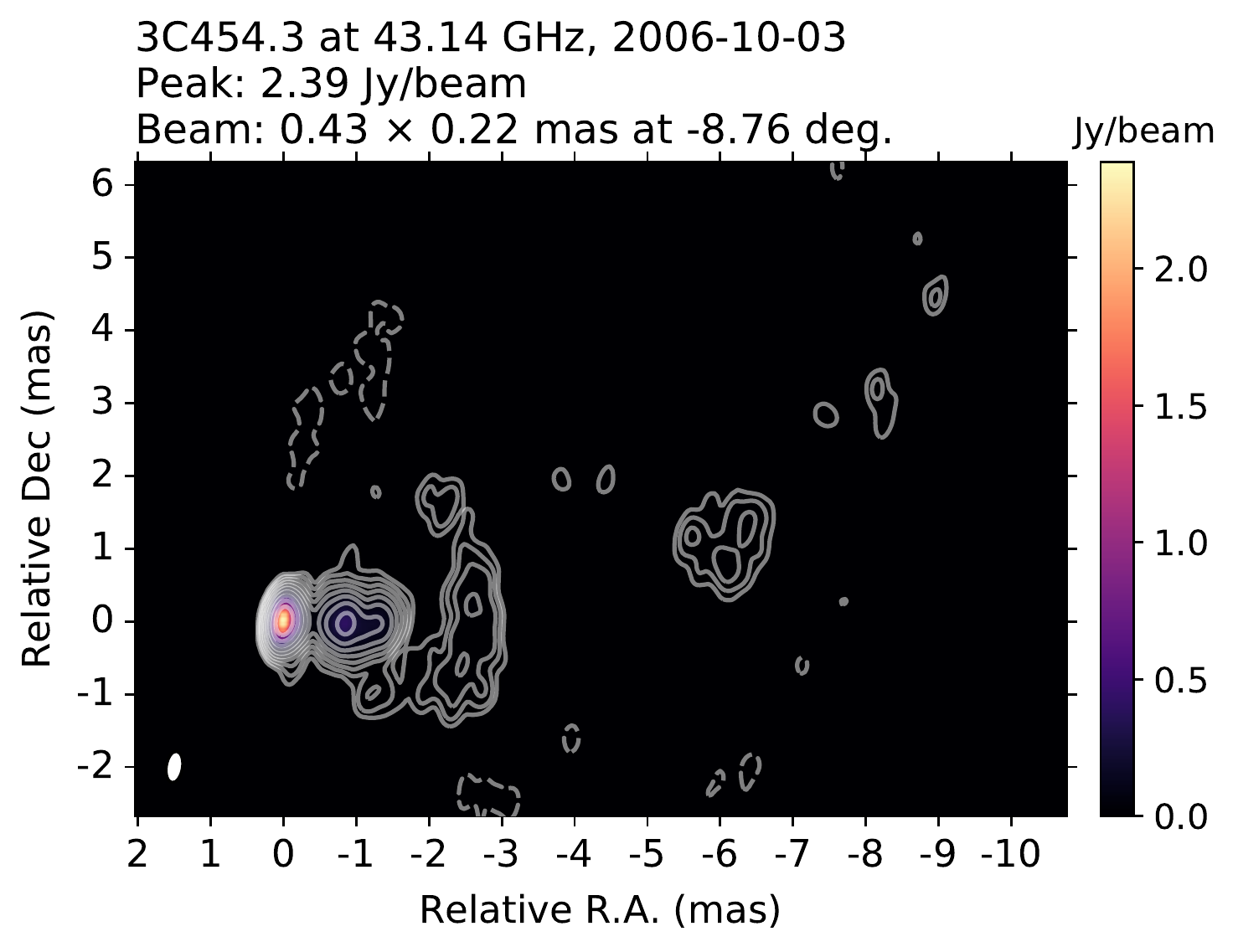}
    }
       \subfigure[]
    {
         \includegraphics[width=0.3\textwidth]{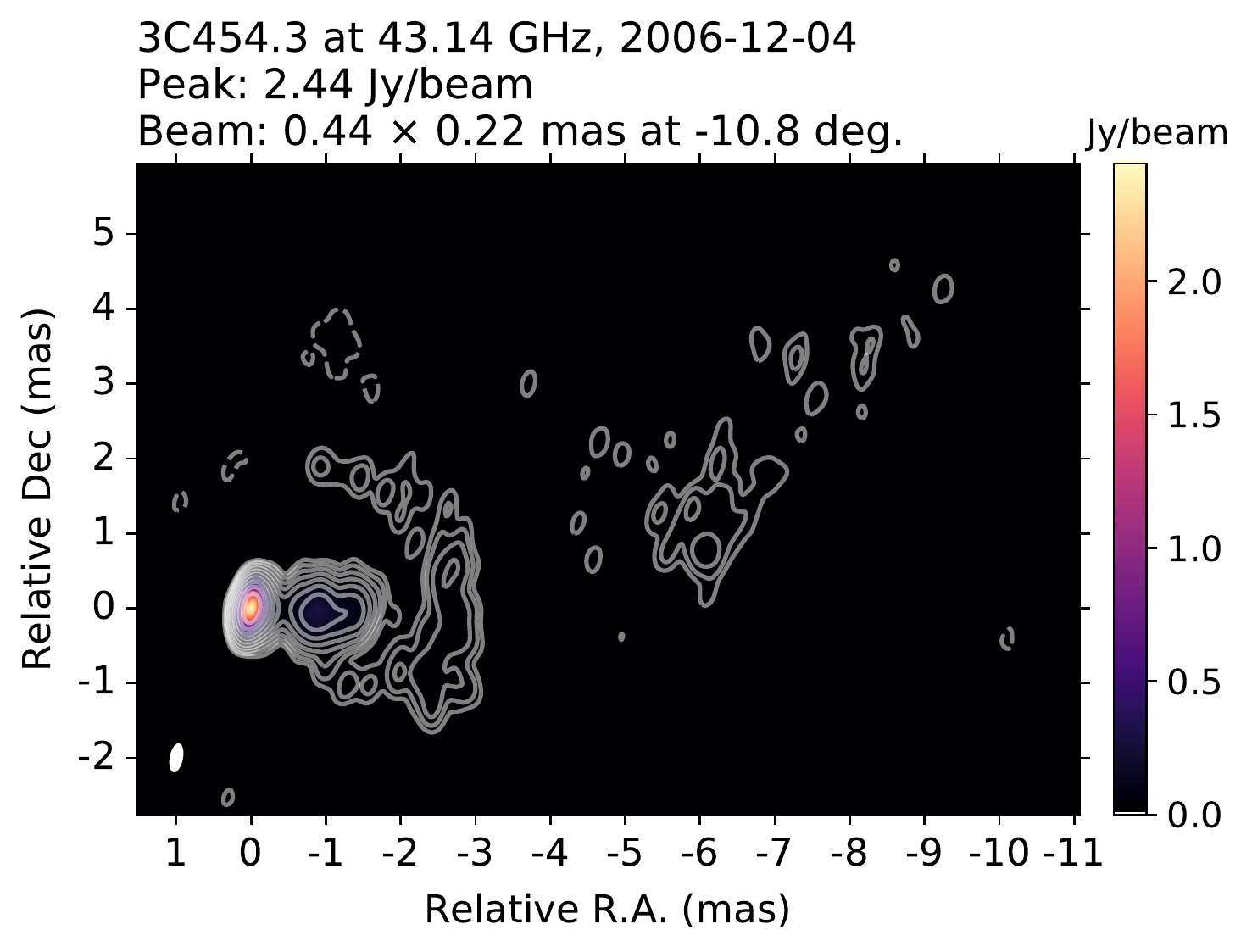}
    }
    \subfigure[]
    {
        \includegraphics[width=0.3\textwidth]{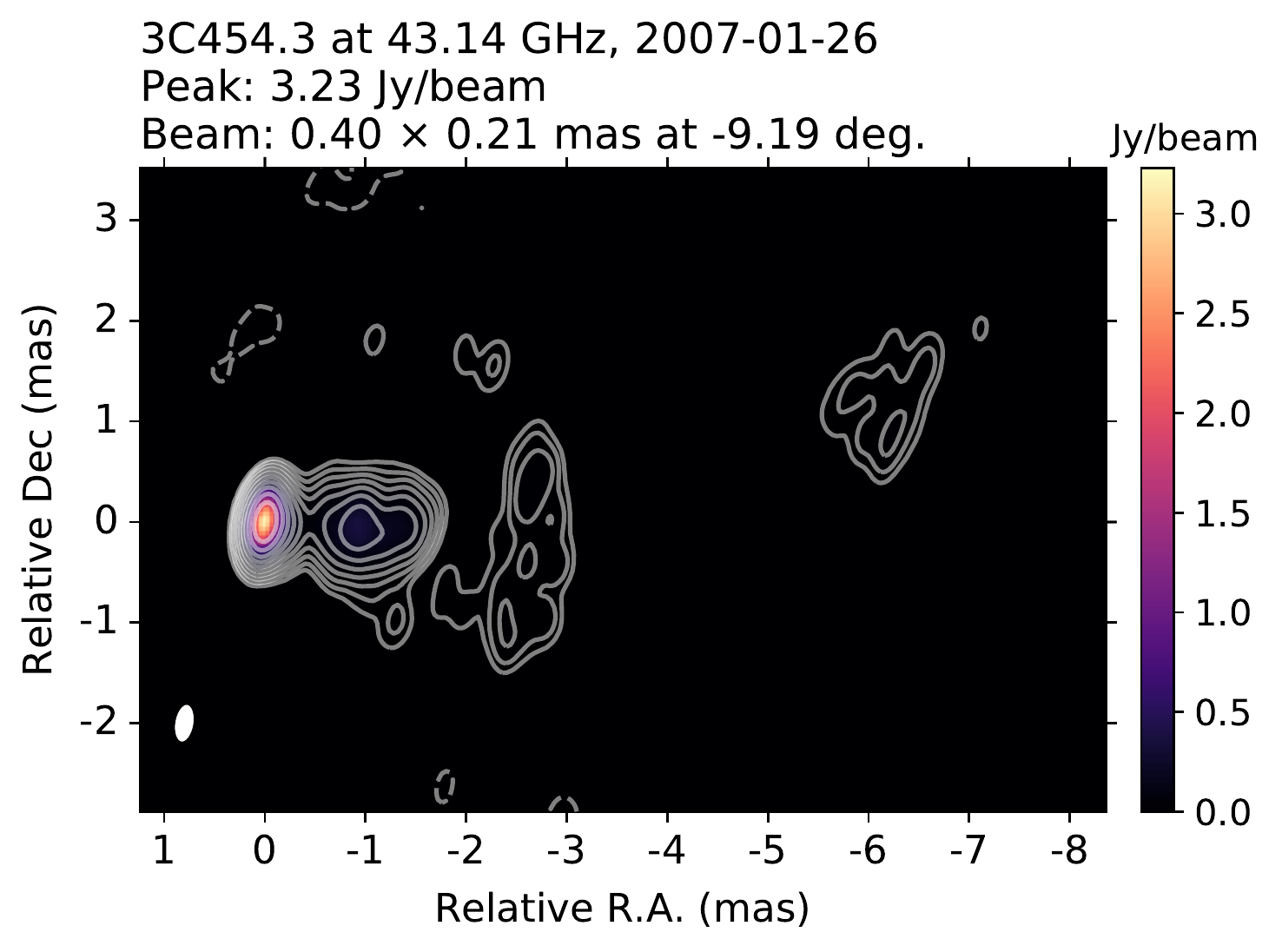}
    }
    \subfigure[]
    {
         \includegraphics[width=0.3\textwidth]{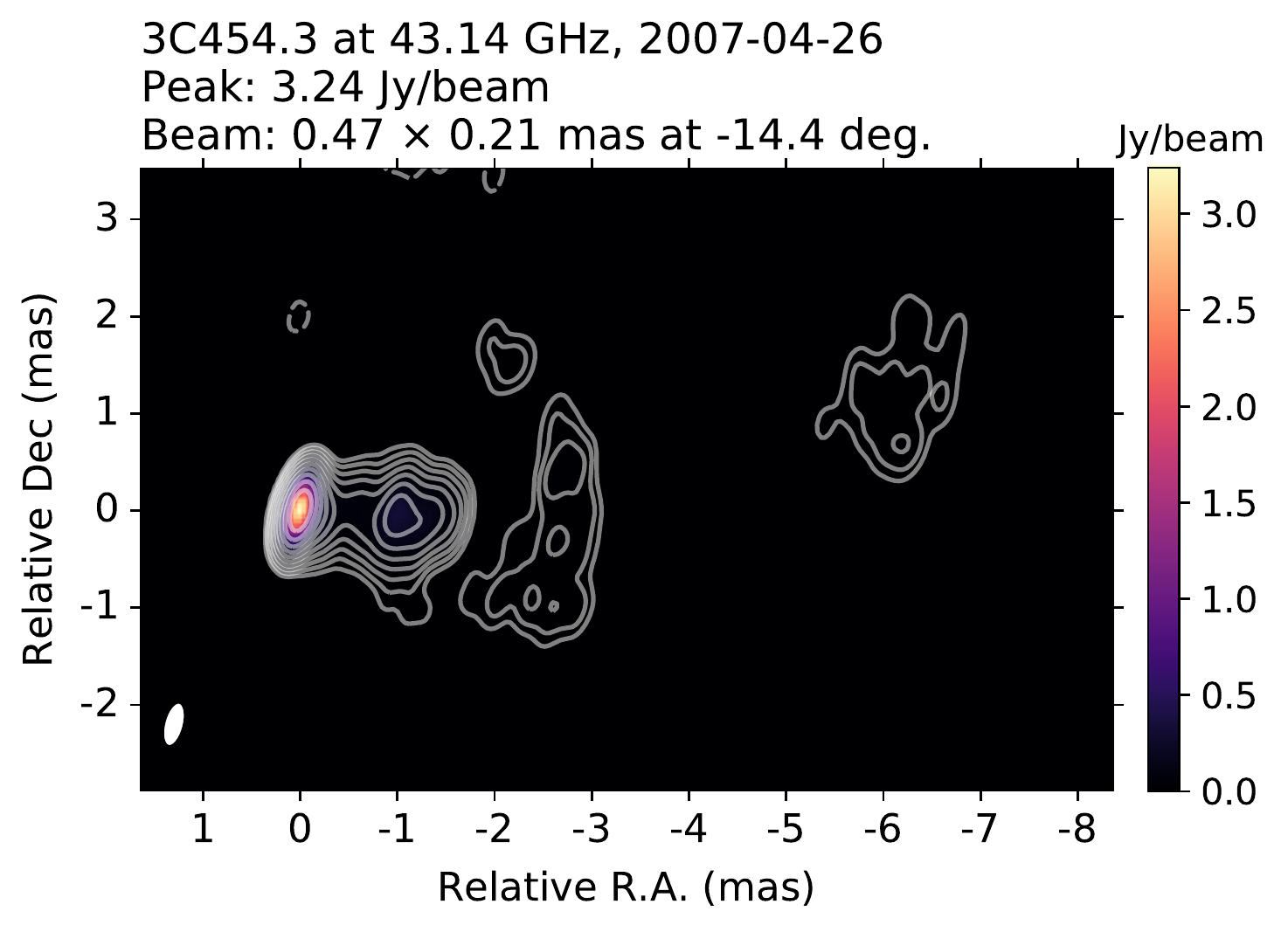}
    }
        \subfigure[]
    {
         \includegraphics[width=0.3\textwidth]{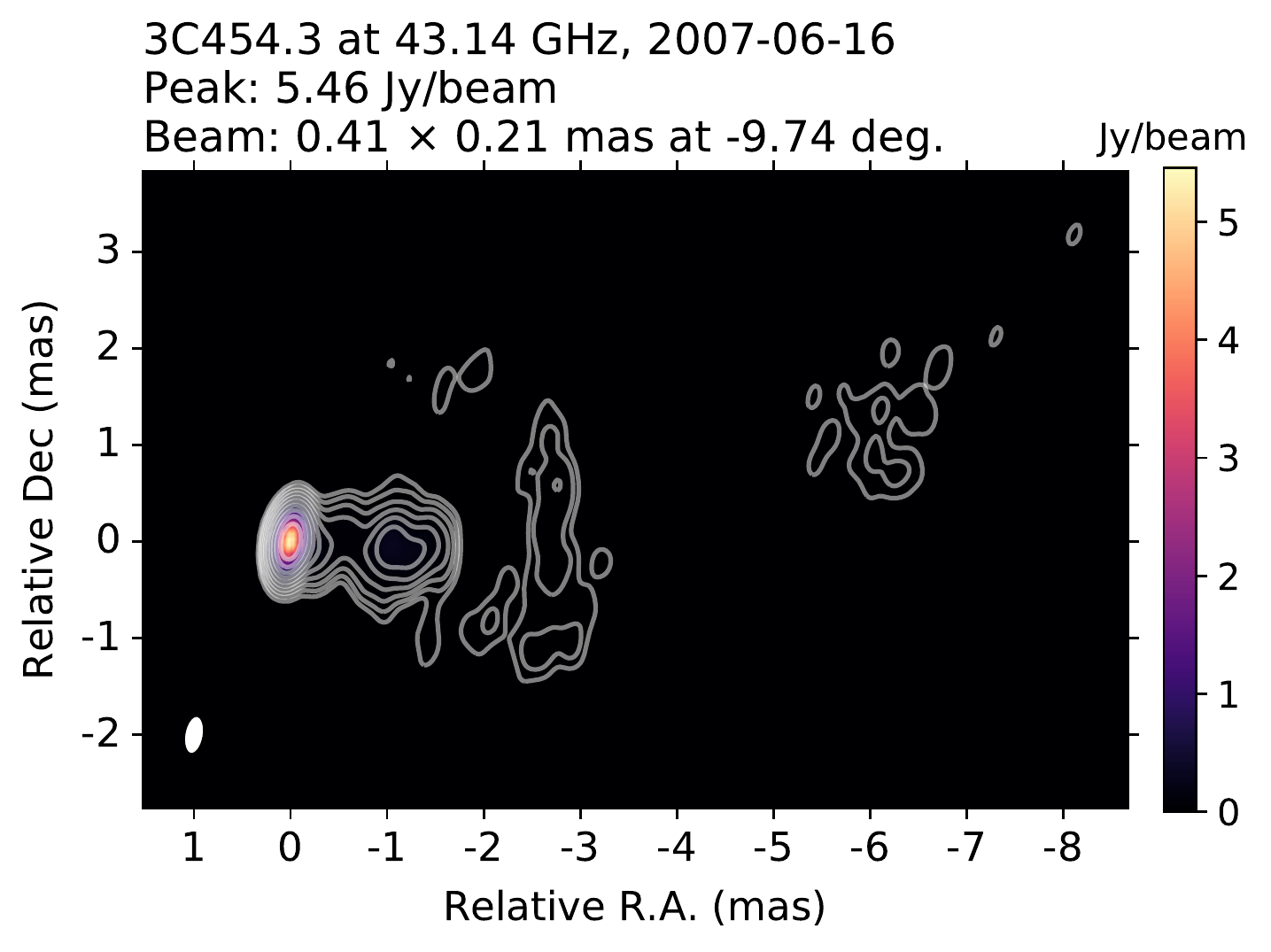}
    }
     \subfigure[]
    {
         \includegraphics[width=0.3\textwidth]{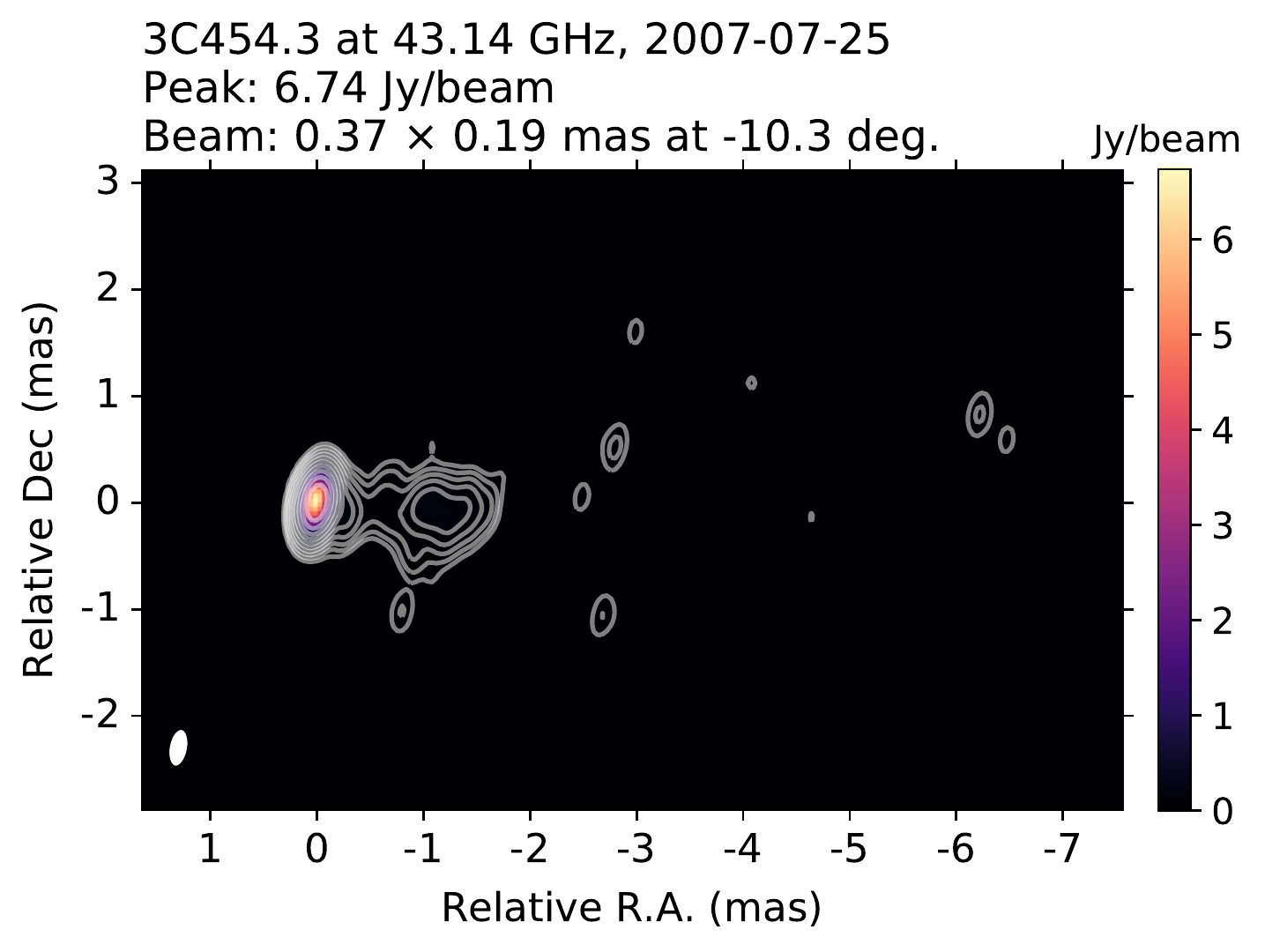}
    }    
         \subfigure[]
    {
        \includegraphics[width=0.3\textwidth]{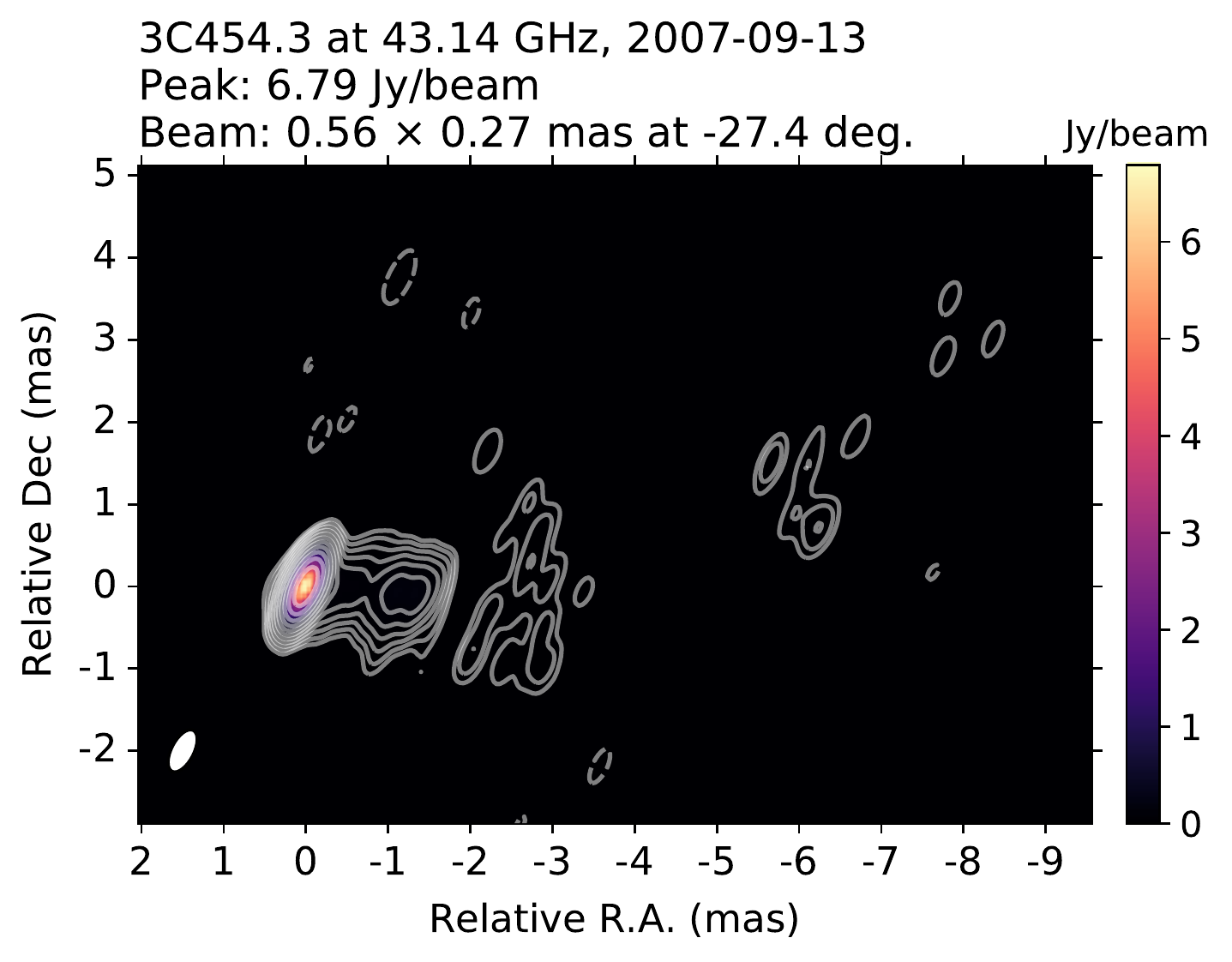}
    }   
     \caption{Q-band (43\,GHz) CLEAN images of 3C454.3 from 2005-05-19 until 2007-09-13 with contours at -0.1\%, 0.1\%, 0.2\%, 0.4\%, 0.8\%, 1.6\%, 3.2\%, 6.4\%, 12.8\%, 25.6\%, and 51.2\% of the peak intensity at each image. For September 13, 2007, the contours are at -0.05\%, 0.05\%, 0.1\%, 0.2\%, 0.4\%, 0.8\%, 1.6\%, 3.2\%, 6.4\%, 12.8\%, 25.6\%, and 51.2\% of the peak intensity. }
    \label{Qbandimagesp1}
\end{figure*}

\begin{figure*}[]
\centering
    \subfigure[]
    {
         \includegraphics[width=0.3\textwidth]{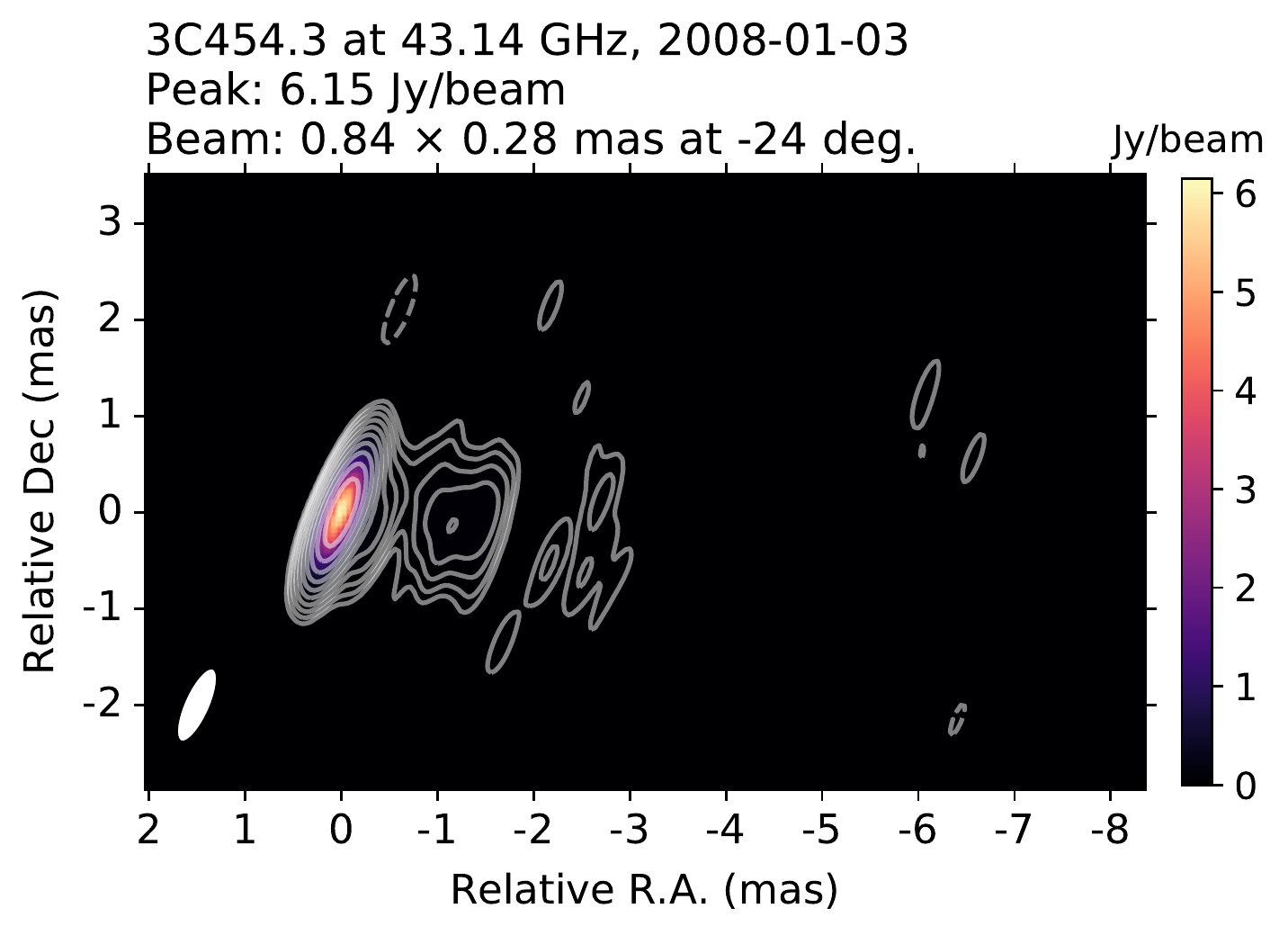}
    }
    \subfigure[]
    {
         \includegraphics[width=0.3\textwidth]{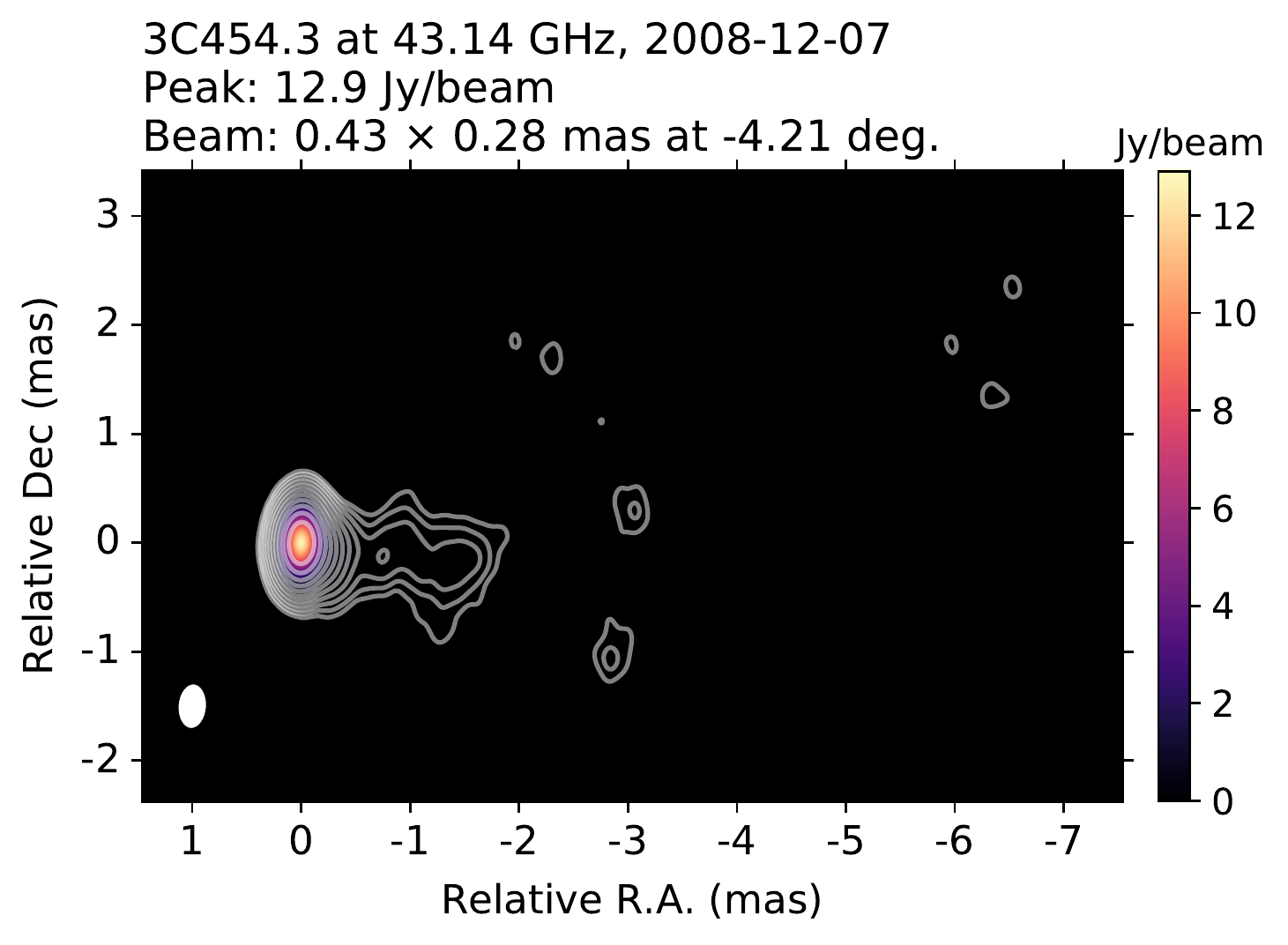}
    }
        \subfigure[]
    {
         \includegraphics[width=0.3\textwidth]{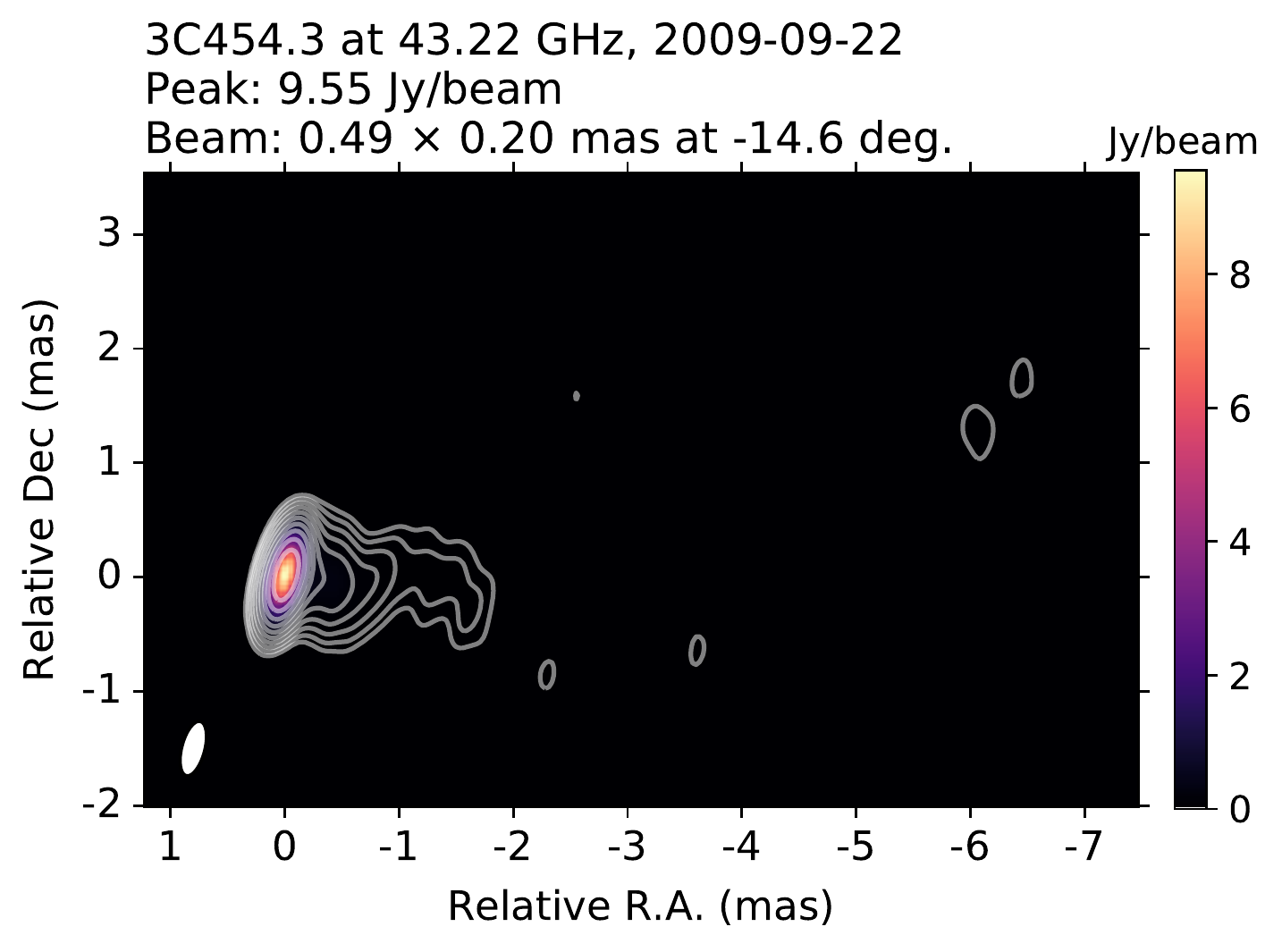}
    }
     \subfigure[]
    {
         \includegraphics[width=0.3\textwidth]{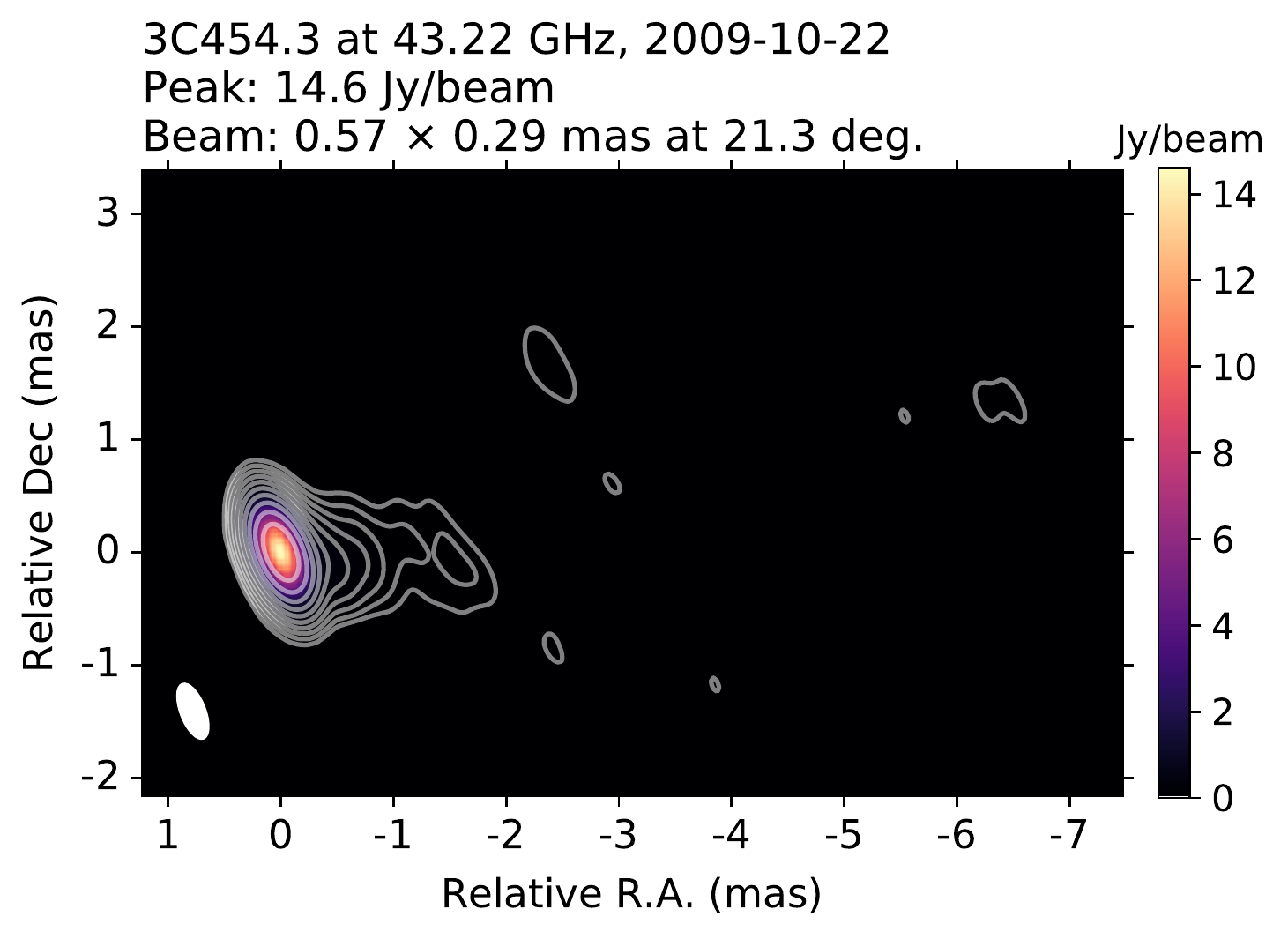}
    }    
         \subfigure[]
    {
        \includegraphics[width=0.3\textwidth]{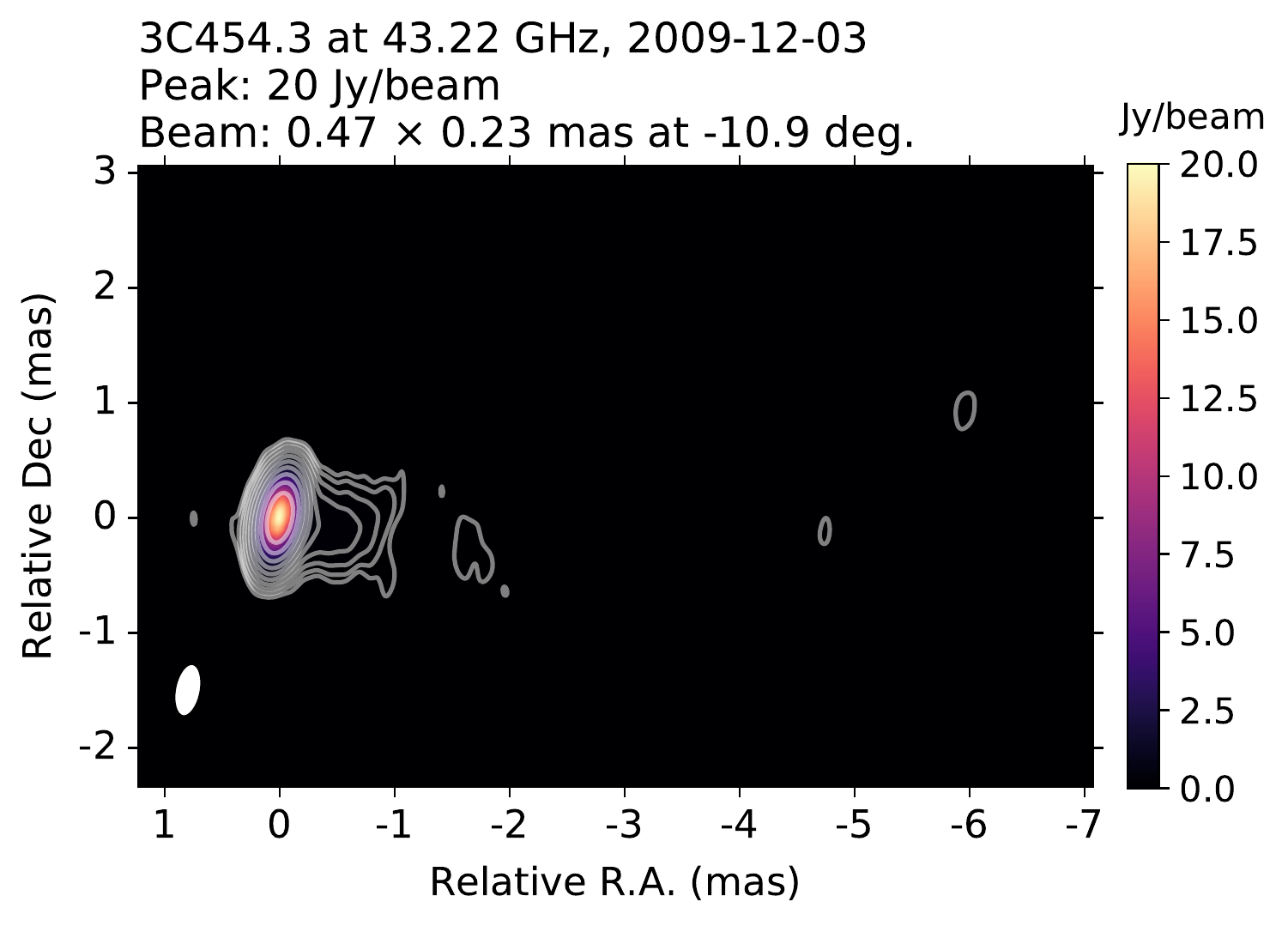}
    }   
         \subfigure[]
    {
        \includegraphics[width=0.3\textwidth]{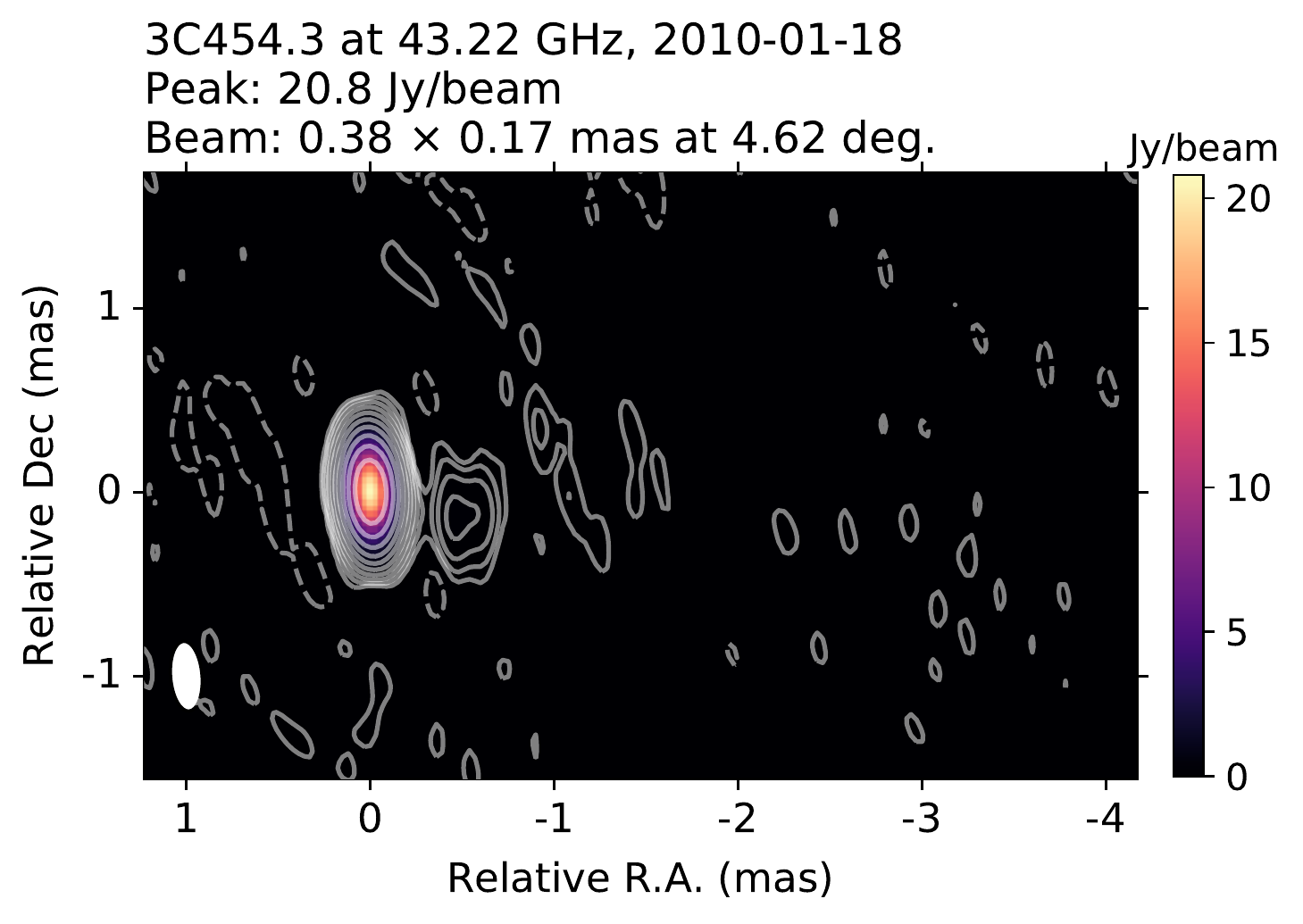}
    }   
       \subfigure[]
    {
        \includegraphics[width=0.3\textwidth]{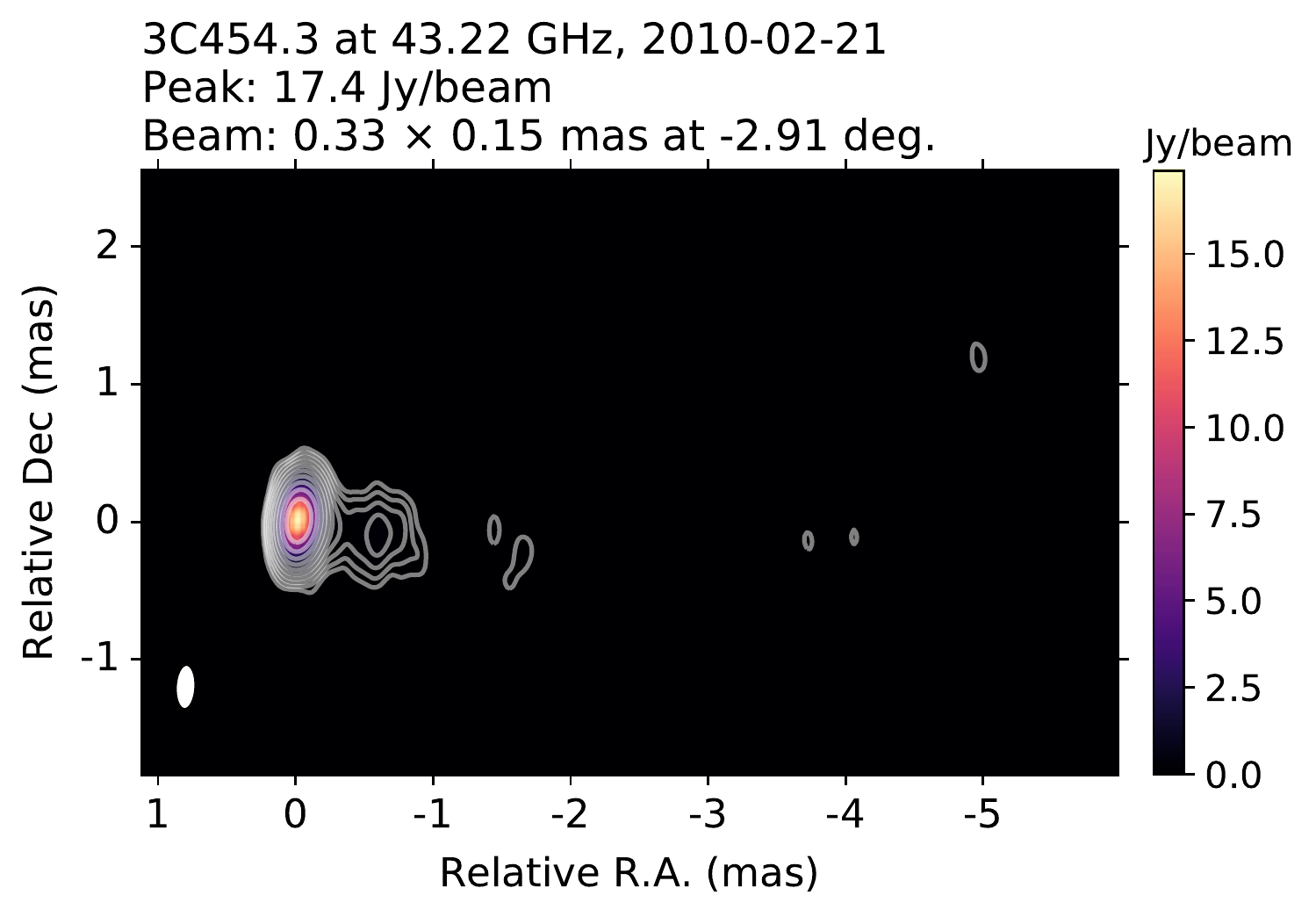}
    }   
     \caption{Q-band (43\,GHz) CLEAN images of 3C454.3 from 2008-01-03 until 2010-02-21. The contours are given at -0.1\%, 0.1\%, 0.2\%, 0.4\%, 0.8\%, 1.6\%, 3.2\%, 6.4\%, 12.8\%, 25.6\%, and 51.2\% of the peak intensity at each image. For December 7, 2008, the contours are at -0.05\%, 0.05\%, 0.1\%, 0.2\%, 0.4\%, 0.8\%, 1.6\%, 3.2\%, 6.4\%, 12.8\%, 25.6\%, and 51.2\% of the peak intensity. }
    \label{Qbandimagesp2}
\end{figure*}

The CLEAN images after self-calibration in amplitudes and in phases are presented here for all epochs. C-band images are displayed in Figures~\ref{Cbandimagesp1} and \ref{Cbandimagesp2}, X-band images in Figures~\ref{Xbandimagesp1} and \ref{Xbandimagesp2}. For the U-band, we present only the epochs that are not published in the MOJAVE website shown if Figure~\ref{Ubandimages}. K-band images are shown in Figures~\ref{Kbandimagesp1} and \ref{Kbandimagesp2}. Finally Q-band images are displayed in Figures~\ref{Qbandimagesp1} and \ref{Qbandimagesp2}.

%%%%%%%%%%%%%%%%%% ONLY C BAND IMAGES %%%%%%%%%%%%%%%%%%%%%%%

%%%%%%%%%%%%%%%%%% ONLY X BAND IMAGES %%%%%%%%%%%%%%%%%%%%%%%

%%%%%%%%%%%%%%%%%% ONLY U BAND IMAGES %%%%%%%%%%%%%%%%%%%%%%%

%%%%%%%%%%%%%%%%%% ONLY K BAND IMAGES %%%%%%%%%%%%%%%%%%%%%%%

%%%%%%%%%%%%%%%%%% ONLY Q BAND IMAGES %%%%%%%%%%%%%%%%%%%%%%%

\section{Spectral index maps}
\label{fullSImaps}
Spectral index maps are obtained after alignment using matched common-uvrange images per frequency pair at all epochs. CX spectral index maps are displayed in Figures~\ref{siCXp1} and \ref{siCXp2}. XU spectral index maps are displayed in Figures~\ref{siXUp1} and \ref{siXUp2}. UK spectral index maps are displayed in Figures~\ref{siUKp1} and \ref{siUKp2}, and KQ spectral index maps are displayed in Figures~\ref{siKQp1} and \ref{siKQp2}.
%%%%%%%%%%%%%%%%%% CX si maps %%%%%%%%%%%%%%%%%%%%%%%

\begin{figure*}[]
\centering
    \subfigure[]
    {
         \includegraphics[width=0.3\textwidth]{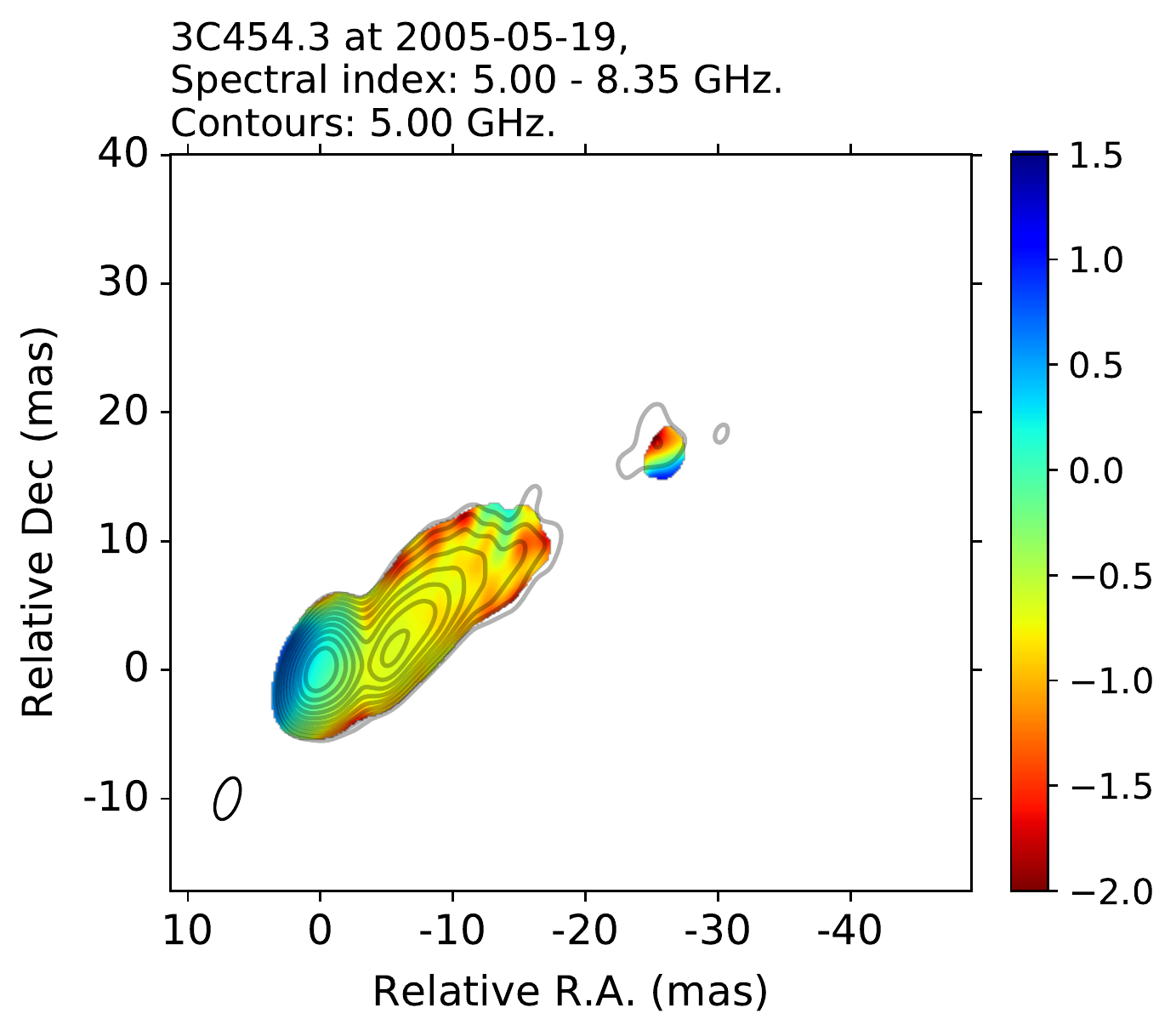}
    }
    \subfigure[]
    {
         \includegraphics[width=0.3\textwidth]{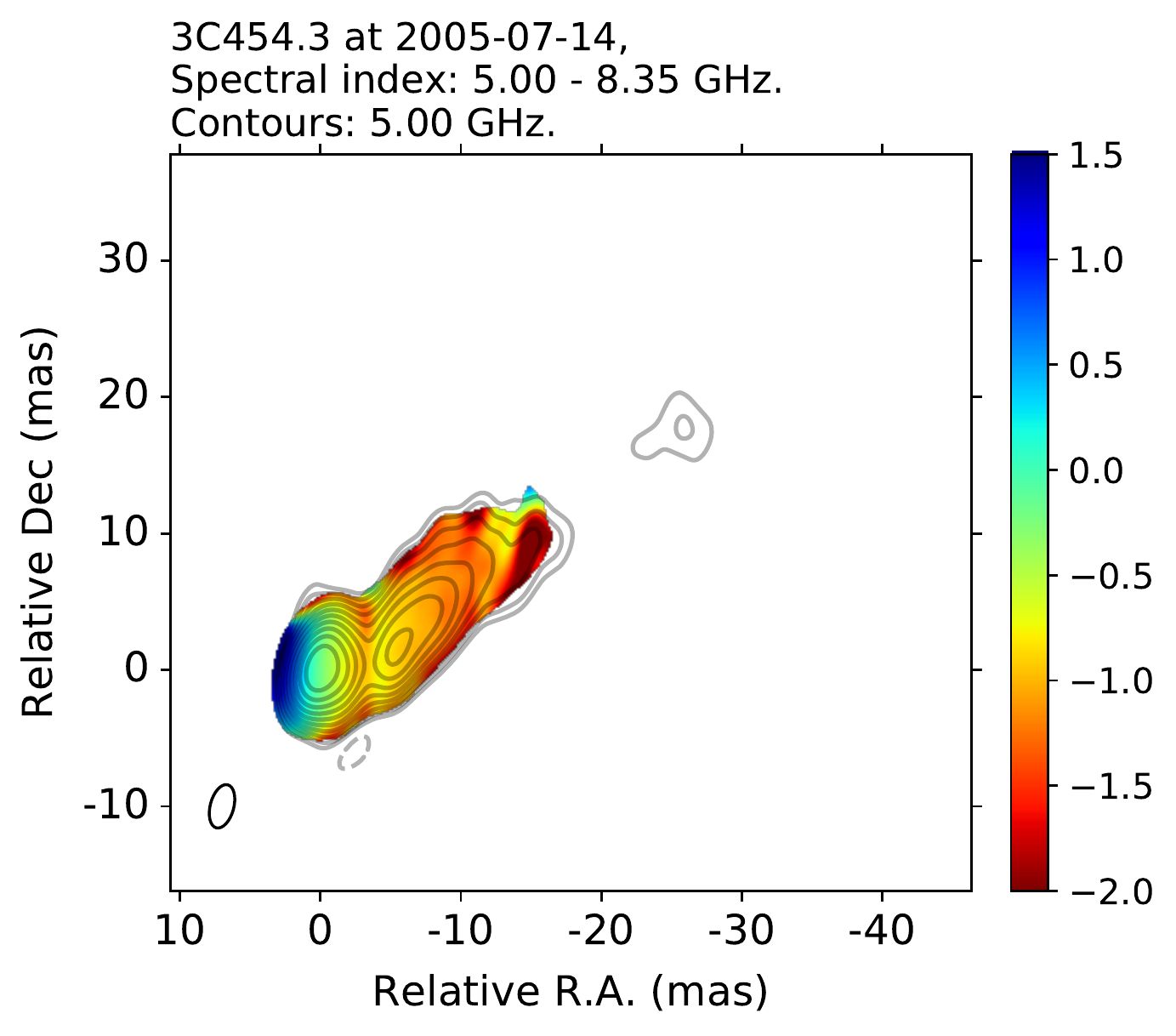}
    }
    \subfigure[]
    {
         \includegraphics[width=0.3\textwidth]{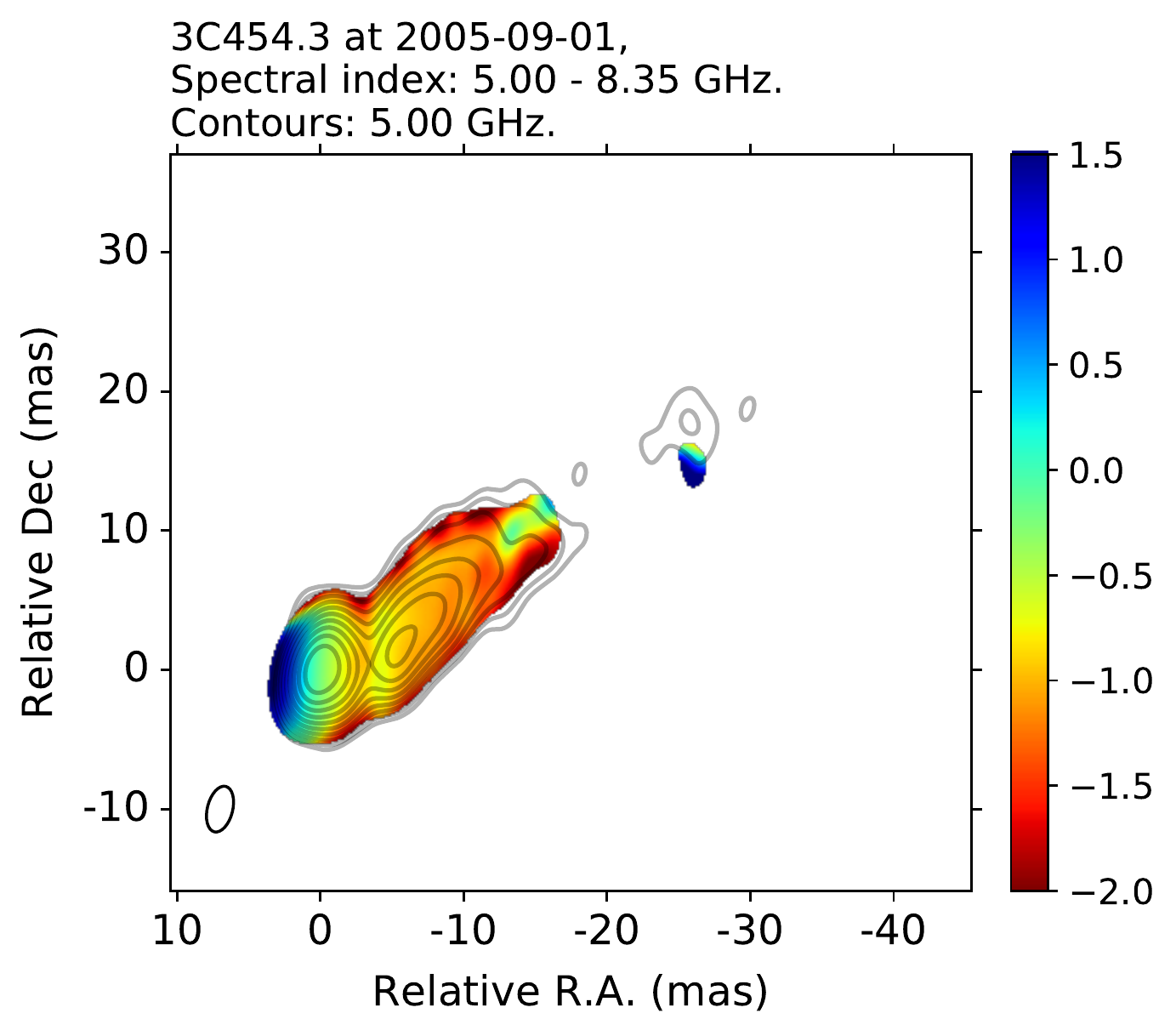}
    }
    \subfigure[]
    {
         \includegraphics[width=0.3\textwidth]{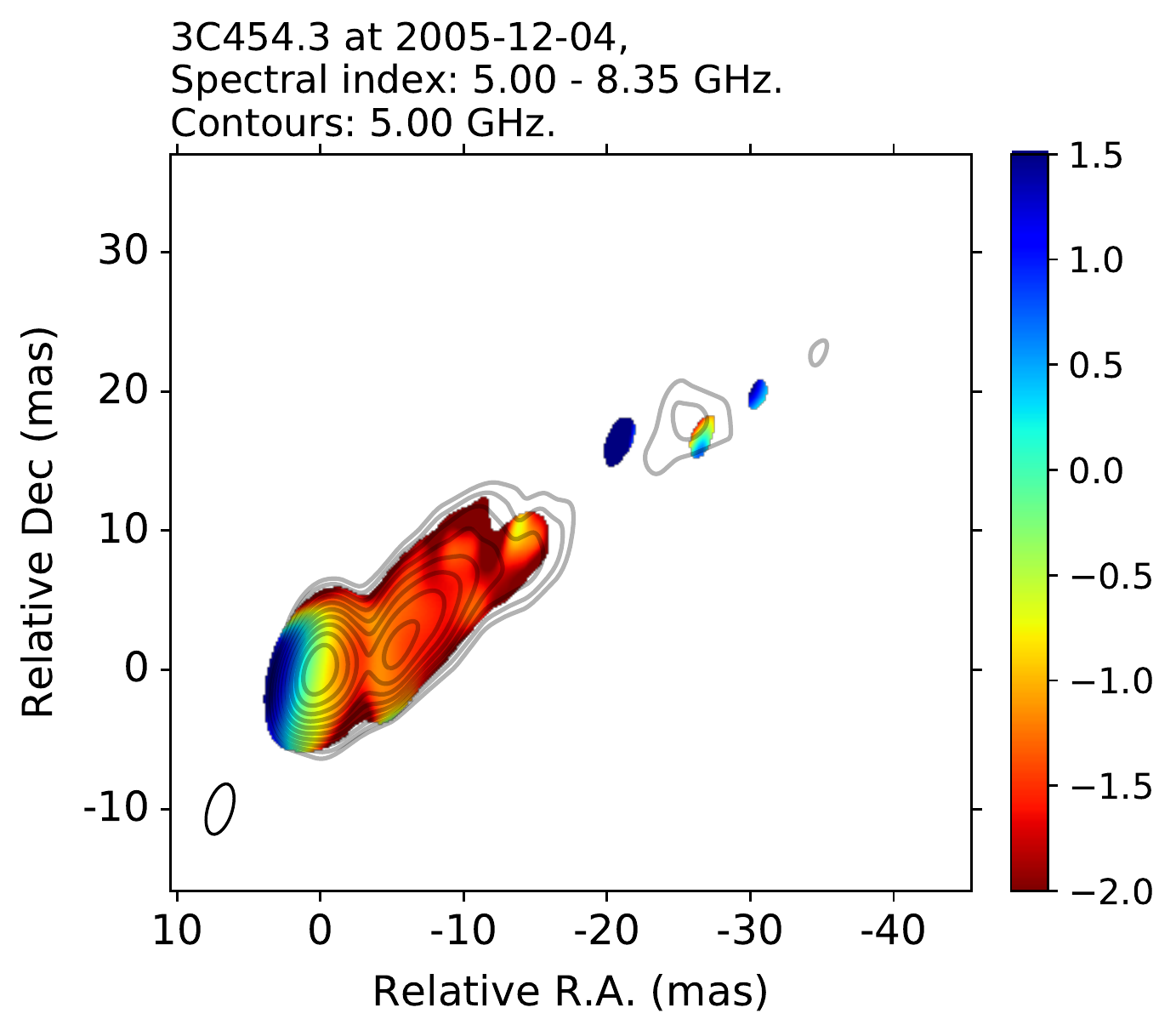}
    }
    \subfigure[]
    {
         \includegraphics[width=0.3\textwidth]{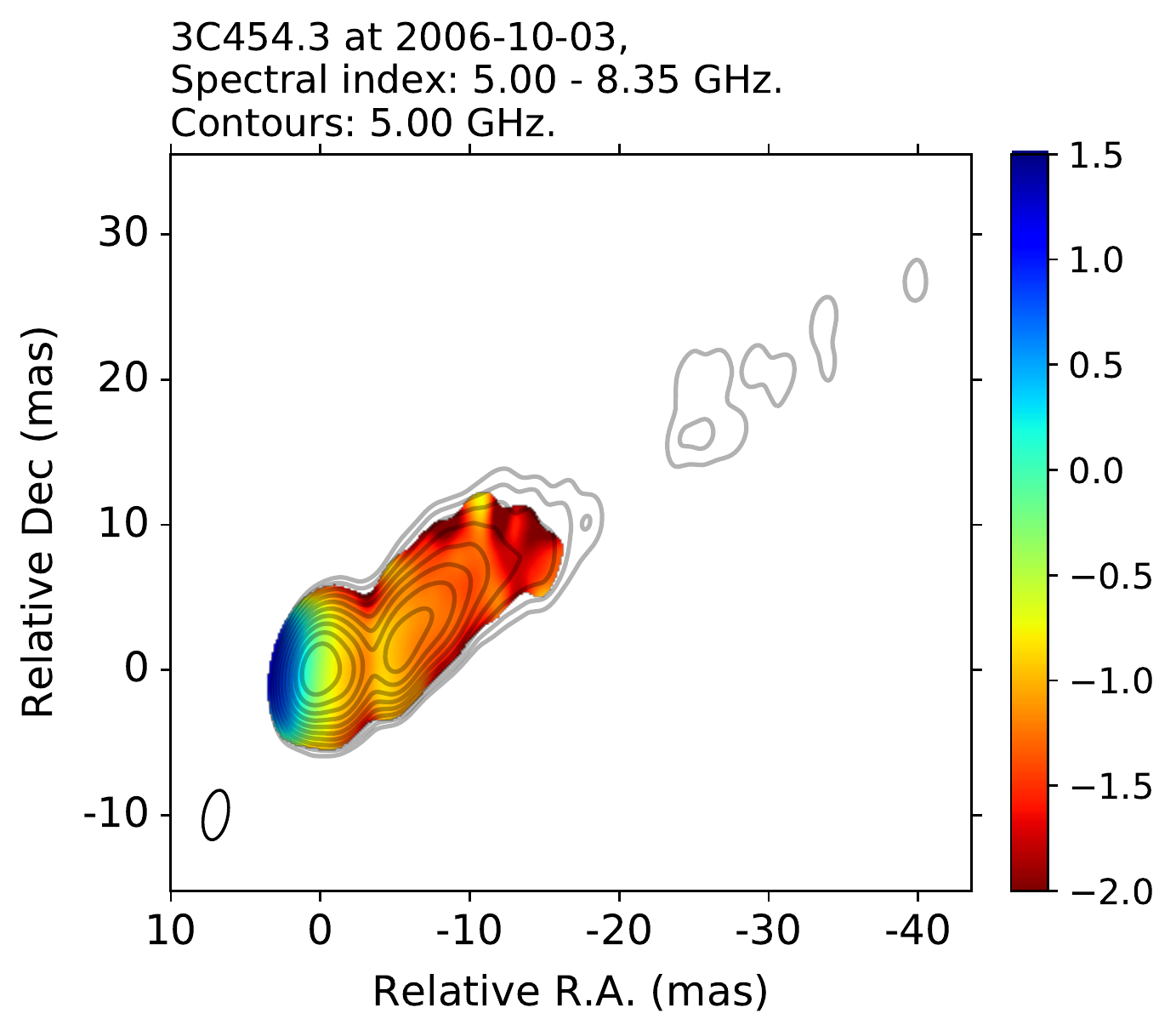}
    }   
    \subfigure[]
    {
         \includegraphics[width=0.3\textwidth]{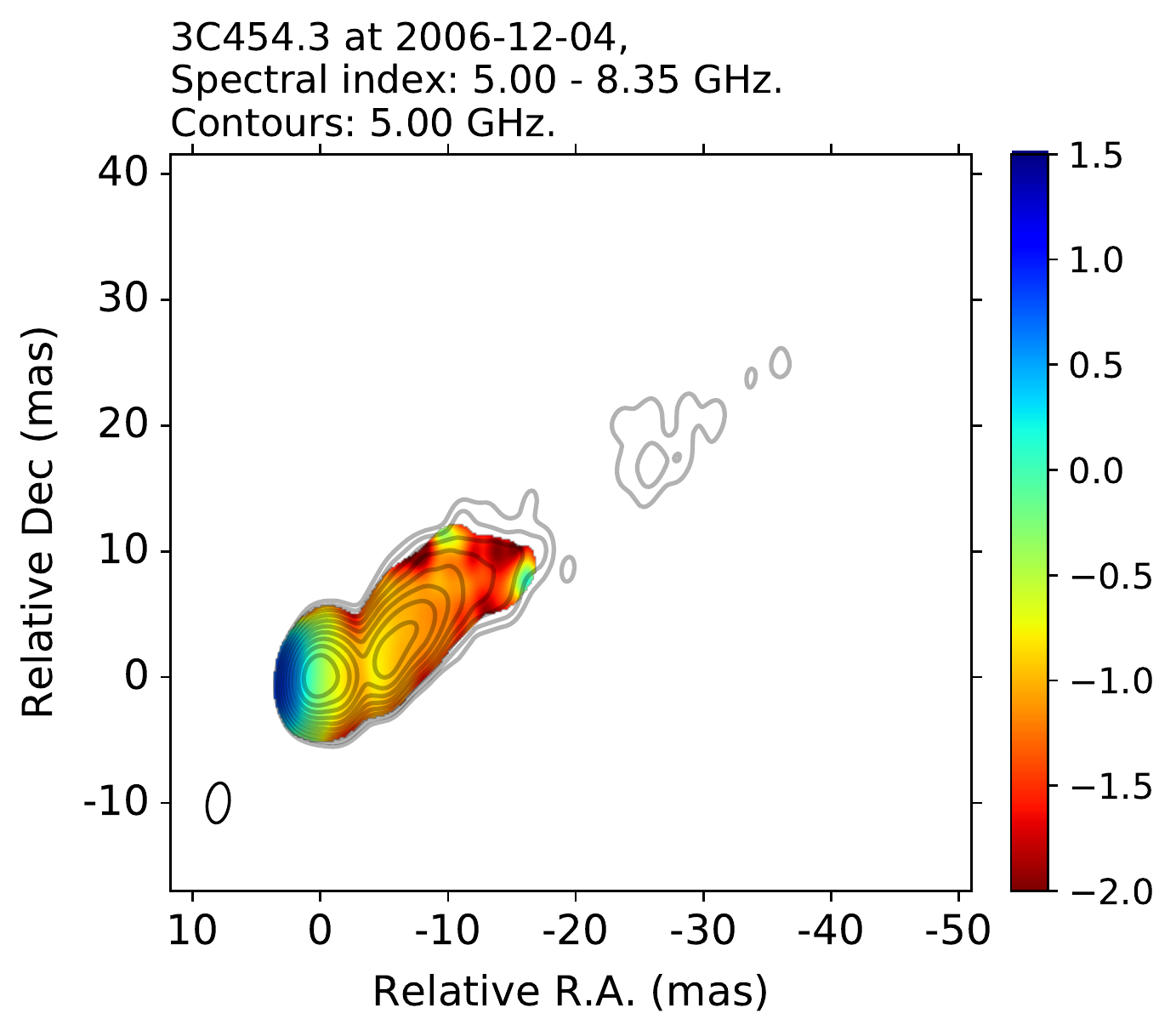}
    }    
    \subfigure[]
    {
         \includegraphics[width=0.3\textwidth]{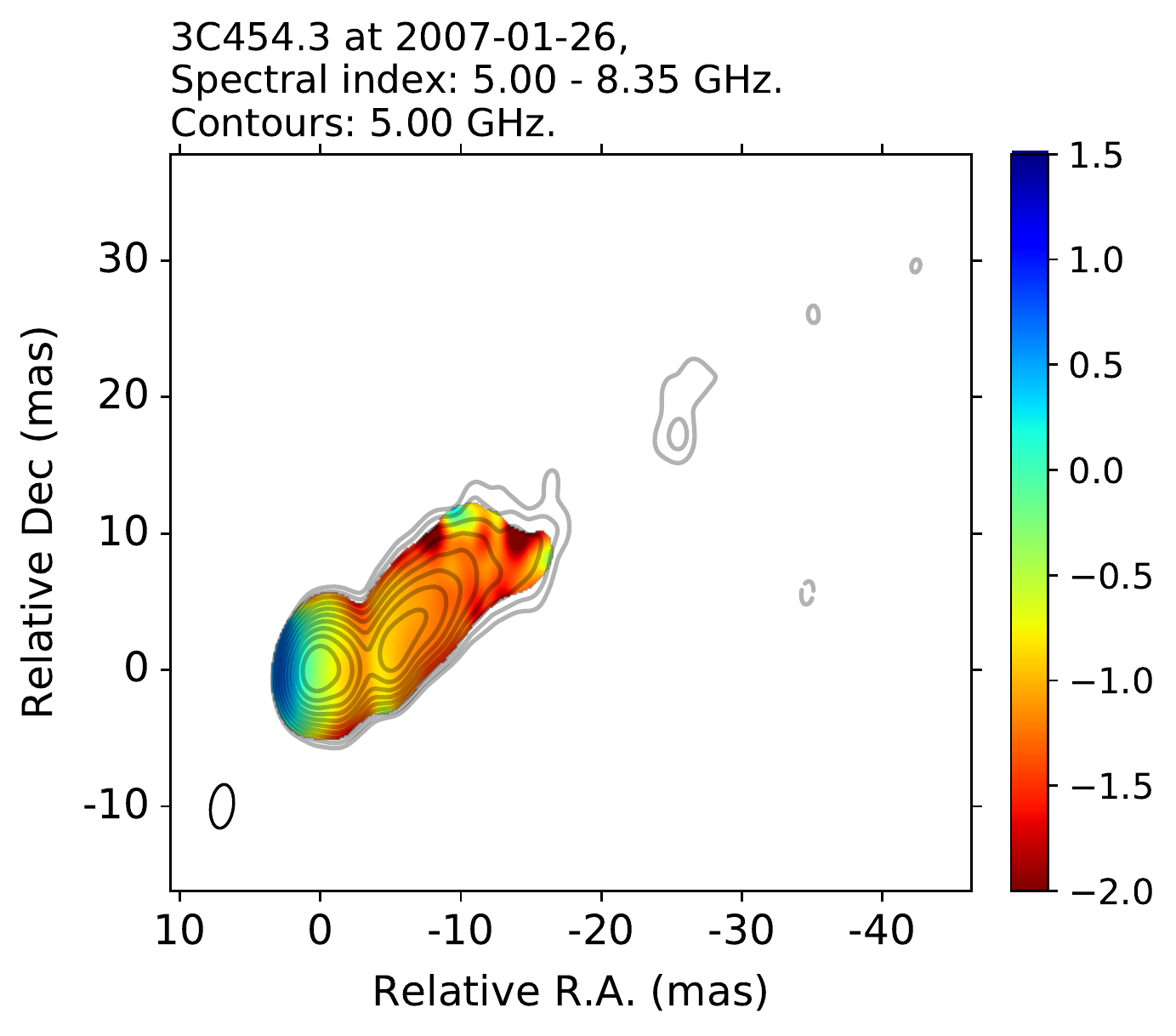}
    }   
    \subfigure[]
    {
         \includegraphics[width=0.3\textwidth]{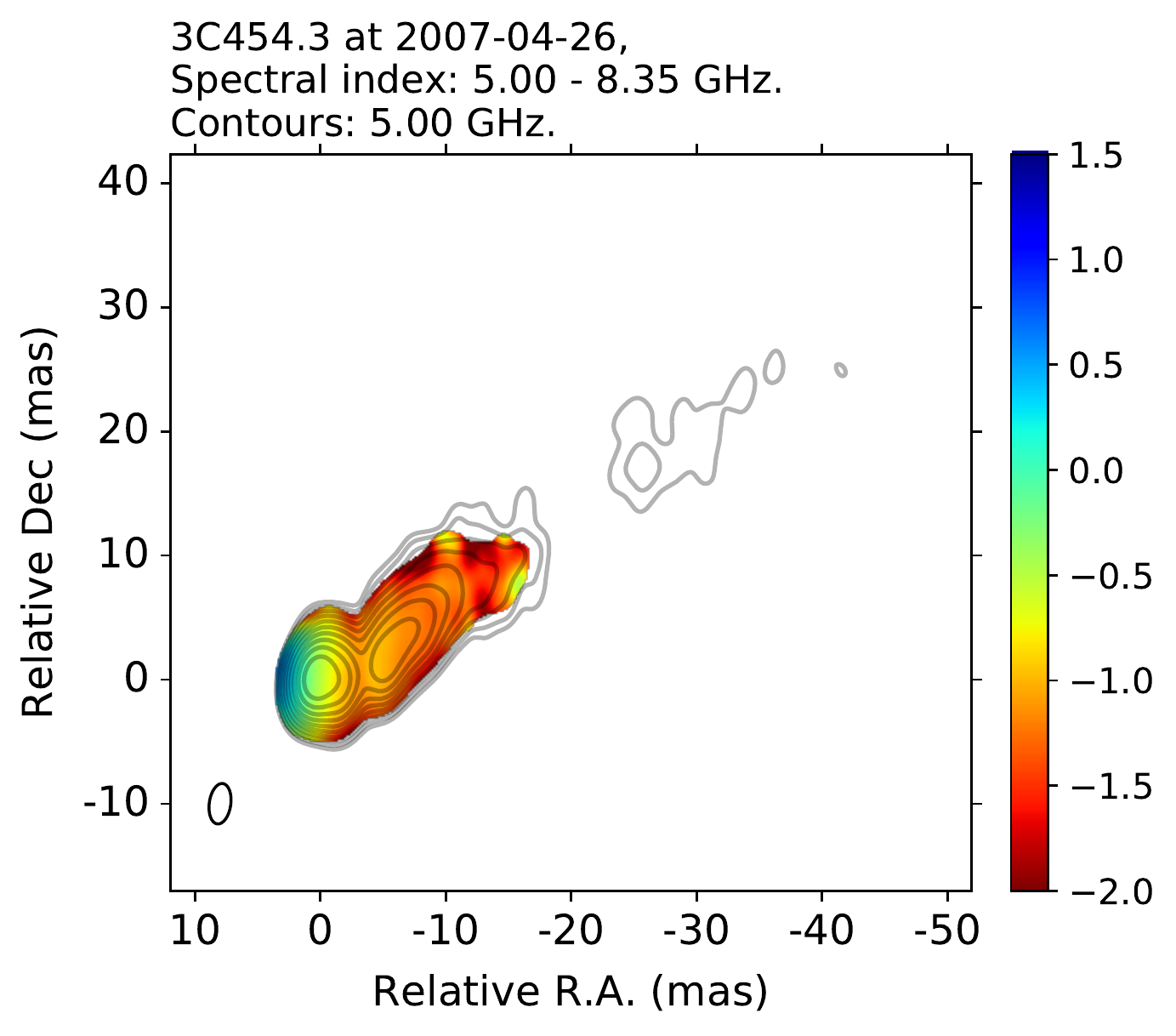}
    }    
        \subfigure[]
    {
         \includegraphics[width=0.3\textwidth]{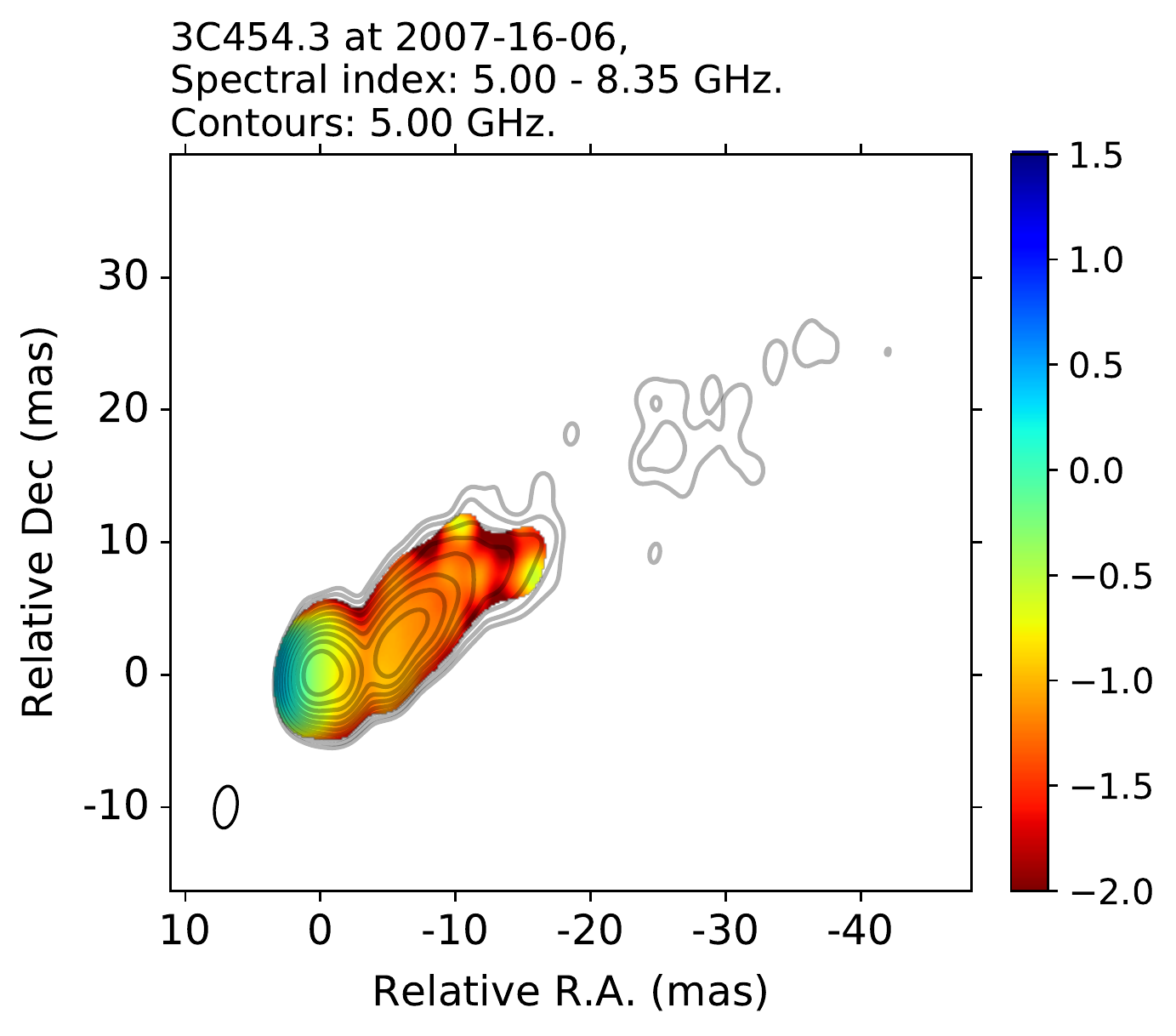}
    }    
        \subfigure[]
    {
         \includegraphics[width=0.3\textwidth]{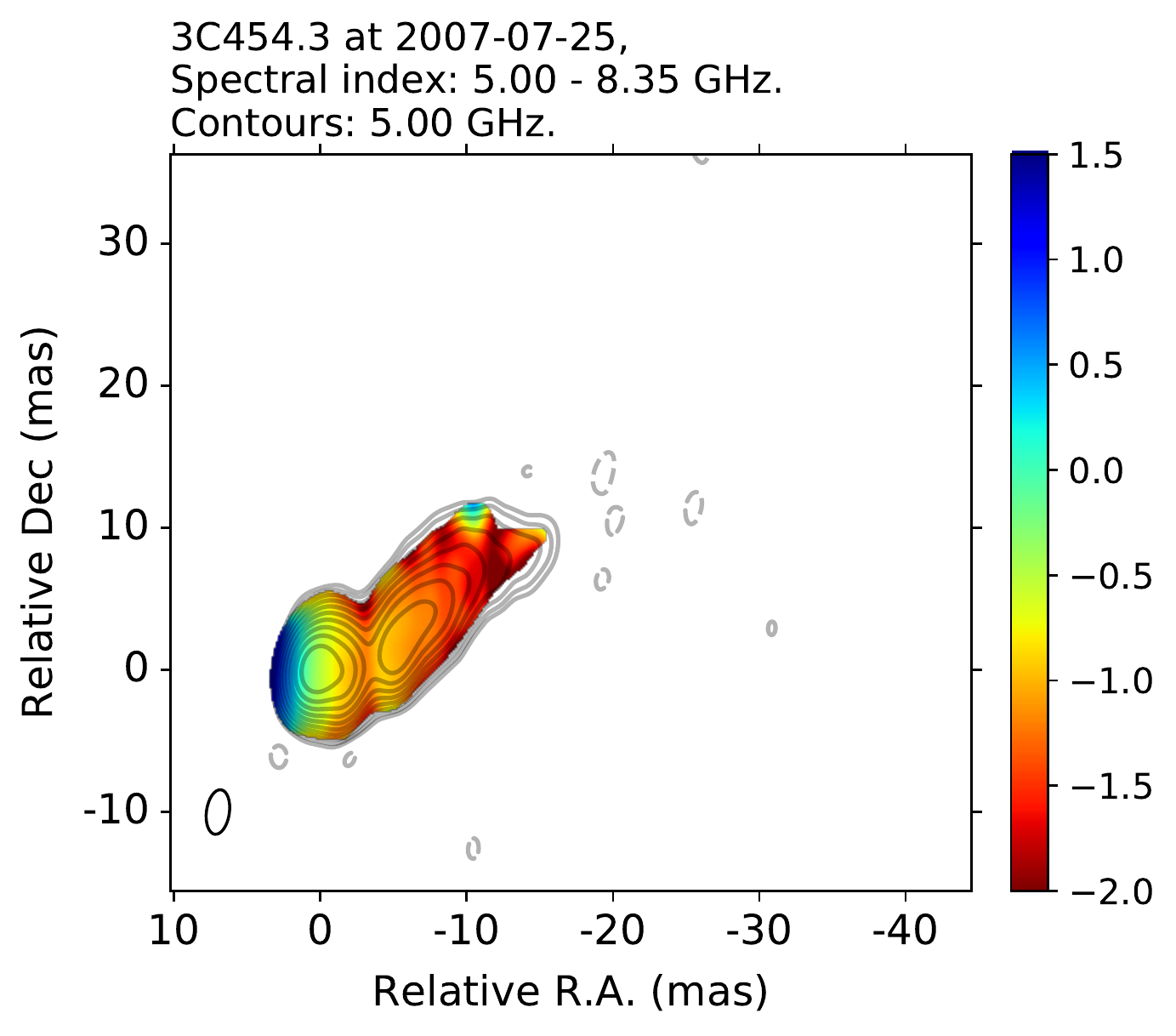}
    } 
            \subfigure[]
    {
         \includegraphics[width=0.3\textwidth]{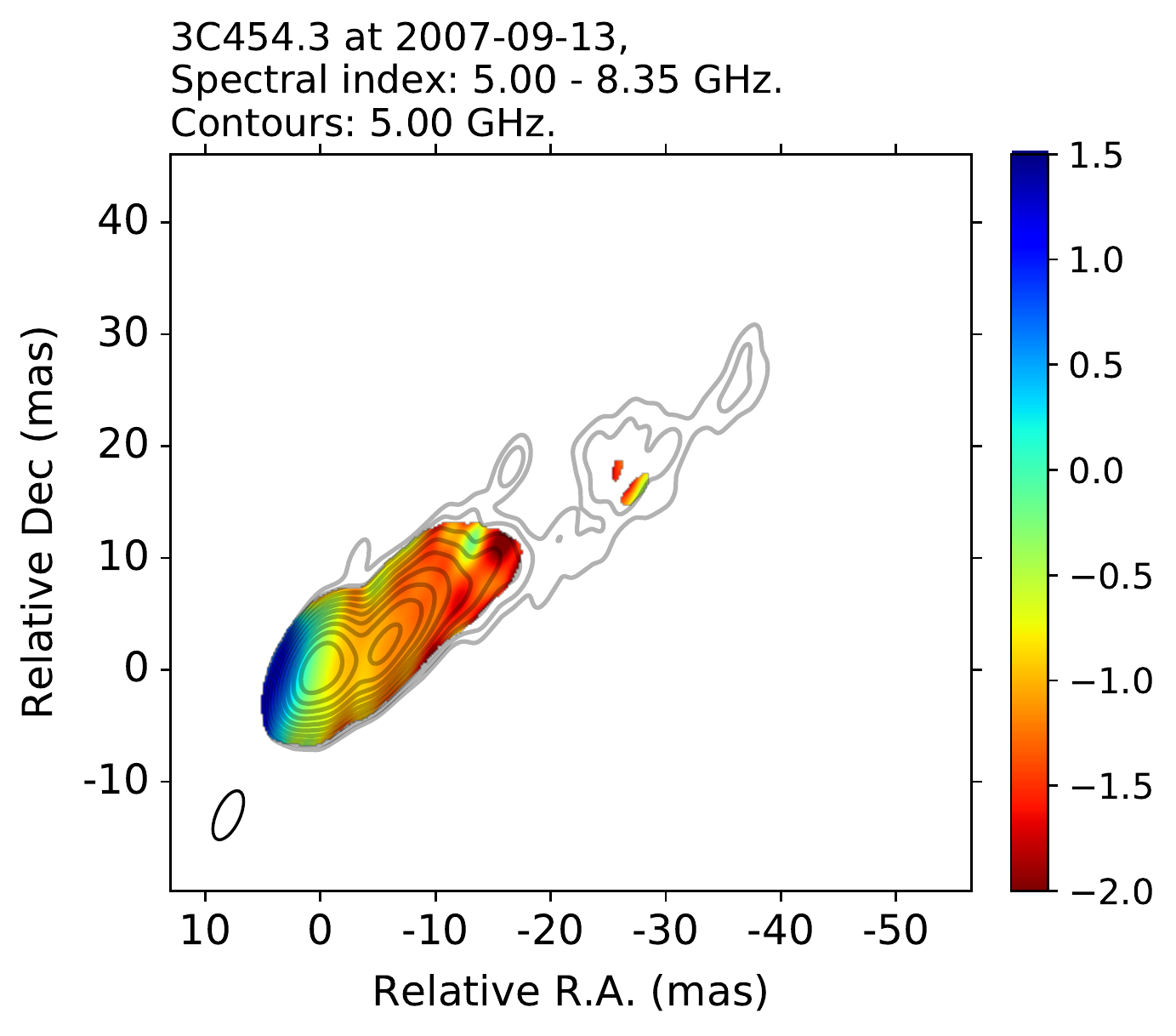}
    } 
            \subfigure[]
    {
         \includegraphics[width=0.3\textwidth]{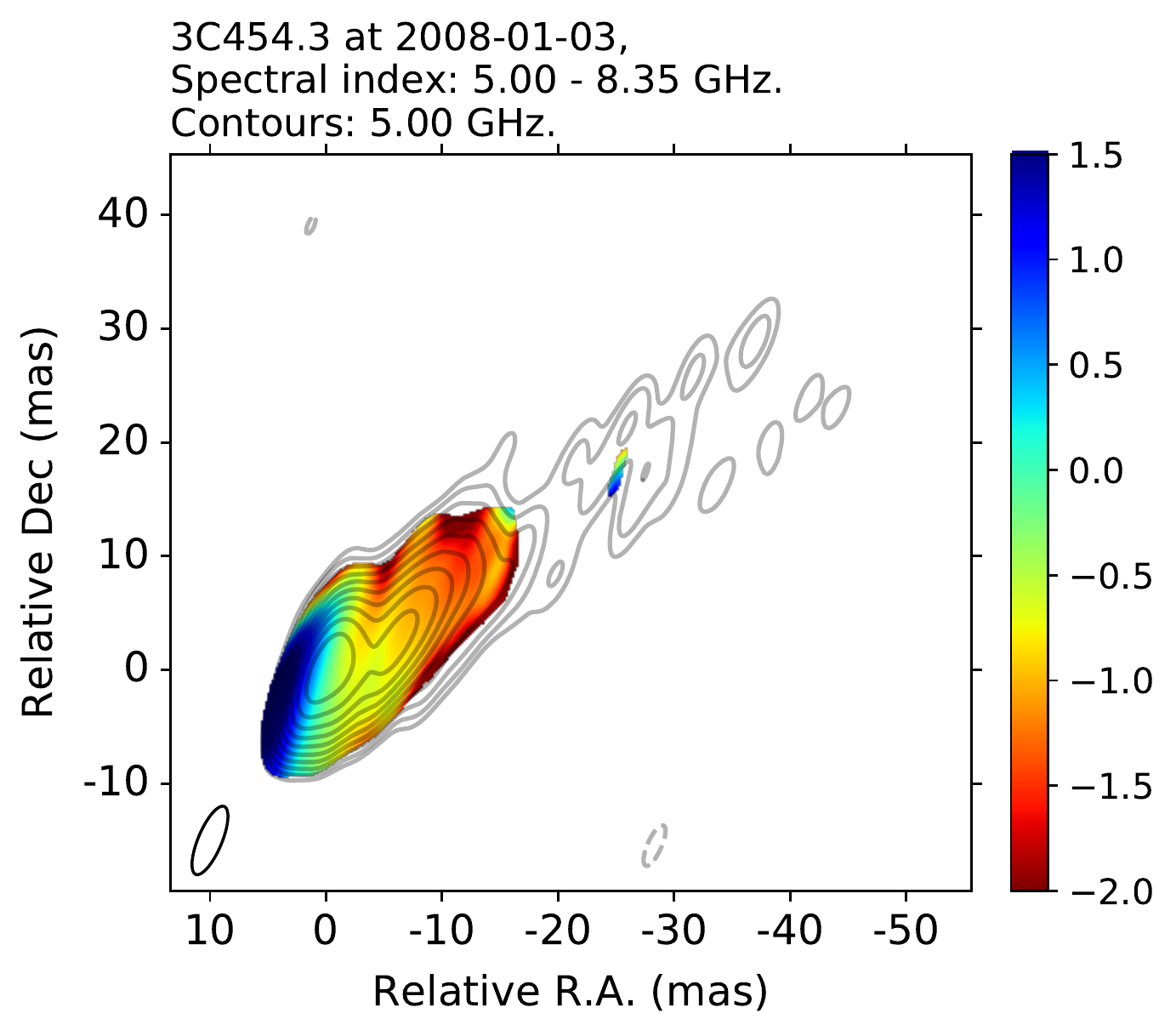}
    } 
     \caption{Spectral index maps for the frequency pair CX (5 - 8\,GHz). The colorbar indicates the spectral index. The ellipse on the bottom left corner represents the interferometric beam. The contour lines are given at   -0.1\%, 0.1\% 0.2\%, 0.4\%, 0.8\%, 1.6\%, 3.2\%, 6.4\%, 12.8\%, 25.6\%, and 51.2\% of the peak intensity at each image. }
    \label{siCXp1}
\end{figure*}

\begin{figure*}[]
\centering
    \subfigure[]
    {
         \includegraphics[width=0.3\textwidth]{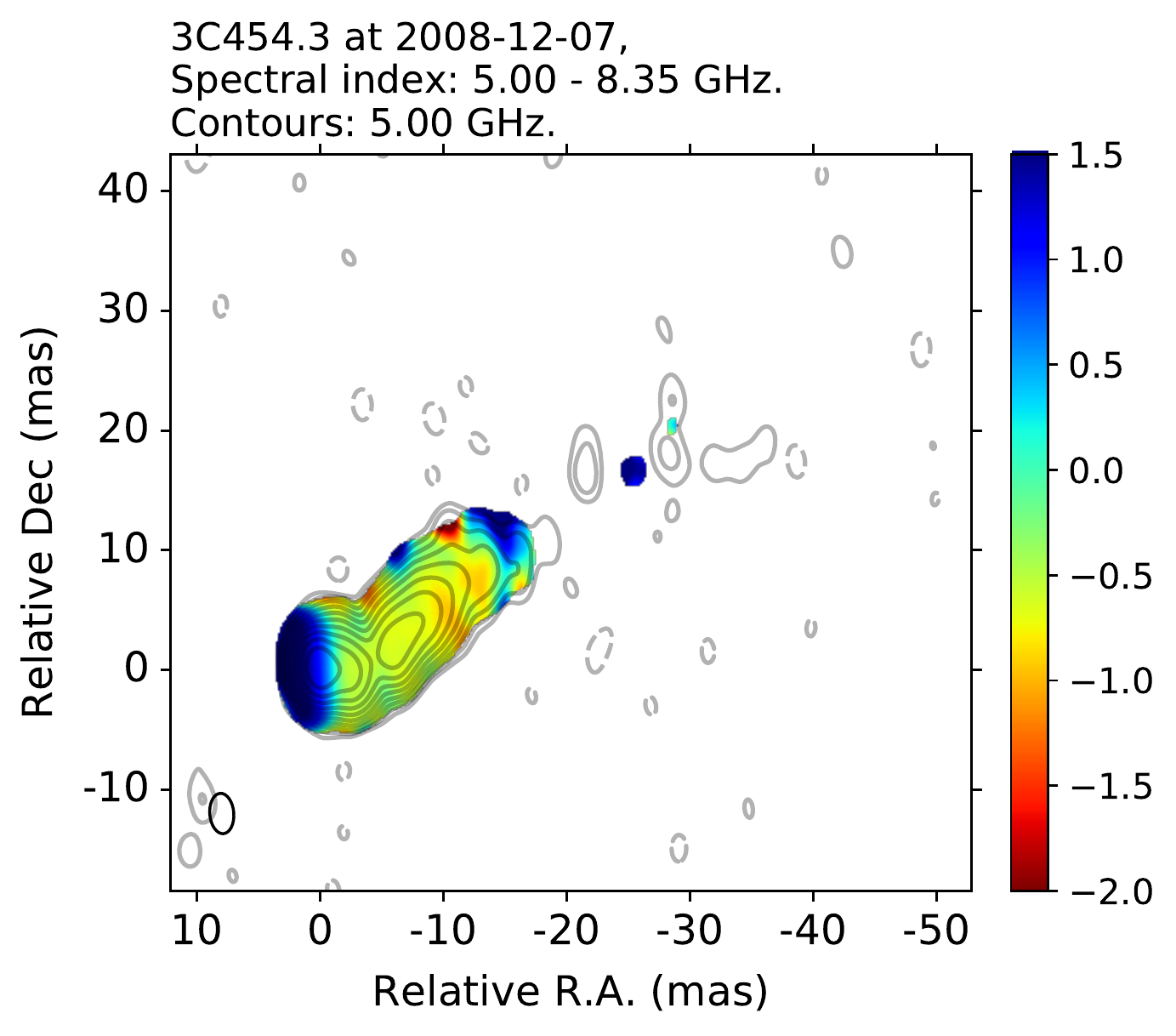}
    }
    \subfigure[]
    {
         \includegraphics[width=0.3\textwidth]{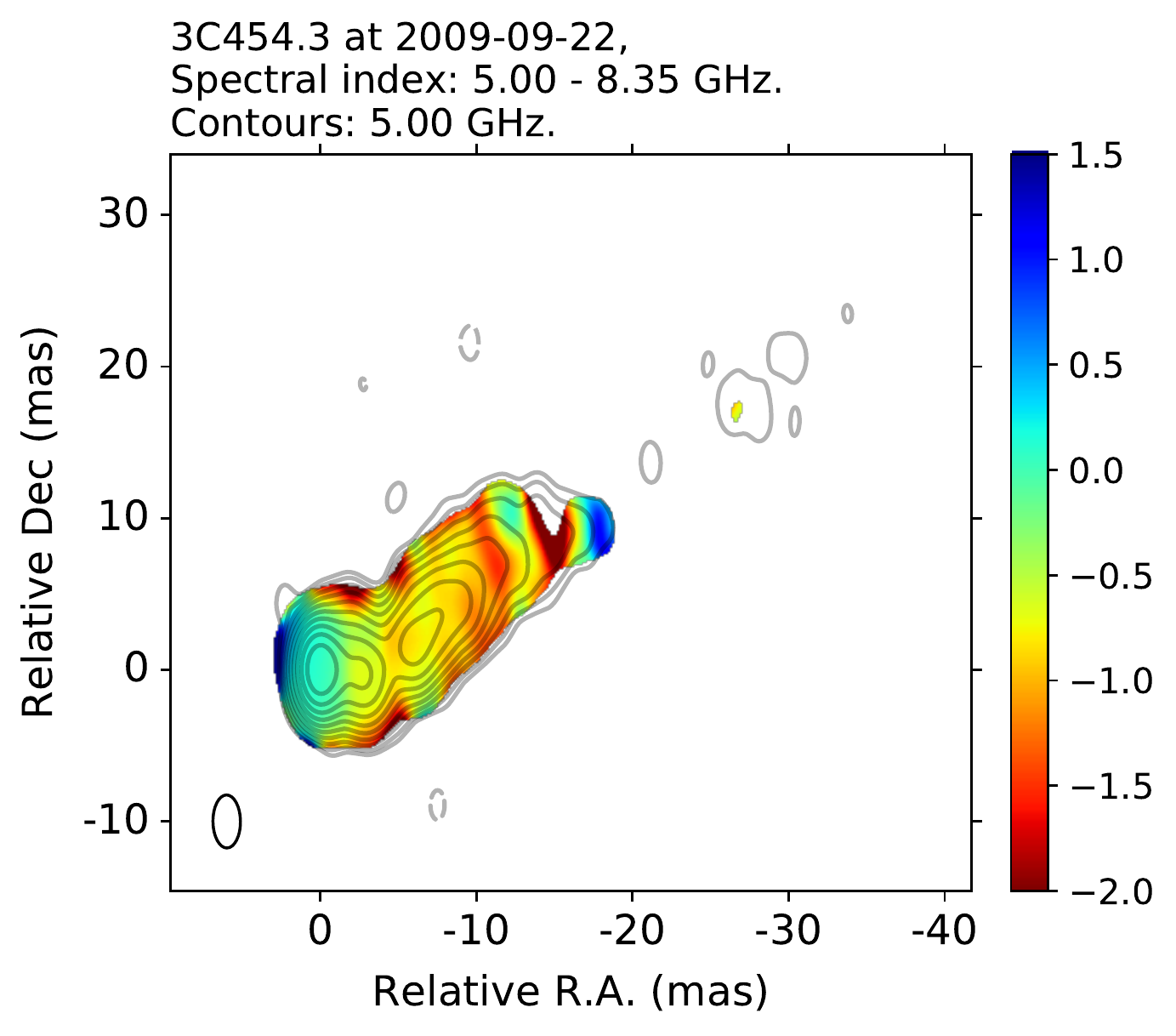}
    }
    \subfigure[]
    {
         \includegraphics[width=0.3\textwidth]{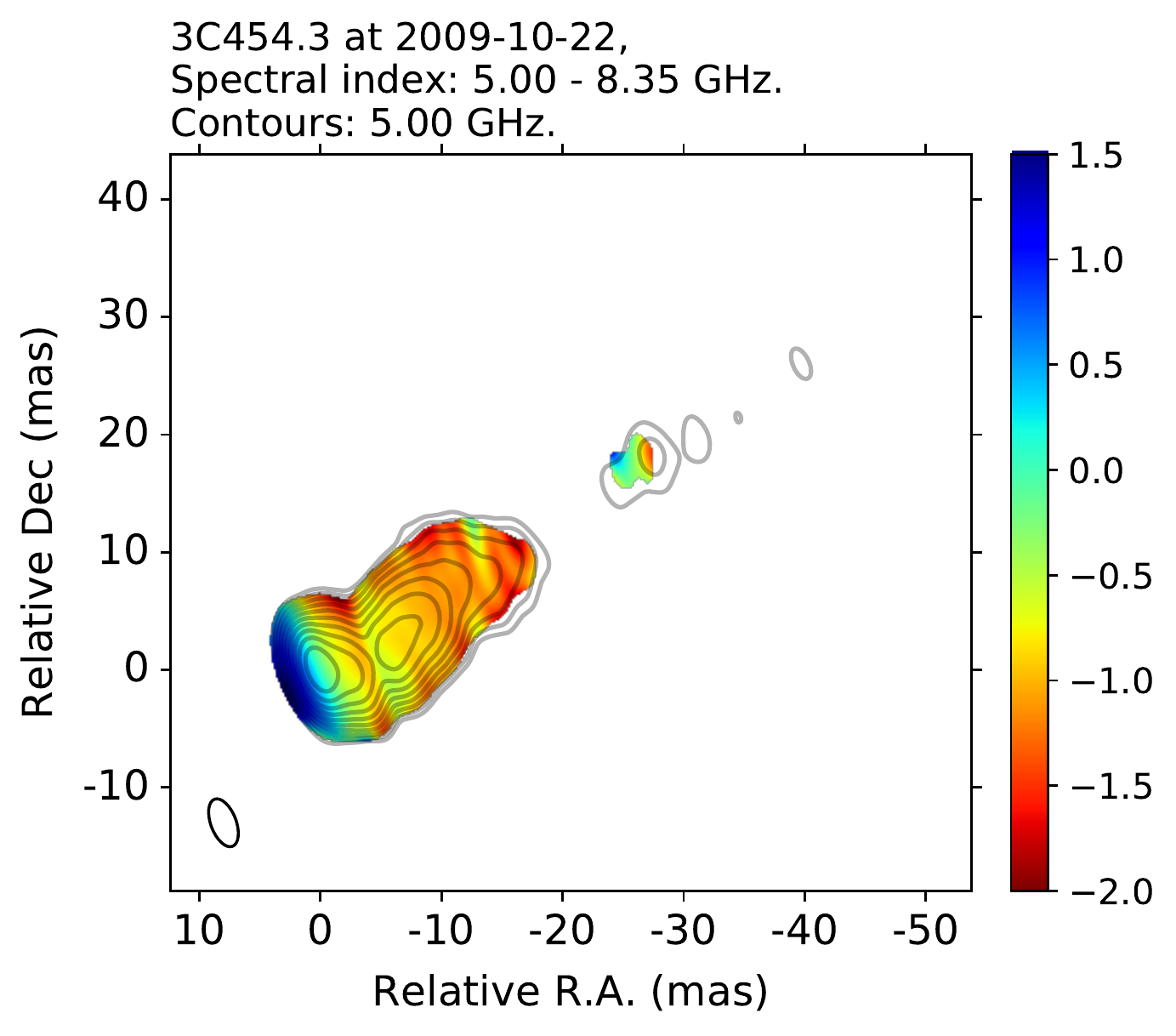}
    }
    \subfigure[]
    {
         \includegraphics[width=0.3\textwidth]{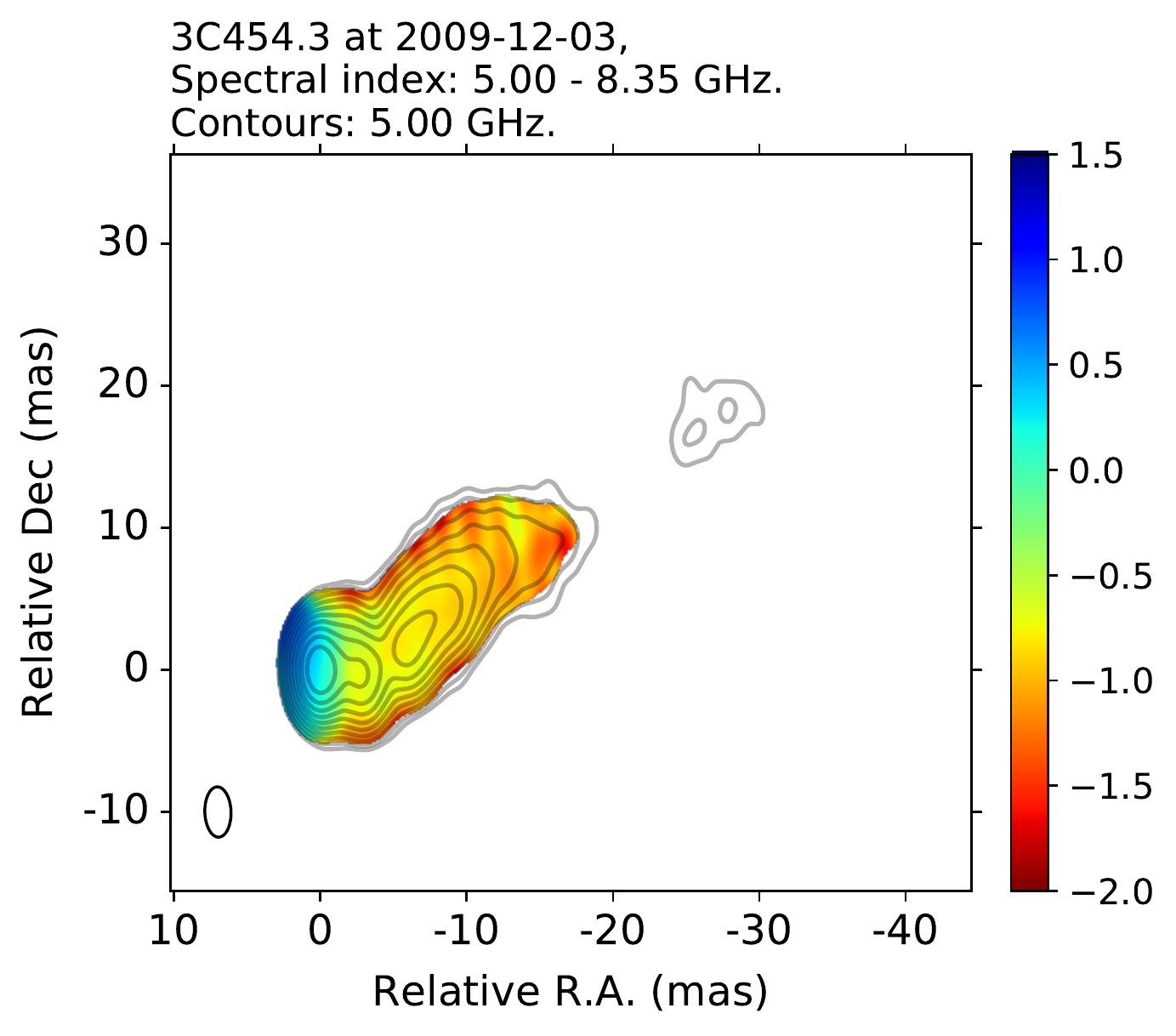}
        
    }
    \subfigure[]
    {
         \includegraphics[width=0.3\textwidth]{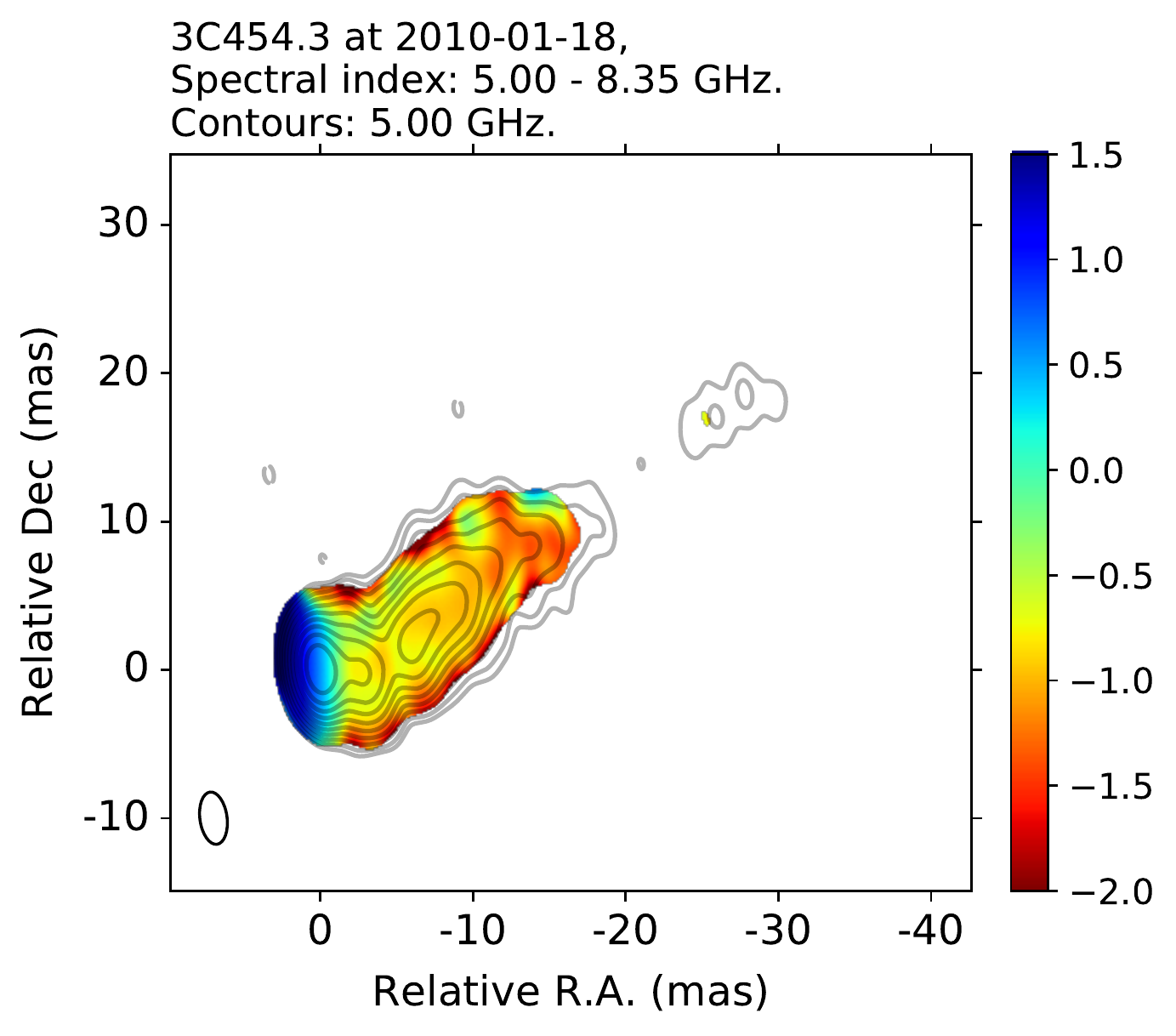}
    }   
    \subfigure[]
    {
         \includegraphics[width=0.3\textwidth]{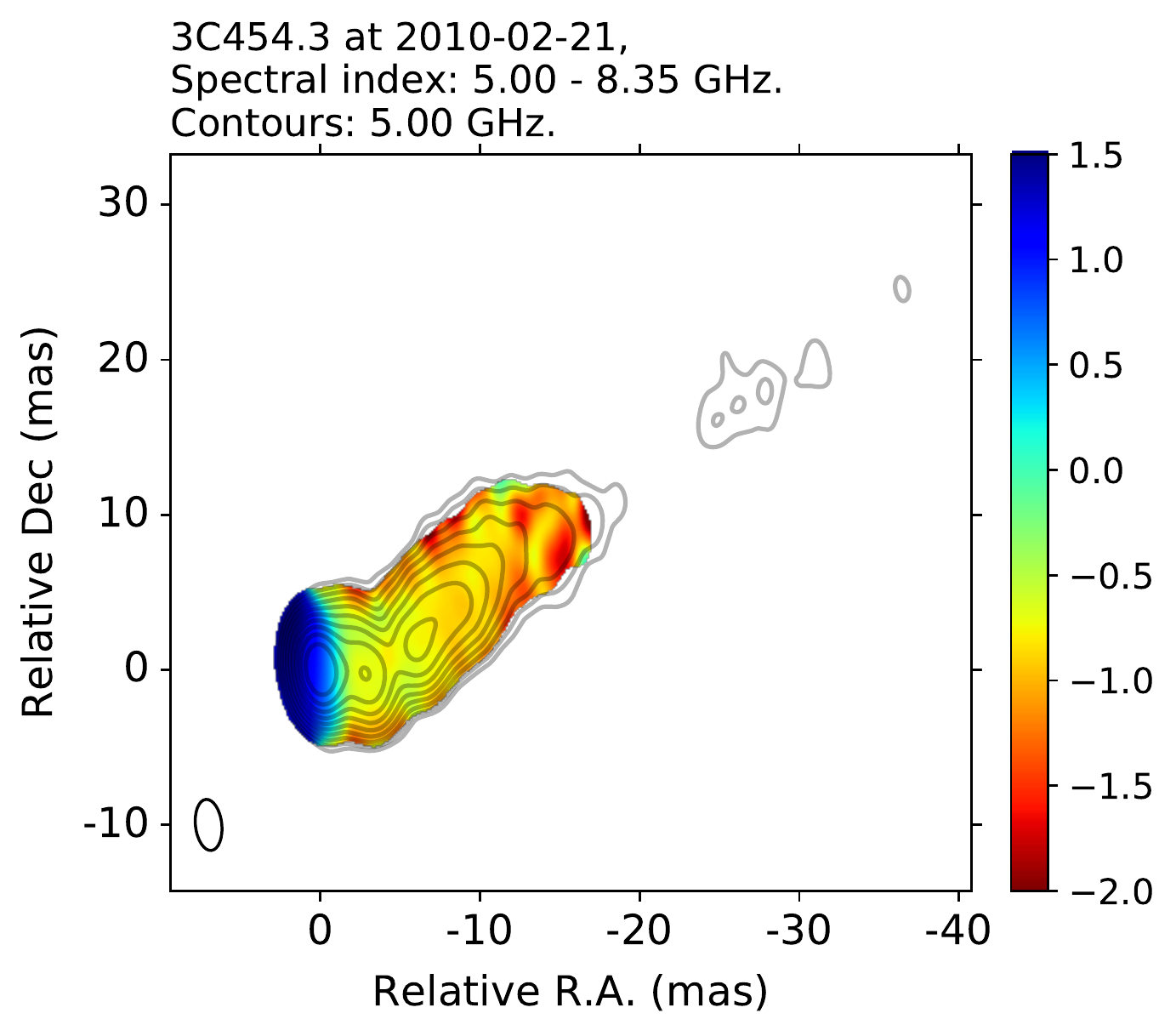}
    }    
        \caption{Continuation of Figure~F.1.}
    \label{siCXp2}
\end{figure*}

%%%%%%%%%%%%%%%%%% XU si maps %%%%%%%%%%%%%%%%%%%%%%%
\begin{figure*}[]
\centering
    \subfigure[]
    {
         \includegraphics[width=0.3\textwidth]{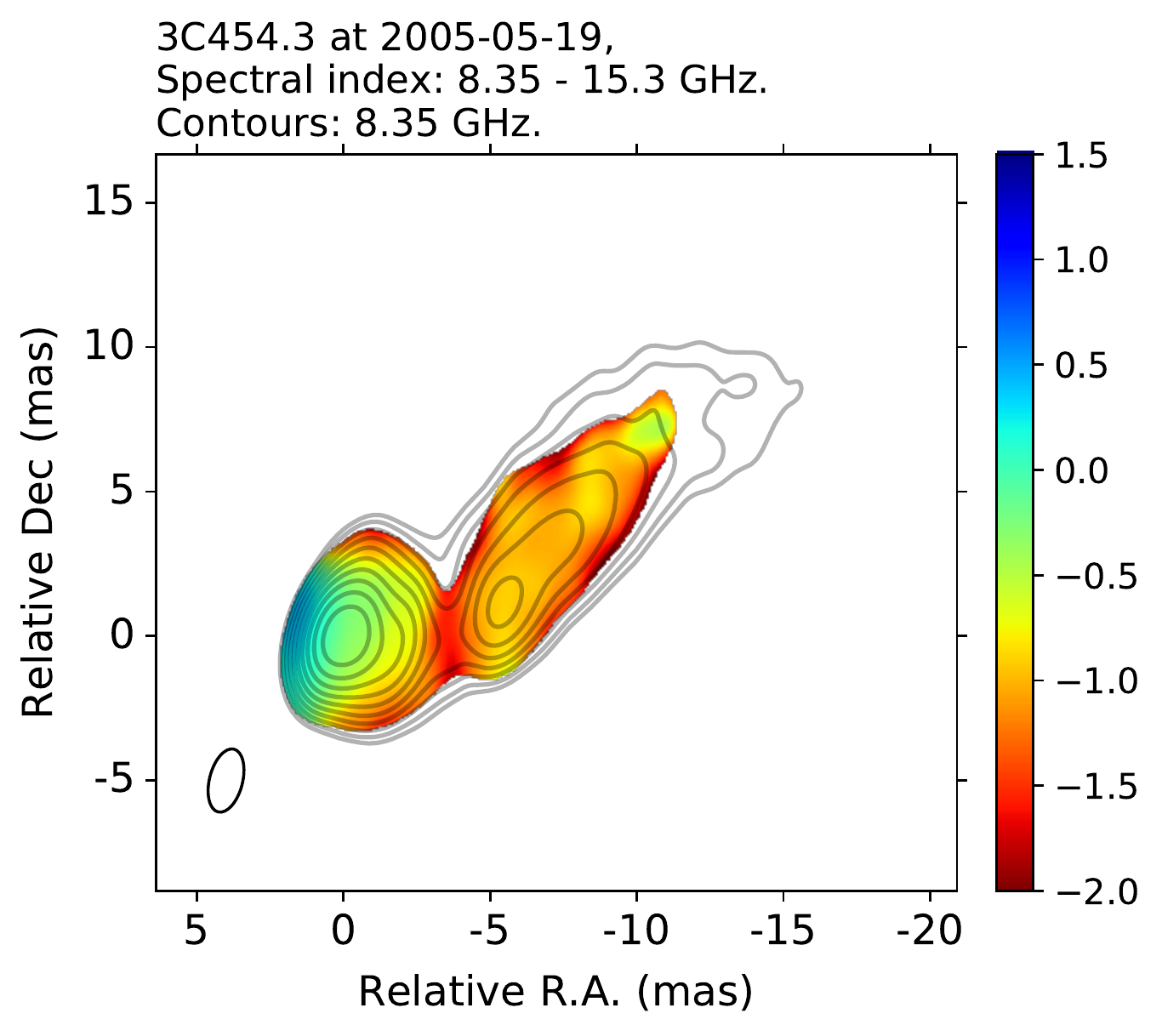}
    }
    \subfigure[]
    {
         \includegraphics[width=0.3\textwidth]{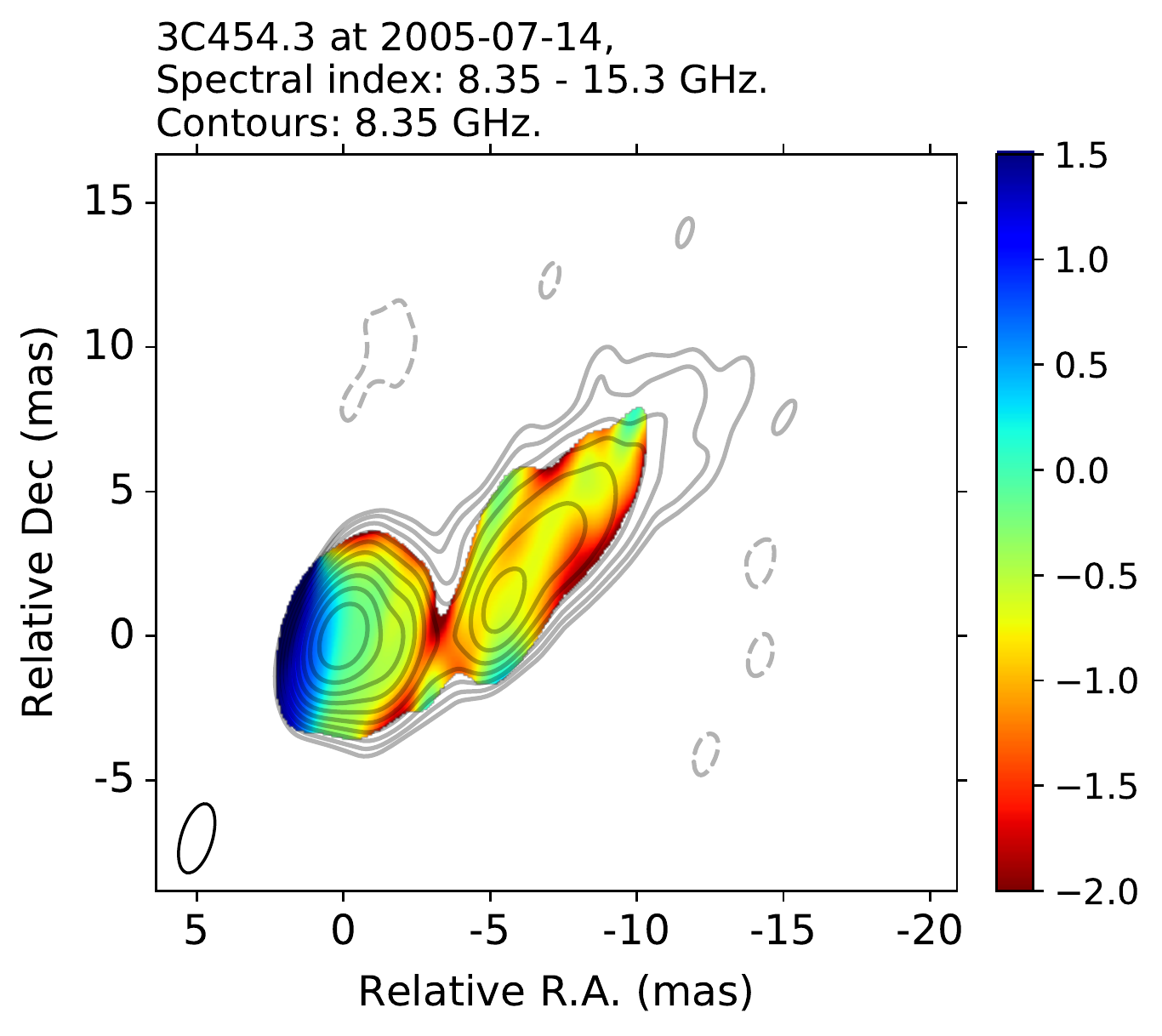}
    }
    \subfigure[]
    {
         \includegraphics[width=0.3\textwidth]{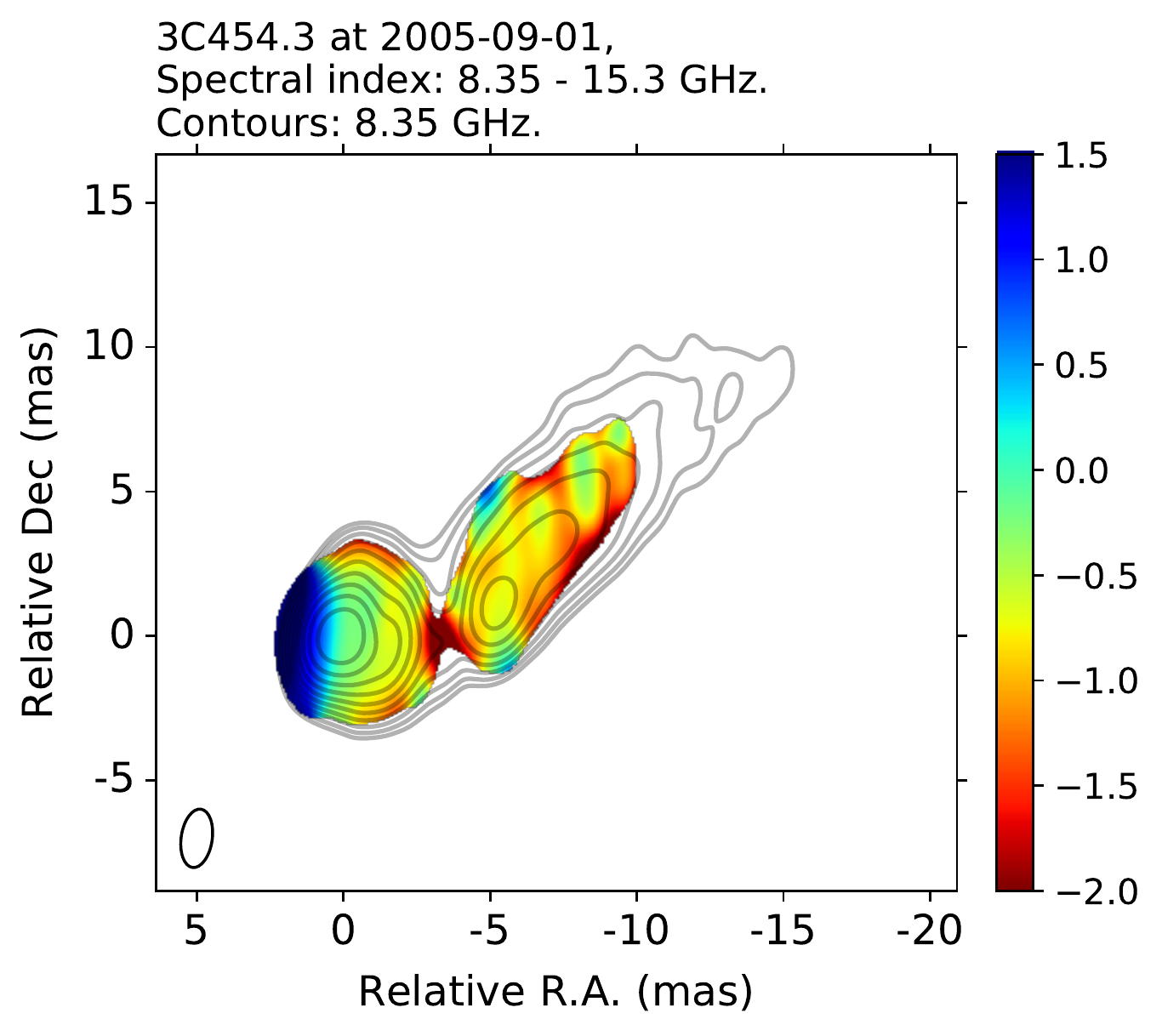}
    }
    \subfigure[]
    {
         \includegraphics[width=0.3\textwidth]{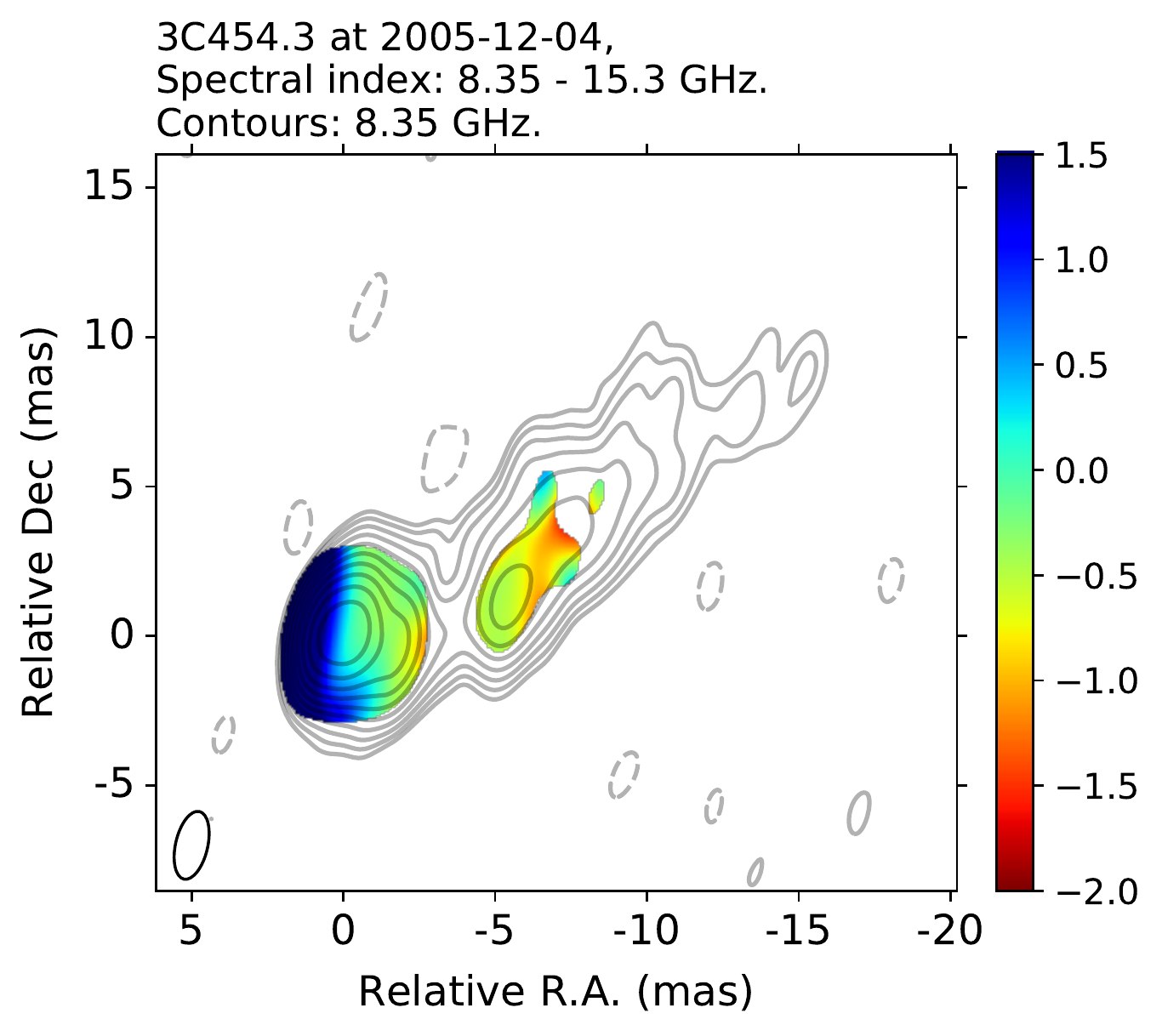}
     }
    \subfigure[]
    {
         \includegraphics[width=0.3\textwidth]{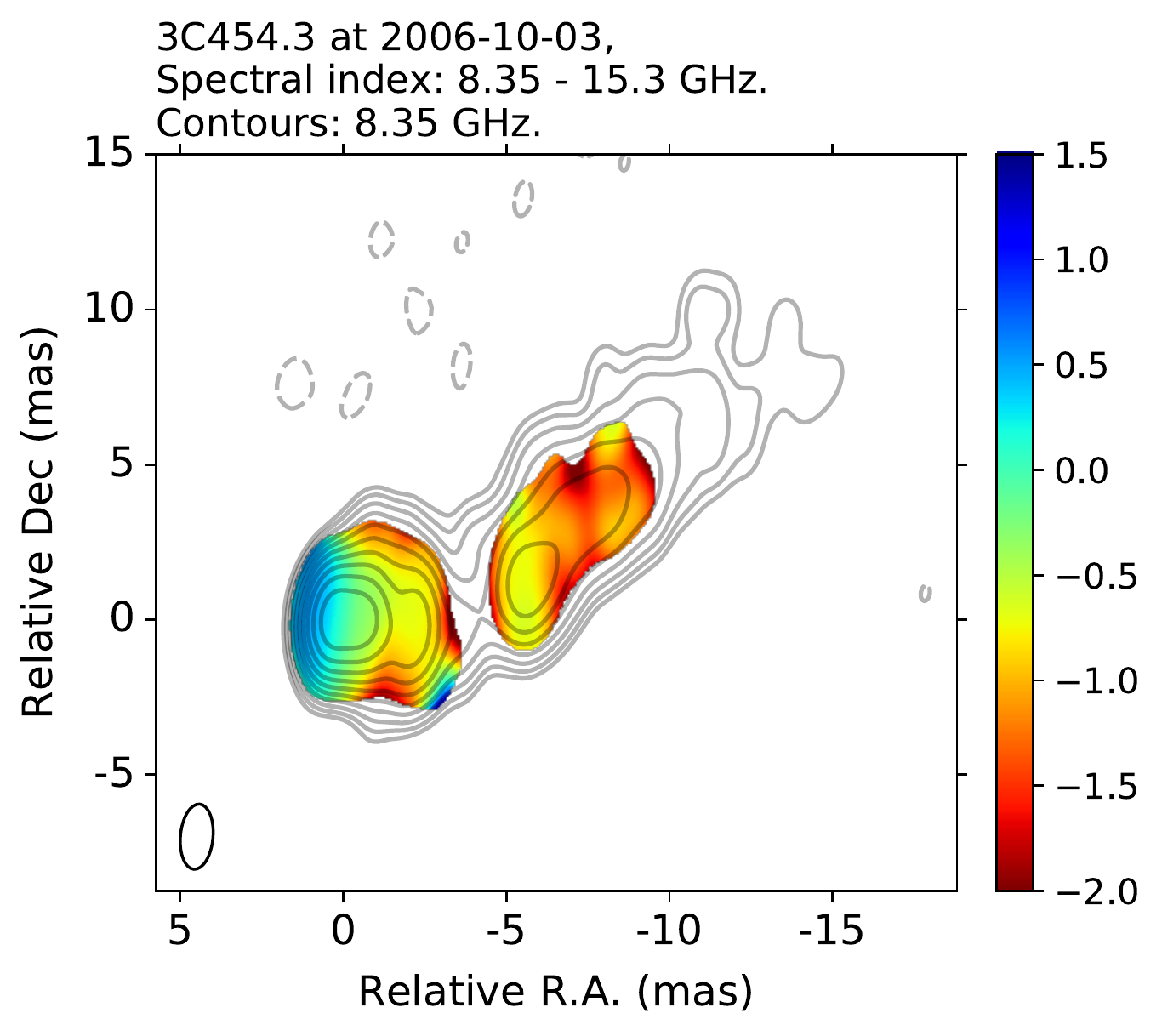}
    }   
    \subfigure[]
    {
         \includegraphics[width=0.3\textwidth]{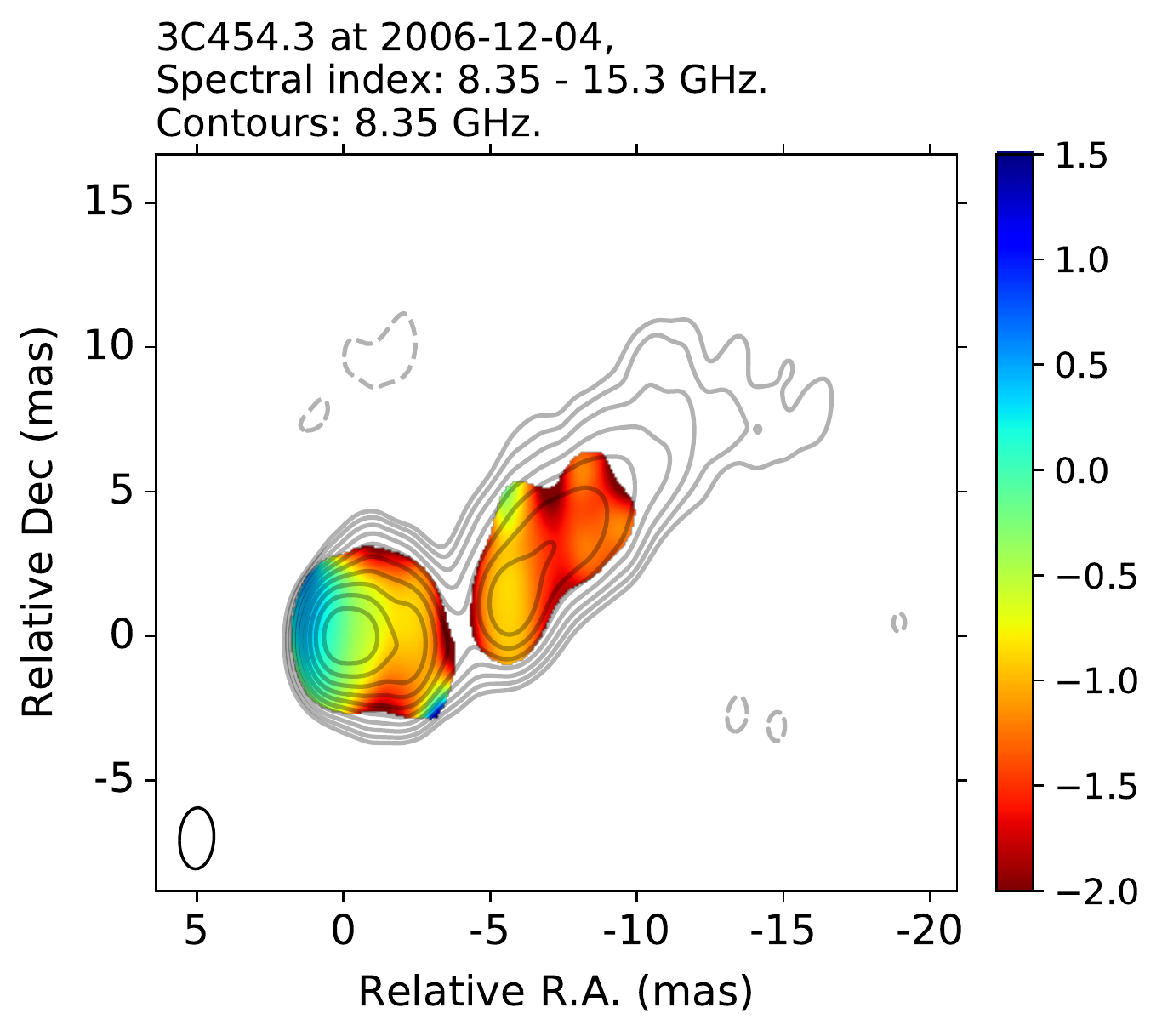}
    }    
    \subfigure[]
    {
         \includegraphics[width=0.3\textwidth]{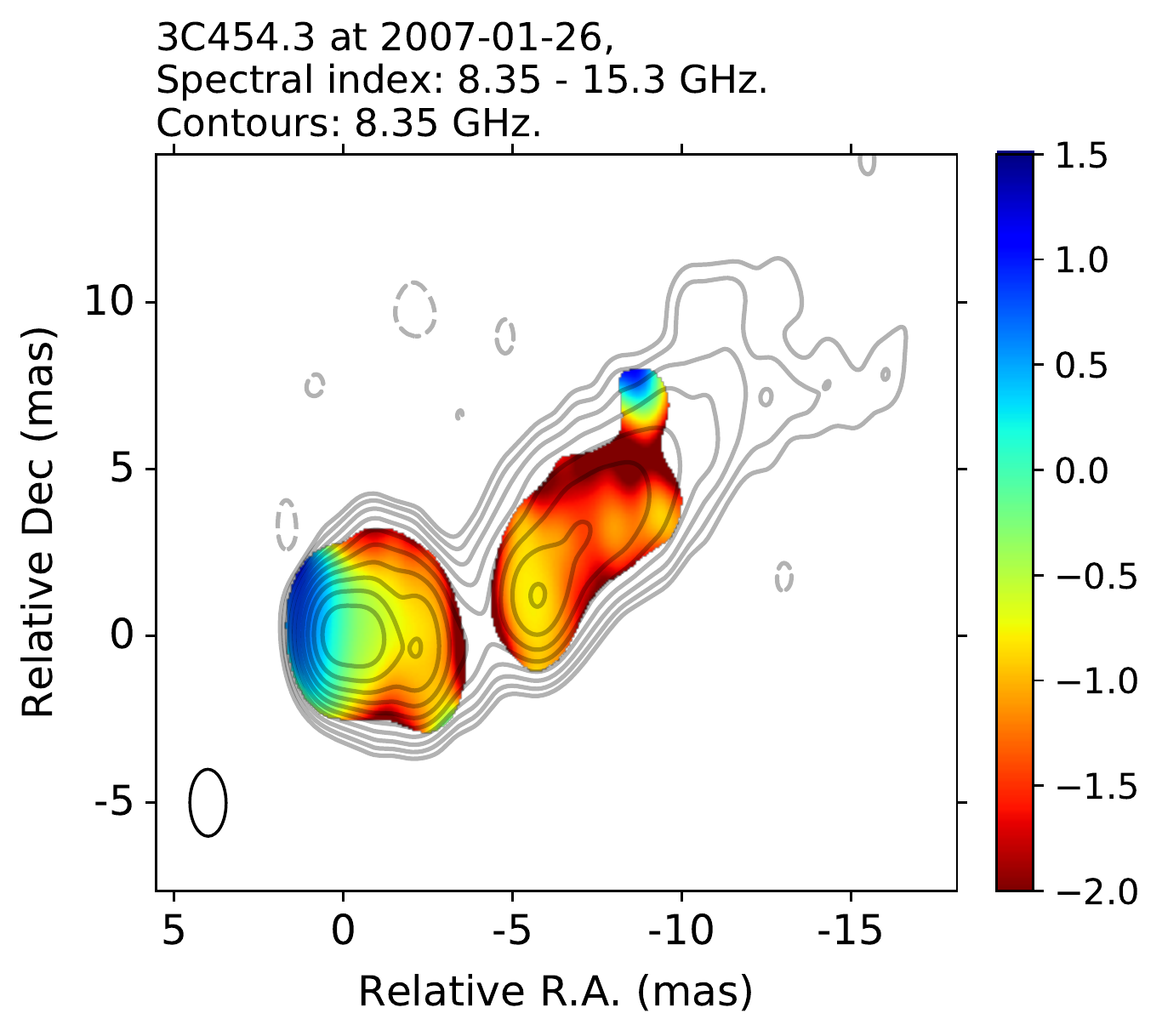}
    }   
    \subfigure[]
    {
         \includegraphics[width=0.3\textwidth]{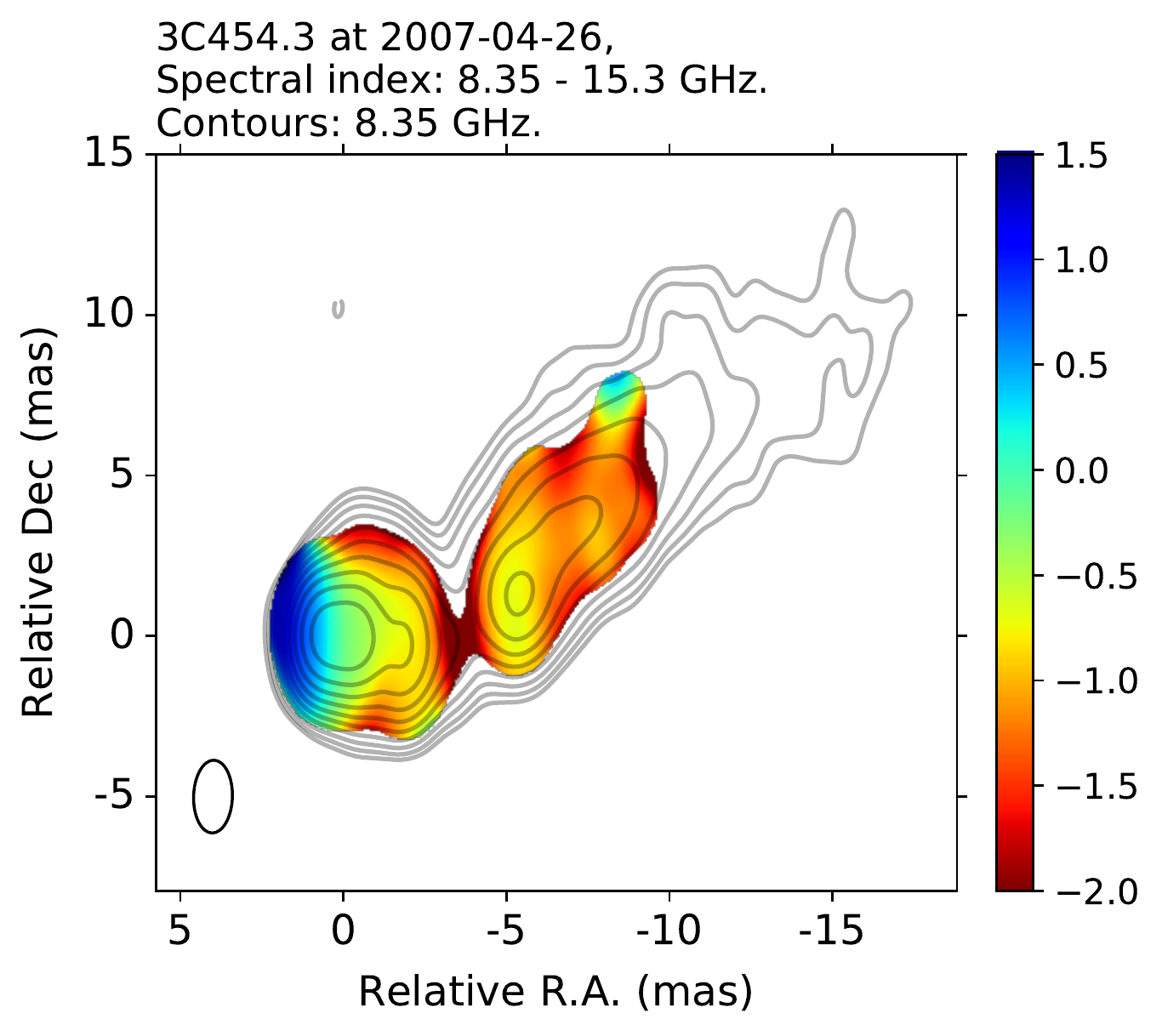}
    }    
        \subfigure[]
    {
         \includegraphics[width=0.3\textwidth]{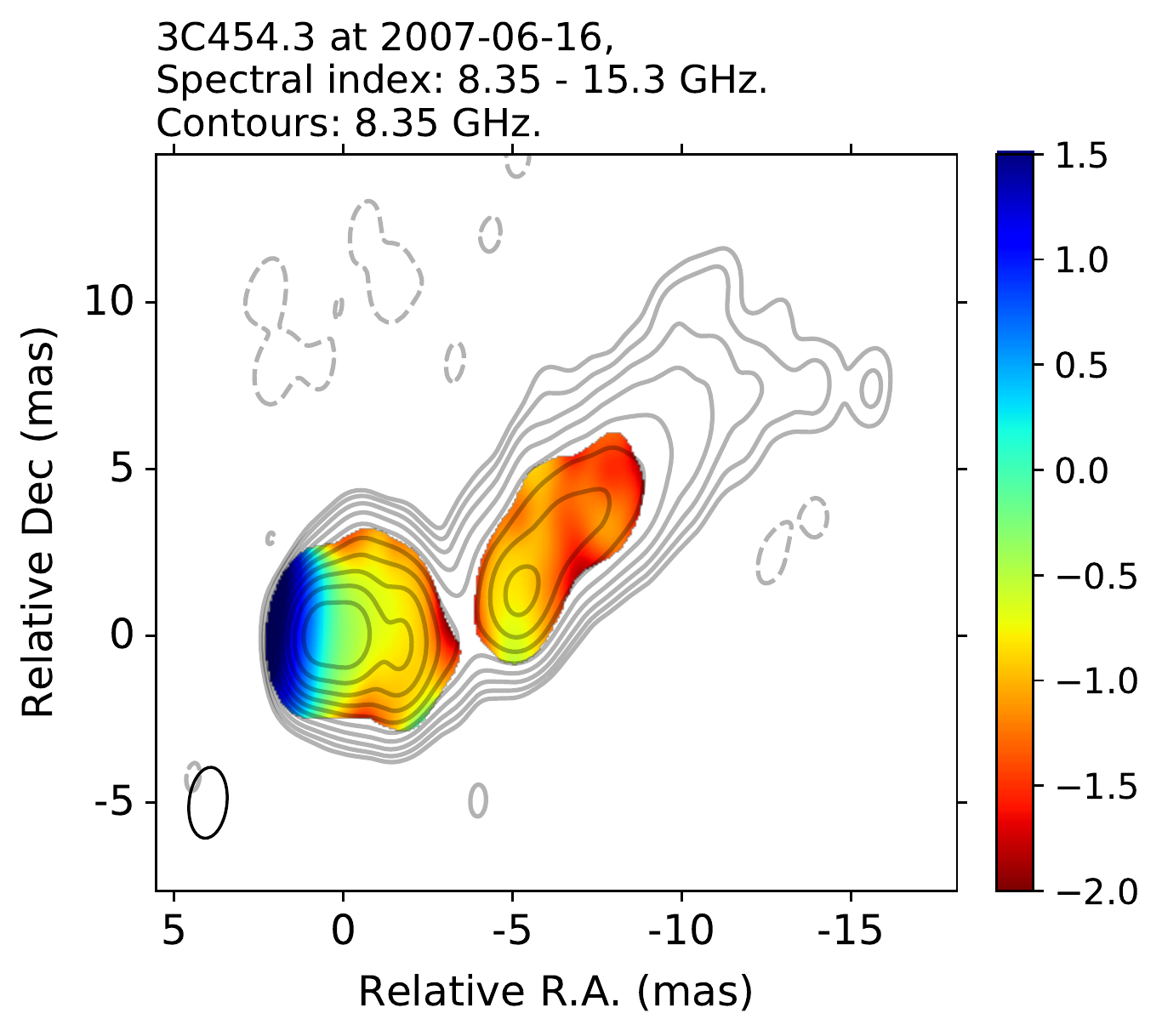}
    }    
        \subfigure[]
    {
         \includegraphics[width=0.3\textwidth]{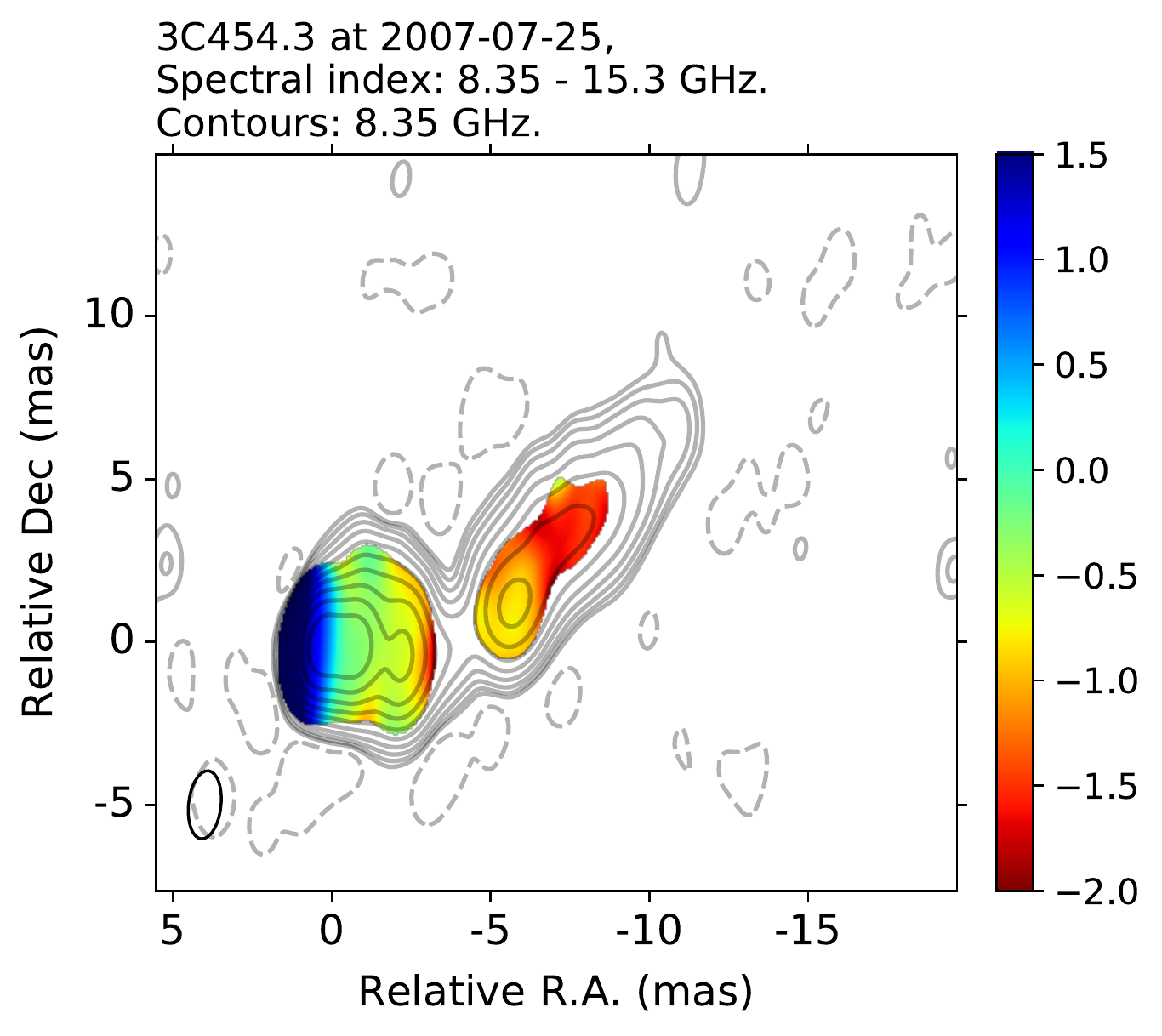}
    } 
            \subfigure[]
    {
         \includegraphics[width=0.3\textwidth]{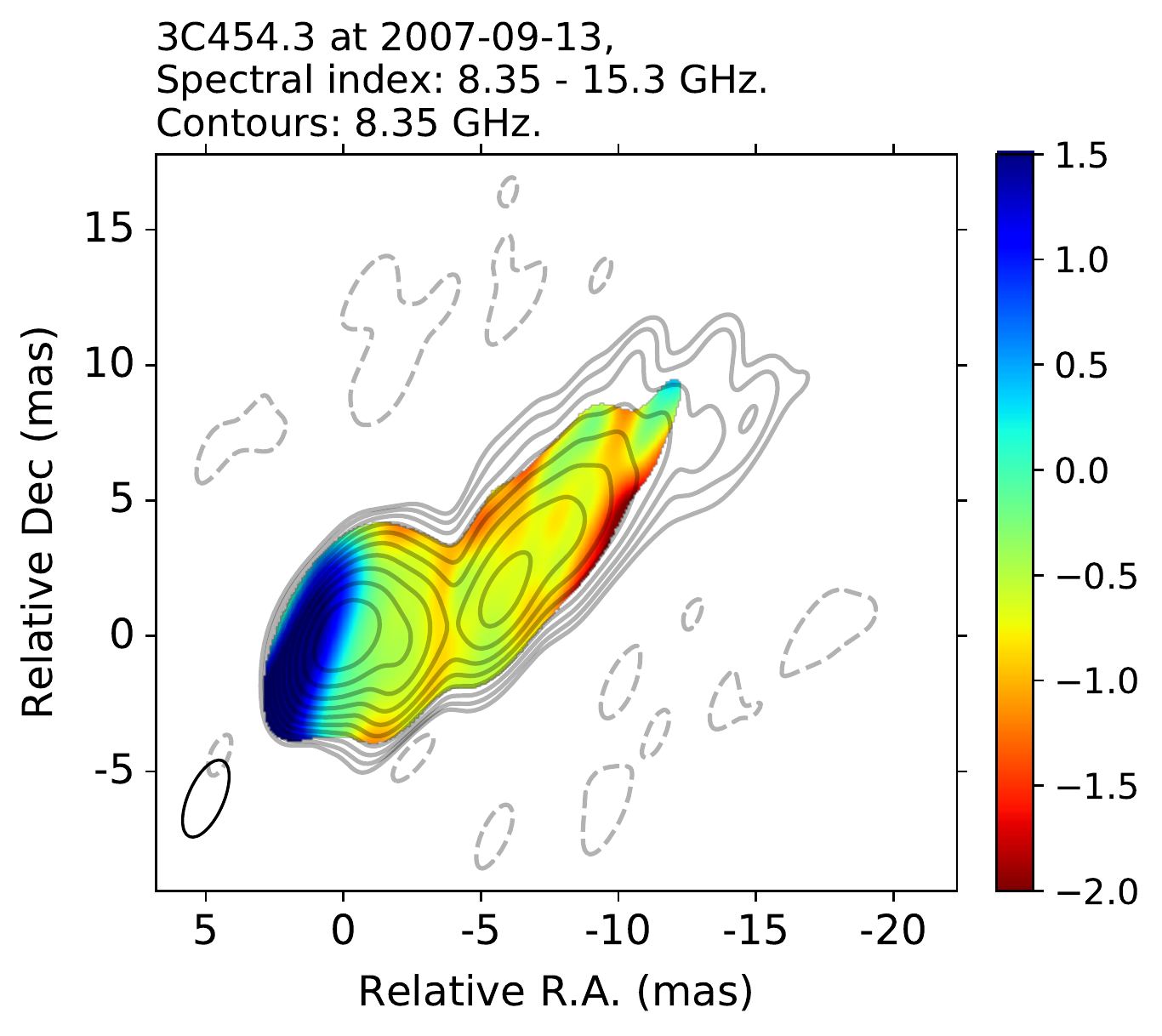}
    } 
            \subfigure[]
    {
         \includegraphics[width=0.3\textwidth]{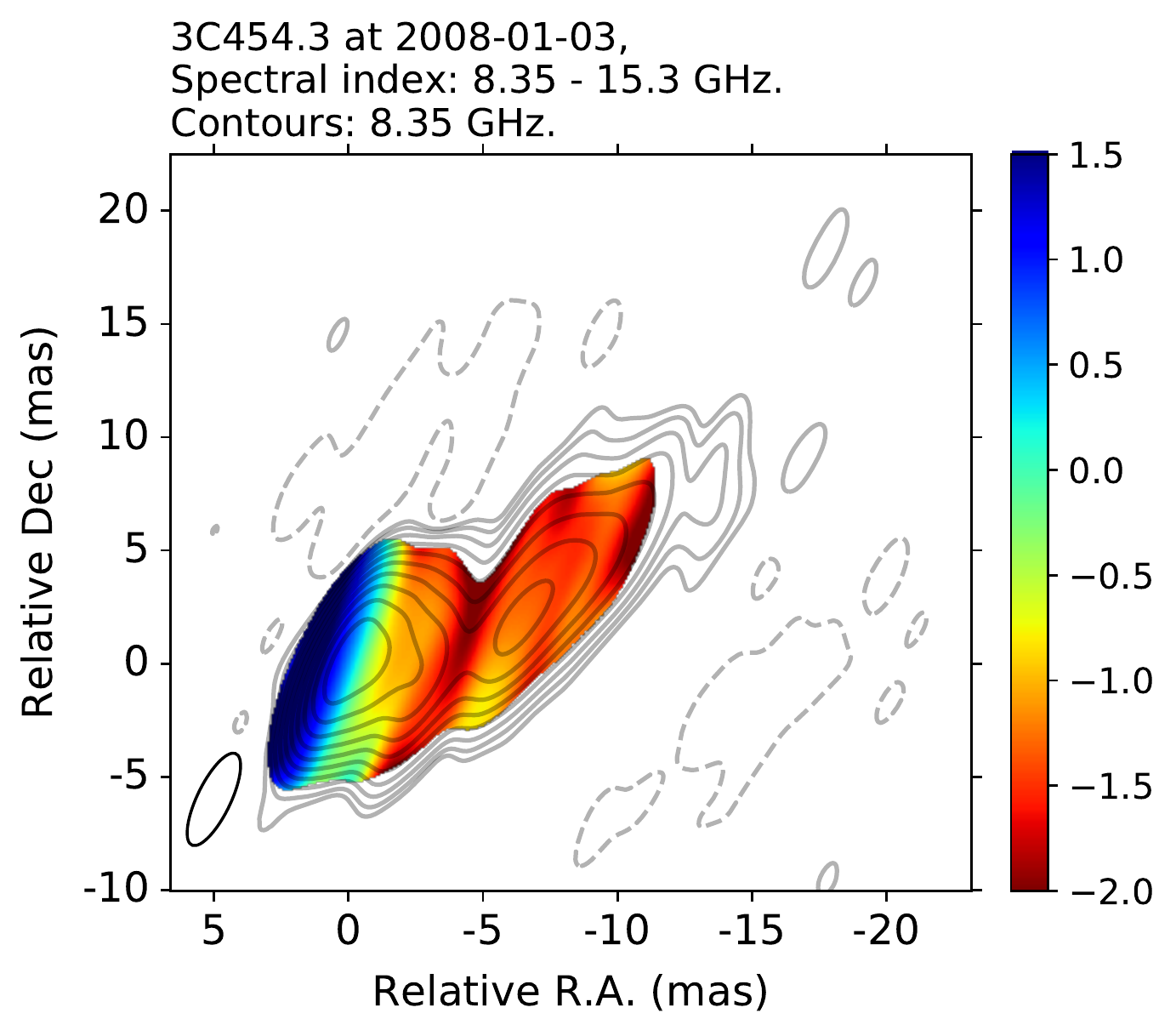}
    } 
     \caption{Spectral index maps for the frequency pair XU (8 - 15\,GHz). The colorbar indicates the spectral index. The ellipse on the bottom left corner represents the interferometric beam. The contour lines are given at   -0.1\%, 0.1\% 0.2\%, 0.4\%, 0.8\%, 1.6\%, 3.2\%, 6.4\%, 12.8\%, 25.6\%, and 51.2\% of the peak intensity at each image. }
    \label{siXUp1}
\end{figure*}

\begin{figure*}[]
\centering
    \subfigure[]
    {
         \includegraphics[width=0.3\textwidth]{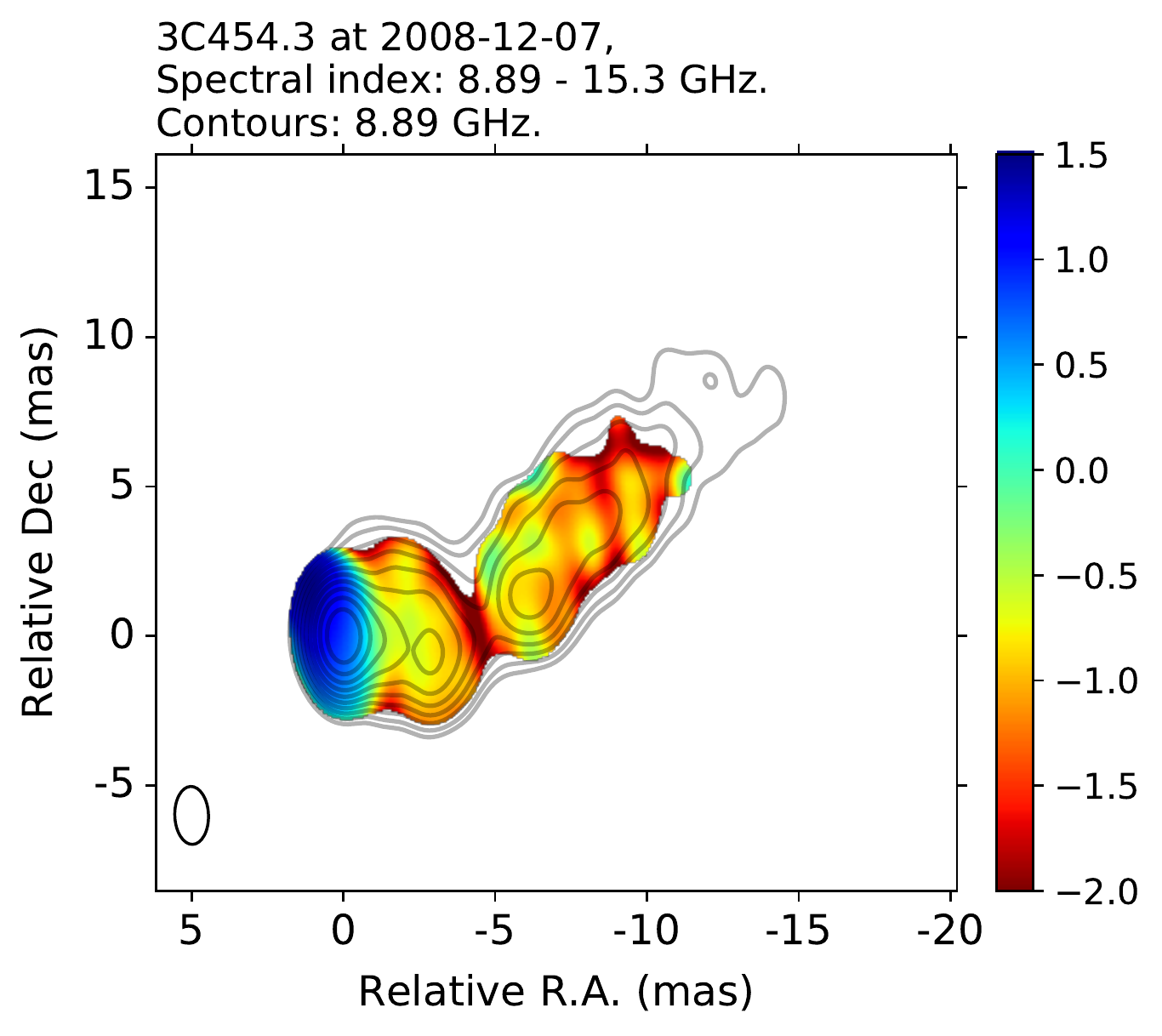}
    }
    \subfigure[]
    {
         \includegraphics[width=0.3\textwidth]{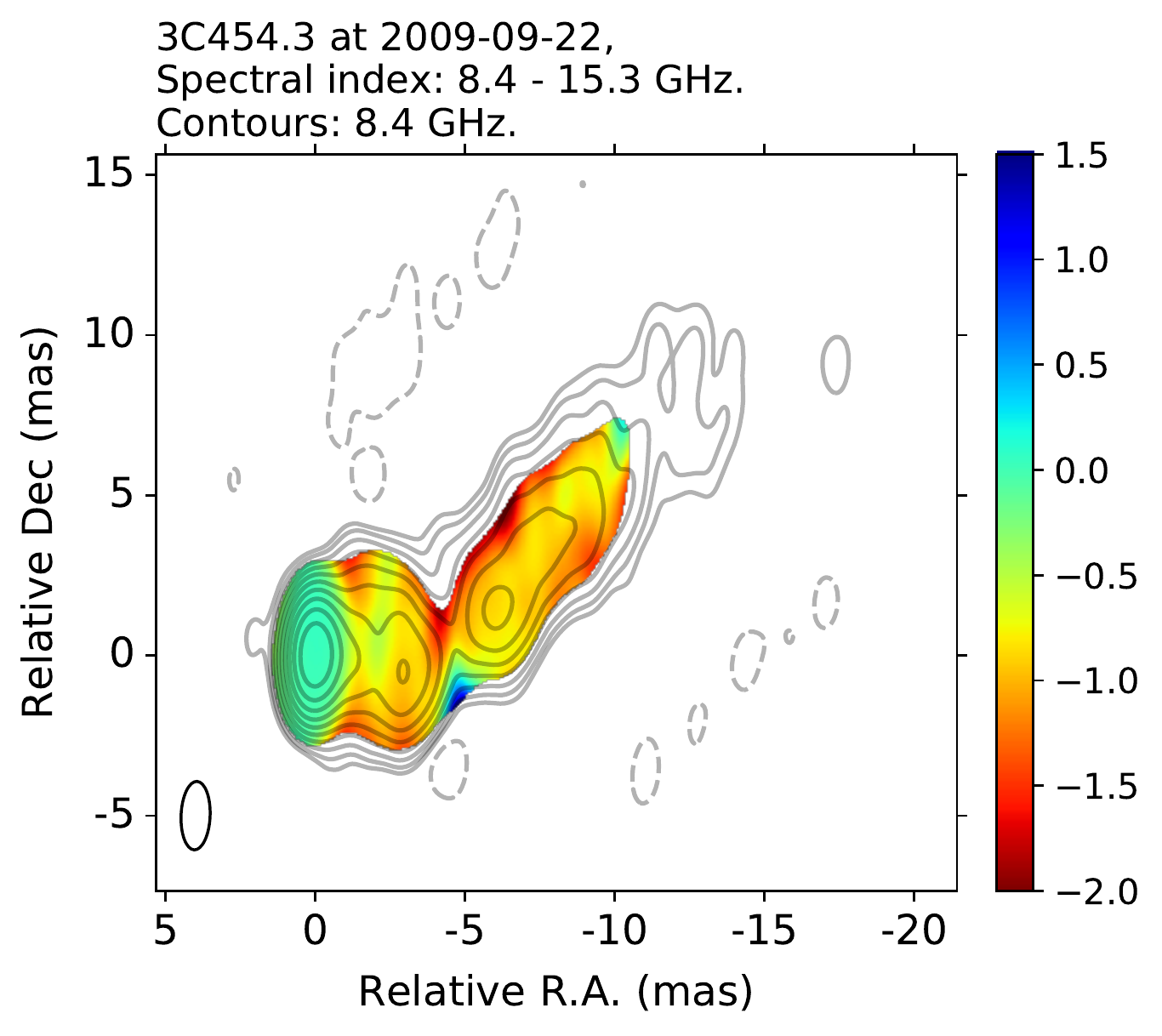}
    }
    \subfigure[]
    {
         \includegraphics[width=0.3\textwidth]{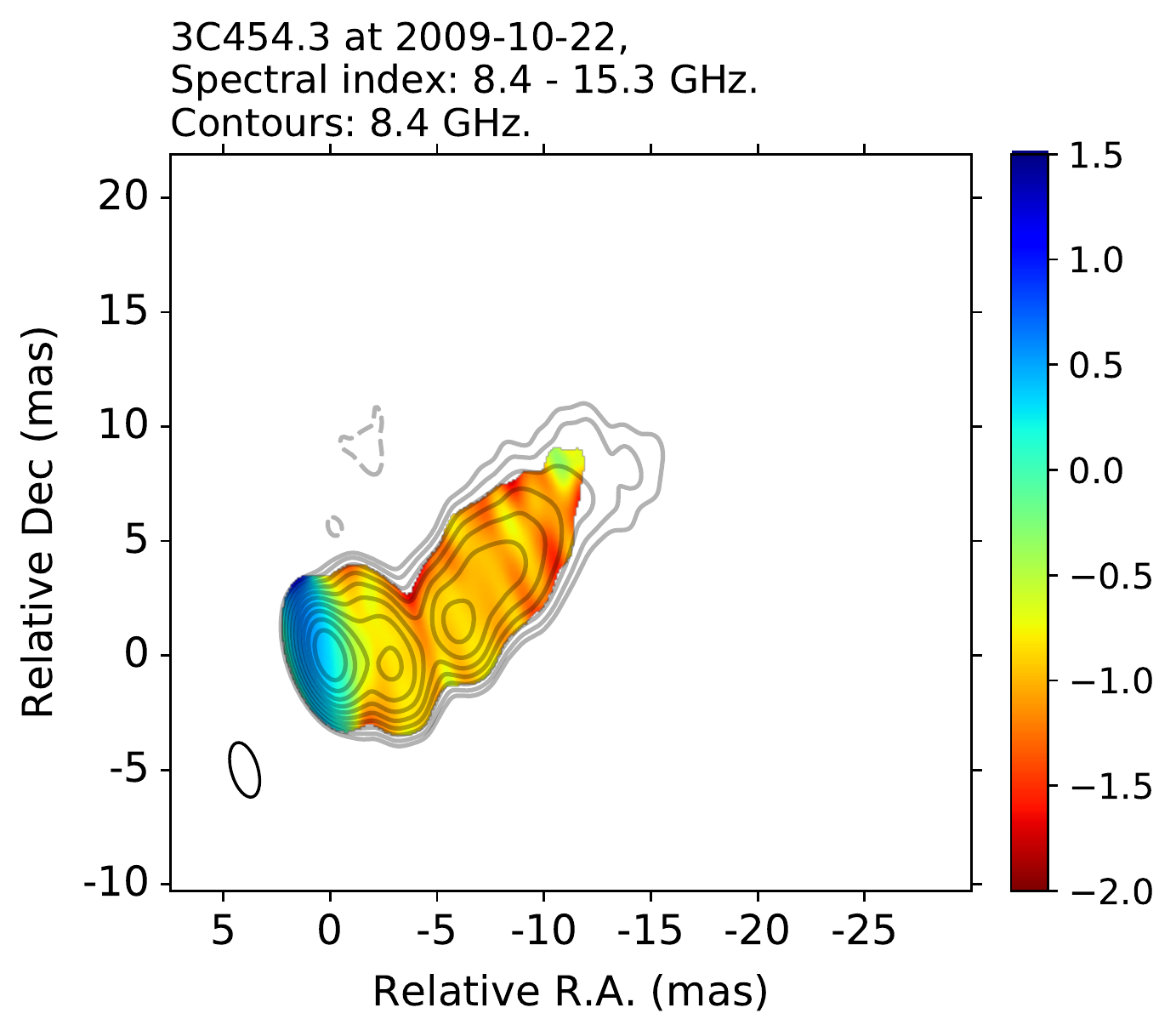}
    }
    \subfigure[]
    {
         \includegraphics[width=0.3\textwidth]{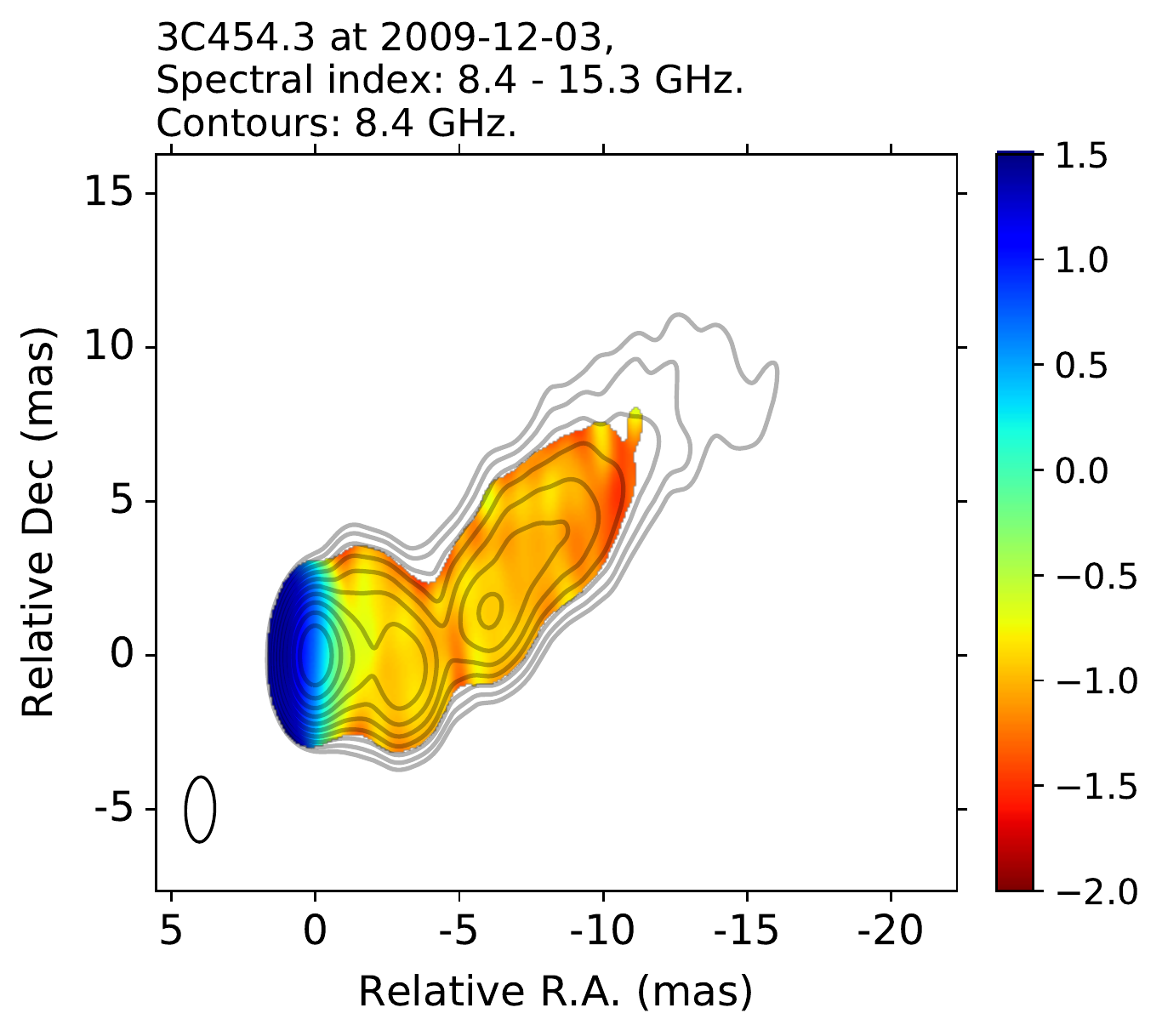}
    }
    \subfigure[]
    {
         \includegraphics[width=0.3\textwidth]{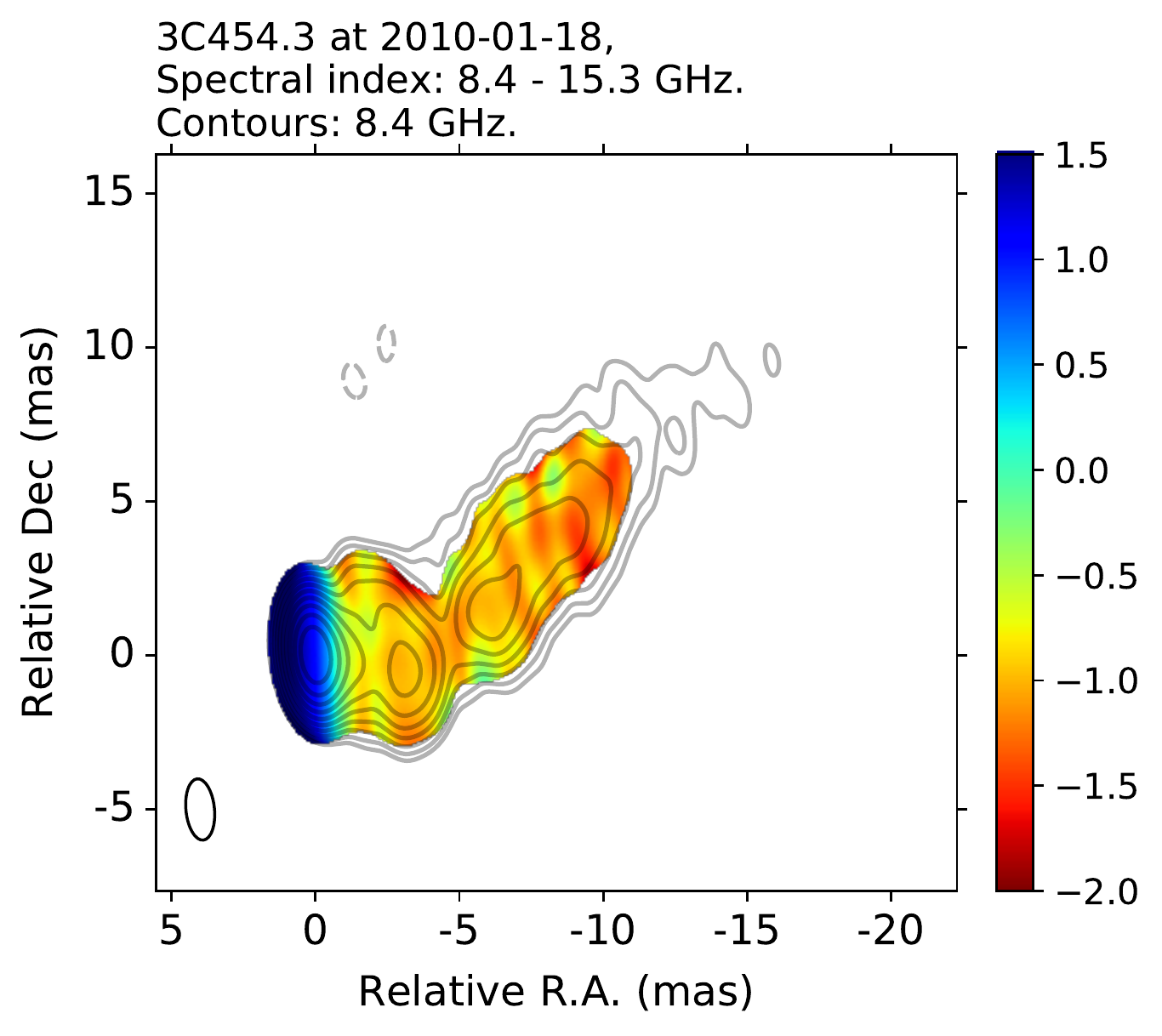}
    }   
    \subfigure[]
    {
         \includegraphics[width=0.3\textwidth]{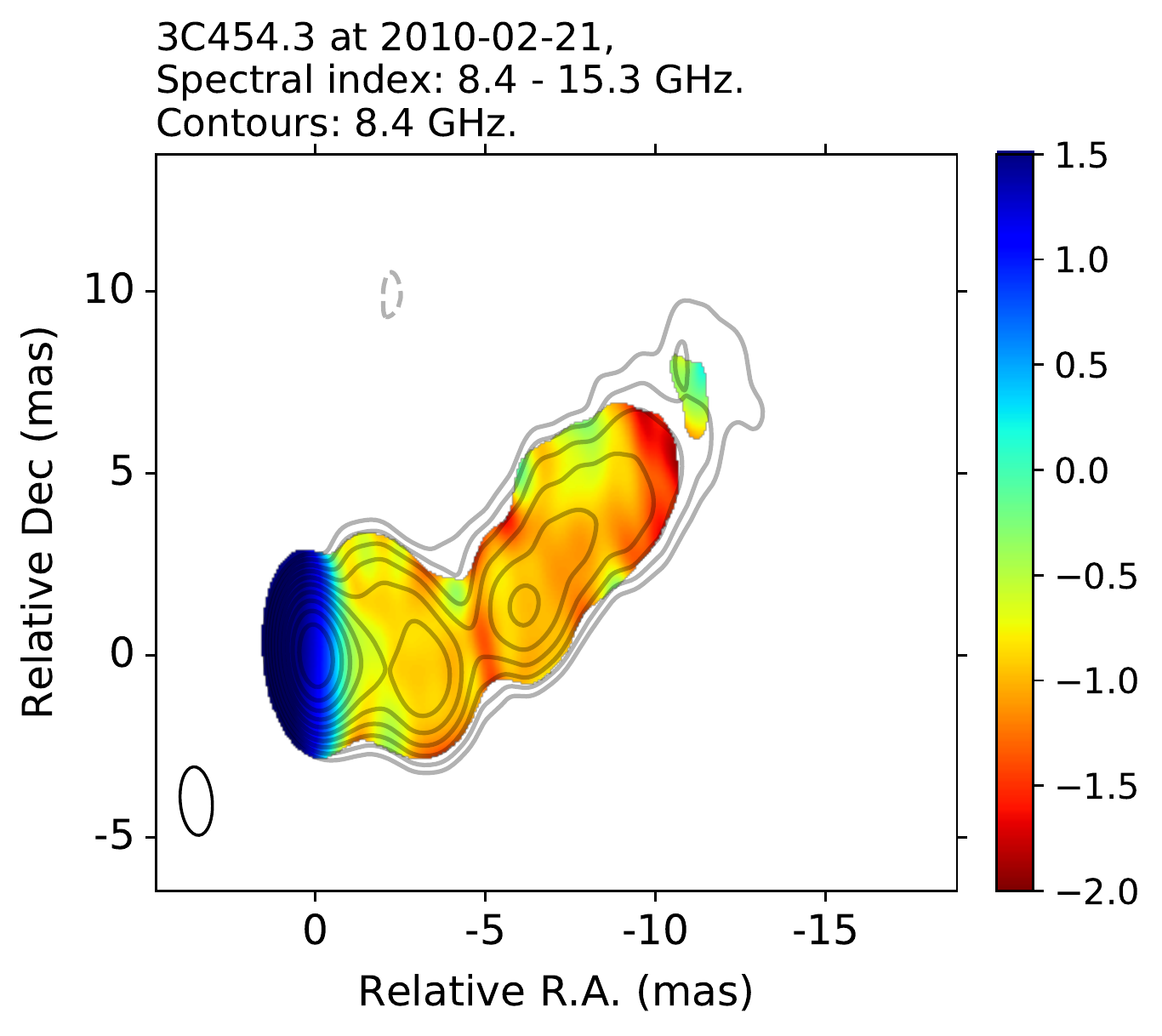}
    }    
        \caption{Continuation of Figure~F.3.}
    \label{siXUp2}
\end{figure*}

%%%%%%%%%%%%%%%%%% UK si maps %%%%%%%%%%%%%%%%%%%%%%%
\begin{figure*}[]
\centering
    \subfigure[]
    {
         \includegraphics[width=0.3\textwidth]{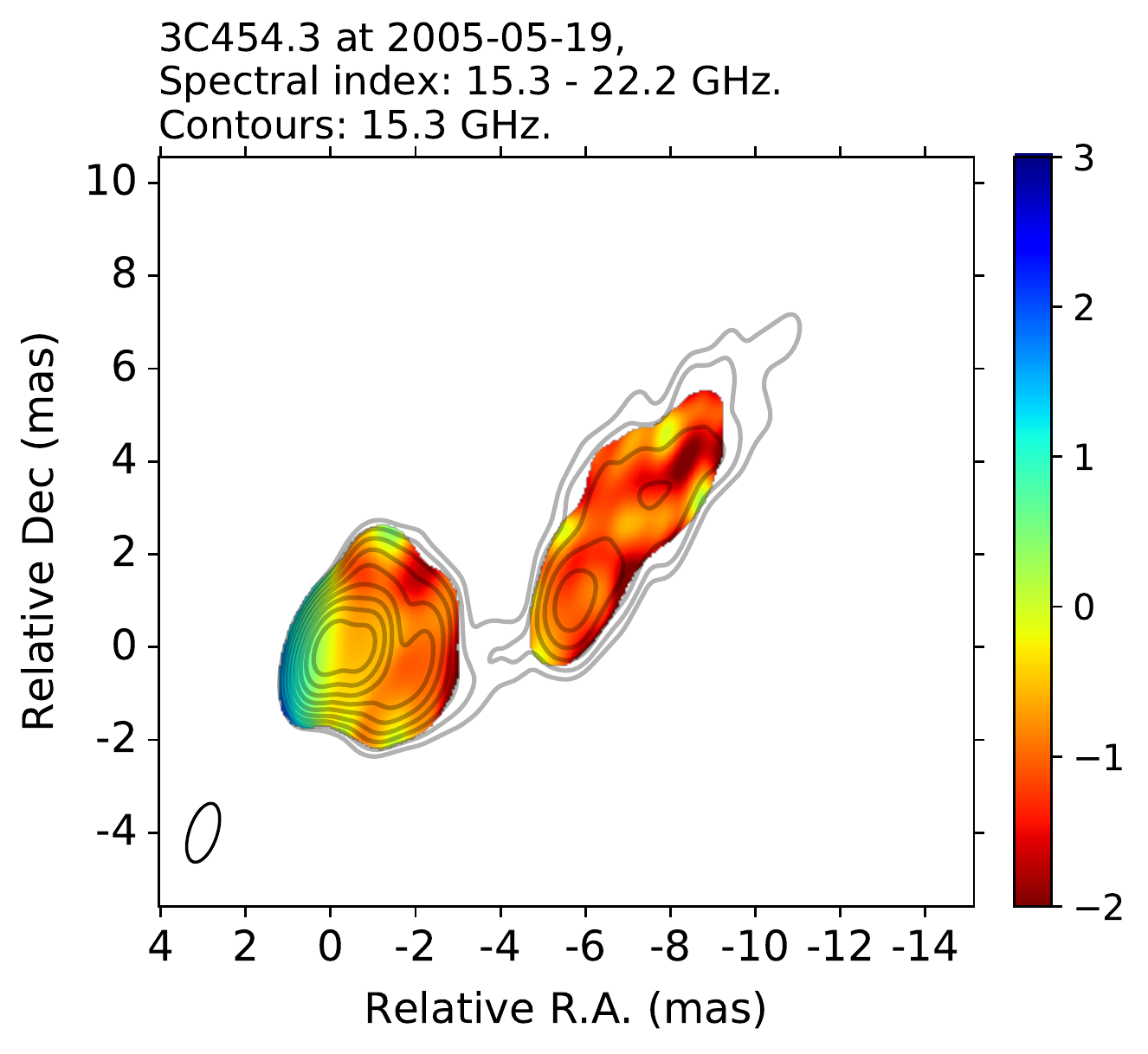}
    }
    \subfigure[]
    {
         \includegraphics[width=0.3\textwidth]{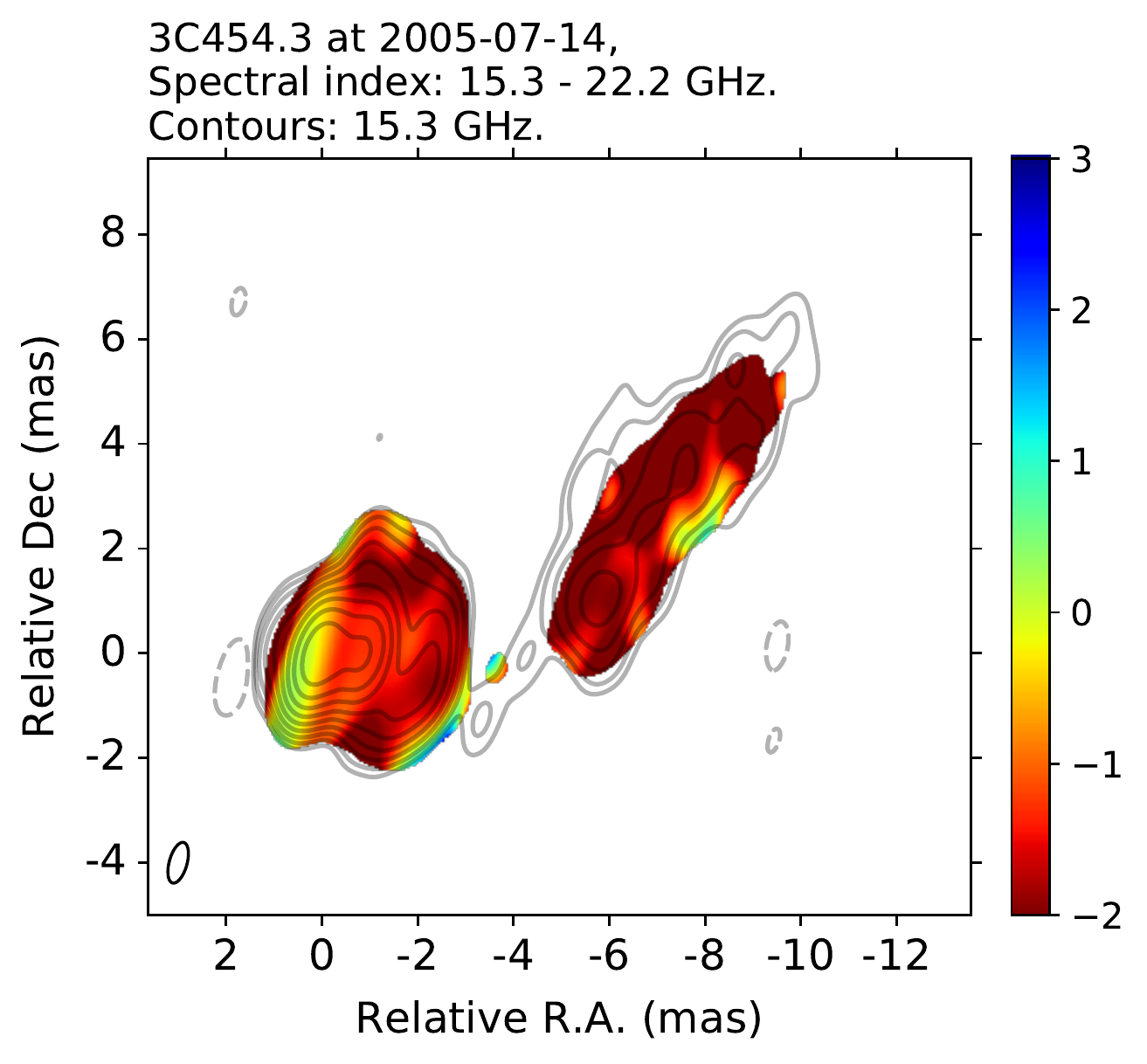}
        
    }
    \subfigure[]
    {
         \includegraphics[width=0.3\textwidth]{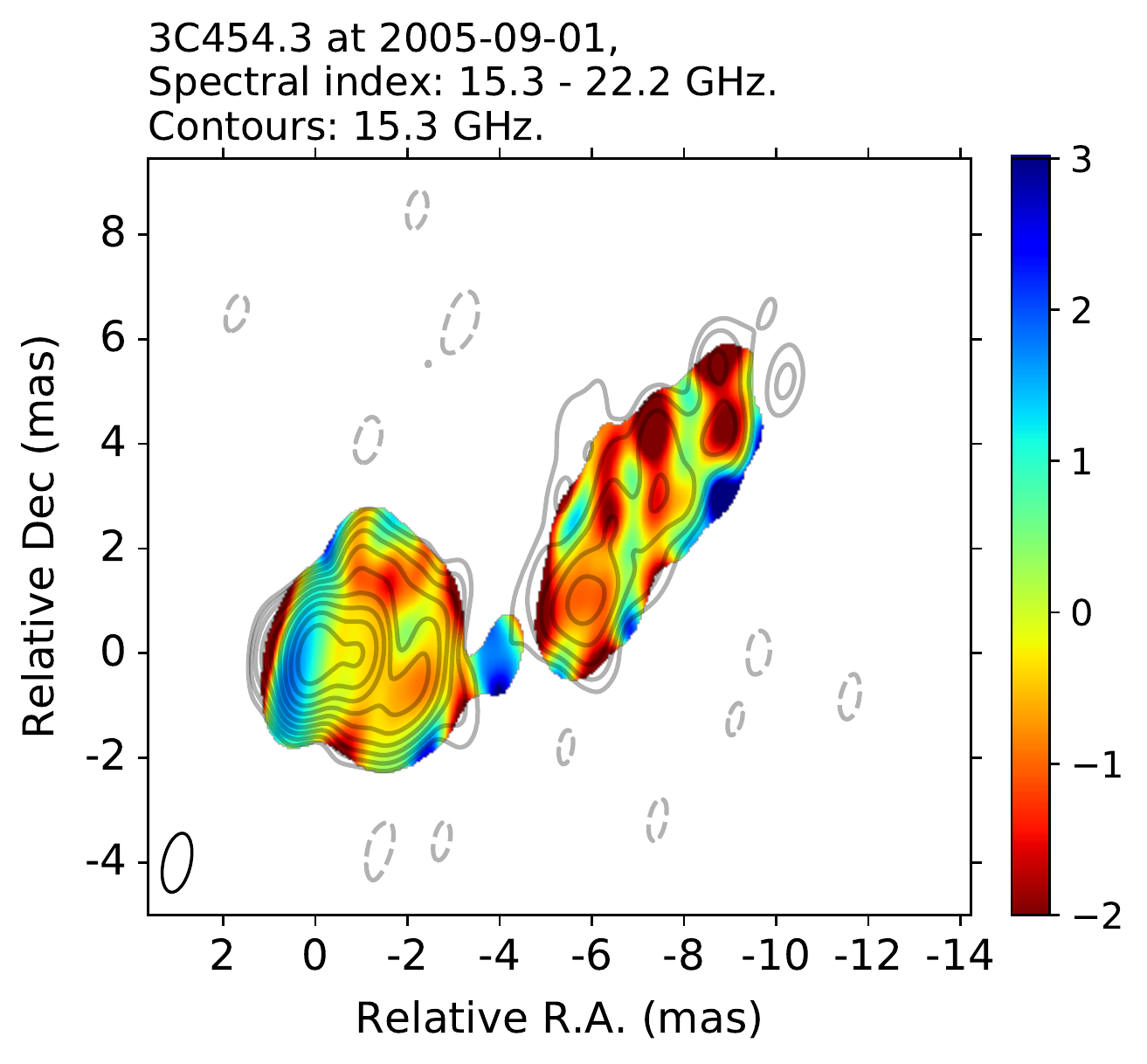}
    }
    \subfigure[]
    {
         \includegraphics[width=0.3\textwidth]{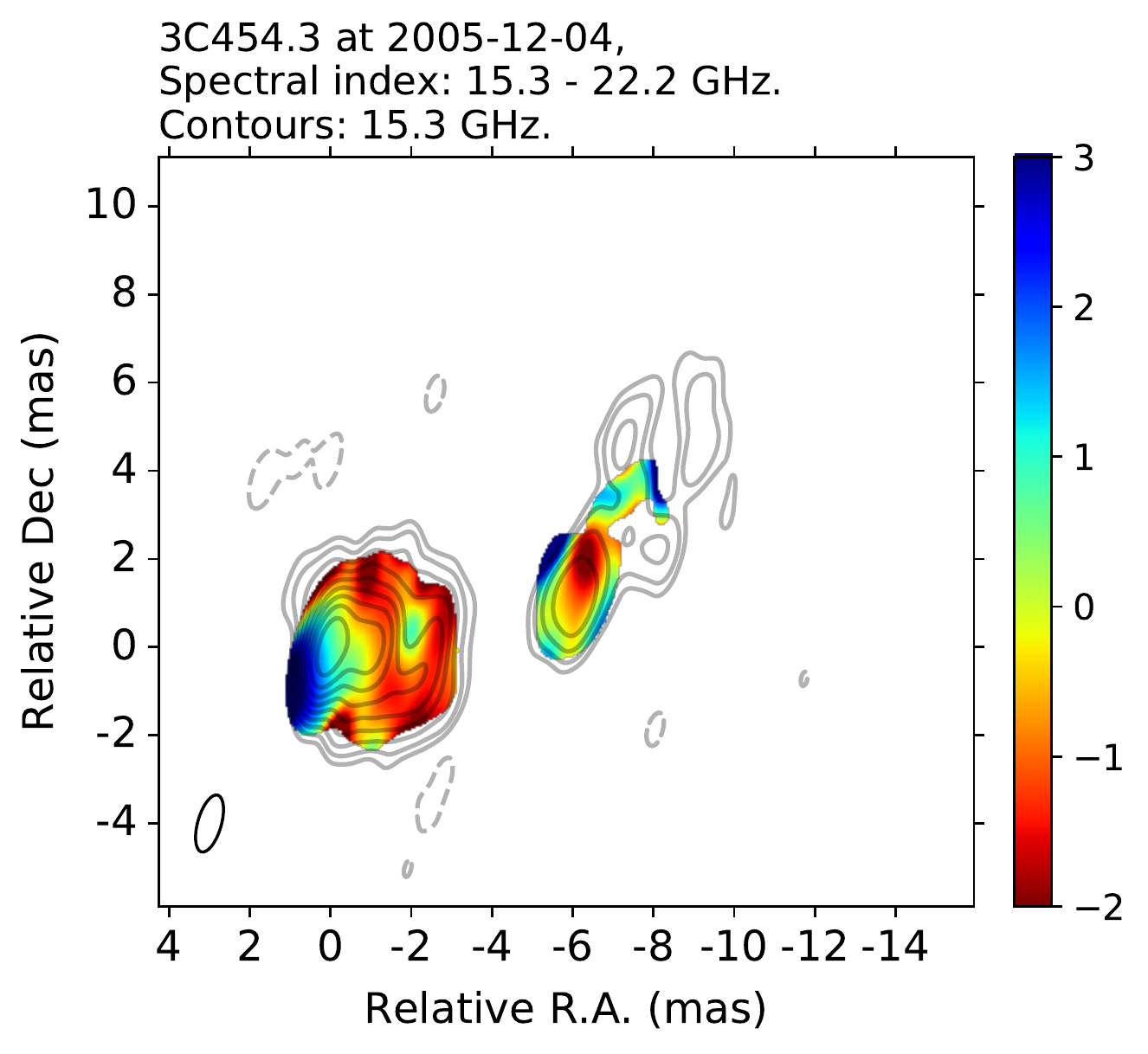}
    }
    \subfigure[]
    {
         \includegraphics[width=0.3\textwidth]{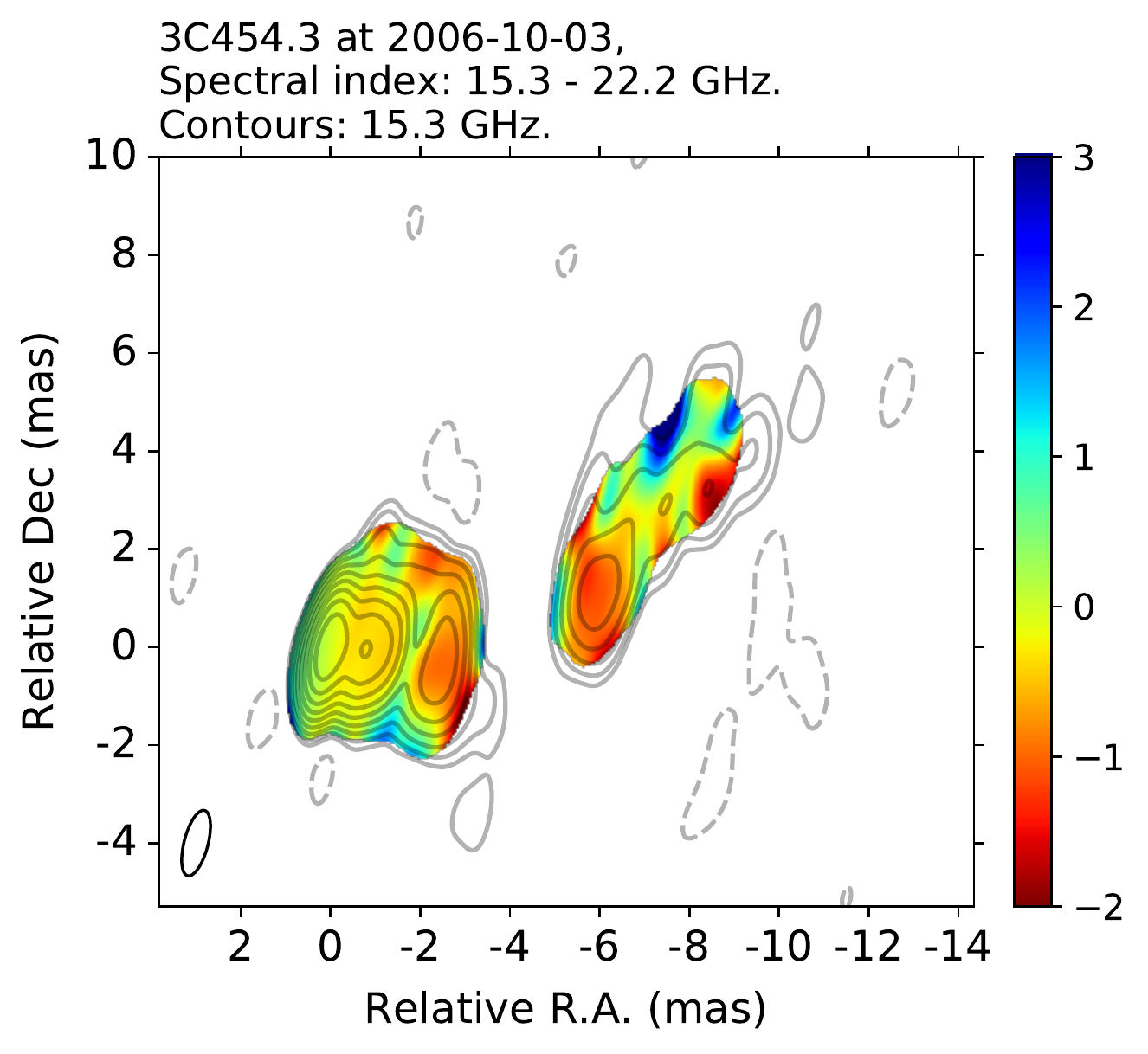}
    }   
    \subfigure[]
    {
         \includegraphics[width=0.3\textwidth]{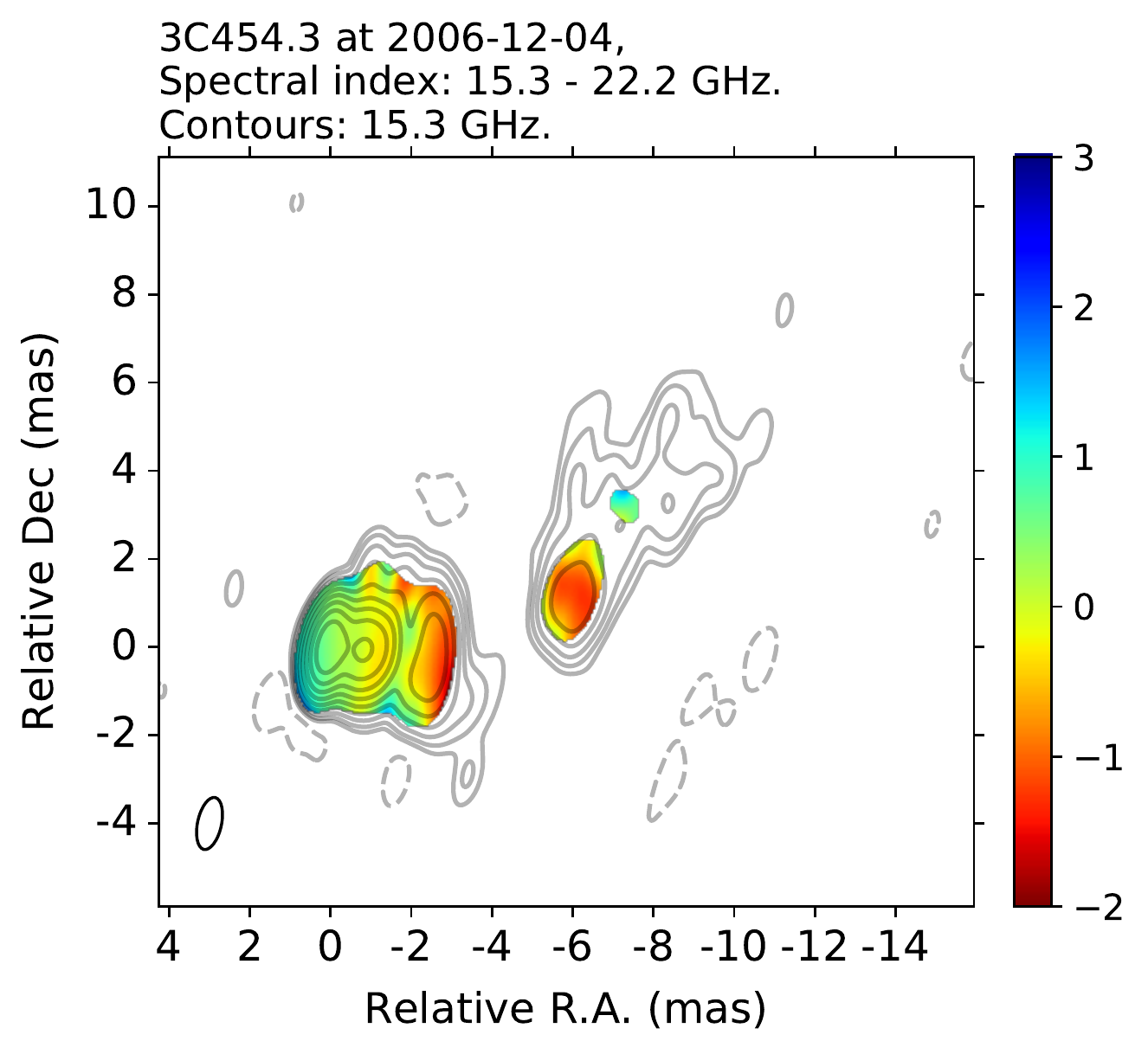}
    }    
    \subfigure[]
    {
         \includegraphics[width=0.3\textwidth]{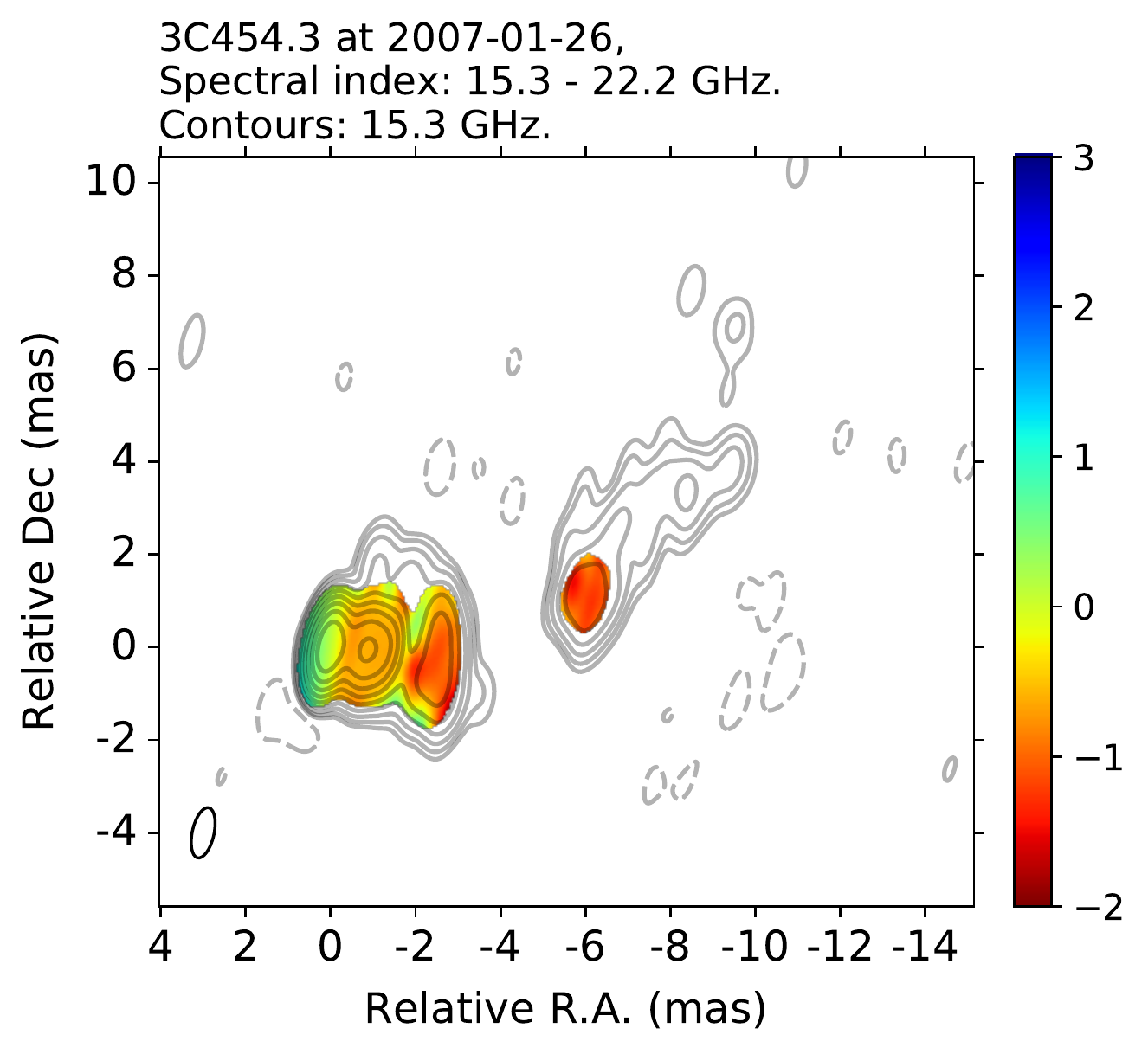}
    }   
    \subfigure[]
    {
         \includegraphics[width=0.3\textwidth]{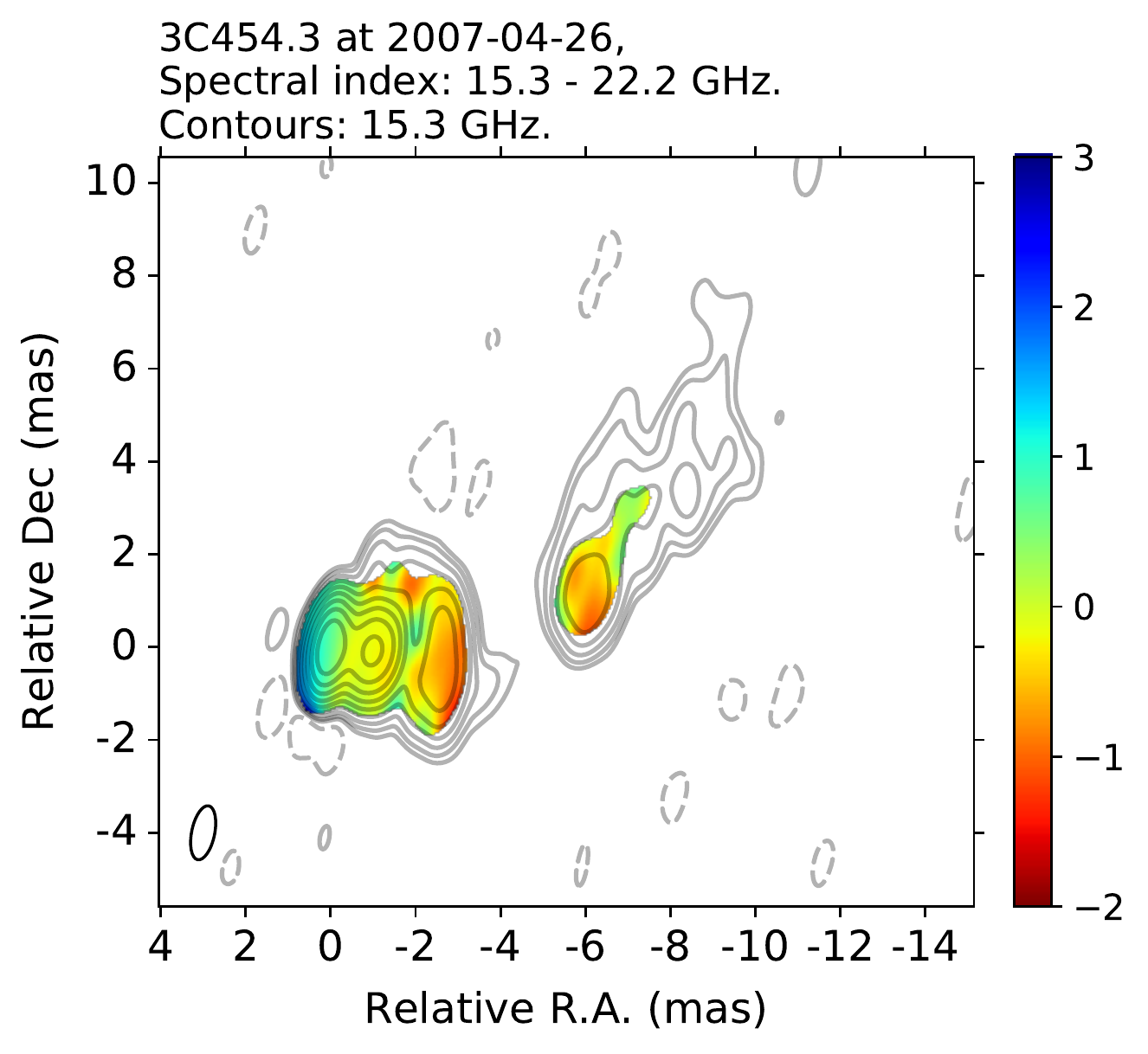}
    }    
        \subfigure[]
    {
         \includegraphics[width=0.3\textwidth]{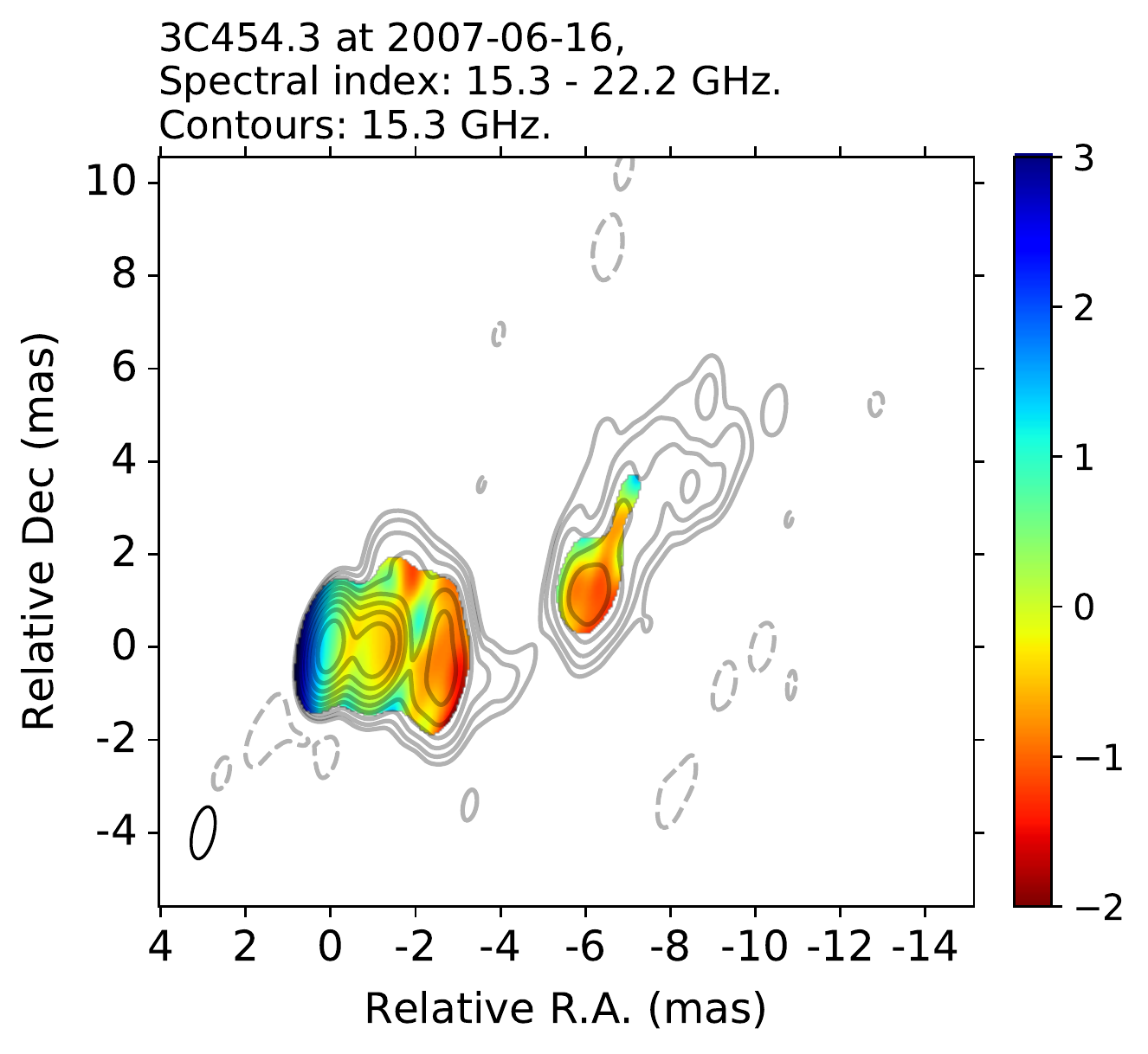}
    }    
        \subfigure[]
    {
         \includegraphics[width=0.3\textwidth]{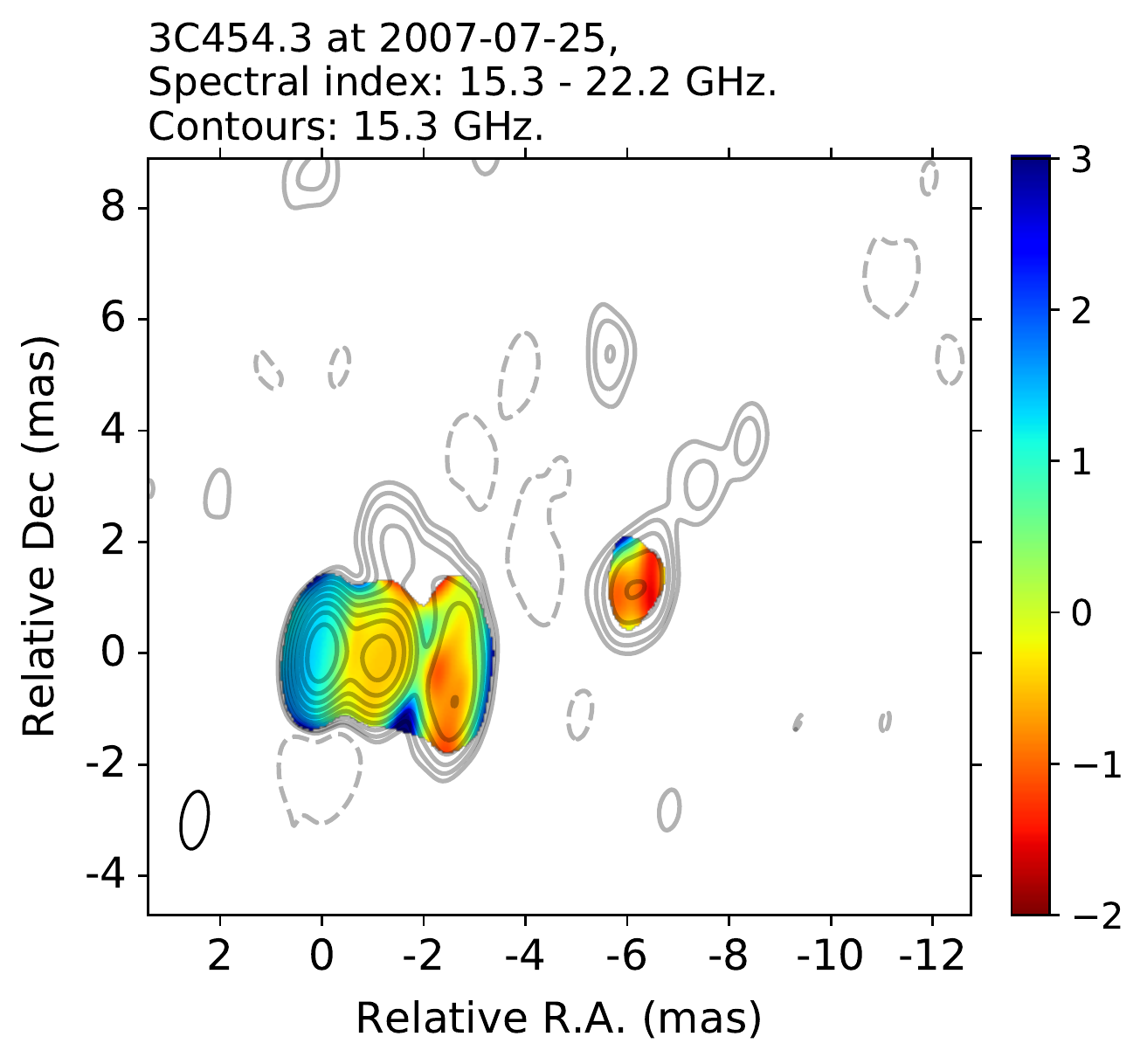}
    } 
            \subfigure[]
    {
         \includegraphics[width=0.3\textwidth]{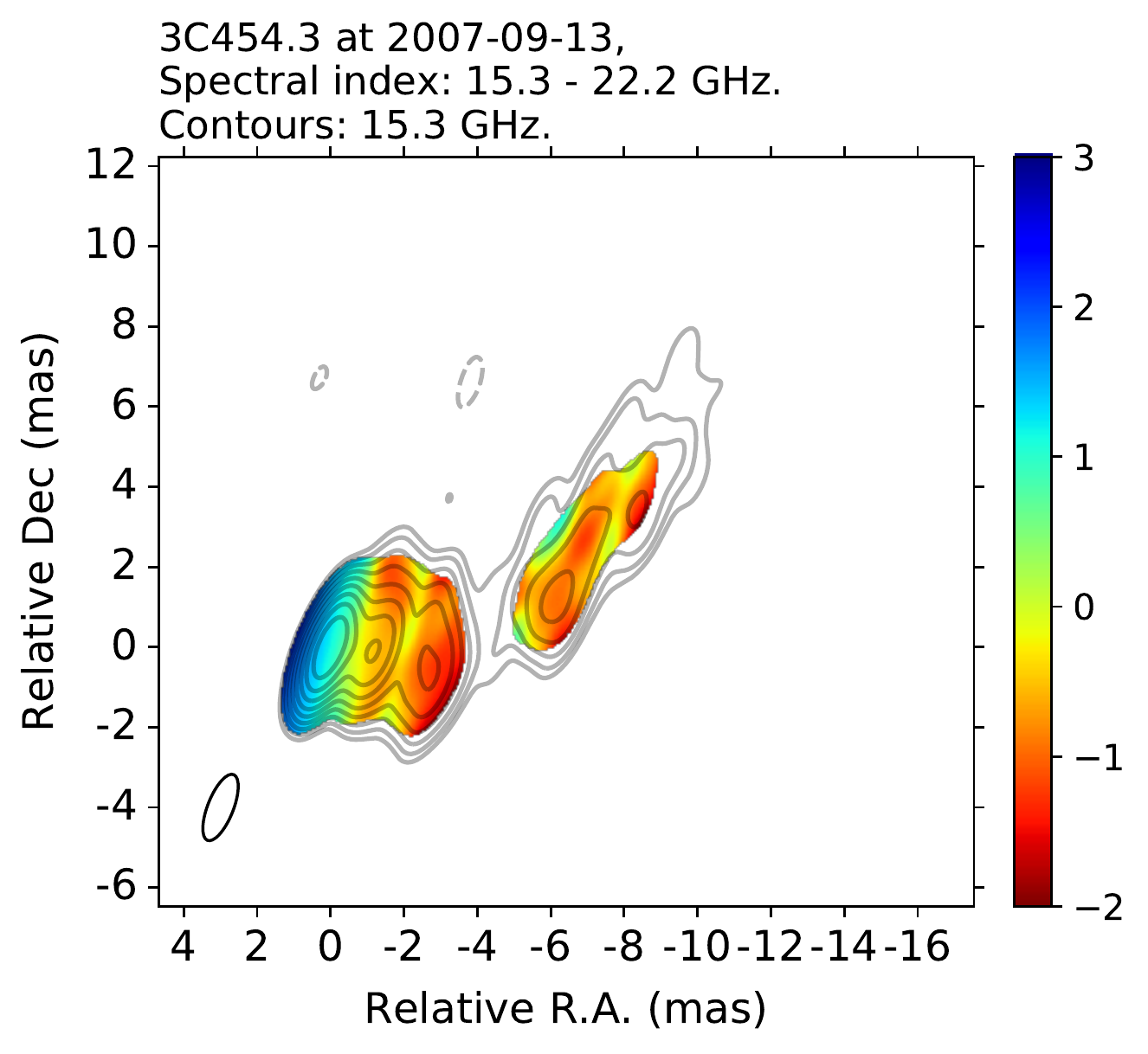}
    } 
            \subfigure[]
    {
         \includegraphics[width=0.3\textwidth]{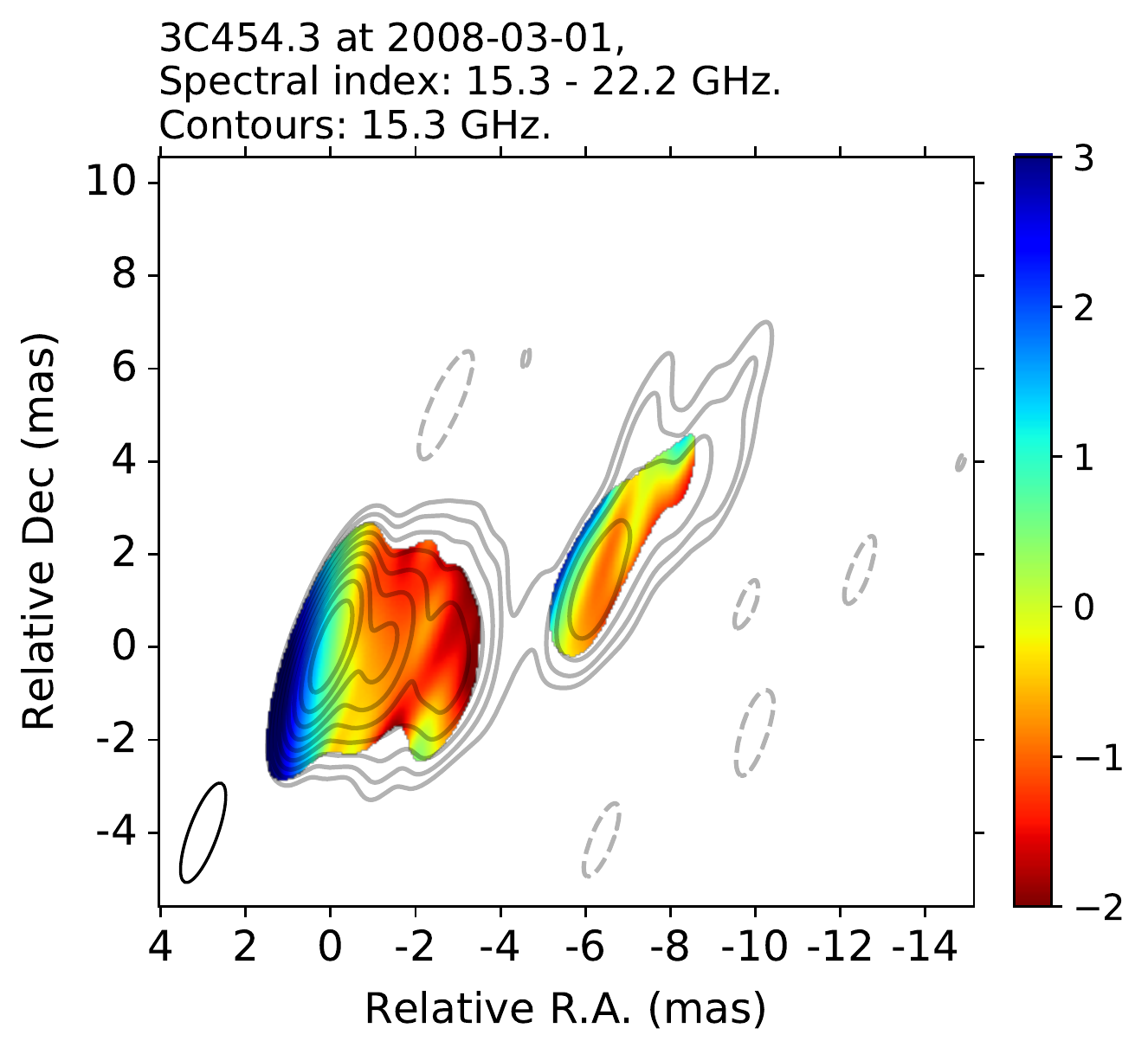}
    } 
     \caption{Spectral index maps for the frequency pair UK (15 -22$-$24)\,GHz. The colorbar indicates the spectral index. The ellipse on the bottom left corner represents the interferometric beam. The contour lines are given at  -0.1\%, 0.1\% 0.2\%, 0.4\%, 0.8\%, 1.6\%, 3.2\%, 6.4\%, 12.8\%, 25.6\%, and 51.2\% of the peak intensity at each image. }
    \label{siUKp1}
\end{figure*}

\begin{figure*}[]
\centering
    \subfigure[]
    {
         \includegraphics[width=0.3\textwidth]{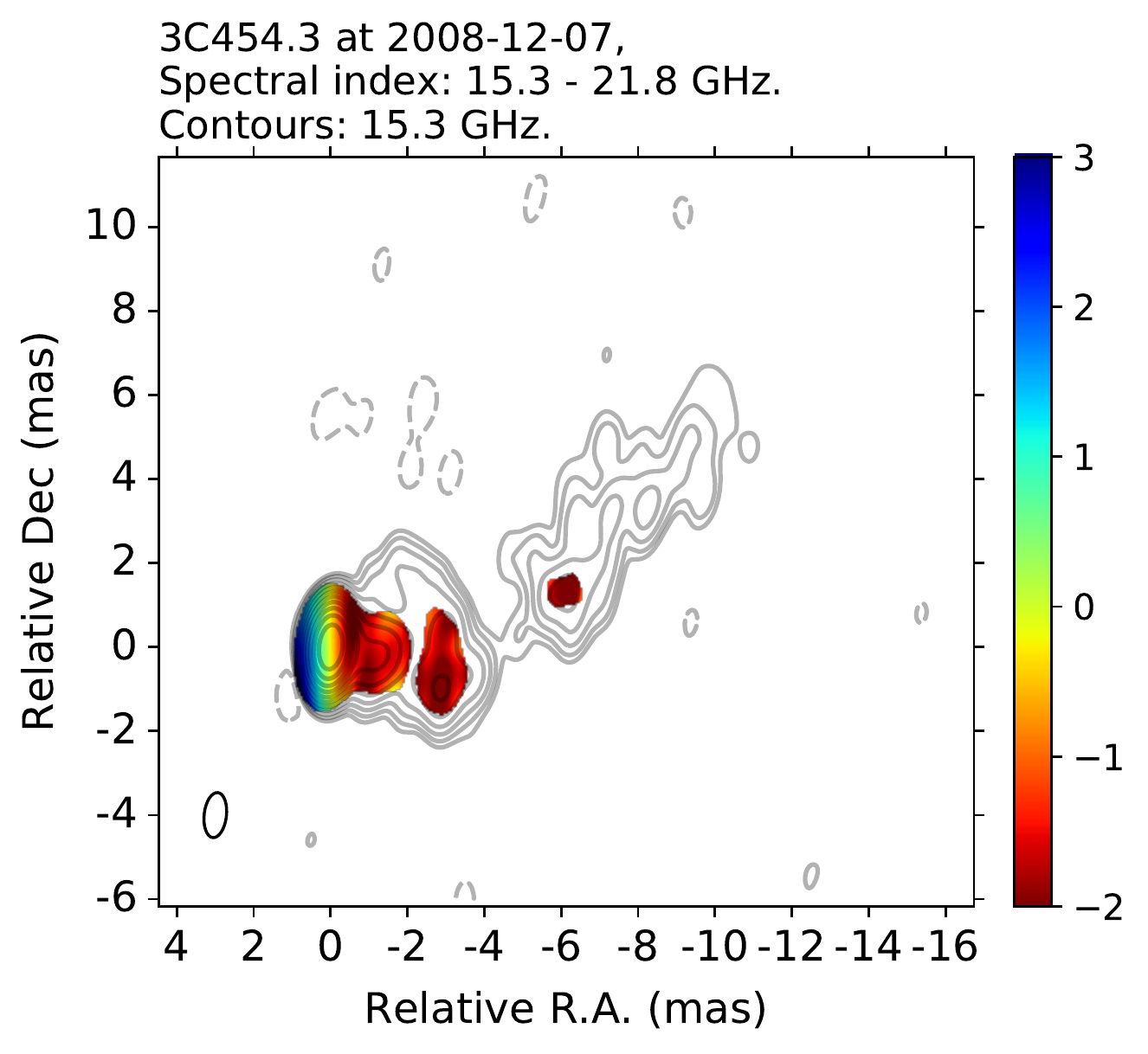}
    }
    \subfigure[]
    {
         \includegraphics[width=0.3\textwidth]{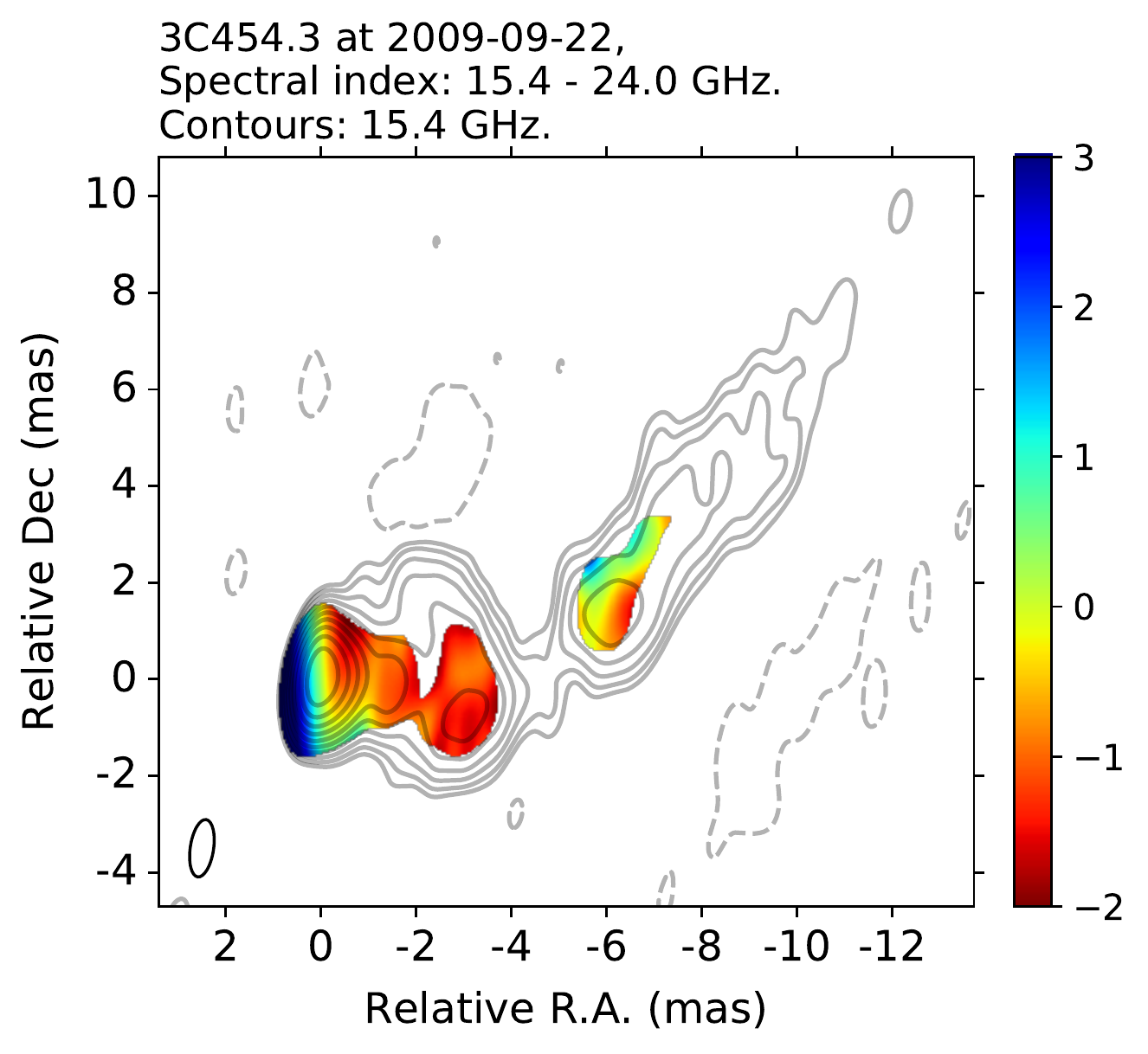}
    }
    \subfigure[]
    {
         \includegraphics[width=0.3\textwidth]{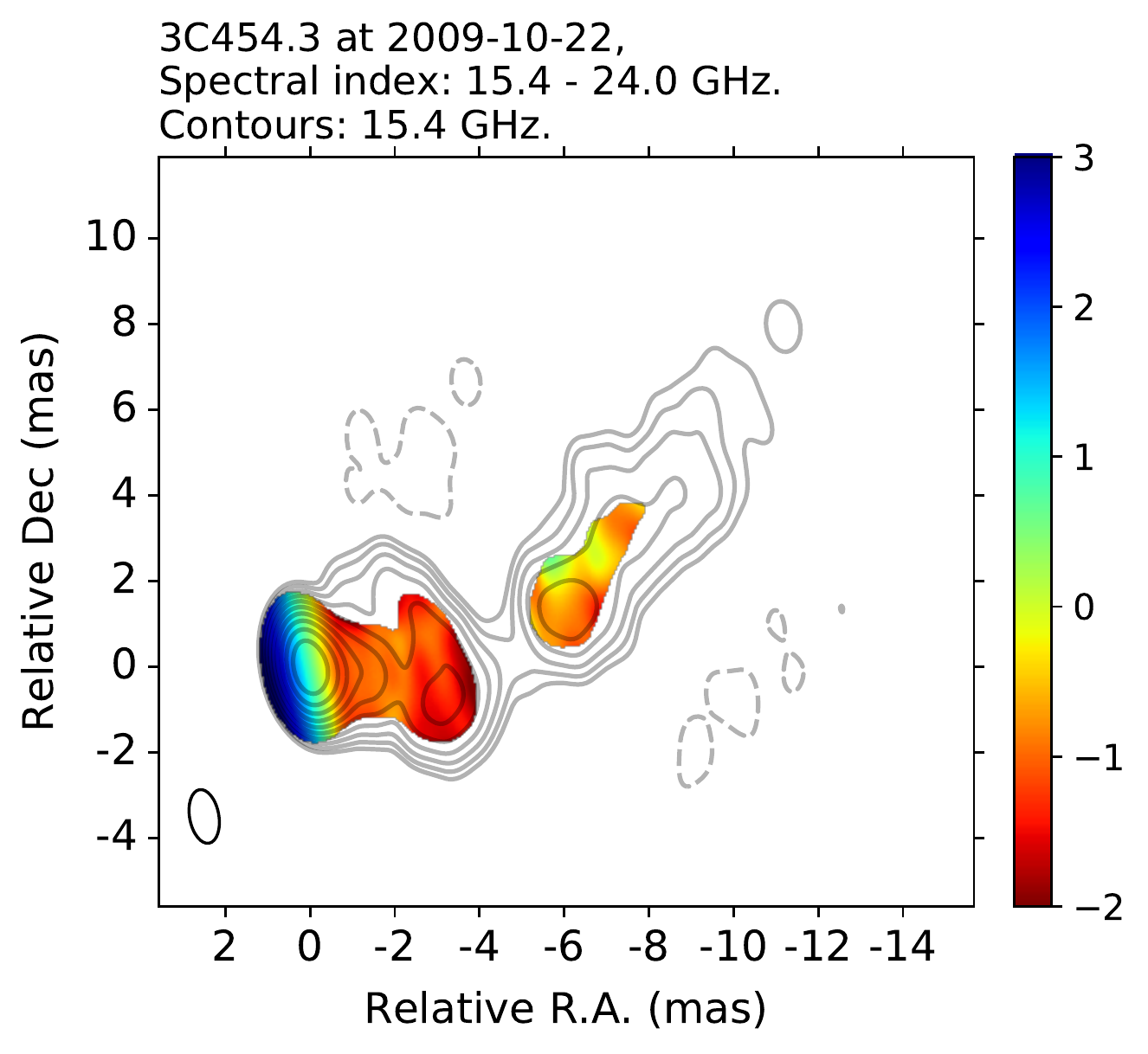}
    }
    \subfigure[]
    {
         \includegraphics[width=0.3\textwidth]{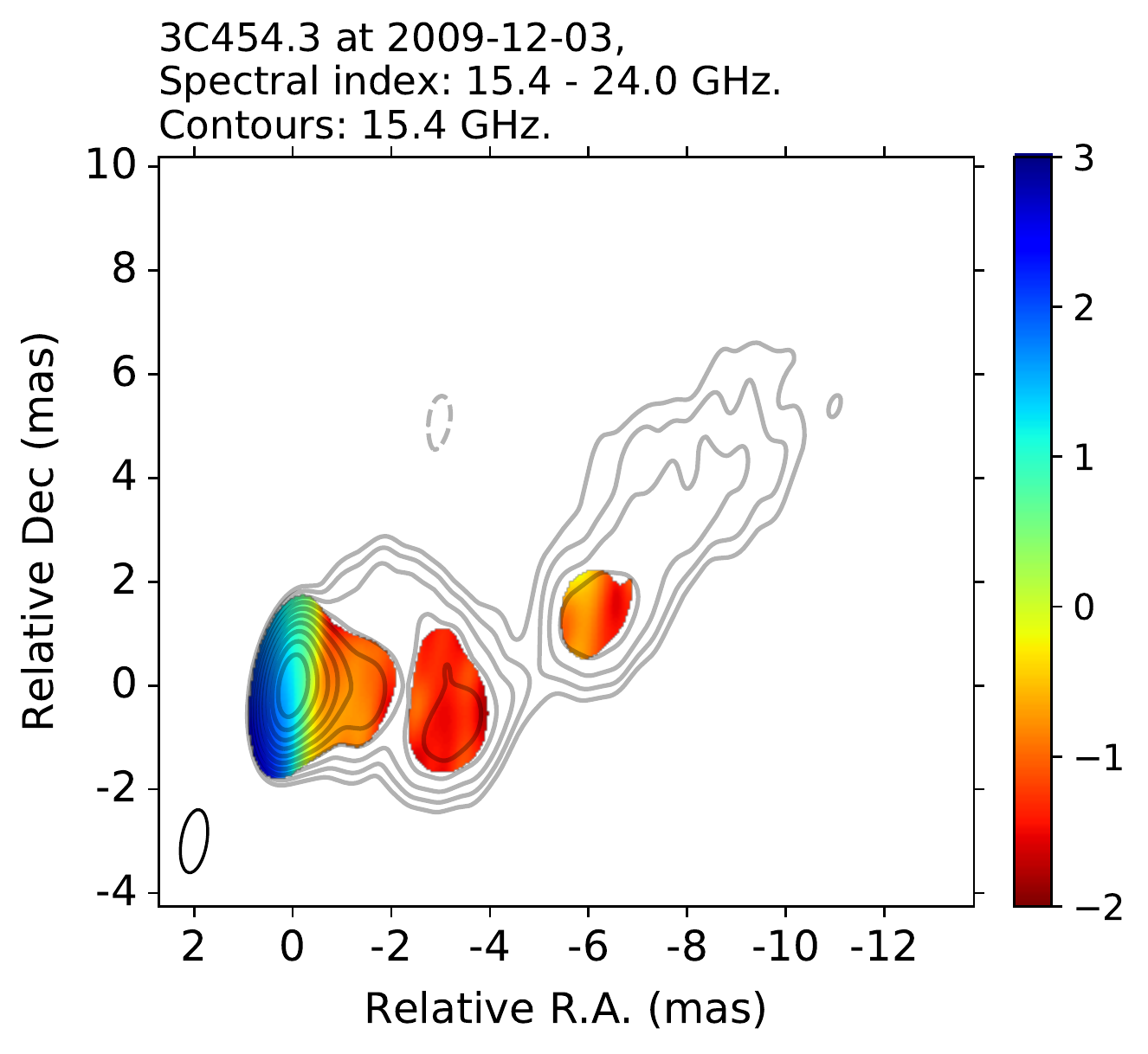}
    }
    \subfigure[]
    {
         \includegraphics[width=0.3\textwidth]{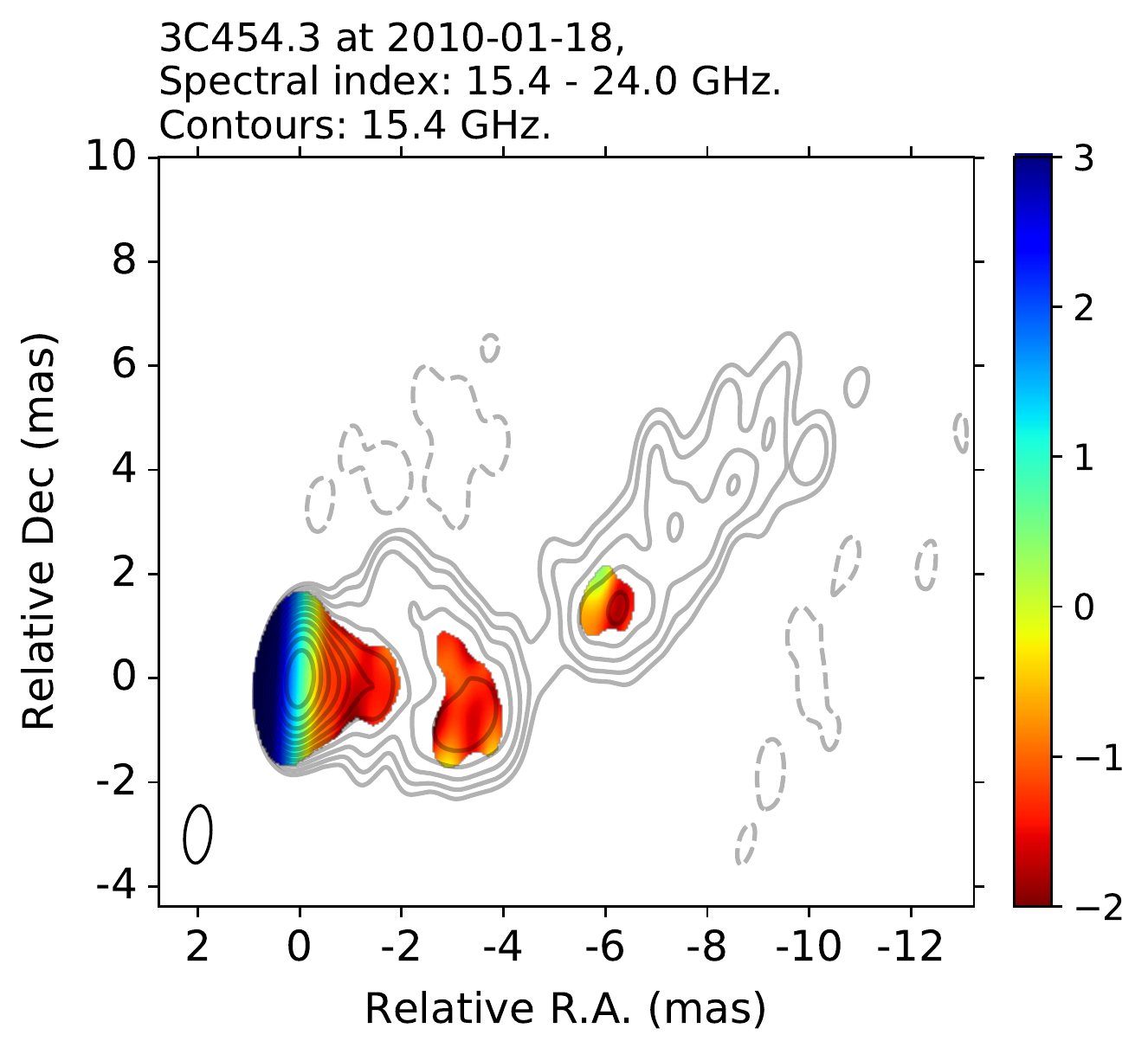}
    }   
    \subfigure[]
    {
         \includegraphics[width=0.3\textwidth]{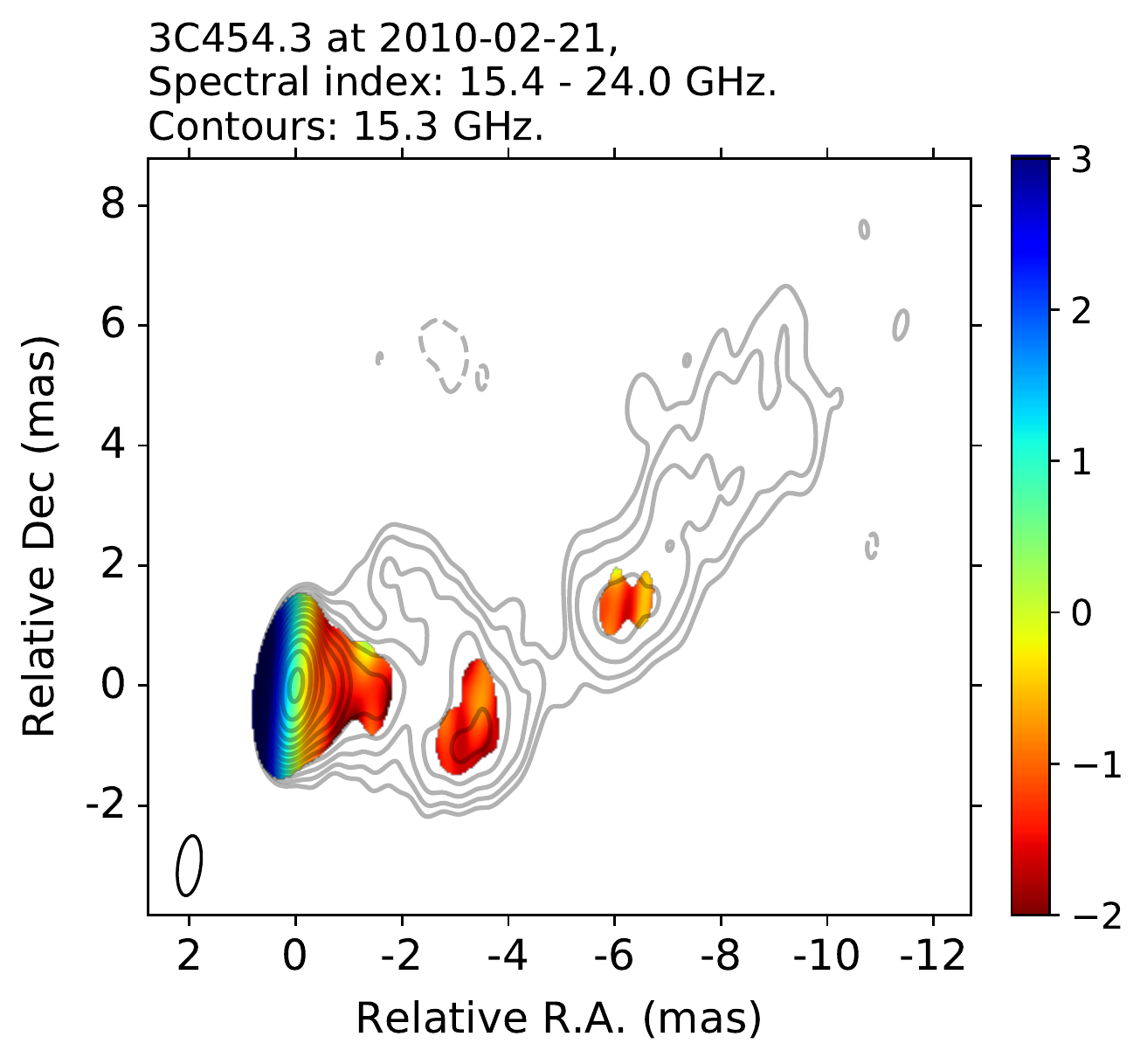}
    }    
        \caption{Continuation of Figure~F.5.}
    \label{siUKp2}
\end{figure*}

%%%%%%%%%%%%%%%%%% KQ si maps %%%%%%%%%%%%%%%%%%%%%%%
\begin{figure*}[]
\centering
    \subfigure[]
    {
         \includegraphics[width=0.3\textwidth]{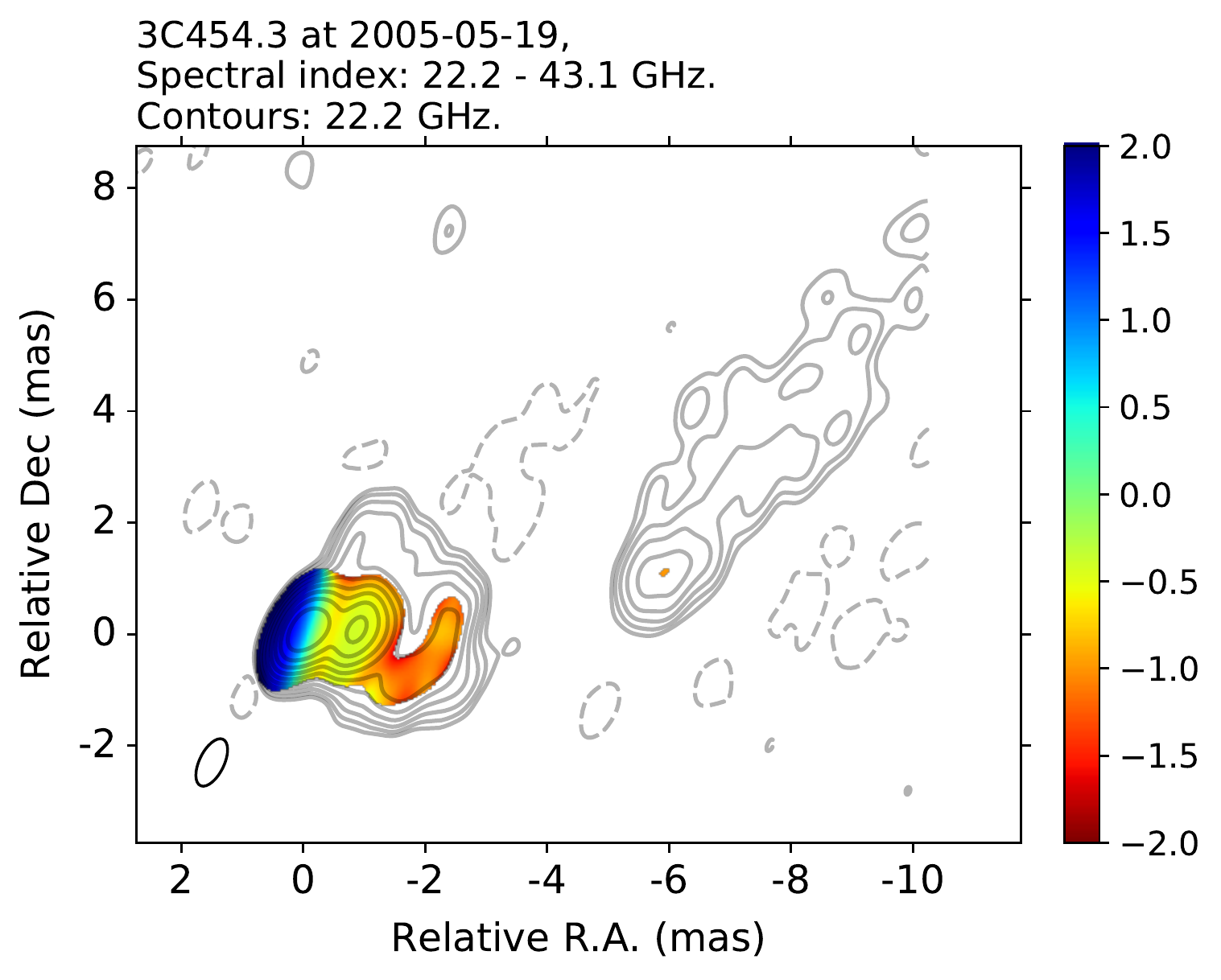}
    }
    \subfigure[]
    {
         \includegraphics[width=0.3\textwidth]{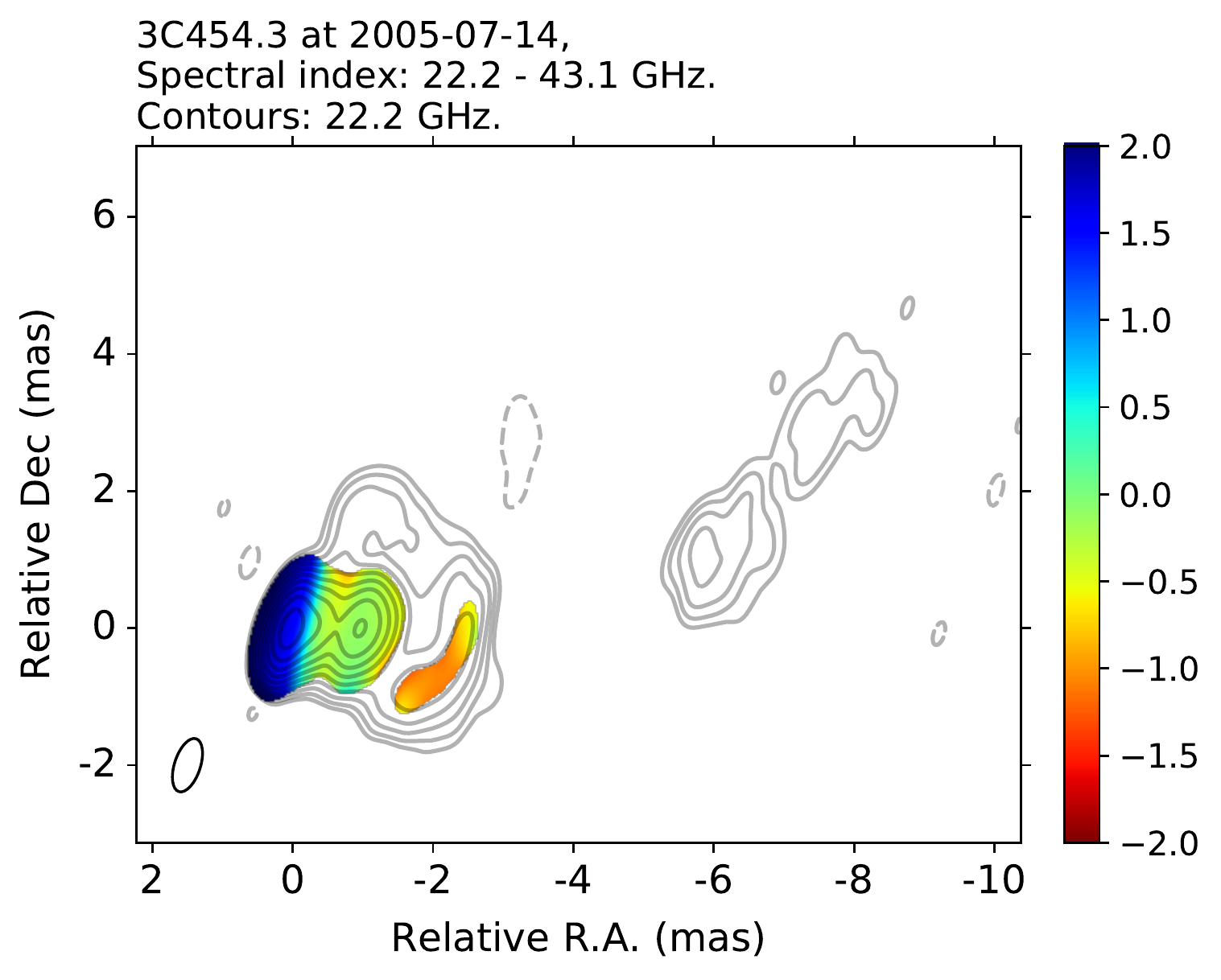}
    }
    \subfigure[]
    {
         \includegraphics[width=0.3\textwidth]{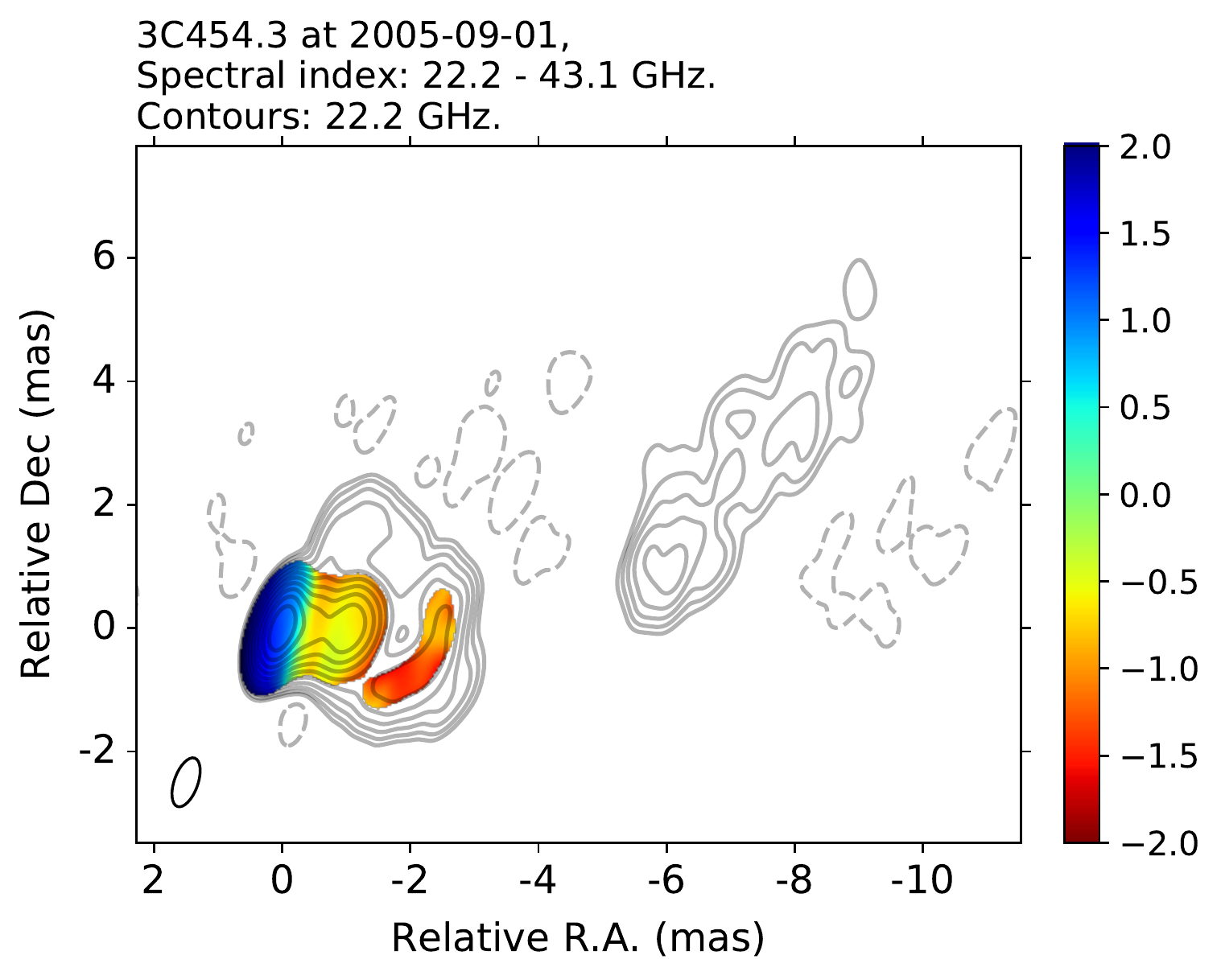}
    }
    \subfigure[]
    {
         \includegraphics[width=0.3\textwidth]{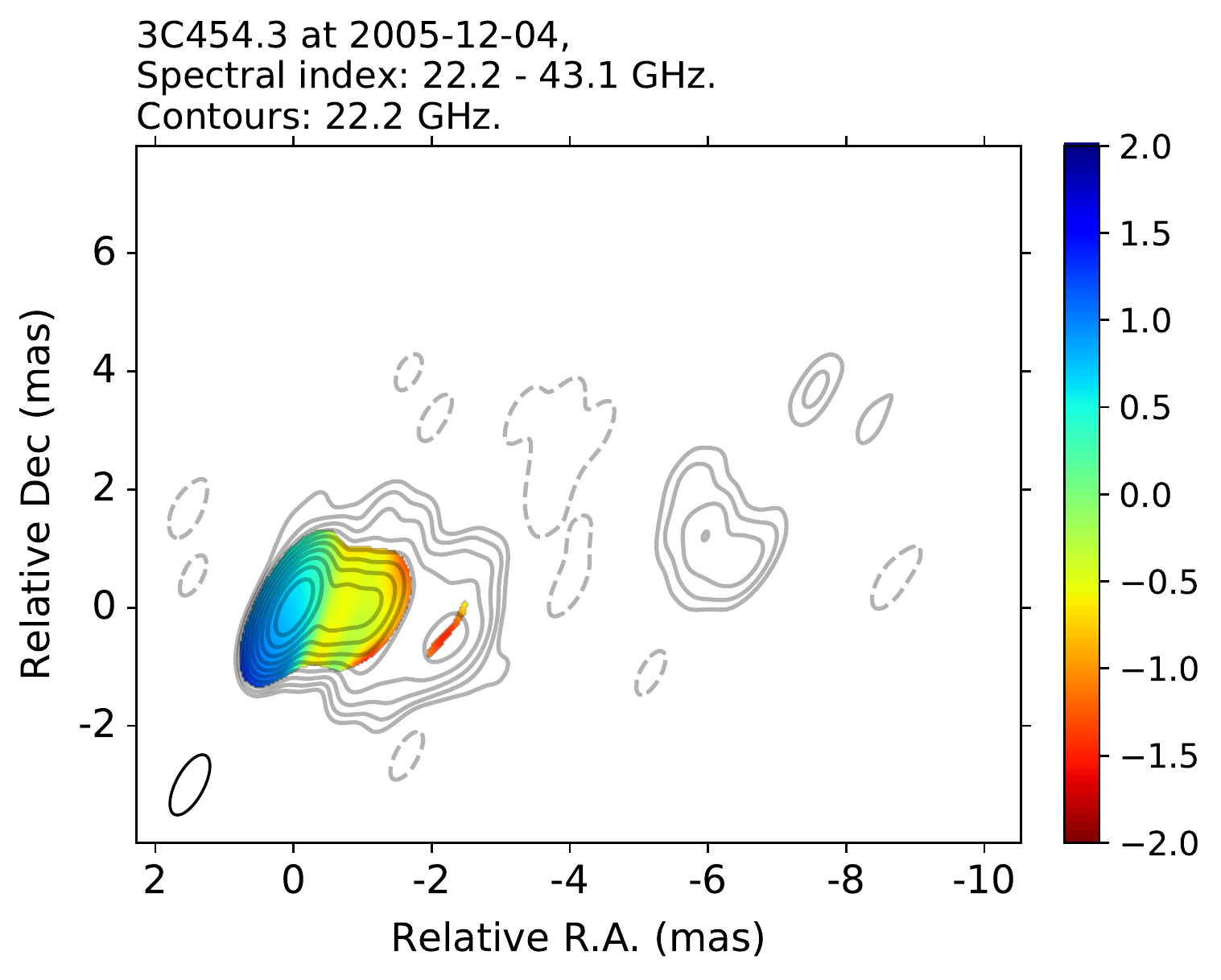}
    }
    \subfigure[]
    {
         \includegraphics[width=0.3\textwidth]{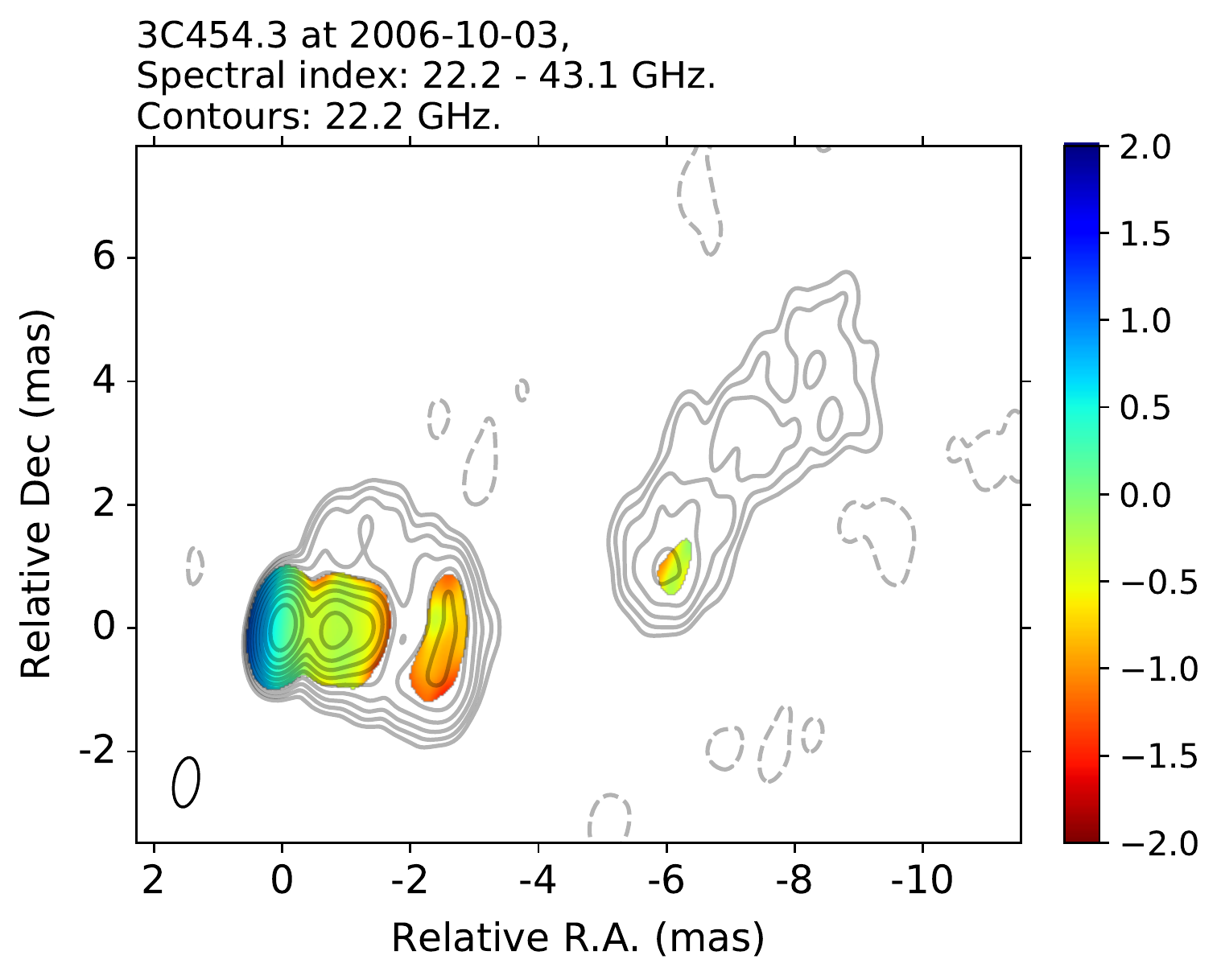}
    }   
    \subfigure[]
    {
         \includegraphics[width=0.3\textwidth]{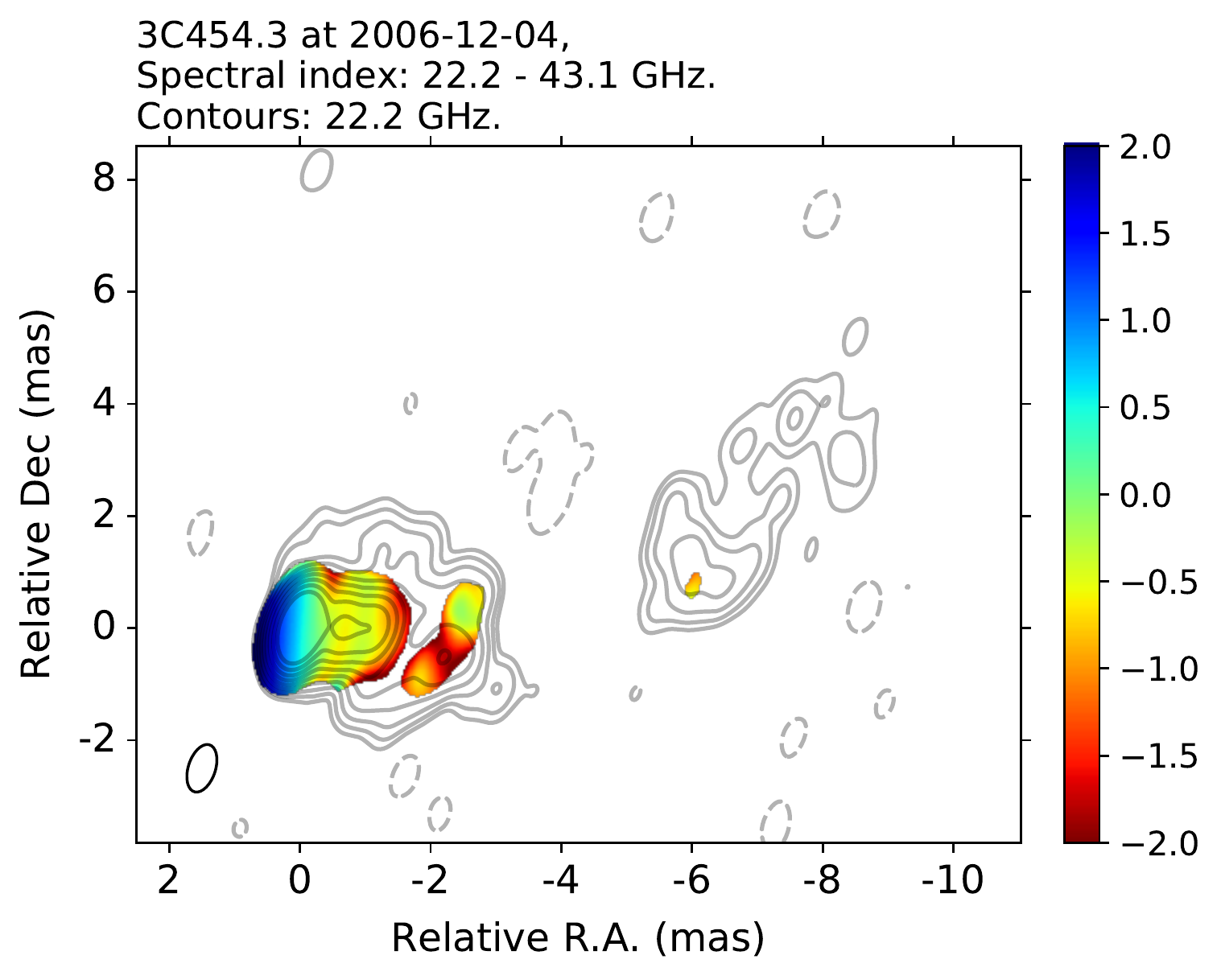}
    }    
    \subfigure[]
    {
         \includegraphics[width=0.3\textwidth]{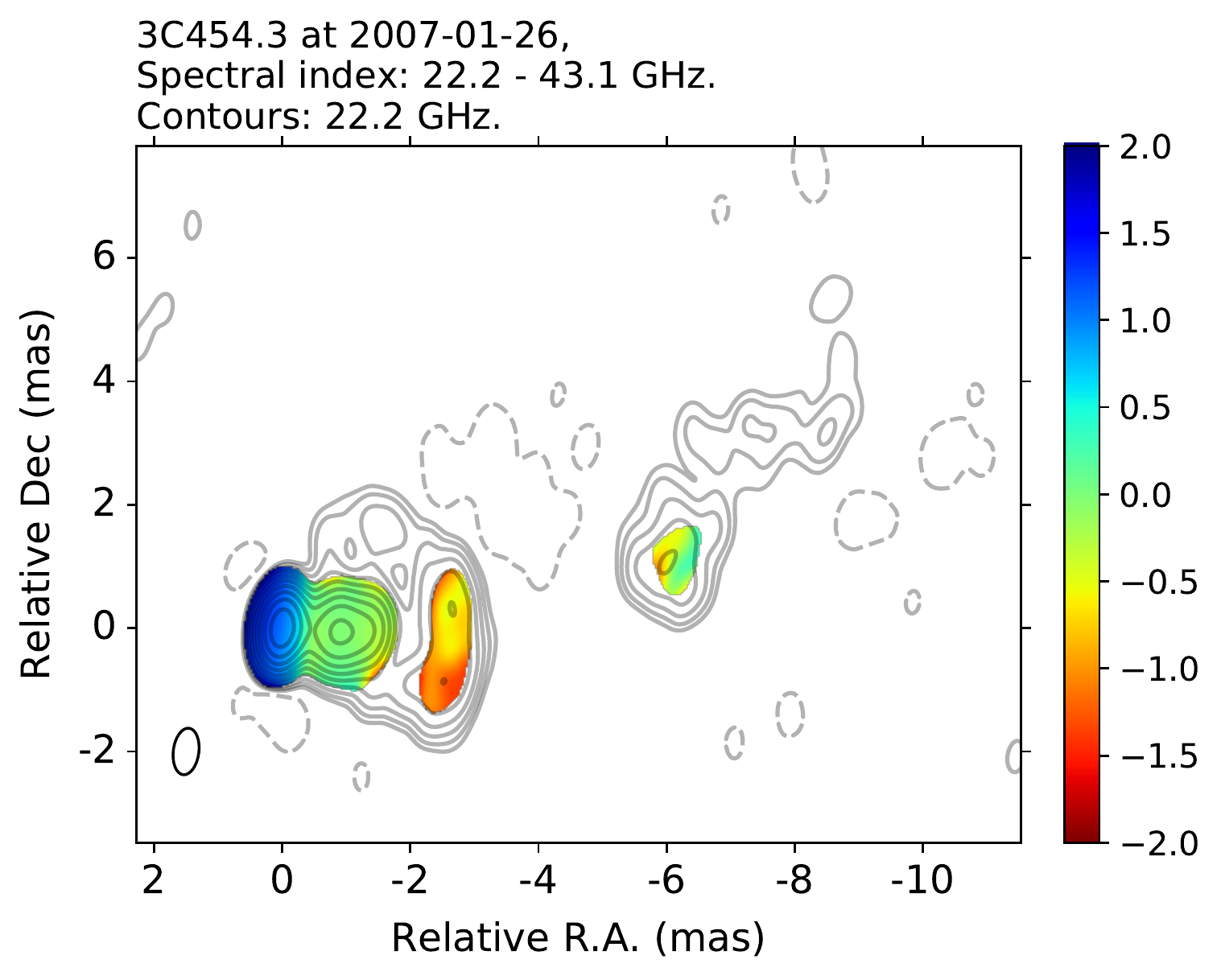}
    }   
    \subfigure[]
    {
         \includegraphics[width=0.3\textwidth]{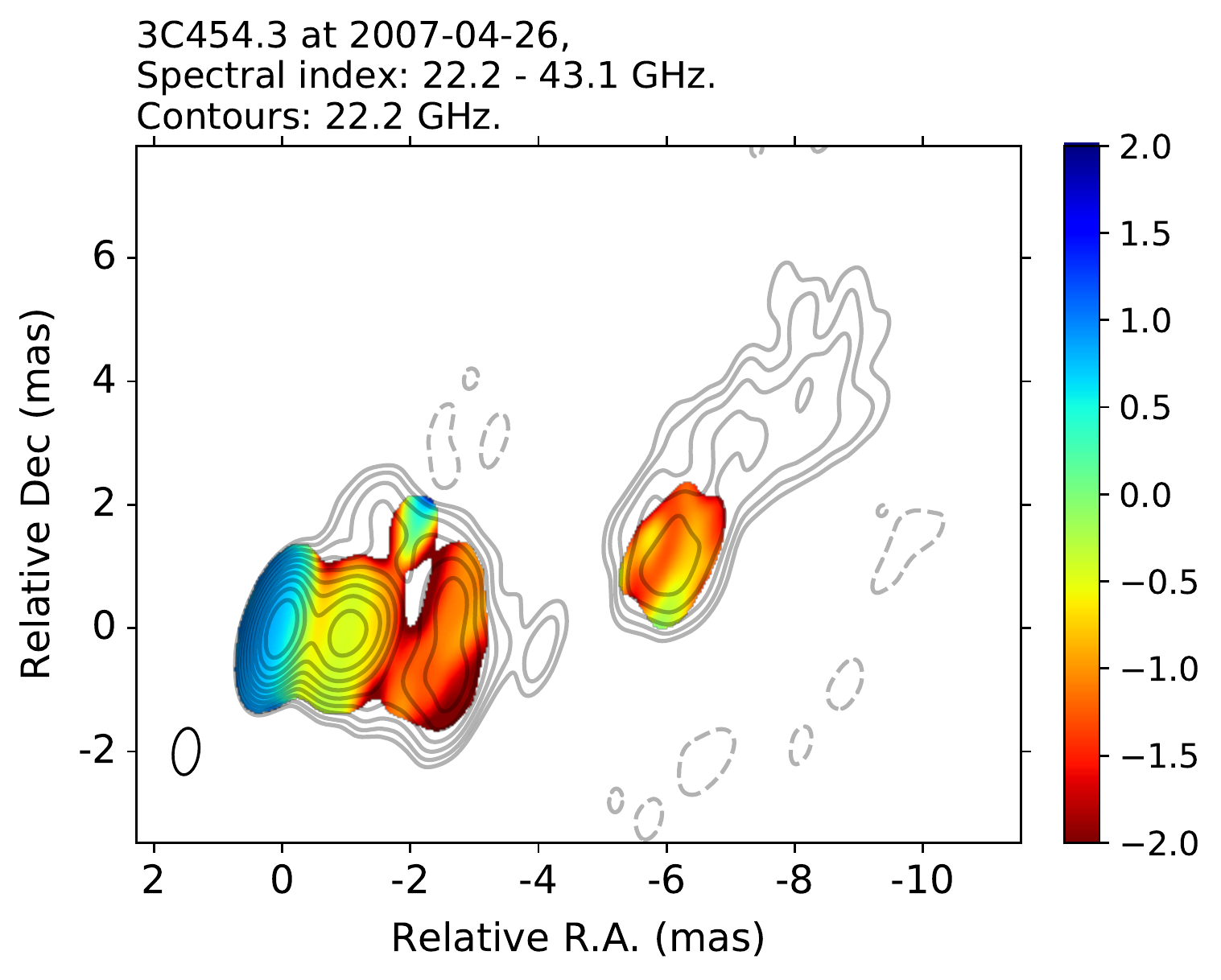}
    }    
        \subfigure[]
    {
         \includegraphics[width=0.3\textwidth]{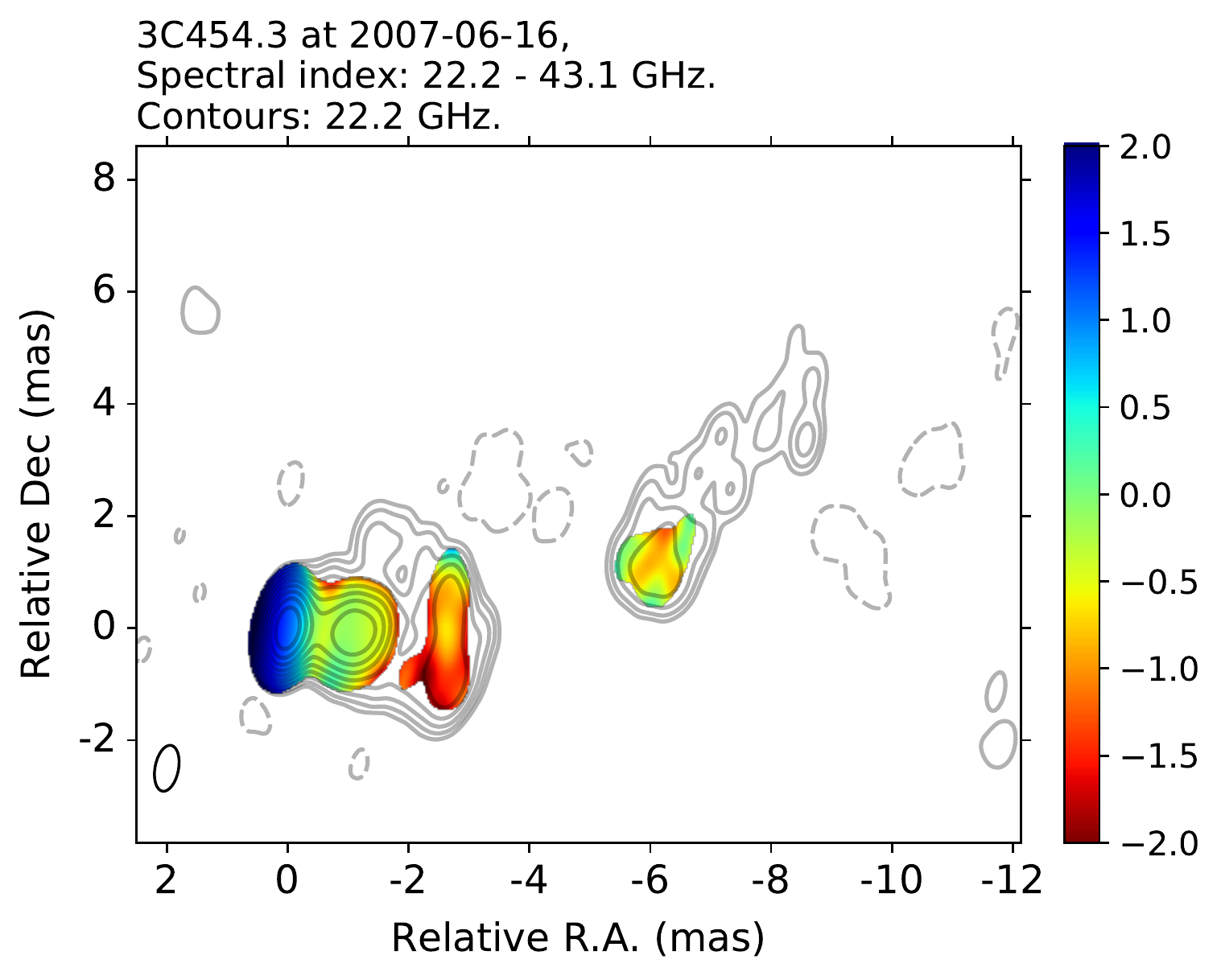}
    }    
        \subfigure[]
    {
         \includegraphics[width=0.3\textwidth]{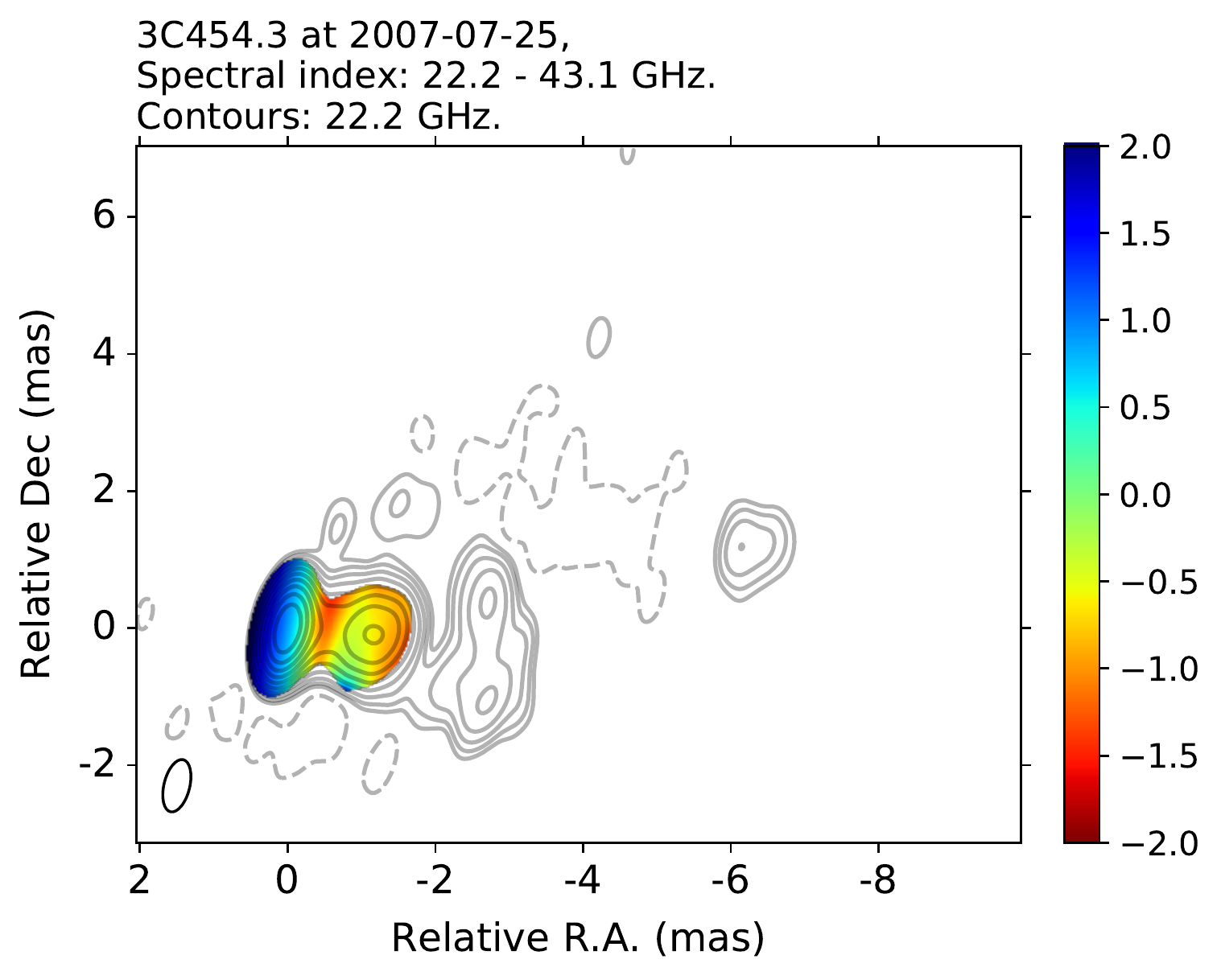}
    } 
            \subfigure[]
    {
         \includegraphics[width=0.3\textwidth]{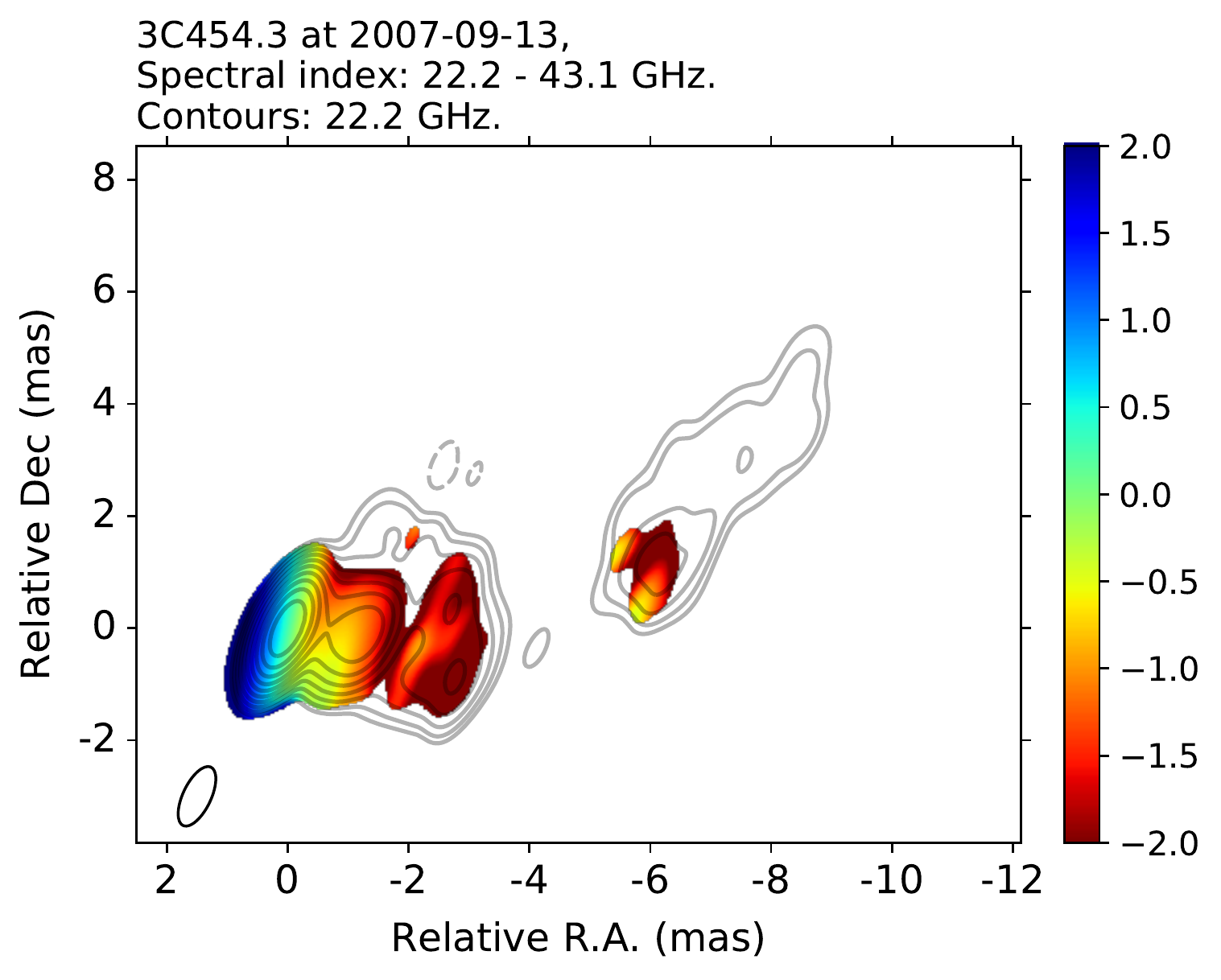}
    } 
            \subfigure[]
    {
         \includegraphics[width=0.3\textwidth]{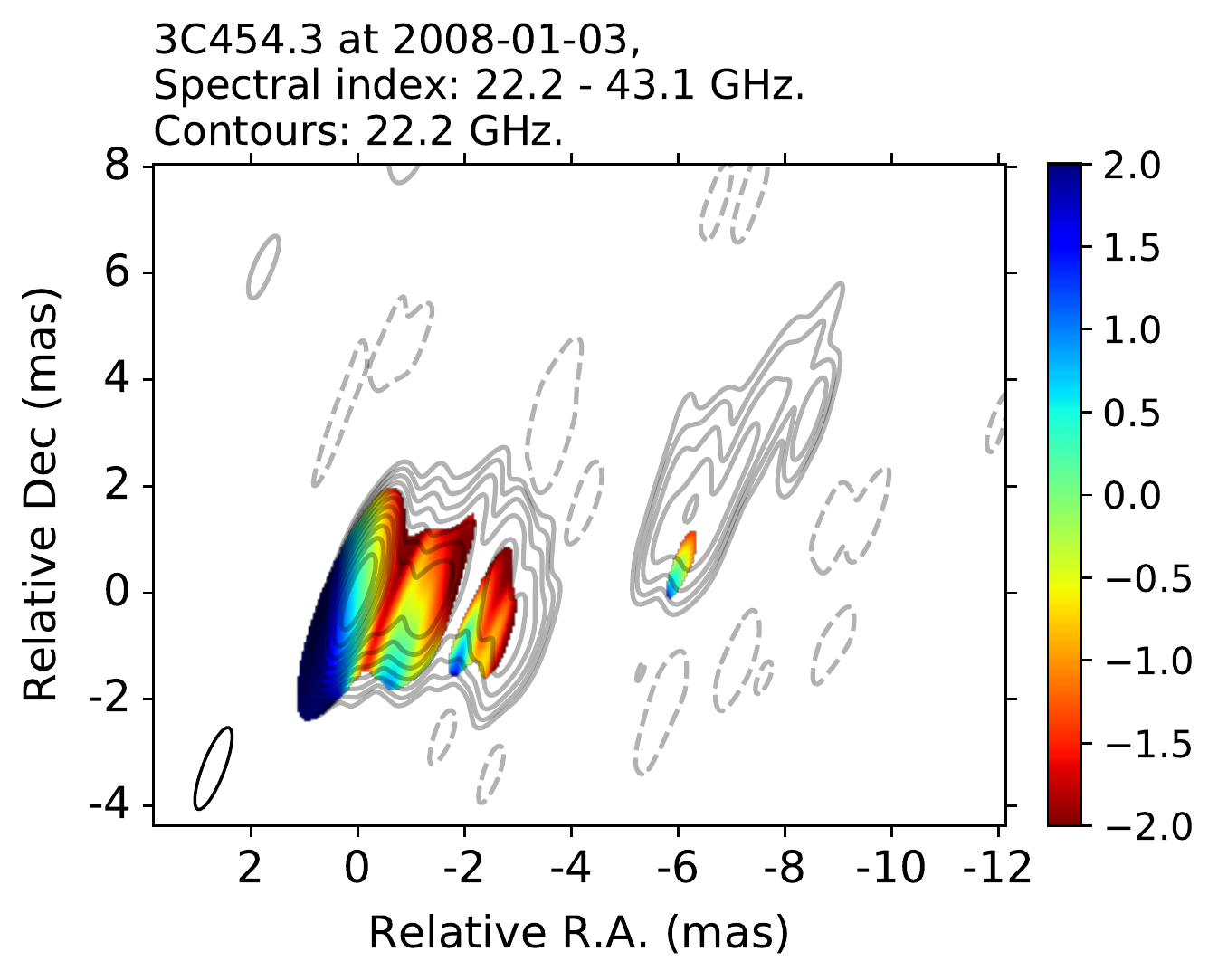}
    } 

     \caption{Spectral index maps for the frequency pair KQ (22$-$24 - 43\,GHz). The colorbar indicates the spectral index. The ellipse on the bottom left corner represents the interferometric beam. The contour lines are given at  -0.1\%,  0.1\% 0.2\%, 0.4\%, 0.8\%, 1.6\%, 3.2\%, 6.4\%, 12.8\%, 25.6\%, and 51.2\% of the peak intensity at each image. }
    \label{siKQp1}
\end{figure*}

\begin{figure*}[]
\centering
    \subfigure[]
    {
         \includegraphics[width=0.3\textwidth]{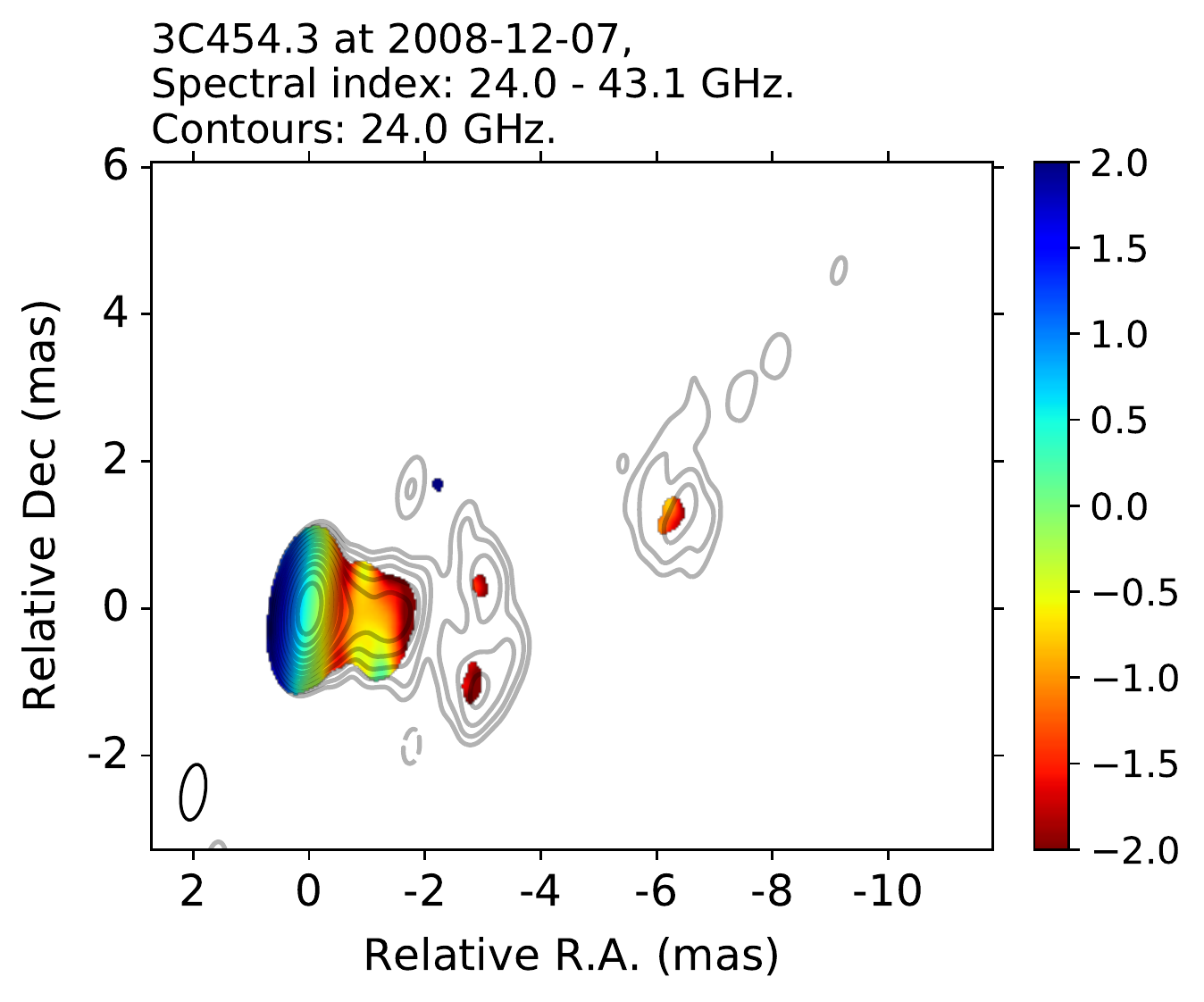}
    }
    \subfigure[]
    {
         \includegraphics[width=0.3\textwidth]{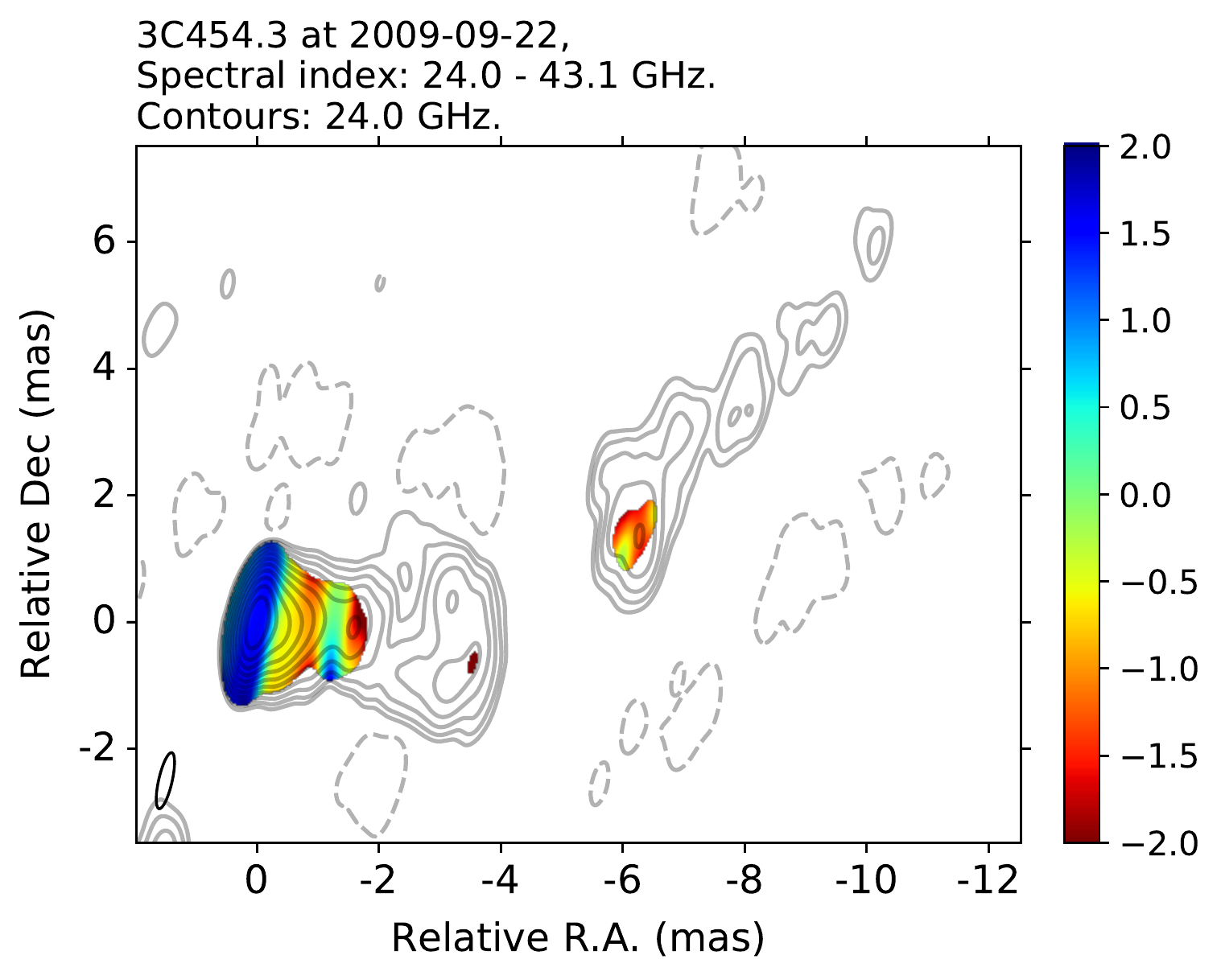}
    }
    \subfigure[]
    {
         \includegraphics[width=0.3\textwidth]{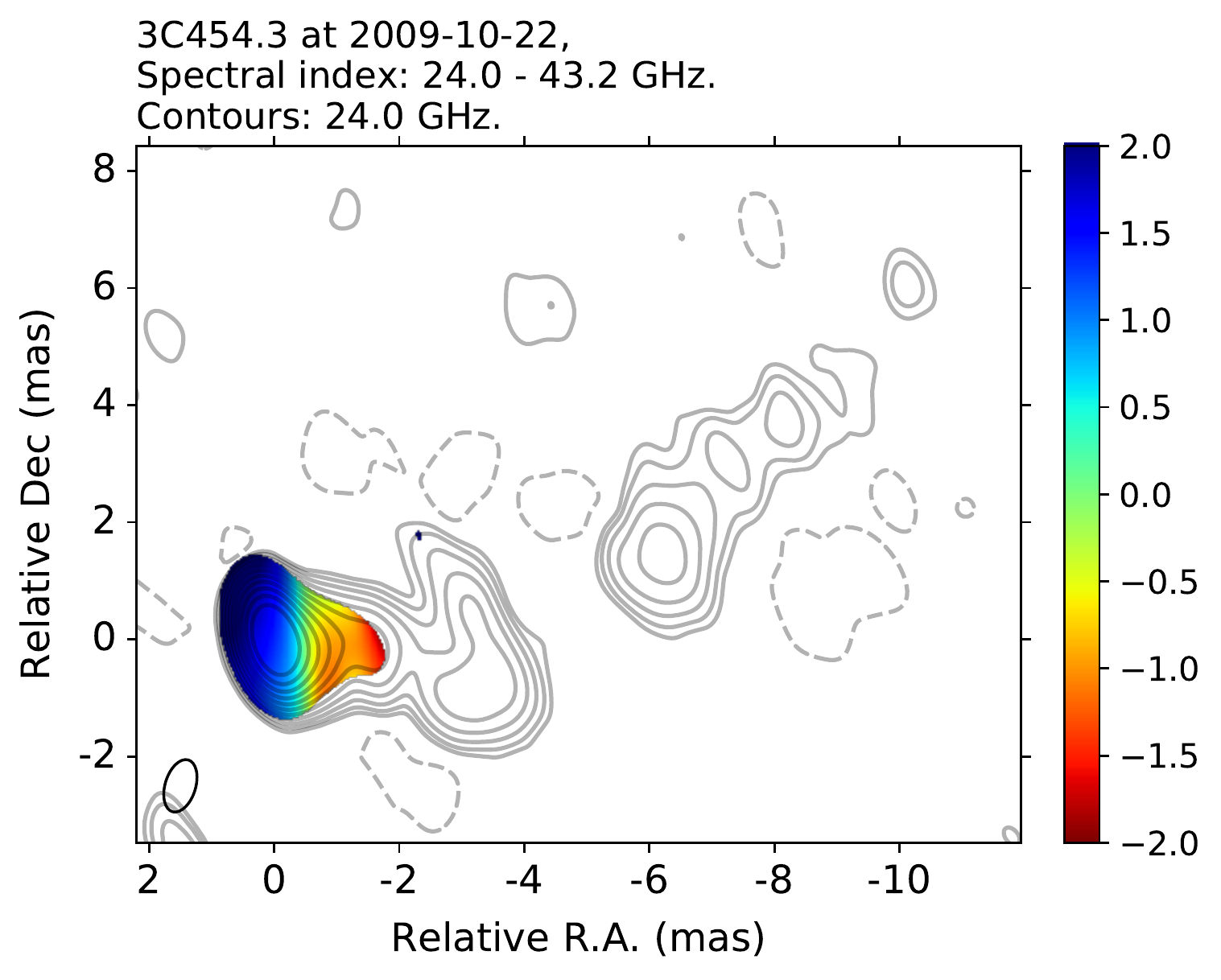}
    }
    \subfigure[]
    {
         \includegraphics[width=0.3\textwidth]{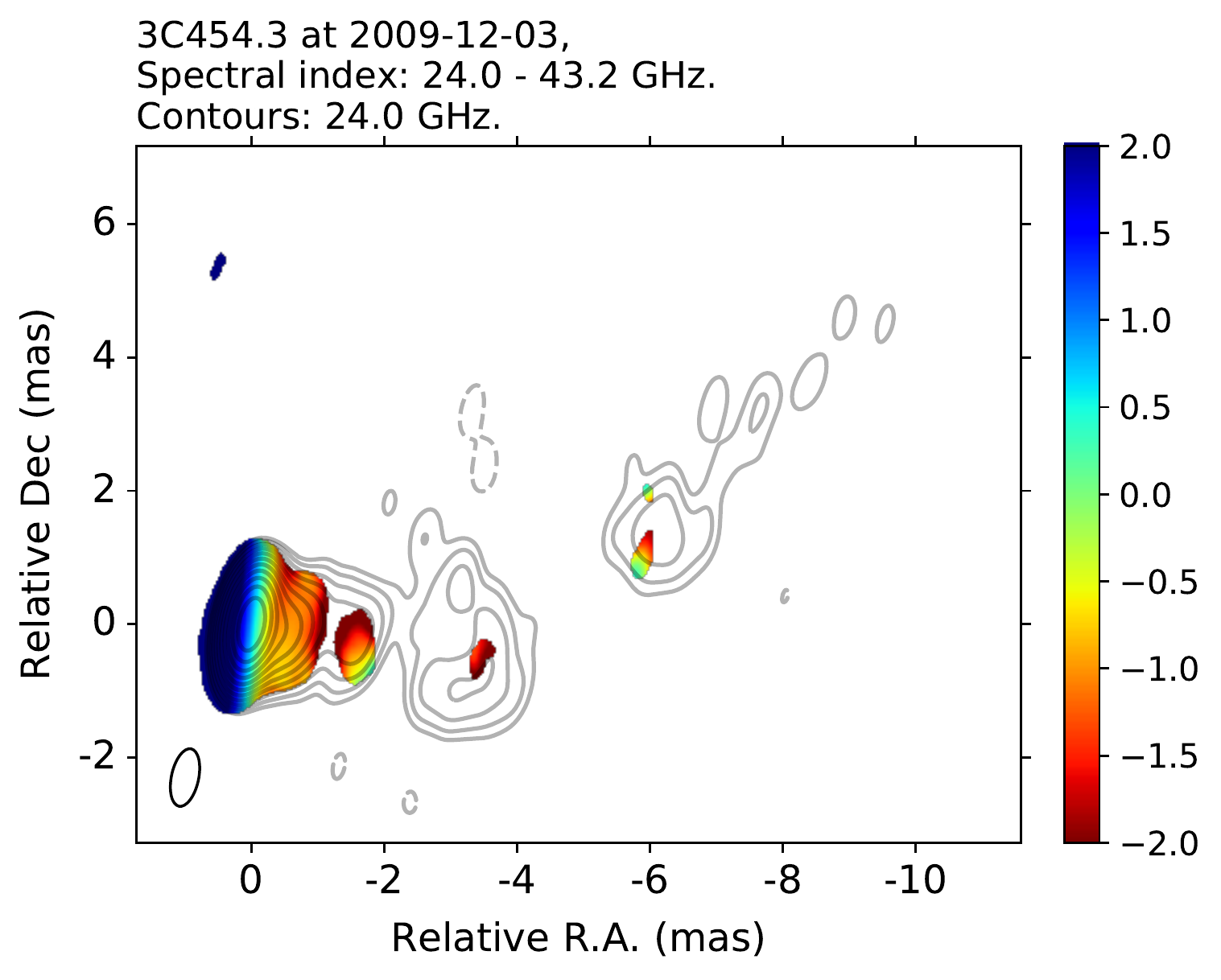}
    }
    \subfigure[]
    {
         \includegraphics[width=0.3\textwidth]{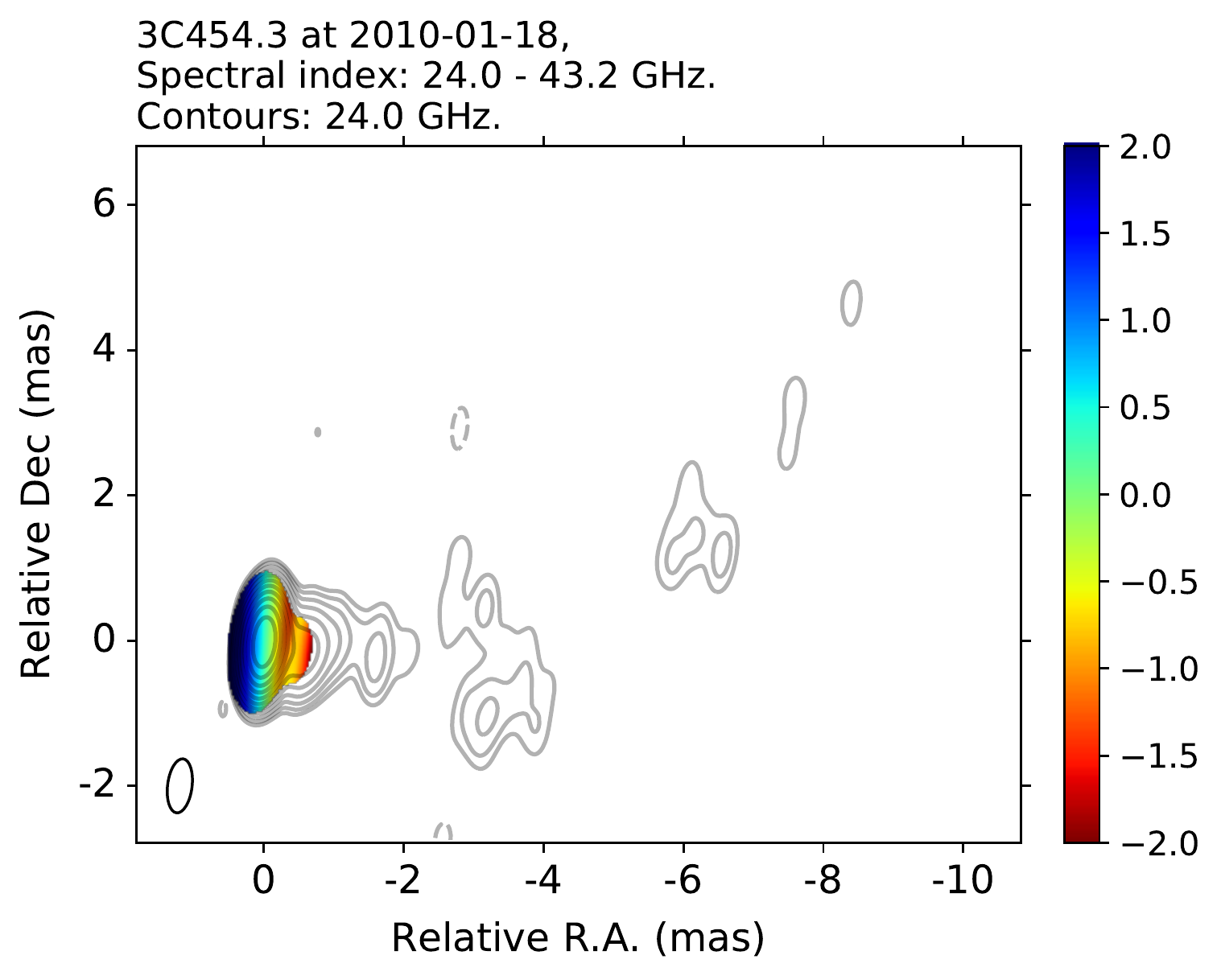}
    }   
    \subfigure[]
    {
         \includegraphics[width=0.3\textwidth]{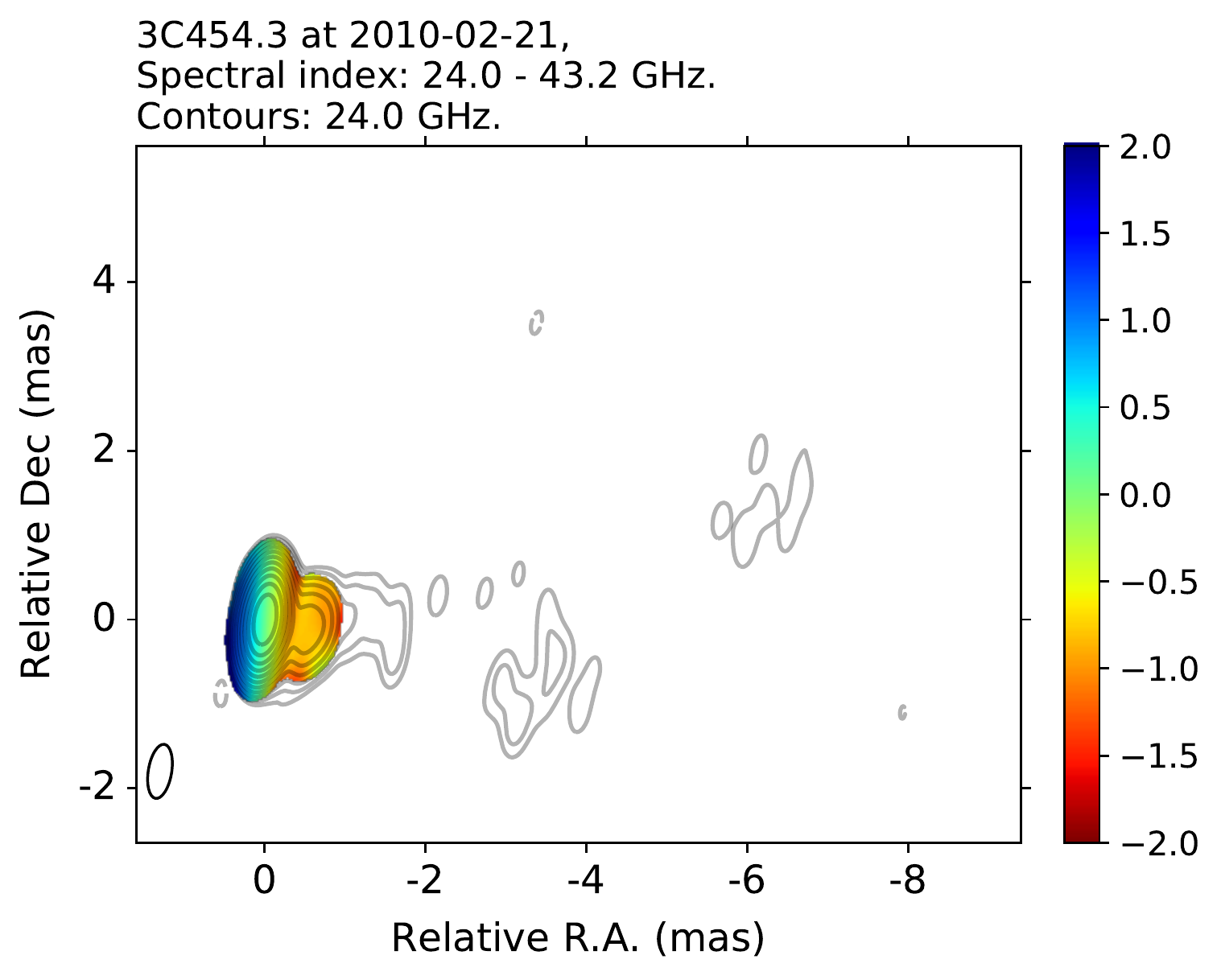}
    }    
        \caption{Continuation of Figure~F.7.}
    \label{siKQp2}
\end{figure*}

\end{appendix}

\end{document}